\documentclass[twocolumn,aps,prd,amsmath,amssymb,longbibliography]{revtex4-1}

\makeatletter                           
\@ifclasswith{revtex4-1}{draft}{        
    \usepackage[notref,notcite]{showkeys}
    
}{}
\makeatother

\usepackage[utf8]{inputenc}
\usepackage{amsmath} 
\usepackage{amsfonts} 
\usepackage{amssymb}
\usepackage{bbm}
\usepackage{epsfig}
\usepackage{graphics}
\usepackage{graphicx}\graphicspath{{figs/}}
\usepackage{titlesec}
\usepackage{mathtools}
\usepackage{environ}
\usepackage{dsfont}
\usepackage[english]{babel}
\usepackage{enumitem}
\newlist{inlinelist}{enumerate*}{1}
\setlist*[inlinelist,1]{ label=(\arabic*)} 
\usepackage{tikz}
\usetikzlibrary{positioning}
\usetikzlibrary{arrows}

\usepackage[colorlinks=true]{hyperref}
\hypersetup{urlcolor=black,linkcolor=black,citecolor=black}
\usepackage[toc,page]{appendix}

\makeatletter
\renewcommand*\env@matrix[1][c]{\hskip -\arraycolsep
  \let\@ifnextchar\new@ifnextchar
  \array{*\c@MaxMatrixCols #1}}
\makeatother

\titleformat{\section}[hang]{\huge\bfseries}{\thesection}{2ex}{}
\titleformat{\subsection}[hang]{\Large\bfseries}{\thesubsection}{2ex}{}
\titlespacing{\subsection}{0pt}{20pt}{10pt}[0pt]
\titleformat{\subsubsection}[hang]{\large\bfseries}{\thesubsubsection}{1ex}{}
\titlespacing{\subsubsection}{0pt}{20pt}{10pt}[0pt]
\titlespacing{\paragraph}{0pt}{15pt}{5pt}[0pt]
\titleformat{\paragraph}[block]{\normalsize\bfseries}{\theparagraph}{1ex}{}

\renewcommand{\thesection}{\arabic{section}}
\renewcommand{\thesubsection}{\thesection.\arabic{subsection}}
\renewcommand{\thesubsubsection}{\thesubsection.\arabic{subsubsection}}
\renewcommand{\theparagraph}{\thesubsubsection.\arabic{paragraph}}

\makeatletter
\renewcommand{\p@subsection}{}
\renewcommand{\p@subsubsection}{}
\makeatother

\usepackage{natbib}

\usepackage[mathscr]{euscript}
\usepackage{braket}
\usepackage{bm}

\usepackage{makecell,tabularx}
\setcellgapes{3pt}

\usepackage{todonotes} 

\newcommand{\gl}{\big{(}}
\newcommand{\gr}{\big{)}}
\newcommand{\im}{{i\mkern1mu}}
\newcommand{\Id}{\bm{1}}
\newcommand{\tr}{\mathrm{tr}}
\newcommand{\ua}{\uparrow}
\newcommand{\da}{\downarrow}
\newcommand{\cD}{\mathcal{D}}
\newcommand{\cL}{\mathcal{L}}
\newcommand{\cB}{\mathscr{B}}
\newcommand{\cA}{\mathcal{A}}
\newcommand{\cK}{\mathscr{K}}
\newcommand{\cM}{\mathcal{M}}
\newcommand{\cW}{\mathcal{W}}
\newcommand{\cS}{\mathcal{S}}
\newcommand{\cN}{\mathcal{N}}
\newcommand{\p}{\partial}
\newcommand{\dif}{\,\mathrm{d}}

\newcommand{\tp}{\text{T}}
\newcommand{\del}{\partial}

\newcommand{\WI}{W_\mathrm{I}}
\newcommand{\WR}{W_\mathrm{R}}
\newcommand{\HI}{H_\mathrm{I}}
\newcommand{\HR}{H_\mathrm{R}}
\newcommand{\JI}{J_\mathrm{I}}
\newcommand{\JR}{J_\mathrm{R}}
\renewcommand{\epsilon}{\varepsilon}
\newcommand{\eps}{\epsilon} 
\newcommand{\Np}{\bar{N}} 
\newcommand{\vp}{\varphi}

\DeclareMathOperator{\spec}{spec}
\DeclareMathOperator{\diag}{diag}

\let\Re\relax
\DeclareMathOperator{\Re}{Re}

\DeclareMathOperator{\TO}{T}

\renewcommand{\d}{\mathop{}\!\mathrm{d}}

\newcommand{\bel}[1]{\begin{equation}\label{#1}}
\newcommand{\ee}{\end{equation}}
\newcommand{\nn}{\nonumber}

\definecolor{refkey}{rgb}{0,0,1}
\definecolor{labelkey}{rgb}{0,1,0}

\numberwithin{equation}{subsection}

\usepackage{mathtools}

\begin{document}
\title[ ]{\Huge The probabilistic world}
\author{C. Wetterich}
\affiliation{Institut  f\"ur Theoretische Physik\\
Universit\"at Heidelberg\\
Philosophenweg 16, D-69120 Heidelberg}

\begin{abstract}
We propose a formulation of physics based on probabilities as
fundamental entities of a mathematical description. Expectation values of
observables are computed according to the classical statistical rule. The
overall probability distribution for one world covers all times from the past to
the future, similar to the functional integral in quantum field theory.

In a general discrete setting time emerges as an ordering structure among
observables which permits to formulate the overall probability distribution as a
product of local factors which only connect variables at neighboring
``time-hypersurfaces''. The time structure induces naturally the quantum
formalism once one focuses on the transport of the time-local probabilistic
information from one hypersurface to a neighboring one. Wave functions or the
density matrix allow the formulation of a general linear evolution law for
classical statistics, which amounts to a generalized Schrödinger- or von-Neumann
equation. In general, no similar law exists for the time-local probability
distribution. The density matrix formulation for time-local subsystems in
classical statistics is a powerful tool which allows us to implement for
generalized Ising models concepts as basis transformations, the momentum
observable and the associated Fourier representation, or the definition of
subsystems by subtraces of the density matrix. The association of operators to
observables permits the computation of expectation values in terms of the
density matrix by the usual quantum rule.

We show that probabilistic cellular automata are quantum systems in a
formulation with discrete time steps and real wave functions. With a complex
structure related to the particle-hole transformation one obtains a Hamiltonian
formulation for the continuous time evolution of a complex wave function.
The Hamilton operator can be expressed in terms of fermionic creation and
annihilation operators. We construct simple automata which are equivalent to
two-dimensional interacting fermionic quantum field theories in a specific
discrete lattice regularization. This demonstrates that at least certain
quantum systems can be described by classical statistics. For these automata it
remains to be shown that the Lorentz symmetry of the naive continuum limit is
indeed realized in the true continuum limit.

The time-local probabilistic information amounts to a subsystem of the overall
probabilistic system which is correlated with its environment consisting of the
past and future. Within classical statistics the correlation with the
environment induces new properties for generic subsystems. Subsystems typically
involve probabilistic observables for which only a probability distribution for
their possible measurement values is available. A characteristic feature is
incomplete statistics which does not permit to compute classical correlation
functions for arbitrary subsystem-observables. This is a consequence of the fact
that large equivalence classes of observables of the overall probabilistic
system are mapped to the same observable of the subsystem. For the time-local
subsystem many overall observables are mapped to the same operator. Incomplete
statistics is the reason why Bell's inequalities are not
generally applicable for measurement correlations in quantum systems.
Furthermore, new statistical observables measure properties
of the probabilistic information, somewhat analogous to temperature. Statistical
observables have no definite values in the states of the overall probabilistic
system and classical correlation functions are not defined. While this work
remains in the context of theoretical physics, the
concepts developed here apply to a wide area of science. 
\end{abstract}

\maketitle
\vspace*{16mm}

\newpage
\clearpage

\tableofcontents
\clearpage
\addtocontents{toc}{\protect\null\vspace*{2mm}}
\section[Fundamental probabilities]{Fundamental\\probabilities}\label{sec:introduction}

The advent of quantum mechanics has opened a probabilistic view on fundamental
physics. It has come, however, with new concepts and rules as wave functions,
non-commuting operators and the rules to associate these quantities to
observations. The unusual probabilistic features have opened many debates on
their interpretation, as well as on suggestions for extensions of quantum
mechanics. Examples are a postulated basic role of observers and measurements
\cite{HEI}, hidden variables to cure an alleged incompleteness \cite{EPR},
attempts to give an observable meaning to the wave function \cite{BOHM}, the
many world hypothesis \cite{EVE}, non-linear quantum mechanics~\cite{TBLSMP,
WEINBERG1989336, GISIN19901, PhysRevLett.66.397} or the incoherent histories
approach \cite{Griffiths:1984rx,omnes1987interpretation,Gell-Mann:1991zrl}. For
many researchers quantum mechanics continues to have mysterious properties.

In the present work we propose that the fundamental physics description of our
world is based entirely on the classical probabilistic concepts of probability
distributions, observables and their expectation values.
Probabilities are considered as the fundamental mathematical concept, not merely
as a measure of lack of knowledge.

In every day life probabilities are often associated to a lack of knowledge. A
girl may be waiting for her boyfriend. When the bell rings she may think: ``it's
most likely, say 90\%, that's him''. This probability concerns her knowledge
rather than being a property of the boyfriend or the person standing on the
other side of the door. Either her boyfriend is already standing there or not.
Once she opens the door she sees the person and the ``observer probability''
that it's the boyfriend changes to either one or zero (or at least very close to
this).

The situation is different if before throwing a dice somebody states that the
probability that the dice will show the face with the number two is one sixth.
She will say this because for a well manufactured dice the probabilities for
showing a given face should all be equal. In principle, she may verify this by
throwing the dice a very large number of times and recording if in a sixth of
the cases it shows the two. The six probabilities for showing the six faces of
the dice may be associated with a property of the dice. Deviations from 1/6
characterize possible small asymmetries. It is very simple to characterize such
possible asymmetries by the probability distribution for the different faces,
much simpler than giving details of geometry, material distribution and so on.

These more intrinsic probabilities which characterize the object should be
distinguished from the observer probability. An observer's knowledge about the
outcome changes from before to after throwing the dice. Once the dice has
stopped the observer knows which face is up - for her the probability to find
the face two jumps from about 1/6 to one or zero. The intrinsic probabilities
for the dice have not changed, however. They do not depend on the observer,
since another observer may throw the same dice, or even the dice has never been
thrown after being manufactured. This simple example demonstrates that
probabilities need not be associated with observers and lack of knowledge. They
can be useful mathematical concepts for describing the state of the dice - and
more generally the state of the world. 

In this work we propose a fundamental description of the world in terms of
``classical'' probabilities. Our starting point will be the simple ``classical''
laws for probability distributions and expectation values of observables. We
will find that rich structures develop from the simple axioms for classical
probabilities. Our basic setting is rather different from classical Newtonian
physics based on differential evolution equations, and corresponding
generalizations to classical field theories. It also differs from the conceptual
setting of quantum mechanics where new mathematical objects and axioms are
introduced beyond classical probabilities. Both classical field theory and
quantum mechanics arise in our approach as particular special cases.

For a formulation of the laws of nature in terms of fundamental probabilities
the basic ingredients are a set of ``classical states'' or ``variables'' $\tau$,
a probability distribution that associates to each $\tau$ a probability $p(\tau)
\geq 0$, $\sum_\tau p(\tau) = 1$, and observables that take real values
$A(\tau)$ for a given variable $\tau$. Expectation values of observables are
computed according to the basic rule of classical statistics
\begin{equation}\label{I1}
\braket{A} = \sum_\tau p(\tau) A(\tau).
\end{equation}
Here $\tau$ may be discrete or continuous variables, with an appropriate
interpretation of the sum as integrals. Examples for the variables~ $\tau$ are
configurations of Ising spins, or scalar fields $\varphi(x)$ for an euclidean
quantum field theory. The description of the world typically uses infinitely
many continuous variables. A scalar field $\varphi(x)$ is of this type since it
amounts to a continuous variable for every point $x$ in space or spacetime.
Other names for the variables $\tau$ are ``basis events", or ``classical
states".

The basic setting of a fundamental probabilistic description of the world is not
based on some evolution equation in time as for classical mechanics, classical
field theory or quantum mechanics. The ``overall probability distribution''
$\lbrace p( \tau)\rbrace$ covers all times - past, present and future - and all
space. Time and space should be considered as derived quantities. Time emerges
as a linear ordering structure among classes of observables. This ordering
property is still quite general, applying similarly to ordering in space or some
other property. Any such ordering can be associated with the concept of
evolution. Physical time is characterized by periodic evolution. This defines
clocks and systems of clocks in a natural way. A continuum limit allows the
transition from the discrete ticking of clocks to continuous time. Finally,
there is an "arrow of time" from the past to the future. Probabilistic physical
time has to realize these three basic properties - ordering, clocks and
periodicity, and arrow of time. 

Out of these three conditions we discuss in this part of the work the first two.
The arrow of time is related to focusing properties of solutions of
time-evolution equations~\cite{VARG} and will be discussed in a later part. We
first establish how in our classical probabilistic formulation time emerges as
an ordering structure for observables \cite{CWPT}. 
A simple setting associates a number of Ising spins $s_\gamma(t)$,
$s_\gamma^2(t) =1$, $\gamma = 1,...,M$, to each discrete time point $t$. The
ordering of the Ising spins according to the label $t$ induces an ordering for
local observables $A(t)$ which can be constructed from $s_\gamma(t)$. For a
simple formulation the probability distribution $p(\tau)$
features some type of locality in time. Examples are local chains for which
$p(\tau)$ is a product of factors $\mathscr{K}(t)$, which each involves only
neighboring variables constructed from $s_\gamma (t)$ and $s_\gamma (t
+\epsilon)$.

Time-local subsystems concentrate on the local probabilistic information at a
given time $t$. The question how the local probabilistic information at a
neighboring time $t + \epsilon$ is related to the one at $t$ reveals the
presence of structures familiar from quantum mechanics. A simple evolution law
for the transport of probabilistic information from $t$ to $t + \epsilon$ needs
local probabilistic information in the form of a classical density matrix. For
suitable (factorizing) boundary conditions one may use a pair of classical wave
functions instead of the classical density matrix. For these ``pure classical
states" the change with $t$ obeys a linear evolution law, described by a
generalized Schrödinger equation, while for the general case one has a
generalized von-Neumann equation. The evolution operator is related to the
transfer matrix. Local observables are associated to operators that typically do
not commute with the evolution operator. This provides for a ``Schrödinger
picture" for the transport of information \cite{CWIT,CWQF}, supplementing the
transfer matrix formalism \cite{TM,MS,FU} which can be seen as a ``Heisenberg
picture". All these structures emerge directly from the minimal setting of
classical
statistics with basic law \eqref{I1}. No additional fundamental concepts need to
be introduced.

Quantum systems are time-local subsystems with the particular property that the
evolution of the classical density matrix with $t$ is unitary, or more generally
orthogonal in a formulation without complex structure. For a unitary evolution
the information is not lost as $t$ increases. One finds that the probabilistic
information available in the subsystem is incomplete. There are local
observables $A(t)$, $B(t)$ for which the expectation values $\braket{A(t)}$ and
$\braket{B(t)}$ can be computed from the probabilistic information of the
time-local subsystem, while the classical correlation function $\braket{A(t)
B(t)}_{cl}$ is either not defined or not accessible with the restricted local
probabilistic information of the subsystem.

In our view the Universe is described by an overall ``classical'' probability
distribution covering the past, present and future. Quantum mechanics arises by
a focus and the time-local subsystem.
Using the embedding of the quantum subsystem in the overall probability
distribution all the quantum rules can be derived from the basic rule \eqref{I1}
of classical statistics. This concerns both the formula for the computation of
expectation values and the association of possible measurement values with the
spectrum of eigenvalues of the quantum operator. We hope that this simple
finding can contribute to a demystification of quantum mechanics. The absence of
conflict with Bell's inequalities \cite{BELL2} for classical correlation
functions finds a simple explanation in the incompleteness of the quantum
subsystem. 
The classical correlations are not relevant for observations or ideal
measurements in quantum subsystems.

The simplest classical probabilistic systems that admit a unitary time evolution
are probabilistic cellular automata. While the sequential updating rule for bit
configurations is deterministic, the crucial probabilistic aspect enters by a
probability distribution over initial bit configurations. If the cells of the
automaton can be labeled by positions in some D-dimensional space, the
probabilistic automata are equivalent to quantum field theories for fermions in
D space and one time-dimension. The causal structure of quantum field theories
with light-cones, the concept of particles whose propagation depends on the
properties of the vacuum, the complex structure and the presence of
antiparticles, or the duality between position and momentum space emerge all in
a very simple, explicit and straightforward way. It is striking that in our
setting the most basic and simplest structures are quantum field theories, while
single-particle quantum mechanics or the quantum mechanics for a few qubits
arises only on the level of appropriate subsystems.

For quantum field theories a close relation to classical statistics has been
exploited for a long time, for example in lattice gauge theories \cite{WIL,
Creutz:1983njd, MLCS, HRLGT, GATLA}. The functional integral for the thermal
equilibrium state at temperature $T$ is directly associated to a probability
distribution $p(\tau)$,
\begin{align}\label{I2}
p(\tau) = Z^{-1} \exp(-S(\tau)), && Z = \sum_\tau \exp(-S(\tau)),
\end{align} 
with $S(\tau)$ the ``classical action", as given by the energy of a state or
field configuration $\tau$ divided by the temperature. This extends to the
vacuum for $T \to 0$. In contrast, the dynamics of a quantum field theory with a
non-trivial evolution in time, in particular for processes of scattering and
decay of particles, employs a complex functional integral where $\exp (-S)$ is
replaced by $\exp ( i\bar{S}_M$). This functional integral with ``Minkowski
signature " defines no longer a probability distribution. One can employ
analytic continuation from the euclidean functional integral \eqref{I2} to
Minkowski signature. In the process of analytic continuation one looses,
however, the property of $p(\tau)$ as a probability distribution, as $\exp
(-S(\tau))$ is replaced by $\exp(i\bar{S}_M(\tau))$. 
The analytic continuation of $S$, namely $-i\bar{S}_M$, is typically complex,
and often purely imaginary.
Probability distributions in euclidean quantum field theories allow for powerful
numerical methods, as Monte-Carlo simulations. These methods do not apply, at
least not in a direct way, to the phase factor $\exp (i\bar{S}_M(\tau))$ for
Minkowski signature. On the other hand, the Minkowski signature is directly
related to the unitary time evolution in quantum mechanics. It can lead to
oscillating behavior of correlation functions, while for euclidean functional
integrals the correlation functions often decay for large distances.

The present work indicates a reconciliation of these seemingly contradictory
aspects. We formulate an underlying functional integral that constitutes a
probability distribution even for the full dynamics of the quantum field theory.
The standard functional integral with Minkowski signature could either be
equivalent to this underlying probability distribution, or be a representation
of the partial information contained in a subsystem. While we have not yet
achieved all steps of this program for realistic quantum field theories, we
already provide examples for simple cases as interacting fermions in two
dimensions.
For suitable probabilistic systems we derive the notion of particles and their
interaction. We explain the origin of the particle-wave duality by discrete
possible measurement values of observables and the continuity of the
probabilistic information contained in the wave function, density matrix or
probability distribution.

Even though the analogy to quantum field theory may suggest a ``preexisting
spacetime'' we introduce neither time nor space as an ``a priori" concept.
Space, spacetime and geometry emerge as structures among observables in our
classical probabilistic setting.
Spacetime and geometry express relations between observables. There is no
``spacetime without matter", where ``matter" includes photons or the
gravitational field.
In particular, a metric can be related to the behavior of the connected
correlation function for suitable observables \cite{CWGEO}.

Starting from a quantum field theory the way to quantum mechanics is, in
principle, straightforward. For a given vacuum one can define single-particle
excitations which obey a one-particle Schrödinger or von Neumann equations, with
generalization to systems of a few particles. If only certain discrete
properties of the particles play a role and position or momentum can be
neglected, one arises at systems for a few qubits. Examples are the spin or
energy eigenstates of atoms. These qubits can, however, not fully be described
by classical probability distributions for only a few variables. They typically
envolve infinitely many classical degrees of freedom, as inherited from the
infinitely many degrees of freedom of the underlying quantum field theory.
Already a single quantum spin needs infinitely many classical bits for a
classical probabilistic description. There are an infinite number of yes/no
decisions for the discrete values of the spin observables in arbitrary
directions.

While conceptually an infinite number of classical bits is required for the
description of a finite number of qubits, it is often sufficient in practice to
employ only a finite number of classical bits. This is analogous to the
representation of real numbers by a finite number of bits in a computer. If one
does not insist on infinite precision for all observables, one can find
interesting systems of a reasonably small number of classical bits whose
classical probability distribution accounts for quantum systems of a certain
number of qubits.


The embedding of quantum mechanics in classical statistics opens new
perspectives for quantum computing \cite{CWQCCB,PEME,PW}, see also refs
\cite{AOR,SFN,AAHE,SLL,MNI,BH} for some related ideas. The heart of quantum
computing \cite{BEN,MAN,FEY,DEU} are quantum correlations relating many parts
(qubits) of a system, such that a change in one qubit affects many others. Such
quantum correlations and related quantum operations can be realized in
``classical systems" as artificial neural networks
\cite{LPLS,SNSSW,CARLTRO,CATRO,KIQB,JBC}, neuromorphic computing
\cite{PBBSM,PJTM,JPBSM,BBNM,ASM,FSGH,DBKB} or even for neurons in the brain,
without the need of low temperatures and very well isolated systems of qubits.
While it is not clear if exact quantum operations can be scaled to a large
number of qubits, which would be required for efficient quantum algorithms, the
conceptual implications of ``classical'' realizations of entangled systems of a
few qubits may open the door to investigations of new forms of ``correlated
computing".

The present paper constitutes a first part of this work and is devoted to the
basic probabilistic concepts and the emergence of quantum field theory. A second
part concentrates on quantum mechanics, with focus on a small number of qubits.

Subsystems of the overall probabilistic system play an important role in this
paper as well as for quantum mechanics. These subsystems are typically
correlated with their environment, and characterized by incomplete statistics.
Together with a focus on conditional probabilities this explains many
``mysteries'' and ``paradoxes'' of quantum mechanics, as the reduction of the
wave function, the violation of Bell's inequalities, entanglement, or the
Einstein-Podolski-Rosen paradox.

In chapter~\ref{sec:Fundamental_probabilism} we start by discussing conceptional
issues for a probabilistic formulation of fundamental physics in terms of the
classical statistical concepts based on a probability distribution. This sets
the stage for the following discussion and gives a first overview of basic ideas
underlying this work. In chapter~\ref{sec:probabilistic_time} we turn to the
concept of time emerging from the overall probabilistic system as an ordering
structure among observables. We formulate basic properties as evolution and
predictivity, using simple examples describing clocks or free particles. This
section also shows how concepts familiar from quantum mechanics, as wave
functions and operators, appear in a natural way in the formalism for evolution
in classical statistical systems. We discuss the continuum limit for time and
clock systems as a basis for physical time. 

In chapter~\ref{sec:free_fermions_in_two_dimensions} we demonstrate the
emergence of quantum field theory from very simple ``classical'' probabilistic
systems.  A particular two-dimensional general Ising model realizes a quantum
field theory for free fermions. It shows the emergence of Lorentz symmetry. It
serves as an example how special relativity and more generally, the concept of
reference frames is a natural consequence of our setting of probabilistic time.
We discuss the emergence of a complex structure which is characteristic for the
``phases" in quantum mechanics. We investigate the concept of different vacua
and the dependence of particle properties on the vacuum properties in the most
simple setting. Already for the very simple generalized Ising model the quantum
concept of momentum-position duality, conserved quantities and symmetries or
particles and antiparticles are shown to be very useful for the understanding of
the dynamics.

 In chapter~\ref{sec:probabilistic_and_deterministic_evolution} we compare
classical probabilistic systems with or without a unitary evolution. Several
simple examples of quantum systems that are realized as subsystems of
``classical'' probabilistic systems illustrate that there is no conceptual
boundary between classical statistics and quantum statistics.  The quantum
formalism with wave functions and operators for observables applies to arbitrary
``classical" probabilistic systems. The comparison with classical systems that
do not follow a unitary evolution, as Markov chains or static memory materials,
reveals the important particular features associated to the unitary evolution of
quantum systems.

Chapter~\ref{sec:quantum_field_theory} deepens the understanding how quantum
field theory emerges naturally from classical probabilistic systems with a
unitary evolution. We remain in the simple setting of probabilistic cellular
automata describing fermionic quantum field theories in two dimensions. The
discussion of chapter~\ref{sec:free_fermions_in_two_dimensions} is extended by
investigating systems with interactions. Exploiting the possibility to perform
Fourier transformations to momentum space we discuss the ``particle physics
vacuum'' for which both particle and antiparticle excitations have positive
energy. We investigate spontaneous symmetry breaking resulting in non-zero
particle masses. We emphasize the importance to perform a true continuum limit
which involves the renormalization flow of couplings characteristic for quantum
field theories.

Chapter~\ref{sec:subsystems} is devoted to a general discussion of subsystems.
Subsystems that are correlated with their environment show already many
conceptual features familiar from quantum mechanics. In particular, we discuss
the relation between observables of the subsystem and the operators in the
associated quantum formalism. This sheds light on the general non-commuting
operator structures. The non-commutativity characteristic for quantum mechanics
extends to many other types of subsystems of ``classical'' probabilistic
systems. We specify the conditions under which such subsystems obey all the
rules of quantum mechanics. Observables of quantum subsystems are typically
probabilistic observables for which only a probability distribution for their
possible measurement values is available for a given state of the subsystem.
This holds even if this observable has fixed values in every state of the
overall probabilistic system. We discuss rather general families of correlated
subsystems for which the overall probability distribution does not factorize
into a part for the subsystem and another part for its environment.  We draw
conclusions for this part in chapter~\ref{sec:Discussion}.

\section[Fundamental probabilism]{Fundamental\\probabilism}
\label{sec:Fundamental_probabilism}

The starting point of the present work assumes that the fundamental description
of our world is probabilistic \cite{CWGENS,CWGEO,CWICS,CWIS}. The basic objects
for this description are probability distributions and observables.
Deterministic physics arises as an approximation for particular cases. Our
description of probabilities remains within the standard setting of classical
statistics. No separate laws for quantum mechanics will be introduced. They
follow from the classical statistical setting for particular classes of
subsystems.

The present section presents basic considerations for a probabilistic
description of Nature. A more systematic mathematical treatment will start in
the next section. We employ units where $\hbar=1$ and the summation convention
where a sum over double indices is implied if not stated otherwise.

\subsection[Probabilistic description of Nature]{Probabilistic description of \\
Nature}
\label{sec:probabilistic_description_of_nature}

Let's look out in the rain. How would a physicist describe raindrops falling
through the atmosphere? She may state that each drop is composed of a very large
number of water molecules. Next she would like to specify how likely it is to
find a number of molecules sufficiently far above the average at a given time
$t$ and given position $\vec{x}$. If the likelihood at $x = (t,\vec{x})$ is high
enough, she would say that it is likely to find a raindrop at time $t$ and
position $\vec{x}$. If it is low, it is unlikely that a drop is at $\vec{x}$. If
for $t_2$ near $t_1$ the high concentration of molecules moves from the position
$\vec{x}_1$ to the position $\vec{x}_2$, and so on, she could construct a
trajectory $\vec{x}(t)$ for a given droplet.
This probabilistic description of the rain already involves many key elements of
our basic approach that we highlight in the following.

\paragraph*{Probability distribution}

The key concept for a probabilistic description of raindrops is the probability
$p[N(x)]$ for finding $N$ water molecules in a volume element around $\vec{x}$
and a time interval around $t$.
%
%
The variables or basis events $\tau = N[x]$ are the molecule distributions over
space and time. Two different molecule distributions in space and time
correspond to two different basis events. To each basis event $\tau = N[x]$ one
associates a probability $p_\tau = p[N(x)]$.
 
Probabilities are real numbers between zero and one, $0 \leq p[N(x)] \leq 1$. If
the probability equals one, an ``event" is certain, while for probability zero
one is certain that an event does not occur. Probabilities are normalized such
that the sum over the probabilities for all possible basis events equals one.

To be more specific, we may divide time into intervals with size $\epsilon$, and
space into cubes with volume $\epsilon^3$. The variables $t$, $x_1$, $x_2$,
$x_3$ are then discrete points on a four-dimensional hypercubic lattice with
lattice distance $\epsilon$. For example, $t$ may take the discrete values $t =
m\epsilon$ with $m$ being an integer. Similarly, the cartesian space coordinates
$x_k$ are given by discrete points $x_k = n_k \epsilon$, with integer $n_k$. A
given distribution of water molecules $N(x)$ specifies how many molecules $N$
are inside the time interval between $t-\epsilon/2$ and $t +\epsilon/2$ for each
point on the lattice $x = (t,x_1,x_2,x_3)$, and similarly in the volume element
given by positions within the intervals $x_k -\epsilon/2$ and $x_k +
\epsilon/2$. These intervals can be visualized as little four-dimensional cubes
around each lattice point. The ensemble of the probabilities for all events
$N(x)$ is called the probability distribution.

A given event or molecule distribution is a (discrete) function $N(x)$. It
specifies the precise number of molecules for each point $x$ by associating to
each lattice point $x$ a positive integer $N(x) \geq 0$. Probability one for a
given function $N(x)$ means that one is certain to find precisely the number of
molecules given by one particular distribution $N(x)$ at each given time $t$ and
each given position $\vec{x}$. For the rain, this situation is not given. Let us
choose $\epsilon$ much smaller than the typical size of a drop such that we can
resolve it, but large enough such that inside a drop we still have large numbers
of molecules in the cubes of size $\epsilon^4$. For a given $(t,\vec{x})$ with
$\vec{x}$ inside a drop at a given time $t$ we may consider the probability to
find $N_1$ molecules at $(t,\vec{x}_1)$ and $N_2$ molecules at an neighboring
position $(t,\vec{x}_2) = (t,\vec{x}_1 + \vec{\delta}_x)$, say $\vec{\delta}_x =
(\epsilon,0,0).$
This probability is expected to be almost equal to the probability to find $N_1
+ 1$ molecules at $(t,\vec{x}_1)$ and $N_2 -1$ molecules at $(t, \vec{x}_2)$. No
experiment or observation, and no dynamical evolution can differentiate between
two situations where (at least) one molecule is rather in one or the other of
the two neighboring volume elements. Given the normalization of the
probabilities we conclude $p[N(x)] < 1$ for all distributions $N(x)$. There is
no way to know certainly that precisely one particular molecule distribution
$N(x)$ is realized. The description of the raindrops is genuinely probabilistic.
No event or molecule distribution in space and time occurs with certainty. Our
physicist decides that she better uses a probabilistic description of raindrops.

\paragraph*{Subsystems}

As a second crucial point she observes that for the understanding of a single
raindrop she has to view it as a subsystem of a larger system that comprises at
least the drop and the atmosphere surrounding it, perhaps the whole region and
duration of the rain, or even further. Some water molecules may move out of the
drop into the atmosphere, some others may move in. Furthermore, the water
molecules inside the drop interact with the ones outside. While a given droplet
appears as a rather well localized separate entity, its properties cannot be
understood without the surrounding ``environment". The environment determines
the pressure and the temperature that are crucial for the behavior of raindrops.
A physicist realizes that a system of water molecules exhibits a first order
phase transition between water and vapor in a certain range of temperature and
pressure. It is this phase transition that is responsible for the presence of
well separated droplets with a surface tension ``holding them together". It is
the underlying reason why molecule distributions with strong local enhancements
of $N(x)$ in certain regions of space and time intervals have comparatively
large probabilities. These concentrations of molecules are the falling droplets.

\paragraph*{Time evolution}

Finally, our physicist may try to understand the system of raindrops by
formulating some type of evolution law. This goes beyond a pure description for
all times and positions by the ``all time probability distribution" or ``overall
probability distribution" $p[N(x)]$. An evolution law could concentrate on
``time-local probabilities" $p[t;N(\vec{x})]$. At every time $t$ one
investigates the probabilities $p[N(\vec{x})]$ to find a distribution of
molecules $N(\vec{x})$ in space. Since one looks at a fixed time, the ``events"
of the local probability distribution are now molecule distributions in space
$N(\vec{x})$ and the $t$-label for the molecule distributions is no longer
needed. In a certain sense the local probability distributions $p[t;N(\vec{x})]$
are snapshots of the rain at given times $t$. Since the raindrops are falling,
the local probability distribution will depend on the time $t$. A droplet
concentrated around $\vec{x}_1$ at $t_1$ is typically concentrated around
another position $\vec{x}_2$ at a subsequent time $t_2$. An evolution law
typically relates the local probabilities at time $t + \epsilon$ to the ones at
time $t$. More formally, it is a relation between $p[t+\epsilon ; N(\vec{x})]$
and $p[t;N(\vec{x})]$. If such a relation exists, a physicist can predict
properties at time $t+\epsilon$, knowing the properties at time $t$. These
properties are probabilistic both at time $t$ and $t + \epsilon$.

We will discuss the precise relation between the overall probability
distribution $p[N(x)]$ and the local probability distribution $p[t;N(\vec{x})]$,
and derive the emergence of evolution laws later in this work. At the present
stage we only observe that the evolution laws typically require additional local
probabilistic information at a time $t$. For the example of droplets one
typically needs the average velocities $\vec{v}(\vec{x})$ of the molecules in
the volume element around $\vec{x}$ at the given time $t$. The local probability
distribution is then given by $p[t,N(\vec{x}),\vec{v}(\vec{x})]$, with events
specified by $N(\vec{x})$ and $\vec{v}(\vec{x})$ simultaneously.
With $N(\vec{x})$ proportional to the particle density this is a type of
probabilistic hydrodynamic description.

\paragraph*{Probabilistic fields}

The quantities $N(\vec{x})$ and $\vec{v}(\vec{x})$ are fields. To every point
$\vec{x}$ one associates a scalar quantity $N(\vec{x})$ and a vector quantity
$\vec{v}(\vec{x})$. The local probability distribution
$p[t,N(\vec{x}),\vec{v}(\vec{x})]$ specifies probabilities for field
configurations. We are dealing with a probabilistic field theory. The overall
probability distribution $p[N(x)]$ specifies probabilities for four-dimensional
fields $N(x)$. This already shows many analogies to quantum field theories. We
will also find that the time-local probability distribution is typically not
sufficient for the formulation of an evolution law for classical probabilistic
systems. The local probabilistic information necessary for the formulation of an
evolution law often involves probability amplitudes or wave functions
$q[t;N(\vec{x}),\vec{v}(\vec{x})]$ or, more generally, a density matrix. These
objects contain local probabilistic information beyond the local probabilities
$p[t;N(\vec{x}),\vec{v}(\vec{x})]$. They point to strong similarities with
quantum mechanics.

For a simple raindrop the physicists description already turns out to be rather
complex. Shouldn't one rather start with pointlike particles, as Newton did for
approximations of planets in the solar system? The reason why we are starting
with the raindrop is that it shares many features of generic physical systems.
First of all, it is a probabilistic system. Second, it is a subsystem of a
larger system. Third, the time evolution concerns the time evolution of a
probability distribution. What is simple depends on the point of view and the
basic physical setting. Basic ingredients in our world are atoms.
Their description and time evolution are probabilistic. Modern physics is
typically described by a quantum field theory, for which atoms are well isolated
subsystems surrounded by a complicated vacuum. Atoms are composed from
elementary particles that are themselves considered probabilistic ``excitations"
of the vacuum.
The raindrops are much closer to fundamental physics than are pointlike planets.
Our short introductory discussion of the probabilistic description of raindrops
shows many analogies to how the concept of particles should be seen in a
fundamental theory.

\paragraph*{Deterministic and probabilistic description of \\ Nature}

If one would attempt a deterministic description of the raindrop based on
Newton's laws one needs to specify at any given time $t$ for each water molecule
labeled by $i$ the position $\vec{x}_i (t)$ and the velocity $\vec{v}_i(t)$, or
the associated momentum $\vec{p}_i(t)$. (In the non-relativistic limit one has
$\vec{p}_i(t) = m \vec{v}_i(t)$, with $m$ the mass of the molecules.)
With a total number of molecules $N_{tot}$ of the order of Avogadro's number
$N_{av} \approx 6 \cdot 10^{23}$, or even much larger, the size of $6 N_{tot}$
real numbers already exceeds any computers storage you may imagine by far. The
positions and momenta would have to be known with extremely high precision,
since two closely neighboring particle trajectories separate from each other
exponentially as time goes on. Furthermore, one would need to store additional
fields as electric and magnetic fields, again with an extremely high precision.
These fields carry memory of the positions and momenta of molecules in the past,
as well as information about the environment outside the droplet. Since water
molecules have a dipole moment, electromagnetic fields directly influence the
trajectories of the molecules. Already the storage of a snapshot of the
situation at a given time $t$ requires information far beyond the one available
in our whole observable universe if a bit is stored in every volume of size
$l_p^3$, with $l_p = 1.6 \cdot 10^{-35}$ meters the Planck length. It is rather
obvious, that this is an idealization that has little to do with a physicist
describing and understanding the real world.

The probabilistic description of raindrops is much simpler. Even though a
sufficiently accurate storage of the local probability distribution
$p[t;N(\vec{x}), \vec{v}(\vec{x})]$ may be a challenge for practical computing,
it is typically a smooth function of the fields $N(\vec{x})$ and
$\vec{v}(\vec{x})$. Also the relevant fields $N(\vec{x})$, $\vec{v}(\vec{x})$
are typically smooth, even though they show strong variations at the boundaries
of droplets. Present computer power can handle the time evolution of raindrops
in the probabilistic approach sketched here. A formulation in terms of
continuous functions often admits, at least partially, an analytic treatment,
helping the understanding greatly.
We note that a given probability distribution $p[t;N(\vec{x}),\vec{v}(\vec{x})]$
can describe many raindrops at once, including processes where two droplets
merge or a given drop separates into smaller droplets.
Again, this shows analogies with fundamental particle physics where particle
numbers are not conserved and may particles can be described at once.

\paragraph*{Observables}

Another important advantage of the probabilistic approach is the formulation of
simple observables that can both be measured and computed. For example, one may
imagine a detector that measures if at least one droplet is in a given detection
volume or not. The corresponding observable equals one if the total number of
molecules in the detection volume $N_{det}$ exceeds a threshold value $N_{th}$,
and it equals minus one if $N_{det}$ remains below $N_{th}$,
\begin{align}
s &= 1 && &\text{for} && N_{det} &\geq N_{th} \nonumber\\
s &= -1 && &\text{for} && N_{det} &< N_{th}.
\end{align} 
The observable $s$ is a two-level observable or an Ising spin, with possible
measurement values $s = \pm 1$. It is associated to a yes/no-decision e.g. a
number above threshold or not. If $N_{th}$ is chosen to be somewhat below the
typical number of molecules in a droplet, one may say that at least one droplet
is inside the detection volume if $s = 1$, and no droplet is within this volume
if $s=-1$. The measurement of $N_{det}$ and therefore of $s$ could be done with
a system of lasers, using reflections at ensembles of molecules.

At any given time the probability $p_+(t)$ for finding $s=1$ at time $t$ can be
computed from the local probability distribution
$p[t;N(\vec{x}),\vec{v}(\vec{x})]$. We first define $p[t;N(\vec{x})]$ by summing
all probabilities with given $N(\vec{x})$, but arbitrary velocities
$\vec{v}(\vec{x})$. The number of molecules $N_{det}$ in the detection volume is
given by the sum over all $N(\vec{x})$ for points $\vec{x}$ inside the detection
volume. The probability $p_+(t)$ is then obtained by summing the probabilities
$p[t;N(x)]$ for all those molecule distributions for which $N_{det}$ exceeds the
threshold. 
Similarly, the sum over all probabilities for $N_{det} < N_{th}$ yields
$p_-(t)$, the probability to find $s=-1$. Since $p_+(t) + p_-(t)$ is the sum
over all probabilities for arbitrary fields and therefore equals one, there is
actually only one independent probability $p_+(t)$, with $p_-(t) = 1 - p_+(t)$.

For a given evolution law the probability $p_+(t)$ can be computed for some time
$t$, given ``initial conditions" at an initial time $t_{in}$. The probability
$p_+(t)$ is ``predicted" for these initial conditions. Comparing with a
measurement of $s$ at time $t$ one can extract information if the evolution law
is valid or not. The measurement will yield either $s=1$ or $s = -1$. If $p_+$
is close to one, say $p_+ > 0.9999$, and the measurement finds $s=-1$, it seems
unlikely that the evolution law is correct. On the other hand, if $s=1$ is
found, the observation is compatible with the prediction of the evolution law.
If the prediction for $p_+$ is far from one or zero, say $p_+ = 0.6$, it will be
difficult to draw any conclusion based on a single measurement.

\paragraph*{Conditional probabilities}

For tests of an evolution law it is therefore preferable to find observables
whose values can be predicted with almost certainty, e.g. $p_+ > 0.9999$. This
can often be achieved by a combination of observables. Assume that the evolution
law is stating that the raindrops fall with a constant velocity $\vec{v}_0$ in
the $z$-direction or 3-direction, $\vec{v}_0 = (v_1,v_2,v_3) = (0,0,-\bar{v})$.
A droplet with center at $\vec{x}_1 = (x_1,x_2,x_3)$ at $t_1$ will then have its
center at $\vec{x}_2 = \vec{x}_1 + \vec{v}_0 (t_2 - t_1) =
(x_1,x_2,x_3-\bar{v}(t_2-t_1))$ at $t_2$. We can now perform a sequence of two
measurements at $t_1$ and $t_2$. For the second measurement at $t_2$ we displace
the detection volume by a vector $(0,0,-\bar{v}(t_2-t_1))$ as compared to the
detection volume of the first measurement at $t_1$.
Since the detector moves with the same velocity as the droplets, any droplet
found in the detector at $t_1$ should also be found in the detector at $t_2$. If
$s(t_1) = 1$ this simple evolution law predicts a probability $p_+(t_2)$ for
$s(t_2) = 1$ to be very close to one. 

We can formulate this in terms of a ``correlation" $s(t_1)s(t_2)$. This
correlation is again a two-level observable or Ising spin. It takes the value
one if both $s(t_1)$ and $s(t_2)$ have the same sign, and the value minus one if
the signs are opposite. In other words, one has $s(t_1)s(t_2) = 1$ if either
there are droplets in the detector both at $t_1$ and $t_2$, or if there is no
droplet in the detector, neither at $t_1$ nor at $t_2$. For a free homogeneous
fall with constant $\vec{v}_0$ it is very unlikely to have a droplet in the
detector at $t_1$ and no droplet at $t_2$, or to have no droplet at $t_1$ and to
find a droplet at $t_2$.
We conclude that the probability $\bar{p}_-$ to find the value minus one for the
observable $s(t_1)s(t_2)$ must be tiny. It is not expected to be exactly zero,
however, since even for the simple fall with constant velocity on expects some
fluctuations.
In turn, for the correlation observable $s(t_1)s(t_2)$ one predicts a
probability $\bar{p}_+$ very close to one, such that this observable can be used
for a test of the evolution law.

Two important lessons can be drawn from this simple discussion. First, the use
of probabilities for the description of a physical situation does not need a
repetition of identical experiments as often assumed. Measuring one given
rainfall can be enough to draw important conclusions.
It is sufficient to concentrate on observables for which the predicted
probability for a given value is very close to one. If many such observables are
available, rather substantial information can be extracted. A fundamental
probabilistic setting is compatible with predictivity.

For the general case the precise relations between sequences of measurements
involves the notion of ``conditional probabilities".
For most purposes a physicist is not interested in the overall probability for
an event. The focus is on the conditional probability that asks how likely is an
event $A$ after an event $B$ has been measured. One does not want to know how
likely it is in the overall history of the Universe that at a given
$(t_2,\vec{x}_2)$ there is a high concentration of water molecules. The
corresponding probability is tiny, since it requires a planet with water at this
place and so on. The relevant question concerns the conditional probability to
find a high concentration of water molecules at $(t_2,\vec{x}_2)$, given that a
high concentration at $(t_1,\vec{x}_1)$ has been observed. We will discuss later
in more detail the rather complex nature of conditional probabilities.
Also the relation between probabilities and series of identical measurements can
be derived later, but needs not to be postulated a priori.

Second, the observables often have discrete values, while the probability
distribution is continuous, with probabilities assuming arbitrary real values
between zero and one. The combination of discrete possible measurement values
and continuous probabilistic information resembles an important aspect in
quantum mechanics, namely particle-wave duality. The particle aspect corresponds
to the discrete possible values of suitable observables, as the Ising spin
associated to the question if a particle is within a certain volume or not. We
will later introduce an operator associated to this observable. It has
eigenvalues $\pm 1$. The wave aspect concerns the continuous behavior of the
probabilistic information. 
For quantum mechanics it is encoded in a wave function or probability amplitude,
rather than in a probability distribution. We will understand the connection
between these different ways of accounting for the relevant time-local
probabilistic information later. 
Finally, our discussion also highlights the important role of simple two-level
observables or Ising spins. 

\paragraph*{Probabilistic particles}

For a probabilistic description of Nature the basic law for the motion of a
classical particle with mass $m$ in a potential $V(\vec{x})$ is the Liouville
equation,
\begin{equation}\label{PR2}
\frac{\del}{\del t} w(t;\vec{x},\vec{p}) = - \frac{p_k}{m} \frac{\del}{\del x_k}
w(t;\vec{x},\vec{p}) + \frac{\del V}{\del x_k} \frac{\del}{\del p_k}
w(t;\vec{x},\vec{p}).
\end{equation}
Here $\vec{x}$ and $\vec{p}$ denote the position and momentum of the particle,
and the time-local probability distribution is denoted by $w$. The particle has
no precise position or momentum. 
We rather deal with the probabilities $w(t;\vec{x},\vec{p})$ to find it at time
$t$ at a position $\vec{x}$ with momentum $\vec{p}$. Eq.~\eqref{PR2} is a
partial differential equation for a function $w$ depending on seven real
variables $(t,x_k,p_k)$. At first sight it looks more complicated than the
deterministic description by Newtons equations
\begin{align}\label{PR3}
\frac{\del}{\del t} x_k(t) = \frac{1}{m} p_k(t), && \frac{\del}{\del t} p_k (t)
= - \frac{\del V(\vec{x})}{\del x_k}(t).
\end{align} 
In eq.~\eqref{PR3} the variables are the sharp position and momentum $\vec{x}$
and $\vec{p}$ of the particle, such that we deal with a partial differential
equation for six functions which depend on $t$. The r.h.s. of the second
equation is, in general, a non-linear function of the position $\vec{x}$.

The Liouville equation \eqref{PR2} and Newton's equation \eqref{PR3} are
related, however. Newton's equation obtains from the Liouville equation in the
limit of a sharp probability distribution which vanishes for all $\vec{x}$ and
$\vec{p}$ that differ from particular sharp values $\vec{x}^{(0)}(t)$ and
$\vec{p}^{(0)}(t)$,
\begin{equation}\label{PR4}
w(t;\vec{x},\vec{p}) = \delta^3(\vec{x}-\vec{x}^{(0)}(t))
\delta^3(\vec{p}-\vec{p}^{(0)}(t)). 
\end{equation}
The sharp values $\vec{x}^{(0)}(t)$ and $\vec{p}^{(0)}(t)$ obey Newton's
equation and define the trajectory of a pointlike particle. In the other
direction, Liouville's equation has been derived from Newton's equation by
assuming a probability distribution for the initial conditions of particle
trajectories.

The advantage of the probabilistic formulation is the possibility to go beyond
the approximation of pointlike particles. For example, individual raindrops may
be associated with isolated particles. This involves approximations, since
processes as merging and splitting of drops are neglected for isolated
particles. We may associate $\vec{X}$ with the center of mass of a droplet and
$\vec{P}$ with its total momentum. At any given time $t$ the one-particle
probability $w(t;\vec{X}, \vec{P})$ can be computed from
$p[t;N(\vec{x}),\vec{v}(\vec{x})]$.
We will describe its possible construction in some detail -- not because it is
actually needed for this work, but rather in order to illustrate the steps from
the water droplets to particles only characterized by position and momentum.

For this purpose we have to employ some definition which positions $\vec{x}$
belong to the droplet. We restrict the functions $N(\vec{x})$ and
$\vec{v}(\vec{x})$ to the region of the droplet. For any given function
$N(\vec{x})$ this defines the position $\vec{x}_0$ of the center of mass, such
that $\vec{x}_0[N(\vec{x})]$ is a function of the particle distribution
$N(\vec{x})$. For the total momentum $\vec{p}_0$ of the drop we start from the
distribution of momenta within the volume element around $\vec{x}$,
$\vec{p}(\vec{x}) = m N(\vec{x}) \vec{v}(\vec{x})$, with $m$ the mass of the
molecules. One obtains the total momentum $\vec{p}_0$ by summing
$\vec{p}(\vec{x})$ over all positions $\vec{x}$ that belong to the drop. Thus
$\vec{p}_0 [ N(\vec{x}), \vec{v}(\vec{x})]$ is a function of $N(\vec{x})$ and
$\vec{v}(\vec{x})$. In turn, the one-particle probability distribution
$w(\vec{X}, \vec{P})$ is found by summing all probabilities for which $\vec{x}_0
[N(\vec{x})] = \vec{X}$ and $\vec{p}_0[N(\vec{x}),\vec{v}(\vec{x})] = \vec{P}$,
\begin{align}
w(\vec{X}, \vec{P}) = Z^{-1} \sum_{N(\vec{x}),\vec{v}(\vec{x})} & p[N(\vec{x}),
\vec{v}(\vec{x})] \delta^3 (\vec{x}_0[N(\vec{x})] - \vec{X}) \nonumber \\
& \cdot \ \delta^3 (\vec{p}_0 [N(\vec{x}), \vec{v}(\vec{x})]-\vec{P}).
\end{align} 
The sum is only over the functions $N(\vec{x})$ and $\vec{v}(\vec{x})$
restricted to $\vec{x}$ inside the drop. This needs a reweighing of the
probability distribution according to
\begin{equation}
Z = \sum_{N( \vec{x}),\vec{v}(\vec{x})} p[N(\vec{x}),\vec{v}(\vec{x})],
\end{equation}
which guarantees the normalization
\begin{equation}
\int \dif^3 X \dif^3 P\ w(\vec{X}, \vec{P}) = 1.
\end{equation}
We label the center of mass $\vec{X}$ and the total momentum $\vec{P}$ of the
droplet by big letters in order to distinguish them from the positions $\vec{x}$
and momenta $\vec{p}$ of volume elements. Nevertheless, $w(\vec{X},\vec{P})$ is
precisely the type of object that appears in the Liouville equation.

The computation of the one-particle probability distribution
$w(\vec{X},\vec{P})$ can be done at every time $t$. For a moving drop the region
of $\vec{x}$ which belongs to the drop will depend on $t$. It may be defined by
the fast fall-off of $N(\vec{x})$ far from the center of the drop, e.g. by
defining some threshold value for $N(\vec{x})$ that needs to be exceeded for
$\vec{x}$ inside the drop. If we know an evolution law for $p[t;
N(\vec{x}),\vec{v}(\vec{x})]$ we can find an expression for $\partial_t
w(t;\vec{X},\vec{P})$. In general, this expression is not a function of
$w(t;\vec{X},\vec{P})$ alone.
Only a suitable approximation determines this expression as a function (more
precisely functional) of $w(t;\vec{X}, \vec{P})$, such that the evolution
equation for the one-particle probability distribution is closed
\begin{equation}\label{PR8}
\frac{\del}{\del t} w(t;\vec{X},\vec{P}) = F[w(t;\vec{X},\vec{P})].
\end{equation}
This form is a generalization of the Liouville equation for non-pointlike
particles.
For raindrops the effective ``probabilistic equation of motion" will be rather
different from the Liouville equation in a gravitational potential. Effects due
to interactions between molecules in the drop and the environment, as friction
and the adaptation of the shape of the drop, play an important role. In contrast
to free falling pointlike particles, raindrops typically reach a maximal
velocity.

Let us discuss the deterministic Newtonian limit in the light of this setting.
The planet Jupiter can be considered as an extended drop. Instead of water
molecules it consists of gases as hydrogen, helium, ammonia, sulfur, methane and
vapor, which become liquids in its central region. The molecules are held
together by gravity. In principle, its one-particle probability distribution
follows a generalization of the Liouville equation of the type \eqref{PR8}. It
turns out that the Liouville equation holds to a very good approximation. On
length scales much larger than the size of Jupiter a pointlike approximation
becomes accurate, and Newton's law follows. The reason for the high accuracy of
the pointlike approximation to the Liouville equation is the important role of
gravity. It holds the planet together. Far away from the planet only its mass
matters.

There is no reason why the Liouville equation should hold for microscopic
particles as atoms. For these objects gravity plays no dominant role and other
criteria may determine the form of eq. \eqref{PR8}. It has been found
\cite{CWQP,CWQPCG,CWQPPS} that for a particular form of $F[w]$ eq.~\eqref{PR8}
leads to the evolution of a quantum particle in a potential, as usually
described by the Schrödinger equation.
This includes all characteristic quantum effects as tunneling or interference.
Furthermore, a family of functionals $F[w]$ interpolates continuously between a
quantum particle on one side and a classical particle on the other side.
Particles with dynamics between quantum physics and classical mechanics have
been called ``zwitters" \cite{CWZWI} and could be relevant for certain
experiments \cite{COFO,EFMC}.
It is an interesting question to find out why Nature prefers a form of
eq.~\eqref{PR8} that describes quantum mechanics for atoms.

\paragraph*{Change of paradigm}

The shift from a fundamentally deterministic setting to an approach where the
fundamental description is probabilistic is a change of paradigm. In the
deterministic view the probabilistic description is an effective approach for
complex situations for which the knowledge of an observer is insufficient to
grasp all details that are, in principle available. For the probabilistic
approach probabilities are fundamental. The deterministic physics is a good
approximation for special situations. Quantum mechanics has already made the
step to a genuinely probabilistic setting, while for classical physics the
deterministic view prevails so far. Quantum mechanics has, however, introduced
new concepts and laws that are believed to go beyond a probability distribution
that is sufficient for the computation of expectation values of observables. We
will stick to the classical statistical concept of probability distributions in
this work. Both classical and quantum systems are described in these terms. The
formulation of quantum mechanics will emerge for particular types of subsystems.

We may paraphrase the change of paradigm by stating that not the pointlike
particles as the planets are fundamental, but rather the probabilistic
description of systems as the rain.
There is no logical inconsistency of a deterministic description of the world.
In practice, however, physicists will always have to deal with subsystems. These
subsystems are necessarily probabilistic. Probabilistic subsystems are the
generic case even for a deterministic description of the world. We find it more
economical and much closer to what a physicist can do to start with a
probabilistic description of the world. There is simply no need for fundamental
determinism with all the problems described above for the raindrop. The observed
deterministic features in our world arise for well understood particular
subsystems, as for the planets and many objects in everyday life.
Thermodynamics is another good example how deterministic aspects arise from the
genuine probabilistic description of gases, liquids, solids and so on.

\subsection{Probabilistic realism}\label{sec:probabilistic_realism}

Probabilistic realism is a basic conceptual, ``philosophical" view of the world.
We highlight here its central ingredients. We discuss properties of a
fundamental overall probability distribution which should describe the whole
Universe. Several criteria for the selection of such a distribution point
towards structures familiar in quantum field theories as local gauge symmetries.

\paragraph*{Reality}

One world exists. Physics describes it by an overall probability distribution
and a selection of observables. The overall probability distribution covers the
whole Universe, for all times. It can even include situations for which time and
space loose their meaning. The world is real, and the probability distribution
with associated observables is a description or a picture of this reality. By
creating humans and scientists the world produces an incomplete picture of
itself as a part of itself. The world comprises everything -- there is nothing
``outside" the world, neither in time, nor in space, nor in any other category.
There is only one world. This contrasts to the ``many world interpretation of
quantum mechanics". This conceptional approach of one real world described by
probabilistic laws has been called \cite{CWICS, CWIS, CWQM} ``probabilistic
realism".

Instead of the whole world the same concepts can be applied to subsystems. This
requires that a subsystem admits a closed description that does not depend
explicitly on the environment of the subsystem. In the most general terms the
environment consists of all the probabilistic information in the world beyond
the one employed for the definition of the subsystem. A subsystem may be
spatially isolated as a single atom. It may be local in time as the concept of
the ``present" in distinction to the future and past. It could also be a
subsystem in the space of correlation functions or some other closed part of the
probabilistic information. 
We will see that for subsystems new probabilistic elements beyond the
probability distribution appear. 

Probabilities that are not very close to one or zero do not lead to definite
conclusions what the outcome of an observation will be. We may employ the notion
of ``certainty" or ''restricted reality" if the probability for a given possible
measurement value of an observable is very close to one. How close is a matter
of definition and may depend on the circumstances. Parameterising with $\Delta =
1-p$ the remaining ``uncertainty", one may sometimes be satisfied with a
``threshold of certainty" $\Delta_c = 10^{-4}$, while in another context a much
smaller value as $\Delta_c = 10^{-100}$ may be appropriate. Requiring $\Delta_c
= 0$ is, in general, too strong since it is never realized in practice. An event
is ``real" in the sense of restricted reality or ``certain" if the probability
for this event exceeds the threshold value, $p>1-\Delta_c$. This restricted
notion of reality coincides with the concept of reality used by Einstein,
Podolski and Rosen \cite{EPR}. As we have seen in the previous discussion of the
rain falling with constant velocity, the restricted reality may concern a
correlation. This important aspect is missing in the discussion of
ref.~\cite{EPR}. We will address this issue in more detail later.

Concerning the philosophical notion of reality the one world is real. Restricted
reality or certainty rather concerns the question for which observables definite
statements can be made about their values.

\paragraph*{Fundamental probabilities}

In our approach probabilities are fundamental. They are not associated to the
lack of knowledge of some observer. They are simply the mathematical objects,
obeying some simple axioms, in terms of which a physicist's picture of the world
is formulated. Let us illustrate the difference between fundamental
probabilities and ``observer probabilities'' that are often associated to
probabilistic settings and should not be confounded.

For this purpose we consider a ball and a cube, one red and the other green. We
compare two different settings. Let us first assume a situation where the
fundamental probability that the ball is green equals one. An observer may have
incomplete information, knowing only that one object is red and the other green.
Without additional knowledge the ``observer probability'' that the ball is green
is only one half. The fundamental probability and the observer probability do
not coincide. The difference between the fundamental probability and the
observer probability is due to the lack of knowledge of a given observer. It
depends on the information available to a particular observer. It can change if
more information becomes available for the observer - for example if she can
have a look and resolve both the shape and the color of the object. If the
fundamental probability for a green ball is close to one there exists typically
a possible setting for which an observer can make the prediction that the ball
is green with high certainty. 

In contrast, consider a second setting where the fundamental probability for a
green ball and a red cube is one half, and the probability for a green cube and
a red ball is again one half. If the fundamental probability for a green ball is
0.5, there exists no observation for which the color of the ball can be
predicted close to certainty. 

We observe that for both settings the two parts of the system are correlated.
Whenever one object is green, the other one is red. Whenever one object is a
ball, the other one is a cube. An observer who has seen a green ball can predict
that the other object is a red cube. The fundamental probability that the two
objects are a green ball and a green cube, or a green ball and a red ball,
vanishes. For the second setting there is no situation for which the color of
the ball can be predicted with certainty. Only the correlations are predicted
with certainty. Once an observer sees a green ball she knows that the other
object is a red cube.

Let us assume that no signals can be exchanged between the two parts of the
system. If a first observer has seen a green ball she can predict with certainty
that an observation of the second object finds a red cube, no matter who
observes it and when the observation is done. In the absence of an exchange of
signals one could erroneously conclude that the observer probability for the
second object being a red cube is one, and therefore the fundamental probability
for a red cube should be one from the beginning, as for our first setting.

This obviously contradicts the distribution of fundamental probabilities for our
second setting. The error arises from the artificial separation of the system
into two isolated objects which is not allowed in the presence of correlations.
Correlations need no exchange of signals to be realized.

One could invoke that the fundamental probability distribution may change after
the first observation due to the action of the observer. This concept is often
not appropriate for fundamental probabilities, however. For example, the
intrinsic probabilistic properties of correlations in the polarization of the
cosmic microwave background do not change if somewhere some intelligent life
observes them. What has changed is the observer probability for a subsequent
observation. On the level of fundamental probabilities this is encoded in the
notion of conditional probabilities for sequences of observations, asking what
is the probability of an outcome $b_n$ for an observable $B$ under the condition
that for $A$ a previous observation has found the value $a_m$. We will discuss
these issues in more detail in the second part on quantum mechanics.

The fundamental probabilities typically refer to the overall probability
distribution of the whole world. They are properties of reality and need no
observer. Probabilistic laws for the emission of the cosmic microwave background
are valid independently of the question if humans or other intelligent life
develops instruments to observe it. Fundamental probabilities are maximal in the
sense that the observer probability before the observation of an event can never
exceed the fundamental probability. For subsystems the fundamental probabilistic
information is no longer available completely. 

\paragraph*{Fundamental probability distribution for the \\ Universe}

For a fundamental probabilistic theory describing our Universe over all times
one needs criteria for the selection of the probability distribution. One way
is, of course, to be guided by observation. One may try to construct an overall
probability distribution that encodes the elementary particles and local gauge
symmetries of the fundamental interactions according to the standard model of
particle physics coupled to quantum gravity, or extensions thereof. Coming from
the other side, one may also ask if there exist general criteria for the
selection of a fundamental probability distribution which naturally induce the
structures of local gauge theories. We will argue that this seems indeed to be
the case.

The issue of general criteria for a fundamental probability distribution gets
more complicated by the fact that the probability distribution itself is not an
observable quantity. Its properties become observable only in connection with a
given association of the values $A_\tau$ of formal observables with given
physical observations. One can perform variable transformations in the space of
events $\tau$ which change both the probabilities $p_{\tau}$ and the values
$A_\tau$, while leaving the expectation values $\langle A\rangle$ invariant. On
a deep conceptual level physical statements and predictions concern only
probabilistic structures between observables. 

While the quest for a fundamental probability distribution constitutes an
important goal, we emphasize that our probabilistic approach is not limited to
fundamental physics. Our discussion of correlated subsystems, time-and space
structures or the possible embedding of quantum mechanics in classical
probabilistic systems remains valid for arbitrary classical probabilistic
systems characterized by a probability distribution $p_\tau$ and observables
$A_{\tau}$. All conclusions are only based on the law~\eqref{I1} for expectation
values, for appropriate selections of $\tau$, $p_{\tau}$ and $A_\tau\,$.

\paragraph*{Structures between observables}
The overall probability distribution is not sufficient to describe the ``state
of the world". It has to be supplemented by a set of observables. The ``state of
the world" can be seen as the ensemble of probabilities for the values of a set
of ``basis observables".
The possible values of these basis observables can be used to specify the set of
basis events $\tau$. An example are Ising spins as basis observables and spin
configurations as basis events. The overall probability distribution associates
to each combination of values
of basis observables 
a probability $p(\tau)$.
This probabilistic information should be sufficient to determine probabilities
for the values of all observables of interest and to make predictions for the
outcomes of measurements. The notion of ``observables of interest" remains
somewhat vague here. It reflects the limitations of a physicist picture of the
world which is necessarily incomplete.

The issue that only a combination of a choice of basis observables and the
probability distribution can yield a description of reality has been addressed
in ``general statistics" \cite{CWGENS}. Consider a probability distribution
$p(\tau)$ depending on a real variable $\tau \in \mathbb{R}$,
\begin{equation}
\int_{-\infty}^\infty \dif \tau\ p(\tau) = \int_\tau p(\tau) = 1.
\end{equation}
Observables are functions $A(\tau)$, with expectation values
\begin{equation}\label{PR10}
\braket{A} = \int_\tau p(\tau) A(\tau).
\end{equation}
One is typically interested in normalizable observables for which
\begin{equation}\label{PR11}
\braket{A^2} = \int_\tau p(\tau) A^2(\tau)
\end{equation}
exists, such that $p^{1/2} A$ is a square integrable function. The spectrum of
possible measurement values of $A(\tau)$ corresponds to the range of possible
values of $A(\tau)$. It is typically, but not necessarily, continuous.

Consider next an invertible variable transformation $\tau \to \tau' =
f^{-1}(\tau)$, or
\begin{equation}
\tau = f(\tau').
\end{equation}
Expressed in the new variables $\tau'$ the observable $A$ reads
\begin{equation}
A(\tau) = A(f(\tau')) = A'(\tau').
\end{equation}
The transformation of the probability distribution involves a Jacobian
$\hat{f}$,
\begin{align}
p'(\tau') = \hat{f} p(f(\tau')), && \hat{f} = |\del f/\del \tau'|,
\end{align}
such that we can continue to use the expression \eqref{PR10} with the new
variables
\begin{equation}\label{PR15}
\braket{A} = \int_{-\infty}^{\infty} d\tau'\ p'(\tau') A'(\tau').
\end{equation}
Omitting the primes the variable transformation amounts to a simultaneous
transformation of observables and probability distribution, which can both be
taken now as functions of a fixed variable $\tau$,
\begin{align}\label{PR16}
A(\tau) \to A(f(\tau)), && p(\tau) \Rightarrow \hat{f}(\tau) p(f(\tau)), &&
\hat{f} = \left| \frac{\del f}{\del \tau}\right|.
\end{align}
Expectation values of observables, as well as the spectrum of their possible
measurement values, are invariant under the variable transformation.

With respect to variable transformations the probability distribution transforms
as a density. For this reason it is often called a ``probability density". For a
suitable choice of $f(\tau)$ any given probability density $p_1(\tau)$ can be
transformed into any other arbitrary probability density $p_2(\tau)$
\cite{CWGENS}. In other words, for two arbitrary probability distributions
$p_1(\tau)$ and $p_2(\tau)$ there always exists an invertible variable
transformation such that
\begin{equation}
p_2(\tau) = \left| \frac{\del f}{\del \tau} \right| p_1(f(\tau)).
\end{equation}
This statement generalizes to $N$ variables, where $\tau$ and $f$ become
$N$-component vectors $\tau_u$, $f_w$, $u,w = 1...N$, and $\hat{f} = |\det(\del
f_w / \del \tau_u)|$.
Invertibility requires $\hat{f} > 0$. Taking sequences with $N \to \infty$ one
finds that every probability distribution of infinitely many real variables can
be transformed into any other probability transformation \cite{CWGENS}.
Infinitely many real variables are a generic case for the description of the
world. Already a real scalar field $\varphi(x)$ corresponds to infinitely many
real variables, one for each point $x$.

This observation has a far reaching consequence. All probability distributions
for infinitely many real variables are equivalent with respect to variable
transformations. A given probability density only specifies a coordinate choice
in the space of variables. If humans use a probability distribution $p_1(\tau)$
for the description of the world, intelligent life on some other planet in the
Universe may use a different probability distribution $p_2(\tau)$. The
conclusion on possible observables will be the same, provided one associates an
observation with the observable $A(f(\tau))$ on the other planet when ever it is
described by $A(\tau)$ on earth.
Only the combination of observables with the probability distribution allows for
statements or predictions for observations. The probability distribution alone
has no physical meaning. 

On a conceptual level statements about observations are related to structures
among observables \cite{CWGENS,CWGEO}. These structures remain invariant under
variable transformations. They do not depend on the choice of variables and the
associated choice of the probability distribution. In this view the physicists
understanding of the world is the unveiling of structures among observables, and
probabilistic statements about possible observations related to these
structures.

In practice it is convenient to work with fixed variables and to make a choice
of the probability distribution. For this choice one associates particular
observables to possible observations. While in principle the choice of the
probability distribution is arbitrary, in practice it should be chosen such that
important structures among observables as time, space and symmetries find an
expression in terms of simple observables. 
This criterion of simplicity, together with a criterion of robustness discussed
below, greatly restricts the choice of the probability distribution. In the
following we mainly will choose a fixed probability distribution for which
important structures among observables find a simple representation. It is
understood that at the end only the structures among observables are related to
statements about possible observations.

\paragraph*{Time and space}

Time and space should be understood as ordering structures among observables.
Time corresponds to a linear order of a class of observables, assigning to every
pair of observables in this class $A_1$ and $A_2$ one of the three relations:
$A_2$ is before, after or simultaneous as compared to $A_1$. This defines
equivalence classes of observables, labeled by time $t$. We can denote
observables in the ordered class by $A(t)$, indicating explicitly the
equivalence class to which they belong. This notion of ``probabilistic time"
\cite{CWPT} does not introduce time as an a priori concept. The notion of time
is meaningful only to the extent that the ordering of observables can be
formulated. By far not all observables can be ordered in time. Simple examples
where this is not possible are correlations of observables at different times as
the products $A(t_1) A(t_2)$. The basic concept of ordering is typically not
unique. Many different time structures can be introduced in this way. One has to
find out which structure can be used to define a type of physical or universal
time.

It is advantageous to use basis observables which belong to the ordered class of
observables. Such basis observables $s(t)$ are labeled by time $t$.
Eq.~\eqref{PR11} requires that $\braket{s^2(t)}$ is defined
\begin{equation}\label{PR18}
\braket{s^2(t)} = \int_{\tau} p(\tau) s^2(t).
\end{equation}
Using the freedom of variable transformations it is useful to concentrate on
probability distributions that have simple properties with respect to the time
structure. In practice we will choose a formulation in the other direction. We
will assume an ordering of the basis observables $s(t)$ and discuss simple types
of probability distributions as local chains. We will then describe the notion
of time emerging for this ``choice of coordinates in field space", and discuss
the question if the time defined in this way can be associated with ``physical
time" as used in observations. On the most fundamental level we will consider
Ising spins $s(t)$ for which eq.~\eqref{PR18} is obeyed trivially since $s^2(t)
= 1$ implies $\braket{s^2(t)} =1$.

Space and geometry can be introduced as structures among observables of a
similar type \cite{CWGEO}. In this case one employs a family of observables
$A(\vec{x})$ that depend on a label $\vec{x}$ which is a point in some subspace
of $\mathbb{R}^D$, with $D$ the dimension of space. One also could use
discretized versions for $\vec{x}$. If the connected correlation function,
\begin{equation}
\braket{A(\vec{x}) A(\vec{y})}_c = \braket{A(\vec{x}) A(\vec{y})} -
\braket{A(\vec{x})}\braket{A(\vec{y})},
\end{equation}
obeys certain (rather mild) conditions \cite{CWGEO}, it can be used to introduce
a distance. The basic idea is, that $\braket{A(\vec{x})A(\vec{y})}_c$ decreases
as the distance increases, and vice versa. If the observables $A(\vec{x})$ are
differentiable with respect to $\vec{x}$ one can extract a metric on a patch of
$\mathbb{R}^D$ from the connected correlation function. Geometry and topology
follow as concepts induced by a structure among observables \cite{CWGEO}.

For a discussion of the structure of spacetime a possible convenient choice for
variables are fields, $\tau = \varphi(x)$, $x = (t,\vec{x})$. A classical state
is then given by the value of $\varphi$ for every spacetime point $x$. We may
consider real variables $\varphi$ as configurations for Ising spins in the limit
$N \to \infty$, similar to the representation of real numbers by bits. One may
also choose discrete fields as Ising spins $s_{\gamma}(x)$. Furthermore, one
could start with discrete points $x$ and take a continuum limit.
The distribution of molecules $N(x)$ for the description of the rain is an
example for a discrete field that can be promoted to a continuous field in the
limit of large $N$. If the variables are fields, the probability distribution
$p[\varphi(x)]$ or the action $S[\varphi(x)]$ is a functional of the fields
$\varphi(x)$. To every field configuration $\varphi(x)$ one associates a real
number $p$ or $S$. Choices for probability distributions that permit a simple
discussion of the structures of space and time correspond to local actions. In
this case the probability distribution $p$ is a product of ``local factors" at
$x$ that each involves fields only in a neighborhood of $x$. We observe that the
variables $\varphi(x)$ can be identified with a particular set of basis
observables.

\paragraph*{Symmetries}

Symmetries are variable transformations \eqref{PR16} that leave the probability
distribution invariant,
\begin{equation}
\hat{f}(\tau) p(f(\tau)) = p(\tau).
\end{equation}
This extends to the case of discrete variables $\tau$ for which the Jacobian
typically equals one, $\hat{f} = 1$. Two observables related by a symmetry
transformation $f(\tau)$ have the same expectation value. For $A'(\tau) =
A(f(\tau))$ one has $\braket{A'} = \braket{A}$,
\begin{equation}
\int_\tau p(\tau) A(f(\tau)) = \int_\tau p(\tau) A(\tau).
\end{equation}
This follows directly from eq.~\eqref{PR15} , using $p' = p$.

For $\tau \in \mathbb{R}^N$ the general symmetry group is $sgen_N$, the group of
$N$-dimensional general coordinate transformations that leave a given
probability density invariant. The structure of this group is independent of the
choice of $p(\tau)$ \cite{CWGENS}. The group $sgen_N$ is a huge group, in
particular if we consider the limit of infinitely many degrees of freedom $N\to
\infty$. Most of the symmetry transformations are, however, complicated
non-linear transformations and not very useful in practice. In particular, many
of those general symmetry transformations are not compatible with the structures
for time and space. A generic variable transformation does not respect the
ordering structure of basis observables $s(t)$ in time, or the concepts of
neighborhood for the observables $A(x)$ for the structures of space and
spacetime.

A class of useful symmetry transformations are those that respect the structures
of space and time. The simplest case are local symmetry transformations that
transform at each point of spacetime $x$ the variables $\varphi(x)$ into
variables at the same location
\begin{equation}
\varphi(x) \to f(x;\varphi(x)).
\end{equation}
Particularly simple are linear local transformations acting on multi-component
fields $\varphi_\gamma(x)$ as
\begin{equation}\label{PR23}
\varphi_\gamma (x) \to B_{\gamma \delta}(x) \varphi_\delta(x).
\end{equation}
These are local gauge symmetries. It seems advantageous to use variables which
realize the local gauge symmetries \eqref{PR23} in a simple way.
This often requires a connection which transforms inhomogeneously, as well as
transformations involving derivatives $\del_\mu \varphi_\gamma(x)$. It is
possible to formulate local gauge theories uniquely in terms of fields that show
the transformation law \eqref{PR23}. The connection then arises as a composite
object \cite{CWSLGT}.

The marking of observables with a spacetime label $x$ should not depend on the
choice of coordinates for the positions of observables. Only the notion of
infinitesimal neighborhood of observables should matter. This induces a symmetry
of general coordinate transformations in $d$-dimensions,
with $d$ the dimension of spacetime. Again, it is advantageous to employ
variables or coordinates in field space for which the probability distribution
is invariant under a simple linear realization of this symmetry. In this case
diffeomorphism symmetry is realized in a standard way.

\paragraph*{Simplicity and robustness}

If we start with a simple representation of spacetime and local symmetries by
choosing a probability distribution depending on local fields $\varphi(x)$, and
perform a general non-linear variable transformation \eqref{PR15}, the result
will be a rather complicated probability distribution for which the structures
of spacetime and local symmetries are not easily visible. Also simple local
observables in the first ``optimal picture'' will be mapped to complicated
non-local observables in the transformed picture. Given our experience that the
structures of spacetime and symmetries are useful for the understanding of the
world, it seems rather obvious that the first simple picture is superior to the
complicated second picture of the same structures. In practice the choice of
coordinates in field space matters - similarly to the advantage of a choice of
coordinates in spacetime that is well adapted to a given problem. The invariance
of the observable structures under general variable transformations \eqref{PR15}
is important conceptually -- in practice the criterion of simplicity for a fixed
set of variables, probability distribution and observables matters. There is a
good reason why local quantum field theories with local gauge symmetries and
diffeomorphism invariance are well adapted to the description of the world.

Simplicity is already a powerful criterion for the selection of the probability
distribution. It is, however, not sufficient. Many local gauge theories are
possible, and the criterion of simplicity does not seem to favor a particular
model. 
Another powerful criterion for the selection of an efficient description is
robustness \cite{CWGEO}. Let us assume that for a given probability distribution
we employ a certain set of local observables $A_i(x)$ for the description of
observations.
The resulting conclusion should not depend strongly on the use of observables
$A_i(x)$ or of closely neighboring observables $A_i(x) + \delta_i(x)$. This
statement is based on the fact that no observation can be infinitely precise.
There will always be very close observational settings which have to be
described by neighboring observables $A_i + \delta_i$. Since there is no way to
decide if measurements are related to $A_i$ or $A_i + \delta_i$, any realistic
description should require insensitivity with respect to the choice of $A_i$ or
$A_i + \delta_i$.

This simple ``criterion of robustness" has important consequences for the choice
of the overall probability distribution for the Universe. Two neighboring
probability distributions should give similar expectation values for relevant
observables. Consider two probability distributions $p(\varphi)$ and $p(\varphi)
+ \delta p(\varphi)$ in close vicinity to each other. 
The distribution $p + \delta p$ can be mapped to $p$ by a transformation
\eqref{PR16} close to the identity, $f(\varphi) = \varphi + \delta f(\varphi)$.
In turn, this transformation will map the observables $A_i (\varphi)$ for the
probability density $p + \delta p$ to observables $A_i + \delta A_i$ for the
probability density $p$. The criterion of robustness tells us that observables
$A_i$ for $p+\delta p$ should lead to a very similar description as the use of
$A_i$ for $p$. The robustness criterion therefore implies ``robustness for
probability distributions".
Two closely neighboring probability distributions should lead to closely similar
outcomes for a given set of observables $A_i(x)$ related to possible
measurements. This should hold at least for the ``relevant observables" that are
used in practice for the description of observations.

The robustness criterion favors a description in terms of ``renormalizable
theories" with a large separation of the length scales of ``microphysics", where
the fundamental overall probability distribution is formulated, and
``macrophysics" where observations a made. In renormalizable theories most of
the details of the microscopic probability distribution are ``forgotten" by the
renormalization flow to larger distances. The macroscopic observations are only
sensitive to the universality class and to the few renormalizable couplings of a
given universality class. This is clearly a great step towards robustness, since
many neighboring probability distributions lead to the same macrophysics.

It is not yet known to which extent the criteria of simplicity and robustness
restrict the possible observations in the macrophysical world. The restrictions
could be so strong that no free parameters remain for the macrophysical
predictions. They may also be weaker. In any case, the seemingly high
arbitrariness in the choice of probability distributions and observables for our
description of the world is highly reduced. It becomes possible for humans to
make meaningful predictions and to test them by observation.

\subsection{Basic concepts}

In this section we formulate the basic concepts used in this work. No other
concepts beyond probabilities, observables and expectation values are employed
for the fundamental definitions. The basic concept of a probabilistic
description can be formulated in an axiomatic approach \cite{KOL}.

\subsubsection{Probabilities}\label{sect:probabilities}

We start with the concept of probabilities. As discussed above, they are treated
as fundamental concepts of a description, rather than as derived quantities
describing a lack of knowledge or properties of sequences of measurements under
identical conditions.
We explain here how fundamental probabilities can be connected to observations
on a basic level, without sequences of repeated measurements.

\paragraph*{Ising spins}

The simplest observables are two-level observables or Ising spins
\cite{LENZ,ISI,KBI}. They correspond to yes/no questions that may be used to
classify possible observations. For example, a researcher may investigate the
activity of neurons. A neuron fires if it sends a pulse with intensity above a
certain threshold. In this case the answer is ``yes" and the Ising spin takes
the value $s = 1$. If not, the answer ``no" corresponds to $s=-1$. The
observable has precisely two possible measurement values, namely $s = \pm 1$. A
yes/no question deciding between mutually excluding alternatives can have only
two possible answers and nothing in-between. Consider three different neurons
corresponding to three Ising spins $s_k$, $k = 1...3$, each obeying $s_k^2 = 1$.
In this simple system a ``basis event" or ``classical state" $\tau$ is a
configuration of the three Ising spins. There are eight different states, $\tau
= 1,...,8$ as (yes, yes, yes), (yes, yes, no) etc.. A given basis event tells
which ones of the three neurons fire. 

It is obvious that Ising spins can be useful observables even in rather complex
situations. Many important properties of a probabilistic description can be
understood in a simple way by the investigation of Ising spins. For this reason
we will use Ising spins as a starting point of our general probabilistic
description in the next section. One may even assert that any practical
observation uses a finite (perhaps large) number of yes/no decisions in the end.
Furthermore, Ising spins offer direct connections to information theory
\cite{SHA}. They can be associated with bits in a computer. We can take $s = 1$
if the bit is one, and $s= -1$ if it is zero.

\paragraph*{Probabilities and predictions}

Our researcher may have developed a model that all three neurons fire
simultaneously if the brain detects the picture of a cat. She casts the outcome
of her model in the form of probabilities $p_\tau$ for the different events
$\tau$. These are positive numbers $p_\tau \geq 0$, normalized such that the sum
equals one, $\sum_\tau p_\tau = 1$. Her model may yield $p_{+++} = 0.95$ for the
event $s_1 = s_2 = s_3 = +1$ or (yes, yes, yes), whenever a cat is shown. The
other seven states have small probabilities, that sum up to $0.05$ by virtue of
the normalization. The ensemble of the eight probabilities $\{p_\tau\}$ is the
``probability distribution", which will be a central concept of this work. 
For a confrontation of theory and experiment, our researcher may show a picture
of a cat within a time interval $\Delta t$, and record the firing of the three
neurons during the same time interval. If all three neurons fire she may take
this as a good start, but if less than three fire she might get worried. The
probability of less than three firing is a small number 0.05, but she may think
that occasionally this could happen. 

In order to improve, she may show a cat a second time. She may label the yes/no
questions with a ``time index" $t$, e.g. $t = 1$ for the firing during the time
interval of the first showing, and $t = 2$ for the time interval of the second
showing. She now has six two-level observables $s_k(t)$, $k=1,...,3$, $t = 1,2$.
Correspondingly, the number of possible basis events is given by $N = 2^6 = 8^2
= 64$. Her model has to yield information about the 64 probabilities $p_\tau$
for the 64 possible basis events.

Our researcher may not be interested in the details of the ``wrong outcomes".
She may define a new ``coarse grained" Ising spin that takes the value
$\bar{s}(t) = 1$ if $s_1(t) = s_2(t) = s_3(t) = 1$ and $\bar{s}(t) = -1$ for all
other seven configurations of the Ising spins $s_k(t)$. 
There remain four possibilities for the coarse grained Ising spins, namely
$(++)$ if $\bar{s}(t_1) = \bar{s}(t_2) = 1$, $(+-)$ for $\bar{s}(t_1) = 1$,
$\bar{s}(t_2) = -1$, $(-+)$ for $\bar{s}(t_1) = -1$, $\bar{s}(t_2) = 1$, and
$(--)$ for $\bar{s}(t_1) = \bar{s}(t_2) = -1$.
The probability $p_{++}$ is the probability that both at $t_1$ and at $t_2$ all
three neurons fire, while $p_{+-}$ is the probability for the event where at
$t_1$ three neurons fire and at $t_2$ less than three neurons fire.

Assume now, that the model tells that the probability for three neurons firing
simultaneously is $0.95$, independently of $t$. The probability $\bar{p}_1$ of
three neurons firing at $t_1$ sums over the different possible events at time
$t_2$, and similarly for $\bar{p}_2$ at $t_2$,
\begin{align}
\bar{p}_1 &= p_{++} + p_{+-} = 0.95, \nonumber \\
\bar{p}_2 &= p_{++} + p_{-+} = 0.95.
\end{align} 
One may compute the probability that either only at $t_1$ or only at $t_2$ or
both at $t_1$ and $t_2$ one finds three neurons firing,
\begin{equation}
\tilde{p} = p_{++} + p_{+-} + p_{-+} = 0.95 + \frac{1}{2}(p_{+-} + p_{-+}),
\end{equation}
where $p_{+-} = p_{-+}$. This is closer to one than $\bar{p_1}$ or $\bar{p_2}$
-- how close needs additional information.

The only rule for probabilities that we use here concerns the grouping of basis
events. If one groups two or more basis events together, the probabilities of
the two basis events add. Since basis events are mutually exclusive, the
grouping of two basis events defines a new combined event, namely that either
the first or the second basis event happens. The probability for the combined
event is the sum of the probabilities for the two basis events that are grouped
into the combined event.

For a computation of $\tilde{p}$ one needs further information about $p_{+-}$.
If the model additionally predicts that the firing at $t_1$ and the firing at
$t_2$ is uncorrelated (see below), one infers
\begin{align}
p_{+-} = p_{-+} = 0.0475, && \tilde{p} = 0.9975.
\end{align} 
This probability is substantially closer to one than $\bar{p}_1$ and
$\bar{p}_2$. Finding less than three neurons firing both at $t_1$ and at $t_2$
casts serious doubts on the correctness of the model, since the probability for
this event, $1-\tilde{p}=0.0025$, is already quite small. The experiment can be
extended to many showings of cats. If the cat is shown ten times, the
probability, that for more than half of the cases all three neurons fire
simultaneously is already extremely close to one. If the observation shows that
in less than half of the cases all three neurons fire simultaneously, our
researcher may consider her model as definitely falsified.

\paragraph*{Conceptual status}

The formulation of the model and the strategy for comparing the model to
observation are entirely formulated in terms of probabilities. No other concepts
enter. Two observations are important in this context. First, probabilities are
not related to any ``lack of knowledge" of the observer. No underlying
``deterministic reality" of eyes and the brain is assumed, for which the
observer would have only limited knowledge and therefore employ probabilities.
Probabilities are not related to or derived from other more basic concepts. They
are the fundamental mathematical objects of the description of the world. In our
case the firing of neurons at different times is also independent of the issue
if they are recorded or not. Our model formulates probabilities for all times.
We will later discuss subsystems, for example for the possible observations
after a certain outcome of the first observation. The probability distribution
for this type of subsystem will depend on the outcome of the first observation.

Second, for the comparison with experiment we do not use the often employed
setting of many uncorrelated identical experiments. Such a setting is an
idealization for a rather particular setting of uncorrelated events. It cannot
be realized in practice for most of the situations. A model may compute the
probability of a big asteroid to hit the earth in the coming hundred years.
There is now way of having ``identical experiments" in this case -- either a big
asteroid hits or not -- there is only one ``experiment". Nevertheless, we would
be much more scared by a probability of 0.1 for this event, rather than
$10^{-9}$. Probabilities have a meaning without the possibility of identical
experiments. We will employ the notion of ``certain events" which have a
probability so close to one that the event is predicted with ``certainty". The
issue is then the construction of such events, which may be combinations of
simpler events as in the example above.

\subsubsection{Axiomatic setting}\label{subsec:axiomatic_setting}

For an axiomatic setting of a theory or model of the world, or some part of it,
we need first to define a ``sample set" of possible outcomes of observations. We
begin by considering a finite ``basis set" of yes/no decisions or Ising spins
$s_\gamma$ with values $\pm 1$, $\gamma = 1,...,N$. A ``basis event" $\tau$ is
an ordered sequence of values for all $N$ Ising spins. There are $2^N$ basis
events which are all mutually exclusive. Two different sequences of yes and no
cannot be realized simultaneously -- possible observations give either the one
or the other outcome. 

\paragraph*{Axioms}

The sample space $\Omega$ is the set that contains all basis events. The set $F$
of all events $E$ is the set of all subsets of $\Omega$, including the empty set
$\emptyset$ and $\Omega$ itself. The union of two events is again an event. 
Two different basis events $\tau_1$ and $\tau_2$ can be grouped together to form
a new event $\tau_1 \cup \tau_2$. 
This can be iterated by grouping $\tau_1 \cup \tau_2$ with another basis element
$\tau_3$ to form the event $\tau_1 \cup \tau_2 \cup \tau_3$, and so on. Adding
$\emptyset$, all events can be constructed by grouping basis events.

A measure space is constructed by assigning to every event $E$ a probability
$p(E)$. The probabilities obey Kolmogorov's axioms \cite{KOL}. The first states
that $p(E)$ is a positive semidefinite real number
\begin{align}
(1) && p(E)\in \mathbb{R},\quad p(E) \geq 0.
\end{align}
The second states that the probability for the set of all basis events equals
one,
\begin{align}
(2) && p(\Omega) =1.
\end{align}
Finally, the third axiom states that the probabilities for the union of two
disjoint elements $E_1$ and $E_2$ is the sum of the probabilities for the events
$E_1$ and $E_2$,
\begin{align}\label{AX3}
(3) && p(E_1 \cup E_2) = p(E_1) + p(E_2).
\end{align}
Disjoint events have no common basis event.

From eq.~(3) one concludes $p(\emptyset) = 0$, since the empty set $\emptyset$
is disjoint from any basis event $E_i$, and $E_i \cup \emptyset = E_i$. Since
the number of events is finite, we can infer for arbitrary mutually disjoint
events $E_i$ the property
\begin{equation}\label{AX4}
p(\bigcup_{i=1}^\infty E_i) = \sum_{i=1}^\infty p(E_i).
\end{equation}
This follows from the axiom (3). 
Eq.~\eqref{AX4} is the usual general formulation of Kolmogorov's third axiom. 

The proof of eq.~\eqref{AX4} uses that only a finite number of elements $E_i$
can be different from the empty set $\emptyset$, and $p(\emptyset)=0$. This
reduces the sum in eq.~\eqref{AX4} to a finite sum. Eq.~\eqref{AX3} applies to
the grouping of two basis events since they are disjoint. Furthermore, the union
$\tau_{a_1} \cup \tau_{a_2} \cup ... \cup \tau_{a_M}$ is disjoint from all other
basis events $\tau_b$ not belonging to the union, $b\neq a_1, a_2,...,a_M$. (It
is not disjoint from $\tau_b$ if $b$ belongs to the list $a_1,...,a_M$.)
Eq.~\eqref{AX3} implies that for any event that is the union of $M$ different
basis events $\tau_{a_1}, \tau_{a_2}, ... , \tau_{a_M}$ one has 
\begin{equation}\label{AX5}
p(\tau_{a_1} \cup \tau_{a_2}\cup ... \cup \tau_{a_M}) = \sum_{i=1}^M
p_{\tau_{a_i}}.
\end{equation}
Every event except $\emptyset$ corresponds to the union of a certain number of
different basis events. Two disjoint events correspond to two different unions
of basis events, with no common basis event. The union of the two disjoint
events is again a union of basis events, now comprising all basis events
belonging to either one of the two disjoint events. The probability of the union
event is the sum of the probabilities of all basis events contained in the union
\eqref{AX5}. This can be continued iteratively until all non-empty disjoint
elements in the sum \eqref{AX4} are included. 

On the other hand, eq.~\eqref{AX3} is part of eq.~\eqref{AX4} if the union
includes no more than two non-empty events. For our setting with a finite number
of basis events eqs.~\eqref{AX3} and \eqref{AX4} are equivalent.
For a finite basis set of yes/no decisions defining the sample set the three
Kolmogorov axioms are the only basis for our probabilistic description of the
world. The central object of the description is the ``probability distribution",
which is defined as the set of probabilities for the basis events. 

We will consider all other cases, as for example a probability distribution for
real variables, as suitable limits $N \to \infty$ for the number of basis
observables. If the limits are well defined this extends the axiomatic setting
to these cases. 

A given description and given probability distribution depend on the selected
set of basis observables. The same observations may be described by a different
set of basis observables. Variable transformations relate two different
descriptions of the same reality. As the basis observables or variables are
transformed, the same also holds for other observables, that are typically
expressed as functions of basis observables.
This is one more aspect of the general discussion of variable transformations in
sect.~\ref{sec:probabilistic_realism}.

\paragraph*{Probabilities and observations}

We still need a connection between a model which specifies a probability
distribution and predictions for the outcome of measurements. We do not use the
idealization of repeated identical experiments for this purpose, since there is
no practical realization for this in most cases. We rather use the notion of
``certain events". A certain event is an event for which the probability is
larger than a threshold probability very close to one,
\begin{equation}
p(E) > 1-\Delta.
\end{equation}
The value of the small parameter $\Delta$ may be adapted to the purpose of the
prediction. A model is considered falsified if a measurement finds the
complement of a certain event $\Omega\setminus E$. Models are considered as
valid as long as they are not falsified. Useful models are valid models that
have not been eliminated by a multitude of different tests. In principle, there
is a notion of human judgment reflected in the choice of $\Delta$. In many
circumstances $\Delta$ can be chosen extremely small and its precise value plays
no role.

Useful quantities for constructing certain events are combined Ising spins. For
two Ising spins $s_1$ and $s_2$ one may define the product $s_1 s_2$. It has the
possible values $\pm 1$ and is therefore again an Ising spin. Out of the four
basis events $(++)$, $(+-)$, $(-+)$ and $(--)$ it groups the two basis events
$(++)$ and $(--)$ to an event $E_1 = \{(++)\cup (--)\}$ and another event $E_2 =
\{(+-) \cup (-+)\}$. The two events $E_1$ and $E_2$ are disjoint. For the event
$E_1$ one has the combined Ising spin $s_1 s_2 = 1$, while for $E_2$ one finds
$s_1 s_2 = -1$. Two other mutually disjoint events are $\bar{E}_1 = \{(++) \cup
(+-) \cup (-+)\}$ and $\bar{E}_2 = \{(--)\}$. The combined Ising spin $\bar{s}$
for this pair equals $+1$ for $\bar{E}_1$ and $-1$ for $\bar{E}_2$. This
combined Ising spin is given by 
\begin{equation}
\bar{s} = \frac{1}{2} (1+s_1+s_2-s_2s_2).
\end{equation}
More generally, a combined Ising spin can be associated to every pair of
mutually disjoint events $E_1$, $E_2$ if $E_1 \cup E_2 = \Omega$.
As we have argued in our example in sect.~\ref{sect:probabilities}, combined
Ising spins are a powerful tool for the construction of ``certain events".


\subsubsection{Observables}

An observable has a fixed value $A_\tau$ or $A(\tau)$ for every classical state
or basis event $\tau$. This value is real, such that observables are maps from
the set of basis events to $\mathbb{R}$. The values $A_\tau$ are the possible
measurement values of the observable $A$. The ensemble of possible measurement
values is the ``spectrum" of $A$. An idealized observation could find one of the
states $\tau$ and therefore the value $A_\tau$ of the observable in this state.
We will later find subsystems for which there no longer are fixed values of all
observables in a state of the subsystem. This occurs rather genuinely if a state
of a subsystem involves basis events with different values $A_\tau$. For
subsystems the observables may become ``probabilistic observables". For the
basic formulation of the overall probabilistic system, however, we employ
observables with fixed values $A_\tau$ in every state $\tau$. This is the
setting of classical statistics, and we therefore call such observables
``classical observables". Beyond the classical observables we will later
encounter further ``statistical observables'' which measure properties of the
probability distribution, without having fixed values for a given basis state
$\tau$. A common example is temperature, which does not have a fixed value in a
given microstate of the system.

\paragraph*{Algebra of observables}

The classical observables form an algebra. Linear combinations of two
observables $A$ and $B$ form new observables $D = \alpha A + \beta B$, for which
the values in every state $\tau$ are given by
\begin{equation}
D_\tau = (\alpha A + \beta B)_\tau = \alpha A_\tau + \beta B_\tau.
\end{equation} 
The classical product of two observables $A$ and $B$ defines an observable $C$
with possible measurement values given by the product of the possible
measurement values of $A$ and $B$,
\begin{equation}
C_\tau = (AB)_\tau = A_\tau B_\tau.
\end{equation}
The classical product is associative and commutative. We will later encounter
other non-commutative product structures for observables.

\paragraph*{Correlation basis}

For a finite number of Ising spins we can construct a ``correlation basis" by
using products of Ising spins. Consider three Ising spins $s_k$. We first have
the three ``basis observables" $s_k$. Second, we can form three products of two
different Ising spins $s_1 s_2$, $s_1 s_3$ and $s_2 s_3$. Finally, there is the
product $s_1 s_2 s_3$. Together with the unit observable we have eight
``correlation-basis observables". Every possible observable can be constructed
as a linear combination of the eight correlation-basis observables. 

This follows from the fact that there are eight basis events $\tau$. We can
construct eight ``projection observables" $P^{(\tau)}$ out of linear
combinations of the correlation-basis observables. The projection observables
take the value one in the state $\tau$ and zero in all other states $\rho \neq
\tau$,
\begin{equation}
P^{(\tau)}_\rho = \delta_\rho^\tau.
\end{equation}
For example, the projection observable for the state $(+++)$ is given by
\begin{equation}
P^{(+++)} = \frac{1}{8} (1 + s_1 +s_2 + s_3 + s_1 s_2 + s_1s_3 + s_2s_3 +
s_1s_2s_3),
\end{equation}
or for the state $(+--)$ one has 
\begin{equation}
P^{(+--)} = \frac{1}{8} (1 + s_1 - s_2 - s_3 - s_1 s_2 - s_1s_3 + s_2s_3 +
s_1s_2s_3).
\end{equation}
Each of the correlation-basis observables appears with a factor $+1/8$ or
$-1/8$, where the signs are chosen such that all contributions are positive for
the particular state $\tau$. For all other states $\rho \neq \tau$ the number of
positive terms equals the number of negative terms. Any observable $A$ with
possible measurement values $A_\tau$ is obviously a linear combination of the
projection observables 
\begin{equation}
A = \sum_\tau A_\tau P^{(\tau)}.
\end{equation}
Since the projection observables are linear combinations of the
correlation-basis observables this proves our statement.

The correlation-basis can be formulated for an arbitrary finite number of Ising
spins. For four Ising spins we have four basis observables, six products of two
Ising spins, four products of three Ising spins and one product of four Ising
spins. Together with the unit observable this makes a total of $1+4+6+4+1 = 16 =
2^4$ correlation-basis observables that form a complete basis. The
generalization of the projection observables is straightforward.

\subsubsection{Correlations}
\label{sec:correlations}

Correlations are a key element for a probabilistic description of the world.
They tell us how different parts of a system are related. They are the
mathematical expression of the deep philosophical insight that the whole is more
than the sum of its parts.

\paragraph*{Correlations and separability}

Consider two Ising spins $s_1$ and $s_2$ and a probability distribution
$p_{++}=p_{--}=0$, $p_{+-}=p_{-+}=1/2$. The product $\overline{s}=s_{1}s_{2}$ is
again an Ising spin.
The probability for $\overline{s}=1$ is given by $p_{++}+p_{--}=0$, while the
probability for $\overline{s}=-1$ amounts to $p_{+-}+p_{-+}=1$.
In this case one is certain that the two spins are opposite. This property makes
only sense for the combined system of the two spins. It cannot be associated to
properties of individual spins. For the individual spins we have no certain
knowledge, since the probability for $s_{1}=1$ is given by $p_{++}+p_{+-}=1/2$,
and similar for $s_{2}$. Thus restricted reality is realized only for a property
of the combined systems or the ``whole'', and not for properties of the
individual spins or the ``parts''. If one tries to assign reality to some
property of the individual parts one runs into contradictions. Our example is
close to a typical Einstein-Rosen-Podolski~\cite{EPR} situation of a spinless
particle decaying into two fermions with spin one half. The ``certainty'' or
``reality'' concerns the property that the total spin of the two fermions
vanishes. This is only a property of the combined system of the two fermions, no
matter how far the fermions are apart after the decay. The alleged
``incompleteness of quantum mechanics''~\cite{EPR} arises from an attempt to
assign reality to properties of the individual fermions.

For a setting with fundamental probabilities the product Ising spin
$\overline{s}=s_{1}s_{2}$ has the same status as the individual spins $s_1$ and
$s_2$. Its expectation value is the classical correlation function of the
individual spins
\begin{equation}\label{XA}
\braket{\overline{s}}=\sum_{\tau}p_{\tau}\overline{s}_{\tau}=\sum_{\tau}p_{\tau}s_{1,\tau}s_{2,\tau}=\braket{s_{1}s_{2}}_{cl}.
\end{equation}
For our example one finds maximal anticorrelation, $\braket{s_{1}s_{2}}=-1$.

For a system that can be completely separated into two independent parts the
probability distribution of the whole factorizes into a product of probabilities
for the parts, which we denote by $p_{\pm}^{(1)}$ and $p_{\pm}^{(2)}$ for spin
one or two being positive or negative,
\begin{align}\label{XB}
&p_{++}=p_{+}^{(1)}p_{+}^{(2)}\;,\quad
p_{+-}=p_{+}^{(1)}p_{-}^{(2)}\;,\nonumber \\
&p_{-+}=p_{-}^{(1)}p_{+}^{(2)}\;,\quad
p_{--}=p_{-}^{(1)}p_{-}^{(2)}\;.\quad
\end{align}
For this factorized form the properties of spin one are independent of the
probability distribution for spin two,
\begin{equation}\label{XC}
\braket{s_1}=p_{++}+p_{+-}-p_{-+}-p_{--}=p_{+}^{(1)}-p_{-}^{(1)}.
\end{equation}
The distribution~\eqref{XB} yields for the correlation function
\begin{align}\label{XD}
\braket{s_{1}s_{2}}_{cl}&=p_{++}+p_{--}-p_{+-}-p_{-+}\nonumber\\
&=p_{+}^{(1)}p_{+}^{(2)}+p_{-}^{(1)}p_{-}^{(2)}-p_{+}^{(1)}p_{-}^{(2)}-p_{-}^{(1)}p_{+}^{(2)}\nonumber\\
&=(p_{+}^{(1)}-p_{-}^{(1)})(p_{+}^{(2)}-p_{-}^{(2)})=\braket{s_1}\braket{s_2}.
\end{align}
The connected correlation function,
\begin{equation}\label{XE}
\braket{s_{1}s_{2}}_{c}=\braket{s_{1}s_{2}}_{cl}-\braket{s_1}\braket{s_2},
\end{equation}
is a measure for the ``inseparability'' of the systems.
A separation into independent parts is only possible if the connected
correlation function vanishes. This extends to higher connected correlation
functions.

\paragraph*{Expectation values and correlation functions}

For a classical observable $A$ the expectation value $\braket{A}$ is defined by
the basic rule of classical statistics,
\begin{equation}
\braket{A} = \sum_\tau p_\tau A_\tau. 
\end{equation}
For the moment, this is simply a definition, and the relation between
expectation values and observables has to be established subsequently. The
expectation value of a product of two classical observables is called a
``classical correlation",
\begin{equation}
\braket{AB}_{cl} = \sum_\tau p_\tau A_\tau B_\tau.
\end{equation}
Expectation values of products of $n$ classical observables are called
``$n$-point functions" or ``$n$-point correlations", e.g.
\begin{equation}
\braket{ABCD}_{cl} = \sum_\tau p_\tau A_\tau B_\tau C_\tau D_\tau.
\end{equation}
As stated before, there is no conceptual difference between the observable $A$
and the product observable $ABCD$.

Our simple example with two Ising spins demonstrates that certain values of
correlation functions permit important conclusions about the system and may lead
to predictions. Expectation values and correlation functions are meaningful
objects without the notion of ``repeated identical experiments''. For example,
the correlation functions of Ising spins at different times can encode
properties of an evolution law.

\paragraph*{Correlation functions and probability distribution}

For a finite number $N$ of Ising spins there is a one to one map between the
$2^N$ probabilities $p_\tau$ and the $2^N$ expectation values of the
correlation-basis observables. The linear map from the probabilities $p_\tau$ to
the expectation values is invertible. Indeed, two different correlation-basis
observables cannot have the same expectation value for all states $\tau$. This
follows directly from the observation that two different correlation-basis
observables have different measurement values in at least one state $\tau$. As a
consequence, the probabilistic information encoded in the probability
distribution $\{p_\tau\}$ is the same as the one encoded in the ensemble of
``basis correlations" $\{\braket{B^{(\rho)}}\}$,
with $\braket{B^{\rho}}$ the expectation value of the correlation-basis
observable $B^{(\rho)}$. At this point the ensemble of basis correlations can be
seen simply as a different way to express the probabilistic information of the
system. If one has a procedure for measuring correlation functions one can
extract information about the probability distribution and use this for testing
a model.

Consider now a large number of Ising spins $N$. Expressing the probabilistic
information in terms of basis correlations involves expectation values of
products of Ising spins with up to $N$ factors. The correlation functions of
very high order are usually not accessible for measurements. For $N=10^6$ one
would need a million-point correlation function. We conclude that the
probability distribution is a basic object for the formulation of the theory,
but usually only some parts and aspects of it are accessible to realistic
observation. There is simply no way to resolve $2^N$ probabilities for $N=10^6$.
What is often accessible are expectation values and low-order correlations of
selected observables. In this sense we may state that for a large number of
variables only expectation values and correlations are observable. The emphasis
of a model for a probabilistic description of the world will therefore be on the
computation of expectation values and correlations.

The classical product of two observables is not the only way to define a product
of two observables. Correspondingly, the classical correlation function is not
the only way to define a correlation function as the expectation value of a
product of observables. We will find that for sequences of measurements
correlation functions different from the classical correlation functions often
play a central role. 

Already at the present stage the emphasis on expectation values and correlation
functions constitutes an important bridge between classical statistics and
quantum mechanics. At first sight these two probabilistic theories seem to have
a very different structure in which the probabilistic information is encoded. A
probability distribution and commuting observables for classical statistics, and
wave functions and operators for quantum mechanics.
Concerning measurements and observation, however, the central quantities for
both approaches are expectation values and correlations.

\paragraph*{Correlations for continuous variables}

The emphasis on correlations is also visible for continuous variables. A
continuous variable $\varphi$ is a real number and needs infinitely many bits or
yes/no decisions for its precise determination. In the language of Ising spins
it corresponds to a limit $N \to \infty$. The probability distribution becomes a
normalized real positive function of $\varphi$,
\begin{align}
p(\varphi) \geq 0, && \int_{-\infty}^\infty \dif \varphi\ p(\varphi) = 1.
\end{align} 
An arbitrarily accurate resolution of a function $p(\varphi)$ by any finite
number of measurements is impossible.

In practice, one typically encodes the available information about $p(\varphi)$
in an approximation with a finite number of parameters. For example, a
probability distribution centered around a definite value $\varphi_0$ may be
approximated by
\begin{equation}\label{EX6}
p(\varphi) = Z^{-1} \exp(-S(\varphi)), \quad Z = \int d\varphi
\exp(-S(\varphi)),
\end{equation}
with
\begin{equation}\label{EX7}
S(\varphi) = \frac{(\varphi - \varphi_0)^2}{2 \Delta^2} + \frac{a_3}{6} (\varphi
-\varphi_0)^3 + \frac{a_4}{24}(\varphi -\varphi_0)^4.
\end{equation}
The probability distribution is characterized by its maximum at $\varphi =
\varphi_0$, a typical width $\Delta$, an asymmetry around the maximum encoded in
$a_3$ and a parameter $a_4$ resolving more of the tail. For $a_3 = a_4 = 0$ this
is a Gaussian probability distribution. 

For the approximation \eqref{EX6} one can compute the expectation value
\begin{equation}
\braket{\varphi} = Z^{-1}\ \int d \varphi \varphi \exp(-S(\varphi)),
\end{equation}
and the connected two point function
\begin{equation}
\braket{\varphi^2}_c = \braket{\varphi^2}-\braket{\varphi}^2 = \braket{(\varphi
- \braket{\varphi})^2}.
\end{equation}
As a third quantity one may employ the connected three point function
\begin{equation}
\braket{\varphi^3}_c = \braket{\varphi^3} - 3 \braket{\varphi^2}
\braket{\varphi} + 2 \braket{\varphi}^3,
\end{equation}
and similarly for $\braket{\varphi^4}_c$. The parameters $\varphi_0$,
$\Delta^2$, $a_3$ and $a_4$ are in one to one correspondence with the
correlation functions $\braket{\varphi}$, $\braket{\varphi^2}_c$,
$\braket{\varphi^3}_c$ and $\braket{\varphi^4}_c$. For a Gaussian probability
distribution $(a_3 = a_4 = 0)$ one has 
\begin{equation}
Z = \sqrt{2\pi \Delta^2}, \quad \braket{\varphi} = \varphi_0, \quad
\braket{\varphi^2}_c = \Delta^2,
\end{equation}
with all higher connected $n$-point functions vanishing. The correlation
functions $\braket{\varphi^3}_c$ and $\braket{\varphi^4}_c$ are therefore a
measure for the deviations from a Gaussian distribution. For many practical
problems the approximation \eqref{EX6} covers the available information,
demonstrating the focus on the low correlation functions. This issue generalizes
to fields $\varphi(x)$, where the connected correlation functions involve field
values at different positions.

\section{Probabilistic time}\label{sec:probabilistic_time}

Time is a fundamental concept in physics. It is the first structure among
observables that we will discuss. 
Rather than being postulated as an ``a priori concept'' with physics formulated
in a pregiven time and space, probabilistic time is a powerful concept to order
and organize observables. There is no time outside the correlations for the
observables of the statistical system.
Introducing time as an ordering structure for observables generates directly the
concepts of locality in time and time-local subsystems that only involve
probabilistic information at some ``present'' time $t$. In turn, this leads to
the concept of evolution, namely the question how the probabilistic information
at some neighboring subsequent time $t+\varepsilon$ is related to the
probabilistic information at time $t$. Understanding the laws of evolution makes
predictions for future events possible. The description of the probabilistic
information for the time-local subsystem and its evolution involves a formalism
similar to quantum mechanics. The necessary local probabilistic information for
a simple evolution equation is encoded in probability amplitudes (wave
functions) or a density matrix. The evolution law is a generalized Schrödinger
or von Neumann equation.
The present chapter deals with the formalism necessary for the understanding of
evolution and presents a few simple instructive examples.

In sect.~\ref{sec:classical_statistics} we first recall
our setting of classical statistics. We adapt the choice of the probability
distribution in order to permit a simple implementation of
the structure of ``time-ordering'' for the basis observables and associated
local observables. We discuss general forms of the overall
probability distribution as unique jump chains or local chains.
For all these classical statistical systems the transfer
matrix and operators representing observables are a central piece of the
formulation. This gives a first glance on non-commutative 
structures in classical statistics.

In sect.~\ref{sec:time_and_evolution} we introduce time as an ordering structure
for a class of observables, and the associated concept of evolution. Time 
defines an equivalence class of observables. Two members of an equivalence class
are two observables ``at the same time $t$''.
The equivalence classes can be ordered according to $t$. Evolution describes how
the probabilistic information at two
neighboring times $t$ and $t+\epsilon$ is connected, such that knowledge at $t$
permits predictions for $t+\epsilon$.
The problem of ``information transport" between two layers of time introduces
the concept of classical wave functions and the classical
density matrix into classical statistics. As an example, we discuss simple clock
systems. 

In sect.~\ref{sec:physical_time} we extend the discrete ticking of clocks to
continuous time. This yields differential evolution equations.
We
introduce ``physical time'' by counting the number of oscillations, and show how
basic concepts of special and general relativity emerge
from our setting of ``probabilistic time'' \cite{CWPT}.

In sect.~\ref{sec:free_fermions_in_two_dimensions} we discuss an overall
classical probability distribution which describes a quantum field theory for
free fermions in one space and one time dimension. This simple two-dimensional
quantum field theory provides a first example for the emergence of quantum
mechanics from classical statistics.

\subsection{Classical statistics}\label{sec:classical_statistics}

We first discuss the classical statistics of the overall probability
distribution for the whole world. Classical statistics is often
associated with commuting structures and a decay of correlations for large
distances, in contrast to quantum statistics with 
its non-commutative structure and oscillatory behavior. This view is too narrow.
We show here explicitly the importance of non-commutative
structures in classical statistics, and give simple examples of oscillatory
behavior.

\subsubsection{Observables and
probabilities}\label{sec:observables_and_probabilities}

In order to permit a self-contained presentation of this chapter we begin by a
summary of classical statistics, adapted to our purpose. It partly recapitulates
in a short form some aspects already discussed in chapter
\ref{sec:Fundamental_probabilism}.

\paragraph*{Two postulates for classical statistics}

A basic concept for a description of the world are ``observables''. They are
denoted by $A$, $B$ etc.. Observables can take different values $A_\tau$, which
can be discrete or continuous real numbers. We assume that the values $A_\tau$
are the possible outcomes of measurements of $A$. We do not enter at this stage
the rather complicated topic how measurements are actually done in real physical
situations and how ``ideal measurements'' are selected. 
We will turn to this issue later.
The characterization of observables by a set of values $A_\tau$ is taken here as
a first postulate or axiom of a probabilistic description of the world or
``classical statistics''.

A ``state'' $\tau$ of classical statistics can be characterized by the values of
a suitable set of observables. Two states $\tau$ and $\rho$ differ if two values
$A_\tau$ and $A_\rho$ differ for at least one observable. On the other hand,
$A_\tau = A_\rho$ does not imply $\tau = \rho$ since some other observable may
have different values in the two states, $B_\tau \neq B_\rho$. We are interested
in situations where a state $\tau$ can be characterized by a set of ``basis
observables'' that we call ``variables''. Then the set of values of the
variables $(A_\tau,\, B_\tau,\, \dots)$ specifies the state $\tau$. Other
observables can be constructed from the basis observables, as the linear
combinations $\alpha A + \beta B$, $\alpha, \beta \in \mathbb{R}$. The value of
the observable $\alpha A + \beta B$ in the state $\tau$ is given by $\alpha
A_\tau + \beta B_\tau$. We can also construct product observables as $A B$ with
values $A_\tau B_\tau$ in the state $\tau$, or function observables $f(A, B)$
with values $f(A_\tau, B_\tau)$. The set of basis observables is 
assumed to be
``complete'' in the sense that all classical observables can be constructed as
functions of the basis observables. 
For a finite number of states $\tau$ all classical correlations as $A_\tau
B_\tau C_\tau$ are well defined for a complete set of basis observables.
Classical statistics is complete in this sense.
For a setting where only functions of the basis observables are considered, two
states $\tau$ and $\rho$ differ if at least one variable takes different values
in the two states. They are taken to be equal if \textit{all} variables or basis
observables $A, B, \dots$ have the same values $A_\tau = A_\rho$.

Our second postulate or axiom of classical statistics associates to every state
$\tau$ a real number $p_\tau$, the ``probability'' of the state $\tau$. It obeys
two basic requirements,
\begin{equation}\label{eq:OP1}
p_\tau \geq 0\, , \quad \sum_\tau p_\tau = 1\, .
\end{equation}
The probabilities obey the axioms discussed in
sect.~\ref{subsec:axiomatic_setting}.
The probabilities are continuous real numbers. They are typically not in the set
of observables -- in general probabilities cannot be measured or observed. For
example, the possible measurement values of the basic observables may be
discrete, say occupation numbers or bits that only take the values $1$ and $0$.
The probabilities $p_\tau$ are continuous real numbers in the interval $[0,1]$.
This ``duality'' between discrete values of observables and continuous
probabilities will be found to be at the root of particle-wave duality in
quantum mechanics.

An observable is called ``discrete'' if the ``spectrum'' of its values $A_\tau$
is discrete. Here the spectrum is the ensemble of the values $A_\tau$ in the
different states $\tau$. For a finite number of discrete variables the states
$\tau$ form a finite discrete ensemble. The sum over the states $\sum_\tau$ in
eq.~\eqref{eq:OP1} is then well-defined. We will define continuous variables as
suitable limits of an infinite set of discrete variables. This is similar to the
``binning of real numbers'' by representing them by an infinite number of bits.
We can then define $\sum_\tau$ for an infinite number of states by a suitable
limit. For continuous variables the ensemble of states is continuous. For
continuous states $\tau$ the sum $\sum_\tau$ becomes an integral over states,
corresponding to the limit procedure.

\paragraph*{Expectation values}

We define the expectation value $\langle A \rangle$ of an observable by the
basic relation of classical statistics
\begin{equation}\label{eq:OP2}
\langle A \rangle = \sum_\tau p_\tau A_\tau\, .
\end{equation}
This includes ``composite observables'', as the classical correlation function
$\langle A B \rangle_{cl}$,
\begin{equation}\label{eq:OP3}
\langle A B \rangle_{cl} = \sum_\tau p_\tau A_\tau B_\tau\, .
\end{equation}
Even though properly speaking the relation~\eqref{eq:OP2} is only a definition
one may call it the third axiom of classical statistics.
All results of this work will be based only on the existence of observables with
values $A_\tau$, the existence of a ``probability distribution'' $\{p_\tau\}$,
which is the ensemble of probabilities for the states $\tau$ obeying
eq.~\eqref{eq:OP1}, and the basic definition of expectation values
\eqref{eq:OP2}. In particular, no new axioms will be introduced for quantum
mechanics. The axioms of quantum mechanics will be \textit{derived} from the
three axioms of classical statistics.

At this stage the two axioms only postulate the existence of the basic objects
of classical statistics, namely the values of observables $A_\tau$ and the
probabilities $p_\tau$. Neither a connection between probabilities and the
outcome of a series of measurements, nor an interpretation of probabilities as a
lack of knowledge for deterministic systems, is assumed here. Probabilities are
simply a basic concept for the formulation of a physical theory. We may later
add postulates about ``ideal measurements''. One such postulate could be that
the possible outcomes of an ideal measurement of the observable $A$ are only the
values $A_\tau$ in its spectrum. 
We will often implicitly assume this postulate by calling $A_\tau$ the
``possible measurement values" of the observable $A$.
Another postulate could be that a sequence of ``identical ideal measurements''
results in an outcome for which the mean over all measurements converges towards
the expectation value $\langle A \rangle$ as the number of such measurements
goes to infinity. We emphasize that for the structural relations developed in
this work the explicite connection to measurements is not needed. It may be
added later as a ``physical interpretation'' of the structures found.

\paragraph*{Weight distribution}

It is often convenient to cast the probabilistic information into a ``weight
function'' or ``weight distribution''. For classical statistics the weights
$w_\tau$ for the states $\tau$ are positive real numbers,
\begin{equation}\label{eq:OP4}
w_\tau \geq 0\, ,
\end{equation}
but the weight distribution is not necessarily normalized. We define the
``partition function'' $Z$ by a sum over all weights
\begin{equation}\label{eq:OP5}
Z = \sum_\tau w_\tau\, .
\end{equation}
A weight distribution defines a probability distribution by
\begin{equation}\label{eq:OP5A}
p_\tau = Z^{-1} w_\tau\, ,
\end{equation}
such that $\{ p_\tau \}$ obeys the criteria \eqref{eq:OP1}. Expectation values
are given in terms of the weight function by
\begin{equation}\label{eq:OP6}
\langle A \rangle = Z^{-1} \sum_\tau w_\tau A_\tau\, .
\end{equation}

\subsubsection{Ising spins, occupation numbers or classical bits}\label{sec:Ising_spins_occupations_numbers_or_classical_bits}

The simplest type of variables are Ising spins. An Ising spin $s$ can only take two values, $s=1$ and $s=-1$. It corresponds to some type of yes/no decision for characterizing some property, $s=1$ for yes and $s=-1$ for no. It can be a macroscopic variable corresponding, for example, to the decision if a neuron fires or not, if a particle hits a detector or not, if some observable quantity is above a certain threshold or not.

Ising spins may also be the fundamental microscopic quantities on which more complex macroscopic structures are built. One may take the attitude that everything that is observable must admit some type of discrete description. If we say that a particle has a position $\bm{x}$, with $\bm{x}$ a continuous variable, we imagine detectors that are able to specify if the particle is in a certain region around $\bm{x}$ or not -- again a yes/no decision. Within a given range and precision a real number can be described by a certain number of yes/no decisions. We use this for the bit representation of real numbers in computers. If the range extends to infinity, or the precision approaches zero, the number of bits needed goes to infinity. Admitting an infinite number of Ising spins the formulation in terms of discrete variables is actually not a restriction.

\paragraph*{Ising variables}

We  will base our general treatment of probabilistic theories on Ising spins as fundamental building blocks. They are the variables or basis observables whose possible values specify the state. There is no need to specify if these are macroscopic variables or the most fundamental microscopic variables. Since we only use probability distributions for some number of Ising spins -- this number may be infinite -- the methods and results will not depend on the physical meaning of these Ising spins. As an important advantage of the  formulation in terms of Ising spins, the discreteness of possible measurement values is built in from the beginning.

Ising spins can be directly associated to bits or fermionic occupation numbers $n$ that can only take the values one or zero,
\begin{equation}\label{eq:IS1}
n = \frac{s + 1}{2} \, , \quad s = 2n-1\, .
\end{equation}
Our probabilistic models will include a probabilistic treatment of classical computing. Deterministic changes of bit configurations will appear as limiting cases of a more general probabilistic approach. For deterministic operations the transition probabilities from one bit configuration to the next one are either one or zero. Computational errors induce transition probabilities that are not exactly one or zero. The association of Ising spins to bits will also allow for an information-theoretic interpretation of the structures that we will find\,\cite{SHA}.

Fermions are a basic building block for elementary particle physics and quantum physics. In quantum field theory or many-body physics they are characterized by occupation numbers that can only take the values $n = 1, 0$. Our treatment of Ising spins can be viewed as a treatment of fermions in the occupation number basis. Probability distributions for Ising spins can be mapped to integrals over Grassmann variables. This very general bit-fermion map \cite{CWFCS,CWQFFT,CWFIM} will allow us to recover the properties of models for fermions based on Grassmann functional integrals. In this sense fermions are not particular ``quantum objects''. They can be taken as the basic building blocks of classical statistical models. 

\paragraph*{Classical states}

The probabilistic description of a single Ising spin involves two ``classical states'' $\tau$, $\tau =1,2$, with $\tau = 1$ denoting $s=1$ or $n=1$, and $\tau = 2$ labeling the state with $s=-1$ or $n=0$. Out of the two positive probabilities $p_1$ and $p_2$ only one is independent since the normalization implies $p_2 = 1 - p_1$. For two Ising spins $s_k$, $k=1,2$, one has four states $\tau=1, \dots, 4$. We may take for $\tau = 1$ the state where both spins are ``up'', i.e. $s_1 = s_2 = 1$ or $\ket{\ua\ua}$. The state with $s_1 = 1$, $s_2 = -1$ or $\ket{\ua\da}$ is labeled by $\tau=2$, while $s_1 = -1$, $s_2=1$ or $\ket{\da\ua}$ corresponds to $\tau = 3$. Finally, $\tau_4$ denotes the state with both spins down or $\ket{\da\da}$. This type of labeling can be extended to the $2^M$ states $\tau = 1, \, \dots, 2^M$ for $M$ spins $s_\gamma$, $\gamma = 1, \dots, M$. In terms of occupation numbers for three spins the labeling of the eight states is shown in table~\ref{tab:1}. 
\begin{table}[t!]
	\centering
	\caption[]{labeling of states for three occupation numbers}
	\label{tab:1}
	\makegapedcells
	\setlength\tabcolsep{3pt}
	\vspace{3mm}
	\begin{ruledtabular}
		\begin{tabular}{ c | c  c  c  c  c  c  c  c }
			$\tau$ & $1$ & $2$ & $3$ & $4$ & $5$ & $6$ & $7$ & $8$ \\ \hline
			$(n_1,\, n_2, \, n_3)$ & $111$ & $110$ & $101$ & $100$ & $011$ & $010$ & $001$ & $000$ 
			\\ \hline
			$N$ & $7$ & $6$ & $5$ & $4$ & $3$ & $2$ & $1$ & $0$ \\
		\end{tabular}
	\end{ruledtabular}
\end{table}
There we also show the integer $N$ that can be associated to the sequences of three bits in the usual binary basis. This generalizes to arbitrary $M$, with $\tau = 2^M - N$. If we consider $2^M$ integers in the interval $[0,\, 2^M -1]$, the first bit is related to the question if the number is in the upper half of this interval, with $s_1 = 1$ if yes. The second bit divides each of the two half-intervals again into two intervals, and so on. Adding additional bits permits an extension of the range or, alternatively, a finer and finer resolution. The labeling is, of course, an arbitrary convention.

The number of states grows very rapidly with $M$. Already a modest $M$, say $M=64$, can account for very large integers or real numbers with a precision that will be sufficient for most purposes. In practice the limit of infinitely many spins, that we will often encounter, can be realized by large finite $M$ with a reasonable size.

Instead of labeling the states by $\tau$ we often use directly the spin configurations $\{ s_\gamma \} = \{ s_1,\, s_2,\, \dots s_M\}$. A spin configuration is an ordered set of values for the spins $s_\gamma$, expressed by $M$ numbers $1$ or $-1$. Each spin configuration corresponds to a possible classical state or a given label $\tau$. For example, for $M=3$ and $\tau=3$ the spin configuration is $\{ s_\gamma \} = \{ 1,\, -1,\, 1 \}$. The corresponding configuration of occupation numbers reads $\{n_\gamma\} =  \{ 1,\, 0,\, 1  \}$, cf. table \ref{tab:1}. We will equivalently use the notations
\begin{equation}\label{eq:IS2}
\tau \; \leftrightarrow \; \{s_\gamma \} \; \leftrightarrow \; \{n_\gamma \}
\end{equation}
and for the probabilities
\begin{equation}\label{eq:IS3}
p_\tau \; \leftrightarrow \; p[s] \; \leftrightarrow \; p[n]\, .
\end{equation}
Here we use a notation familiar from functional integrals, i.e. $p[s] \equiv p(\{ s_\gamma\})$ associates to each spin configuration a probability. In this notation $p[s]$ can be viewed as a function of $M$ discrete variables $s_\gamma$. The sum over configurations is denoted by
\begin{equation}\label{eq:IS4}
\sum_\tau = \int \mathcal{D}s = \prod_{\gamma=1}^M \left( \sum_{s_\gamma = \pm 1} \right)\, .
\end{equation}
Again, the notation resembles functional integrals. We will later define functional integrals as limits of sums over spin configurations for an infinite number of spins.

\paragraph*{Observables for Ising spins}

Possible observables take a real value $A_\tau$ in every state $\tau$. We can write them as real functions $A[s]$ of the discrete variables $s_\gamma$. In this language the expectation value reads
\begin{equation}\label{eq:IS5}
\langle A \rangle = \sum_\tau p_\tau A_\tau = \int \mathcal{D}s\; p[s]\, A[s]\, .
\end{equation}
Similarly, the classical correlation function for two observables $A$ and $B$ reads
\begin{equation}\label{eq:IS6}
\langle AB \rangle = \int \mathcal{D}s\, p[s]\, A[s]\, B[s]\, .
\end{equation}

For a finite number $M$ of Ising spins any observable $A[s]$ is a finite polynomial. This follows from the relation $s_\gamma^2 = 1$. For every term in the polynomial each given Ising spin can either be present or absent. An observable can be written as a linear combination of basis observables in the correlation basis. The basis observables are the possible products of Ising spins. There are $2^M$ different basis observables (including unity), in one-to-one correspondence with the $2^M$ probabilities $p_\tau$.
\subsubsection{Unique jump chains}\label{sec:unique_jump_chains}

We will next discuss simple probabilistic systems for Ising spins. We will label Ising spins with $s_k(m)$ with $m$ an integer
used to order the Ising spins partially. We discuss probability distributions consisting of factors which involve each only 
Ising spins of neighboring layers $m$ and $m+1$. This choice will later permit a simple realisation of time as a structure between
observables. We start with a particularly simple example, the unique jump chains. They can be associated to automata or 
a deterministic evolution.

\paragraph*{Local factors}
Let us consider $\mathcal{M}+1$ Ising spins $s(m)$ on a chain labeled by integers $0 \leq m \leq \mathcal{M}$. We start with a very simple probability distribution
\begin{equation}\label{eq:UJ1}
p[s] = p[\{s(m)\}] = Z^{-1} w[s]\, ,
\end{equation}
with 
\begin{equation}\label{eq:UJ2}
w[s] = \left( \prod_{m=0}^{\mathcal{M}-1} \delta \big( s(m+1) - s(m) \big) \right)
\mathscr{B}(s_f,\, s_{in})\, .    
\end{equation}
The $\delta$-function,
\begin{align}\label{eq:UJ3}
\delta\big( s(m+1) - s(m) \big) &= \delta_{s(m+1),\, s(m)} \notag \\
&= 
\begin{cases}
1 \text{ for } s(m+1) = s(m) \\
0 \text{ for } s(m+1) \neq s(m)
\end{cases} ,
\end{align}
implies that $p[s]$ only differs from zero if $s(m+1)$ equals $s(m)$ for all $m$, such that all Ising spins $s(m)$ must be equal. The boundary term $\mathscr{B}(s_f,\, s_{in}) \geq 0$ only involves the ``initial spin'' $s_{in} = s(m=0)$ and the ``final spin'' $s_f = s(m=\mathcal{M})$ on the chain. The partition function is trivial since only configurations with all $s(m)$ equal contribute,
\begin{equation}\label{UJ4}
Z = \int \dif s_{in}\; \mathscr{B} (s_{in},\, s_{in}) = \sum_{s_{in} = \pm 1} 
\mathscr{B}(s_{in},\, s_{in})\, .
\end{equation}
In general, the final and initial spins can be in four combinations $(s_f,\, s_{in}) = (+,+)$, $(+,-)$, $(-,+)$, $(-,-)$, with associated boundary coefficients $\mathscr{B}_{++}$, $\mathscr{B}_{+-}$, $\mathscr{B}_{-+}$, and $\mathscr{B}_{--}$. The coefficients $\mathscr{B}_{+-}$ and $\mathscr{B}_{-+}$ do not matter, and $Z = \mathscr{B}_{++} + \mathscr{B}_{--}$. All spins are up with the probability $\mathscr{B}_{++}/Z$, and down with probability $\mathscr{B}_{--}/Z$. Expectation values of observables are easily computed with this information. Only states with all spins equal contribute in the configuration sum. 

The  probability distribution \eqref{eq:UJ1},~\eqref{eq:UJ2} can be expressed as a product of ``local factors'' $\mathscr{K}(m)$ which depend only on the spins $s(m)$ and $s(m+1)$,
\begin{equation}\label{eq:UJ5}
w[s] = \prod_{m=0}^{\mathcal{M}-1} \mathscr{K}(m)\, \mathscr{B}\, ,
\end{equation}
with 
\begin{equation}\label{UJ6}
\mathscr{K}(m) = \delta\big( s(m+1) - s(m)\big)\, .
\end{equation}
The boundary term $\mathscr{B}$ appears as an additional factor. One can write the local factor as
\begin{equation}\label{eq:UJ7}
\mathscr{K}(m) = \lim_{\beta \to \infty} \exp  \big\{ \beta\big( s(m+1)\, s(m) - 1 \big) \big\}\, .
\end{equation}
If $s(m+1)$ equals $s(m)$ the exponent is zero and $\mathscr{K}(m)=1$, while for $s(m+1)$ different from $s(m)$ one has $\lim_{\beta\to \infty} \exp(-2\beta) = 0$ .

\paragraph*{Local action}
Since $\mathscr{K}(m) \geq 0$ for all $m$, we can write the probability distribution in terms of an action $\mathcal{S}$,
\begin{equation}\label{eq:UJ8}
w[s] = \exp\big\{ - \mathcal{S}[s] \big\}\, \mathscr{B}\, ,
\end{equation}
with
\begin{equation}\label{eq:UJ9}
\mathcal{S}[s] = \sum_{m=0}^{\mathcal{M}-1} \mathcal{L}(m)\, ,
\end{equation}
and
\begin{equation}\label{eq:UJ10}
\mathcal{L}(m) = - \lim_{\beta\to\infty} \beta\big( s(m+1)\, s(m) - 1\big)\, .
\end{equation}
Since only two neighboring spins are connected, this is called a ``next neighbor interaction''. For next neighbor interactions the action is ``local''.

We may consider a different probability distribution with opposite sign of the next neighbor interaction,
\begin{equation}\label{eq:UJ11}
\mathcal{L}(m) = \lim_{\beta\to\infty} \beta \big( s(m+1)\, s(m) + 1\big)\, .
\end{equation}
The local factor is now given by 
\begin{equation}\label{eq:UJ12}
\mathscr{K}(m) = \delta\big( s(m+1) + s(m) \big) \, .
\end{equation}
Nonzero probabilities arise only for a small subset of the possible spin configurations: whenever the spin $s(m)$ is positive, the neighboring spin $s(m+1)$ has to be negative. The spins have to flip from one side to the next one. The ``allowed configurations'' with nonzero probabilities can be characterized by a ``propagation of spins''. A given spin at site $m$ has only a unique possibility to propagate to the site $m+1$: it has to change its sign. Probability distributions where for every spin configuration at $m$ the neighboring spin configuration at $m+1$ is uniquely determined are called ``unique jump chains''. Here chain refers to the ordering of $m$.

\paragraph*{Unique jump chains for three Ising spins}
More possibilities arise if one places more than a single spin on every site $m$. As an example, we may consider three Ising spins at every site $m$, $s_k(m) = \pm 1$, $k=1,2,3$. For $\mathcal{L}(m)$ one may consider
\begin{align}\label{eq:UJ13}
\mathcal{L}_H(m) = \lim_{\beta\to\infty} \beta \big\{ &s_2(m+1)\, s_2(m) - s_3(m+1)\, s_1(m) \notag \\
& - s_1(m+1)\, s_3(m) + 3 \}\, .
\end{align}
For this unique jump chain the spin $s_2$ has to change its sign when moving from $m$ to $m+1$, and the two spins $s_1$ and $s_3$ are exchanged,
\begin{equation}\label{eq:UJ14}
V_H\, : \quad s_2 \rightarrow - s_2\, , \quad s_1 \rightarrow s_3\, , \quad s_3 \rightarrow s_1\, .
\end{equation}
Only spin configurations that obey the rule \eqref{eq:UJ14} for neighboring sites contribute to expectation values of observables. We will later associate this unique jump operation with the Hadamard gate in a quantum subsystem for which the three Ising spins $s_k$ are associated to the three cartesian directions of a single qubit or quantum spin. 

Another choice could be 
\begin{align}\label{eq:UJ15}
\mathcal{L}_{12}(m) = \lim_{\beta\to\infty} \beta \big\{& s_1(m+1)\, s_2(m) - s_2(m+1)\, s_1(m) \notag \\
&- s_3(m+1) \, s_3(m) +3 \big\}\, .
\end{align}
The unique jump corresponds to a rotation between the spins $s_1$ and $s_2$, leaving $s_3$ invariant,
\begin{equation}\label{eq:UJ16}
V_{12}\, : \quad s_1 \rightarrow s_2\, , \quad s_2 \rightarrow - s_1\, , \quad s_3 \rightarrow s_3\, ,
\end{equation}
which stands for $s_1(m+1) = -s_2(m)$, $s_2(m+1) = s_1(m)$, $s_3(m+1) = s_3(m)$.

\paragraph*{Products of unique jumps}
There is no need that the action $\mathcal{S}$ in eq.~\eqref{eq:UJ9} has for every $m$ the same $\mathcal{L}(m)$. For example, we may consider a situation where $\mathcal{L}(m) = \mathcal{L}_H(m)$ for $m$ even, and $\mathcal{L}(m) = \mathcal{L}_{12} (m)$ for $m$ odd. Starting from some even $m$, the propagation of spins undergoes first the transformation $V_H$, and subsequently the transformation $V_{12}$. The combined propagation from $m$ to $m+2$ corresponds to 
\begin{equation}\label{eq:UJ17}
V_{12}V_H\, : \quad s_1 \rightarrow s_3\, , \quad s_2 \rightarrow s_1\, , \quad s_3 \rightarrow s_2\, .
\end{equation}
Correspondingly, we can define a combined local factor
\begin{align}\label{eq:UJ18}
\hat{\mathscr{K}} (m) &= \int \dif s (m+1)\, \mathscr{K}(m+1)\, \mathscr{K}(m) \notag \\
&= \delta\big( s_3(m+2) - s_1(m)\big)\, \delta\big( s_1 (m+2) - s_2(m) \big) \notag \\
& \quad \times \delta \big( s_2(m+2) - s_3(m) \big) \, .
\end{align}
It involves the spins at even sites $m$ and $m+2$, while the spins at the intermediate odd site $m+1$ is ``integrated out'', with
\begin{equation}\label{eq:UJ19}
\int \dif s (m+1) = \prod_k \sum_{s_k (m+1) = \pm 1}\, .
\end{equation}

\paragraph*{Coarse graining}
On the level of the action this sequence of two unique jumps amounts to a combined term $\hat{\mathcal{L}} (m)$, defined by
\begin{align}\label{eq:UJ20}
\exp \big\{ - \hat{\mathcal{L}}(m) \big\} = \int &\dif s(m+1) \, \exp \big\{ - (\mathcal{L}_H (m) \notag \\
& + \mathcal{L}_{12} (m+1)) \big\} \, .
\end{align}
Indeed, evaluating explicitly the r.h.s. of eq.~\eqref{eq:UJ20} yields
\begin{align}\label{eq:UJ21}
& \exp \big\{ - \hat{\mathcal{L}}(m) \big\} = \int \dif s(m+1) \notag \\
& \quad  \exp \big\{ - \beta \big[ s_1(m+2) \, s_2(m+1) - s_2(m+2)\, s_1(m+1) \notag \\
& \qquad \quad \, \, \, -s_3(m+2) \, s_3(m+1) + s_2(m+1)\, s_2(m) \notag \\
& \qquad \quad \, \, \, - s_3(m+1)\, s_1(m) - s_1 (m+1)
\, s_3(m) + 6 \big] \big\} \notag \\
&= \exp \big\{ 2\beta\, \big[ s_3 (m+2)\, s_1(m) + s_1(m+2)\, s_2(m) \notag \\
& \qquad \qquad \quad  + s_2(m+2)\, s_3(m)
-3 \big] \big\} \, .
\end{align}
Here we use for $\beta\to\infty$ the identity for Ising spins 
\begin{equation}\label{eq:UJ22}
\sum_{s' = \pm 1} \exp \big\{ - \beta [ (s'' + s) s' +2 ] \big\} 
= \exp \big\{ 2\beta [ s'' s - 1] \big\} \, .
\end{equation}
The factor $\hat{\mathscr{K}}(m) = \exp \{ - \hat{\mathcal{L}}(m)\}$ accounts indeed for the propagation \eqref{eq:UJ17}. 

For an even number $\mathcal{M}$ of sites on the chain we can integrate out all spins on odd sites. The action is a sum over $\hat{\mathcal{L}}(m)$ at all even sites $m$. The propagation of spins from an even site to the next even site is given by the operation $V_{12} V_H$ in eq.~\eqref{eq:UJ17}. This procedure amounts to a ``coarse graining'' of the action and the associated probability distribution. We can define a new ``coarse-grained'' probability distribution that depends only on the spins at even sites
\begin{equation}\label{eq:UJ22A}
\hat{p}[s] = Z^{-1}\, \hat{w}[s]\, , \quad Z = \int \mathcal{D} s_{\text{even}} \hat{w}[s]\, ,
\end{equation}
with
\begin{equation}\label{eq:UJ22B}
\hat{w}[s] = \exp ( - \hat{\mathcal{S}}[s] )\, \mathscr{B},
\end{equation}
and
\begin{equation}\label{eq:UJ22C}
\hat{\mathcal{S}}[s] = \sum_m \hat{\mathcal{L}}(m)\, .
\end{equation}
All coarse-grained quantities (with a hat) depend only on the spins at even sites, and the configuration sum $\int \mathcal{D} s_{\text{even}}$ sums only over configurations for this restricted set of spins. Formally, the coarse-grained weight function $\hat{w}(s)$ obtains by a sum (or ``integration'') $\int \mathcal{D} s_{\text{odd}}$ over the configurations of spins at odd sites,
\begin{equation}\label{eq:UJ22D}
\hat{w}[s] = \int \mathcal{D} s_{\text{odd}} w[s]\, ,
\end{equation}
such that 
\begin{equation}\label{eq:UJ22E}
Z = \int \mathcal{D}s\, w(s) = \int \mathcal{D} s_{\text{even}}\int \mathcal{D} s_{\text{odd}}\, w[s] = \int \mathcal{D} s_{\text{even}}\, \hat{w} [s]\, .
\end{equation}
The expectation values of observables that involve only spins at even sites can be computed from the coarse-grained probability distribution.

\paragraph*{Non-commutativity}
The operations $V_{12}$ and $V_H$ do not commute. Indeed, one finds for the other order
\begin{equation}\label{eq:UJ23}
V_H V_{12}\, : \quad s_1 \rightarrow - s_2\, ,\quad s_2 \rightarrow -s_3 \, , \quad
s_3 \rightarrow s_1 \, .
\end{equation}
This clearly differs from the transformation $V_{12} V_H$ in eq.~\eqref{eq:UJ17}. The importance of the order of transformations gives a first glance at the presence of non-commutative aspects in classical statistics. The operations $V_{12}$ and $V_H$ have an inverse. From $V_H^2 = 1$ one finds $V_H^{-1} = V_H$. The inverse transformation of $V_{12}$ is given by 
\begin{equation}\label{eq:UJ23A}
(V_{12})^{-1}\, : \quad s_1 \rightarrow - s_2\, , \quad s_2 \rightarrow s_1 \, ,
\quad s_3 \rightarrow s_3\, .
\end{equation}
Products are defined by composition, as for example $V_H V_{12}$ or
\begin{equation}\label{eq:UJ23B}
(V_H V_{12})^2\, : \quad s_1 \rightarrow s_3\, , \quad s_2 \rightarrow - s_1 \, 
, \quad s_3 \rightarrow
- s_2\, ,
\end{equation}
and
\begin{equation}\label{eq:UJ23C}
(V_H V_{12})^3 = 1\, ,
\end{equation}
with $1$ the unit transformation. The spin transformations form the non-commutative group of permutations of three elements $P_3$, augmented by sign changes of the spins.

Unique jump chains can represent transformations beyond the permutations and sign changes of spins. Denoting by $\rho$ the $2^M$ configurations of $M$ spins $s_\gamma(m)$, and by $\tau$ the $2^M$ configurations of spins $s_\gamma (m+1)$, any transformation $\rho \to \tau(\rho)$ maps each configuration at $m$ to a configuration at $m+1$. The invertible unique jump operators form the group of permutations of $2^M$ elements. General unique jump transformations contain conditional transformations. For our example of three spins $s_k(m)$ the transformations
\begin{align}\label{eq:UJ23D}
\tau(1)&=1\, , \quad \tau(2) = 3\, , \quad \tau(3) = 2\, , \quad \tau(4) = 4\, , \notag \\
\tau(5) &= 5\, , \quad \tau(6) = 6\, , \quad \tau(7) = 7\, , \quad \tau(8) = 8
\end{align}
corresponds to an exchange of $s_2$ and $s_3$ if $s_1 = 1$, while under the condition $s_1 = -1$ all spins remain invariant.

\paragraph*{Probabilistic automata and deterministic\\computing}

The unique jumps describing the propagation of spins from one site to the next are completely deterministic. They describe automata. In particular, together with certain locality properties in some other quantity as space, they realize cellular automata\,\cite{WOLZ,TH,TH2,EL, Banks:2020wul, Pizzi_2022}.
Cellular Automata can be realized by classical computers or quantum systems\,\cite{LTP,CWEL,NP,ALL,ICJ, Banchi_2017}.
Our formulation of the unique jumps in terms of an action permits us to deal with the statistics of automata. Probabilistic aspects are only introduced by the boundary term $\mathscr{B}(s_f, s_{in})$, while the propagation of every individual spin configuration to sites with larger $m$ is purely deterministic. For the boundary term we may take a direct product form
\begin{equation}\label{eq:UJ24}
\mathscr{B}(s_f, s_{in}) = \mathscr{B}_f(s_f) \, \mathscr{B}_{in} (s_{in})\, .
\end{equation}
With open boundary condition at the final site, $\mathscr{B}_f = 1$, the relative probabilities of the different spin configurations are determined by $\mathscr{B}_{in}$. For the three initial spins $s_{k, \, in} = s_k (0)$ there are eight possible configurations $\tau$. Each initial configuration propagates to larger $m$ according to the deterministic rule of the cellular automaton. We only need the probabilities $p_\tau$ for the different initial spin configurations. Using the $\delta$-functions in the local factors $\mathscr{K}(m)$ we can integrate out all spins at sites $m \geq 1$. The probabilities for the initial configurations are then determined by 
\begin{equation}\label{eq:UJ25}
p_\tau = Z^{-1} \mathscr{B}_\tau\, , \quad Z = \sum_\tau \mathscr{B}_\tau\, .
\end{equation}
Here $\mathscr{B}_\tau$ is the value that $\mathscr{B}_{in}(s_{in})$ takes for the different initial configurations $\tau$.

Standard deterministic computing is a particular case of automata for which the initial spin configuration is uniquely fixed. The bits with values $n=1, 0$ are directly related to the Ising spins by $n= (s+1)/2$. The initial configuration of the three spins $s_{k,\, in}$ is uniquely fixed by $\mathscr{B}_{\tau_0}$ = 1 for the given initial configuration $\tau_0$, and $\mathscr{B}_\tau = 0$ for $\tau \neq \tau_0$. With $\mathscr{B}_f = 1$ the final spin configuration $\{ s_{k,\, f} \}$ is the result of the processing of the initial spin configuration $\tau = \{ s_{k,\, in}\}$. With observables placed at the final site $m=\mathcal{M}$ one can read out the result of the computation.

The advantage of the probabilistic description of automata is that many methods of statistical physics can be implemented directly, as coarse graining or the systematic investigation of correlations and the associated generating functionals. Furthermore, the formalism is easily extended to non-perfect computations for which a unique jump is only performed with a certain error. This occurs for finite $\beta$, a case to which we will turn next.
For a very large number of initial spins deterministic initial conditions are no longer realistic. One rather has to deal with an initial probability distribution for the configurations of initial spins. This is the case if we want to describe the Universe by a cellular automaton -- the number of initial spins is infinite. In such a description the Universe would be a \textit{probabilistic} cellular automaton.

\subsubsection{Local chains}\label{sec:local_chains}

For local chains the weight function can be written as a product of local factors similar to eq.\,\eqref{eq:UJ5}. These local factors $\mathscr{K}(m)$
only involve Ising spins at neighboring layers $m$ and $m+1$. They form the basis of our discussion of probabilistic systems. Local
chains describe a very large class of systems. For most of the developments in this work there will be no need to go beyond the 
setting of local chains.

\paragraph*{Ising chain}

The one-dimensional Ising model or Ising chain is one of the best known and understood models in classical statistics. Originally  developed for the understanding of magnetic properties, it has found wide applications in various branches of science. The probability distribution is given by an action with next-neighbor interactions,
\begin{equation}\label{eq:LC1}
\mathcal{S} = \sum_{m=0}^{\mathcal{M} - 1} \mathcal{L}(m)\, , \quad 
\mathcal{L}(m) = \beta\, \big(\kappa\, s(m+1)\, s(m) + 1 \big)\,,
\end{equation}
as
\begin{equation}\label{eq:LC2}
p[s] = Z^{-1}\, w[s]\, , \quad w[s] = \exp( -\mathcal{S} ) \, \mathscr{B}\, ,
\end{equation}
with boundary term $\mathscr{B}$ depending on $s_{in}$ and $s_f$ and $Z = \int\mathcal{D} s \, w[s]$. We take $\beta > 0$ and choose a normalization such that $\kappa = \pm 1$. For $\kappa= -1$ the interaction is ``attractive'' and configurations with aligned spins are favored, similar to ferromagnets. The ``repulsive interaction'' for $\kappa = + 1$ yields higher probabilities if spins at neighboring sites have opposite signs, resembling antiferromagnets. For $\beta\to \infty$ we recover the trivial unique jump chain \eqref{eq:UJ1} - \eqref{eq:UJ10} for $\kappa = -1$, and the alternating unique jump chain \eqref{eq:UJ11}, \eqref{eq:UJ12} for $\kappa = 1$. For the Ising model the local factors are given by
\begin{equation}\label{eq:LC3}
\mathscr{K}(m) = \exp\big( - \mathcal{L}(m)\big)\, ,
\end{equation}
with $w[s]$ given by eq.~\eqref{eq:UJ5}.

\paragraph*{Local chains}

We want to generalize the Ising model to general ``local chains'', which are defined by
\begin{equation}\label{eq:LC4}
w[s] = \prod_{m=0}^{\mathcal{M} - 1} \mathscr{K} (m)\, \mathscr{B}\, ,
\end{equation}
with $\mathscr{K}(m)$ depending on spins $s_\gamma (m+1)$ and $s_\gamma(m)$ and $\mathscr{B}$ depending on $s_{\gamma,\, in} = s_\gamma(0)$ and $s_{\gamma,\, f} = s_\gamma (\mathcal{M})$. The local factors $\mathscr{K}(m)$ and the boundary term $\mathscr{B}$ have to be chosen such that $w[s] \geq 0$ for all spin configurations $\{ s_\gamma(m)\}$. For $M$ spins $s_\gamma$ at a given site, $\gamma=1, \dots, M$, and $\mathcal{M}+1$ sites on the chain, $m=0, \dots, \mathcal{M}$, the total number of Ising spins $\mathcal{N}$ is
\begin{equation}\label{eq:LC4a}
\mathcal{N} = M\,(\mathcal{M} + 1)\, ,
\end{equation} 
and the total number of configurations amounts to $2^\mathcal{N}$. The configuration sum or ``functional integral'' reads
\begin{equation}\label{eq:LC5}
\int \mathcal{D}s = \prod_{m=0}^\mathcal{M}\, \prod_{\gamma=1}^M \, \sum_{s_\gamma(m) = \pm 1}\, .
\end{equation}

In this generality local chains do not only cover one-dimensional systems. Two-dimensional systems on a square lattice are described by sites with two integer coordinates $(m_1,\, m_2)$, $m_i = 0, \dots, \mathcal{M}$. Taking $m=m_1$ and $\gamma = m_2 + 1$, $M = \mathcal{M} + 1$, the two-dimensional system takes the form of a local chain if $w[s]$ is of the form \eqref{eq:LC4}. This requires that $w[s]$ can be written as a product of factors that only involve $s(m_1 + 1,\, m_2)$ and $s(m_1,\, m'_2)$. 
The ``internal label'' $\gamma$ becomes the discrete coordinate $m_2$.
There is at this stage no restriction on the dependence on $m_2$, $m'_2$ -- the next-neighbor property is only required in the direction of $m_1$. A system is also local in the $m_2$-direction if $w[s]$ can be written as a product of factors $\mathscr{K}(m_1, m_2)$ involving only $s(m_1 + 1,\, m_2 +1)$, $s(m_1 + 1,\, m_2)$, $s(m_1,\, m_2 + 1)$, and $s(m_1,\, m_2)$. In this case we could equivalently define the chain in the $m_2$-direction. A chain is selected by a choice of a sequence of hypersurfaces. For our example the hypersurfaces are at fixed $m_1$ for the chain in the $m_1$-direction, and at fixed $m_2$ for a chain in the $m_2$-direction. The choice of the hypersurfaces or chain direction is not necessarily determined by properties of the probability distribution. It may be merely a matter of convenience. A generalization to higher-dimensional systems, or more than a simple species of spins at every site, is straightforward by a suitable range of the index $\gamma$.

\paragraph*{Two-dimensional Ising models}

As an example we may consider the two-dimensional Ising model with next neighbor interactions,
\begin{equation}\label{eq:LC6}
\mathcal{S} = \sum_{m_1, \,m_2} \mathcal{L} (m_1, \,m_2)\, ,
\end{equation}
with
\begin{align}\label{eq:LC6A}
\mathcal{L}(m_1, m_2) = \beta\, \big\{ & \kappa\, \big[ s(m_1+1,\, m_2)\, s(m_1,\, m_2) 
\notag \\
& + s(m_1,\, m_2 + 1)\, s(m_1,\, m_2) \big] + 2 \big\}\, ,
\end{align}
and 
\begin{equation}\label{eq:LC7}
w[s] = \text{e}^{-\mathcal{S}}\, \mathscr{B} \, .
\end{equation}
The boundary term should only involve spins on the boundary, i.e. $m_1 = 0,\,\mathcal{M}$ or $m_2 = 0,\, \mathcal{M}$,
\begin{equation}\label{eq:LC8}
\mathscr{B} = \mathscr{B}_1 \big[ s(0,m_2),\, s(\mathcal{M},m_2) \big]\,
\mathscr{B}_2 \big[ s(m_1, 0),\, s(m_1,\, \mathcal{M})\big] \, .
\end{equation}

One could take periodic boundary conditions, e.g.
\begin{equation}\label{eq:LC9}
\cB_2 = \exp \Big\{ - \beta\, \sum_{m_1} \big[ \kappa\, s(m_1, \mathcal{M}) \, 
s(m_1, 0) + 1 \big] \Big\} \, ,
\end{equation}
and similarly for $\cB_1$. In this case one has
\begin{equation}\label{eq:LC10}
w[s] = \prod_{m_1 = 0}^\cM \, \prod_{m_2 = 0}^\cM \, \cK (m_1, m_2)\, ,
\end{equation}
with local factors
\begin{equation}\label{eq:LC11}
\cK (m_1, m_2) = \exp \big\{ - \cL (m_1, m_2) \big\}\, ,
\end{equation}
and $s(\cM +1, m_2) \equiv s(0, m_2)$, $s(m_1, \cM + 1) \equiv s(m, 0)$. For every ``link'' connecting two neighboring sites the weight function contains a factor $\exp \{ - \beta\, (\kappa + 1) \}$ if the two spins at the end of the link are equal, and a factor $\exp \{ \beta\, (\kappa - 1)\}$ if they are different. For $\kappa = \pm 1$ these factors equal either $1$ or $\exp( - 2\beta)$. Thus the probabilities for links with the ``wrong sign'' of the spins at the ends are suppressed.

The Ising model \eqref{eq:LC6A} has only next-neighbor interactions. Diagonal interactions may be described by adding to $\cL (m_1, m_2)$ a term
\begin{align}\label{eq:LC12}
\cL_d(m_1, m_2) &= \beta\, \big\{ \kappa_d\, \big[ s(m_1 + 1,\, m_2 +1)\, 
s(m_1,\, m_2) \notag \\
& \quad + s(m_1,\, m_2 +1)\, s(m_1+1,\, m_2) \big] + 2 \big\}\, .
\end{align}
Already at this stage a solution of the model becomes rather complex. The complexity is enhanced for more general boundary conditions, say for $\cB_1$. It should be clear at this stage that local chains cover a very large variety of probability distributions. They should not be considered as a specific model, but rather as a general setting.

\paragraph*{Generalized Ising chains}

A simple way of ensuring the positivity of the probability distribution is to take all local factors $\cK(m)$ positive, as well as a positive boundary term
\begin{equation}\label{eq:LC13}
\cK(m) \geq 0\, , \quad \cB \geq 0\, .
\end{equation}
There should be at least one configuration for which all $\cK(m)$ and $\cB$ differ from zero, such that $Z > 0$. Local chains with these properties are called ``generalized Ising chains''.

Due to the positivity of $\cK(m)$ and $\cB$ the weight function $w[s]$ can be written in the form of an action
\begin{equation}\label{eq:LC14}
w[s] = \exp \big\{ - \cS - \cS_\cB \big\}\, ,
\end{equation}
with
\begin{equation}\label{eq:LC15}
\cS = \sum_{m=0}^{\cM - 1} \cL (m)\, , \quad \cK(m) = \exp \big\{ - \cL(m) \big\}\, ,
\end{equation}
and
\begin{equation}\label{eq:LC16}
\cB = \exp ( - \cS_\cB ) \, .
\end{equation}
If $\cK(m)$ vanishes for some spin configuration, $\cL(m)$ diverges for this configuration. This configuration does not contribute to the weight function. With zero probability, it is effectively excluded from the configuration sum. Diverging $\cL(m)$ is a convenient way to ``forbid'' certain configurations and to restrict the space of ``allowed configurations''. The same holds for the boundary term, where vanishing $\cB$ for some configuration is realized by diverging $\cS_\cB$.

Generalized Ising chains cover again a very wide range of probability distributions. This holds, in particular, if one considers local factors that differ for different $m$. One can implement ``selection rules'' by diverging factors $\cL_m$. In particular, for the unique jump chains or cellular automata the selection rules are so strong that only one particular spin configuration is allowed for a particular initial boundary configuration. Cellular automata and classical computing are of the type of generalized Ising chains. 
Nevertheless, generalized Ising chains are not the most general local chains. The condition \eqref{eq:LC13} is not necessary for obtaining a positive weight distribution.

Also local chains defined by eq.~\eqref{eq:LC4} are not the only possibility for realizing a locality property. In Appendix~\ref{app:matrix chains} we discuss matrix chains where $\cK (m)$ is replaced by an $n\times n$-matrix, similarly for $\cB$, and a trace is taken. For a particular structure of $2\times 2$-matrices this realizes a complex action, close to Feynman's path integral. Matrix chains can be found by coarse graining of local chains.
\subsubsection{Transfer matrix}\label{sec:transfer_matrix}

In our discussion of the deterministic unique jumps of cellular automata in sect.\,\ref{sec:unique_jump_chains} we have encountered the operations $V_H$ or $V_{12}$ in eqs.\,\eqref{eq:UJ14}, \eqref{eq:UJ16}. They describe how a spin configuration at site $m$ is transformed into a spin configuration at site $m+1$. More generally, this concerns the question how the probabilistic information about the spins $s_\gamma (m)$ is passed over to probabilistic information about the spins $s_\gamma (m+1)$.

We aim for a generalization of this concept to truly probabilistic or non-deterministic systems. This leads us to the transfer matrix\,\cite{TM,MS,FU}
and the step evolution operator\,\cite{CWIT}.
With these concepts the non-commutative structures in classical statistics become very apparent. The transfer matrix can be formulated for arbitrarily local chains.

Since we want to deal with explicit matrices, it is convenient to introduce a basis for functions that depend on discrete spin variables. We will concentrate here on the occupation number basis\,\cite{CWIT,CWQF}.
For other possible choices of basis functions, 
see ref.\,\cite{CWQF}.

\paragraph*{Occupation number basis}

For the choice of a basis it is convenient to switch to occupation numbers $n_\gamma = (s_\gamma + 1)/2$. They can take the values one and zero and obey the relation
\begin{equation}\label{eq:TS1}
n_\gamma^2 = n_\gamma\, .
\end{equation}
For a function  $f[n]$ depending on a single occupation number we define two basis functions
\begin{equation}
h_1[n] = n\, , \quad h_2 [n] = 1 - n\, .
\end{equation}
Due to $n^2 = n$, any arbitrary real function $f[n]$ is linear in $n$ and can be written as
\begin{equation}\label{eq:TS3}
f[n] = q_1\, h_1[n] + q_2\, h_2[n] = q_\tau\, h_\tau [n]\, ,
\end{equation} 
with real coefficients $q_1$ and $q_2$. The basis function $h_1[n]$ equals one in the state $\tau = 1$, and zero in the state $\tau = 2$. Similarly, $h_2[n]$ equals one in the state $\tau = 2$ and zero in the state $\tau = 1$. In view of this close correspondence we label the basis functions $h_\tau [n]$ with the same label as the states $\tau$. (A confusion of $\tau$ labeling basis functions or states should be easily avoided from the context.)

For functions depending on two occupation numbers $n_1$, $n_2$ the four basis functions are defined as
\begin{align}\label{eq:TS4}
h_1[n] = n_1\, n_2\, , \quad &  h_2 [n] = n_1\, (1 - n_2)\, , \notag \\
h_3 [n] = (n_1 - 1)\, n_2\, , \quad & h_4[n] = (1 - n_1)\, (1 - n_2)\, .
\end{align}
They have the property (no sum over $\tau$ here)
\begin{equation}\label{eq:TS5}
n_\gamma\, h_\tau [n] = (n_\gamma)_\tau\, h_\tau [n]\, ,
\end{equation}
with $(n_\gamma)_\tau$ the value of $n_\gamma$ in the state $\tau$. In other words, multiplication of $h_\tau$ by $n_\gamma$ ``reads out'' the value that $n_\gamma$ has in the state $\tau$. This system of basis functions is easily extended to an arbitrary number $M$ of occupation numbers. Every basis function $h_\tau[n]$ is a product of $M$ factors $f_\gamma$, where each factor is either $f_\gamma = n_\gamma$ or $f_\gamma = 1 - n_\gamma$. The factor $n_\gamma$ occurs for all $\tau$ for which $n_\gamma = 1$, and a factor $(1 - n_\gamma)$ is present for those $\tau$ for which $n_\gamma = 0$. We call this system the ``occupation number basis''.

An arbitrary function $f[n]$ can be expanded in these basis functions
\begin{equation}\label{eq:TS6}
f[n] = q_\tau\, h_\tau[n]\, .
\end{equation}
Due to the relation $n_\gamma^2 = n_\gamma$, an arbitrary function $f[n]$ is a sum of terms where each term either contains a given $n_\gamma$ or not. The number of such terms is $2^M$, and they have arbitrary coefficients. This can be reordered to a sum of terms that either contain a factor $n_\gamma$ or a factor $(1 - n_\gamma)$, according to  $an + b = (a + b) n + b\, (1 - n)$. After this reordering the relation \eqref{eq:TS6} is obvious.

The basis functions $h_\tau [n]$ obey several important relations. The multiplication rule,
\begin{equation}\label{eq:TS7}
h_\tau[n]\, h_\rho [n] = \delta_{\tau \rho} \, h_\tau [n]\, ,
\end{equation}
follows from $n_\gamma\, (1 - n_\gamma) = 0$. As a consequence, $h_\tau h_\rho$ can differ from zero only if all factors $f_\gamma$ in $h_\tau$ and $h_\rho$ are the same. The summation rule,
\begin{equation}\label{eq:TS8}
\sum_\tau h_\tau [n] = 1\, ,
\end{equation}
results from the identity $n_\gamma + (1 - n_\gamma) = 1$. For $\gamma = M$ the basis functions can be divided into pairs. For each pair the factors $f_\gamma$ are equal for all $\gamma < M$. Out of the two members of a given pair one has a factor $n_M$, and the other a factor $(1 - n_M)$. Taking for each pair the sum of the two members one remains with a system of basis functions for $M-1$ occupation numbers. This can be done consecutively for $\gamma = M-1$ and so on, proving eq.~\eqref{eq:TS8} by iteration.

Furthermore, we have the integration rule
\begin{equation}\label{eq:TS9}
\int \cD n \, f_\tau [n] = 1\, .
\end{equation}
It follows from the simple identities
\begin{equation}\label{eq:TS10}
\sum_{n=0,1} n = 1\, , \quad \sum_{n=0,1} (1 - n) = 1\, .
\end{equation}
Since every $f_\tau$ has for each $\gamma$ either a factor $n_\gamma$ or $1 - n_\gamma$ and
\begin{equation}\label{eq:TS11}
\int \cD n = \prod_\gamma \, \sum_{n_\gamma = 0,1}\, ,
\end{equation}
eq.~\eqref{eq:TS9} follows by use of eq.~\eqref{eq:TS10} for every $n_\gamma$. We can combine eqs~\eqref{eq:TS7} and \eqref{eq:TS9} in order to establish the orthogonality relation
\begin{equation}\label{eq:TS12}
\int \cD n \, h_\tau [n]\, h_\rho [n] = \delta_{\tau \rho}\,.
\end{equation}
The orthogonality relation implies for any function $f[n]$ according to eq.~\eqref{eq:TS6} the relation
\begin{equation}\label{eq:TS13}
q_\tau = \int \cD n \, h_\tau [n]\, f[n]\,.
\end{equation}

Finally, the completeness relation reads
\begin{equation}\label{eq:TS14}
h_\tau [n] \, h_\tau[n'] = \delta [n - n']\, ,
\end{equation}
where $\delta[n - n']$ equals one if two configurations $\{ n_\gamma\}$ and $\{ n'_\gamma \}$ coincide, and is zero otherwise. For any given configuration $\{ n_\gamma \}$ a given basis function $h_\tau [n]$ takes either the value one or zero, depending on whether $\tau$ coincides with $\{ n_\gamma \}$ or not. Two different configurations $\{ n_\gamma\}$ and $\{ n'_\gamma \}$ differ in at least one occupation number $\bar{n}_\gamma$. Thus for every $\tau$, $h_\tau [n]$ and $h_\tau [n']$ cannot both equal one. One concludes $h_\tau[n]\, h_\tau[n'] = 0$ for all $\{ n _\gamma \} \neq \{ n'_\gamma \}$. For $\{ n_\gamma \} = \{ n'_\gamma\}$ one has $h_\tau^2 [n] = h_\tau [n]$ according to eq.\eqref{eq:TS7}, and $h_\tau [n]\, h_\tau [n'] = \sum_\tau h_\tau[n] = 1$ according to eq.~\eqref{eq:TS8}, establishing the relation \eqref{eq:TS14}.

\paragraph*{Local states and local occupation number basis}

We define ``local occupation numbers'' as occupation numbers $n_\gamma (m)$ at a given site $m$. Locality refers here to the position on the chain. With $\gamma = 1, \dots,\, M$ we will denote by ``local states'' $\tau$, $\tau = 1, \dots,\, 2^M$, those states that can be constructed from the local occupation numbers at a given $m$. In the following $\tau$ will typically label local states, not to be confused with the $2^{M\, (\cM + 1)}$ overall states or spin configurations of the system.
(The use of the same symbol $\tau$ for denoting the local states and the states of the overall probability distribution should not give
rise to confusion. We use $\tau$ for a general labeling of probabilistic states. In general the meaning is clear from the context,
and will be recalled if necessary.)
Functions of local occupation numbers can be expanded in the ``local occupation number basis''. The basis functions $h_\tau [n(m)]$ involve then the occupation number $n(m)$ at given $m$. A ``local function'' $f[n(m)]$ depends only on occupation numbers on a given site $m$. It can be expanded as
\begin{equation}\label{eq:TS14A}
f[n(m)] = q_\tau (m)\, h_\tau [n(m)]\, .
\end{equation}
The relations \eqref{eq:TS5} -- \eqref{eq:TS14} hold now for a given $m$.

\paragraph*{Transfer matrix for local chains}

The local factor $\cK (m)$ for local chains depends on two sets of occupation numbers $\{ n_\gamma (m)\}$ and $\{ n_\gamma (m+1)\}$. We can use a double expansion
\begin{equation}\label{eq:TS15}
\cK (m) = \hat{T}_{\tau \rho} (m) \, h_\tau [n (m+1)]\, h_\rho [ n(m)]\, .
\end{equation}
The coefficients $\hat{T}_{\tau\rho} (m)$ are the elements of the ``transfer matrix'' $\hat{T}(m)$ at the site $m$. The double expansion \eqref{eq:TS15} uses a separate expansion in basis functions for each site $m$. We will employ the shorthands $h_\tau (m) \equiv h_\tau [ n(m)]$ in order to indicate that the basis functions are functions of the occupation numbers on the site $m$.

Consider the integration over the product of two neighboring local factors
\begin{align}\label{eq:TS16}
&\int \cD n (m+1)\, \cK (m+1) \, \cK (m) \notag \\
&\quad =  \big( \hat{T} (m+1)\, 
\hat{T} (m) \big)_{\tau \rho}\, h_\tau (m+2) \, h_\rho (m)\, .
\end{align}
It involves the matrix product of two neighboring transfer matrices
\begin{equation}\label{eq:TS17}
\big( \hat{T}(m+1)\, \hat{T}(m) \big)_{\tau \rho} = \hat{T}_{\tau\sigma} (m+1) 
\hat{T}_{\sigma \rho}(m)\, ,
\end{equation}
and basis functions at the sites $m+2$ and $m$, e.g. depending on occupation numbers $\{ n_\gamma (m+2)\}$ and $\{ n_\gamma (m)\}$. The relation \eqref{eq:TS16} obtains by expanding both $\cK (m+1)$ and $\cK (m)$ and using the orthogonality relation \eqref{eq:TS12} for the basis functions $h(m+1)$,
\begin{align}\label{eq:TS18}
& \int \cD n \, (m+1)\, \cK (m+1)\, \cK(m) \notag \\
& \quad = \int \cD n(m+1) \, \hat{T}_{\tau\mu} (m+1)\, h_\tau(m+2)\,
h_\mu (m+1) \notag \\
& \qquad\quad \times \hat{T}_{\sigma \rho} (m)\, h_\sigma (m+1)\, h_\rho (m) \notag \\[4pt]
& \quad = \hat{T}_{\rho\sigma} (m+1) \, \hat{T}_{\sigma\rho} (m) \, h_\tau (m+2) \,
h_\rho (m)\, .
\end{align}

The local factors $\cK (m)$ and $\cK (m+1)$ are functions of occupation numbers and their order plays no role,
\begin{equation}\label{eq:TS19}
\cK (m+1) \, \cK (m) = \cK (m)\, \cK (m+1)\, .
\end{equation}
Ordering them with increasing $m$ towards the left is a pure matter of convenience. In contrast, the transfer matrices do not commute, in general, if the local factors at different $m$ are different. The order of the matrices plays now a role. It is such that the transfer matrix at site $m+1$ stands on the left of the one for the site $m$ in the matrix multiplication \eqref{eq:TS17}. This extends the non-commutative structure that we have discussed above for unique jump chains to arbitrary local chains. The appearance
of the matrix product in the identity \eqref{eq:TS16} is at the root of noncommuting structures in classical statistics. We will find below 
similar matrix structures or operators associated to observables.

We can continue the procedure of multiplying neighboring local factors and integrating out the common occupation numbers. For example, the relation
\begin{align}\label{eq:TS20}
& \int \cD n(m+2)\, \int \cD n(m+1) \, \cK (m+2) \, \cK (m+1) \, \cK (m) \notag \\
& \quad = \left( \hat{T}(m+2)\, \hat{T}(m+1) \, \hat{T}(m) \right)_{\tau\rho}
h_\tau (m+3)\, h_\rho (m)
\end{align}
involves the ordered multiplication of three transfer matrices and the basis functions at sites $m+3$ and $m$. If we expand also the boundary term
\begin{equation}\label{eq:TS20A}
\cB = \hat{B}_{\tau\rho} \, h_{\tau}(n_{in}) \, h_\rho (n_f)
\end{equation}
we can derive for the partition function the relation
\begin{equation}\label{eq:TS21}
Z = \tr \left\{ \hat{T} (\cM - 1) \, \hat{T}(\cM - 2) \dots \hat{T}(1)\, \hat{T}(0)\, \hat{B} \right\} \, .
\end{equation}
It expresses the partition function as the trace of a chain of transfer matrices, multiplied with the boundary matrix $\hat{B}$. For the particular case where all $\hat{T}(m)$ are equal and $\hat{B} = \hat{T}$ one has
\begin{equation}\label{eq:TS22}
Z = \tr \left\{ \hat{T}^{\cM + 1} \right\}\, .
\end{equation}
This formula has been used for exact solutions of the Ising model \cite{KBI}. We emphasize that this expression for the partition function
holds for all local chains. 

\paragraph*{Transfer matrix for generalized Ising chains}

The occupation number basis is a convenient tool for the explicit construction of the transfer matrix for 
generalized Ising models.
Generalized Ising chains, defined in terms of the action,
\begin{equation}\label{eq:TS22A}
\cS = \sum_{m=0}^{\cM - 1} \cL (m)\, ,
\end{equation}
involve local factors
\begin{equation}\label{eq:TS22B}
\cK (m) = \exp ( - \cL (m) )\, ,
\end{equation}
with real $\cL(m)$. For next-neighbor interactions $\cL (m)$ depends on neighboring occupation numbers $n (m+1)$ and $n(m)$.

The occupation number basis offers a simple way for the computation of the transfer matrix. With
\begin{equation}\label{eq:TS22C}
\cL (m) = \cL_{\tau\rho} (m)\, h_\tau (m+1)\, h_\rho (m)\,,
\end{equation}
one finds the simple relation 
\begin{align}\label{eq:TS22D}
\cK (m) &= \hat{T}_{\tau\rho} (m) \, h_\tau (m+1) \, h_\rho(m) \notag \\
&= \exp \{-\cL_{\tau\rho}(m)\} \, h_\tau (m+1)\, h_\rho (m)\, .
\end{align}
The $(\tau\rho)$-element of the transfer matrix is given by
\begin{equation}\label{eq:TS22E}
\hat{T}_{\tau\rho}(m) = \exp \left\{ - \cL_{\tau\rho} (m) \right\}\, .
\end{equation}
In consequence, the transfer matrix $\hat{T}$ is a nonnegative matrix in the sense that all its elements obey $\hat{T}_{\tau\rho} \geq 0$.

The proof of the simple relation \eqref{eq:TS22E} employs the properties of the basis functions $h_\tau$. We can expand
\begin{align}\label{eq:TS22F}
& \exp \left\{ - \cL_{\tau\rho} (m) \, h_\tau (m+1)\, h_\rho(m) \right\} 
\notag \\[4pt]
& \quad =  1 - \cL_{\tau\rho} (m) \, h_\tau (m+1) \, h_\rho(m) \notag \\
& \qquad + \frac{1}{2} \cL_{\tau\rho}(m)\, h_\tau(m+1) \,h_\rho (m) \,     
\cL_{\sigma\omega} (m) \, h_\sigma (m+1) \, h_\omega (m) \notag \\
& \qquad - \dots \notag \\
& \quad = 1 + \sum_{\tau,\rho} h_\tau (m+1) \, h_\rho (m) \, \left( - \cL_{\tau\rho} + 
\frac{1}{2} \cL^2_{\tau\rho} - \dots \right)\, .
\end{align}
Here we use the multiplication rule for basis functions \eqref{eq:TS7} which implies
\begin{align}\label{eq:TS22G}
& \cL_{\tau\rho} (m)\, \cL_{\sigma \omega} (m)\, h_\tau (m+1) \, h_\sigma (m+1)\, h_\rho (m) \, h_\omega (m) \notag \\
& \quad = \sum_{\tau,\rho} \cL_{\tau\rho} (m)\, \cL_{\sigma \omega} (m) \, h_\tau (m+1)\, 
h_\rho (m)\, \delta_{\tau\sigma} \, \delta_{\rho\omega} \notag \\
& \quad = \sum_{\tau,\rho} \left( \cL_{\tau\rho} (m) \right)^2\, h_\tau (m+1) \, 
h_\rho (m)\, .
\end{align}
Similar relations hold for the higher terms in the series expansion of the exponential function, such that
\begin{align}\label{eq:TS22H}
\cK (m) &= \exp \big\{ - \cL_{\tau\rho} (m) \, h_\tau (m+1) \, h_\rho (m) \big\} \notag \\
&= 1 + \sum_{\tau, \rho} \big( \exp ( - \cL_{\tau\rho} ) -1 \big) \, h_\tau (m+1) \, 
h_\rho (m) \notag \\
&= \sum_{\tau, \rho} \exp (-\cL_{\tau\rho} ) \, h_\tau (m+1) \, h_\rho (m)\, .
\end{align}
For the last identity we use the sum rule \eqref{eq:TS8} for
\begin{equation}\label{eq:TS22I}
\sum_{\tau, \rho} h_\tau (m+1) \, h_\rho (m) = 1\, .
\end{equation}
This establishes eqs~\eqref{eq:TS22D}, \eqref{eq:TS22E}.

\paragraph*{Transfer matrix for the Ising chain}

It is straightforward to apply these results for a computation of the transfer matrix for the Ising chain. For this purpose one has to expand $\cL(m)$ in the occupation number basis. For the Ising chain one has
\begin{align}\label{eq:TS23}
&\cL (m) = \beta \{ \kappa\, s(m+1)\, s(m) + 1 \} \notag \\[4pt]
& \, = \beta \big\{ (1 + \kappa) \big[ n (m+1)\, n(m) \notag \\
& \qquad + (1 - n(m+1))\,(1 - n(m)) \big] \notag \\
& \qquad + (1 - \kappa) \big[ n(m+1) \, (1 - n(m)) \notag \\
& \qquad + ( 1 - n(m+1))\, n(m) \big] \big\} \\[4pt]
& \, = \beta \big\{ (1+ \kappa)\, \big[ h_1 (m+1)\, h_1(m) + h_2 (m+1)\, 
h_2 (m) \big] \notag \\
& \qquad + (1 - \kappa) \big[ h_1 (m+1)\, h_2 (m) + h_2(m+1) \, h_1(m) \big] \big\} \, . \notag
\end{align}
One infers for the transfer matrix
\begin{equation}\label{eq:TS24}
\hat{T} = \begin{pmatrix}
\text{e}^{-\beta\, (1 + \kappa)} & \text{e}^{-\beta\, (1 - \kappa)} \\
\text{e}^{-\beta\, (1 - \kappa)} & \text{e}^{-\beta\, (1 + \kappa)}
\end{pmatrix}\, .
\end{equation}
For the attractive interaction, $\kappa = -1$, this yields
\begin{equation}\label{eq:TS25}
\hat{T} = \begin{pmatrix}
1 & \text{e}^{-2\beta} \\
\text{e}^{-2\beta} & 1
\end{pmatrix} \, ,
\end{equation}
with
\begin{equation}\label{eq:TS26}
\hat{T}^2 = \begin{pmatrix}
1 + \text{e}^{-4\beta} & 2 \text{e}^{-2\beta} \\
2 \text{e}^{-2\beta} & 1 + \text{e}^{-4\beta}
\end{pmatrix}\, .
\end{equation}
For the repulsive interaction, $\kappa = 1$, one finds
\begin{equation}\label{eq:TS27}
\hat{T} = \begin{pmatrix}
\text{e}^{-2\beta} & 1 \\
1 & \text{e}^{-2\beta}
\end{pmatrix}\, ,
\end{equation}
which yields the same expression \eqref{eq:TS26} for $\hat{T}^2$. For periodic Ising chains with an even number of sites or $\cM$ odd the partition function
\begin{equation}\label{eq:TS28}
Z = \tr \left( \hat{T}^{\cM + 1} \right)
\end{equation}
is therefore independent of the sign $\kappa$.

For the evaluation of the partition function we compute the eigenvalues $\lambda_\pm$ of the transfer matrix $\hat{T}$. For $\kappa = -1$ one finds
\begin{equation}\label{TS29}
\lambda_\pm = 1 \pm \text{e}^{-2\beta} \, ,
\end{equation}
while for $\kappa = 1 $ one has
\begin{equation}\label{eq:TS30}
\lambda_\pm = \pm \left( 1 \pm \text{e}^{-2\beta} \right)\, .
\end{equation}
We can express the partition function \eqref{eq:TS28} as the sum of the eigenvalues of $\hat{T}^{\cM + 1}$, and therefore compute it in terms of $\lambda_\pm$. For odd $\cM$ this yields for both values of $\kappa$
\begin{align}\label{eq:TS31}
Z &= (\lambda_+)^{\cM + 1} + (\lambda_-)^{\cM + 1} \notag \\
&= \left( 1 + \text{e}^{-2\beta} \right)^{\cM + 1} + \left( 1 - 
\text{e}^{-2\beta} \right)^{\cM + 1} \, .
\end{align}
For finite $\beta > 0$ only $\lambda_+$ contributes in the limit $\cM \to \infty$, since $\lambda_-/\lambda_+ < 1$ and $\lim_{\cM \to \infty} (\lambda_- / \lambda_+)^{\cM} = 0$. In this limit one obtains
\begin{equation}\label{eq:TS32}
\ln Z = (\cM + 1) \, \ln \left( 1 + \text{e}^{-2\beta} \right) \, .
\end{equation}
This is the well known result for the Ising chain, with $\cM + 1$ playing the role of the volume. 

The $\beta$-dependence of $Z$ can be used to compute the classical correlation function between neighboring spins
\begin{align}\label{eq.TS33}
\frac{\p \ln Z}{\p \beta} &= - \sum_m \left( \kappa \left\langle s(m+1)\, s(m) 
\right\rangle + 1 \right) \notag \\
&= - (\cM + 1) \left( \kappa \left\langle s(m+1) \, s(m) \right\rangle + 1 \right) 
\notag \\
&= - 2(\cM + 1)\, \text{e}^{-\beta}\, (\text{e}^{\beta} + \text{e}^{-\beta} ) ^{-1}\, , 
\end{align}
where the second line uses that $\langle s(m+1)\, s(m) \rangle$ does not depend on $m$ and the third line takes $\cM \to \infty$. One infers the next-neighbor correlation in the infinite-volume limit
\begin{equation}\label{eq:TS34}
\langle s(m+1) \, s(m) \rangle = -\kappa\, \tanh (\beta) \, .
\end{equation}

\paragraph*{Transfer matrix for unique jump chains}

For a unique jump chain a given configuration $\rho$ of spins at $m$ transforms uniquely to a configuration $\tau(\rho)$ at $m+1$. The column $\rho$ of the transfer matrix has therefore a one for $\tau = \tau (\rho)$, and zero for all other entries in the column,
\begin{equation}\label{eq:TS35}
\hat{T}_{\tau\rho} = \delta (\tau,\, \tau(\rho)) \, .
\end{equation}
This has to hold for each $\rho$. If the map $\rho \to \tau(\rho)$ is invertible, there can be only a unique entry one in each row $\tau$. Otherwise, two different $\rho_1 \neq \rho_2$ would be mapped to the same $\tau$. Transfer matrices for a unique jump chain with invertible transformation have in each column and row precisely one element one, and all other elements zero.

Unique jump chains can be considered as limiting cases for generalized Ising models with $\beta \to \infty$. They are given by
\begin{equation}\label{eq:TS36}
\cL (m) = \beta \, \sum_{\tau, \rho} (1 - \delta(\tau, \, \tau(\rho)) \, h_\tau (m+1)  
\, h_\rho (m)\, .
\end{equation}
The limits $\beta \to \infty$ of the Ising chains \eqref{eq:TS25}, \eqref{eq:TS27} are unique jump chains, with $\tau(\rho) = \rho$ or $\tau(\rho) = 3 -\rho$, respectively. Also the unique jump chains \eqref{eq:UJ13} or \eqref{eq:UJ15} have transfer matrices of this type. In this case the transfer matrices are $(8\times 8)$-matrices that follow from eqs~\eqref{eq:UJ13} and \eqref{eq:UJ15} by expanding $\cL_H(m)$ and $\cL_{12} (m)$ in the occupation number basis. The spin transformations $V_H$ and $V_{12}$ in eqs~\eqref{eq:UJ14}, \eqref{eq:UJ16} generate the map of configurations $\rho \to \tau(\rho)$.

\paragraph*{Weight function in terms of chains of transfer matrices}

We first consider real factors $\cK(m)$ for local chains. Using the product rule \eqref{eq:TS7} for the basis functions one obtains
\begin{align}\label{eq:TS45}
\cK (m+1)\, \cK (m) &= \sum_{\tau, \sigma, \rho} \hat{T}_{\tau\sigma} (m+1)\, \hat{T}_{\sigma \rho} (m) \notag \\
& \quad \times h_\tau (m+2)\, h_\sigma (m+1) \, h_\rho (m) \, .
\end{align}
This extends to the product \eqref{eq:LC4}, and we can express the weight function in terms of transfer matrices and basis functions,
\begin{align}\label{eq:TS46}
w[n] = \sum_{\rho_0,\, \rho_1,\, \dots,\, \rho_{\cM}} w_{\rho_0\rho_1 \cdots 
	\rho_{\cM}} \,h_{\rho_0} (0)\, h_{\rho_1} (1) \cdots h_{\rho_{\cM}}(\cM)
\end{align}
with
\begin{align}\label{eq:TS47}
w_{\rho_0 \rho_1 \cdots \rho_{\cM}} &=  \hat{T}_{\rho_{\cM}\, \rho_{\cM - 1}} (\cM - 1) 
\cdots \, \hat{T}_{\rho_2 \rho_1}(1) \notag \\
& \quad \times \hat{T}_{\rho_1 \rho_0} (0)\, \hat{B}_{\rho_0 \rho_{\cM}} \, .
\end{align}
We observe that on the r.\,h.\,s.\ of eq.~\eqref{eq:TS47} every index appears twice but there is no summation in this expression, and therefore no matrix multiplication.

The basis functions take the values one or zero for all configurations and are therefore positive. The configurations of all occupation numbers on the chain can be labeled by specifying the local configuration $\rho_m$ at every site $m$, i.e. by $(\rho_0, \rho_1, \dots , \rho_{\cM})$. The sum in eq.~\eqref{eq:TS46} can be seen as a sum over all configurations of occupation numbers on the chain. For each value of the vector  $(\rho_0, \rho_1, \dots , \rho_{\cM})$ there exists indeed precisely one particular configuration of occupation numbers for which the product of basis functions in eq.~\eqref{eq:TS46} equals one. For this particular configuration all other products of basis functions in eq.~\eqref{eq:TS46} vanish, such that $w$ is given by $w_{\rho_0 \rho_1 \cdots \rho_{\cM}}$ for the corresponding values $(\rho_0, \rho_1, \dots , \rho_{\cM})$. 

The positivity of $w[n]$ for all configurations therefore requires
\begin{equation}\label{eq:TS48}
w_{\rho_0 \rho_1 \cdots \rho_{\cM}} \geq 0
\end{equation}
for all values of the multi-index $(\rho_0, \rho_1, \dots , \rho_{\cM})$. We can interpret
\begin{equation}\label{eq:TS49}
p_{\rho_0 \rho_1 \cdots \, \rho_{\cM}} = Z^{-1} \, w_{\rho_0 \rho_1 \cdots\, \rho_{\cM}}
\end{equation}
as the probability to find the particular configuration of occupation numbers $(\rho_0, \rho_1, \dots , \rho_{\cM})$. 
We discuss in appendix~\ref{app:positivity_of_overall_probability_distribution} the conditions for the positivity of the weight distribution.
 It is obvious that the weight function is positive if all matrix element of the transfer matrix are positive, provided all matrix element of the boundary matrix are positive as well. There are many more possibilities to obtain $w\geq 0$, however. We conclude that a positive overall probability distribution can be obtained also for transfer matrices with negative elements provided the conditions of appendix~\ref{app:positivity_of_overall_probability_distribution} are obeyed. For positive $w_{\rho_0 \rho_1 \cdots\,
	\rho_{\cM}}$ the overall probability distribution for configurations of occupation number can be inferred directly from eq.~\eqref{eq:TS46}.

The partition function obeys
\begin{equation}\label{eq:TS50}
Z = \sum_{\rho_0,\, \rho_1,\, \dots ,\, \rho_{\cM}} w_{\rho_0 \rho_1 \cdots\,
	\rho_{\cM}}\, .
\end{equation}
With the summation over all $\rho_m$, the relation \eqref{eq:TS47} turns into a trace over a matrix product, identical to eq.~\eqref{eq:TS21}.

\paragraph*{Use of the transfer matrix}

So far we have seen that the transfer matrix can be employed for the computation of the partition function -- this has been its
main historical use. We also have established an explicit expression of the probability distribution in terms of the transfer matrix.
For local chains, including local matrix chains, the ensemble of transfer matrices at all $m$ contains the same probabilistic information
as the probability distribution. One can be computed from the other. 

We will next see how the transfer matrix can be employed for the computation of expectation values of local observables. This formalism
in classical statistics is the equivalent of the Heisenberg formalism in quantum mechanics. The transfer matrix plays for classical
statistical systems a similar role as the evolution operator for quantum mechanics. In an appropriate normalization as the step evolution
operator\,\cite{CWIT}
it will be the generator of evolution for classical statistical systems. We will also develop for classical statistics an analogue of the Schr\"odinger picture in quantum mechanics. This involves wave functions and a classical density matrix. For particular types of subsystems
-- the quantum subsystems -- the step evolution operator will be related directly to the Hamiltonian of the corresponding quantum system. 
\subsubsection{Operators for local observables}\label{sec:operators_for_local_observables}

Quantum mechanics is formulated in terms of operators. One associates to each observable a suitable operator, usually represented
as a matrix or differential operator. This structure is also present for classical statistical systems. Constructing an operator
associated to a local observable we can express the expectation values of this observable by a trace over appropriate products 
of the corresponding operator and powers of the transfer matrix.

\paragraph*{Local observables}

Local observables are constructed from Ising spins at a given position $m$ on the chain, or in its neighborhood. 
The concept of ``locality'' refers here to locality on the chain, e.g. locality in $m$. With a narrow definition, ``local observables'' are those observables that can be constructed from the occupation numbers or Ising spins at a single site $m$. Local observables at $m$ are therefore functions of $n_\gamma (m)$. An example are the occupation numbers $n_\gamma (m)$ themselves, or products of occupation numbers at the same $m$, as $n_\gamma (m)\, n_\delta (m)$. We will later also employ an extended definition of local observables where we only require that local observables depend on occupation numbers in a neighborhood of $m$. At present, we stick to the narrow definition above. Local observables are then local functions that can be expanded in the local occupation number basis
\begin{equation}\label{eq:LO1}
A[n(m)] = A_\tau(m)\, h_\tau [n(m)] = A_\tau (m) \, h_\tau(m)\, .
\end{equation}
The coefficients $A_\tau (m)$ are the values that the observable takes in the local state $\tau$.

\paragraph*{Expectation values of local observables}

The expectation value of a local observable can be written in the form
\begin{align}\label{eq:LO2}
\langle A(m) \rangle &= Z^{-1} \int \cD n \, w[n] \, A[n(m)] \notag \\
&= Z^{-1} \int \cD n (m) \notag \\
& \quad\times\int \cD n (m' > m)\, \prod_{m' \geq m} \cK (m')\, 
A[n(m)] \notag \\
& \quad\times \int \cD n (m' < m)\, \prod_{m' < m} \cK (m') \, \cB\, .
\end{align}
Here we have split the configuration sum into a sum over local configurations $n(m)$, a sum over all configurations of occupation numbers $n(m')$ with $m' > m$, and a similar sum over configurations with $m' < m$. We have further ordered the product of local factors $\cK(m')$ in a corresponding way. 

Insertion of the expression of local factors in terms of transfer matrices \eqref{eq:TS15} and use of products with integration over intermediate $n$ in eq.~\eqref{eq:TS16} yields
\begin{align}\label{eq:LO3}
& \langle A(m) \rangle = Z^{-1} \int \dif n_f \int \dif n_{in} \, 
h_{\rho_{\cM}} (\cM)\, h_{\rho_0} (0) \notag \\
& \quad \times\hat{T}_{\rho_{\cM}\, \rho_{\cM - 1}} (\cM - 1) \cdots 
\hat{T}_{\rho_{m+1} \rho_m} (m) \notag \\
& \quad \times \hat{T}_{\rho'_m \rho_{m-1}} (m-1) \cdots 
\hat{T}_{\rho_1 \rho_0} (0) \,
\cB(n_f,\, n_{in}) \, \hat{A}_{\rho_m \rho_{m'}}\, ,
\end{align}
where $\hat{A}$ collects the parts depending on $n(m)$, 
\begin{equation}\label{eq:LO4}
\hat{A}_{\rho_m \rho_{m'}} (m) = \int \dif n(m) \, h_{\rho_m} (m) \, 
h_{\rho_{m'}} (m)\, h_\tau (m) A_\tau (m) \, .
\end{equation}

Performing integrations over $n_f$ and $n_{in}$ we can write $\langle A(m) \rangle$ as a matrix trace,
\begin{align}\label{eq:LO5}
\langle A(m) \rangle &= Z^{-1}\, \tr \big\{ \hat{T} (\cM - 1) \, \hat{T}(\cM - 2) 
\cdots\notag \\
& \qquad \times \hat{T}(m) \, \hat{A} (m) \, \hat{T}(m-1) \cdots\, \hat{T} (0) \, 
\hat{B} \big\} \, .
\end{align}
This expression differs from eq.~\eqref{eq:TS21} for $Z$ by the insertion of the matrix $\hat{A}(m)$ in the chain of transfer matrices at a position between $\hat{T} (m)$ and $\hat{T}(m-1)$. Since the matrices do not commute the position matters.

\paragraph*{Local operators}

The matrix $\hat{A}_{\rho_m \rho_{m'}}(m)$ is the ``local operator'' representing the local observable $A(m)$. In the occupation number basis we can easily perform the remaining integration over $n(m)$ and find
\begin{equation}\label{eq:LO6}
\hat{A}_{\rho_m \rho_{m'}} (m) = \sum_\tau A_\tau (m) \delta_{\rho_m \tau} 
\delta_{\rho_{m'} \tau} \, .
\end{equation}
The local operator is a diagonal matrix $\sim \delta_{\rho_m \rho_{m'}}$, with diagonal elements given by $A_\tau$. The possible values $A_\tau (m)$ of the observable correspond to the eigenvalues of the operator $\hat{A}(m)$, or the ``spectrum'' of the operator $\hat{A}(m)$. The fact that the operators for local observables are diagonal is a property of the local occupation number basis. It will not hold for an arbitrary basis. The relation \eqref{eq:LO5} and the definition \eqref{eq:LO4} hold in an arbitrary complete basis. The general operator expression of $A(m)$ in eq.\,\eqref{eq:LO4} involves an integration over a product of three basis functions\,\cite{CWQF}.

If the transfer matrices are all diagonal the position of $\hat{A} (m)$ in eq.~\eqref{eq:LO5} does not matter. The expectation value $\langle A(m) \rangle$ is then independent of $m$ if at every $m$ the observable has the same local values $A_\tau (m)$. This extends to the more general case where $\hat{A} (m)$ commutes with all transfer matrices. For the case where all $\hat{T}(m)$ are equal, and we consider an observable with the same $A_\tau (m)$ for every $m$, one has
\begin{equation}\label{eq:LO7}
\langle A(m) \rangle = Z^{-1}\, \tr \big\{\hat{T}^{\cM - m} \, \hat{A} \,
\hat{T}^m \hat{B} \big\}\, .
\end{equation}
If $\hat{A}$ commutes with $\hat{T}$ the observable $A$ is a ``conserved quantity'' with $\langle A(m) \rangle$ independent of $m$,
\begin{equation}\label{eq:LO8}
[ \hat{A},\, \hat{T} ] = 0 \quad \Rightarrow \quad \langle A(m) \rangle \;
\text{independent of }m\, .
\end{equation}

The relation $[\hat{A},\, \hat{T}]=0$ will not hold for arbitrary $A$ unless $\hat{T}$ is a diagonal matrix. For $[\hat{A},\ \hat{T}]\neq 0$ the position of the operator in the chain \eqref{eq:LO7} matters, and the expectation value $\langle A(m) \rangle$ will depend on the position $m$. This will be the generic case. These simple arguments show that non-commutative structures play a central role in classical statistics. The widespread opinion that quantum mechanics is non-commutative and classical statistics commutative is an unfounded prejudice.
\subsection{Time and evolution}\label{sec:time_and_evolution}

The concept of ``probabilistic time''\,\cite{CWPT} 
does not employ time as an a priori entity that exists per se and can be
used for the formulation of a theory. It rather views time as a particular structure between observables, conceptually 
similar to many other possible structures. Structures among observables do not depend on a particular choice for the probability
distribution. We have stated in sect. \ref{sec:probabilistic_realism} that for infinitely many variables any probability distribution
can be transformed into any other probability distribution by a suitable variable transformation. One can use this freedom to bring
the probability distribution into a form most suitable for the discussion of a particular structure. For the structure of time this form corresponds to the
local chains discussed in sect. \ref{sec:classical_statistics}.

Together with the concept of time arises the concept of locality in time. Events can happen at the same time or at a neighboring time 
before or after. We will see that for an understanding of the behavior of expectation values of local observables in a certain time
region only local probabilistic information is necessary. This is only a small part of the overall probabilistic information that
describes the world for all times. 

The notion of locality in time induces the concept of evolution. Given the local probabilistic information at some time $t_1$, one
asks if one can make statements about the probabilistic information at some later time $t_2 > t_1$. For local chains the answer is
positive and can be cast into the form of an evolution law. Knowing this evolution law permits to make predictions. For a given
probabilistic information at some ``present time" $t_1$ one can compute the probabilistic information at $t_2 > t_1$ and therefore
make statements about the probability of events in the future. In this section we develop the formalism for this setting. A simple
linear evolution law is formulated in terms of classical wave functions or a classical density matrix.

It is striking that for general classical statistical systems the time evolution law cannot be formulated in terms of the time-local probability distribution. More extended time-local probabilistic information is needed for casting the evolution from one time-layer to the next into a simple law. This time-local probabilistic information is encoded in a density matrix, whose diagonal elements are the time-local probabilities. For the density matrix the evolution law is linear, in the form of a generalized von Neumann equation. The close analogy to evolution in quantum mechanics is remarkable. 

The basic concepts of time and evolution apply to a large variety of classical statistical systems. ``Time'' needs not to be physical time. It can be some arbitrary linear order of layers, for example hypersurfaces in a multidimensional geometry. For a concept of ``physical time'' we will require additional criteria, relating it to the notion of clocks. In this sense physical time is closely connected to oscillatory phenomena. It is typically defined by a counting of the number of oscillations.
We discuss in this section simple classical statistical systems that show an oscillatory behavior and can be used as clocks. In the following sections we argue that the  choice
of a time structure is not unique -- different time structures define different clock systems. In this way basic features of special
and general relativity arise directly from the concept of probabilistic time. 

\subsubsection{Time as ordering structure}\label{sec:time_as_ordering_structure}

In this section we discuss the ordering relation between observables of a
certain class, which includes the basis observables. We 
define equivalence classes of observables according to this ordering and discuss
the associated concept of locality in time.

\paragraph*{Ordering relation}

What is time? In its most general form time is an ordering structure among
observables. Consider a set of basis observables or variables $\{ B_i\}$. A
linear ordering relation defines for each pair of variables $(B_i, \, B_j)$ if
$B_i$ is ``before'' $B_j$, ``after'' $B_j$ or ``simultaneous with'' $B_j$. We
denote by ``$<$'', ``$>$'' and ``$=$'' the relations before, after and
simultaneous with, respectively. The ordering relation for time has the property
that for any additional third variable $B_k$ one has
\begin{align}\label{eq:TOS1}
& B_i \;\text{``}=\text{''}\; B_j \; \text{and} \; B_j\, \text{``}=\text{''}\;
B_k \quad
\Rightarrow \quad B_i \; \text{``}=\text{''}\; B_k\, , \notag \\
& B_i\, \text{``}<\text{''}\; B_j \; \text{and} \; B_j \;\text{``}<\text{''}\;
B_k \quad
\Rightarrow \quad B_i \text{``}<\text{''} B_k \notag\, , \\
& B_i\, \text{``}>\text{''}\; B_j \; \text{and} \; B_j\, \text{``}>\text{''}\;
B_k \quad
\Rightarrow \quad B_i \text{``}>\text{''} B_k \, ,
\end{align}
as well as
\begin{align}\label{eq:TOS2}
& B_i \;\text{``}=\text{''}\; B_j \; \text{and} \; B_j \;\text{``}<\text{''}\;
B_k \quad
\Rightarrow \quad B_i \text{``}<\text{''} B_k\, , \notag \\
& B_i \;\text{``}=\text{''}\; B_j \; \text{and} \; B_j\, \text{``}>\text{''}\;
B_k \quad
\Rightarrow \quad B_i\, \text{``}>\text{''} \,B_k\, , \notag \\
& B_i\, \text{``}<\text{''}\; B_j \; \text{and} \; B_j \;\text{``}=\text{''}\;
B_k \quad
\Rightarrow \quad B_i \text{``}<\text{''} B_k\, , \notag \\
& B_i \;\text{``}>\text{''}\; B_j \; \text{and} \; B_j \;\text{``}=\text{''}\;
B_k \quad
\Rightarrow \quad B_i \;\text{``}>\text{''}\; B_k \, .
\end{align}
The relations \eqref{eq:TOS1}, \eqref{eq:TOS2} hold for arbitrary triplets
$B_i$, $B_j$ and $B_k$, and therefore for arbitrary permutations of the indices
$i$, $j$ and $k$. 

We assume that the ordering relation for time is defined for all basis
observables or variables of the system.
In a general context this amounts to a selection of variables compatible with
the time structure, or a ``coordinate choice in field space'' adapted to the
time structure.

\paragraph*{Equivalence classes for observables}

The presence of a linear ordering relation permits to order all variables into
equivalence classes labeled by $m$. If two variables are simultaneous, they
belong to the same equivalence class. We denote the basis observables in a given
equivalence class $m$ by $B_\alpha (m)$. The first relation \eqref{eq:TOS1}
implies that all members of an equivalence class are simultaneous with each
other, i.e. for all $\alpha$, $\beta$ one has
\begin{equation}\label{eq:TOS3}
B_\alpha (m)\; \text{``}=\text{''}\; B_\beta (m)\, .
\end{equation}
Furthermore, if some variable $B_\gamma (m')$ is before some other variable
$B_\alpha (m)$, all members of the equivalence class $m'$ are before all members
of the equivalence class $m$,
\begin{equation}\label{eq:TOS4}
B_\gamma (m') \; \text{``} < \text{''}\; B_\alpha (m) \quad \Rightarrow \quad
B_\delta (m') \; \text{``}< \text{''} B_\beta (m)\, , 
\end{equation}
for all $\delta$ and $\beta$, see eqs~\eqref{eq:TOS1}, \eqref{eq:TOS2}. The
ordering relation among variables implies an ordering relation between
equivalence classes, such that eq.~\eqref{eq:TOS4} can also be written in the
short form
\begin{equation}\label{eq:TOS5}
m' \; \text{``} < \text{''} m\, .
\end{equation}
The same holds for the relation ``$>$''.

The labels $m$ are strictly ordered. Without loss of generality we can use
integers to label the equivalence classes since they have the same strict order
relation, $m\in \mathbb{Z}$. The interval of integers chosen for $m$ is
arbitrary. The integers $m$ are a measure for discrete time. In these units time
is dimensionless. We will later introduce a factor $\varepsilon$ with dimension
``time'', such that 
\begin{equation}\label{eq:TOS5A}
t = \varepsilon\,(m - m_0)\, ,
\end{equation}
with suitably chosen $m_0$. In the limit of infinitely many sites and
$\varepsilon\to 0$ the time variable $t$ becomes continuous (see
sect.\,\ref{sec:continuous_time}).

\paragraph*{Local chains for probabilistic time}

If the variables are Ising spins or occupation numbers we can label all
variables by $s_\gamma(m)$ or $n_\gamma(m)$.
Using general variable transformations we can choose probability distributions
for the overall probabilistic system that are adapted to the ordering structure.
These are the local chains of Ising spins discussed in
sect.~\ref{sec:classical_statistics}. 
In the following we will assume that the overall probability distribution takes
the form of a local chain. More precisely, we assume that by suitable variable
transformations the overall probability distribution can be brought to the form
of a local chain for Ising spins $s_{\gamma}(m)$. These Ising spins are then
considered as the basis observables or variables with a natural ordering in
discrete time $m$. This fixes our ``coordinates" for the pair (observables +
probability distribution). From there on we admit only variable transformations
which preserve the structure of the local chain.

Before we have sometimes assumed implicitly that the number of spins at each
site $m$ is given by $M$ independent of $m$. This is not a necessity for the
concept of time, which can be implemented also if the number of variables $M(m)$
in the different equivalence classes varies with $m$. For example, this may play
a role if the variables are neurons in an artificial neuronal network and $m$
denotes the layers. For most of our discussion we will take $M$ independent of
$m$, however.

\paragraph*{Locality}

The ordering relation allows us to define the concept of locality, more
precisely locality in time or time-locality. We define as ``classical local
observables'' at time $m$ all observables that can be constructed from the
variables in the equivalence class $m$. For our local chains they are functions
of the local occupation numbers $n(m)$ and are denoted by $A(m)$. We can also
define local probabilities $\{ p_\tau (m)\}$. They obtain from the overall
probability distribution by ``integrating out'' all variables $n_\gamma (m')$
for $m' \neq m$. For the purpose of a clear distinction we denote in the
following the states of the overall probabilistic system for all times by
$\omega$, and the states of the time-local subsystem by $\tau$.

More in detail, the overall weight distribution over states~$\omega$
\begin{equation}\label{eq:TOS6}
w[n] = \{ w_\omega \}
\end{equation}
defines the partition function $Z$
\begin{equation}\label{eq:TOS6A}
Z = \int \cD n\, w[n] = \prod_{m'} \int \cD n(m')\, w[n]
\end{equation}
and the overall probability distribution
\begin{equation}\label{eq:TOS7}
p[n] = Z^{-1}\, w[n]\, , \quad p_\omega = Z^{-1}\, w_\omega\, .
\end{equation}
The local weight distribution obtains as
\begin{equation}\label{eq:TOS8}
w_1 [n(m)] = \{ w_\tau \} = \prod_{m' \neq m} \, \int \cD n(m')\, w[n]\, ,
\end{equation}
where the index 1 stands for a single layer.
It depends on the local occupation numbers $n(m)$ or local states $\tau$. The
partition function can be expressed in terms of the local weight distribution by
integrating over the local occupation numbers $n(m)$, 
\begin{equation}\label{eq:TOS9}
Z = \int \cD n (m) \, w_1 [n(m)]\, .
\end{equation}
Indeed, insertion of eq.~\eqref{eq:TOS8} for $w_1[n(m)]$ directly yields
eq.~\eqref{eq:TOS6A}.
The local probability distribution reads
\begin{equation}
\label{eq:TOS10}
p_1[n(m)] = \{ p_{\tau}(m)\} = Z^{-1}\, w_1 [n(m)]\,.
\end{equation}
It is related to the overall probability distribution by integrating out the
variables $n(m')$ for all $m' \neq m$,
\begin{equation}
p_1[n(m)] = \prod_{m' \neq m} \int \cD n(m') \, p[n]\, .
\label{eq:TOS10b}
\end{equation}
If confusion is unlikely we will drop the index $1$ for $w_1$ and $p_1$.

\paragraph*{Local observables}

The expectation values of classical local observables can be computed from the
local probability distribution at a given time $m$,
\begin{equation}\label{eq:TOS11}
\langle A(m) \rangle = \int \cD n(m) \, p[n(m)]\, A[n(m)]\, .
\end{equation}
This follows from the basic definition of expectation values \eqref{eq:OP6}
\begin{align}\label{eq:TOS12}
\langle A(m) \rangle &= \prod_{m'} \int \cD n(m') \, p[n]\, A[n(m)] \notag \\
&= \int \cD n(m) \, \prod_{m' \neq m} \int \cD n(m') \, p[n]\, A[n(m)] \notag \\
&= \int \cD n(m) \, p[n(m)]\, A[n(m)]\, .
\end{align}
No detailed information about the behavior of the overall probability
distribution for occupation numbers $n(m')$ at sites $m' \neq m$ is needed. This
is an enormous reduction of the probabilistic information needed for a
description of the expectation values of 
local observables. 

We will later discuss more general local observables, cf. sect.
\ref{sec:observables_and_operators}.
For these more general local observables we can define extended equivalence
classes, for which each equivalence class at a given $m$ comprises all local
observables at $m$. This equivalence class includes the local basis observables
or variables at the given $m$. Nevertheless, the order relation for time is not
defined for arbitrary observables of the system. As an example we may consider
two correlation functions $n(1)\, n(4)$ and $n(2)\, n(3)$. There is no
well-defined relation ``before'' or ``after'' for this pair of observables. In
short, the time $m$ or $t$ orders all local observables which only depend on
occupation numbers at a single site $m$, while ``non-equal time'' correlation
functions or observables which involve occupation numbers at different sites
are, in general, not ordered.

Our definition of time as an ordering of observables does not yet address the
issue of the arrow of time. Indeed, at the present stage the ordering of
equivalence classes can be done with ascending or descending $m$. We will turn
much later to the emergence of an arrow of time as an attractor property of a
family of solutions of field equations~\cite{VARG}. We also have not dealt yet
with the notion of clocks and the observed ``uniqueness of time'' for many
physical situations. For the moment, one can imagine many different orderings of
the variables. We will address this issue of ``physical time'' in sect.
\ref{sec:properties_of_physical_time}.

\paragraph*{Probabilities for histories}

The overall configuration of Ising spins $\omega$ can be seen as a sequence of
time-local configurations $\tau(m)$, 
\bel{PHI1}
\omega=\big{(}\tau(0),\, \tau(1), \, \dots \, \tau(m)\, \dots \,
\big{)}=\big{\lbrace}\tau(m)\big{\rbrace}\ .
\ee
For a given $m$ we can interpret $\tau(m)$ as a ``classical event at time $m$".
Thus $\omega$ denotes an ordered time sequence of classical events or a
``classical history". The probabilities $p_\omega$ of the overall probability
distribution are probabilities for classical histories. In turn, the probability
to find for a classical local observable $A(m)$ one of its possible measurement
values $\lambda_A$ equals the sum of all probabilities for classical histories
for which $A(m)$ takes the value $\lambda_A$. We first infer from
eq.~\eqref{eq:TOS10b} that the local probabilities $p_{\tau}(m)$ are the sum of
all probabilities for histories for which at $m$ the configuration $\tau(m)$ is
realized. In turn, we can group together all local configurations $\tau(m)$ for
which $A(m)$ takes the value $\lambda_A$ and sum the respective probabilities.

At this point one may observe a certain analogy of our probabilistic setting
with the incoherent history approach to quantum
mechanics~\cite{Griffiths:1984rx, omnes1987interpretation,
Gell-Mann:1991zrl},~\cite{Banks:2020wul},~\cite{HALLIWELL_1995,
Griffiths:1984rx, PhysRevLett.70.2201, PhysRevD.47.3345, PhysRevD.44.3173,
OMNES1989157, RevModPhys.64.339, Di_si_1995}.
The question becomes more complex, however, once we turn to sequences of
measurements. First of all, a sequence of measurements is based on conditional
probabilities which depend on the way how measurements are performed. At best
for some type of idealized measurements a sequence of classical events can be
identified with a sequence of measurements for classical local observables. More
importantly, we will discuss more general local observables which are not
functions of the local occupation numbers $n(m)$. An example is the momentum
observable in a quantum field theory. If some of the measurements in the
sequence concern this extended set of local observables the simple reasoning
above does not apply. Defining a consistent probability for the outcome of such
a general sequence of measurements is a rather complicated story.

\subsubsection{Evolution}\label{sec:evolution}

Evolution as a central idea in science is directly related to the concept of locality in time. In its most general form it asks how
the local probabilistic information at two different times $t_1$ and $t_2$ is related. Together with the concept of ordering in time
one can deal with the question how a given initial state, specified by the local probabilistic information at $t_\text{in}$, evolves
to a different state at later times $t > t_\text{in}$. We will see that the local probabilistic information necessary for an understanding
of evolution exceeds, in general, the probabilistic information in the local probability distribution \eqref{eq:TOS10}, \eqref{eq:TOS10b}. Similar to
quantum mechanics, the local probabilistic information necessary for the evolution in classical statistical systems involves a classical
density matrix or classical wave functions

\paragraph*{Evolution of local observables}

Consider a family of local observables $A(m)$, with time or site $m$ parameterizing the members of this family. Each member of the family is the same function of the local occupation numbers $n_\gamma(m)$. Examples are $A(m) = n_\gamma (m)$ or $A(m) = n_\gamma (m)\, n_\delta (m)$ for given $\gamma$ and $\delta$. The different members of this family are only distinguished by the index $m$. One calls the change of the expectation value $\langle A (m) \rangle$ as a function of $m$ the ``evolution'' of the observable $A(m)$. In the particular case that $\langle A(m) \rangle$ is independent of $m$, the observable $A(m)$ is called an invariant observable, e.g. it is invariant with respect to translations in time or translations on a chain.

With eq.~\eqref{eq:TOS11} the evolution of local observables is due to the dependence of the local probability distribution $p_1 [n(m)]$ on the time $m$. While the value of the observable $A_\tau$ in a local state $\tau$ is the same, the observable evolves due to the time dependence of the probabilities $p_\tau (m)$ for the states $\tau$. For understanding the time evolution of expectation values of local observables we have to understand the time evolution of the local probability distribution.

For the definition of the family of local observables we have assumed implicitly that the number of occupation numbers $M$ at a site $m$ does not depend on $m$. We could add an ``explicit time dependence'' of the local observables by having the values of $A_\tau$ dependent on $m$. In this case we could also consider an $m$-dependent number $M(m)$. We will not consider these possible generalizations here. We will often speak about ``time evolution'', but it should be clear that our general concept of evolution does not depend on an identification of $m$ with discrete time. We could interpret everything equivalently as an evolution in space, with $m$ labeling sites on a one-dimensional chain or hypersurfaces in a space with more than one dimension. At the present stage, there is no distinction between space and time. The label $m$ could also denote steps in a computation, layers in neural networks or generations in biological
evolution. The concept of evolution developed here is very general for all probabilistic systems that admit an ordering structure for
observables.

\paragraph*{Local evolution}

So far our notion of time and evolution is very general. We will now concentrate on a more restricted ``local evolution''. Intuitively, for a local evolution the change of a system at time $t$ should only depend on the physics at the time $t$. This concept is at the basis of differential evolution equations common to many areas of science. Local evolution addresses the evolution between two neighboring sites $m$ and $m+1$. As a central feature of local evolution, the evolution of local observables between neighboring times or sites involves only ``local probabilistic information'', rather than knowledge of the overall probability distribution for all sites of the chain. The local probabilistic information at a given $m$ is processed by some type of ``evolution law'' in order to obtain the local probabilistic information at higher $m' > m$.

We have to specify what we mean by local probabilistic information. The simplest form of local probabilistic information is the local probability distribution,
\begin{equation}\label{eq:EV1}
p_1(m) = p_1[n(m)] = \prod_{m' \neq m}\, \int \cD n(m')\, p[n]\, .
\end{equation}
It is sufficient for a computation of the expectation values of all local observables $A(m)$. If an evolution law relating $p_1(m+1)$ to $p_1(m)$ exists, one can immediately compute the evolution from $\langle A(m)\rangle $ to $\langle A(m+1) \rangle$. Such an evolution law exists for particular cases as unique jump chains or Markov chains. An evolution law based only on local probabilities is, however, not the general case. For example, the Ising chain does not admit such a simple evolution law.

In general, higher order local probabilistic information is necessary for the formulation of a local evolution law. An example for local probabilistic information beyond $p_1$ is the ``two-site probability distribution'',
\begin{equation}\label{eq:EV2}
p_2(m) = p_2[n(m+1),\, n(m) ] = \prod_{m' \neq m,\, m+1} \cD n(m')\, p[n]\, .
\end{equation}
The two-site probability distribution is a function of occupation numbers at two neighboring sites $m$ and $m+1$. It is sufficient in order to compute the expectation values of local observables $A(m)$ and $A(m+1)$, as well as all observables $A[n(m+1),\, n(m)]$ that can be expressed as functions of $n(m)$ and $n(m+1)$. Indeed, one has
\begin{align}\label{eq:EV3}
& \langle A[n (m+1), \, n(m)] \rangle = \int \cD n(m+1) \, \int \cD n(m) \notag \\
& \qquad \times p_2[n(m+1),\, n(m)]\, A[n(m+1),\, n(m)]\, .
\end{align}
Such observables include, in particular, products of local observables $A(m+1)\, B(m)$, whose expectation values are $\langle A(m+1)\, B(m) \rangle$. 

The local probabilities obtain from the two-site probability distribution by integrating out one set of variables,
\begin{align}\label{eq:EV4}
& p_1(m) = \int \cD n(m+1)\, p_2 (m)\, , \notag \\
& p_1(m+1) = \int \cD n(m)\, p_2(m)\, .
\end{align}
The two-site probability distribution contains, however, local probabilistic information beyond $p_1(m)$ and $p_1(m+1)$. 
For example, correlation functions for Ising spins at neighboring sites can be directly computed from the two-site probability 
distribution.

The local chains discussed in sect.~\ref{sec:local_chains}, or the matrix chains in Appendix~\ref{app:matrix chains}, are systems with a local evolution. The evolution law needs, in general, local probabilistic information beyond the local probability distribution. Its formulation involves concepts such as classical wave functions, which play the role of generalized probability amplitudes similar to quantum mechanics, or the classical density matrix.

For local chains, we will find that a simple evolution law can be formulated for the two-site density matrix $\rho_2 (m)$,
\begin{equation}\label{eq:EV5}
\rho_2(m) = Z^{-1}\, \int \cD n (m' \neq m,\, m+1) \, \prod_{m' \neq m} \, \cK(m) \, 
\cB\, .
\end{equation}
It depends on the occupation numbers at sites $m$ and $m+1$. Up to a missing factor $\cK(m)$ this is the two-site probability distribution $p_2(m)$. The latter obtains, in turn, as
\begin{equation}\label{eq:EQ6}
p_2 (m) = \cK (m) \, \rho_2 (m) \, .
\end{equation}
Thus $\rho_2(m)$ contains the necessary local probabilistic information for all observables whose expectation values can be computed from $p_2 (m)$.
The inverse, a computation of $\rho_2(m)$ from $p_2(m)$, is not always possible since $\cK(m)$ may vanish for certain spin configurations.
The classical density matrix will be closely related to $\rho_2(m)$.

\paragraph*{Unique jump chains}

Unique jump chains are a simple example for a local evolution based only on the local probability distributions $p_1(m)$ and $p_1(m+1)$. The evolution law specifies $p_1(m+1)$ as a function of $p_1(m)$. Consider an invertible unique jump for which a configuration $\rho$ at $m$ is mapped to a configuration $\tau (\rho)$ at $m+1$. This maps each local probability $p_\rho (m)$ to the same local probability at $m+1$ for the state $\tau (\rho)$, i.e.
\begin{equation}\label{eq:EV7}
p_{\tau (\rho)} (m+1) = p_\rho (m)\, .
\end{equation}
The set of local probabilities $p_\tau (m+1)$ at $m+1$ is the same as the set of $p_\rho (m)$ at $m$. The only change concerns the state $\tau$ or $\rho$ to which a given probability is associated. In other words, if at $m$ the probability to find a given sequence of bits $\rho$ is $p_\rho (m)$, and this sequence is mapped uniquely and invertibly to a new bit sequence $\tau(\rho)$ at $m+1$, the probability to find at $m+1$ the sequence $\tau(\rho)$ must be the same as $p_\rho (m)$. Invertibility of the unit jump map $\tau(\rho)$ is important for this circumstance. If two different spin configurations $\rho_1$ and $\rho_2$ at $m$ would be mapped to the same $\tau$ at $m+1$, the probability $p_\tau (m+1)$ would be given by the sum of the probabilities $p_{\rho_1} (m)$ and $p_{\rho_2} (m)$. A simple formal proof of the intuitively clear property \eqref{eq:EV7} will be given in sect.~\ref{sec:step_evolution_operator}, based on the step evolution operator.

\paragraph*{Markov chains}

Markov chains are characterized by ``transition probabilities'' $W_{\tau\rho}$ that obey the relation
\begin{equation}\label{eq:EV8}
W_{\tau\rho} \geq 0 \, , \quad \sum_\tau W_{\tau \rho} = 1\, .
\end{equation}
A state $\rho$ at $m$ is transformed with transition probability $W_{\tau\rho}$ to a state $\tau$ at $m+1$, in the sense that the local probabilities obey
\begin{equation}\label{eq:EV9}
p_{1\tau} (m+1) = W_{\tau \rho} (m)\, p_{1\rho} (m)\, .
\end{equation}
The relations \eqref{eq:EV8} guarantee the properties of a local probability distribution at $m+1$, i.e. $p_{1 \tau} (m+1) \geq 0$ and $\sum_\tau p_{1 \tau} (m+1) = 1$, provided that $p_1 (m)$ is a probability distribution. The local probabilistic information for the evolution law \eqref{eq:EV9} involves the local probability distributions $p_1 (m)$ and $p_1 (m+1)$. 

Unique jump chains are a special limiting case of Markov chains, with the property
\begin{equation}\label{eq:EV10}
W_{\tau\rho} = \delta_{\tau(\rho),\, \rho}\, .
\end{equation}
The probability of a transition from $\rho$ to $\tau (\rho)$ equals one, while the transition probabilities to all other states $\tau \neq \tau (\rho)$ vanish. 
In contrast to probabilistic automata, generic Markov chains have a probabilistic evolution. The probabilistic aspects do not only concern the boundary term. We will discuss in sect.~\ref{sec:markov_chains} under which circumstances Markov chains arise from our setting of an overall probability distribution for all times.
\subsubsection{Classical wave functions}
\label{sec:classical_wave_functions}

The classical wave functions are the objects which permit a simple formulation
of a local evolution law for all local chains
with factorizing boundary conditions.
They contain the relevant local probabilistic information. Similar to quantum
mechanics, they are a type of probability amplitude. The local probability
distribution is a bilinear in the classical wave functions. In contrast to
quantum mechanics there are in general two different classical wave functions.
The conjugate wave function is not related directly to the wave function by some
operation as complex conjugation. As a consequence, the wave function and the
conjugate wave function are not normalized individually. A classical density
matrix can be constructed as a bilinear in the classical wave functions. Similar
to quantum mechanics, its diagonal elements are the local probabilities. We will
see that for the particular case of a unique jump chain the two wave functions
can be identified, leading to discrete quantum mechanics in a real formulation.

\paragraph*{Pure state boundary condition}

Classical wave functions can be formulated for ``pure state boundary
conditions''. For this type of boundary condition the boundary term $\cB
(n_{in}, \, n_f)$ in eq.~\eqref{eq:LC4} is a product of an initial boundary term
$\tilde{f}_{in}(n_{in})$ and a final boundary term $\bar{f}_f (n_f)$,
\begin{equation}\label{eq:CWF1}
\cB (n_{in},\, n_f) = \tilde{f}_{in} (n_{in})\, \bar{f}_f (n_f)\, .
\end{equation}
Probability distributions with pure state boundary conditions describe
``classical pure states''. Similar to mixed states in quantum mechanics, general
boundary conditions can be formulated later as weighted sums of pure state terms
\eqref{eq:CWF1}.

\paragraph*{Normalization of local factors}

It is convenient for a discussion of general properties to normalize the local
factors $\cK(m)$ and the boundary term $\cB$ in eq.~\eqref{eq:LC4} such that the
partition function equals one,
\begin{equation}\label{eq:CWF2}
Z = 1\, .
\end{equation}
For positive weights the weight distribution coincides with the probability
distribution in this case. The normalization \eqref{eq:CWF2} can be achieved by
multiplying first each local factor $\cK(m)$ by a constant factor. We will
discuss a suitable normalization of $\cK(m)$ in
sect.~\ref{sec:step_evolution_operator}. For a given normalization of $\cK(m)$,
the boundary term $\cB$ can be multiplied by a constant factor in order to
realize eq.~\eqref{eq:CWF2}. For simple systems these normalization operations
can be done explicitly. If this is no longer possible for more complex systems,
the existence of such operations is still useful for the conceptual discussion.
If $Z=1$ cannot be realized explicitly, one has to multiply the definition of
the classical wave functions and other objects discussed in the following by
appropriate powers of $Z$ or similar partial normalization factors. In the
following we assume $Z=1$ if not indicated differently.

\paragraph*{Classical wave functions}

The classical wave function $\tilde{f}(m)$ or $\tilde{f}(t)$ is a local object
at a given site $m$ or time $t$. It is a function of the local occupation
numbers $n_\gamma (m)$. It obtains by integrating out the variables with $m' <
m$, as
\begin{equation}\label{eq:CWF3}
\tilde{f}(m) = \tilde{f}[n(m)] = \prod_{m' = 0}^{m - 1} \int \cD n(m')
\prod_{m' = 0}^{m-1} \cK (m')\, \tilde{f}_{in}\, .
\end{equation}
Indeed, the integrand on the r.h.s. of eq.~\eqref{eq:CWF3} involves only
occupation numbers at sites $m' \leq m$. Performing the partial configuration
sum over all $n(m')$ with $m' < m$, the integral only depends on the local
occupation numbers at the site $m$. 

The conjugate wave function $\bar{f}(m)$ or $\bar{f}(t)$ is constructed by
integrating over the occupation numbers at $m' > m$, 
\begin{equation}\label{eq:CWF4}
\bar{f}(m) = \bar{f}_f \,\prod_{m' = m+1}^{\cM} \, \int \cD n(m')\, 
\prod_{m'=m}^{\cM - 1} \, \cK (m')\, .
\end{equation}
It again depends only on the local occupation numbers at the site $m$. The
product $\bar{f} (m)\, \tilde{f}(m)$ integrates over all variables at sites $m'
\neq m$. It contains all local factors $\cK (m')$ in eq.~\eqref{eq:LC4}. Also
the full boundary term is contained in this product. As a result, the product
$\bar{f} (m) \, \tilde{f} (m)$ integrates out all variables at sites $m' \neq m$
in the overall probability distribution \eqref{eq:LC4}. The result is the local
probability distribution $p_1(m)$ in eq.~\eqref{eq:TOS10}, \eqref{eq:TOS10b},
\begin{equation}\label{eq:CWF5}
\bar{f}(m) \, \tilde{f}(m) = p_1 (m)\, .
\end{equation}
The local probability distribution is a bilinear in the classical wave
functions, involving both the classical wave function $\tilde{f}(m)$ and the
conjugate classical wave function $\bar{f}(m)$. In turn, the classical wave
functions are a type of ``probability amplitudes", with some analogy to the wave
function in quantum mechanics.

The simplicity of the local evolution law is based on the observation that
$\tilde{f}(m+1)$ obtains from $\tilde{f}(m)$ by adding a local factor $\cK (m)$
and integrating out the variables $n(m)$
\begin{equation}\label{eq:CWF6}
\tilde{f}(m+1) = \int \cD n(m) \, \cK (m) \, \tilde{f}(m)\, .
\end{equation}
As it should be, $\tilde{f}(m+1)$ depends on the local occupation numbers at the
site $m+1$. Similarly, one finds for the conjugate wave function
\begin{equation}\label{eq:CWF7}
\bar{f} (m-1) = \int \cD n(m) \, \bar{f}(m) \, \cK (m-1)\, .
\end{equation}
Again, $\bar{f} (m-1)$ depends on the local occupation numbers at $m-1$.
We will see in sect.~\ref{sec:step_evolution_operator} that the simple relations
\eqref{eq:CWF6}, \eqref{eq:CWF7} result in linear
evolution laws for the classical wave functions in the occupation number basis.

\paragraph*{Classical wave functions in occupation number \\basis}

The classical wave functions can be expanded in the local occupation number
basis
\begin{align}\label{eq:CW8}
& \tilde{f}(m) = \tilde{q}_\tau (m) \, h_\tau (m) \, , \notag \\
& \bar{f}(m) = \bar{q}_\tau (m)\, h_\tau (m) \, .
\end{align}
The coefficients $\tilde{q}_\tau (m)$ and $\bar{q}_\tau (m)$ are real vectors
with $2^M$ components. They constitute the classical wave function and the
conjugate wave function in the occupation number basis, and have a status
analogue to the wave function in quantum mechanics in a given basis.

Using $h_\tau (m)\, h_\rho (m) = \delta_{\tau\rho}\, h_\tau (m)$ one obtains for
the bilinear
\begin{equation}\label{eq:CW9}
\bar{f}(m) \, \tilde{f}(m) = \sum_\tau \bar{q}_\tau (m) \, q_\tau (m) \, h_\tau
(m)\, .
\end{equation}
This may be compared with the expansion of the local probability distribution in
the occupation number basis
\begin{equation}\label{eq:CW10}
p_1 (m) = \sum_\tau p_\tau (m)\, h_\tau (m) \, .
\end{equation}
One infers from eq.~\eqref{eq:CWF5} that the local probability $p_\tau(m)$ for a
state $\tau$ at $m$ is a bilinear in the wave functions (no sum over $\tau$
here)
\begin{equation}\label{eq:CW11}
p_\tau (m) = \bar{q}_\tau (m)\, \tilde{q}_\tau (m)\, .
\end{equation}
Even though the notation is similar to quantum mechanics, we recall that in
general $\bar{q}_\tau (m)$ is not related to $\tilde{q}_\tau (m)$. The two wave
functions depend on separate initial and final boundary factors.

For generalized Ising chains, with positive local factors $\cK (m) \geq 0$ and
positive boundary factors $\bar{f}_f \geq 0$, $\tilde{f}_{in} \geq 0$, the
classical wave functions are both positive, obeying for all configurations $\{
n(m)\}$
\begin{equation}\label{eq:CW11A}
\tilde{f}(m) \geq 0\, , \quad \bar{f} (m) \geq 0\, .
\end{equation}
This implies that all components are positive,
\begin{equation}\label{eq:CW11B}
\tilde{q}_\tau (m) \geq 0\, , \quad \bar{q}_\tau (m) \geq 0 \, .
\end{equation}
Indeed, if one component would be negative, say $\tilde{q}_{\tau_0} < 0$, the
configuration $\tau_0$ would lead to negative $\tilde{f}(m)$. For this
particular configuration only $h_{\tau_0}$ differs from zero, such that
$\tilde{f}(m) = \tilde{q}_{\tau_0} (m) \, h_{\tau_0} (m)$. Since $h_{\tau_0}$ is
positive ($h_{\tau_0} = 1$ for the configuration $\tau_0$), this leads to
$\tilde{f}(m) < 0$, proving eq.~\eqref{eq:CW11B} by contradiction. At first
sight the requirement of positivity \eqref{eq:CW11B} seems to be rather
restrictive. We will see later\,\cite{CWQF} 
that the sign of the wave functions is partly a matter of conventions. It can be
changed by gauge transformations, and eq.~\eqref{eq:CW11B} corresponds to a
particular gauge fixing in the formulation of the probability distribution for
local chains.

\paragraph*{Expectation value of local observables}

For a local observable $A[n(m)]$, with associated operator $\hat{A}(m)$ given by
eq.~\eqref{eq:LO6}, the expectation value has a simple expression as a bilinear
of the wave functions. For local chains it reads
\begin{align}\label{eq:CW19}
\langle A[n(m)]\rangle & = \langle \bar{q}(m) | \hat{A} (m) | \tilde{q} (m)
\rangle 
\notag \\
&= \bar{q}_\tau (m) \, \hat{A}_{\tau\rho} (m) \, \tilde{q}_\rho (m)\, ,
\end{align}
where we use the bra-ket formulation familiar from quantum mechanics. Indeed,
eq.~\eqref{eq:CW19} is rather similar to the definition of expectation values in
quantum mechanics, except for the presence of two different classical wave
functions. In contrast to quantum mechanics we do not need to postulate this
rule as an axiom. It follows directly from the definition of expectation values
in classical statistics \eqref{eq:OP2}.

Indeed, for $Z=1$ we can write
\begin{align}\label{eq:CW20}
\langle A[n(m)]\rangle &= \int \cD n\, A[n(m)] \, \prod_{m' = 0}^{\cM -1} \,
\cK(m) 
\, f_{in} \, \bar{f}_f \notag \\
&= \int \cD n(m) \, A[n(m)]\, \bar{f}(m)\, \tilde{f}(m)\, .
\end{align}
Here we have used the definitions \eqref{eq:CWF3}, \eqref{eq:CWF4} of the wave
functions in order to absorb all local factors $\cK(m)$ and the boundary terms
$f_{in}$, $\bar{f}_f$, together with the integration over all variables $n(m')$
except for $m' = m$. Equivalently, one may use the expression for the local
probability distribution $p_1(m)$ in eq.~\eqref{eq:CWF5}, and 
\begin{equation}\label{eq:CW21}
\langle A[n(m)]\rangle = \int \cD n(m)\, A[n(m)]\, p_1(m)\, .
\end{equation}

The expansion of the wave functions \eqref{eq:CW8}, together with the expansion
\begin{equation}\label{eq:CW22}
A[n(m)] = A_\tau (m) \, h_\tau (m)\, ,
\end{equation}
yields
\begin{align}\label{eq:CW23}
\langle A [n(m)] \rangle &= \int \cD n(m) \, \bar{q}_\tau (m) \, A_\sigma (m) \, 
\tilde{q}_\rho (m) \notag \\
& \qquad \times h_\tau (m)\, h_\sigma (m) \, h_\rho (m) \notag \\
&= \bar{q}_\tau (m) \, \hat{A}_{\tau\rho} (m) \, \tilde{q}_\rho(m)\, ,
\end{align}
with operator
\begin{align}\label{eq:CW24}
\hat{A}_{\tau\rho} (m) &= \int \cD n(m)\, A_\sigma (m)\, h_\tau (m)\,
h_\sigma(m) \,
h_\rho (m) \notag \\
&= A_\tau (m) \, \delta_{\tau\rho} \, .
\end{align}
This proves eq.~\eqref{eq:CW19}. We recall that the coefficients $A_\tau (m)$
are the values the observable takes in the state $\tau$ at $m$.

The expression of the expectation value of local observables in terms of wave
functions and operators is similar to the Schr\"odinger
picture in quantum mechanics. It is equivalent to the expression \eqref{eq:LO5},
which corresponds to the Heisenberg picture. The
operators associated to local observables are identical. The important advantage
of the use of classical wave functions is that only
the local probabilistic information is involved. Once the classical wave
functions are computed, no further information from the
overall probability distribution is needed. The classical wave functions give
access to all expectation values of local observables. 

\subsubsection{Step evolution operator}\label{sec:step_evolution_operator}

The step evolution operator is the analogue of the evolution operator in quantum mechanics for discrete time steps. It is related
to the transfer matrix by an appropriate normalization.
The step evolution operator will be a key quantity in this work. It encodes the ``dynamics'' of the probabilistic system.

\paragraph*{Normalization of local factors}

Multiplication of a local factor $\cK (m)$ by a constant $c(m)$ does not change the overall probability distribution. The multiplication of the weight distribution $w[n]$ by $c(m)$ is canceled by the multiplication of the partition function $Z$ by the same factor $c(m)$. In principle, one could take $c(m)$ to be arbitrary real or even complex numbers different from zero. We take here real positive $c(m)$. We employ the freedom of this multiplication in order to normalize the transfer matrix conveniently. Indeed, by virtue of eq.\,\eqref{eq:TS15} the transfer matrix is multiplied by the same factor $c(m)$ as for $\cK(m)$. 

We use this freedom for a normalization where the largest absolute value of the eigenvalues of the transfer matrix equals one. There may be a single ``largest eigenvalue'' $\lambda$ with $|\lambda| = 1$, or there could be multiple largest eigenvalues $\lambda_i$, with $|\lambda_i| = 1$. With this normalization the transfer matrix is called the ``step evolution operator'', denoted by $\hat{S}$. For local chains with the appropriate normalization of the local factors one has
\begin{equation}\label{eq:SE1}
\cK (m) = \hat{S}_{\tau\rho} (m) \, h_\tau (m+1) \, h_\rho (m)\, .
\end{equation}
The step evolution operator obeys all the identities \eqref{eq:TS16}--\eqref{eq:TS21} for the transfer matrix. The condition on the largest eigenvalue of $\hat{S}$ fixes the normalization of all local factors. The condition $Z=1$ is then realized by a suitable multiplicative normalization of the boundary matrix $\hat{B}$ in eq.~\eqref{eq:TS21},
\begin{equation}\label{eq:SE2}
\tr \{ \hat{S} (\cM - 1) \cdots \, \hat{S}(1)\, \hat{S}(0)\, \hat{B} \} = 1\, .
\end{equation}

\paragraph*{Evolution law for classical wave functions}

With eq.~\eqref{eq:SE1} we can translate the evolution law \eqref{eq:CWF6} for the classical wave function to the occupation number basis
\begin{equation}\label{eq:SE3}
\tilde{q}_\tau (m+1) = \hat{S}_{\tau\rho} (m) \, \tilde{q}_\rho (m)\, .
\end{equation}
This is a simple linear evolution law, similar to the evolution law of the wave function in quantum mechanics. 
It multiplies the vector $\tilde{q}$ by a matrix $\hat{S}$. This matrix is the step evolution operator, with a status similar to the evolution operator in quantum mechanics.
In particular, the superposition principle for possible solutions of the evolution equation holds: if $\tilde{q}_1$ and $\tilde{q}_2$ are two solutions of the evolution equation \eqref{eq:SE3}, also $\alpha\,\tilde{q}_1 + \beta\, \tilde{q}_2$ is a possible solution. The evolution law \eqref{eq:SE3} is the generalization of a discrete Schrödinger equation to classical statistics. For quantum mechanics $\tilde{q}$ is replaced by the complex wave function $\psi$, and $\hat{S}$ is replaced by the unitary step evolution operator $U (t + \varepsilon,\, t)$.

The derivation of the linear evolution law \eqref{eq:SE3} is straightforward, using for the basis functions the orthogonality relation \eqref{eq:TS12}
\begin{align}\label{eq:SE4}
\tilde{f}(m+1) & = \tilde{q}_\tau (m+1) \, h_\tau (m+1) \notag \\
&= \int \cD n(m) \, \cK (m) \, \tilde{f}(m) \notag \\
&= \int \cD n(m) \, \hat{S}_{\tau\sigma} (m) \, h_\tau (m+1) \notag \\
& \qquad \times h_\sigma (m) \,\tilde{q}_\rho (m) \, h_\rho (m) \notag \\[4pt]
&= \hat{S}_{\tau\sigma} (m) \, \delta_{\sigma\rho}\, \tilde{q}_\rho (m) \, h_\tau (m+1) 
\notag \\[4pt]
&= \hat{S}_{\tau\rho} (m) \, \tilde{q}_\rho (m) \, h_\tau (m+1)\, ,
\end{align}
and comparing the coefficients of $h_\tau (m+1)$.

Analogously, one obtains the evolution law for the conjugate wave function from eq.~\eqref{eq:CWF7},
\begin{equation}\label{eq:SE5}
\bar{q}_\tau (m-1) = \bar{q}_\rho (m)\, \hat{S}_{\rho\tau} (m-1)\, .
\end{equation}
or
\begin{equation}\label{eq:SE6}
\bar{q}_\tau (m) = ( \hat{S}^\tp )_{\tau\rho} (m) \, \bar{q}_\rho (m+1)\, ,
\end{equation}
with $\hat{S}^\tp$ the transpose of the step evolution operator. We will assume that the step evolution operator $\hat{S}$ is a regular matrix, such that the inverse $\hat{S}^{-1}$ exists. This is the case for almost all overall probability distributions discussed in this work. In this case we can invert eq.~\eqref{eq:SE6},
\begin{equation}\label{eq:SE7}
\bar{q}_\tau (m+1) = ( \hat{S}^\tp )^{-1}_{\tau\rho} (m) \, \bar{q}_\rho (m)\, .
\end{equation}
We observe that the evolution operator for the wave function $\tilde{q}$ and the conjugate wave function $\bar{q}$ differ unless the step evolution operator is an orthogonal matrix.

\paragraph*{Evolution of local probability distribution}

With eq.~\eqref{eq:CW11} we can compute the evolution of the local probabilities (no sum over $\tau$)
\begin{align}\label{eq:SE8}
p_\tau (m+1) &= \bar{q}_\tau (m+1) \, \tilde{q}_\tau (m+1) \notag \\
&= \sum_{\rho,\sigma} \bar{q}_\rho (m) \, \hat{S}_{\rho\tau}^{-1} (m) \, 
\hat{S}_{\tau\sigma} (m) \, \tilde{q}_\sigma (m)\, .
\end{align}
For general wave functions $\tilde{q}$ and $\bar{q}$ this can be written as an evolution law for local probabilities only if $\hat{S}$ obeys a particular condition, namely if for each $\tau$, $\rho$, $\sigma$ it satisfies
\begin{equation}\label{eq:SE9}
\hat{S}_{\rho\tau}^{-1} (m) \, \hat{S}_{\tau\sigma} (m) = W_{\tau\rho}^{(M)} (m) \, 
\delta_{\rho\sigma}\, .
\end{equation}
If the expression on the l.\,h.\,s. does not vanish for $\rho \neq \sigma$, the r.h.s. of eq.\,\eqref{eq:SE8} contains a combination of wave functions that cannot be expressed by local probabilities,
in general. 

If eq.\,\eqref{eq:SE9} holds one recovers
for arbitrary wave functions
the evolution law for Markov chains \eqref{eq:EV9}, with $W_{\tau\rho} = W^{(M)}_{\tau\rho}$ the transition probabilities, provided that $W_{\tau\rho} \geq 0$. The normalization condition \eqref{eq:EV8} for the transition probabilities $W_{\tau\rho}$ is obeyed automatically by taking a sum over $\tau$ in eq.\,\eqref{eq:SE9}. 
We conclude that
for the general case
Markov chains can arise only for particular probabilistic states for which the products $\bar{q}_\rho \tilde{q}_\sigma$ for $\rho \neq \sigma$ can be expressed in terms of local probabilities.

In general, the condition \eqref{eq:SE9} is not realized. There exists then no general local evolution law that can be formulated in terms of the local probabilities alone. The pair of classical wave functions contains local probabilistic information beyond the one contained in the local probability distribution. (See sect.\,\ref{sec:time_local_subsystems} for a more detailed discussion.) This additional local probabilistic information is needed for the evolution of the local probability distribution. 
This holds for generic systems, as we will see below explicitly for the Ising model.

We conclude that a description of the evolution of expectation values of local observables needs more information than available from the local probability distribution alone. This is the central reason for the use of classical wave functions or the associated classical density matrix.
The observation that the local probability distribution $p_1(m)$ is sufficient for the computation of expectation values for all local observables has often led to the misconception that $p_1(m)$ contains all the relevant local probabilistic information. As we have seen, however, the local probability distribution is insufficient for the formulation of an evolution law.

\paragraph*{Step evolution operator for generalized Ising chains}

For generalized Ising chains the local factor can be written as $\cK(m) = \exp \{ - \cL (m) \}$. A multiplicative renormalization $\cK(m) \to c(m)\, \cK(m)$ corresponds to a shift of $\cL (m)$ by a constant
\begin{equation}\label{eq:SE10}
\cL (m) \rightarrow \cL (m) - \ln c(m)\, .
\end{equation}
After this shift the elements of the step evolution operator obey
\begin{equation}\label{eq:SE11}
\hat{S}_{\tau\rho} = \exp \{ - \cL_{\tau\rho}\}\, ,
\end{equation}
with $\cL_{\tau\rho}$ given by eq.~\eqref{eq:TS22C}. The shift \eqref{eq:SE10} corresponds for all elements $\cL_{\tau\rho}$ to the same constant shift
\begin{equation}\label{eq:SE12}
\cL_{\tau\rho} \rightarrow \cL_{\tau\rho} - \ln c(m)\, ,
\end{equation}
according to
\begin{equation}\label{eq:SE13}
\int \cD n(m+1) \, \int \cD n(m) \, h_\tau (m+1) \, h_\rho (m) = 1\, .
\end{equation}
Thus all elements of the transfer matrix are multiplied by the constant $c(m)$, as expected, and we can employ the shift \eqref{eq:SE10} in order to achieve $| \lambda | = 1 $ for the largest eigenvalue of $\hat{S}$. We conclude that for generalized Ising models the step evolution operator is a nonnegative matrix with largest eigenvalue $|\lambda| = 1$. We will further assume that it is a regular matrix. These conditions lead to important restrictions for the properties of the step evolution operator, that we will discuss in sect.~\ref{sec:orthogonal_and_unitary_step_evolution_operators}.

For the particular case of the Ising chain with action
\begin{equation}\label{eq:SE13AA}
\cS = \beta\, \sum_m \big( \kappa\, s(m+1)\, s(m) + 1\big)\,,
\end{equation}
the step evolution operator reads
\begin{align}\label{eq:SE13A}
& \hat{S}_+ = \frac{1}{2\cosh\beta} \begin{pmatrix}
\text{e}^{-\beta} & \text{e}^\beta \\
\text{e}^\beta & \text{e}^{-\beta}
\end{pmatrix}
\quad \text{ for } \kappa = 1\, , \notag \\
& \hat{S}_- = \frac{1}{2\cosh\beta} \begin{pmatrix}
\text{e}^\beta & \text{e}^{-\beta} \\
\text{e}^{-\beta} & \text{e}^\beta
\end{pmatrix}
\quad \text{ for } \kappa = -1\, .
\end{align} 
It obtains from the transfer matrices \eqref{eq:TS27} and \eqref{eq:TS25} by multiplication with
\begin{equation}\label{eq:SE13B}
c = \frac{\text{e}^\beta}{2\cosh \beta}\,,
\end{equation}
or by adding to $\cL (m)$ in eq.~\eqref{eq:TS23} the constant $-\ln c$. The eigenvalues of $\hat{S}_+$ are
\begin{equation}\label{eq:SE13C}
\lambda_+ = ( 1,\, - \tanh \beta )\, ,
\end{equation}
while one finds for $\hat{S}_-$ the eigenvalues
\begin{equation}\label{eq:SE13D}
\lambda_- = (1,\, \tanh\beta ) \, .
\end{equation}
For $\beta \to \infty$ the second eigenvalue approaches $-1$ for $\lambda_+$ and $+1$ for $\lambda_-$. As it should be, the largest eigenvalue obeys $|\lambda| = 1$.

One can verify directly that the condition \eqref{eq:SE9} is not obeyed for finite $\beta$. A Markov chain can at best become an approximation to the evolution in the Ising chain, with wave functions obeying particular conditions that we will discuss in sect.\,\ref{sec:markov_chains}.

\paragraph*{Markov chains}

Without restrictions on the wave function general step evolution operators do not lead to Markov chains. 
In the other direction, general
Markov chains cannot be realized by step evolution operators
of local chains with the same number of Ising spins. This is another facet of the observation that in general the probabilistic information
in the local probability distribution $p_1(m)$ is insufficient for the formulation of an evolution law. 

The condition \eqref{eq:SE9} for Markov chains is only obeyed for a special class of step evolution operators. In particular, one needs for arbitrary $\tau$, $\rho$, $\sigma$ the relation
\begin{equation}\label{eq:SE14}
\hat{S}_{\rho\tau}^{-1} \, \hat{S}_{\tau\sigma} = 0 \quad \text{ for } \rho \neq 
\sigma\, .
\end{equation}
Basic issues can be understood for $(2 \times 2)$-matrices, that we parametrize by
\begin{equation}\label{eq:SE15}
\hat{S} = \begin{pmatrix}
a & c \\ d & b
\end{pmatrix}\, , \quad
\hat{S}^{-1} = \frac{1}{ab - cd} \begin{pmatrix}
b & -c \\ -d & a
\end{pmatrix}\, .
\end{equation}
We assume here regular matrices $\hat{S}$ such that $ab - cd \neq 0$. For $a = \hat{S}_{11} \neq 0$ one needs $\hat{S}_{21}^{-1} = 0$ or $d = 0$. By similar considerations one finds
\begin{align}\label{eq:SE16}
& a \neq 0 \,\Rightarrow \, d = 0\, , \quad b \neq 0\,\Rightarrow\, c = 0\, , \notag \\
& c \neq 0 \, \Rightarrow \, b=0\, , \quad d \neq 0\, \Rightarrow \, a = 0\, . 
\end{align}
The only possible regular matrices obeying these conditions are
\begin{align}\label{eq:SE17}
\hat{S}_1 = \begin{pmatrix}
a & 0 \\ 0 & b
\end{pmatrix}\, , \quad
\hat{S}_2 = \begin{pmatrix}
0 & c \\ d & 0
\end{pmatrix}\, .
\end{align}

If the condition \eqref{eq:SE14} holds, the transition probabilities are given by (no index sums)
\begin{equation}\label{eq:SE18}
W_{\tau\rho}^{(M)} = ( \hat{S}^{-1} )_{\rho\tau} \, \hat{S}_{\tau\rho} = \hat{S}_{\tau\rho}\,
(\hat{S}^\tp)^{-1}_{\tau\rho} \, .
\end{equation}
The transition probabilities corresponding to $\hat{S}_1$ and $\hat{S}_2$ are given, respectively, by
\begin{equation}\label{eq:SE19}
W_1 = \begin{pmatrix}
1 & 0 \\ 0 & 1
\end{pmatrix}\, , \quad
W_2 = \begin{pmatrix}
0 & 1 \\ 1 & 0
\end{pmatrix}\, .
\end{equation}
The result \eqref{eq:SE19} holds independently of the normalization of the step evolution operator. Normalization requires for $\hat{S}_1$ either $| a | = 1$, $|b| \leq 1$ or $|b|=1$, $|a| \leq 1$, while $\hat{S}_2$ needs $|cd| = 1$, such that the eigenvalues for real $\hat{S}_2$ are $\lambda = \pm 1$ for $cd = 1$ and $\lambda = \pm \im$ for $cd = -1$.

Several conclusions can be drawn from this simple finding. First, local chains or generalized Ising chains are, in general, not Markov chains. The step evolution operator \eqref{eq:SE13A} for the Ising chain is for finite $\beta$ not of the form \eqref{eq:SE17} and does not obey the condition \eqref{eq:SE14}. For generic local chains the formulation of a local evolution law has to employ local probabilistic information beyond the probability distributions. 
Only for $\beta \to\infty$ the step evolution operator 
$\hat{S}_+$ in eq.\,\eqref{eq:SE13A}
is of the form $\hat{S}_2$, 
while $\hat{S}_-$
approaches $\hat{S}_1$, with $c=d=1$ or $a=b=1$, respectively.

Second, the transition probabilities $W_1$ and $W_2$ have the same form as for the change of local probabilities by unique jumps. For $\hat{S}_1$ the local probabilities do not change, while for $\hat{S}_2$ the local probabilities $p_1$ and $p_2$ are switched at every step. The general form of a Markov chain for $M=1$ is given by 
\begin{equation}\label{eq:SE20}
W^{(M)} = \begin{pmatrix}
A & 1 - B \\
1 - A & B
\end{pmatrix}\, ,
\end{equation}
with $0 \leq A \leq 1$, $0 \leq B \leq 1$. Only the limits $A=B=1$ and $A = B = 0$ are realized by the evolution of local chains. Thus generic Markov chains cannot be realized by the evolution of local probabilities in local chains.

The structure above generalizes to larger matrices for the transition probabilities. For example, for $(4 \times 4)$-matrices the step evolution operator
\begin{equation}\label{eq:SE20A}
\hat{S} = \begin{pmatrix}
0 & a & 0 & 0 \\
0 & 0 & b & 0 \\
0 & 0 & 0 & c \\
d & 0 & 0 & 0 
\end{pmatrix}\, , \quad
\hat{S}^{-1} = \begin{pmatrix}
0 & 0 & 0 & d^{-1} \\
a^{-1} & 0 & 0 & 0 \\
0 & b^{-1} & 0 & 0 \\
0 & 0 & c^{-1} & 0
\end{pmatrix}
\end{equation}
obeys the condition \eqref{eq:SE14}. The corresponding matrix $W$ according to eq.~\eqref{eq:SE18} reads
\begin{equation}\label{eq:SE20B}
W^{(M)} = \begin{pmatrix}
0 & 1 & 0 & 0 \\
0 & 0 & 1 & 0 \\
0 & 0 & 0 & 1 \\
1 & 0 & 0 & 0
\end{pmatrix}\, .
\end{equation}
This corresponds again to the transition probabilities for a unique jump chain. This property holds for all matrices $\hat{S}$ for which each row and each column contains only a single nonzero element.

We will see in sect.\,\ref{sec:markov_chains} that often Markov chains do not describe the evolution for arbitrary local probabilistic states. They are typically realized only for wave functions obeying constraints. In this case the condition \eqref{eq:SE9} is weakened.

\paragraph*{Unique jump chains}

For a unique jump chain the step evolution operator takes a very simple form. Every state $\rho$ at site $m$ is mapped uniquely to a state $\tau(\rho)$ at site $m+1$. The overall probability distribution vanishes if the configuration of occupation numbers $\{ n(m) \}$ corresponds to the state $\rho$, while $\{n(m+1)\}$ differs from the one corresponding to the state $\tau (\rho)$. This property is realized by a local factor
\begin{equation}\label{eq:SE21}
\cK (m) = \sum_\rho h_{\tau(\rho)} (m+1) \, h_\rho (m)\, .
\end{equation}
Each term for a given state $\rho$ at $m$ vanishes for all states $\tau$ at $m+1$ that differ from $\tau(\rho)$. Indeed, $h_\rho (m)$ equals one precisely for the configuration of occupation numbers corresponding to $\rho$, and vanishes for all other configurations, and similar for $h_{\tau(\rho)} (m+1)$. If the configuration at $m+1$ corresponds to $\tau(\rho)$ one has $h_{\tau(\rho)} (m+1)\, h_\rho (m) = 1$.

Writing eq.~\eqref{eq:SE21} as
\begin{align}\label{eq:SE22}
\cK (m) &= \sum_{\tau, \rho} \delta_{\tau,\, \tau(\rho)} \, h_\tau (m+1) \, h_\rho (m) \notag \\
&= \sum_{\tau, \rho} \hat{S}_{\tau\rho}\, h_\tau (m+1) \, h_\rho (m)\, ,
\end{align}
we can extract the step evolution operator
\begin{equation}\label{eq:SE23}
\hat{S}_{\tau\rho} = \delta_{\tau,\, \tau(\rho)}\, .
\end{equation}
This matrix has a one in each column $\rho$ in the row $\tau(\rho)$. It is invertible if there is only a single one in each row. We will restrict the discussion to invertible unique jump chains for which there is precisely one element $1$ in each row and column of $\hat{S}$, while all other matrix elements are $0$. 

The step evolution operators for invertible unique jump chains are called ``unique jump operators''. The unique jump operators are orthogonal matrices, as follows from
\begin{align}\label{eq:SE24}
\hat{S}^\tp_{\sigma\tau} \hat{S}_{\tau\rho} &= \sum_\tau \hat{S}_{\tau\sigma} \hat{S}_{\tau\rho} \notag \\
&= \sum_\tau \delta_{\tau,\, \tau(\sigma)} \delta_{\tau,\, \tau(\rho)} = 
\delta_{\sigma\rho}\, .
\end{align}
As for all orthogonal matrices, the eigenvalues of $\hat{S}$ obey all
\begin{equation}\label{eq:SE25}
|\lambda_i| = 1\, .
\end{equation}

For invertible unique jump chains the evolution of the classical wave functions is very simple. Both $\tilde{q}$ and $\bar{q}$ have the same evolution law, since $(\hat{S}^\tp)^{-1} = \hat{S}$ for the conjugate wave function. One finds
\begin{equation}\label{eq:SE25A}
\tilde{q}_\tau (m+1) = \hat{S}_{\tau\rho} (m) \, \tilde{q}_\rho = \sum_\rho 
\delta_{\tau,\, \tau(\rho)} \tilde{q}_\rho (m) = \tilde{q}_{\rho(\tau)} (m)\, ,
\end{equation} 
and the same for $\bar{q}$. Here $\rho(\tau)$ is given by the inverse of the map $\tau(\rho)$. In other words, the $\tau$-component of $\tilde{q}(m+1)$ equals precisely the particular component of $\tilde{q}(m)$ that is mapped to $\tau$. 

Since both $\bar{q}$ and $\tilde{q}$ follow the same evolution, one obtains for the local probabilities (no sum over $\tau$)
\begin{align}\label{eq:SE26}
p_\tau (m+1) &= \bar{q}_\tau (m+1) \, \tilde{q}_\tau (m+1) \notag \\
& = \bar{q}_{\rho(\tau)} (m) \, \tilde{q}_{\rho(\tau)} (m) = p_{\rho(\tau)} (m)\, ,
\end{align}
which is equivalent to
\begin{equation}\label{eq:SE27}
p_{\tau(\rho)} (m+1) = p_\rho (m)\, .
\end{equation}
This proves the relation \eqref{eq:EV7}. We can also write eq.~\eqref{eq:SE26} as
\begin{equation}\label{eq:SE28}
p_\tau (m+1) = \hat{S}_{\tau\rho} (m)\, p_\rho (m)\, .
\end{equation}
Thus unique jump chains are particular limiting cases of Markov chains, with $M_{\tau\rho} = \hat{S}_{\tau\rho}$.

For unique jump chains the structure of the overall probability distribution is very simple. For every initial configuration $\rho$ at $m=0$ we can define a trajectory in configuration space. It obtains by the invertible map $\tau(\rho)$ applied at every time step. Every allowed trajectory may be labeled by its initial configuration $\rho$. The family of these trajectories defines the ``allowed overall configurations''. For all overall configurations not belonging to the family of allowed trajectories the overall probability distribution vanishes. The non-vanishing overall probabilities are given by the probabilities for the different allowed trajectories. They can be labeled by the probabilities for the initial configurations $p_{\rho}(0)=\overline{q}_{\rho}(0)\tilde{q}_{\rho}(0)$. Since the probabilities are simply transported to any other $m>0$ by virtue of eq.~\eqref{eq:SE27} we can equally define the probability for a given trajectory by $p_{\overline{\tau}}(m)=p_{\rho}(0)$, where the configuration $\overline{\tau}$ at $m$ is on the same trajectory as a given $\rho$ at $m=0$.

These simple features imply important properties for the classical wave functions of unique jump chains. We may specify the overall probabilities for the allowed trajectories by the time-local probabilities at the final boundary (no sum over $\tau$)
\begin{equation}\label{UJC1}
p_{\tau}(\cM )=\overline{q}_{\tau}(\cM)\tilde{q}_{\tau}(\cM )\, .
\end{equation}
Since these probabilities need to be positive we conclude that $\overline{q}_{\tau}(\cM)$ needs to have the same sign as $\tilde{q}_{\tau}(\cM )$. (We will discuss this more formally in sect.~\ref{sec:free_particles_in_two_dimensions} where a concrete example may help the intuitive understanding.)
With this condition the overall probabilities are positive for arbitrary signs of $\tilde{q}_{\tau}(\cM )$ and therefore also for arbitrary signs of $\tilde{q}_{\rho}(0 )$ or $\tilde{q}_{\overline{\tau}}(m)$.
If $\tilde{q}_{\tau}(\cM )=0$ for some particular configuration $\tau$ we may set $\overline{q}_{\tau}(\cM)=0$ for this configuration as well since non-zero values will not change the vanishing probability $p_{\tau}(\cM)=0$ for to the corresponding trajectory. For nonzero $\tilde{q}_{\tau}(\cM)$ we can take $\overline{q}_{\tau}(\cM)=c_{\tau}^{2}\tilde{q}(m)$, $c_{\tau}>0$.
Since $\tilde{q}_{\tau}(m)$ and $\overline{q}(m)$ obey the same evolution law one has the same relation for all $m$
\begin{equation}\label{ULJ2}
\overline{q}_{\tau}(m)=c_{\tau}^{2}\tilde{q}_{\tau}(m)\, .
\end{equation}

Without changing the overall probability distribution we can rescale the wave functions in the initial and final boundary terms, replacing $\tilde{q}_{\tau}(0)$ by $\tilde{q}_{\tau}'(0)=q_{\tau}(0)$ and modifying the final boundary term such that also $\overline{q}_{\tau}(0)$ changes to $\overline{q}_{\tau}'(0)=q_{\tau}(0)$ (no sum over $\tau$), 
\begin{align}
\label{UJC3}
q_{\tau}(0)&=c_{\tau} \tilde{q}_{\tau}(0)=\frac{1}{c_{\tau}}\overline{q}_{\tau}(0)\, ,\notag \\
p_{\tau}(0)=q_{\tau}^{2}(0)&=\overline{q}_{\tau}(0)\tilde{q}_{\tau}(0)=c_{\tau}^{2}\tilde{q}_{\tau}^{2}(0)\, .
\end{align}
With these rescalings one has for all m
\begin{equation}
\label{UJC4}
\overline{q}_{\tau}'(m)=\tilde{q}_{\tau}'(m)=q_{\tau}(m)\, .
\end{equation}

We conclude that for unique jump chains the choice $\overline{q}(m)=\tilde{q}(m)=q(m)$ does not require a particular selection of the overall probability distribution. It rather amounts to a suitable distribution of the probabilistic information to the boundary factors which leaves the overall probability distribution unaffected. For pure classical states of unique jump chains only a single wave function $q(m)$ is needed. The local probabilistic information is substantially reduced as compared to the general classical statistical case with two distinct wave functions $\tilde{q}(m)$ and $\overline{q}(m)$. Still, the local probabilistic information exceeds the one stored in the time-local probability distribution since it also contains the signs $q_{\tau}=s_{\tau}\sqrt{p_{\tau}}$, $s_{\tau}=\pm 1$.
The signs $s_{\tau}$ do not affect the expectation values of local observables which can be formed as functions of occupation numbers at $m$. They will play a role, however, for new sets of probabilistic observables that we will discuss later. For unique jump chains the choice of signs for the wave function $q(m)$ does not affect the overall probability distribution. Every allowed overall configuration picks up a factor $q_\tau$ from the initial boundary term and another factor $q_\tau$ from the final boundary term.

\subsubsection{Influence of boundary conditions}\label{sec:influence_of_boundary_conditions}

We have set up the local chains with boundaries at $m=0$ and $m=\cM$, and a ``bulk'' for the values of $m$ in between. We want to investigate how the probabilistic information contained in the boundary term $\cB$ ``propagates'' into the bulk. In other words, we want to understand how the expectation values of observables inside the bulk depend on the boundary term. 

The response of bulk expectation values to boundary conditions characterizes different types of
time evolution. Imagine that far inside the bulk, for large values of the distances from the boundary, given by $m$ and $\cM - m$, the probability distribution reaches a unique equilibrium distribution. In this case the boundary information is ``lost'' or ``forgotten'' as the evolution proceeds inside the bulk. On the other hand, for unique jump chains the boundary information is not lost arbitrarily far inside the bulk. The initial configuration propagates in a processed way to arbitrarily high $m$. While the concept of time can be formulated for both cases and the formalism is the same, some notion of ``physical time'' is rather associated to the second case where a nontrivial evolution takes on forever. It is our task to find the general conditions for boundary information to propagate into the bulk. 

The classical wave functions are a very convenient tool for an investigation of the "boundary problem". The boundary problem is also
of relevance for many practical problems  -- for example the storage of information in some memory. Our formalism with classical wave
functions offers a systematic way for its solution.

\paragraph*{Initial value problem}

The linear evolution law \eqref{eq:SE3}, \eqref{eq:SE5} for the pair of classical wave functions formulates the boundary problem in terms of some type of ``discrete differential equation''. The ``initial value'' for $\tilde{q}$ is given by $\tilde{q}(0) = \tilde{q}_{in}$, which encodes the ``initial boundary  factor'' $\tilde{f}_{in} (n_{in})$ according to eq.~\eqref{eq:CW8}, by using the expression
\begin{equation}\label{eq:BC1}
\tilde{f}_{in} (n(m=0)) = \tilde{f}(0) = \tilde{q}_\tau (0)\, h_\tau(0)\, . 
\end{equation}
Using eq.~\eqref{eq:SE3} we can compute $\tilde{q}(m)$ for increasing $m$. Similarly, the boundary value for the conjugate wave function is given by the ``final value'' $\bar{q}(\cM) = \bar{q}_f$, resulting from the expansion of the ``final boundary factor'' $\bar{f}_f(n_f)$,
\begin{equation}\label{eq:BC2}
\bar{f}_f(n(m = \cM)) = \bar{f}(\cM) = \bar{q}_\tau(\cM)\, h_\tau (\cM)\, .
\end{equation}
Starting from there, one can use eq.~\eqref{eq:SE5} to compute $\bar{q}(m)$ for decreasing $m$. We observe that the boundary conditions for $\tilde{q}$ and $\bar{q}$ are given at the opposite ends of the chain.

The stepwise evolution is the discrete version of a first order differential evolution equation. We will obtain in sect.~\ref{sec:continuous_time} such a differential evolution equation in the limit $t= \varepsilon\, (m - m_0)$, $\varepsilon\to 0$, for chains with an infinite number of sites. Similar to a first-order differential equation the value of the function at the boundary is sufficient to determine the solution for all $m$ uniquely, provided that the boundary values $\tilde{q}_{in}$ and $\bar{q}_f$ of the classical wave functions are given. For a regular step evolution operator $\hat{S}$ the solutions $\tilde{q}(m)$ and $\bar{q}(m)$ exist for all $m$. In a matrix notation, with $\tilde{q}$ and $\bar{q}$ vectors, the solution reads formally
\begin{align}\label{eq:BC3}
\tilde{q}(m) &= \hat{S} (m-1)\, \hat{S}(m-2) \, \cdots \, \hat{S}(0)\, \tilde{q}(0)\, ,
\notag \\
\bar{q}^\tp (m) &= \bar{q}^\tp (\cM)\, \hat{S}(\cM - 1)\, \hat{S}(\cM - 2) \, \cdots
\, \hat{S}(m)\, . 
\end{align}
Solving the evolution equation for $\tilde{q}(m)$ and $\bar{q}(m)$ we can compute the expectation values of local observables 
$A(m)$ in the bulk for arbitrary boundary conditions. This is the formal solution of the boundary problem.

\paragraph*{Boundary value problem for the Ising chain}

As an example we solve the boundary value problem for the Ising chain by use of the classical wave functions. We follow here closely ref.~\cite{CWIT} and discuss the solution in some detail. This demonstrates the procedure for more general cases. We consider the attractive Ising chain with $\kappa = -1$ in eq.~\eqref{eq:SE13A}, for which the step evolution operator
\begin{equation}\label{eq:BC4}
\hat{S} = \frac{1}{2\cosh\beta} \begin{pmatrix}
\text{e}^\beta & \text{e}^{-\beta} \\
\text{e}^{- \beta} & \text{e}^\beta
\end{pmatrix}
\end{equation}
has the eigenvalues
\begin{equation}\label{eq:BC5}
\lambda_1 = 1\, , \quad \lambda_2 = \tanh\beta\, .
\end{equation}

The equilibrium wave functions are the eigenfunctions to the eigenvalue $\lambda_1 = 1$,
\begin{equation}\label{eq:BC6}
\tilde{q}_{eq} = \frac{1}{\sqrt{2}} \begin{pmatrix}
1 \\ 1
\end{pmatrix}\, , \quad  \bar{q}_{eq} = \frac{1}{\sqrt{2}} \begin{pmatrix}
1 \\ 1
\end{pmatrix}\, .
\end{equation}
They obey
\begin{equation}\label{eq:BC7}
\hat{S}\, \tilde{q}_{eq} = \tilde{q}_{eq}\, , \quad \bar{q}_{eq}^\tp\, \hat{S} = 
\bar{q}_{eq}^\tp\, .
\end{equation}
For $\tilde{q}_{in} = \tilde{q}_{eq}$ one has therefore $\tilde{q}(m) = \tilde{q}_{eq}$ for all $m$. Similarly $\bar{q}_f = \bar{q}_{eq}$ implies $\bar{q}(m) = \bar{q}_{eq}$ for all $m$. If both $\tilde{q}$ and $\bar{q}$ are given by the equilibrium wave function one finds for the local probabilities
\begin{align}\label{eq:BC8}
& p_1 (m) = \bar{q}_1 (m)\, \tilde{q}_1 (m) = \frac{1}{2}\, , \notag \\
& p_2 (m) = \bar{q}_2 (m)\, \tilde{q}_2 (m) = \frac{1}{2}\, .
\end{align}
For the equilibrium the probability to find $s(m) = 1$ is the same as for $s(m) = -1$, and therefore $\langle s \rangle = 0$.

There are various ways to set up different boundary conditions. For example, we may consider open final boundary conditions where no spin direction of $s(\cM)$ is preferred by the ``final'' boundary term. This amounts to $\bar{q}(\cM) = \bar{q}_{eq}$. We may then fix $s(0)$, or the expectation value $\langle s(0)\rangle$. Alternatively, we may impose boundary conditions at both ends of the chain by specifying both $\langle s(0) \rangle$ and $\langle s(\cM) \rangle$.

For an investigation of the boundary problem it is convenient to parameterize
\begin{align}\label{eq:BC9}
& \tilde{q}_{in} = \tilde{q}(0) = c\, (\tilde{q}_{eq} + \delta\tilde{q}_{in} )\, ,     
\notag \\
& \bar{q}_f = \bar{q} (\cM) = \bar{c}\, (\bar{q}_{eq} + \delta \bar{q}_f )\, ,
\end{align}
with $\delta \tilde{q}$ and $\delta \bar{q}$ the eigenvectors of the second eigenvalue of $\hat{S}$,
\begin{equation}\label{eq:BC10}
\hat{S}\, \delta \tilde{q}(m) = \lambda_2 \, \delta\tilde{q}(m)\, ,\quad \delta 
\bar{q}^\tp (m)\, \hat{S} = \lambda_2\, \delta \bar{q}^\tp (m)\, .
\end{equation}
These eigenvectors are given by
\begin{equation}\label{eq:BC11}
\delta \tilde{q} (m) = a (m) \begin{pmatrix}
1 \\ -1
\end{pmatrix}
\, , \quad \delta \bar{q}^\tp (m) = \bar{a} (m) \left( 1, -1\right) \, .
\end{equation}
The two eigenvectors evolve separately, and one finds the general solution
\begin{align}
\tilde{q} (m) &= c\, ( \tilde{q}_{eq} + \lambda_2^m\, \delta \tilde{q}_{in} ) \, , 
\notag \\
\bar{q} (m) &= \bar{c}\, \big( \bar{q}_{eq} + \lambda_2^{(\cM - m)}\, \delta \bar{q}_f 
\big) \, .
\end{align}
For any finite $\beta$ one has $\lambda_2 < 1$. The deviations $\delta \tilde{q}$, $\delta \bar{q}$ from the equilibrium wave functions shrink as one moves into the bulk.

We can associate $\lambda_2$ to the ``correlation time'' or ``correlation length'' $\xi$ of the Ising chain. With
\begin{equation}\label{eq:BC13}
t = \varepsilon m \, , \quad t_f = \varepsilon \cM\,,
\end{equation}
we write
\begin{equation}\label{BC14}
\lambda_2^m = \exp \left\{ - \frac{t}{\xi} \right\}\, , \quad \lambda_2^{(\cM - m)} =
\exp \left\{ - \frac{(t_f - t)}{\xi}  \right\} \,.
\end{equation}
The correlation length is identified by
\begin{align}\label{eq:BC15}
& \xi^{-1} = - \frac{\ln \lambda_2}{\varepsilon}\, , \notag \\
& \xi = \frac{\varepsilon}{\ln (1/\lambda_2)} = \frac{\varepsilon}{\ln (\coth    
	\beta)}\, .
\end{align}
Here $\varepsilon$ only plays the role of setting units for time or length. In these units the general solution reads
\begin{align}\label{eq:BC16}
\tilde{q}(t) &= c \left(\tilde{q}_{eq} + \exp \left( - \frac{t}{\xi} \right) \delta 
\tilde{q}_m \right)\, , \notag \\
\bar{q}(t) &= \bar{c} \left( \bar{q}_{eq} + \exp \left( - \frac{t_f - t}{\xi} \right) 
\delta \bar{q}_f \right)\, .
\end{align}

We next compute the local probabilities, which directly yield the expectation value of $s(m)$. With
\begin{equation}\label{eq:BC17}
\delta \tilde{q}_{in} = \frac{a_{in}}{\sqrt{2}} \begin{pmatrix}
1 \\ -1
\end{pmatrix}\, , \quad
\delta \bar{q}_f = \frac{a_f}{\sqrt{2}}
\begin{pmatrix}
1 \\ -1
\end{pmatrix}\,,
\end{equation}
one finds for the local probabilities
\begin{align}\label{eq:BC18}
p_1 (m) &= \bar{q}_1(m)\, \tilde{q}_1 (m) = \frac{\bar{c} c}{2} 
( 1 + \lambda_2^{(\cM - m)}\, a_f ) ( 1 + \lambda_2^m\, a_{in} )\, , \notag \\
p_2(m) &= \bar{q}_2 (m) \, \tilde{q}_2 (m) = \frac{\bar{c} c}{2} 
( 1 - \lambda_2^{(\cM - m)}\, a_f ) ( 1 - \lambda_2^m\, a_{in} )\, .
\end{align}
The normalization condition
\begin{equation}\label{eq:BC19}
p_1 (m) + p_2 (m) = \bar{c} c\, ( 1 + \lambda_2^{\cM}\, a_f\, a_{in} ) = 1
\end{equation}
holds independently of $m$. It fixes $\bar{c} c$ as a function of $a_f\, a_{in}$. The expectation value of the local spin at $m$ is given by
\begin{align}\label{eq:BC20}
\langle s(m) \rangle &= p_1 (m) - p_2 (m) \notag \\[4pt]
&= \bar{c} c\, \big( \lambda_2^{(\cM - m)}\, a_f + \lambda_2^m\, a_{in} \big) 
\notag \\[4pt]
&= \frac{\lambda_2^{(\cM - m)}\, a_f + \lambda_2^m\, a_{in}}
{1 + \lambda_2^{\cM}\, a_f\, a_{in}} \, .
\end{align}
This is the general solution of the boundary problem.

We may express $a_{in}$ and $a_f$ by the expectation values of the initial and final spin
\begin{align}\label{eq:BC21}
\langle s(0) \rangle &= s_0 = \frac{a_{in} + \lambda_2^{\cM}\, a_f}
{1 + \lambda_2^{\cM}\, a_f\, a_{in}}\, , \notag \\[4pt]
\langle s(\cM) \rangle &= s_f = \frac{a_f + \lambda_2^{\cM}\, a_{in}}
{1 + \lambda_2^{\cM}\, a_f \, a_{in}}\, .
\end{align}
For open final boundary conditions one has $a_f = 0$, $\langle s(0) \rangle = a_{in}$ and therefore
\begin{align}
\langle s(m) \rangle &= \lambda_2^m \, \langle s(0) \rangle = (\tanh \beta)^m \, 
\langle s(0) \rangle \, , \notag \\
\langle s(t) \rangle &= \exp \left\{ - \frac{t}{\xi} \right\}\, \langle s(0) \rangle
\, .
\end{align}
For any finite $\beta$ or finite correlation length $\xi$ this describes an exponential approach to the equilibrium state, or an exponential ``loss of initial information'', as one moves away from the boundary further inside the bulk. For $\beta\to \infty$ one has $\xi \to \infty$. This limit corresponds to a unique jump chain. The initial $\langle s(0) \rangle$ is preserved, $\langle s(m) \rangle = \langle s(0) \rangle$.

For general boundary conditions with both $a_{in}$ and $a_f$ different from zero we consider first a large length of the chain as compared to the correlation length, or the limit
\begin{equation}\label{eq:BC23}
\lambda_2^{\cM} \ll 1\, .
\end{equation}
In this case one has
\begin{equation}\label{eq:BC24}
\langle s(0) \rangle = a_{in}\, , \quad \langle s(\cM)\rangle = a_f\, .
\end{equation}
The spin expectation value in the bulk decays exponentially with the distance from both boundaries,
\begin{equation}\label{eq:BC25}
\langle s(m) \rangle = \lambda_2^m \, \langle s(0) \rangle + \lambda_2^{\cM - m}\,
\langle s(\cM)\rangle\, ,
\end{equation}
or
\begin{equation}\label{eq:BC26}
\langle s(t) \rangle = \exp \left( - \frac{t}{\xi} \right)\, s_0 + 
\exp \left( - \frac{t_f - t}{\xi} \right)\, s_f\, .
\end{equation}

For the general case we parameterize
\begin{equation}\label{eq:BC27}
\lambda_2^{\cM} = 1 - \delta\, , \quad 0 \leq \delta \leq 1\, ,
\end{equation}
such that
\begin{align}\label{eq:BC28}
s_{in} + s_f &= \frac{(2 - \delta) (a_{in} + a_f)}{1 + a_{in}\, a_f - \delta\, 
	(a_{in}\, a_f)}\, , \notag \\
s_{in} - s_f &= \frac{\delta\, (a_{in} - a_f)}
{1 + a_{in}\, a_f - \delta\, a_{in}\, a_f}\, .
\end{align}
Inverting this relation yields $a_{in}$ and $a_f$ as a function of $s_{in}$ and $s_f$. Eq.~\eqref{eq:BC20} yields then the
expectation value $\braket{s(m)}$ in the bulk as a function of expectation values of the boundary spins.

As an example, we may discuss opposite spin expectation values at the boundaries, $s_f = - s_{in}$. In this case one has
\begin{equation}\label{eq:BC29}
s_f = - s_{in} \; \Rightarrow \; a_f = - a_{in} = - a\,,
\end{equation}
and therefore
\begin{equation}\label{eq:BC30}
s_{in} = \frac{\delta a}{1 - a^2\, (1 - \delta)}\, .
\end{equation}
One infers
\begin{align}\label{eq:BC31}
\langle s(m) \rangle &= \frac{s_{in}}{\delta} \, ( \lambda_2^m - \lambda_2^{\cM - m} )
\notag \\
&= \frac{s_{in}}{\delta} \, 
\left[ (1 - \delta)^{\frac{m}{\cM}} - (1 - \delta)^{\frac{\cM - m}{\cM}} \right]\, .
\end{align}
In the limit $\delta \to 0$ this yields a linear decrease of $\langle s(m) \rangle$ from $s_{in}$ to $s_f$,
\begin{equation}\label{eq:BC32}
\langle s (m) \rangle = s_{in}\, \left( 1 - \frac{2m}{\cM} \right)\, .
\end{equation}
From eq.~\eqref{eq:BC30} we infer that for $\delta \to 0$ the initial wave function is given by
\begin{equation}\label{eq:BC33}
a = 1 - \frac{1 - s_{in}}{2\, s_{in}}\, \delta\, .
\end{equation}
We conclude that the solution of the boundary value problem is rather simple in terms of the classical wave functions.

We could solve the boundary value problem equivalently by use of the transfer matrix. For this purpose one uses the trace identity \eqref{eq:LO5} for the expectation value of the local spin operator. Transforming to a basis where $\hat{T}$ is diagonal yields the same result. In a certain sense the transfer matrix formalism corresponds to the Heisenberg picture in quantum mechanics, while the use of the classical wave functions constitutes the associated Schrödinger picture. Depending on the problem one or the other of these equivalent formulations may be more convenient. The use of wave functions has the advantage that the local probabilistic information is directly available for every site $m$.
This is of great help if the step evolution operator depends on $m$, or for many cases where approximations are needed.

\paragraph*{Loss of memory}

We have found systems with different qualitative behavior of the solutions of the boundary problem. For the Ising chain with finite
$\beta$ the boundary information is lost far inside the bulk. For unique jump chains the memory of the boundary information is
preserved inside the bulk. We will generalize these cases and discuss situations inbetween the extremes. This issue is crucial
for physical time since it decides on the possibility of oscillating behavior.

We concentrate for this discussion on step evolution operators $\hat{S}$ which do not depend on $m$.
Let us consider infinite chains, $\cM \to \infty$, or chains with very large $\cM$. For Ising chains with a given finite $\beta$, and therefore a second eigenvalue $\lambda_2$ of the step evolution operator $\hat{S}$ obeying $|\lambda_2| < 1$, this implies $\lambda_2^{\cM} \to 0$. For values of  $m$ far enough inside the bulk one also has $\lambda_2^m \to 0$, $\lambda_2^{\cM - m} \to 0$. For this part of the chain the classical wave functions have reached the equilibrium values $\tilde{q}_{eq}$ and $\bar{q}_{eq}$, and therefore also the local probability distribution becomes the equilibrium distribution. Far enough inside the bulk the system has ``lost memory'' of the boundary conditions. More precisely, the boundary term plays no longer a role when the distances $\Delta t_{in} = \varepsilon\, m$ from the initial time, and $\Delta t_f = \varepsilon\, (\cM - m)$ from the final time, both exceed the correlation time $\xi$ by many factors. Boundary effects are then suppressed exponentially by $\exp ( -\Delta t / \xi )$.

This generalizes to the ``bulk weight distribution'' or ``bulk probability distribution''. The bulk weight distribution is defined by integrating out the spins for a sufficient number of sites close to the boundaries,
\begin{equation}\label{eq:BC34}
w[n_{\text{bulk}} ] = \prod_{m' = 0}^{\bar{m} - 1}\, \int \cD n (m')\
\prod _{m'' = \cM - \bar{m} + 1}^{\cM} \cD n (m'')\, w[n]\, .
\end{equation}
It depends on the occupation numbers $n(m)$ for $\bar{m} \leq m \leq \cM - \bar{m}$. The integration over the products of local factors in the ranges of $m'$ between $0$ and $\bar{m} - 1$, or $m''$ between $\cM - \bar{m}$ and $\cM - 1$, can be represented as products of step evolution operators. For large enough $\bar{m}$, such that $\lambda_2^{\bar{m}} \approx 0$, only the eigenvectors to the largest eigenvalue $\lambda_1 = 1$ survive for the wave functions. This results in a reduced bulk system for which the ``boundary wave functions'' are now given by the equilibrium wave function, $\tilde{q} (\bar{m}) = q_{eq}$, $\bar{q}(\cM - \bar{m}) = \bar{q}_{eq}$. If one ``waits long enough'', the system is found in its equilibrium state. 

Not only the expectation values of local observables, but also correlations of observables at different $m$, $m'$ inside the bulk
can be calculated for the equilibrium state. This is a typical situation for a canonical ensemble in thermal equilibrium, with 
$\beta$ proportional to the inverse temperature. Indeed, we can compute the probability distribution for all configurations
$\{s(m')\}$ with $m'$ inside the bulk. It replaces the overall probability distribution for the corresponding range of $m'$. It is
the same as for the overall probability distribution with equilibrium boundary conditions. For equilibrium boundary conditions the size
of the bulk does not matter. One has $\hat{S}\tilde{q}_\text{eq} = \tilde{q}_\text{eq}$ and similar for $\bar{q}_\text{eq}$ for all $m$.

This ``loss of memory of boundary information'' generalizes immediately to more complex systems. Assume that the step evolution operator $\hat{S}$ has a unique eigenvalue $\lambda_1 = 1$, and all other eigenvalues $\lambda_k$ obey $|\lambda_k| \leq 1 - g$, with $g>0$ a finite ``gap'' between the largest eigenvalue and the smaller eigenvalues $\lambda_k$. Let us denote the second largest eigenvalue by $\lambda_2$, with $|\lambda_2| = 1 - g$. Sufficiently far inside the bulk, $\lambda_2^{\bar{m}} \approx 0$, the boundary information is again lost. The bulk system is in a unique equilibrium state, for which at $\bar{m}$ and $\cM - \bar{m}$ the effective boundary wave functions are the equilibrium wave functions $\tilde{q}_{eq}$ and $\bar{q}_{eq}$, obeying
\begin{align}\label{eq:BC35}
& \hat{S}\, \tilde{q}_{eq} = \tilde{q}_{eq}\, , \quad \bar{q}^\tp_{eq}\, \hat{S} = 
\bar{q}^\tp_{eq}\, , \notag \\ 
& \tilde{q} (\bar{m}) = \tilde{q}_{eq}\, , \quad  \bar{q}(\cM - \bar{m}) = \bar{q}_{eq}\, .
\end{align}
If there are several different smaller eigenvalues the loss of boundary information may proceed in steps. Each smaller eigenvalue $\lambda_k$ defines a ``partial correlation time'' $\xi_k$. For $\Delta t \gg \xi_k$ the boundary information concerning the eigenvectors to the eigenvalue $\lambda_k$ is lost.

The situation is very different for unique jump chains. In this case all eigenvalues of the step evolution operator obey $|\lambda_i | = 1$. No loss of boundary information occurs. The evolution is a rotation of the wave functions. Since unit jump operators are orthogonal matrices, the norm of the classical wave functions, $|\tilde{q}|^2 = \sum_\tau \tilde{q}_\tau^2 (m)$, $|\bar{q}|^2 = \sum_\tau \bar{q}_\tau^2 (m)$, is independent of $m$ and therefore ``conserved''.

One may also consider mixed situations where a certain number $\bar{N}$ of eigenvalues of $\hat{S}$ obey $|\lambda_i| = 1$, $i=1,\, \dots ,\, \bar{N}$, while the remaining $2^{M} - \bar{N}$ eigenvalues $\lambda_k$ are ``smaller eigenvalues'', $|\lambda_k| < 1$. If there is a gap, $|\lambda_k| \leq 1 - g$, the boundary information concerning the eigenvectors to $\lambda_k$ will be lost inside the bulk. Inside the bulk one encounters a system for which the boundary information concerning all eigenfunctions of the largest eigenvalues $\lambda_i$ is still available. Within this subspace of eigenfunctions the wave function can have a nontrivial evolution, given by a rotation in this subspace. 

Let us denote by $\tilde{q}_{(i)}$, $\bar{q}_{(i)}$ the eigenfunctions to the eigenvalues $\lambda_i$, $|\lambda_i| = 1$,
\begin{equation}\label{eq:BC36}
\hat{S}\, \tilde{q}_{(i)} = \lambda_i\, \tilde{q}_{(i)}\, , \quad
\bar{q}^\tp_{(i)}\, \hat{S} = \lambda_i\, \bar{q}^\tp_{(i)}\, .
\end{equation}
The bulk system \eqref{eq:BC34} will be characterized by effective boundary terms
\begin{equation}\label{eq:BC37}
\tilde{q}(\bar{m}) = a_i(\bar{m}) \, \tilde{q}_{(i)}\, , \quad \bar{q}(\cM - \bar{m}) = 
b_i\, (\cM - \bar{m} )\, \bar{q}_{(i)}\, .
\end{equation}
The coefficients $a_i$ and $b_i$ typically depend on $\bar{m}$ or $\cM - \bar{m}$. 
They contain all the remaining probabilistic information for the bulk. The memory of boundary information stored in eigenfunctions to the eigenvalues $|\lambda_k|\leq 1-g$ is no longer available.

For sites $m$ in the range $\bar{m} \leq m \leq \cM - \bar{m}$ the solution for the wave function will be of the type
\begin{equation}\label{eq:BC38}
\tilde{q} (m) = a_i (m) \, \tilde{q}_{(i)}\, , \quad \bar{q}(m) = \bar{a}_i (m) \, 
\bar{q}_{(i)}\, .
\end{equation}
For a nontrivial solution the coefficients $a_i(m)$, $\bar{a}_i (m)$ will depend on $m$.
They follow an evolution described by a reduced step evolution operator $\hat{S}_{ij}$,
\begin{equation}
a_i(m+1) = \hat{S}_{ij}a_j(m),\quad \bar{a}_i(m+1) = (\hat{S}^\mathrm{T})^{-1}_{ij} \bar{a}_j(m).
\label{eq:SUSYA}
\end{equation}
Formally, the reduced step evolution operator can be found by a similarity transformation that makes $\hat{S}$ block diagonal in the large eigenvalues $|\lambda_i|=1$ and small eigenvalues $|\lambda_k|<1-g$.
The evolution in the bulk is either a rotation or static (e.g. trivial ``rotation"). In the subspace of eigenvectors to eigenvalues
$|\lambda_i| = 1$ the length of the wave function vectors cannot shrink or increase, such that $(\tilde{q}_\tau \tilde{q}_\tau)$
and $(\bar{q}_\tau \bar{q}_\tau)$ remain constant. This permits an oscillatory evolution. 

Physical time for realistic systems 
involves oscillating expectation values. Oscillation periods that are large as compared to $\varepsilon$
become possible for unique jump chains or the mixed case with $\bar{N} \gg 1$.  For $\cM \rightarrow\infty$ the  original boundaries $m=0$, $m=\cM$ may correspond to the infinite past or the infinite future, respectively. Boundary information associated to eigenfunctions of the smaller eigenvalues $\lambda_k$, $|\lambda_k| \leq 1 - g$, is no longer available for the ``present epoch'', which corresponds to a part of the chain sufficiently far away from the boundaries. If there is a complex evolution in the present epoch, the number $\bar{N}$ of largest eigenvalues has to be sufficiently large. 
This also holds for a periodic evolution with a large period.

Finally, we may consider the case where the gap $g$ goes continuously to zero. Very small nonzero $g$ typically requires some tuning of parameters if a model has a finite number $M$ of local Ising spins. For infinite $M$, as realized for many realistic systems, one often encounters a continuous spectrum of eigenvalues without a gap. The two limits $\cM \to \infty$ at fixed $g$ (typically fixed finite $M$), and $g\to 0$ (typically $M \to \infty$) and fixed $\cM$, are different. They do not commute even for the case where at the end both
limits $g \to 0$ and $\cM \to \infty$ are taken. For any arbitrarily large but finite size of chain $\cM$, a continuous spectrum without gap implies that there will always be eigenvalues for which $|\lambda_k|^{\cM}$ is not small, while $|\lambda_k| < 1$. Boundary information concerning their eigenfunctions is still available in the bulk. The issue of loss of memory is then more complex. Typical physical systems with such a behavior are critical phenomena for phase transitions of second order, for which the correlation length diverges.
\subsubsection{Classical density matrix}\label{sec:classical_density_matrix}

The classical density matrix $\rho'$ is for classical probabilistic systems the analogue of the density matrix $\rho$ for quantum systems. It is defined for every site of a local chain as a $(2^M\times 2^M)$-matrix $\rho'(m)$, with generalizations to matrix chains. The elements $\rho'_{\tau\rho} (m)$ contain all the local probabilistic information that is necessary for the formulation of a local evolution law. Its diagonal elements $\rho'_{\tau\tau}(m)$ are the local probabilities $p_\tau (m)$. In general, the evolution law also involves the off-diagonal elements of $\rho'$, and can therefore not be described by the local probability distribution alone. We denote the classical density matrix $\rho'$ with a prime in order to stress that it does not share all the properties of a density matrix in quantum mechanics. In the basis used here it is a real matrix.

For factorizing boundary conditions the classical density matrix can be constructed as a bilinear in the classical wave function 
and the conjugate classical wave function. It is more general, however. Similar to quantum mechanics, it can also account for 
mixed state boundary conditions. In particular, it is suitable for formulating the probabilistic information for subsystems. 
Even if the overall probabilistic system has pure state boundary conditions the classical density matrix for a subsystem is often no longer of the pure state type.
The classical density
matrix is the central object for the description of time-local physics in probabilistic systems.

\paragraph*{Density matrix for pure classical states}

For pure-state boundary conditions \eqref{eq:CWF1} the classical density matrix is defined as the product of the classical wave function $\tilde{q}$ and the conjugate wave function $\bar{q}$
\begin{equation}\label{eq:DM1}
\rho'_{\tau\rho} (m) = \tilde{q}_\tau (m)\, \bar{q}_\rho (m)\, .
\end{equation}
From eq.~\eqref{eq:CW11} one infers that the diagonal elements are the local probabilities (no sum over $\tau$)
\begin{equation}\label{eq:DM2}
p_\tau (m) = \rho'_{\tau\tau} (m)\, .
\end{equation}
In consequence, $\rho'$ is normalized according to
\begin{equation}\label{eq:DM3}
\tr \rho' = 1\, .
\end{equation}
For pure classical states one has the relation
\begin{equation}\label{eq:DM4}
\rho'^2 = \rho'\, .
\end{equation}
This follows from
\begin{equation}\label{eq:DM5}
\rho'_{\tau\sigma} \rho'_{\sigma\rho} = \tilde{q}_\tau \bar{q}_\sigma \tilde{q}_\sigma
\bar{q}_\rho = \tilde{q}_\tau \bar{q}_\rho\, .
\end{equation}
Thus $\rho'$ can only have eigenvalues one and zero. From $\tr \rho' = 1$ one infers that one eigenvalue equals one, and all other eigenvalues equal zero. The eigenvector to the eigenvalue one is given by $\tilde{q}$,
\begin{equation}\label{eq:DM6}
\rho'\, \tilde{q} = \tilde{q}\, ,
\end{equation}
according to
\begin{equation}\label{eq:DM7}
\tilde{\rho}'_{\tau\rho} \tilde{q}_\rho = \tilde{q}_\tau \bar{q}_\rho \tilde{q}_\rho = 
\tilde{q}_\tau\, .
\end{equation}
In general, $\rho'$ is not a symmetric matrix, an exception being $\bar{q} = a\, \tilde{q}$, see below.

\paragraph*{Functional integral for classical density matrix}

We can also represent the classical density matrix as a function of occupation numbers
\begin{equation}\label{eq:DM8}
\rho' (m;\, n(m),\, \bar{n}(m)) = \rho'_{\tau\rho} (m)\, h_\tau [n(m)]\, h_\rho [\bar{n}(m)]\, .
\end{equation}
Here we use for the site $m$ two different sets of occupation numbers $\{ n(m)\}$ and $\{ \bar{n}(m)\}$. In this representation, $\rho' (m)$ is a product of classical wave functions as functions of occupation numbers \eqref{eq:CWF3}, \eqref{eq:CWF4},
\begin{equation}\label{eq:DM9}
\rho' (m;\, n(m),\, \bar{n}(m)) = \tilde{f}[n(m)]\, \bar{f}[\bar{n}(m)]\, .
\end{equation}
The equivalence of the expressions \eqref{eq:DM9} and \eqref{eq:DM1} is easily seen by expanding eq.~\eqref{eq:DM9} in the occupation number basis. 

Insertion of configuration sums for $\tilde{f}[n(m)]$ and $\bar{f}[\bar{n}(m)]$ yields the ``functional integral expression'' for the classical density matrix
\begin{align}\label{eq:DM10}
& \rho' (m;\, n[m],\, \bar{n}[m]) = Z^{-1} \int \cD n(m'\neq m)\, \cB
\prod_{m' \neq m-1,m} \cK (m') \notag \\
& \, \times \cK (m; n(m+1), \bar{n}(m))\, \cK (m-1; n(m), n(m-1))\, .
\end{align}
This expression integrates the weight function over all occupation numbers except $n(m)$, with a replacement $n(m) \to \bar{n}(m)$ in the local factor $\cK(m)$. We have so far discussed the classical wave functions with a normalization of $w[n]$ where $Z = 1$. We have written the normalization factor $Z^{-1}$ explicitly in eq.~\eqref{eq:DM10}, such that this definition remains valid for weight functions without a particular normalization. We can check the correct normalization of the density matrix by noting
\begin{align}\label{eq:DM11}
\tr \rho' &= \int \cD n(m)\, \cD\bar{n}(m)\, \delta(n(m) - \bar{n}(m))\, 
\rho'(m;\, n(m),\, \bar{n}(m)) \notag \\
&= \int \cD n(m)\, \rho' (m;\, n(m),\, n(m)) = Z^{-1} \int \cD n\, w[n] = 1\, .
\end{align}
For a compact notation we use
\begin{equation}\label{eq:DM12}
\rho' (m) = Z^{-1} \int \cD n(m' \neq m)\, C(m)\, w[n]\, ,
\end{equation}
where the cut-operation $C(m)$ replaces $n \to \bar{n}$ in $\cK (m)$,
\begin{equation}\label{eq:DM13}
C(m)\, \cK (m;\, n(m+1),\, n(m)) = \cK (m;\, n(m+1),\, \bar{n}(m))\, .
\end{equation}

\paragraph*{General boundary conditions}

A general boundary term $\cB$ can be written as a linear combination of pure-state boundary terms
\begin{equation}\label{eq:DM14}
\cB = \sum_\alpha \bar{w}_\alpha\, \cB^{(\alpha)}\, ,
\end{equation}
with $\cB^{(\alpha)}$ denoting pure-state boundary conditions,
\begin{equation}\label{eq:DM15}
\cB^{(\alpha)} = \tilde{f}^{(\alpha)}_{in} (n_{in})\, \bar{f}^{(\alpha)}_f(n_f)\, .
\end{equation}
We denote the weight function with pure state boundary term $\cB^{(\alpha)}$ by $w^{(\alpha)} [n]$. Since the weight function $w^{(\alpha)}$ is linear in $\cB^{(\alpha)}$, the weight function for the general boundary condition \eqref{eq:DM14} obeys
\begin{equation}\label{eq:DM16}
w[n] = \sum_\alpha \bar{w}_\alpha\, w^{(\alpha)} [n]\, .
\end{equation}
If all $w^{(\alpha)}$ are positive semidefinite for all configurations of occupation numbers, and the coefficients $\bar{w}_\alpha$ obey
\begin{equation}\label{eq:DM17}
\bar{w}_\alpha \geq 0\, ,
\end{equation}
the positivity of $w[n]$ is guaranteed. Boundary conditions obeying the condition \eqref{eq:DM17} with at least two different $\bar{w}_\alpha$ different from zero are called ``mixed-state boundary conditions'', in close analogy to mixed states in quantum mechanics. 

The classical density matrix for arbitrary boundary conditions is defined by eq.~\eqref{eq:DM12}. We may denote by $\rho'^{(\alpha)}(m)$ the classical density matrix corresponding to the pure state boundary term $\cB^{(\alpha)}$, with partition function $Z_{(\alpha)}$,
\begin{equation}\label{eq:DM18}
\rho'^{(\alpha)} (m) = Z_{(\alpha)}^{-1} \, \int \cD n(m' \neq m)\, C(m)\, 
w_\alpha [n]\, .
\end{equation}
The linearity of the functional integral expression of $\rho'$ in terms of $w$ implies for general boundary conditions
\begin{align}
\label{eq:DM18A} \rho'(m) &= Z^{-1} \int \cD n(m' \neq m)\, C(m) \, w[n] \\
&= Z^{-1} \sum_\alpha \bar{w}_\alpha \int \cD n(m' \neq m)\, C(m)\, w^{(\alpha)} [n] 
\notag \\
 &= \sum_\alpha \frac{\bar{w}_\alpha Z_{(\alpha)} }{Z}\, \rho^{(\alpha)} (m)\, .\nonumber
\end{align}

With
\begin{equation}\label{eq:DM20}
Z = \sum_\alpha \bar{w}_\alpha Z_{(\alpha)} \, ,
\end{equation}
we may interpret the expressions
\begin{equation}\label{eq:DM21}
\bar{p}_\alpha = \frac{\bar{w}_\alpha Z_{(\alpha)}}{Z}\, , \quad \sum_\alpha 
\bar{p}_\alpha = 1\, , \quad \bar{p}_\alpha \geq 0\, ,
\end{equation}
as the probabilities that a mixed-state system is a pure-state system with $w^{(\alpha)} [n]$. The mixed-state probability distribution can be seen as a weighted sum of pure-state probability distributions, with weights given by the probabilities $\bar{p}_\alpha$,
\begin{equation}\label{eq:DM22}
\bar{p}[n] = \sum_\alpha \bar{p}_\alpha\, p^{(\alpha)} [n]\, , \quad p^{(\alpha)}[n] = 
Z_{(\alpha)}^{-1} w^{(\alpha)}[n]\, .
\end{equation}
Correspondingly, the mixed-state classical density matrix is a weighted sum of density matrices for pure classical states
\begin{equation}\label{eq:DM23}
\rho'(m) = \sum_\alpha \bar{p}_\alpha \rho'^{(\alpha)} (m)\, .
\end{equation}
The relations \eqref{eq:DM12}, \eqref{eq:DM13} between the diagonal elements of the density matrix and the local probabilities hold for arbitrary boundary conditions.

For generalized Ising chains the elements of pure-state density matrices obey in our basis
\begin{equation}\label{eq:DM23A}
\rho'_{\tau\rho} (m) \geq 0\, .
\end{equation}
This follows directly from eq.~\eqref{eq:CW11B}. Since this condition holds for every $\rho'^{(\alpha)}$ in eq.~\eqref{eq:DM23}, it also holds for the mixed-state boundary conditions with $\bar{p}_{\alpha} \geq 0$.

\paragraph*{Expectation value of local observables}

The expectation values of local observables at the site $m$ can be computed directly from the classical density matrix $\rho' (m)$ as
\begin{equation}\label{eq:DM34}
\langle A(m) \rangle = \tr \big\{ \hat{A} (m)\, \rho' (m) \big\}\, .
\end{equation}
This formula is familiar from quantum mechanics. It involves the operator $\hat{A}(m)$ associated to the observable $A(m)$ by eq.~\eqref{eq:CW24}. For pure classical states eq.~\eqref{eq:DM34} follows directly from eq.~\eqref{eq:CW19},
\begin{equation}\label{eq:DM35}
\langle A(m) \rangle = \hat{A}_{\tau\rho} (m)\, \tilde{q}_\rho (m)\, \bar{q}_\tau (m)\, 
.
\end{equation}
Since the relation between $\langle A \rangle$ and $\rho'$ is linear, the relation \eqref{eq:DM34} extends to mixed-state boundary conditions. 

We could also employ directly the defining relation for expectation values in terms of the overall probability distribution, which yields for local chains and local observables
\begin{align}\label{eq:DM36}
\langle A (m) \rangle &= \int \cD n(m)\, A_\sigma (m)\, h_\sigma (m)\, h_\tau (m) 
\notag \\
& \qquad \times \rho'_{\tau\rho} \big( m;\, n(m),\, n(m)\big) \, h_\rho (m)\, .
\end{align}
Here we employ the relation (cf. eq.~\eqref{eq:DM10})
\begin{equation}\label{eq:DM37}
\rho' \big( m;\, n(m),\, n(m)\big) = Z^{-1} \int \cD n(m' \neq m)\, w[n] = p_1 (m)\, .
\end{equation}
With eq.~\eqref{eq:CW24} the relation \eqref{eq:DM36} equals eq.~\eqref{eq:DM34}. 

The relation \eqref{eq:DM34} demonstrates again the close similarities between local evolution in classical statistical systems and the formalism of quantum mechanics. The analogues become even stronger if we consider later observables for which the associated operators are not diagonal.

\paragraph*{Evolution of density matrix}

A central advantage of the use of the classical density matrix in classical probabilistic systems is the existence of a simple evolution law which permits to compute $\rho'(m+1)$ from $\rho'(m)$, namely 
\begin{equation}\label{eq:DM38}
\rho'(m+1) = \hat{S}(m)\, \rho' (m) \, \hat{S}^{-1} (m)\, .
\end{equation}
It is linear in $\rho'$, implying the superposition principle for possible solutions. The step evolution operator $\hat{S}$ acts as a similarity transformation, preserving the eigenvalues of $\rho'$. We observe again the close analogy to the formalism of quantum mechanics, with step evolution operator $\hat{S}$ replacing the unitary step evolution operator between $t$ and $t + \varepsilon$ for quantum mechanics. We will see later that eq.~\eqref{eq:DM38} \textit{is} actually the quantum evolution law if $\hat{S}$ belongs to a unitary subgroup of the orthogonal transformations $SO(2^M)$, and if $\rho'$ is compatible with an appropriate complex structure. 

For pure-state boundary conditions the evolution law \eqref{eq:DM38} follows directly from the evolution of the wave functions \eqref{eq:SE3}, \eqref{eq:SE5},
\begin{align}\label{eq:DM39}
\rho'_{\tau\rho} (m+1) &= \tilde{q}_\tau (m+1)\, \bar{q}_\rho (m+1) \notag \\
&= \hat{S}_{\tau\sigma} (m)\, \tilde{q}_\sigma (m) \, \bar{q}_\mu (m)\, 
\hat{S}_{\mu\rho}^{-1} (m) \notag \\
&= \hat{S}_{\tau\sigma} (m) \, \rho'_{\sigma\mu} (m)\, \hat{S}_{\mu\rho}^{-1} (m)\, .
\end{align}
For general boundary conditions we employ that eq.~\eqref{eq:DM23} holds for all $m$. Insertion of eq.~\eqref{eq:DM38} for each pure-state density matrix $\rho'^{(\alpha)}$ yields the evolution \eqref{eq:DM38} for $\rho'$ with arbitrary boundary factor. 

The classical density matrix contains local probabilistic information beyond the local probability distribution. This additional information is encoded in the off-diagonal elements of $\rho'$. We can compute $\rho'$ as function of occupation numbers $n(m)$ and $\bar{n}(m)$ from the two-site density matrix $\rho_2$ in eq.~\eqref{eq:EV5} as
\begin{align}\label{eq:DM33}
& \rho' (m;\, n(m),\, \bar{n}(m)) = \int  \cD n(m+1)\notag \\
& \qquad\times \cK (m,\, \bar{n}(m),\, n(m+1))\, 
\rho_2(m)\, .
\end{align}

At this point we note that the two-site density matrix $\rho_2$ in eq.~\eqref{eq:EV5} obeys a similar linear evolution law. Expanding
\begin{equation}\label{eq:DM40}
\rho_2 (m) = (\rho_2)_{\tau\rho} (m)\, h_\tau (m)\, h_\rho (m+1)\, ,
\end{equation}
the evolution law for the matrix $\rho_2$ in the occupation number basis reads
\begin{equation}\label{eq:DM41}
\rho_2 (m+1) = \hat{S}(m)\, \rho_2 (m)\, \hat{S}^{-1} (m+1)\, .
\end{equation}
For pure-state boundary conditions we can write
\begin{align}\label{eq:DM42}
& \rho_2 (m) = f(m)\, \bar{f}(m+1)\, , \notag \\
& (\rho_2)_{\tau\rho}(m) = \tilde{q}_\tau (m)\, \bar{q}_\rho(m+1)\, ,
\end{align}
such that eqs.~\eqref{eq:SE3}, \eqref{eq:SE5} imply the evolution law \eqref{eq:DM41}. The close relation between $\rho'$ and $\rho_2$ shows that the local information necessary for the formulation of a simple evolution law concerns two neighboring sites. This also suggests that $\rho' (m)$ may contain enough local probabilistic information to compute the expectation values of certain observables beyond the local observables at $m$. We will see in sect.~\ref{sec:observables_and_operators} that this is indeed the case.

In general, the off-diagonal elements of the classical density matrix matter for the evolution of the local probabilities. This can be seen from
\begin{equation}\label{eq:DM43}
p_\tau (m+1) = \rho'_{\tau\tau}(m+1) = \hat{S}_{\tau\sigma} (m) \, \rho'_{\sigma\mu} (m)
\, (\hat{S}^{-1})_{\mu\tau} (m)\, .
\end{equation}
Only if the condition \eqref{eq:SE14} holds, the r.h.s. can be expressed in terms of the diagonal elements $\rho'_{\sigma\sigma} (m)$ only. In this case the evolution is given by a Markov chain. For general step evolution operators this condition is not obeyed. For the generic case we may state that a local evolution law for the local probabilistic information requires information for two neighboring sites, rather than a single site. This is the basic reason for the appearance of an object as the classical density matrix, which contains more local probabilistic information than the local probability distribution alone. 

\paragraph*{Properties of the classical density matrix}

The classical density matrix $\rho'$ shows many similarities with the density matrix in quantum mechanics. For $Q$ quantum spins
or qubits the quantum density matrix is a hermitian normalized positive $2^Q \times 2^Q$-matrix. Combining the real and imaginary 
parts of the complex quantum wave function into a real $2^{Q+1}$-component vector, the quantum density matrix becomes a real symmetric
$2^{Q+1} \times 2^{Q+1}$-matrix. The evolution law of the quantum system is given in this real basis by eq.~\eqref{eq:DM38}, with
$\hat{S}$ an orthogonal matrix belonging to the unitary $U(2^Q)$-subgroup of $SO(2^{Q+1})$. The hermitian operators for observables
in quantum systems are symmetric matrices in the real formulation. We conclude that both the classical density matrix and the quantum
density matrix in the real formulation share several common features:
\begin{inlinelist}
\item They obey the same normalization. \item For pure states both obey the relation $\rho'^2 = \rho'$, with eigenvalues
of $\rho'$ being one or zero. \item The formula \eqref{eq:DM34} for the computation of expectation values is the same. 
\item The formula \eqref{eq:DM38} for the evolution is the same.
\end{inlinelist}

Despite these many common features, there are also important differences. In general, the classical density matrix $\rho'$ is not
a symmetric matrix, and the step evolution operator is not an orthogonal matrix. These differences are rooted in the fact that
for general local chains one has a pair of two distinct wave functions $\tilde{q}$ and $\bar{q}$, whereas in quantum mechanics the
conjugate wave function is related to the wave function by complex conjugation. We will discuss the properties of the classical density
matrix in more detail in the following.

A central difference of the classical density matrix 
as compared to the density matrix in quantum mechanics is the lack of symmetry for generic classical density matrices. We may decompose $\rho'$ into its symmetric and antisymmetric parts,
\begin{equation}\label{eq:DM24}
\rho' = \rho'_S + \rho'_A \, , \quad \rho'^\tp_S = \rho'_S \, , \quad \rho'^\tp_A = 
- \rho'_A\, .
\end{equation}
One may be tempted to identify the symmetric part $\rho'_S$ with the density matrix in quantum mechanics. In quantum 
mechanics the positivity of the density matrix plays an important role. All eigenvalues of the density matrix are positive
semidefinite. Let us therefore investigate the issue of positivity for the symmetric part of the classical density matrix.

The symmetric part defines a bilinear for the classical wave functions,
\begin{equation}\label{eq:DM25}
\langle \tilde{q}^{(1)} | \,\tilde{q}^{(2)} \rangle_\rho = \tilde{q}^{(1)}_\tau 
(\rho'_S)_{\tau\rho} \,\tilde{q}^{(2)}_\rho\, .
\end{equation}
In general, the property
\begin{equation}\label{eq:DM26}
\langle \tilde{q}\,  |\,  \tilde{q} \rangle_\rho \geq 0
\end{equation}
is not realized, however, for arbitrary symmetric $\rho'_S$ and arbitrary $\tilde{q}$. This can be seen by simple examples of $(2\times 2)$-matrices. In quantum mechanics in a real basis the inequality \eqref{eq:DM26} holds and the bilinear \eqref{eq:DM25} can serve
as a type of scalar product.

For the particular case of symmetric pure-state density matrices the inequality \eqref{eq:DM26} holds, since the eigenvectors to the eigenvalues one and zero can be used as an ON-basis, with left- and right-eigenvalues being identical
\begin{equation}\label{eq:DM27}
\rho' \, \tilde{q}^{(k)} = \lambda_k \, \tilde{q}^{(k)} \; \Rightarrow \; (\tilde{q}^{(k)})^\tp \rho' = \lambda_k (\tilde{q}^{(k)})^\tp\, .
\end{equation}
Denoting $\lambda_1 = 1$, $\lambda_{k > 1} = 0$ and expanding an arbitrary vector $\tilde{q}$ in terms of $\tilde{q}^{(k)}$,
\begin{equation}\label{eq:DM28}
\tilde{q} = \sum_k c_k \tilde{q}^{(k)}\, ,
\end{equation}
one infers
\begin{align}\label{eq:DM29}
\tilde{q}^\tp \rho' \tilde{q} = \sum_{k,l} c_k c_l (\tilde{q}^{(k)})^\tp \rho' 
\tilde{q}^{(l)} = c_1^2 (\tilde{q}^{(1)})^\tp \tilde{q}^{(1)} \geq 0\, .
\end{align}

This generalizes to mixed-state boundary conditions, where
\begin{equation}\label{eq:DM30}
\rho'_S = \sum_\alpha \bar{w}_\alpha \rho'^{(\alpha)}_S\, , \quad \bar{w}_\alpha \geq 0\,.
\end{equation}
If all $\rho'^{(\alpha)}$ are symmetric, $\rho'^{(\alpha)}_A = 0$, or if all $\rho'^{(\alpha)}_S$ have a representation as a pure 
state density matrix with appropriate $\tilde{q}^{(\alpha)}_S,\bar{q}^{(\alpha)}_S$, the sum \eqref{eq:DM30} extends over positive 
matrices with positive coefficients. In this case
one has for arbitrary $\tilde{q}$
\begin{equation}\label{eq:DM31}
\langle \tilde{q} \,|\, \tilde{q} \rangle_\rho = \tilde{q}^\tp \rho'_S \tilde{q} \geq 0 \, .
\end{equation}
As a consequence, all eigenvalues $\lambda_i$ of $\rho'_S$ obey $\lambda_i \geq 0$.
This establishes the positivity of $\rho'$. 

Furthermore, for an orthogonal transformation
\begin{equation}\label{eq:DM32}
\rho'_{(O)} = O\, \rho' \, O^\tp\, , \quad O\, O^\tp = 1
\end{equation} 
the transformed matrix $\rho'_{(O)}$ remains symmetric and all its diagonal elements are positive, $(\rho'_{(O)})_{\tau\tau} \geq 0$. All these are known properties of the hermitian density matrix in quantum mechanics. We conclude that symmetric pure-state classical density matrices have the same positivity properties as the density matrix in quantum mechanics. This extends to mixed-state density matrices \eqref{eq:DM30} that are sums of symmetric pure-state density matrices.

We observe that the symmetry of classical density matrices and the orthogonality of the step evolution operator are closely related.
If the evolution is not orthogonal, a symmetric density matrix does not stay symmetric in the course of the evolution. On the other
hand, for an orthogonal evolution the split \eqref{eq:DM24} into an antisymmetric and symmetric part is preserved by the evolution. For 
observables represented by symmetric operators, in particular the diagonal operator for local observables, the antisymmetric part 
$\rho'_A$ does not contribute in the formula\eqref{eq:DM34}. It plays no role in this respect and may be omitted. This suggests that 
classical probabilistic systems with orthogonal evolution behave in the same way as quantum systems. For a positive density matrix
$\rho'_S$ and in the presence of a complex structure all aspects of evolution and expectation values are identical to quantum 
mechanics. We will come back later to this issue.

\paragraph*{Symmetric density matrices for unique jump chains}

For unique jump chains the step evolution operator is orthogonal. For suitable boundary conditions these systems share the
properties of discrete quantum mechanics. We show this first for classical pure states and generalise later to mixed states.

For orthogonal step evolution operators
\begin{equation}\label{eq:DM44}
\hat{S}^\tp(m)\, \hat{S}(m) = 1\, ,
\end{equation}
the evolution equation \eqref{eq:DM38} transforms the symmetric and antisymmetric parts of $\rho'$ separately,
\begin{align}\label{eq:DM45}
& \rho'_S (m+1) = \hat{S} (m)\, \rho'_S (m)\, \hat{S}^{-1} (m)\, , \notag \\
& \rho'_A (m+1) = \hat{S} (m)\, \rho'_A (m)\, \hat{S}^{-1} (m)\, .
\end{align}
If we choose a boundary term such that $\rho'$ is a symmetric matrix at some given site $m$, it will be a symmetric matrix for all sites $m$.

A symmetric density matrix is realized for pure classical states by pairs of wave functions obeying
\begin{equation}\label{eq:DM46}
\bar{q}_\tau (m) = a(m)\, \tilde{q}_\tau (m)\, .
\end{equation}
For a pure-state boundary term the condition \eqref{eq:DM46} is necessary for realizing $\rho'^\tp = \rho'$. From
\begin{equation}\label{eq:DM47}
\tilde{q}_\tau \bar{q}_\rho = \tilde{q}_\rho \bar{q}_\tau
\end{equation}
one infers by multiplication with $\bar{q}_\rho$ and summation over $\rho$ that each component $\bar{q}_\tau$ is proportional to the $\tau$-component of $\tilde{q}$,
\begin{equation}\label{eq:DM48}
\bar{q}_\tau = \sum_\rho (\bar{q}_\rho)^2\, \tilde{q}_\tau = a\,\tilde{q}_\tau\, .
\end{equation}

Orthogonal transformations do not change the length of the vectors $\tilde{q}$ and $\bar{q}$, $\sum_\rho(\tilde{q}_\rho)^2 = \text{const}$, $\sum_\rho(\bar{q}_\rho)^2 = \text{const}$, such that $a$ in eq.~\eqref{eq:DM46} is a factor independent of $m$. A constant $a$ can be absorbed by a multiplicative renormalization  of the boundary factors that leave $\cB$ invariant, $f \to\sqrt{a}f$, $\bar{f} \to \bar{f}/\sqrt{a}$. In this normalization symmetric density matrices require for all $m$
\begin{equation}\label{eq:DM49}
\bar{q}(m) = \tilde{q}(m) = q(m)\,,
\end{equation}
with
\begin{equation}\label{eq:DM50}
p_\tau (m) = (q_\tau(m))^2\, .
\end{equation}
There are no longer two separate classical wave functions, but a unique wave function that we denote by $q(m)$. We recover the finding in eq.~\eqref{UJC3} that for a positive overall probability distribution we can distribute the probabilistic information to the two boundary factors such that $\overline{q}(m)=\tilde{q}(m)=q(m)$ for all $m$. With this choice the density matrix for unique jump chains is symmetric.

The irrelevance of the distribution of the probabilistic information on the boundary terms is underlined by the observation that a possible antisymmetric part of the density matrix $\rho'_A$ does not affect the expectation values of local observables. The operator $\hat{A}(m)$ associated to a local observable is symmetric,
\begin{equation}\label{eq:DM51}
\hat{A}_{\tau\rho} (m) = \hat{A}_{\rho\tau} (m)\, .
\end{equation}
Only the symmetric part of $\rho'$ contributes to the trace \eqref{eq:DM34}. Since $\rho'_A$ does neither influence the evolution of $\rho'_S$ nor contribute to the expectation values of observables, we can simply omit it by choosing $\cB$ such that the classical density matrix is symmetric.

Mixed states can be constructed by choosing boundary conditions for which for all $\alpha$ in eq.\,\eqref{eq:DM23}the matrices
$\rho'^{(\alpha)}$ are pure state density matrices with $\bar{q}^{(\alpha)} = \tilde{q}^{(\alpha)}$. In this case $\rho'$ is a positive
symmetric matrix.

For orthogonal step evolution operators the picture is even closer to quantum mechanics than for the general case. For pure classical states we only need a single wave function, and the expectation values of observables are bilinear in this wave function,
\begin{equation}\label{eq:DM52}
\langle A(m) \rangle = q^\tp (m)\, \hat{A}(m)\, q(m)\, .
\end{equation}
The classical density matrix is symmetric, with properties analogous to the hermitean density matrix in quantum mechanics. The orthogonal evolution of the density matrix and the wave functions is the analogue of the unitary evolution in quantum mechanics. For unique jump chains the step evolution operator $\hat{S}$ is orthogonal. We will see that in the presence of an appropriate complex structure the unique jump chains do indeed realize quantum mechanics for discrete evolution steps.

\subsubsection{Independence from the future}\label{sec:independence_from_the_future}

In our common concept of time we have some information about the past, but we do not know the future. The use of differential equations for predictions of future events is set as an initial value problem, where the state of the system at some time $t_0$ selects among the different possible solutions. With initial conditions at $t_0$ we predict expectation values of observables for $t > t_0$. No direct information about the future is used in this process. We would like to set up the evolution for local chains in a similar spirit, such that no input from the future is necessary. The generalized von Neumann equation~\eqref{eq:DM38} is already of this type. With initial conditions set by $\rho'(t_0)$ we can compute $\rho'(t)$ without the use of future information for $t'>t$.

The independence from the future is accompanied by a similar independence from the past. All the relevant boundary information and its evolution to $t_0$ is encoded in the density matrix $\rho'(t_0)$. The local probabilistic information in $\rho'(t_0)$ is only a very small part of the information on what happened in the past or will happen in the future. The future and the past have an influence on the present only via the initial density matrix $\rho'(t_0)$. This influence can be important, however, by restricting the allowed values of $\rho'(t_0)$. This is connected to the loss of memory in the bulk discussed above. Only for particular systems (or subsystems) without information loss restrictions on $\rho'(t_0)$ are absent. Only in this case the common view that the future is irrelevant for the present is realized.

A functional integral formulation of the initial value problem leads to the concept of ``double chains'' by which expectation values of observables at $t > t_0$ can be computed from ``initial conditions'' set by the classical density matrix at $t_0$. This resembles the Schwinger-Keldyish formalism~\cite{JSS},~\cite{KEL} for quantum field theories.

\paragraph*{Initial density matrix}

The evolution equation for the classical density matrix \eqref{eq:DM38} permits us to formulate the influence of boundary terms as an initial value problem. If at some time $t_0$ the density matrix $\rho' (t_0)$ is known, we can compute $\rho'(t_0 + \varepsilon)$ by eq.~\eqref{eq:DM38} and proceed to arbitrary higher $t$. We can also invert eq.~\eqref{eq:DM38} in order to proceed to lower $t$,
\begin{equation}\label{eq:IF1}
\rho'(m-1) = \hat{S}^{-1}(m-1)\, \rho' (m)\, \hat{S}(m-1)\, .
\end{equation}
No arrow of time is selected at this stage. Once the density matrix is given at some ``reference time'' $t_0$, or some site $m_0$, the solution of the evolution law determines $\rho'(m)$ for all sites. From there the expectation values of all local observables can be computed. 

We can take $m_0 = 0$ and start with the initial density matrix
\begin{equation}\label{eq:IF2}
\rho'(0) = \rho'_{in}\, .
\end{equation}
The solution of the ``discrete differential equation'' \eqref{eq:DM38} with given $\rho'_{in}$ solves the ``boundary value problem'' how boundary terms influence observables in the bulk. In this formulation, the boundary value problem is turned to an ``initial value problem''. One only needs probabilistic information at one site, namely $\rho'(m_0)$. Information at some ``present time'' $t_0$ is indeed sufficient to compute the behavior in the future in terms of a local evolution law. As we have argued, this is the way how we set up the problem of evolution for most situations in physics. We usually do not employ information about the future. For the evolution law \eqref{eq:DM38} one also does not need to know what happens in the past. If $\rho'(t_0)$ is known at some ``present time'' $t_0$, we can compute both the future and the past if we are able to solve the evolution equation.

The original boundary value problem has boundary conditions at both ends of the chain, involving both functions of $n(0)$ and $n(\cM)$. We have to relate this to the initial value problem by a determination of the initial density matrix $\rho'_{in}$. The issue can be understood most easily for pure-state boundary conditions. In this case the initial density matrix involves both wave functions at $m=0$, 
\begin{equation}\label{eq:IF3}
(\rho'_{in})_{\tau\rho} = \tilde{q}_\tau (0)\, \bar{q}_\rho (0)\, .
\end{equation}
While $\tilde{q}(0)$ obtains directly from the initial boundary term $f_{in}$, the conjugate wave function involves the final boundary term $\bar{f}(\cM)$, corresponding to $\bar{q}(\cM)$. For a determination of $\bar{q}(0)$ we need to solve the evolution equation for $\bar{q}$, starting with $\bar{q}(\cM)$. One may not care what $\bar{q}(\cM)$ is precisely and take the attitude that $\bar{q}(0)$ is as good as a boundary condition as $\bar{q}(\cM)$. With $\tilde{q}(0)$ and $\bar{q}(0)$ we have initial values at $m=0$ from which we can compute the evolution towards larger $m$. Since there is an invertible map between $\bar{q}(\cM)$ and $\bar{q}(0)$ one is indeed free to choose which quantity is used for a specification of the solution of the evolution equation.

\paragraph*{Restrictions on initial density matrix}

While this argument is valid in principle, there may be a problem from the choice of the range of $\bar{q}(0)$ which is compatible with the evolution in local chains. A large range of values of $\bar{q}(\cM)$ may be mapped to a very restricted range of possible values for $\bar{q}(0)$. This becomes particularly relevant for large values of $\cM$, for which the map $\bar{q}(\cM) \to \bar{q}(0)$ may have strong focus properties. An example is the Ising chain for finite values of $\beta$. With
\begin{equation}\label{eq:IF4}
\bar{q}(\cM) = \begin{pmatrix}
a + b \\ a - b
\end{pmatrix}
\end{equation}
one has 
\begin{equation}\label{eq:IF5}
\bar{q}(0) = \begin{pmatrix}
a + \lambda_2^{\cM}\, b \\ a - \lambda_2^{\cM} \,b
\end{pmatrix}
= \begin{pmatrix}
a + (\tanh\beta)^{\cM}\, b \\
a - (\tanh\beta)^{\cM}\, b
\end{pmatrix}\, .
\end{equation}
(Here $a = \bar{c}/\sqrt{2}$, $b = \bar{c}\, a_f/\sqrt{2}$ in the notation of eqs~\eqref{eq:BC9}, \eqref{eq:BC17}.) The difference,
\begin{equation}\label{eq:IF6}
\bar{q}_1(0) - \bar{q}_2(0) = (\tanh\beta)^{\cM}(\bar{q}_1(\cM) - \bar{q}_2(\cM))\,,
\end{equation}
can be much smaller than the corresponding difference at the site $\cM$. For finite $\beta$ and finite $\bar{q}_1(\cM) - \bar{q}_2 (\cM)$ it vanishes for $\cM \to\infty$.
In this limit the map from $\bar{q}(\cM)$ to $\bar{q}(0)$ is no longer invertible.

We will not go into details of the possible restrictions for the initial density matrix $\rho'_{in}$ and refer for this issue to ref.~\cite{CWQF}. One simple condition for the allowed $\rho'_{in}$ is that the corresponding solution has to guarantee for all $m$ and all $\tau$ the bound $0 \leq \rho'_{\tau\tau}(m) \leq 1$. We rather give examples for the absence or presence of restrictions by discussing unique jump chains and the Ising chain.

\paragraph*{Initial value problem for unique jump chains}

For unique jump chains there are no restrictions on the initial value of the density matrix. In this case the influence of the future can be eliminated completely. The situation is similar to quantum mechanics. The absence of restrictions is directly related to the fact that no boundary information is lost. For pure states we have already seen that a natural choice of boundary conditions leads to $\bar{q}(t)=\tilde{q}(t)$. For a given $\tilde q(t_{in})$ one can follow its evolution to $t_f$. At first sight the choice $\bar q(t_f)=\tilde q(t_f)$ at the final boundary seems to be a special condition. We have seen, however, that this choice is possible for arbitrary overall probability distributions by employing the scaling~\eqref{UJC3}. With a free choice of the single wave function $\overline{q}(t)=\tilde{q}(t)=q(t)$ arbitrary symmetric classical density matrices $\rho'(t)$ can be realized.

For arbitrary mixed states the situation remains similar. Consider an arbitrary boundary term~\eqref{eq:DM14}
\begin{align}
\label{XAX}
\cB=&\sum_\alpha\bar w_\alpha \tilde f_{in}^{(\alpha)}(n_{in})\bar f_f^{(\alpha)}(n_f)\nonumber\\
=&\sum_\alpha\bar w_\alpha \tilde q_\tau^{(\alpha)}(t_{in})\bar q_\rho^{(\alpha)}(t_f)h_\tau(t_{in})h_\rho(t_f)\nonumber\\
=&\hat{B}_{\tau\rho}(t_{in},t_f)h_\tau(t_{in})h_\rho(t_f)\ .
\end{align}
The boundary matrix $\hat{B}(t_{in},t_f)$ is the same as in eq.~\eqref{eq:TS47}, with $\hat T$ replaced by $\hat S$. With $\rho'(t_{in})$ given by the functional~\eqref{eq:DM18A} for $m=0$ we can integrate out all $n(m>0)$. Integrating out first only $n(t_f)$ leads to a restricted wave function with $t_f$ replaced by $t_f-\epsilon$. The corresponding boundary matrix $\hat{B}(t_{in},t_f-\epsilon)$ obtains from eq.~\eqref{XAX} by replacing $\bar q^{(\alpha)}(t_f)\to\bar q^{(\alpha)}(t_f-\epsilon)$, $h_\rho(t_f)\to h_\rho(t_f-\epsilon)$. Here $\bar q^{(\alpha)}(t_f-\epsilon)$ obtains from $\bar q^{(\alpha)}(t_f)$ by use of the evolution law. Performing these integrations stepwise yields
\begin{equation}\label{XBX}
\rho'_{\tau\rho}(t_{in})=\sum_\alpha\bar w_\alpha\tilde q^{(\alpha)}_\tau(t_{in})\bar q_\rho^{(\alpha)}(t_{in})\ .
\end{equation}
General $\rho'(t_{in})$ can be found for appropriate $\hat{B}(t_{in},t_f)$, which is simply found from eq.~\eqref{XBX} by replacing $\bar q_\rho^{(\alpha)}(t_{in})$ by $\bar q_\rho^{(\alpha)}(t_f)$.

\paragraph*{Boundary problem for the Ising chain}

For pure classical states we have already solved the boundary value problem for the Ising chain in terms of the classical wave functions. This can be taken over to the corresponding pure-state classical density matrices. We complete the discussion here for mixed-state boundary conditions by discussing solutions of the evolution equation \eqref{eq:DM38} for $\rho'$. For more details see ref.~\cite{CWQF}. We demonstrate the appearance of restrictions by a discussion of possible static solutions.

The most general ``static solution'' with $\rho'$ independent of $m$ is given by 
\begin{equation}\label{eq:IF7}
\rho' = \frac{1}{2} \begin{pmatrix}
1 & a \\ a & 1
\end{pmatrix}\, , \quad
|a| \leq 1\, .
\end{equation}
General static solutions obey 
\begin{equation}\label{eq:IF8}
\hat{S}\, \rho'\, \hat{S}^{-1} = \rho'\, ,
\end{equation}
such that all $\rho'$ commuting with $\hat{S}$ are static,
\begin{equation}\label{eq:IF9}
[\hat{S},\, \rho'] = 0\, .
\end{equation}
For the attractive Ising chain ($\kappa = -1$) the step evolution operator has the form
\begin{align}\label{eq:IF10}
\hat{S} = \frac{\text{e}^\beta}{2\cosh\beta}\, \bm{1}_2 + \frac{\text{e}^{-\beta}}{2\cosh\beta} \,\tau_1\, , \quad \tau_1 = \begin{pmatrix}
0 & 1 \\ 1 & 0
\end{pmatrix}\, ,
\end{align}
and the general solution for the commutation relation \eqref{eq:IF9} is easily established.

We note that $a=1$ corresponds to a pure-state density matrix with 
\begin{align}\label{eq:IF11}
\tilde{q} = c\, \tilde{q}_{eq} = \frac{c}{\sqrt{2}} \begin{pmatrix}
1 \\ 1
\end{pmatrix}\, , \quad
\bar{q} = \bar{c}\,\bar{q}_{eq} = \frac{\bar{c}}{\sqrt{2}} \begin{pmatrix}
1 \\ 1
\end{pmatrix}\, , \quad \bar{c}\, c = 1\, .
\end{align}
Also $a = -1$ is a pure-state density matrix, $\rho'^2 = \rho'$. The corresponding wave functions are the eigenfunctions of the eigenvalue $\lambda_2$,
\begin{equation}\label{eq:IF12}
\tilde{q}(m) = \frac{d(m)}{\sqrt{2}} \begin{pmatrix}
1 \\ -1
\end{pmatrix}\, , \quad
\bar{q}(m) = \frac{\bar{d}(m)}{\sqrt{2}} \begin{pmatrix}
1 \\ -1
\end{pmatrix}\, , \quad
\bar{d}\, d = 1\, .
\end{equation}
The coefficients $d(m)$, $\bar{d}(m)$ obey
\begin{equation}\label{eq:IF13}
d(m) = \lambda_2^m\, d(0)\, , \quad \bar{d}(m) = \lambda_2^{\cM - m}\, \bar{d}(\cM)\, ,
\end{equation}
such that the static density matrix with $a=-1$ requires
\begin{equation}\label{eq:IF14}
\lambda_2^{\cM}\, d(0)\, \bar{d}(\cM) = 1\, .
\end{equation}
For finite $\beta$ and $\cM \to \infty$ the product of coefficients $d(0)\, \bar{d}(\cM)$ has to diverge as $\lambda_2^{-\cM}$. If the values of $d(0)$ and $\bar{d}(\cM)$ are bounded, this static solution cannot be realized. This is an example for restrictions on the possible values of the initial density matrix $\rho'_{in}$.

The local probabilities $p_1(m) = p_2(m) = 1/2$ are independent of $a$. Also the expectation values of local observables are independent of $a$. The solution for general boundary conditions approaches the static solution with $a=1$ as one moves from the boundary into the bulk if $\lambda_2 < 1$. This is a simple example of ``syncoherence''\,\cite{CWQM} where a mixed state evolves towards a pure state. We will give in sect.\,\ref{sec:continuous_time} a few more details in a formulation with continuous time. For $\beta\to\infty$ every initial density matrix generates a static solution. Since the coefficient of $\tau_1$ in eq.\,\eqref{eq:IF10} vanishes, every density matrix is static.

\paragraph*{Double chains}

For local chains the initial value problem can be formulated directly as a functional integral. This leads to the concept of ``double chains''. These double chains are equivalent to the original local chains. The boundary conditions appear here directly in the form of the initial classical density matrix $\rho_{in}$. This formulation renders the predictivity of the model independent of the future and the distant past. Only the density matrix at some reference time $t_0$ is needed for predictions on local observables for $t > t_0$.

Double chains formulate a probability distribution or ``functional integral" that is equivalent to the one for local chains for all
``classical observables" of the local chain. The boundary conditions appear, however, in a different form, involving now the initial
density matrix. It is a matter of taste which one of the formulations is considered as fundamental. One may either consider the double
chains as a reformulation of the local chains, or take them as the basic overall probability distribution.

We construct the double chain first for local chains with pure-state boundary conditions in a normalization with $Z=1$. The basis is the general expression of $\bar{q}(0)$ in terms of $\bar{q}(\cM)$,
\begin{equation}\label{eq:IF15}
\bar{q}(0)_\tau^\tp = \bar{q}(\cM)^\tp_\rho \big[ \hat{S}(\cM - 1)\, \hat{S}(\cM - 2)
\cdots \, \hat{S}(1)\, \hat{S}(0) \big]_{\rho\tau}\, .
\end{equation}
For invertible step evolution operators we can invert this equation and express $\bar{q}(\cM)$ in terms of $\bar{q}(0)$,
\begin{equation}
\label{eq:IF16}
\bar{q}_\tau (\cM) = \big[ (\hat{S}^\tp)^{-1}(\cM -1)\cdots\, (\hat{S}^\tp)^{-1} (0) \big]_{\tau\rho} \, \bar{q}_\rho (0)\, .
\end{equation}
We introduce occupation numbers $\bar{n}(0)$, different from $n(0)$, in order to define a ``conjugate initial boundary term''
\begin{equation}
\label{eq:IF17}
\bar{f}_{in} = \bar{q}_\sigma (0)\, h_\sigma [\bar{n}(0)]\, .
\end{equation}
The final boundary term can then be expressed in terms of $\bar{f}_{in}$ as
\begin{align}\label{eq:IF18}
\bar{f}_f &= \bar{q}_\tau (\cM)\, h_\tau (\cM) \notag \\
&= \int \cD \bar{n}(0)\, h_\tau(\cM)
\big[ (\hat{S}^\tp)^{-1} (\cM -1)\cdots\, (\hat{S}^\tp)^{-1} (0) \big]_{\tau\rho}
\notag \\
& \quad \times h_\rho[\bar{n}(0)]\bar{f}_{in}\, ,
\end{align}
using
\begin{equation}\label{eq:IF19}
\int \cD \bar{n}(0)\, h_\rho [\bar{n}(0) ]\, \bar{f}_{in} = \bar{q}_\rho(0)\, .
\end{equation}

Next we define ``conjugate local factors'' $\bar{\cK}(m)$ by
\begin{equation}\label{eq:IF20}
\bar{\cK}(m) = \bar{h}_\tau (m+1) \,(S^\tp)^{-1}_{\tau\rho}\, \bar{h}_\rho(m)\,,
\end{equation}
with $\bar{h}_\rho (m)$ basis functions in terms of $\bar{n}(m)$,
\begin{equation}\label{eq:IF21}
\bar{h}_\rho (m) = h_\rho [\bar{n}(m)]\, .
\end{equation}
The conjugate local factor $\bar{\cK}(m)$ therefore depends on the occupation numbers $\bar{n}(m)$ and $\bar{n}(m+1)$. These variables are different from the variables $n(m)$ and $n(m+1)$ that are used for the local factors $\cK(m)$. The final boundary term $\bar{f}_f$ in eq.~\eqref{eq:IF18} can be written as a product of conjugate local factors
\begin{align}\label{eq:IF22}
\bar{f}_f = \prod_{m' = 0}^{\cM} \int \cD \bar{n}(m')\, \delta 
\big( \bar{n}(\cM) - n(\cM)\big) \prod_{m' = 0}^{\cM - 1} \bar{\cK} (m')\,
\bar{f}_{in}\, .
\end{align}
This employs the product \eqref{eq:TS16}
\begin{align}\label{eq:IF23}
& \int \cD \bar{n} (m+1)\, \bar{\cK}(m+1)\, \bar{\cK}(m) = \bar{h}_{\rho_{m+2}}(m+2) \notag 
\\
& \qquad\times  \big[ (\hat{S}^\tp)^{-1} (m+1)\, (\hat{S}^\tp)^{-1} (m) 
\big]_{\rho_{m+2}\rho_m} \bar{h}_{\rho_m}(m)\, .    
\end{align}
The $\delta$-function,
\begin{equation}\label{eq:IF24}
\delta(\bar{n}(\cM) - n(\cM)) = \bar{h}_\sigma (\cM)\, h_\sigma (\cM)\, ,
\end{equation}
identifies $\bar{n}(\cM)$ and $n(\cM)$, according to $h_\tau(\cM)$ in eq.~\eqref{eq:IF18} depending on $n(\cM)$.

Insertion of eq.~\eqref{eq:IF22} into the weight function $w[n]$ of the local chain \eqref{eq:LC4} yields
\begin{equation}\label{eq:IF25}
w[n] = \prod_{m'=0}^{\cM} \int \cD \bar{n}(m')\, w[n,\, \bar{n}]\, ,
\end{equation}
with
\begin{equation}\label{eq:IF26}
w[n,\, \bar{n}] =  \delta (\bar{n}(\cM) - n(\cM)) \prod_{m' = 0}^{\cM - 1} 
\big( \bar{\cK}(m')\,\cK (m') \big) \, \bar{f}_{in}\, f_{in}\, .
\end{equation}
Eq.~\eqref{eq:IF26} defines the weight function $w[n,\, \bar{n}]$ of the ``double chain''. It involves two sets of occupation numbers $\{ \bar{n}(m)\}$ and $\{ n(m)\}$ -- the origin of the naming. In general, the conjugate local factor $\bar{\cK}(m)$ differs from the local factor $\cK(m)$. The double chain is not simply the product of two independent local chains since the factor $\delta(\bar{n}(\cM) - n(\cM))$ identifies occupation numbers at the endpoints of the chain. There are no longer any ``final boundary terms''. The boundary factors $\bar{f}_{in}$ and $f_{in}$ depend only on ``initial occupation numbers'' $\bar{n}(0)$ and $n(0)$, respectively.

\paragraph*{Integrating out the future}

The double chain has the same partition function as the original local chain
\begin{equation}\label{eq:IF27}
Z = \int \cD \bar{n} \,\cD n\, w[n,\, \bar{n}] = \int \cD n \, w[n]\, .
\end{equation}
We observe the identity 
\begin{align}\label{eq:IF28}
& \int\cD \bar{n}(m+1)\,\cD n(m+1)\, \delta (\bar{n}(m+1) - n(m+1)) \notag \\
& \quad \times \bar{\cK}(m)\, \cK(m) \notag \\
&  = \delta(\bar{n}(m) - n(m))\,.
\end{align}
It follows from
\begin{align}\label{eq:IF29}
& \int \cD \bar{n}(m+1)\, \cD n (m+1) \, \bar{h}_\sigma (m+1)\, h_\sigma (m+1) 
\notag \\
& \quad \times \bar{h}_\tau (m+1) (\hat{S}^\tp(m))^{-1}_{\tau\alpha} \bar{h}_\alpha
(m) \, h_\rho(m+1)\, \hat{S}_{\rho\beta}(m)\, h_\beta (m) \notag \\[4pt]
& = (\hat{S}^\tp(m))^{-1}_{\sigma\alpha}\, \hat{S}_{\sigma\beta}(m)\,
\bar{h}_\alpha (m)\, h_\beta (m) = \bar{h}_\alpha (m) \, h_\alpha (m)\, .
\end{align}
We can employ this identity for integrating out in eq.~\eqref{eq:IF27} subsequently the pairs of occupation numbers $\bar{n}(m')$ and $n(m')$, starting from $m' = \cM$. This yields
\begin{align}\label{eq:IF30}
Z &= \int \cD \bar{n}(0)\, \cD n(0)\, \delta(\bar{n}(0) - n(0)) \, \bar{f}_{in}
[\bar{n}(0)]\, f_{in} [n(0)] \notag \\
&= \int \cD \bar{n}(0) \, \cD n(0)\, \bar{h}_\sigma (0) \, h_\sigma (0) \,
\bar{q}_\tau (0)\, \bar{h}_\tau (0)\, \tilde{q}_\rho (0) \, h_\rho (0) \notag \\
&= \bar{q}_\sigma (0)\, \tilde{q}_\sigma (0) = 1\, .
\end{align}

We observe that the normalization of the step evolution operator does actually not matter for the weight distribution $w[n,\,\bar{n}]$ defined by eq.~\eqref{eq:IF26}. Replacing $\hat{S}(m)$ by the unnormalized transfer matrix $\hat{T}(m)$ results in a multiplicative renormalization $\cK (m) \to \alpha(m)\,\cK(m)$. The conjugate wave function is renormalized by the inverse factor, $\bar{\cK}(m) \to \alpha^{-1}(m)\, \bar{\cK}(m)$, such that the product $\bar{\cK}(m)\,\cK(m)$ is not affected. For the normalization $Z=1$ it is sufficient to normalize the initial factors $\bar{f}_{in}$ and $f_{in}$ such that $\bar{q}_\tau (0)\, q_\tau (0) = 1$.

Computing the expectation value of a local observable $A(m)$ from the double chain we can integrate out all pairs of $\bar{n}(m')$ and $n(m')$ with $m' > m$,
\begin{equation}\label{eq:IF31}
\langle A(m) \rangle = \prod_{m'=0}^m \int \cD \bar{n}(m')\, \cD n(m')\, A(m)\, 
w[m;\, n,\, \bar{n}]\, ,
\end{equation}
with 
\begin{equation}\label{eq:IF32}
w[m;\, n,\, \bar{n}] = \prod_{m' = 0}^{m-1} \bar{\cK}(m')\, \cK(m')\, \bar{f}_{in}\, 
f_{in}\, .
\end{equation}
The probability distribution $w[m;\, n,\, \bar{n}]$ of the ``restricted double chain'' only involves the occupation numbers $\bar{n}(m')$ and $n(m')$ for $m' \leq m$. No knowledge of the local factors $\cK(m')$ for $m' > m$ is needed. Local physics becomes independent of the future!

\paragraph*{Density matrix as boundary condition}

All these properties can be directly extended to arbitrary boundary conditions. One simply replaces the product of initial factors $\bar{f}_{in}\, f_{in}$,
\begin{equation}\label{eq:IF33}
\bar{f}_{in}\, f_{in} \; \rightarrow \; \sum_\alpha \bar{w}_\alpha\, \bar{f}_{in}^{(\alpha)}\, f_{in}^{(\alpha)} = \cB_{in} [\bar{n}(0),\, n(0)]\, .
\end{equation}
With 
\begin{equation}\label{eq:IF34}
f_{in}^{(\alpha)} = \tilde{q}_\tau^{(\alpha)}(0)\, h_\tau (0)\, ,\quad 
\bar{f}_{in}^{(\alpha)}[\bar{n}(0)] = \bar{q}_\rho^{(\alpha)} (0)\, 
\bar{h}_\rho (0)\, ,
\end{equation}
one has
\begin{align}\label{eq:IF35}
\cB_{in} &= h_\tau (0) \, \sum_\alpha \big( \bar{p}_\alpha\, \tilde{q}_\tau^{(\alpha)}(0)\, \bar{q}_\rho^{(\alpha)} (0)\big)\,\bar{h}_\rho (0) \notag \\
&= h_\tau (0)\, \rho'_{\tau\rho} (0)\, \bar{h}_\rho (0) = h_\tau (0)\, 
(\rho'_{in})_{\tau\rho}\, \bar{h}_\rho (0)\, .
\end{align}
The probability distribution for the double chain is directly formulated in terms of the initial classical density matrix $\rho'_{in}$,
\begin{align}\label{eq:IF36}
w[n,\, \bar{n}] &= \delta(\bar{n}(\cM) - n(\cM)) \prod_{m'=0}^{\cM - 1} \big(
\bar{\cK}(m')\, \cK(m')\big)\, \notag \\
& \quad \times h_\tau (0)\, (\rho'_{in})_{\tau\rho}\, \bar{h}_\rho (0)\, .
\end{align}
This makes it suitable for the investigation of the initial value problem. 

As we have seen, we can choose $\cM$ freely, provided it is larger or equal to the maximal value of $m$ for which we want to compute expectation values of observables. The expectation values are independent of $\cM$ in this case. This statement extends beyond local observables. Let us call $m_{max}$ the largest value of $m$ for which occupation numbers $n(m)$ influence the values of the observables of interest. For $\cM \geq m_{max}$ the expectation values of all those observables are independent of $\cM$. We may also call $m_{min}$ the smallest value of $m$ which affects the observables of interest. It is then not necessary to know all the ``past'' of the double chain for $n(m')$ and $\bar{n}(m')$ with $m' < m_{min}$. We can define the probability distribution for the ``bounded double chain'' by
\begin{align}\label{eq:IF37}
& w[m_{up},\, m_{low};\, n,\, \bar{n}] = \delta(\bar{n}(m_{up}) - n(m_{up})) 
\notag \\ 
& \qquad\qquad \times \prod_{m' = m_{low}}^{m_{up} -1} 
\big( \bar{\cK}(m')\, \cK(m')\big) \, \rho'(m_{low})\, ,
\end{align}
with
\begin{equation}\label{eq:IF38}
\rho'(m_{low}) = h_{\tau}(m_{low}) \, \rho'_{\tau\rho} (m_{low})\, 
\bar{h}_\rho(m_{low})\, .
\end{equation}
It depends on the occupation numbers $n(m')$ and $\bar{n}(m')$ in the range $m_{low} \leq m' \leq m_{up}$. For $m_{low} \leq m_{min}$ and $m_{up} \geq m_{max}$ the expectation values of the observables of interest do not depend on the choice of $m_{up}$ or $m_{low}$.

The independence of $m_{low}$ follows from the possibility to integrate out in the double chain weight distribution the occupation numbers with $m' < m_{low}$. By use of the identity
\begin{equation}\label{eq:IF39}
\int \cD n(m) \, \cD \bar{n}(m)\, \bar{\cK}(m)\, \cK (m)\, \rho' (m) = \rho'(m+1)
\end{equation}
this integration simply transports $\rho'(0)$ to higher values of $m$, ending at $m_{low}$. In turn, the identity \eqref{eq:IF39} follows by expanding all quantities in the occupation number basis, 
\begin{align}\label{eq:IF40}
& \int \cD n(m)\, \cD \bar{n}(m)\, f_\tau (m+1)\, \hat{S}_{\tau\alpha} (m)\, 
f_\alpha (m) \notag \\
& \qquad \times \bar{f}_\rho (m+1)\, (S^\tp)^{-1}_{\rho\beta} (m)\, 
\bar{f}_\beta (m) f_\gamma (m)\, \rho'_{\gamma\delta}(m)\, \bar{f}_\delta (m)         
\notag \\[4pt]
& \quad = f_\tau (m+1)\, \hat{S}_{\tau\alpha}(m)\, \rho'_{\alpha\beta}(m) 
(\hat{S}^\tp)^{-1}_{\rho\beta}(m)\, \bar{f}_\rho (m+1) \notag \\[4pt]
& \quad = f_\tau (m+1)\, \rho'_{\tau\rho}(m+1)\, \bar{f}_\rho (m+1)\, ,
\end{align}
where the last line uses the evolution law for $\rho'$. The identity \eqref{eq:IF39} does not make any assumption about the precise form of the step evolution operator. The step evolution operators for $m' < m_{low}$ may be a complicated chain of different operators for different $m'$. All this detailed information is not needed. The only information from the past is summarized in the density matrix $\rho'(m_{low})$.

\paragraph*{Time-local physics}

These findings reveal an important structure for local chains: local physics needs only local information! If one is interested in expectation values and relations of observables with support in a time range $\Delta t$ between $t_{min}$ and $t_{max}$, no information about the future for $t > t_{max}$ is needed. Furthermore, all information needed from the past is summarized in the classical density matrix $\rho'(t_{min})$. With this information, and the specification of the model by the local factors $\cK(t)$ in the range $t_{min} \leq t \leq t_{max}$, all quantities of interest are, in principle, computable.

The formulation as a double chain needs the conjugate local factors $\bar{\cK}(m)$. For given $\cK(m)$ they may be difficult to compute, since the step evolution operator $\hat{S}(m)$ must be found and inverted. A much simpler structure arises if the step evolution operator is an orthogonal matrix. In this case one has $(\hat{S}^\tp)^{-1} = \hat{S}$, and $\bar{\cK}(m)$ equals $\cK(m)$ up to the replacement of $n(m+1)$ and $n(m)$ by $\bar{n}(m+1)$ and $\bar{n}(m)$. Furthermore, for a symmetric initial density matrix $\rho'_{in} = (\rho'_{in})^\tp$, the density matrix remains symmetric for all $m$. With symmetric $\rho'$ the double chain has an important symmetry: it is invariant under a simultaneous exchange of occupation numbers $n(m) \; \leftrightarrow \; \bar{n}(m)$ for all $m$. For orthogonal $\hat{S}$ the double chain is the analogue of the Schwinger-Keldysh formalism \cite{JSS,KEL} for quantum mechanics and quantum field theory.

For unique jump chains the step evolution operator is orthogonal. The local factors $\cK(m)$ are $\delta$-functions that transport every configuration $\rho$ at $m$ to a configuration $\tau(\rho)$ at $m+1$. This holds in parallel for the configurations $\{ n(M)\}$ and $\{\bar{n}(m)\}$. For a given density matrix $\rho'_{\rho\alpha} (m)$ the density matrix at $m+1$ reads
\begin{equation}\label{eq:IF41}
\rho'_{\tau\beta} (m+1) = \rho'_{\rho(\tau),\,\alpha(\beta)} (m)\, ,
\end{equation}
with $\rho(\tau)$ the inverse of $\tau(\rho)$, and correspondingly $\alpha(\beta)$ the inverse of $\beta(\alpha)$. The initial value problem can be solved by deterministic updating  of the density matrix in consecutive steps.
\subsubsection{Clock systems}\label{sec:clock_systems}

Unique jump chains are the simplest setting where initial information is not lost. The orthogonal step evolution operator resembles the unitary evolution in quantum mechanics.
One of the simplest systems is the clock system. It consists of a unique jump chain of $N$ states that are reached sequentially as time progresses. The evolution is periodic, with time period given by $\varepsilon N$.
Periodic evolution and clocks are basic ingredients for the concept of ``physical time" discussed in sect. \ref{sec:properties_of_physical_time}.

\paragraph*{Deterministic and probabilistic clocks}

\begin{figure}[t!]
	\includegraphics[scale=.75]{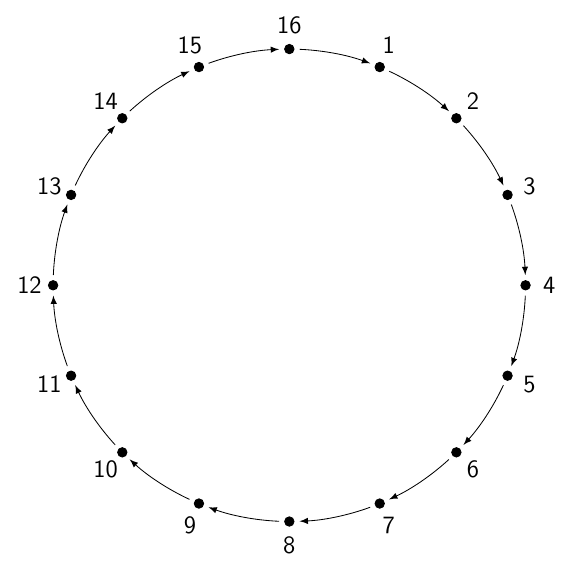}
	\caption{Clock system for $N=16$.}\label{fig:1} 
\end{figure}

With $\tau = 1,\, \dots,\, N$ the step evolution operator for a clock system is given by the unique jump operator
\begin{equation}\label{eq:CS1}
\hat{S}_{\tau\rho}(t) = \delta_{\tau,\,\rho+1}\, .
\end{equation}
Whenever the system is in the state $\rho$ at $t$, it is necessarily in the state $\rho +1$ at $t + \varepsilon$. Otherwise the probability distribution vanishes. For $\rho = N$ we identify the state $\rho = N+1$ with $\rho = 1$. The evolution is therefore periodic. For sharp initial states, where the probability at $t = 0$ is one only for one particular state $\rho_0$, the clock system is a simple type of cellular automaton, depicted in Fig.~\ref{fig:1}. The drawing associates the evolution with the motion of a pointer in a clock.

We consider here probabilistic boundary conditions. For pure classical states they are encoded in the classical wave functions $\tilde{q}_\tau(t_{in})$ and $\bar{q}_\tau (t_f)$. We do not take time to be periodic. Typically $(t_f - t_{in})/\varepsilon$ is taken much larger than $N$, such that starting from $t_{in}$ the clock can rotate for many periods before $t_f$ is reached. We may take $(t_f - t_{in})/(\varepsilon N) \to \infty$ at the end, such that ``the clock rotates forever''. With step evolution operator \eqref{eq:CS1} the evolution of the wave functions obeys
\begin{align}\label{eq:CS2}
& \tilde{q}_\tau (t + \varepsilon) = \hat{S}_{\tau\rho}\,\tilde{q}_\rho (t) = 
\tilde{q}_{\tau - 1} (t)\, , \notag \\
& \bar{q}_\tau (t + \varepsilon) = \bar{q}_{\tau - 1} (t)\, ,
\end{align}
where we employ that the unique jump operator \eqref{eq:CS1} is an orthogonal matrix. In turn, this implies for the local probabilities
\begin{equation}\label{eq:CS3}
p_\tau (t+\varepsilon) = p_{\tau - 1} (t)\, .
\end{equation}
For the special case $p_\tau (t=0) = \delta_{\tau,\, \tau_0}$ the pointer starts at $t=0$ at the position $\tau_0$, at $t+\epsilon$ it is at the position $\tau_0 + 1$, and so on. This is a deterministic clock. The probabilistic clock is simply a probability distribution over deterministic clocks, with $p_\tau(t=0)$ the probability for an initial pointer position at $\tau$.

Alternatively, we may specify the system by the initial classical density matrix $\rho'(t=0)$. The evolution of the density matrix obeys the simple law
\begin{equation}\label{eq:CS4}
\rho'_{\tau\rho} (t+\varepsilon) = \rho'_{\tau - 1,\, \rho-1} (t)\, .
\end{equation}
In particular, the diagonal elements $\rho'_{\tau\tau} = p_\tau$ obey eq.~\eqref{eq:CS3}. As for every unique jump chain, a clock system is a limiting case of a Markov chain. Only knowledge of the local probabilities $p_\tau$ at $t$ is needed for the computation of $p_\tau(t+\varepsilon)$.

\paragraph*{Continuous rotation}\label{sec:continuous_rotation}

We may build a clock from $M$ Ising spins or bits, with $N=2^M$. We can use the bits in the same way as the representation of integers in a binary basis, such that $\tau = 1,\, \dots,\, N$ is directly given by the representation of the integer $\tau - 1$ in terms of a sequence of $M$ bits. We can also associate $\tau$ with an angle $\alpha_\tau$,
\begin{equation}\label{eq:CS5}
\alpha_\tau = \frac{2\pi\,\tau}{N}\, .
\end{equation}
The periodicity $N + \tau = \tau$ is then mapped to the periodicity of the angle with period $2\pi$. We may represent the wave functions as functions of the discrete angle $\alpha_\tau$,
\begin{equation}\label{eq:CS6}
\tilde{q}_\tau \equiv \tilde{q}(\alpha_\tau)\, .
\end{equation}
Taking $M \to \infty$ at fixed periodicity for $\alpha$, the angular variable $\alpha$ becomes continuous. Wave functions and local probabilities are in this limit functions of a continuous angular variable $\alpha$, e.g. $\tilde{q}(\alpha)$, $\bar{q}(\alpha)$ and $p(\alpha)$. The classical density matrix depends on two angles, $\rho'(\alpha,\, \alpha')$.

As long as $M$ remains finite, the angular step $\Delta\alpha$ performed during a time step $\varepsilon$ or unit step on the local chain is given by
\begin{equation}\label{eq:CS7}
\Delta\alpha = \frac{2\pi}{2^M}\, ,
\end{equation}
such that
\begin{equation}\label{eq:CS8}
\tilde{q}(t+\varepsilon,\, \alpha) = \tilde{q}(t,\, \alpha - \Delta\alpha)\, .
\end{equation}
The general solution is given by
\begin{equation}\label{eq:CS8A}
\tilde{q}(t,\,\alpha) = \tilde{q} \Big(0,\, \alpha - \frac{t\,\Delta\alpha}
{\varepsilon} \Big)\, ,
\end{equation}
and similar for $\bar{q}$, $p$ and $\rho'$. We can define a discrete derivative
\begin{equation}\label{eq:CS9}
\p_t \tilde{q}(t,\, \alpha) = \frac{1}{2\varepsilon} \big[ 
\tilde{q}(t + \varepsilon,\, \alpha) - \tilde{q}(t-\varepsilon,\, \alpha) \big]\,.
\end{equation}
It obeys
\begin{equation}
\p_t \tilde{q}(t,\, \alpha) = -\frac{1}{2\varepsilon} \big[ \tilde{q} (t,\, \alpha + \Delta\alpha) - 
\tilde{q}(t,\, \alpha - \Delta\alpha) \big]\, .
\end{equation}
Defining similarly
\begin{equation}\label{eq:CS10}
\p_\alpha \tilde{q} (t,\, \alpha) = \frac{1}{2\Delta\alpha} \big[ 
\tilde{q}(t + \Delta\alpha) - \tilde{q} (t,\, \alpha - \Delta\alpha)\big]\, ,
\end{equation}
the evolution law reads
\begin{equation}\label{eq:CS11}
\p_t \tilde{q}(t,\,\alpha) = -\frac{\Delta\alpha}{\varepsilon} \p_\alpha 
\tilde{q}(t,\, \alpha)\, .
\end{equation}

The time units $\varepsilon$ are arbitrary. We can choose them such that the ratio $\Delta\alpha/\varepsilon$ remains fixed as $M \to \infty$,
\begin{equation}\label{eq:CS12}
\frac{\Delta\alpha}{\varepsilon} = \omega\, .
\end{equation}
In this limit the general solution \eqref{eq:CS8A} is a continuous rotation in time
\begin{equation}\label{eq:CS13}
\tilde{q}(t,\, \alpha) = \tilde{q} (0,\, \alpha - \omega t ) \, ,
\end{equation}
with $\omega$ the angular frequency and time period $2\pi/\omega$. If the initial wave function $\tilde{q}(0,\, \alpha)$ has a particular feature at $\alpha = \alpha_0$, say a maximum or some other feature, this may be associated with the position of the pointer of the clock. At later times the position of the pointer will be at
\begin{equation}\label{eq:CS14}
\alpha(t) = \alpha_0 + \omega t\, .
\end{equation}
The pointer rotates continuously. 

If furthermore the angular dependence of $\tilde{q}(t,\,\alpha)$ is sufficiently smooth, the discrete derivative $\p_\alpha$ in eq.~\eqref{eq:CS10} becomes the standard partial derivative. The evolution is then given by a differential equation,
\begin{equation}\label{eq:CS15}
\p_t \tilde{q} = -\omega\, \p_\alpha \tilde{q}\, .
\end{equation}
The same applies to $\bar{q}$, $p$ and $\rho'$. A differential equation emerges as a limiting case of local chains for Ising spins.
The clock system in the limit $N \to \infty$ is a first example for a continuous evolution in time. We will discuss continuous time
in more detail in sect. \ref{sec:continuous_time}.

\paragraph*{Unique jump chains and clock systems}

Consider unique jump chains with a finite number $M$ of occupation numbers or Ising spins at every layer $t$, and with local factors $\cK(t)$ or associated step evolution operators $\hat{S}(t)$ independent of $t$. All non-trivial unique jump systems with constant $\hat{S}$ are clock systems. They may describe several clocks with different periodicities. The maximal periodicity for a single clock is $2^M$.

For an invertible unique jump operation a given state $\tau_1$ can either remain the same or move to a different state $\tau_2$. In the next step $\tau_2$ either moves back to $\tau_1$, or it reaches a new state $\tau_3$ which is different from $\tau_1$ and $\tau_2$. It cannot stay at $\tau_2$ since this would contradict invertibility. If it moves back to $\tau_1$, the two states $\tau_1$ and $\tau_2$ constitute a clock system with period $N=2$. This logic continues if $\tau_3$ has been reached. In the next step the system either returns to $\tau_1$, such that the states $(\tau_1,\,\tau_2,\,\tau_3)$ realize a clock system with period $N=3$, or else a new state $\tau_4$ must be reached, which has to be different from $\tau_2$ and $\tau_3$ due to invertibility. At some point the chain of unique jumps has to close, at the latest after $2^M$ steps. There is then no new state $\tau_{2^M+1}$ available, such that the only possibility remains a return to $\tau_1$. We conclude that either $\tau_1$ is an invariant state, or it belongs to a clock system with period $N_1$. The maximal value for $N_1$ is $2^M$, but any $N_1 \leq 2^M$ is possible. For an invariant state we take $N_1 = 1$.

For $N_1 = 2^M$ all available states have been reached by the clock. In contrast, for $N_1 < 2^M$ there remain $2^M - N_1$ states that have not been reached by starting at $\tau_1$. Out of the $2^M - N_1$ remaining states we may take a state $\tau'_1$. The evolution of $\tau'_1$ during subsequent steps can never reach one of the states of the first clock, since this would contradict invertibility of the unique jump chain. Either $\tau'_1$ is invariant, or the later steps of the evolution starting at $\tau'_1$ have to form a second clock system with period $N_2$. This period is bounded by $N_2 \leq 2^M - N_1$. For $2^M - N_1 - N_2 > 0$ we continue by considering a state $\tau''_1$ that does not belong to the states of the first two clocks. It is invariant or belongs to a third clock system. Continuing this procedure until all $2^M$ states of the system are used, one finds that the unique jump chain describes a system of clocks with periods $N_j$. The sum of all periods equals the number of states,
\begin{equation}\label{eq:CS16}
\sum_j N_j = 2^M\, .
\end{equation}
This sum includes invariant states with $N_j = 1$. If all $N_j$ equal one the unit jump chain is the trivial chain where all states are invariant. Otherwise, at least one clock contains more than one state and features true periodicity. The order of the states in a given clock is determined by the sequence of unit jump operations realizing the clock. Only the starting point is arbitrary. 

Any system consisting of periodic subsystems is periodic itself. The combined period can be much longer than the largest period of a subsystem. As an example, consider $M=4$, $N=16$, and three subsystems with $N_1=4$, $N_2=5$, $N_3=7$. The total period is $4\cdot 5\cdot 7 = 140$, since all three subsystems have to reach simultaneously the initial state. For large $M$ extremely large periods become possible, much larger than $2^M$. Nevertheless, for any finite $M$ also the period of the whole system remains finite.
We conclude that generic unique jump chains with large $M$ constitute clock systems with many clocks. Clock systems with clocks that
can be synchronized are an essential ingredient for the universality of physical time discussed in sect.~\ref{sec:physical_time}.

\paragraph*{Clock systems and time units}

The choice of the angular frequency $\omega$ in eq.~\eqref{eq:CS12} is arbitrary, since the choice of $\varepsilon$ is arbitrary. A single clock system only sets the units of time. The frequency $\omega$ and the associated time unit is not a measurable quantity. For a system with two clocks the ratio of frequencies,
\begin{equation}\label{eq:CS17}
\frac{\omega_1}{\omega_2} = \frac{\Delta\alpha_2}{\Delta\alpha_1} = \frac{N_1}{N_2}\, 
,
\end{equation}
becomes a measurable quantity since the arbitrary unit $\varepsilon$ drops out. It is given by the inverse ratios of periods. The ``measurement'' consists of counting how many periods the clock two traverses during a period of the clock one. This is the way how time has always been measured. A first periodic system sets the units, for example the day for the rotation of the Earth, or the year for the rotation of the Earth around the Sun. The evolution of other systems can then be measured in time units set by the first clock. Other clocks can be ``gauged'' by determining the ratio \eqref{eq:CS17}. In present days, the time units are set by the periodic evolution in atomic clocks.
\subsection{Physical time}\label{sec:physical_time}

Physical time is based on the counting of oscillations for a periodic evolution. Different oscillations of clocks can be synchronized to clock systems and establish a certain universality of time. The continuum limit induces units for time and energy and introduces Planck's constant as a conversion factor. For an orthogonal step evolution operator and in presence of a complex structure compatible with the evolution we obtain the familiar Schrödinger or von Neumann equation of quantum mechanics. The possibility to use different ordering structures for the definition of time is reflected by different reference frames which measure different time.
\subsubsection{Continuous time}\label{sec:continuous_time}

Continuous time is realized by local chain systems as a limiting process, the ``continuum limit". An example has been given for the clock system in sect.\,\ref{sec:clock_systems}. This limit consists of taking $\varepsilon \rightarrow 0$, while $t_\mathrm{f}-t_\mathrm{in}$ remains finite. It therefore corresponds to the number of sites on the chain or local factors $\mathcal{M} \rightarrow \infty$. A continuum limit is possible for classical wave functions, classical density matrices and observables if their dependence on $t$ is sufficiently smooth. Typically we require that time-derivatives can be defined that keep finite values for $\varepsilon \rightarrow 0$.

Let us consider the classical wave function and the evolution equation $\tilde{q}(t+\varepsilon) = \hat{S}(t)\tilde{q}(t)$. We define the discrete time derivative by
\begin{equation}
\partial_t \tilde{q}(t) 
= \frac{1}{2\varepsilon} \left( \tilde{q}(t+\varepsilon) - \tilde{q}(t-\varepsilon) \right) 
= W(t)\tilde{q}(t),
\label{eq:ct1}
\end{equation}%
where
\begin{equation}
W(t) = \frac{1}{2\varepsilon} \left( \hat{S}(t) - \hat{S}^{-1}(t-\varepsilon) \right).
\label{eq:ct2}
\end{equation}%
For a continuous evolution we may require that $\partial_t \tilde{q}(t)$ remains finite for $\varepsilon \rightarrow 0$, except perhaps for a certain number of ``singular time points".

In sect.~\ref{sec:observables_and_operators} we will discuss time-derivative observables, analogous to velocity observables as time-derivatives of position observables. A continuum limit requires that expectation values of such observables should not depend on the precise microscopic definition of the time-derivative. Extending this to product observables whose expectation values define correlation functions, we will learn that classical observable products and the classical correlation function are not compatible with the continuum limit. Consistency with the continuum limit will be an important criterion for the selection of robust observable products and associated correlations that do not depend on the precise definition of time-derivatives.

\paragraph*{Continuum limit}

Since $\varepsilon$ only sets the time units, with time-distance between two neighboring sites on the local chain defined to be $\varepsilon$, we should specify the meaning of the limit $\varepsilon \rightarrow 0$. What matters are only dimensionless quantities. Let us take some time interval $\Delta t = \varepsilon \tilde{\mathcal{M}}$. The limit $\varepsilon \rightarrow 0$ at fixed $\Delta t$ corresponds to $\tilde{\mathcal{M}} \rightarrow \infty$. It is this type of limit that we have to consider. A continuum limit is realized if $\tilde{q}(t+\Delta t) - \tilde{q}(t)$ remains small for small $\Delta t$, while $\tilde{\mathcal{M}} \rightarrow \infty$.

The continuum limit can be implemented by a sequence of models, as for the clock system. We may keep both $\Delta t$ and the angle $\Delta\varphi = \tilde{\mathcal{M}} \Delta\alpha$ fixed, while increasing $\tilde{\mathcal{M}}$ and simultaneously the number $M$ of Ising spins at a given site, such that $\tilde{\mathcal{M}}/N = 2^{-M} \tilde{\mathcal{M}}$ remains constant. In this limit one finds for the angle $\Delta\varphi$, that the pointer has rotated in the time interval $\Delta t$, the expression
\begin{equation}
\Delta\varphi = \Delta\alpha\tilde{\mathcal{M}}
= (2\pi) \tilde{\mathcal{M}}/N = (2\pi) 2^{-M} \Delta t/\varepsilon.
\label{eq:ct3}
\end{equation}%
This is independent of the limit $\tilde{\mathcal{M}} \rightarrow \infty$, $\tilde{\mathcal{M}}/N$ constant. The pointer rotates by an angle
\begin{equation}
\alpha(t+\Delta t) - \alpha(t) = \Delta\varphi = \omega\Delta t,
\label{eq:ct4}
\end{equation}%
independently of $\varepsilon$.

In practice, it is often sufficient to consider one given model with a very large number $\tilde{\mathcal{M}}$, instead of the formal limit $\tilde{\mathcal{M}} \rightarrow \infty$ for a sequence of models. The criterion for a continuum limit is then that $\alpha$ has changed only by a finite small amount for a suitable ``small" time interval $\Delta t$. Even for the small time interval the number of time points is still very large in this case. In short, a continuum limit amounts to a small change of the wave function for a very large number of time steps.

We emphasize that the continuum limit does not require that the step evolution operator $\hat{S}(t)$ is close to the unit operator, as one may infer too naively from eqs.\,(\ref{eq:ct1})(\ref{eq:ct2}). For the clock system the step evolution operator is a unique jump operator which does not change in the continuum limit. Only the number of points on the circle that corresponds to a given angle $\Delta\varphi$ increases.

The continuum limit for the one particle wave function in sect.\,\ref{sec:free_particles_in_two_dimensions} is of a similar type. For a fixed $\Delta t$ and increasing $\tilde{\mathcal{M}}$ also the number $\tilde{\mathcal{M}}_x$ of $x$-points for a given $\Delta x$ increases. Defining
\begin{equation}
\tilde{\mathcal{M}}_t = \frac{\Delta t}{\varepsilon_t},\quad \tilde{\mathcal{M}}_x = \frac{\Delta x}{\varepsilon_x},
\label{eq:ct5}
\end{equation}%
one has
\begin{equation}
\frac{\Delta x}{\Delta t} = \frac{\tilde{\mathcal{M}}_x \varepsilon_t}{\tilde{\mathcal{M}}_t \varepsilon_x},
\label{eq:ct6}
\end{equation}%
and the continuum limit, $\tilde{\mathcal{M}}_t \rightarrow \infty$, $\tilde{\mathcal{M}}_t / \tilde{\mathcal{M}}_x = \textnormal{const}$, can be taken. The choice of $\tilde{\mathcal{M}}_t / \tilde{\mathcal{M}}_x$ and $\varepsilon_x/\varepsilon_t$ fixes the units for time intervals and space distances. We have chose them in sect.\,\ref{sec:free_particles_in_two_dimensions} such that the velocity of the particles equals one. Again, the step evolution operator is not close to the unit matrix. The $W$-operator in eq.\,(\ref{eq:ct2}) equals the negative $x$-derivative $\partial_x$ defined in eq.\,(\ref{eq:FP19})
\begin{equation}
W = -\partial_x.
\label{eq:ct7}
\end{equation}%
Its explicit form (\ref{eq:CQ9}) has off-diagonal elements neighboring the diagonal which are proportional $\varepsilon^{-1}$.

\paragraph*{Units and Planck's constant}

With $\varepsilon$ having units of time, the $W$-operator (\ref{eq:ct2}) has units of inverse time, since $\hat{S}$ is dimensionless. In natural units $W$ has units of energy, which equal units of inverse time. We could define energy units different from inverse time by introducing a ``conversion constant" $\hbar$,
\begin{equation}
W(t) = \frac{\hbar}{2\varepsilon} \left( \hat{S}(t) - \hat{S}^{-1}(t) \right).
\label{eq:ct8}
\end{equation}%
This conversion constant appears then in the evolution equation
\begin{equation}
\hbar \,\partial_t \tilde{q} = W\tilde{q}.
\label{eq:ct9}
\end{equation}%
If we identify $\hbar$ with Planck's constant we use a conversion between inverse seconds and some energy units as Newtons. It is obvious that $\hbar$ is a purely human invention, adapted to practical historical definitions of seconds and energy units. Its value has no fundamental meaning. Civilizations on other planets would typically use other units, and therefore a different value of $\hbar$. With $\hbar$ a conversion constant used to define units, it does no depend on fields, time, or other quantities. The limit $\hbar \rightarrow 0$ used often for the classical limit in quantum mechanics is actually a limit where the action $\tilde{S}$ of a system is large compared to $\hbar$, corresponding to a divergence of the dimensionless ratio $\tilde{S}/\hbar \rightarrow \infty$. We will adopt natural units and set $\hbar = 1$.

\paragraph*{Evolution of classical density matrix}

The time-evolution equation of the conjugate classical wave function $\bar{q}$ can be expressed as 
\begin{equation}
\partial_t \bar{q}(t) = \frac{1}{2\varepsilon} \left[ \bar{q}(t+\varepsilon) - \bar{q}(t-\varepsilon) \right] = -\tilde{W}^\mathrm{T}(t) \bar{q}(t),
\label{eq:ct10}
\end{equation}%
with
\begin{equation}
\tilde{W}(t) = \frac{1}{2\varepsilon} \left[ \hat{S}(t-\varepsilon) - \hat{S}^{-1}(t) \right].
\label{eq:ct11}
\end{equation}%
The operator $\tilde{W}(t)$ equals $W(t)$ for $\hat{S}$ independent of time. In the type of continuum limit that we consider here the difference between $\tilde{W}(t)$ and $W(t)$ can be neglected, such that the product rule for differentiation applies and the normalization is preserved,
\begin{equation}
\partial_t (\bar{q}_\tau \tilde{q}_\tau) = \bar{q}_\tau\,\partial_t \tilde{q}_\tau + \partial_t \bar{q}_\tau \,\tilde{q}_\tau = 0.
\label{eq:ct12}
\end{equation}%
In turn, the evolution of the classical density matrix obeys
\begin{equation}
\partial_t \rho'(t) = \frac{1}{2\varepsilon} \left( \rho'(t+\varepsilon) - \rho'(t-\varepsilon) \right)
= \left[ W(t), \rho'(t) \right].
\label{eq:ct13}
\end{equation}%
In the last expression we have neglected correction terms similar to the difference between $\tilde{W}$ and $W$. Again, this preserves the normalization, $\partial_t \tr (\rho') = 0$. For $\tilde{W}=W$ it is compatible with the product rule for differentiation for pure classical states where $\rho'_{\tau\sigma} = \tilde{q}_\tau \bar{q}_\sigma$. 

Indeed, the continuum limit applies for slowly varying classical wave functions and density matrices, $\tilde{q}(t+\varepsilon) = \tilde{q}(t) + \mathcal{O}(\varepsilon)$ etc., such that in leading order relations of the type
\begin{equation}
\hat{S}(t)\tilde{q}(t) \approx \hat{S}(t+\varepsilon) \tilde{q}(t+\varepsilon) \approx \hat{S}(t+\varepsilon) \tilde{q}(t)
\label{eq:ct14}
\end{equation}%
lead to
\begin{equation}
W(t) = \tilde{W}(t) = \frac{1}{2\varepsilon} \left( \hat{S}(t) - \hat{S}^{-1}(t) \right).
\label{eq:ct15}
\end{equation}%
In leading order, only the difference between $\hat{S}$ and $\hat{S}^{-1}$ matters, whereas the precise location of the step evolution operators plays a negligible role. For a smooth enough classical density matrix one requires the leading relation
\begin{equation}
\hat{S} \rho' \hat{S}^{-1} - \hat{S}^{-1} \rho' \hat{S} \approx \hat{S}\rho' + \rho'\hat{S}^{-1} - \hat{S}^{-1}\rho' - \rho'\hat{S},
\label{eq:ct16}
\end{equation}%
where the precise position of $\rho'$ and $\hat{S}$ does not matter. This leads to the relation (\ref{eq:ct13}), while the correction terms vanish in this limit.

We are interested in observables for which a continuum limit exists. Their expectation values should vary slowly in time, such that the difference between $\braket{A(t+\varepsilon)}$ and $\braket{A(t)}$ is of the order $\varepsilon$. For sufficiently smooth $\rho'(t)$ this is realized if the associated operators $\hat{A}(t)$ only vary slowly in time. In this case the continuum limit can be taken as
\begin{equation}
\braket{A(t)} = \tr \left\lbrace \rho'(t) \hat{A}(t) \right\rbrace,
\label{eq:ct17}
\end{equation}%
independently of the precise location of $\rho'(t)$ and $\hat{A}(t)$. The continuum limit looses the resolution of time differences of the order $\varepsilon$. Of course, time differences on much larger time scales $\Delta t$ can still occur. With eq.\,(\ref{eq:ct13}) one finds for $\hat{A}(t)$ independent of $t$
\begin{equation}
\partial_t \braket{A(t)} = -\tr \left\lbrace \rho'(t) [W(t),\hat{A}] \right\rbrace.
\label{eq:ct18}
\end{equation}%

The definition of discrete time-derivatives (\ref{eq:ct1}), (\ref{eq:ct10}), (\ref{eq:ct13}) is not unique. One may consider other discrete derivatives as
\begin{equation}
\partial_t^{(+)} \tilde{q} = \frac{1}{\varepsilon} \left( \tilde{q}(t+\varepsilon) - \tilde{q}(t) \right).
\label{eq:ct19}
\end{equation}%
In the continuum limit the difference between $\partial_t \tilde{q}$ and $\partial_t^{(+)} \tilde{q}$ should vanish.
This requirement extends to time-derivative observables and their correlations.

Finally, we note that a continuum limit remains possible also for large $\tilde q(t+\epsilon)-\tilde q(t)$, or for $\hat S(t+\epsilon)$ much different from $\hat S(t)$. It is sufficient that smoothness is realized after a certain number $P$ of $\epsilon$-steps. One replaces in this case $\epsilon$ by $P\epsilon$ in the discussion above.

\paragraph*{Decomposition of the $W$-operator}

The $W$-operator can be decomposed into antisymmetric and symmetric parts,
\begin{equation}
W = \hat{F} + \hat{J},
\label{eq:ct20}
\end{equation}%
with
\begin{equation}
\hat{F}^\mathrm{T} = -\hat{F},\quad \hat{J}^\mathrm{T} = \hat{J}.
\label{eq:ct21}
\end{equation}%
In the continuum limit,
\begin{equation}
W(t) = \frac{1}{2\varepsilon} \left( \hat{S}(t) - \hat{S}^{-1} (t) \right),
\label{eq:ct22}
\end{equation}%
we observe that an orthogonal step evolution operator, $\hat{S}^{-1}(t) = \hat{S}^\mathrm{T}(t)$, leads to an antisymmetric $W$-operator or $\hat{J}=0$. For $\hat{J}=0$ the length of the classical wave function vector $\tilde{q}$ is conserved as appropriate for a rotation, 
\begin{equation}
\partial_t (\tilde{q}_\tau \tilde{q}_\tau) 
= \tilde{q}_\tau \hat{F}_{\tau\rho} \tilde{q}_\rho + \hat{F}_{\tau\rho} \tilde{q}_\rho \tilde{q}_\tau = 0.
\label{eq:ct23}
\end{equation}%
The same holds for the conjugate classical wave function $\bar{q}$. This corresponds to an evolution with conserved norm, similar to the conserved norm of the wave function in quantum mechanics.

There is a one to one correspondence between vanishing $\hat{J}$ and an evolution with orthogonal step evolution operator. In particular, for unique jump chains $\hat{S}$ is orthogonal and therefore $\hat{J}=0$. We will see that in presence of a complex structure $F$ corresponds to a hermitian Hamiltonian, while $J$ introduces a part not present in quantum mechanics.

For an orthogonal evolution the probabilistic information in the boundary terms is not lost as one moves inside the bulk. On the other hand, nonzero $\hat{J}$ is typically associated to a loss of boundary information. As an example we may take the Ising model discussed in sect.\,\ref{sec:influence_of_boundary_conditions}.

\paragraph*{Ising model}

With $\hat{S}$ given by eq.\,(\ref{eq:BC4}) and
\begin{equation}
\hat{S}^{-1} = \frac{1}{2\sinh \beta}
	\begin{pmatrix}
	e^\beta & -e^{-\beta} \\
	-e^{-\beta} & e^\beta
	\end{pmatrix},
\label{eq:ct24}
\end{equation}%
one obtains
\begin{equation}
W = \hat{J} = -\frac{e^{-2\beta}}{\varepsilon(1-e^{-4\beta})} \begin{pmatrix}
	1 & -1 \\
	-1 & 1
\end{pmatrix},
\label{eq:ct25}
\end{equation}%
while the antisymmetric part vanishes, $\hat{F}=0$. A continuum limit can be defined for
\begin{equation}
e^{-2\beta} = \gamma\varepsilon,
\label{eq:ct26}
\end{equation}%
with constant $\gamma$ and $\varepsilon \rightarrow 0$. It results in the evolution equation
\begin{equation}
\partial_t \tilde{q} = -\gamma (1-\tau_1)\tilde{q}.
\label{eq:ct27}
\end{equation}%

The $W$-operator,
\begin{equation}
W = -\gamma(1-\tau_1),
\label{eq:ct28}
\end{equation}%
has one vanishing eigenvalue. The corresponding eigenvector is the equilibrium wave function (\ref{eq:BC6}). For
\begin{equation}
\tilde{q}(t) = \frac{1}{\sqrt{2}} \begin{pmatrix}
1 \\ 1 \end{pmatrix}
+ a(t) \begin{pmatrix}1 \\ -1\end{pmatrix},
\label{eq:ct29}
\end{equation}%
one obtains
\begin{equation}
\partial_t a = -2\gamma a.
\label{eq:ct29a}
\end{equation}%
This amounts to an exponential decay of $a$, or an exponential approach towards the equilibrium wave function,
\begin{equation}
\tilde{q}(t_2) = \tilde{q}_\mathrm{eq} + a(t_1) \exp 
\left( -\frac{(t_2-t_1)}{\xi} \right)
\begin{pmatrix} 1 \\ -1 \end{pmatrix}.
\label{eq:ct30}
\end{equation}%
The correlation length,
\begin{equation}
\xi = \frac{1}{2\gamma} = \frac{\varepsilon}{2e^{-2\beta}},
\label{eq:ct31}
\end{equation}%
agrees with eq.\,(\ref{eq:BC15}) in the continuum limit $\varepsilon \rightarrow 0$, $e^{-2\beta} \rightarrow 0$.

A continuum limit offers a particularly simple way to solve the boundary value problem by a solution of generalized Schrödinger equations for classical wave functions. This can be helpful for more complex systems for which the computation of eigenvalues and eigenvectors of the step evolution operator may be more involved.

\paragraph*{Hamilton operator and Schrödinger equation}

Let us focus on evolution laws that are compatible with a complex structure. This is the case if there exists a basis for which the $W$-operator takes the form 
\begin{equation}
W = \begin{pmatrix}
\WR & -\WI \\
\WI & \WR
\end{pmatrix}
= \WR + I\WI\ ,
\label{eq:ct32}
\end{equation}%
where the matrix $I$ and an involution operator $K_c$ are defined by
\begin{equation}\label{YYA}
I=\begin{pmatrix}0&-1\\1&0\end{pmatrix}\ ,\quad K_c=\begin{pmatrix}1&0\\0&-1\end{pmatrix}\ ,\quad \{I,K_c\}=0\ .
\end{equation}
Here $K_c$ will play the role of complex conjugation, and $I$ operates multiplication by $i$. (For more details on complex structures see sects.~\ref{sec:complex_structure},~\ref{sec:particles_and_holes}.) We split the wave function $\tilde q$ into eigenfunctions of $K_c$, with ``real part'' $\tilde q_\text{R}$ corresponding to positive eigenvalues of $K_c$, $K_c\tilde q_\text{R}=\tilde q_\text{R}$, and ``imaginary parts'' $\tilde q_\text{I}$ corresponding to negative eigenvalues, $K_c\tilde q_\text{I}=-\tilde q_\text{I}$,
\begin{equation}\label{YYB}
\tilde q=\begin{pmatrix}\tilde q_\text{R}\\ \tilde q_\text{I}\end{pmatrix}\ .
\end{equation}
We introduce a complex wave function by
\begin{equation}\label{YYC}
\psi=\tilde q_\text{R}+i\tilde q_\text{I}\ .
\end{equation}

The evolution equation for the complex wave function $\psi$ becomes a complex equation
\begin{equation}
i\,\partial_t\psi = G\psi,
\label{eq:ct33}
\end{equation}%
with
\begin{equation}
G = i\WR - \WI = H + iJ.
\label{eq:ct34}
\end{equation}%
Here we have decomposed the complex matrix $G$ into an hermitean part $H$ and an antihermitean part $iJ$,
\begin{equation}
H^\dagger = H,\quad J^\dagger = J.
\end{equation}%

We may compare with the decomposition (\ref{eq:ct20}). With
\begin{align}
\begin{split}
H &= \frac{1}{2}(G + G^\dagger) = -\frac{1}{2} (\WI + \WI^\tp) + \frac{i}{2} (\WR - \WR^\tp),\\
J &= -\frac{i}{2}(G - G^\dagger) = \frac{1}{2} (\WR + \WR^\tp) + \frac{i}{2} (\WI - \WI^\tp),
\end{split}\label{eq:ct36}
\end{align}%
and
\begin{align}
\begin{split}
\hat{F} &= \frac{1}{2} \begin{pmatrix}
	\WR-\WR^\tp & -(\WI+\WI^\tp) \\[1mm]
	\WI+\WI^\tp & \WR-\WR^\tp
\end{pmatrix}
	= -I\hat{H}, \\[2mm]
\hat{J} &= \frac{1}{2} \begin{pmatrix}
	\WR+\WR^\tp & -(\WI-\WI^\tp) \\[1mm]
	\WI-\WI^\tp & \WR+\WR^\tp
\end{pmatrix},
\end{split}\label{eq:ct37}
\end{align}%
one finds that $H$ is the complex representation of $\hat{H}$, and $J$ the complex representation of $\hat{J}$,
\begin{equation}
\hat{H} = \HR + I\HI,\quad \hat{J} = \JR + I\JI\ .
\label{eq:ct38}
\end{equation}%
Here $\hat{H}$ is given by
\begin{equation}
\hat{H} = I\hat{F} = -\frac{1}{2} (\WI + \WI^\tp) + \frac{1}{2} (\WR - \WR^\tp)I.
\label{eq:ct39}
\end{equation}

For an orthogonal evolution with $\hat{J}=0$ the evolution equation in the complex formulation is the Schrödinger equation
\begin{equation}
i\partial_t \psi = H\psi.
\label{eq:ct40}
\end{equation}%
The Hamilton operator $H$ is a hermitean matrix that plays precisely the same role as in quantum mechanics. The real eigenvalues of $H$ can be associated with the energy for the corresponding eigenvectors.

\paragraph*{Quantum systems}

For orthogonal step evolution operators as for unique jump chains the norm $\tilde{q}_\tau\tilde{q}_\tau$ is independent of $t$. We can choose a normalization such that the initial wave function at $t_\mathrm{in}$ obeys $\tilde{q}_\tau(t_\mathrm{in}) \tilde{q}_\tau(t_\mathrm{in}) = 1$. This normalization is preserved in time. In the presence of a complex structure it translates to the normalization
\begin{equation}
\psi^\dagger(t)\psi(t) = 1.
\label{eq:ct41}
\end{equation}%
For unique jumps chains or other systems with $\hat{J}=0$ the presence of a complex structure implies a unitary evolution of the complex wave function. It is precisely the same as in quantum mechanics. We will see later that often complex conjugation is related to a combination of time reversal and some other discrete symmetry.

The close analogy with the evolution in quantum mechanics for $\hat{J}=0$ extends to the conjugate classical wave function $\bar{q}$ and the density matrix $\rho'$. From the continuum evolution equation (\ref{eq:ct10}), (\ref{eq:ct15}),
\begin{equation}
\partial_t \bar{q} = -W^\tp \bar{q} = (\hat{F} - \hat{J})\bar{q},
\label{eq:ct42}
\end{equation}%
one infers that for $\hat{J}=0$ the conjugate wave function $\bar{q}$ follows the same evolution law as the wave function $\tilde{q}$. Let us extend the complex structure to the conjugate wave function
\begin{equation}
\tilde{q} = \begin{pmatrix}
\tilde{q}_\mathrm{R} \\ \tilde{q}_\mathrm{I}
\end{pmatrix},\quad
\bar{q} = \begin{pmatrix}
\bar{q}_\mathrm{R} \\ \bar{q}_\mathrm{I}
\end{pmatrix}.
\label{eq:ct43}
\end{equation}%
If we define
\begin{equation}
\bar{\psi} = \bar{q}_\mathrm{R} - i\bar{q}_\mathrm{I},
\label{eq:ct44}
\end{equation}%
the complex formulation of the evolution equation reads
\begin{equation}
i\partial_t \bar{\psi} = -(H^\tp + iJ^\tp)\bar{\psi} = -(H^* + iJ^*)\bar{\psi}.
\label{eq:ct45}
\end{equation}%
We may compare this with the complex conjugate of the evolution equation (\ref{eq:ct33}) for $\psi$
\begin{equation}
i\partial_t \psi^* = -(H^* - iJ^*)\psi^*.
\label{eq:ct46}
\end{equation}%
We conclude that for $J=0$ the complex formulation of the classical conjugate wave function $\bar{\psi}$ follows the same evolution law as $\psi^*$. In general, this does not yet imply that $\bar{\psi}$ can be identified with $\psi^*$. The wave function $\psi^*$ is determined by the boundary conditions at $t_\mathrm{in}$, while $\bar{\psi}$ depends on the boundary conditions at $t_\mathrm{f}$. For boundary conditions leading to $\bar{q}=\tilde{q}$ one has indeed $\bar{\psi}=\psi^*$. We will often concentrate on this type of boundary condition.

If an observable is represented by an operator $\hat{A}$ which is compatible with the complex structure,
\begin{equation}\label{YYD}
\hat A=\begin{pmatrix}A_\text{R}&-A_\text{I}\\ A_\text{I}&A_\text{R}\end{pmatrix}\ ,
\end{equation}
we can use the associated complex operator
\begin{equation}\label{YYE}
A=A_\text{R}+iA_\text{I}\ ,
\end{equation}
for the computation of the expectation value of this observable
\begin{equation}
\braket{A(t)} = \bar{\psi}^\tp A\psi.
\label{eq:ct51}
\end{equation}%
For boundary conditions with $\bar{\psi}=\psi^*$ this is the rule of quantum mechanics
\begin{equation}
\braket{A(t)} = \psi^\dagger A\psi = \braket{\psi|A|\psi}.
\label{eq:ct52}
\end{equation}%

The time evolution of probabilistic systems for which a continuum limit with $\hat{J}=0$ exists has all properties of the time evolution for quantum systems, provided that a complex structure exists and we choose boundary conditions such that $\bar{\psi}=\psi^*$. For a finite number of components $\psi_\alpha$ the complex wave functions form a finite-dimensional Hilbert space. Arbitrary normalized complex wave functions can be realized by suitable boundary conditions. The superposition principle for solutions of the linear evolution equation guarantees consistency for linear combinations. The same scalar product as in quantum mechanics can be defined. Suitable limits for an infinite number of components $\psi_\alpha$, as for example $\psi(x)$, define an infinite dimensional Hilbert space. Expectation values of local observables that are compatible with the complex structure can be computed according to the quantum rule. The only remaining question is then if all relevant observables in the quantum system find a suitable counterpart in the classical overall probabilistic system. The momentum observable in sect.\,\ref{sec:conserved_quantities_and_symmetries} will provide for a first example of non-commuting operator structures and the presence of observables that are not classical observables of the overall probabilistic system.

\paragraph*{General evolution in presence of a complex structure}

For general classical statistical systems admitting a complex structure the normalization of the local probability distribution $\partial_t(\bar{q}_\tau \tilde{q}_\tau) =0$ corresponds to 
\begin{equation}
\partial_t(\bar{\psi}^\mathrm{T} \psi)=0,
\label{eq:ct47}
\end{equation}%
where we employ
\begin{equation}
i\partial_t \bar{\psi}^\tp = -\bar{\psi}^\tp G.
\label{eq:ct47a}
\end{equation}%
The change of the length of the complex wave function $\psi$ obeys
\begin{equation}
\partial_t (\psi^\dagger\psi) = 2 \psi^\dagger J \psi.
\label{eq:ct48}
\end{equation}%
This demonstrates again that the antihermitean part of $G$ corresponding to $\hat{J}$ acts as a ``generalized damping term'' that can change the norm of the classical wave function. The normalization of the local probability distribution $\bar{q}_\tau(t) \tilde{q}_\tau(t)=1$ amounts to
\begin{equation}
\bar{\psi}^\tp (t) \psi (t)=1.
\label{eq:ct49}
\end{equation}%
For $\hat{J}=0$ we may combine with eq.\,(\ref{eq:ct41}),
\begin{equation}
\left(\bar{\psi}(t) - \psi^*(t)\right)^\tp \psi(t) =0,
\label{eq:ct50}
\end{equation}%
and conclude that the difference between $\bar{\psi}$ and $\psi^*$ is for all $t$ orthogonal to $\psi$.


\paragraph*{Generalized von-Neumann equation}

For a classical pure state admitting a complex structure we define the complex density matrix $\rho$,
\begin{equation}
\rho_{\alpha\beta}(t) = \psi_\alpha(t) \bar{\psi}_\beta(t).
\label{eq:ct53}
\end{equation}%
Its evolution equation is a generalized von-Neumann equation
\begin{equation}
i\partial_t \rho = [G,\rho].
\label{eq:ct54}
\end{equation}%
For $J=0$, $G=H$, this is the von-Neumann equation with hermitean Hamiltonian $H$. Expectation values of observables that can be associated to a complex operator $A$ obey the ``quantum rule''
\begin{equation}
\braket{A(t)} = \tr \{ A\rho(t) \}.
\label{eq:ct55}
\end{equation}%
The precise boundary conditions at $t_\mathrm{f}$ which determine $\bar{\psi}(t)$ are now encoded in the form of $\rho(t)$. For $\bar{\psi}(t) = \psi^*(t)$ the complex density matrix is hermitean, $\rho^\dagger(t) = \rho(t)$. This property is preserved for a unitary evolution, $J=0$, but not for more general evolution laws with $J \neq 0$. As usual, general density matrices obtain by weighted sums of pure state density matrices.

In the presence of a complex structure the real classical pure state density matrix $\rho'$ takes the form
\begin{equation}
\rho' = \begin{pmatrix}
\tilde{q}_\mathrm{R} \bar{q}_\mathrm{R} & \tilde{q}_\mathrm{R} \bar{q}_\mathrm{I} \\
\tilde{q}_\mathrm{I} \bar{q}_\mathrm{R} & \tilde{q}_\mathrm{I} \bar{q}_\mathrm{I}
\end{pmatrix},
\label{eq:ct56}
\end{equation}%
where
\begin{align}
\tilde{q}_\mathrm{R} &= \frac{1}{2} (\psi + \psi^*),\quad \tilde{q}_\mathrm{I} = -\frac{i}{2} (\psi - \psi^*), \\
\bar{q}_\mathrm{R} &= \frac{1}{2} (\bar{\psi} + \bar{\psi}^*),\quad \bar{q}_\mathrm{I} = \frac{i}{2} (\bar{\psi} - \bar{\psi}^*).
\label{eq:ct57}
\end{align}%
(We omit indices $\tilde{q}_{\mathrm{R}\,\alpha}$, $\psi_\alpha$ etc.). In distinction, the real matrix that would be associated to a complex matrix $\rho$ by the complex structure is given by
\begin{equation}
\tilde{\rho} = \begin{pmatrix}
\tilde{\rho}_\mathrm{R} & -\tilde{\rho}_\mathrm{I} \\
\tilde{\rho}_\mathrm{I} & \tilde{\rho}_\mathrm{R}
\end{pmatrix},
\label{eq:ct58}
\end{equation}%
where
\begin{equation}
\tilde{\rho}_\mathrm{R} = \tilde{q}_\mathrm{R}\bar{q}_\mathrm{R} + \tilde{q}_\mathrm{I}\bar{q}_\mathrm{I},\quad 
\tilde{\rho}_\mathrm{I} = \tilde{q}_\mathrm{I}\bar{q}_\mathrm{R} + \tilde{q}_\mathrm{R}\bar{q}_\mathrm{I}.
\label{eq:ct59}
\end{equation}%
The matrix $\tilde{\rho}$ obtains from $\rho'$ as
\begin{equation}
\tilde{\rho} = \rho' - I\rho' I.
\label{eq:ct60}
\end{equation}%
We conclude that the definition of a complex density matrix does not require that $\rho'$ is a matrix that is compatible with a complex structure. It does not need to be of the form~\eqref{YYD}. More generally, the linear map
\begin{equation}
B \to C(B) = \frac{1}{2}(B-IBI)
\label{eq:ct61}
\end{equation}%
projects $B$ to a matrix $C(B)$ which is compatible with the complex structure, with $C(C(B)) = C(B)$.

In summary, the continuum limit of the time evolution of the local probabilistic information in the form of classical wave functions or density matrices has a structure that is similar to the Schrödinger equation or von-Neumann equation in quantum mechanics. In general, the evolution operator $W$ is not antisymmetric, however. Correspondingly, in a complex formulation the evolution operator $G$ is not hermitean. An important exception is an orthogonal evolution with antisymmetric $W$ or hermitean $G=H$. For appropriate boundary conditions the evolution is the same as for quantum mechanics in this case.
\subsubsection{Properties of physical
time}\label{sec:properties_of_physical_time}

The linear ordering structure of variables $s(t)$ or $\varphi(t)$ introduced in
sect.\,\ref{sec:time_as_ordering_structure} is very general. It does not yet
distinguish between time and space. In this section we discuss further criteria
for the selection of a ``physical time" that may be used to order the events of
our world. 
Physical time is based on clocks and clock systems with oscillatory behavior.
We will see in the next section how basic concepts of special and general
relativity emerge naturally from our formulation of probabilistic time.

\paragraph*{Time or space?}
We have encountered already rather different probabilistic systems that admit an
ordering of a class of observables into equivalence classes that we have labeled
by ``time"~$t$.
They include Ising models with next-neighbor interactions in arbitrary
dimension, or the clock systems in sect.\,\ref{sec:clock_systems}. The formalism
for a description of evolution, with wave functions or the classical density
matrix, and observables represented by operators, is the same for all these
systems. It only involves the organization of the overall probability
distribution as a local chain. The behavior of the evolution is rather
different, however.

For Ising models with next-neighbor interaction the boundary information is
gradually lost as one moves inside the bulk. Far enough inside the bulk one
finds the equilibrium density matrix. One may associate this type of evolution
with space rather than time. For the clock systems the evolution is periodic.
This is what one may associate with time. We will base the concept of physical
time on a periodic evolution. Periodicity of the evolution for at least one
observable is a necessary criterion for physical time. It is not sufficient,
however, since one may also find periodic patterns in space.

\paragraph*{Oscillation time}
Any concept of physical time needs to select some particular ordering structure
for which at least one observable shows a periodic behavior of its evolution.
``Oscillation~time" counts the number of oscillations for this observable. This
concept of time has always been used by humans. The oscillatory behavior may be
related to the rotation of the Earth, with oscillation time the number of days,
or to the rotation of the Earth around the Sun, with oscillation time the number
of years. Later, one uses more local clocks as associated to the periodic
evolution of a pendulum. Today the time standards are set by the oscillations of
the electromagnetic field or photon wave function for the radiation emitted for
some particular atomic transition.

Oscillation time is a physical time, being based on observable phenomena which
admit a simple counting. It does not depend on the choice of a time-coordinate
by an observer. It only uses the fact that our world shows oscillatory phenomena
and counts the periods. The concept of oscillation time is an important notion
if one deals with complex questions about the beginning of the
universe~\cite{Rubakov:2022fqk, CWPTBU}.

\paragraph*{Universality of time}
In practice, humans use many different clocks and compare the oscillation time
for one clock to the oscillation time for some other clock. We compare the year
to the time units set by atomic clocks. This allows us to define a type of
universal time. 

For a definition of physical time we require the existence of ``clock systems"
with many different clocks. Two clocks of a clock system are synchronized. This
means that both clocks count the number of oscillations of their respective
periodic evolution, and it is possible to specify how many oscillations of clock
two occur during one oscillation of clock one. This ratio needs not be an
integer since the comparison can extend over many oscillation periods of clock
one. The clocks of a clock system are all synchronized -- this means that every
clock in the system is synchronized with some standard clock. In turn, every
pair of clocks in a clock system is synchronized. Synchronization of a pair of
clocks may happen by use of intermediate clocks. A simple examphe encountered
before are unique jump chains with different periods.

A clock system with many clocks realizes ``universality of physical time". The
same time can be used for all clocks in the clock system. It is given by the
number of oscillations of the ``standard clock" to which all clocks in the
system are synchronized.
The number of oscillations of the standard clock defines a ``universal physical
time" or ``universal oscillation time" for the whole clock system. For example,
the standard clock for some type of cosmic time could be defined by the
oscillations of photons with a given comoving wavelength in the
cosmic microwave background.

For physical time we require periodicity to be as accurate as possible. 
We may, nevertheless, include clocks with a repetition of events that permit
counting, like zeros of amplitudes for processes for which the period depends on
time.
There are other clocks based on a repetition of characteristic features. They
may, however, only be approximately periodic. Examples are the biorhythm of
animals or the seasons in meteorology. Synchronization of such ``imperfect
clocks" with the physical clocks of a clock system is only approximative.

\paragraph*{Clock systems as equivalence classes}

A clock system defines an equivalence class. All clocks that can be synchronized
to the standard clock are equivalent in this sense. We may restrict the notion
of equivalence of two clocks A and B such that a finite number of oscillations
for clock A corresponds to a finite number of oscillations of clock B, and vice
versa. This entails that whenever one clock ticks an infinite number of times
this holds for all clocks in its equivalence class. Concerning physical time we
have to distinguish between two different types of equivalence classes. The
first concerns the time-ordering of observables which defines a time structure.
Two observables belong to the same class if they are simultaneous with respect
to a given ordering. The second is an equivalence of clocks which is used for a
notion of universality of physical time. If two different equivalence classes of
clocks exist, they can define two different ``universal times''. There may be no
notion of a unique universal physical time valid for all phenomena.

\paragraph*{Time units -- from counting to continuous time}
Counting is discrete, and the oscillation time is a dimensionless number. If it
is an integer for the standard clock, it does not need to be an integer for the
other clocks of the clock system. If the conversion factor of the
synchronization is a rational number, clocks synchronized with the standard
clock measure time as a rational number. 

We are used to consider time as a continuous quantity, and to associate a unit
to it. In practice, we measure time in seconds and consider a continuous time
flow. Two steps are to be taken for the transition from oscillation time to the
more standard concepts of time. The first is the transition to continuous time
as discussed in sect.\,\ref{sec:continuous_time}. After this step, the time of a
given clock is mapped to a dimensionless continuous real number which is
proportional to the number of its oscillations. We may define the scale of this
number by mapping a certain, typically very high, fixed number of oscillations
of the standard clock to the real number one. The second step is the
introduction of a specific unit of time. For a given standard clock, say some
atomic clock, we may define the real number one to correspond to one second.
From there on time is a continuous real variable that is measured in seconds.
Starting from the discrete time of a local chain, this procedure defines the
time difference~$\epsilon$ between two neighboring positions on the chain in
terms of seconds. One counts the number of time steps on the chain needed to
perform one oscillation of the standard clock, and then converts it to seconds.
The time variable on the chain is connected in this way to physical time.

\paragraph*{Different time units}
The synchronization of a clock with the standard clock is straight forward for
discrete oscillation time. It just compares two different discrete counts. This
gives a unique result. For continuous time the issue can be more involved. Two
clocks (or two observers) do not need to use the same time units. For a certain
given number of ticks of a standard clock one may assign two different time
intervals $\Delta t_{1}\neq\Delta t_{2}$ to two different clocks, for example to
two clocks in relative motion to each other. 
While the ``physical time" of the two clocks is equal, namely set by the
comparison with the standard clock, the continuous time interval is not.

This freedom in the choice of a continuous time interval can be an important
advantage. It allows, for example, a simple expression for physical laws.
Einstein's development of special relativity is based of the insight that
physical processes in a spacecraft moving with constant velocity relative to
another spacecraft obey the same laws (we neglect here gravitational fields).
This property can be realized in a simple way if the time intervals of clocks in
the two spacecrafts are chosen to be different. Here $\Delta t_1$ and $\Delta
t_2$ both relate to the same time interval $\Delta t$ of oscillation time or the
same number of oscillations of the standard clock.
Assume that both spacecrafts host an identical atomic clock, and both count
(during a fixed time interval $\Delta t$ of universal time) the number of
oscillations $n$ for the photons of some given atomic transition line. One will
find $n_{1}\neq n_{2}$ if the satellites move with different relative velocities
$\vec{v}_{1}\neq\vec{v}_{2}$ with respect to the standard clock. (We neglect
here gravitational fields and the change of velocities during $\Delta t$.) 

It would seem that the laws in the two spacecrafts are different since the
number of observed oscillations during $\Delta t$ is different. Einstein
predicts
\begin{equation}\label{PT1}
\frac{n_{1}}{n_{2}}=\frac{\gamma_{2}}{\gamma_{1}} \quad , \quad
\gamma_{i}=\frac{1}{\sqrt{1-\vec{v_{i}}^2}}.
\end{equation}
Observers in the two satellites may decide to use different time units, 
\begin{equation}\label{PT2}
\Delta t_{i}=\Delta t / \gamma_{i}.
\end{equation}
In this case they find the same number of oscillations per time unit and realize
the universality of physical laws
\begin{equation}\label{PT3}
\frac{n_{i}}{\Delta t_{i}}=\frac{n_{i}\gamma_{i}}{ \Delta
t}=\frac{\bar{n}\gamma_{i}}{\gamma_{i}\Delta t}=\frac{\bar{n}}{\Delta t}.
\end{equation}
This choice of $\Delta t_{i}$ corresponds to proper time.

\paragraph*{Different time structures and reference frames}

\begin{figure}[t!]
\includegraphics{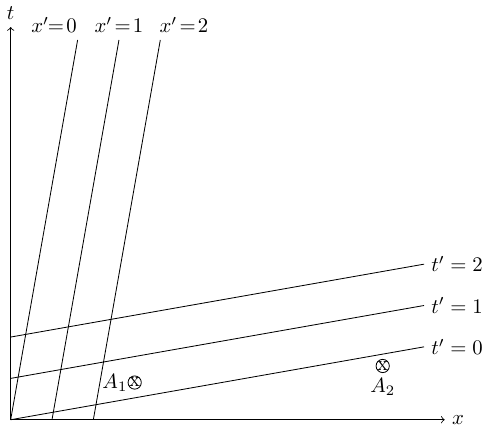}
\caption{Families of hypersurfaces for a time structure labeled by $t'$ and a
space structure labeled by $x'$.}\label{figure:PT1}
\end{figure}

In general, an overall probabilistic system admits different possible time
structures. As an example we consider two-dimensional generalized Ising models
on a quadratic lattice. Different possible time-orders correspond to different
hypersurfaces on the lattice. One possible time-order takes time along one of
the axes of the lattice that we may name the $t$-axis. A second time structure
is shown in Fig.~\ref{figure:PT1}, with time labeled by $t'$. For the $t'$-time
structure the hypersurfaces are given by the straight lines $t=cx+t'$, $c>0$. We
display in Fig.~\ref{figure:PT1} two observables $A_1$ and $A_2$, associated to
a particular point $(x_1,t_1)$ and $(x_2,t_2)$ on the lattice. With respect to
the $t$-ordering $A_1$ is before $A_2$, while with respect to the $t'$-ordering,
it occurs after $A_2$. This mismatch extends to the equivalence classes of
observables that are simultaneous with $A_1$ or $A_2$. The two different time
structures induce different equivalence classes. For example, we may shift $A_1$
in time to $A_1'$ at position $(x_1,t_2)$, such that $A_1'$ is simultaneous with
$A_2$ with respect to the $t$-ordering (same $t$-coordinate). With respect to
$t$-ordering $A_1'$ is in the same equivalence class as $A_2$. It is obvious
that the values of $t'$ of these two points to not coincide, such that $A_1'$
and $A_2$ belong to different equivalence classes with respect to $t'$ ordering.

For each time structure we may define a physical time by an appropriate clock
system. A priori, the time in one time structure may not be related to the time
in a different time structure. The different ordering systems can be seen as
different reference frames in which observers measure time. There may be rules
to relate different reference frames. In the next section we establish for a
particular model the connection to Lorentz transformations between different
inertial frames.

More generally, it may be possible to establish laws that relate particular
possible time structures, as the Lorentz transformations. It is important to
realize that even in this case one deals with different time structures,
involving different orderings of observables. This is not the same as using
different clocks in a clock system of a given time structure, or employing
different units for these clocks. With respect to clocks the two reference
frames corresponding to $t$-ordering and $t'$-ordering in Fig.~\ref{figure:PT1}
can be synchronized. Clocks in both systems belong to the same clock-equivalence
class.

\section[Fermions]{Fermions}\label{sec:free_fermions_in_two_dimensions}

In this section we present a first simple classical statistical model that is
equivalent to a quantum field theory. It is a generalized Ising model which
describes free massless fermions in one time and one space dimension. Even
though this two-dimensional model remains extremely simple, all aspects of the
formalism of quantum mechanics are already visible. This concerns both the
Schrödinger or von Neumann equation for a unitary evolution and the appearance
of non-commuting operators for observables as position or momentum. Concerning
time we will see the emergence of Lorentz symmetry in the continuum limit and
understand the origin of the Lorentz transformation between different reference
frames. At a later stage interactions will be added to this model in
sect.~\ref{sec:Fermionic_quantum_field_theory_with_interactions}.

\subsection[Quantum field theory for free fermions in two dimensions]{Quantum field theory for free\\fermions in two dimensions}
\label{sec:free_particles_in_two_dimensions}

In this section we investigate a particular Ising-type model on a two-dimensional square lattice. It only has diagonal next-neighbor interactions. Those are asymmetric, allowing propagation only along one of the two diagonals. This model describes a simple cellular automaton of right-movers and left-movers. We establish the equivalence with a two-dimensional quantum field theory for free fermions. Quantum mechanics obtains by a concentration on one-particle states.

\paragraph*{Diagonal Ising model in two dimensions}

Consider a two-dimensional lattice, with lattice sites labeled by pairs of integers $(m_1,\, m_2)$ in the range $0 \leq m_1 \leq \cM_1$, $0 \leq m_2 \leq \cM_2$. The variables are given by a single Ising spin $s(m_1,\, m_2)$ at each lattice site. The action of our model has diagonal neighbor interactions in one diagonal direction only,
\begin{equation}\label{eq:FP1}
\cS = -\beta \sum_{m_1 = 0}^{\cM_1 - 1} \sum_{m_2 = 0}^{\cM_2 - 1} \big[ 
s(m_1 +1,\, m_2 +1)\, s(m_1,\, m_2) -1 \big]\, .
\end{equation}
We take $\beta > 0$, such that the interaction is attractive and tends to align spins along diagonal directions. 

The interactions are only among spins on a given diagonal line, without mixing spins on different diagonal lines. We can therefore write the action as a sum of independent pieces $\cS^{(d)}$,
\begin{align}\label{eq:FP2}
\cS = \sum_d \cS^{(d)}\, ,
\end{align}
where $d$ labels different lines in the diagonal direction from low left to high right. For boundary conditions not mixing spins on different lines, our model is simply a collection of independent one-dimensional Ising models, each with spins along a given line $d$. The boundary value problem can then be solved exactly, as in sects~\ref{sec:influence_of_boundary_conditions}, \ref{sec:independence_from_the_future}.

We consider the diagonal Ising model here as a two-dimensional model. This is imposed by probabilistic boundary conditions that mix spins on different lines. For such boundary conditions the model can no longer be decomposed into a collection of one-dimensional models. The boundary conditions induce important correlations. For $\beta\to\infty$ this model is a probabilistic cellular automaton. It describes free fermions in Minkowski space.

We will later consider more elaborate models that cannot be decomposed into independent one-dimensional models for particular boundary conditions. The present model will be an interesting limiting case.

\paragraph*{Time and space}

For $\cM_2 = \cM_1$ the setting is completely symmetric under an exchange of directions.  The transformation $s(m_1,\, m_2) \to s(m_2,\, m_1)$ leaves the action $\cS$ invariant. We will investigate the evolution in one of the two directions. Quite arbitrarily we select $m_1$ and define time variables $t$ and space variables $x$ by
\begin{equation}\label{eq:FP3}
t = \epsilon\left(m_1-\frac{\cM_1}{2}\right)\ ,\quad x=\epsilon\left(m_2-\frac{\cM_2}{2}\right)\ .
\end{equation}
The diagonal Ising model defines a local chain, with hypersurfaces $m_1$ orthogonal to the evolution direction. For the local chain we have $s_\gamma (m_1) \equiv s(m_1,\, m_2)$, e.g. $\gamma = m_2 = 0,\,\dots,\, \cM_2$, $M = \cM_2 + 1$. The generalized Ising chain is specified by
\begin{equation}\label{eq:FP4}
\cL (m_1) = -\beta \sum_{m_2 = 0}^{\cM_2 - 1}  \big[ s(m_1 + 1,\, m_2 + 1)\, 
s(m_1,\, m_2) - 1\big]\, .
\end{equation}
Correspondingly, we fix boundary terms at $m_1 = 0$ and $m_1 = \cM_1$.

The local chain structure in the $m_1$-direction also allows for boundary terms in the $x$-direction, involving $s(m_1,\, 0)$, $s(m_1 + 1,\, 0)$, $s(m_1,\, \cM_2)$ and $s(m_1 + 1,\, \cM_2)$. We take here periodic boundary conditions in the $x$-direction by adding to $\cL(m_1)$ a term
\begin{equation}\label{eq:FP5}
\cL_b (m_1) = - \beta \big( s(m_1+1,\, 0)\, s(m_1,\, \cM_2) - 1 \big)\, .
\end{equation}
This identifies effectively
\begin{equation}\label{eq:FP6}
s(m_1,\, \cM_2 + 1) = s(m_1,\, 0)\, ,
\end{equation}
such that the $x$-variables are discrete points on a circle. With the inclusion of the boundary term, the sum in eq.~\eqref{eq:FP4} extends to $\cM_2$, i.e. over all $m_2$ in the range $0 \leq m_2 \leq \cM_2$.
This boundary term breaks the symmetry between the $t$ and $x-$directions. This asymmetry becomes unimportant in the limit of a very large number of lattice points.

\paragraph*{Unique jump chain}

The local factor $\cK(m_1)$ for the local chain in the $m_1$-direction is given by
\begin{align}\label{eq:FP7}
& \cK(m_1) = \exp \big(-\cL(m_1)\big) \notag \\
& \quad = \exp \bigg\{ \beta \sum_{m_2 = 0}^{\cM_2}
\big( s(m_1 + 1,\, m_2 + 1)\, s(m_1,\, m_2) - 1 \big) \bigg\}\, .
\end{align}
It equals one if for all $m_2$ the diagonally neighboring spins $s(m_1+1,\, m_2+1)$ have the same sign as $s(m_2,\, m_1)$. For each diagonal spin pair with opposite sign, $\cK(m_1)$ picks up a factor $\text{e}^{-2\beta}$. We will consider the limit $\beta\to\infty$. In this limit one finds a unique jump chain, where $\cK(m_1)$ vanishes for all configurations $\{ s(m_1 + 1)\}$ of the spins at $m_1 + 1$ for which $s(m_1 + 1,\, m_2 +1)$ differs from $s(m_1,\, m_2)$ for at least one value of $m_2$. Only those configurations contribute to the weight function for which $\{ s(m_1 + 1)\}$ is an exact copy of $\{s(m_1)\}$, but displaced by one unit to higher $m_2$. For this pair of configurations $\cK(m_1)$ equals one. An example for two neighboring ``allowed configurations'' with $\cK(m_1) = 1$ is given for the associated occupation numbers $n=(s+1)/2$ by
\begin{align}\label{eq:FP7A}
\{n(m_1 + 1)\} \; & \begin{pmatrix}
1 & 0 & 1 & 1 & 0 & 1 & 1 & 0 & 0 & 0 
\end{pmatrix} \quad : \; \tau(\rho) \notag \\
\{n(m_1)\} \; & \begin{pmatrix}
0 & 1 & 1 & 0 & 1 & 1 & 0 & 0 & 0 & 1 
\end{pmatrix} \quad : \; \rho
\end{align}
The unique jump property holds for all $m_1$. As a result, the only overall spin configurations that contribute to the overall probability distribution have all spins aligned on each diagonal line $d$, either all plus one or all minus one. This is, of course, what one could easily infer from the decomposition into independent Ising chains along the diagonal lines $d$.

We switch to occupation numbers $n(m_1,\,m_2) = (s(m_1,\, m_2)+1)/2$, where
\begin{align}\label{eq:FP8}
\cK (m_1) &= \exp \big\{ 2\,\beta\, [ n(m_1 + 1,\, m_2 + 1)\, n(m_1,\, m_2) \notag\\
& + (1 - n\, (m_1 +1,\,m_2 + 1))\,(1 - n(m_1,\, m_2)) - 1 ]\big\}\, .
\end{align}
From eq.~\eqref{eq:FP8} we can read off the corresponding step evolution operator, which takes the form
\begin{equation}\label{eq:FP9}
\hat{S}_{\tau\rho} (m_1) = \delta_{\tau,\, \tau(\rho)}\, .
\end{equation}
Here the map $\rho \to \tau(\rho)$ maps every configuration $\rho$ of occupation numbers at $m_1$ to a configuration $\tau(\rho)$ that is displaced by one unit in the positive $m_2$-direction. We have indicated the map $\tau(\rho)$ in the example \eqref{eq:FP7A}. In other words, the step evolution operator $\hat{S}$ maps uniquely every sequence of occupation numbers $\{ n_0,\, n_1,\, \dots,\, n_{\cM_2 - 1},\, n_{\cM_2}\}$ at $m_1$ to the sequence $\{ n_{\cM_2}, n_0,\, n_1,\, \dots,\, n_{\cM_2 - 1}\}$ at $m_1 + 1$. The inverse step evolution operator $\hat{S}^{-1}$ exists. It corresponds to the displacement of the whole sequence by a shift of one unit in the negative $m_2$-direction. The solution for the classical wave functions for a pure classical state reads
\begin{equation}\label{eq:FP10}
\tilde{q}_\tau (m_1 + 1) = \tilde{q}_{\rho(\tau)} (m_1)\, , \quad 
\bar{q}_\tau (m_1 + 1) = \bar{q}_{\rho(\tau)}(m_1)\, ,
\end{equation}
with $\rho(\tau)$ the inverse map of $\tau(\rho)$.

\paragraph*{Particle interpretation}

A possible particle interpretation of this model states that a fermion is present at time $t=m_1\,\varepsilon$ at the position $x = m_2\,\varepsilon$ whenever $n(m_1,\, m_2) = 1$. No particle is present at this time and location if $n(m_1,\, m_2) = 0$. Since the occupation numbers for our generalized Ising chains can only take the numbers $n = 0, 1$, we interpret the particle as a fermion. With this interpretation our model describes possible multi-fermion states, with a number of fermions between $0$ and $\cM_2+1$. This interpretation assumes implicitly that the vacuum is given by the zero-particle states where all occupation numbers $n(m_1,\, m_2)$ vanish. Other vacua are possible, and the definition of particles has to be adapted correspondingly\,\cite{CWFIM,CWQFTCS,CWCPMW}. We will discuss this point in sects.~\ref{sec:Fermionic_quantum_field_theory_with_interactions},~\ref{sec:fourier_transform_for_cellular_automata} and~\ref{sec:Particles_and_antiparticles}.

The number of ones in a given bit sequence at fixed $m_1$ does not depend on $m_1$. The particle number $\Np(t)$ at a given time is a local observable,
\begin{equation}\label{eq:FP11}
\Np(m_1) = \sum_{m_2 = 0}^{\cM_2} n(m_1,\, m_2)\, .
\end{equation}
Its expectation value is independent of time, i.e. $\langle \Np(m_1)\rangle$ does not depend on $m_1$ for arbitrary boundary conditions.
The particle number $\Np$ is a ``conserved quantity".
The operator associated to $\Np(m_1)$ is given by the matrix $\hat{N}(m_1)$, with matrix elements
\begin{equation}\label{eq:FP12}
(\hat{N}(m_1))_{\tau\rho} = \Np_\rho\, \delta_{\tau\rho}\, .
\end{equation}
Here, $\Np_\rho$ counts the number of ones in a given state $\rho$ at fixed $m_1$. We take the same definition of particle number for all $m_1$, such that the operator $\hat{N}$ does not depend on $m_1$. It commutes with the step evolution operator,
\begin{equation}\label{eq:FP13}
[\hat{S}(m_1),\, \hat{N}] = 0\, .
\end{equation}

We can classify the states or configurations $\rho$ (at given $m_1$) according to the particle number $\Np_\rho$. A zero-particle state has all $n(m_1,\, m_2) = 0$. This is a single configuration. One-particle states have exactly one bit equal to one in the sequence, and all other bits zero. There are $M = \cM_2 + 1$ such bit-configurations, and we label them by the position of the single particle at $x = \epsilon(m_2-\cM_2/2)$. For two-particle states two bits differ from zero. The $M\,(M -1)/2$ configurations of this type can be labeled by the two positions of the particles at $x$ and $y$. There is no distinction between the two particles. Similarly, there are $M\,(M - 1)\,(M -2)/6$ three-particle states, and so on. The totally occupied state with $\Np_\rho = M$ is again a single configuration. As it should be, the total number of states with fixed particle number sums up to $2^{M}$. 

The evolution does not mix configurations with a given particle number. In particular, if a wave function $\tilde{q}(m_1)$ is an eigenstate of $\hat{N}$ with a given eigenvalue $\Np$,
\begin{equation}\label{eq:FP14}
\hat{N}\, \tilde{q}(m_1) = \Np\, \tilde{q}(m_1)\, ,
\end{equation}
it remains an eigenstate for all $m_1$. This follows directly from the commutation relation \eqref{eq:FP13},
\begin{align}\label{eq:FP15}
\hat{N}\, \tilde{q}(m_1 + 1) &= \hat{N}\, \hat{S}(m_1)\, \tilde{q}(m_1) = 
\hat{S}(m_1)\, \hat{N}\, \tilde{q}(m_1)  \notag \\
& = \Np\,\hat{S}(m_1)\, \tilde{q}(m_1) = \Np\,\tilde{q}(m_1 + 1)\, .
\end{align}
The same holds true for the conjugate wave function $\bar{q}$. If both $\tilde{q}$ and $\bar{q}$ are eigenfunctions of $\hat{N}$ with the same $\Np$, we call this a fixed-particle-number pure state. The generalization to a mixed $\Np$-particle state is given by the condition
\begin{equation}\label{eq:FP16}
\hat{N}\, \rho' = \rho'\,\hat{N} = \Np\,\rho'\, .
\end{equation}
The reader should notice that the word ``state'' is used in this work with two different meanings. In one sense it describes configurations of occupation numbers, e.g., $\rho$, $\tau$ etc. In the other sense -- for example for a one-particle state -- it designs the probabilistic information contained in the wave functions of the density matrix. The latter depends on the boundary information. The ambiguity of this wording is historical. We keep this naming, but the conceptual difference should be remembered from the context.

\paragraph*{One-particle states}

\begin{figure}[t!]
\includegraphics[scale=0.45]{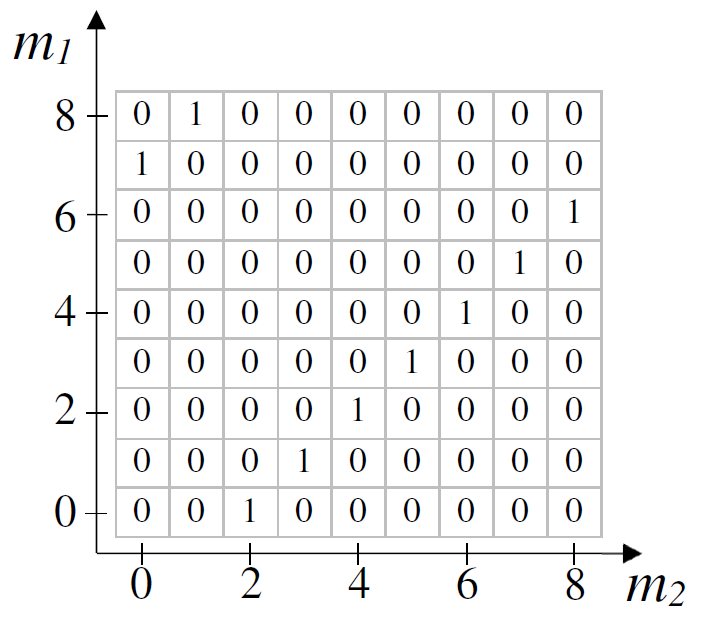}
\caption{One-particle state for a lattice with $\cM_1 = \cM_2 = 8$. At $t=0$ the position of the particle is $x(0) = 2\,\varepsilon$. For increasing $t$ it moves to the right according to $x(t) = 2\,\varepsilon + t$.}\label{fig:2}
\end{figure}

Let us choose boundary conditions such that $\tilde{q}(t=0)$ is the wave function of a one-particle state, $\hat{N}\,\tilde{q}(t=0) = \tilde{q}(t=0)$, and similar for $\bar{q}(t= \varepsilon\,\cM_1)$. We consider here ``symmetric boundary conditions''
for which $\bar{q}(\cM_1)$ is chosen such that $\bar{q}(m_1) = \tilde{q}(m_1) = q(m_1)$ (For a discussion of other boundary conditions see sect.\,\ref{sec:static_memory_materials}.) A very simple such state is a ``sharp position state'' for which $\tilde{q}_\tau (0) = 1$ for one particular bit configuration $\tau$, e.g. a particle located at a given $m_2$, while it vanishes for all other configurations. The evolution of this wave function is very simple, since the particle moves at each time step one position to the right. The resulting overall bit-configuration for an initial wave function of this type is shown in Fig.\,\ref{fig:2}. Of course, configurations where the particle is positioned at $m_1 = 0$ at some other $m_2$ behave in the same way. 

A general one-particle wave function at $m_1 = 0$ may differ from zero for more than one $m_2$. We adopt the notation
\begin{equation}\label{eq:FP17}
q_{in} (m_2) = \tilde{q}(0,\, m_2) = q_{dl}(0,\, x)\, .
\end{equation}
The initial classical wave function is an arbitrary function of the discrete periodic variable $x = \epsilon(m_2-\cM_2/2)$, $x + L = x$, $L = \varepsilon\, (\cM_2 + 1)$. For symmetric boundary conditions one has $\bar{q}(0,\, x) = \tilde{q}(0,\, x) = q_{dl}(0,\, x)$. The wave function is subject to the normalization condition 
\begin{align}\label{eq:FP17A}
\sum_{m_2 = 0}^{\cM_2} q_{dl}^2 (0,\, m_2) = \varepsilon \sum_{m_2 = 0}^{\cM_2} q^2 (0,\, m_2) = \int_x q^2 (0,\, x) = 1\, ,
\end{align}
where the index $dl$ stands for dimensionless and we define
\begin{equation}\label{eq:FP17B}
\int_x = \int_{-(L-\epsilon)/2}^{(L-\epsilon)/2} \dif x = \varepsilon \sum_{m_2 = 0}^{\cM_2}\, .
\end{equation}
The unit $\varepsilon$ is arbitrary. We can connect the normalization of $q(t,x)$ in eq.\,\eqref{eq:FP17A} to the original dimensionless wave function $q_\mathrm{dl}(m_1,m_2)$ either by setting $\varepsilon = 1$ or by employing a relative normalization factor $\sqrt{\varepsilon}$ between dimensionful and dimensionless wave functions,
\begin{equation}\label{eq:FP17C}
q_{dl}(m_1,\, m_2) = \sqrt{\varepsilon} \, q(t,\, x)\, .
\end{equation}
We use the same unit $\varepsilon$ for both the time and space direction. This measures space-distances in ``light seconds'' or sets the ``light velocity'' as $c=1$. 

The classical wave function obeys a discrete Schrödinger-type evolution equation
\begin{equation}\label{eq:FP18}
q(t + \varepsilon,\, x) = q(t,\, x-\varepsilon)\, .
\end{equation}
We may define discrete time- and space-derivatives by
\begin{align}\label{eq:FP19}
\p_t\, q(t,\, x) &= \frac{1}{2\varepsilon} \big[ q(t+\varepsilon,\, x) - 
q(t - \varepsilon, \, x)\big]\, , \notag \\
\p_x\, q(t,\, x) &= \frac{1}{2\varepsilon} \big[ q(t,\, x+\varepsilon) - 
q(t,\, x-\varepsilon) \big]\, .
\end{align}
In terms of these discrete derivatives the evolution equation takes the form
\begin{equation}\label{eq:FP20}
\p_t\, q (t,\, x) = - \p_x\, q(t,\, x)\, .
\end{equation}
The general solution of the initial value problem is simply
\begin{equation}\label{eq:FP21}
q(t,\, x) = q(0,\, x - t) = q_{in} (x - t)\, .
\end{equation}
The wave functions solving the Schrödinger equation \eqref{eq:FP20} only depend on the difference $x-t$.

\paragraph*{One-particle observables}

One-particle observables are observables whose expectation value can be computed from the one-particle wave function (or corresponding density matrix). A simple example is the local occupation number $n(t,\, x)$ at a given position $x$. The associated local operator $\hat{n}(t,\, x)$ is given by
\begin{align}\label{eq:FP22}
& \big( \hat{n}(t,\,x) \big)_{\tau\rho} = \big( \hat{n}(t,\, x) \big)(m'_2,\, m''_2) \notag \\[4pt]
& \quad = \delta \Big( \frac{x}{\varepsilon},\, m'_2\Big)
\, \delta \Big( \frac{x}{\varepsilon},\, m''_2 \Big) = \varepsilon^2\, 
\delta (x - x')\, \delta(x' - x'')\, .
\end{align}
The definition \eqref{eq:FP22} is meaningful only for one-particle states for which $\tau$ and $\rho$ can be characterized by ``particle positions'' $m'_2$ and $m''_2$. The operator $\hat{n}(t,\, x)$ is a matrix in the space of one-particle states. The different local particle number operators for the different locations $x$ are labeled by $x$. The corresponding operators depend on $x$ but are independent of $t$, e.\,g.\ $\hat{n}(t,x)=\hat{n}(x)$. The eigenvalues of these operators are zero or one, as it should be for a fermionic occupation number. Indeed, the Kronecker $\delta\big( \frac{x}{\varepsilon}, \, m'_2\big)$ equals one for $x/\varepsilon = m_2 = m'_2$ and vanishes otherwise,
\begin{equation}\label{eq:FP23}
\sum_{m_2} \delta \Big( \frac{x}{\varepsilon},\, m'_2 \Big) = \sum_{m_2} 
\delta (m_2,\, m'_2) = 1\, .
\end{equation}
We have normalized $\delta(x- x')$ such that
\begin{equation}\label{eq:FP24}
\int_x \delta (x - x') = 1\, .
\end{equation}

The expectation value of the local particle number depends on $t$ according to
\begin{align}\label{eq:FP25}
\langle n(t,\, x) \rangle &= \sum_{m'_2,\, m''_2} q_{dl}(t,\, m'_2)
\big(\hat{n}(t, x)\big) (m'_2, m''_2)\, q_{dl} (t,\, m''_2) \notag \\
&= \frac{1}{\varepsilon} \int_{x'} \int_{x''} q(t,\, x')\, \big(\hat{n}(t,\, n)\big)
(x',\, x'')\, q(t,\, x'') \notag \\
&= \frac{1}{\varepsilon}\, \langle q(t)\, | \, \hat{n}(t,\, x)\, | \, q(t) \rangle = 
\varepsilon\, q^2 (t,\, x)\, . 
\end{align}
Similar to quantum mechanics for one-particle states the number density is given by the square of the wave function
\begin{equation}
\frac{1}{\varepsilon} \braket{n(t,x)} = q^2(t,x),\quad \int_x q^2(t,x) =1.
\label{eq:DIMA}
\end{equation}
For the sum of occupation numbers at all positions one has for a one-particle state
\begin{align}\label{eq:FP26}
\langle \Np \rangle = \sum_{m_2} \langle n( t,\, m_2)\rangle = \frac{1}{\varepsilon} 
\int_x \langle n(t,\, x) \rangle = \int_x q^2 (t,\, x) = 1\, ,
\end{align}
as it should be.

The expectation value $n(t,\, x)$ can, of course, be directly computed from the overall probability distribution, without any restriction to one-particle states. For general wave functions the expression
\begin{equation}\label{eq:FP27} 
\langle n(t,\, x) \rangle = q_\tau (t) \big(\hat{n}(t,\, x)\big)_{\tau\rho}\, q_\rho(t)
\end{equation}
has also contributions from states $\tau$, $\rho$ that correspond to multi-particle states. We may define projectors $P_{\Np}$ on $\Np$-particle states by
\begin{equation}\label{eq:FP28}
(P_{\Np})_{\tau\rho} = \begin{cases}
\delta_{\tau\rho} \quad \text{if } \tau, \rho \text{ are $\Np$-particle 
	configurations} \\
0 \qquad \text{ otherwise}
\end{cases}\, .
\end{equation}
These projectors obey
\begin{align}\label{eq:FP29}
& \sum_{\Np} P_{\Np} = 1\, , \quad P_{\Np}^2 = 1\, , \quad P_{\Np}\, P_{\Np'} = \delta_{\Np\Np'}\, , 
\notag \\[4pt]
& \hat{N}\, P_{\Np'} = \Np\, \delta_{\Np\Np'} \, P_{\Np'}\, .
\end{align}
Inserting unit factors in eq.~\eqref{eq:FP27},
\begin{equation}\label{eq:FP30}
\langle n(t,\, x) \rangle = \langle q(t)\, | \, \Big( \sum_{\Np'} P_{\Np'} \Big)\, 
\hat{n}(t,\, x)\, \Big( \sum_{\Np} P_{\Np} \Big)\, | \, q(t) \rangle\, ,
\end{equation}
and observing the vanishing commutator
\begin{equation}\label{eq:FP31}
[ \,\hat{n}(t,\, x),\, P_{\Np}\, ] = 0\, ,
\end{equation}
one arrives at
\begin{equation}\label{eq:FP32}
\langle n(t,\, x) \rangle = \sum_{\Np} \langle q_{\Np}(t)\, \hat{n}_{\Np} (t,\, x)\, q_{\Np}(t) \rangle\, ,
\end{equation}
with
\begin{align}\label{eq:FP33}
q_{\Np}(t) = P_{\Np}\, q(t)\, , \quad \hat{n}_{\Np} (t,\, x) = P_{\Np}\, \hat{n}(t,\, x)\, P_{\Np}\, .
\end{align}

The operator $\hat{n}(t,\, x)$ in eq.~\eqref{eq:FP22} corresponds to $\hat{n}_1 (t,\, x)$ in eq.~\eqref{eq:FP33} in an appropriate basis for the one-particle states. For general wave functions there will be contributions from $\Np\neq 1$ in eq.~\eqref{eq:FP32}. The contribution from the one-particle sector retains the form \eqref{eq:FP25}. The normalization condition for $q(x)$ changes, however, since $\int_x q^2 (t,\, x) < 1 $ if the wave function $q_\tau (t)$ has nonzero contributions from states with $\Np\neq 1$.

\paragraph*{Particle position}

For one-particle states we can define a position observable $X(t)$,
\begin{equation}\label{eq:FP34}
X(t) = \sum_x x\, n(x,\, t) = \frac{1}{\varepsilon} \int_x x\, n(x,\, t)\, ,
\end{equation}
with associated operator
\begin{equation}\label{eq:FP35}
\hat{X}(t) = \frac{1}{\varepsilon}\int_x x\, \hat{n}(x,\, t) = \varepsilon\, x'\, 
\delta(x' - x'')\, .
\end{equation}
Its expectation value obeys
\begin{equation}\label{eq:FP36}
\langle X (t) \rangle = \frac{1}{\varepsilon} \int_x x\, q^2_{dl} (t,\, x) =
\int_x x\, q^2(t,\, x)\, ,
\end{equation}
similar to the position operator in quantum mechanics.

In a similar way, we can define the observable $X^2(t)$ by
\begin{align}\label{eq:FP37}
& X^2(t) = \frac{1}{\varepsilon} \int_x x^2\, n(x,\, t)\, , \notag \\
& \hat{X}^2(t) = 
\frac{1}{\varepsilon} \int_x x^2\, \hat{n}(x,\, t) = \varepsilon\, x'^2\, 
\delta(x' - x'')\, .
\end{align}
Similar to quantum mechanics, the expectation value reads
\begin{equation}\label{eq:FP39}
\langle X^2 (t) \rangle = \int_x\, x^2\, q^2(x)\, .
\end{equation}
We emphasize that the observable $X^2(t)$ differs from the square of the observable $X(t)$, e.g.
\begin{equation}\label{eq:FP38}
\big(X(t)\big)^2 = \frac{1}{\varepsilon^2} \int_{x, y} x\, y\, n(x,\, t)\, n(y,\, t)\, .
\end{equation}

The observable $X^2(t)$ is an example for a product structure among observables that differs from the classical observable product.
It is related to the operator product for the associated operators,
\begin{equation}
\hat{X}^2(t) = \hat{X}(t)\hat{X}(t)\,.
\end{equation}
This is a product of matrices. (The factors $\varepsilon$ assure the correct matrix multiplication in the discrete basis where
the matrix elements are denoted by $\hat{X}(m_2,m'_2)$.)
We observe that both the classical observable product $\left( X(t) \right)^2$ in eq.~\eqref{eq:FP38} and the observable product
$X^2(t)$ are classical observables whose expectation value can be computed from the overall probability distribution. They simply
correspond to different possible structures defining a multiplication law for observables.

The observable $X^2(t)$ can be used to define the ``dispersion''
\begin{align}\label{eq:FP40}
D(t) &= \langle X^2(t) \rangle - \langle X(t) \rangle^2 \notag \\
&= \int_x x^2\, q^2(x) - \bigg( \int_x x\, q^2 (x) \bigg)^2 \notag \\
&= \int_x \big(x - \bar{x}(t)\big)^2\, q^2(x)\, , 
\end{align}
where we employ the shorthand
\begin{equation}\label{FP41}
\bar{x}(t) = \langle X(t) \rangle\, .
\end{equation}
The dispersion is a measure for the ``width'' of the wave function. It is positive semi-definite and vanishes only for the ``sharp one-particle states'' shown in fig.~\ref{fig:2}. In a discrete notation one has
\begin{align}\label{eq:FP42}
& \bar{x} (t) = \varepsilon\, \sum_{m_2}\left(m_2-\frac{\cM_2}{2}\right)\, q^2_{dl} (t,\, m_2)\, , \notag \\
& \langle X^2(t) \rangle = \varepsilon^2 \sum_{m_2} \left(m_2-\frac{\cM_2}{2}\right)^2 \, q_{dl}^2 (t,\, m_2)\, .
\end{align}
For a sharp one-particle state,
\begin{equation}\label{eq:FP43}
q_{dl}(t,\, m_2) = \delta (m_2,\, m_0+\cM_2/2 + t\, \varepsilon)\, ,
\end{equation}
one has
\begin{align}\label{eq:FP44}
\bar{x}(t) = \varepsilon\, m_0 + t\, , \quad \langle X^2 (t) \rangle = 
(\varepsilon\, m_0 + t)^2\, ,
\end{align}
and therefore $D(t) = 0$.

\paragraph*{Correlations for one-particle states}

For a generic one-particle initial wave function, the spins or occupation numbers on different diagonal lines $d$ are correlated. This correlation implies that the diagonal two-dimensional Ising model can no longer be decomposed as a collection of independent one-dimensional Ising models. Any correlation between the spins in different diagonal lines $d$ forbids to treat these lines as independent systems. The initial correlation between Ising spins at $m_1 = 0$ is transported to similar correlations in the bulk at arbitrary $m_1$.

We define the correlation function
\begin{align}\label{eq:FP45}
G (t;\, x,\, y) &= \langle n(t,\, x)\, n(t,\, y)\rangle - \bar{n}(t,\, x)\, 
\bar{n}(t,\, y) \notag \\
& = \langle (n(t,\, x) - \bar{n}(t,\, x))\, (n(t,\, y) - \bar{n}(t,\, y)) 
\rangle\, ,
\end{align}
with
\begin{equation}\label{eq:FP46}
\bar{n}(t,\, x) = \langle n(t,\, x)\rangle\, .
\end{equation}
Assume that for $x\neq y$ the correlation function does not vanish, $G(t;\, x,\, y) \neq 0$. The spins at $x$ and $y$, for a given $t$, say $t=0$, are then correlated. This implies immediately that the spins on different diagonal lines are correlated. The correlation function can be computed from the wave function as
\begin{align}\label{eq:FP47}
& G(t;\, x,\, y)\notag \\
& \;\;= \sum_{m_2} \delta\Big(\frac{x}{\varepsilon},\, m_2-\frac{\cM_2}{2}\Big)\, 
\delta \Big( \frac{y}{\varepsilon},\, m_2-\frac{\cM_2}{2} \Big)\, q_{dl}^2 (t,\, m_2) \notag \\
& \qquad - \sum_{m_2} \sum_{m'_2} \delta\Big( \frac{x}{\varepsilon},\, m_2-\frac{\cM_2}{2} \Big)
\delta \Big( \frac{y}{\varepsilon},\, m'_2-\frac{\cM_2}{2} \Big)\notag\\
&\qquad\times q_{dl}^2 (t,\, m_2)\, 
q_{dl}^2 (t,\, m'_2)\phantom{\Big{|}} \notag \\
&\;\; =  \delta(x,\, y)\, q_{dl}^2 (t,\, x) - q_{dl}^2 (t,\, x)\, q_{dl}^2(t,\, y) \notag
 \\
&\;\; = \varepsilon^2 [ \delta(x-y)\, q^2 (t,\, x) - q^2(t,\, x)\, q^2(t,\, y) ] 
\, . 
\end{align}
The first term in the difference vanishes for $x\neq y$, reflecting the simple property $n(t,\, x) \, n(t,\, y) = 0$ for $x\neq y$. The second term gives a negative contribution to $G$ for all one-particle states that have a nonvanishing dispersion $D$. We conclude that for all initial states with nonvanishing dispersion, the spins on different diagonal lines are correlated.

For a nonzero dispersion the wave function differs from zero for more than one site $m_2$. The normalization \eqref{eq:FP17A} therefore implies $q_{dl}^2(x) < 1$. In consequence, one finds for $G(x,\, x)$ a positive value,
\begin{equation}\label{eq:FP48}
G(x,\, x) = q_{dl}^2(x) - q_{dl}^4(x)\, .
\end{equation}
This is consistent with the sum rule
\begin{equation}\label{eq:FP49}
\int_y G(t;\, x,\, y) = 0\, .
\end{equation}

For the particular sharp one-particle state \eqref{eq:FP43} one finds for all $x$ and $y$
\begin{equation}\label{eq:FP50}
G(t;\, x,\, y) = 0\, .
\end{equation}
In this case the vanishing for $x\neq y$ follows from the fact that $q_{dl}(t,\, m_2)$ differs from zero only for a single $m_2$, and for $y = x$ one has at this particular point $q_{dl}^4(x) = q_{dl}^2(x) = 1$. Sharp one-particle states can obviously be represented by independent one-dimensional generalized Ising models, with all spins up on one diagonal line, and all spins down on all other diagonal lines. The presence of correlations for all other one-particle states has a simple origin. The probabilistic information for spins on one diagonal line $d_1$ cannot be independent
of the probabilistic information for all other lines $d_2$, since more than one diagonal line contributes to the total particle number $\langle \Np \rangle = 1$. We will see that the presence of correlations gives rise to properties that cannot be found for one-dimensional Ising models.

\paragraph*{Sign of the wave function}

For generalized Ising models the definition \eqref{eq:CWF3}, \eqref{eq:CWF4}, \eqref{eq:CW8} of the wave function $\tilde{q}(t)$ and the conjugate wave function $\bar{q}(t)$ leads to
\begin{equation}\label{eq:FP51}
\tilde{q}_\tau (t) \geq 0\, , \quad \bar{q}_\tau (t) \geq 0\, .
\end{equation}
All local factors $\cK(t)$ obey $\cK(t) \geq 0$ for arbitrary configurations $\{ n(t)\}$ and $\{ n(t+\varepsilon)\}$, and the boundary terms $f_{in}$ and $\bar{f}_f$ are assumed to be positive as well, $f_{in} \geq 0$, $\bar{f}_f \geq 0$. As a result, $f(t)$ and $\bar{f}(t)$ are positive for all configurations $\{ n(t)\}$,
\begin{equation}\label{eq:FP52}
f(t) \geq 0\, ,\quad \bar{f}(t) \geq 0\, .
\end{equation}
All coefficients $\tilde{q}_\tau(t)$, $\bar{q}_\tau(t)$ for the expansion in the occupation number basis therefore obey eq.~\eqref{eq:FP51}. At first sight this seems to be an important difference as compared to quantum mechanics. In quantum mechanics, we can write the complex Schrödinger equation in a real form for the real and imaginary parts of the wave function. This yields a real evolution equation for the double number of components. For general solutions of this real Schrödinger equation, the components can be both positive and negative. The condition \eqref{eq:FP51} seems to eliminate many of the general solutions of the real evolution equation \eqref{eq:FP20}. 

The sign of the classical wave functions and the step evolution operator is, however, partly a matter of conventions \cite{CWQF}. A ``sign transformation'',
\begin{equation}\label{eq:448}
\tilde{q}_\tau (t) \rightarrow \sigma_\tau (t)\, \tilde{q}_\tau (t)\, , \; 
\bar{q}_\tau(t) \rightarrow \sigma_\tau (t)\, \bar{q}_\tau (t)\, , \; 
\sigma_\tau (t) = \pm 1\, ,
\end{equation}
leaves the local probabilities  $p_\tau (t) = \tilde{q}_\tau (t)\, \bar{q}_\tau (t)$ unchanged. As a consequence, the expectation values of local observables are independent of the choice of $\sigma_\tau (t)$. As an example, the one-particle wave function $q(t,x) = \tilde{q}(t, x) = \bar{q} (t, x) = \cos(p(x-t))$ is equivalent to the wave function $q(t,x) = |\cos(p(x-t))|$. From the perspective of the local probabilistic information the sign of the wave function does not constitute a restriction as compared to quantum mechanics. The condition \eqref{eq:CQ6} needs not to be imposed. 

This extends to the overall probability distribution. Changing the sign of the wave function has to be accompanied by corresponding changes of signs of the step evolution operator \cite{CWQF}. Indeed, the overall probability distribution or weight function \eqref{eq:TS46}, \eqref{eq:TS47} (with $\hat{T} \rightarrow \hat{S}$) is not changed by the transformation
\begin{equation}\label{eq:449}
\hat{S}_{\tau\rho} (t) \rightarrow D_{\tau\alpha} (t + \varepsilon)\, 
\hat{S}_{\alpha\beta}(t) \, D_{\beta\rho} (t)\, ,
\end{equation}
where $D(t)$ is a diagonal matrix involving the signs $\sigma_\tau (t)$,
\begin{equation}\label{eq:450}
D_{\tau\rho} (t) = \sigma_\tau (t)\, \delta_{\tau\rho}\, .
\end{equation}
This holds provided that one transforms simultaneously the initial and final wave functions ($D^\tp = D^{-1} = D$) according to
\begin{align}\label{eq:451}
& \tilde{q}(t_{in}) \; \rightarrow \; D(t_{in})\, \tilde{q}(t_{in})\, , \notag \\
& \bar{q}(t_f) \; \rightarrow \; D(t_f)\, \bar{q} (t_f)\, .
\end{align}
The wave functions and the density matrix then transform as
\begin{align}\label{eq:452}
& \tilde{q} (t) \; \rightarrow \; D(t)\, \tilde{q}(t)\, , \notag \\
& \bar{q} (t) \; \rightarrow \; D(t) \, \bar{q}(t)\, , \notag \\
& \rho'(t) \; \rightarrow \; D(t)\, \rho'(t) \, D(t)\, , 
\end{align}
and operators for local observables are changed by multiplication with the matrix $D(t)$ from both sides,
\begin{equation}\label{eq:453}
\hat{A} (t) \; \rightarrow \; D(t)\, \hat{A}(t) \, D(t)\, .
\end{equation}

We may consider the local sign changes encoded in $D(t)$ as local gauge transformations. They change the wave functions, classical density matrix and step evolution operator, while diagonal operators $\hat{A}(t)$ remain the same. Since the weight distribution is invariant, all expectation values of local observables are independent of the choice of gauge. The local sign changes are a particular case of more general similarity transformations \cite{CWQF}, see sect.\,\ref{sec:Change_of_basis_and_similarity_transformations}.

We may next ask the question if for a given choice of step evolution operators, say step evolution operators with only positive elements, arbitrary signs for the wave functions are consistent. The consistency criterion is the positivity of the overall probability distribution. The answer depends on the characteristics of the step evolution operator. For unique jump step evolution operators arbitrary signs for the wave function are consistent, as we have argued already in sect.~\ref{sec:step_evolution_operator}. We employ here the sign transformations in order to establish that for unique jump chains arbitrary signs for the initial wave function $\tilde{q}(t_{\text{in}}) $ lead to a positive overall probability distribution provided that the signs of the final wave function $\overline{q}(t_{\text{f}}) $ are chosen appropriately. This line of argument is an important property of unique jump chains and we will point out the possible differences to arbitrary local chains.

We may start with the diagonal Ising model with nonnegative step evolution operators $\hat{S}$ and consider first the restriction of $\hat{S}$ to the one-particle sector, with $\tau = x/\varepsilon$.
Let us assume arbitrary signs for $\tilde{q}_{\text{in}}(x)=\tilde{q}(t_{\text{in}},x)$.
We can transform this wave function to a positive wave function $\tilde{q}'_{\text{in}}(x)$ by an appropriate choice of signs $\sigma(t_{\text{in}},x)$
\begin{equation}
\label{SCA}
  \tilde{q}'_{\text{in}}(x)=\sigma(t_{\text{in}} , x)\, \tilde{q}_{\text{in}}(x)\, .
\end{equation}
This transformation modifies $\widehat{S}(t_{\text{in}})$ according to eq.~\eqref{eq:449}. By an appropriate choice of signs $\sigma(t_{\text{in}}+\varepsilon , x)$ we can compensate this change and keep $\widehat{S}(t_{\text{in}})$ invariant, shifting the sign change to $\widehat{S}(t_{\text{in}}+\varepsilon)$.
This procedure can be continued. The chain of step evolution operators is left invariant provided we impose for all $t\leq t_{\text{f}}-\varepsilon$ the condition
\begin{equation}\label{eq:454}
\sigma(t + \varepsilon,\, x) = \sigma (t,\, x-\varepsilon)\, .
\end{equation}
At $t_{\text{f}}$ we also transform the final wave function $\overline{q}_{\text{f}}(x)=\overline{q}(t_{\text{f}},x)$ to 
\begin{equation}
\label{SCB}
\overline{q}_{\text{f}}'(x)=\sigma(t_{\text{f}} , x) \, \overline{q}_{\text{f}}(x)\, .
\end{equation}
After the transformation the final boundary term is positive provided $\overline{q}_{\text{f}}'(x)$ is positive for all $x$. For this particular gauge the initial and final boundary terms as well as all step evolution operators are positive, implementing a positive weight distribution. The weight distribution remains invariant if we transform back to the original wave functions $\tilde{q}_{\text{in}}(x)$ and $\overline{q}_{\text{f}}(x)$.

This demonstrates that for arbitrary $\tilde{q}_{\text{in}}(x)$ the weight distribution is positive for a suitable choice of $\overline{q}_{\text{f}}(x)$.
Only the signs of the final wave function $\bar{q}_f(x) = \bar{q}(t_f,\, x)$ have to be correlated to the signs of $\tilde{q}_{in} (x)$. For a given choice of $\tilde{q}_{in} (x)$ the signs $\sigma(t_{in},\, x)$ are fixed. By virtue of eq.~\eqref{eq:454} this fixes the signs for all later $t$, including $t=t_f$, and therefore the signs of $\overline{q}(t_{\text{f}}) $ which lead to positive $\overline{q}'(t_{\text{f}})$.

For the particular choice of boundary conditions implementing $\overline{q}(t)=\tilde{q}(t)$ one finds positive $\overline{q}_{\text{f}}'=\tilde{q}'(t_{\text{f}})$. This results from positive $\tilde{q}'_{\text{in}}$ since the step evolution operators are all positive. The change from $\overline{q}_{\text{f}}'$ to $\overline{q}_{\text{f}}$ by the signs $\sigma(t_{\text{f}})$ is the same as the change from $\tilde{q}'(t_{\text{f}})$ to $\tilde{q}(t_{\text{f}})$.
The boundary condition leading to $\overline{q}(t)=\tilde{q}(t)=q(t)$ therefore imply a positive weight distribution for arbitrary signs of $q(t_{\text{in}})$ and therefore arbitrary signs of $q(t)$. For $\overline{q}(t)=\tilde{q}(t)$ the only restriction on the wave function is the normalization as a unit vector.

Switching from $\overline{q}_{\text{f}}=\tilde{q}(t_{\text{f}}) $ to some other $\overline{q}_{\text{f}}^{(s)}$ by a sign flip for some of the non-vanishing components induces a negative weight function for some particular configurations. Indeed, the negative signs in $\overline{q}_{\text{f}}'^{(s)}$ induce negative signs in the final boundary term in the gauge where $\tilde{q}_{\text{in}}'$ is positive. We conclude that the positivity of the weight function requires the signs of all components of $\overline{q}_{\text{f}}(x)$ to be the same as for $\tilde{q}(t_{\text{f}},x)$.

The above line of arguments generalizes to arbitrary unique jump chains. For periodic boundary conditions the one-particle subsector is equivalent to the clock system in sect.~\ref{sec:clock_systems}.
For arbitrary multi-particle states of the two-dimensional diagonal Ising model or more general unique jump chains characterized by configurations $\tau$ the freedom of choice of signs for the initial wave function generalizes directly. The condition \eqref{eq:454} is extended to
\begin{equation}\label{eq:455}
\sigma_\tau (t+\varepsilon) = \sigma_{\rho(\tau)} (t)\, ,
\end{equation}
where we replace $x$ by $\tau$ and $x-\epsilon$ by $\rho(t)$.

We conclude that for arbitrary unique jump chains there is no restriction on the signs of the initial wave function
$\tilde{q}_{in}$. The signs of the final wave function are restricted for given signs of the initial wave function. They have to be the same as for $\tilde{q}(t_{\text{f}})$. For the time-local probabilistic information a single real wave function $q(t)$ is sufficient. It is a unit vector without restrictions on the signs of its components. Without affecting the positive weight function we can now perform gauge transformations for which $\widehat{S}(t)$ is no longer a non-negative matrix.

The complete freedom in the choice of signs for the wave function does not extend to arbitrary local chains beyond unique jump chains. We can still achieve positive $\tilde{q}_{\text{in}}'$ by appropriate signs $\sigma_{\tau}(t_{\text{in}})$.
What is not guaranteed is the existence of a choice of signs $\sigma_{\tau}(t_{\text{in}} + \varepsilon)$ such that $\widehat{S}(t_{\text{in}}) $ remains invariant. The matrix multiplication $\widehat{S}(t_{\text{in}})\, D(t_{\text{in}}) $ yields simultaneous sign changes for all elements in a given column. Further multiplication with $D(t_{\text{in}}+\varepsilon)$ from the left changes simultaneously the signs of all elements in a given row. Starting from positive $\widehat{S}(t_{\text{in}})$ the matrix $\widehat{S}(t_{\text{in}}) \, D(t_{\text{in}})$ may have different signs in the same row which cannot be removed for $D(t_{\text{in}}+\varepsilon)\,\widehat{S}(t_{\text{in}})\, D(t_{\text{in}})$. Positivity of the overall weight function is then no longer guaranteed.

\paragraph*{Multi-particle states}

The possible two-particle states are characterized by bit sequences with one bit at $x$ equal one and a second bit at $y$ equal one, while all other bits are zero. They describe one particle at $x$ and another particle at $y$. The two particles are indistinguishable, and the locations have to be different, $x\neq y$. A general two-particle wave function depends on two arguments $x$ and $y$ and can be taken to be antisymmetric under the exchange $x \leftrightarrow y$,
\begin{equation}\label{eq:456}
\tilde{q}_2 (t;\, x,\, y) = - \tilde{q}_2 (t;\, y,\, x)\, .
\end{equation}
The antisymmetry guarantees that $\tilde{q}_2$ vanishes for $x = y$. Similar to the familiar wave functions for two fermions in quantum mechanics the two particles are indistinguishable. The same holds for the conjugate wave function
\begin{equation}\label{eq:457}
\bar{q}_2 (t;\, x,\, y) = - \bar{q}_2 (t;\, y,\, x)\, ,
\end{equation}
or, for suitable boundary conditions, for the single wave function $q_2(t;\, x,\, y) = \tilde{q}_2 (t;\, x,\, y) = \bar{q}_2 (t; \, x,\, y)$.

The generalization to states with a fixed particle number $\Np > 2$ is straightforward. The three-particle wave function depends on three positions $\tilde{q}_3 (t;\, x_1,\, x_2,\, x_3)$. It is taken to be totally antisymmetric  under the exchange of particle positions $x_i \leftrightarrow x_j$. Within the subsector of states where precisely three spins $s(t,\, x)$ are positive, and all other spins are negative, the local three-particle probability distribution
\begin{equation}\label{eq:458}
p_3 (t;\, x_1,\, x_2,\, x_3) = \bar{q}(t;\, x_1,\, x_2,\, x_3)\, 
\tilde{q} (t;\, x_1,\, x_2,\, x_3)
\end{equation}
describes the probability to find one spin up at $x_1$, another one at $x_2$ and the third one at $x_3$. It vanishes whenever two locations coincide. It is invariant under the exchange of positions $x_i \leftrightarrow x_j$, reflecting that the particles cannot be distinguished. 

If we continue to increase the number of particles we finally arrive at the totally occupied state with all $M$ spins positive. The wave function $\tilde{q}(t;\, x_1,\, x_2,\, \dots,\, x_M) = \tilde{q}(n;\, m_1,\, m_2,\, \dots ,\, m_{\Np})$ is totally antisymmetric under the exchange of positions $x_i \leftrightarrow x_j$, or $m_i \leftrightarrow m_j$. It is fixed uniquely up to a sign, being proportional to the totally antisymmetric tensor in $M$ dimensions. Also the totally empty or vacuum state with all spins negative has a unique wave function up to a sign.

\paragraph*{Fermionic quantum field theory}

The diagonal Ising model \eqref{eq:FP1} describes a two-dimensional quantum field theory for free massless Majorana-Weyl fermions. The Weyl condition restricts the particle content to ``right movers'', for which the position variable $x$ increases as $t$ increases. The Majorana condition makes the wave functions real, as in our case. If we take boundary conditions for which $\tilde{q}(t) = \bar{q}(t) = q(t)$, there is a one-to-one map of the general wave functions in the diagonal Ising model to the general wave functions in the fermionic quantum field theory. They are superpositions of wave functions with fixed particle numbers. In both models the wave functions obey the same evolution law. If the initial conditions are the same, all expectation values of observables are the same. The two models can therefore be identified. 

There exists a general exact ``bit-fermion map'' between ``functional integrals'' based on local chains for occupation numbers and Grassmann functional integrals \cite{CWFIM}. This maps the diagonal Ising model \eqref{eq:FP1} to a standard Grassmann functional integral for two-dimensional free Majorana-Weyl fermions in a discretized version. The two-dimensional Lorentz symmetry of the continuum limit of the model is manifest in the Grassmann formulation. Also the antisymmetry of the wave functions for fixed particle numbers has a simple root in the anticommuting property of the Grassmann variables. We will investigate the bit-fermion map more in detail in later parts of this work.

The diagonal two-dimensional Ising model is a first, still rather simple, example of a classical statistical system describing a quantum
field theory in Minkowski space. More complex models with a richer structure will be developed as this work goes on. Already at this 
simple level we have found the important structures of wave functions, operators and observable products that do not correspond to the
classical observable product. By a modest generalisation to two different Ising spins per site we will also see further characteristics
of quantum field theory and quantum mechanics emerging in a simple way: the complex structure, the momentum observable and non-commuting
operators for observables.
\subsection{Complex structure}\label{sec:complex_structure}

In quantum mechanics particles are described by complex wave functions. One can
equivalently use a formulation in terms of real wave functions with twice the
number of components. These real wave functions are simply the real and
imaginary part of the complex wave function. In the other direction, a map of
real wave functions to complex wave functions requires particular properties,
called a complex structure. If it exists, a complex formulation is rather
powerful. It allows for a simple description of the momentum observable and
complex basis transformations as the Fourier transform. We discuss in this
section general complex structures. We present a simple example for a single
species of bits. We further investigate several distinct complex structures for
two species of bits. They can be used for two-dimensional generalized Ising
models which describe a quantum field theory for free Weyl fermions. For a
single species of bits a particular complex structure relates complex
conjugation to a mapping between particles and holes. This will be discussed in
sect.~\ref{sec:particles_and_holes}.

\paragraph*{General complex structure}

We have already encountered in sect.~\ref{sec:continuous_time} and
appendix~\ref{app:matrix chains} the map from real $(2\times 2)$-matrices to
complex numbers which is compatible with the multiplication rule for complex
numbers, cf. eqs~\eqref{YYD},~\eqref{YYE} or~\eqref{eq:MC5}--\eqref{eq:MC13}. In
the present part we address complex structures more generally and focus on the
issue of complex wave functions. We consider a real vector $q_\tau$ and ask for
the conditions of a complex formulation.

A general complex structure involves two discrete transformations $K_c$ and $I$
that act as matrices on the vector $q$, 
\begin{equation}\label{eq:CC1}
q'_\tau = (K_c)_{\tau\rho}\, q_\rho\, , \quad q''_\tau = (I)_{\tau\rho}\,
q_\rho\, .
\end{equation}
They have to obey the relations
\begin{equation}\label{eq:CC2}
K_c^2 = 1\, , \quad I^2 = -1\, , \quad \{ K_c, I \} = 0\, .
\end{equation}
In the complex formulation $K_c$ describes the operation of complex conjugation.
From $K_c^2= 1$ one infers that the eigenvalues are $\lambda_c = \pm 1$. Even
functions are eigenvectors to $\lambda_c = +1$ and correspond to real
quantities. Odd functions change sign under the action of complex conjugation,
being eigenfunctions to the eigenvalue $\lambda_c = - 1$. They are associated to
imaginary parts of complex numbers, e.g.
\begin{equation}\label{eq:CC3}
K_c q_R = q_R\, , \quad K_c q_I = - q_I\, .
\end{equation}
The operation $I$ accounts in the complex formulation for the multiplication
with $\im$. From $I^2 = -1$ we infer the eigenvalues $\lambda_I = \pm \im$. The
anticommutation relation $\{K_c, I\} = 0$ implies that the number of components
$q_R$ and $q_I$ must be the same. Since $I$ is a regular matrix, the numbers of
independent components of $I q_R$ and $q_R$ are the same. From $K_c I\, q_R = -
I\, K_c q_R = -I q_R$ we conclude that the number of independent components of
$q_I$ is at least the number of components of $q_R$. With $I q_I$ being even,
the number of components of $q_R$ must be equal or exceed the number of
components of $q_I$. The two inequalities imply the same number of components
for $q_R$ and $q_I$. In particular, the vector $q$ can admit a complex structure
\eqref{eq:CC2} only if the number $N$ of components, $\tau=1,\, \dots,\, N$, is
even.

In the other direction, an equal number of positive and negative eigenvalues of
$K_c$ implies the existence of $I$. The matrix $K_c$ has $N/2$ eigenvalues $+1$
and $N/2$ eigenvalues $-1$. We can choose a basis for which
\begin{equation}\label{eq:CC4}
q = \begin{pmatrix}
q_R \\ q_I
\end{pmatrix}\, , \quad
K_c = \begin{pmatrix}
1 & 0 \\ 
0 & -1
\end{pmatrix}\, ,
\end{equation}
with $q_R$ and $q_I$ vectors with $N/2$ components and $K_c$ involving
$(N/2\times N/2)$-unit matrices. Within this basis we can choose $I$ as
\begin{equation}\label{eq:CC5}
I = \begin{pmatrix}
0 & -1 \\
1 & 0
\end{pmatrix} \, .
\end{equation}
We can map $(q_R,\, q_I)$ to a complex $N/2$-component vector
\begin{equation}\label{eq:CC6}
\psi = q_R + \im q_I\, .
\end{equation}
The action of $K_c$ and $I$ is transferred to a transformation of $\psi$,
\begin{equation}\label{eq:CC7}
K_c(\psi) = \psi^*\, , \quad I(\psi) = \im \psi\, .
\end{equation}
These are the standard operations of complex conjugation and multiplication by
$\im$.

The relations \eqref{eq:CC2} are invariant under similarity transformations. An
arbitrary complex structure can be obtained from the representation
\eqref{eq:CC4}, \eqref{eq:CC5} by a similarity transformation.
We will encounter situations when $K_c$ can be realized, while $I$ is realized
only for a part of $q$. Only this part of $q$ can be described by a genuine
complex wave function. The other parts are either even with respect to $K_c$ and
therefore real, or odd and therefore purely imaginary. Multiplication with $\im$
or more general complex numbers is not defined for the purely real or purely
imaginary parts. This corresponds to situations where both complex and real wave
functions are encountered for suitable subspaces.

\paragraph*{Complex operators}

Since multiplication by real numbers and multiplication by $\im$ is defined for
$\psi$, also the multiplication with arbitrary complex numbers is defined. For
$N/2 > 1$ this extends to multiplication by arbitrary complex $(N/2\times
N/2)$-matrices,
\begin{equation}\label{eq:CC8}
A = \im A_I\, .
\end{equation}
A real $N \times N$-matrix $\hat{A}$ is compatible with the complex structure if
it is of the form
\begin{equation}\label{eq:CC9}
\hat{A} = 1\otimes A_R + \tilde{I}\otimes A_I = A_R + I\, A_I = \begin{pmatrix}
A_R & - A_I \\
A_I & A_R
\end{pmatrix}\ ,
\end{equation}
where $I=\tilde{I}\otimes1$, $A_{R,I}=1\otimes A_{R,I}$ in the second
expression. The multiplication of the real vector $q$ in eq.~\eqref{eq:CC4} by
such a real $(N\times N)$-matrix
results after the map to the complex vector $\psi$ in the complex matrix
multiplication
\begin{equation}\label{eq:CC10}
q'=\hat{A} q \to \psi' = A\, \psi\, .
\end{equation}
Multiplication with complex numbers is realized by $A_R$ and $A_I$ proportional
to unit matrices.

For a symmetric matrix $\hat{A}$ eq.\,\eqref{eq:CC9} one has $A_R^\mathrm{T} =
A_R$, $A_I^\mathrm{T}=-A_I$. In turn, the complex matrix $A$ in
eq.\,\eqref{eq:CC8} is hermitian. In the real formulation observables are
represented by symmetric operators $\hat{A}$. If they are compatible with the
complex structure they will be represented by hermitian operators in the complex
formulation. Not all real symmetric observables are compatible with the complex
structure, however.

A necessary and sufficient condition for $\hat A$ to be compatible with a given
complex structure is that it commutes with $I$,
\bel{3.4.81A}
\big[\hat A,I\big]=0\ .
\ee
This identity follows directly from eq.~\eqref{eq:CC9}. In the other direction,
for arbitrary $K_c$ and $I$ obeying eq.~\eqref{eq:CC2} we can employ a basis
transformation to bring them into a direct product form
\bel{3.4.81B}
K'=DK_cD^{-1}=\tau_3\otimes1\ ,\quad I'=DID^{-1}=\tilde I\otimes1\ ,
\ee
with $\tilde I=\pm i\tau_2$. This can either be achieved by choosing $I$
appropriately as discussed above. Or, for more general $I$, one may employ basis
transformations that leave $K'$ invariant. In the basis~\eqref{3.4.81B} a
general real matrix $\hat A$ can be written in the form
\bel{3.4.81C}
\hat A=1\otimes A_R+\tilde I\otimes A_I+\tau_1\otimes A_1+\tau_3\otimes A_3\ .
\ee
The commutation relation~\eqref{3.4.81A} implies $A_1=0$, $A_3=0$ and therefore
implies the compatibility relation~\eqref{eq:CC9}.

\paragraph*{Complex evolution law}

An evolution law is compatible with the complex structure of the wave function
if the step evolution operator $\hat{S}$ takes the form \eqref{eq:CC9}. In this
case it becomes in the complex language a complex $(N/2\times N/2)$-matrix,
defined by eq.~\eqref{eq:CC8}. Formally we can always introduce a complex
structure for the wave function whenever $N$ is even. It is sufficient to define
matrices $K_c$ and $I$ with the properties \eqref{eq:CC2}. Such a complex
structure is useful, however, only if it is compatible with the evolution law.
The criterion for the possibility of a complex structure requires therefore that
$\hat{S}$ can be brought to the form \eqref{eq:CC9} by a suitable similarity
transformation. In particular, if $\hat{S}$ is block diagonal in the form
\begin{equation}\label{eq:CC11}
\hat{S} = \begin{pmatrix}
A_R & 0 \\
0 & A_R
\end{pmatrix}\, ,
\end{equation}
a complex structure can always be introduced, and $\hat{S}$ remains a real
matrix in the complex formulation.

If the step evolution operator $\hat{S}$ is orthogonal and compatible with the
complex structure, the associated complex evolution operator is unitary. This
follows from the conservation of the norm of the wave function, $q^\mathrm{T}q =
\mathrm{const}$, for orthogonal $\hat{S}$. In turn, the complex bilinear is
conserved as well, $\psi^\dagger \psi = q^\mathrm{T}q = \mathrm{const}$.
Transformations that preserve the norm $\psi^\dagger\psi$ are unitary. In other
words, if $\hat{S}$ belongs to some subgroup of SO($N$) and is compatible with
the complex structure, the associated complex step evolution operator belongs to
a subgroup of U($N/2$).

\paragraph*{Complex structures for single species of bits}

For automata representing free massless fermions in two dimensions there exist
many possible complex structures obeying the general criteria described above.
Which one is most useful depends partly on the form of interactions that will be
introduced later. Since the complex structure is a crucial feature in quantum
field theory and quantum mechanics we will present several complex structures in
detail. This underlines the fact that the complex structure is not unique and
not fixed a priori.

A general way of introducing a complex structure divides the configurations of
occupation numbers $\tau$ into two sets $\tau_1$ and $\tau_2$ of equal size.
This may be done by using an observable $A$ with two possible values $(A_1,A_2)$
for $A_\tau$, with an equal number of configurations for both values. We can
then define $\tau_1$ and $\tau_2$ by the conditions $A_{\tau_1}=A_1$,
$A_{\tau_2}=A_2$. As one of the possibilities we can associate the components
$\tilde q_{\tau_1}$ of the wave function to the real part of the complex wave
function, and components $\tilde q_{\tau_2}$ to its imaginary part. We further
assume an invertible map from the set $\tau_1$ to the set $\tau_2$,
$\tau_1\to\bar\tau_2(\tau_1)$. We can then order the wave function $\tilde q$ in
a two-component form
\bel{NCS1}
\tilde q=\begin{pmatrix}\tilde q_{R,\tau_1}\\ \tilde q_{I,\tau_1}\end{pmatrix}\
,\quad \tilde q_{R,\tau_1}=\tilde q_{\tau_1}\ ,\quad \tilde q_{I,\tau_1}=\tilde
q_{\bar\tau_2(\tau_1)}\ .
\ee
For any particular order $\bar\tau_2(\tau_1)$ the complex structure can be
defined by eqs.~\eqref{eq:CC4},~\eqref{eq:CC5}. The complex wave
function~\eqref{eq:CC6} has components denoted by $\tau_1$.

As an example, we take for the Ising model~\eqref{eq:FP1} an observable based on
the total particle number $\Np$,
\bel{NCS4}
A=\theta\left(\Np-\frac{M}{2}\right)\ .
\ee
Here we assume $M$ odd (e.g. $\mathcal{M}_2=M-1$ even), such that $\Np-M/2$
takes either positive or negative half-integer values. For $\Np>M/2$ one has
$A=A_1=1$, while for $\Np<M/2$ one has $A=A_2=0$. The real part of the wave
function corresponds to the configurations with more than half-filling, i.e.
with more sites occupied than empty. The imaginary part is given by the
configuration with less than half-filling. The configurations $\tau_1$ are
arbitrary multi-particle states, with the restriction $\Np>M/2$. This covers
indeed precisely one half of the total configurations. The step evolution
operator for right-transport commutes with the observable $A$. It therefore does
not mix $\tilde q_R$ and $\tilde q_I$. For suitable maps $\bar\tau_2(\tau_1)$ it
acts in the same way on $\tilde q_R$ and $\tilde q_I$, such that $\hat S_R$ is
block diagonal in the form~\eqref{eq:CC11}. The complex structure is compatible
with the evolution in this case.

An even simpler setting is the addition of a single bit or occupation number
$\overline{n}_{c}(t)$ at every $t$. Its value is not changed by the step
evolution operator. Associating $A$ with $\overline{n}_{c}$ the configurations
$\tau_1$ correspond to $\overline{n}_{c}=1$, while $\tau_2$ is the part for
$\overline{n}_{c}=0$. The real and imaginary part of the wave function
correspond to $\overline{n}_{c}=1$ or $\overline{n}_{c}=0$. This general setting
can be realized whenever one has two sets of real wave functions
$\tilde{q}_{1}(t)$ and $\tilde{q}_{2}(t)$. They can be labeled by the additional
bit $\overline{n}_{c}(t)$, with $\overline{n}_{c}(t)=1$ for $\tilde{q}_{1}(t)$
and $\overline{n}_{c}(t)=0$ for $\tilde{q}_{2}(t)$.

\paragraph*{Complex conjugation as local time reversal}

The two sets of wave functions may correspond to wave functions at neighboring
times, $\tilde{q}_{1}(t)=\tilde{q}(t)$,
$\tilde{q}_{2}(t)=\tilde{q}(t+\varepsilon)$. Instead of the complex
structure~\eqref{eq:CC4},~\eqref{eq:CC5} acting on
$\tilde{q}_{1}$,$\tilde{q}_{2}$ we may use
\begin{equation}\label{CT1}
K_c = \begin{pmatrix}
0 & 1 \\ 
1 & 0
\end{pmatrix}\, ,\quad 
I = \begin{pmatrix}
\phantom{-}0 & 1 \\ 
-1 & 0
\end{pmatrix}\, .
\end{equation}
Complex conjugation acts as local time reversal,
\begin{equation}\label{CT2}
K_c \begin{pmatrix}
\tilde{q}(t) \\ 
\tilde{q}(t+\varepsilon)
\end{pmatrix}=
K_c \begin{pmatrix}
\tilde{q}_{1} \\ 
\tilde{q}_{2}
\end{pmatrix}=
 \begin{pmatrix}
\tilde{q}_{2} \\ 
\tilde{q}_{1}
\end{pmatrix}=
\begin{pmatrix}
\tilde{q}(t+\varepsilon) \\ 
\tilde{q}(t)
\end{pmatrix}\ ,
\end{equation}
exchanging $\tilde{q}(t)$ and $\tilde{q}(t+\varepsilon) $. The complex wave
function $\varphi(t)$ obtains as
\bel{CT3}
\varphi(t)=\frac{1+i}{\sqrt {2}}\tilde{q}(t)+\frac{1-i}{\sqrt
{2}}\tilde{q}(t+\varepsilon)\ .
\ee
It combines information from two neighboring time steps. Correspondingly, the
evolution of the time-local probabilistic information is encoded in a series of
time steps with distance $2\varepsilon$, $\varphi(t)$,
$\varphi(t+2\varepsilon)$, $\varphi(t+4\varepsilon) \, \dots$.

For the time-evolution of the complex wave-function $\varphi(t)$ we have to
relate the pair $\tilde{q}(t+2\varepsilon)$, $\tilde{q}(t+3\varepsilon)$ to the
pair $\tilde{q}(t)$, $\tilde{q}(t+\varepsilon)$.
A step evolution operator which is compatible with this complex structure
requires the relations
\bel{CT4}
\varphi(t+2\varepsilon)=\widehat{S}_{c}(t)\varphi(t)\ ,
\ee
or
\begin{align}
\label{CT5}
\tilde{q}(t+2\varepsilon)=&\widehat{S}_{R}(t)\tilde{q}(t)+\widehat{S}_{I}(t)\tilde{q}(t+\eps)\,\nn\\
\tilde{q}(t+3\varepsilon)=&\widehat{S}_{R}(t)\tilde{q}(t+\varepsilon)-\widehat{S}_{I}(t)\tilde{q}(t)\
,
\end{align}
with
\bel{CT6}
\widehat{S}_{c}(t)=\widehat{S}_{R}(t)+i\widehat{S}_{I}(t)\ .
\ee
For a free right moving particle $\widehat{S}_{c}$ is real, $\widehat{S}_{I}=0$,
with $\widehat{S}_{R}(t)$ the operator for right transport by two units
$2\varepsilon$. 

If eq.\eqref{CT5} holds the evolution law can be written in a complex form. The
imaginary and real parts of $\varphi(t)$ are not independent, however, since
$\tilde{q}(t+\varepsilon)$ can be computed from $\tilde{q}(t)$.
This changes if we replace $\tilde{q}_{2}(t)$ by a different wave function at
$t+\varepsilon$, for example by the particle-hole conjugated wave function
discussed in the next section.

\paragraph*{Complex structure for Weyl fermions}

We next address further possible complex structures for two species of
occupation numbers at each position. We focus first on the ``Weyl complex
structure''. A two-dimensional Weyl fermion can be composed of two independent
Majorana-Weyl fermions. For Weyl fermions the wave functions are complex. We can
describe this by a two-dimensional diagonal Ising model for two species of Ising
spins. Starting from a real formulation we will introduce a complex structure
and complex wave functions.

For every point $(m_1,\, m_2)$ we take two species of Ising spins $s_1(m_1,\,
m_2)$ and $s_2(m_1,\, m_2)$. The action of the two-dimensional Weyl-Ising model
extends eq.~\eqref{eq:FP1} to two species,
\begin{align}\label{eq:459}
\cS = -\beta \sum_{m_1 = 0}^{\cM_1 - 1} \sum_{m_2 = 0}^{\cM_2} \sum_{\alpha =
1}^2 
\big[ s_\alpha(m_1 + 1, m_2 + 1) s_\alpha (m_1,m_2) - 1 \big]
\end{align}
where we keep the periodic structure \eqref{eq:FP6} in the $x$-direction
\begin{equation}\label{eq:460}
s_\alpha (m_1,\, \cM_2 + 1) = s_\alpha (m_1,\, 0)\, .
\end{equation}
The number of Ising spins or occupation numbers at every layer $m_1$ of the
local chain is now given by
\begin{equation}\label{eq:461}
M = 2\, (\cM_2 + 1)\, ,
\end{equation}
such that the number of states $\tau$ equals
\begin{equation}\label{eq:462}
2^{M} = \Big( 2^{(\cM_2 + 1)} \Big)^2 \, .
\end{equation}
We may label them by an index $\tau = (\tau_1, \, \tau_2)$, where $\tau_1$
specifies the positions which are occupied by particles of species $\alpha = 1$,
while $\tau_2$ does the same for particles of species $\alpha = 2$. For every
position $x$ we can have four possibilities for the pair of occupation numbers
$(n_1, n_2)$, e.g. $(0,0)$, $(1,0)$, $(0,1)$ and $(1,1)$. Correspondingly, the
real classical wave functions are labeled as $\tilde{q}_{\tau_1,\, \tau_2}(t)$
and $\bar{q}_{\tau_1,\, \tau_2}(t)$.

A possible involution corresponding to complex conjugation switches the sign of
$\tilde{q}_{\tau_1,\,\tau_2}$ and $\bar{q}_{\tau_1,\,\tau_2}$ whenever $\tau_2$
corresponds to an odd number of particles of species $\alpha = 2$. To every
state $(\tau_1, \tau_2)$ we can associate particle numbers $N_\alpha$, with
$N_\alpha$ counting the number of positions for which $s_\alpha$ is positive.
The involution $K_c$ is given by
\begin{align}\label{eq:463}
& K_c (\tilde{q}_{\tau_1,\,\tau_2}) = (-1)^{N_2}\,\tilde{q}_{\tau_1,\,\tau_2}\,
, \notag \\
& K_c (\bar{q}_{\tau_1,\,\tau_2}) = (-1)^{N_2}\, \bar{q}_{\tau_1,\,\tau_2}\, .
\end{align}
We may represent $K_c$ by a diagonal $(2^M\times 2^M)$-matrix with diagonal
elements $(-1)^{N_2}$. Quantities that are odd with respect to $K_c$ will be
considered as imaginary with respect to the complex structure, while even
quantities are real. For example, the vacuum wave function with $N_1 = N_2 = 0$
is a real constant. We will see below alternative choices of complex conjugation
related to the map between particles and holes. This relation to particle-hole
conjugation is better adopted to quantum field theories. We continue here first
with the choice~\eqref{eq:463} in order to demonstrate the freedom in the choice
of complex structures by a particularly simple example.

\paragraph*{Complex one-particle wave function}

For one-particle wave functions one has either $N_1 = 1$, $N_2 = 0$, or $N_1 =
0$, $N_2 = 1$. The real one-particle wave functions can be written in the form
\begin{equation}\label{eq:464}
\tilde{q}(t,x) = \begin{pmatrix}
\tilde{q}_1(t,x) \\
\tilde{q}_2(t,x)
\end{pmatrix}\, ,
\end{equation}
with operators for particle numbers of type $\alpha$ given by
\begin{equation}\label{eq:465}
\hat{N}_1 = \begin{pmatrix}
1 & 0 \\
0 & 0 
\end{pmatrix}\, , \quad \hat{N}_2 = \begin{pmatrix}
0 & 0 \\
0 & 1
\end{pmatrix} \, .
\end{equation}
For $\tilde{q}_2 (t,x) = 0$ the particle is of type $\alpha=1$, while for
$\tilde{q}_1 (t,x) = 0$ it is of type $\alpha = 2$. For general $\tilde{q}(t,x)$
one has a superposition of states with a single particle being of type one or
two. Eq.~\eqref{eq:464} is the most general eigenstate of the particle number
operator $\hat{N} = \hat{N}_1 + \hat{N}_2$ with eigenvalue one. For the
conjugate wave function we use a form similar to eq.~\eqref{eq:464}.

We can map the two real functions $\tilde{q}_1$, $\tilde{q}_2$ to a complex
function, and similar for $\bar{q}_1$, $\bar{q}_2$, 
\begin{align}\label{eq:466}
& \tilde{\psi}(t,x) = \frac{1}{\sqrt{\varepsilon}} \big( \tilde{q}_1 (t,x) + 
\im \tilde{q}_2 (t,x) \big)\, , \notag \\
& \bar{\psi} (t,x) = \frac{1}{\sqrt{\varepsilon}} \big( \bar{q}_1 (t,x) - 
\im \bar{q}_2 (t,x) \big)\, . 
\end{align}
Since $\tilde{q}_2$ and $\bar{q}_2$ change sign under the complex conjugation
$K_c$ in eq.~\eqref{eq:463}, the involution $K_c$ acts on $\tilde{\psi}$ and
$\bar{\psi}$ as a standard complex conjugation,
\begin{align}\label{eq:467}
& K_c ( \tilde{\psi}(t,x)) = \tilde{\psi}^* (t,x) ,\, \notag \\
& K_c (\bar{\psi}(t,x)) = \bar{\psi}^* (t,x)\, .
\end{align}

The conventions in eq.~\eqref{eq:466} are taken such that
\begin{align}\label{eq:468}
\bar{\psi}(t,x)\, \tilde{\psi}(t,x) &= \frac{1}{\varepsilon} \Big( \bar{q}_1
(t,x)\, 
\tilde{q}_1(t,x) + \bar{q}_2 (t,x)\, \tilde{q}_1 (t,x) \Big) \notag \\
&= p(t,x)\, ,
\end{align}
with $p(t,x)$ the local probability density to find a particle at position $x$,
normalized for a one-particle state according to 
\begin{equation}\label{eq:469}
\int_x p(t,x) = 1\, .
\end{equation}
Choosing boundary conditions for which $\bar{q}_\tau (t) = \tilde{q}_\tau(t)$,
one has the relations similar to quantum mechanics for a single particle,
\begin{equation}\label{eq:470}
\bar{\psi}(t,x) = \psi^* (t,x)\, , \quad \int_x \psi^* (t,x)\, \psi (t,x) = 1\,
.
\end{equation}
This reflects the general property for orthogonal step evolution operators that
the conjugate classical wave function $\bar{q}$ can
be identified with the classical wave function $\tilde{q}$. For our choice of
boundary conditions and complex structure we find
the standard formulation of quantum mechanics with a complex wave function,
including the standard normalization \eqref{eq:470}.

In the real formulation \eqref{eq:464} the involution $K_c$ is given by
multiplying the wave function vector with a diagonal matrix,
\begin{equation}\label{eq:471}
K_c = \begin{pmatrix}
1 & 0 \\
0 & -1
\end{pmatrix}\, , \quad K_c^2 = 1\, .
\end{equation}
We can also define the transformation 
\begin{equation}\label{eq:472}
I = \begin{pmatrix}
0 & -1 \\
1 & 0
\end{pmatrix}\, , \quad I^2 = -1\, , \quad \{ K_c, I \} = 0\, .
\end{equation}
With
\begin{equation}\label{eq:473}
I\, \begin{pmatrix}
\tilde{q}_1 \\ \tilde{q}_2
\end{pmatrix} = \begin{pmatrix}
- \tilde{q}_2 \\ \tilde{q}_1
\end{pmatrix}\, , \quad I \begin{pmatrix}
\bar{q}_1 \\ \bar{q}_2
\end{pmatrix} = \begin{pmatrix}
- \bar{q}_2 \\
\bar{q}_1
\end{pmatrix}\, ,
\end{equation}
one realizes the multiplication with $\im$ in the complex formulation
\eqref{eq:466},
\begin{equation}\label{eq:474}
I(\tilde{\psi}) = \im \tilde{\psi}\, , \quad I(\bar{\psi}) = -\im \bar{\psi}\, .
\end{equation}
For $\bar{q}_\alpha = \tilde{q}_\alpha$, $\bar{\psi} = \tilde{\psi}^*$, this is
compatible with $(\im \tilde{\psi})^* = - \im \tilde{\psi}^*$.

\paragraph*{Schrödinger equation}

In the real basis \eqref{eq:464} the time evolution of the wave function obeys
in the continuum limit
\begin{equation}\label{eq:475}
\p_t \begin{pmatrix}
\tilde{q}_1 \\
\tilde{q}_2
\end{pmatrix} = - \p_x
\begin{pmatrix}
\tilde{q}_1 \\
\tilde{q}_2
\end{pmatrix} \, .
\end{equation}
We can multiply with the matrix $I$,
\begin{equation}\label{eq:476}
I\, \p_t \begin{pmatrix}
\tilde{q}_1 \\
\tilde{q}_2
\end{pmatrix} = - I\, \p_x \begin{pmatrix}
\tilde{q}_1 \\
\tilde{q}_2
\end{pmatrix} \, .
\end{equation}
In consequence, the complex wave function obeys a Schrödinger equation for a
one-particle state
\begin{equation}\label{eq:477}
\im \p_t \tilde{\psi}(t,x) = \hat{P}\, \tilde{\psi}(t,x)\, .
\end{equation}
Here $\hat{P}$ is the momentum operator
\begin{equation}\label{eq:478}
\hat{P} = -\im \p_x\, .
\end{equation}
As usual, the eigenstates of $\hat{P}$ are plane waves,
\begin{equation}\label{eq:479}
\tilde{\psi} = c\, \exp \big( - \im p\, (t-x) \big)\, ,
\end{equation}
corresponding to solutions of eq.~\eqref{eq:477} with boundary condition
\begin{equation}\label{eq:480}
\tilde{\psi}_{in} = \tilde{\psi}(0,x) = c\,\exp(\im p x)\, .
\end{equation}
For boundary conditions with $\bar{q}(t) = \tilde{q}(t)$, $\tilde{\psi}(t) =
\psi (t)$, $\bar{\psi}(t) = \psi^* (t)$, the complex normalization constant $c$
obeys
\begin{equation}\label{eq:481}
|c|^2 = \frac{1}{L}\, , \quad L = \int_x = \varepsilon\, (\cM_2 + 1)\, .
\end{equation}
The formulation of the evolution of one-particle states in the Weyl-Ising model
is precisely the form of quantum mechanics in a discretized version.

\paragraph*{Momentum basis}

We can translate back and investigate the momentum eigenstates in the real
formulation. The initial boundary condition has $\tilde{q}_{\tau_1,\,\tau_2} (0)
= 0$ for all configurations where the total particle number differs from one.
For real $c$ and a superposition of configurations where $s_1(m_2) = 1$ for one
particular site $m_2$, while all other spins equal $-1$, one find the part of
the initial wave function
\begin{equation}\label{eq:482}
\tilde{q}_{in}(m_2) = \frac{1}{\sqrt{\cM_2 + 1}} \cos (\varepsilon\, p\, m_2 )\,
,
\end{equation}
while for a superposition of configurations with $s_2 (m_2) = 1$, and all other
spins negative, one has
\begin{equation}\label{eq:483}
\tilde{q}_{in} (m_2) = \frac{1}{\sqrt{\cM_2 + 1}} \sin (\varepsilon\, p\, m_2)\,
.
\end{equation}
The initial overall spin configuration of our unique jump chain is given for the
plane wave \eqref{eq:479} at $t=0$ by
\begin{align}\label{eq:484}
& \tilde{q}_1 (m_1, m_2) = \bar{q}_1(m_1,m_2) = \frac{1}{\sqrt{\cM_2 +1}} 
\cos \big( \varepsilon p (m_2 - m_1)\big)\, , \notag \\
& \tilde{q}_2 (m_1, m_2) = \bar{q}_2 (m_1, m_2) = \frac{1}{\sqrt{\cM_2 +1}} 
\sin \big( \varepsilon p (m_2 - m_1) \big) \, .
\end{align}
The probability to find the single particle is the same for all $x$ and all $t$,
\begin{equation}\label{eq:485}
p(m_1, m_2) = \tilde{q}_1^2 (m_1, m_2) + \tilde{q}^2_2 (m_1, m_2) = 
\frac{1}{\cM_2 +1} \, .
\end{equation}
What oscillates, however, is the type of particle found. It oscillates from type
one to type two and back, with period given by the inverse of the momentum $p$,
\begin{equation}\label{eq:486}
N_p = \Delta m_1 = \frac{\pi}{\varepsilon p} \, .
\end{equation}

By virtue of the periodic boundary conditions in $x$ the possible values of $p$
are discrete,
\begin{equation}\label{eq:487}
p = \frac{2\pi\, n_p}{\varepsilon\,(\cM_2 +1)} = \frac{2\pi\, n_p}{L}\, , \quad 
n_p = \mathbb{Z}\, .
\end{equation}
Furthermore, the discretization of $x$-points implies that $p + 2\pi
k/\varepsilon$ and $p$ lead to the same wave function for any integer $k$. We
can therefore restrict $p$ to the range
\begin{equation}\label{eq:488}
- \frac{\pi}{\varepsilon} \leq p \leq \frac{\pi}{\varepsilon}\, ,
\end{equation}
with boundaries of the interval identified. Correspondingly, one has
$|n_p|\leq(\cM_2+1)/2$. The total number of $p$-modes equals $\cM_2$ and
coincides with the total number of discrete positions.

We observe that for discrete $x$-points and finite $L$ the operator $\p_x$ is a
finite antisymmetric $(2^M \times 2^M)$-matrix,
\begin{equation}\label{eq:489}
(\p_x)_{m^{}_2 m'_2,\, \alpha \alpha'} = \frac{1}{2\varepsilon} \Big( 
\delta_{m^{}_2,\, m'_2 -1} - \delta_{m^{}_2,\, m'_2 +1} \Big)\,
\delta_{\alpha\alpha'}\, .
\end{equation}
The operator $I\p_x$ is a symmetric matrix in the real representation,
corresponding to $\hat{P}=-\im \p_x$ being hermitian in the complex basis. A
different hermitian momentum operator $\tilde{P}$ with eigenvalues $p$ can be
related to the discrete Fourier transform discussed in
sect.~\ref{sec:fourier_transform_for_cellular_automata}. As familiar from
quantum mechanics the eigenfunctions of $\tilde{P}$ form a complete basis. An
arbitrary complex periodic function $\psi(t,x)$ can be represented by a
(discrete) Fourier series
\begin{equation}\label{eq:490}
\psi (x) = \int_p \hat{\psi}(p)\, \text{e}^{\im p x}\, ,
\end{equation}
where $\int_p$ involves a summation over the discrete momenta in the interval
\eqref{eq:488}.

We conclude that the one-particle state of the Weyl-Ising model obeys all the
properties of a free Weyl fermion in quantum mechanics for one space and one
time dimension. No additional axioms are needed beyond the three axioms of
classical statistics. For discrete space points and finite $L$ the Hilbert space
is finite-dimensional. The continuum limit $\varepsilon \to 0$ can be taken in a
very straightforward way, resulting in an infinite-dimensional Hilbert space.
Also the infinite-volume limit $L \to \infty$ does not pose any major problems. 

\paragraph*{Plane wave solutions and probabilistic clocks}

The plane wave solutions or momentum eigenstates constitute probabilistic
clocks. Indeed, for any given fixed position $x$ the evolution of $\tilde{q}(t)$
is periodic and constitutes a clock with period $N_q = 2 N_p$ given by
eq.~\eqref{eq:486}. It is described by a rotating angle $\tilde{\alpha}(t)$,
\begin{align}\label{eq:491}
\begin{pmatrix}
\tilde{q}_1 \\
\tilde{q}_2
\end{pmatrix} = \frac{1}{\sqrt{\cM_2 + 1}} \begin{pmatrix}
\cos (\tilde{\alpha}(t)) \\
\sin(\tilde{\alpha}(t))
\end{pmatrix}\, ,
\end{align}
with 
\begin{equation}\label{eq:492}
\tilde{\alpha} (t) = p x - \omega t\, , \quad \omega = p\, .
\end{equation}

In contrast to the clock systems in sect.~\ref{sec:clock_systems} the angle
$\tilde{\alpha}(t)$ does not correspond to different states $\tau$. It is purely
a property of the probabilistic information for two states, namely $\gamma = 1$
and $\gamma = 2$ at some given position $x$. The space-local subsystem at a
given $x$ constitutes a simple example for a subsystem with periodic evolution
not described by unique jump operations. Unique jump operations on a two-state
system can have only period two (or one), while the period in our case is given
by
\begin{equation}\label{eq:493}
N_q = \frac{\cM_2 + 1}{n_p}\, ,
\end{equation}
with integer $n_p$. For non-integer $N_q$ the wave function at a given discrete
time position $t$ turns back to some initial value only after a period that is
an integer times $N_q$, such that the product is an integer.

With the initial conditions of a plane wave solution we can integrate out spins
at locations different from the given value $x$. The remaining subsystem is a
two-state system for the two particle species, $\gamma = 1,2$. As we have seen,
the evolution of the two-state subsystem can be computed from the local
probabilistic information of the subsystem alone. This example generalizes the
concept of probabilistic clocks. It is sufficient that the local probabilistic
information shows a periodic evolution. No states $\tau$ need to be associated
to the different steps of the evolution.

The angular step $\Delta\tilde{\alpha}$ performed in one time step $\varepsilon$
is given by
\begin{equation}\label{eq:494}
\Delta \tilde{\alpha} = \omega \varepsilon = p \varepsilon = \frac{2\pi\,
n_p}{\cM_2 +1}\, ,
\quad n_p \in \mathbb{Z}\, .
\end{equation}
For $\cM_2\to\infty$ and fixed $n_p$ the clock performs a continuous rotation.

For every location $x$ we can define a ``local clock'' related to the subsystem
at $x$. The different clocks have all the same frequency. They only differ in
phase. Nevertheless, they are all correlated. Whenever the pointer of one local
clock changes by $\Delta\tilde{\alpha}$, the pointers of all other local clocks
change by the same amount $\Delta\tilde{\alpha}$. The one-particle plane wave
function can be interpreted as a system of correlated local clocks, distributed
over the whole space, and ticking all with the same frequency. We observe that
the period of the clock depends on the boundary term or initial condition for
the overall probabilistic 
system. This boundary term fixes the momentum $p$. 

\paragraph*{Complex one-particle density matrix}

For $\tilde{\psi}(x) = \psi(x)$, $\bar{\psi} = \psi^*(x)$ the pure-state density
matrix in the one-particle sector is hermitian,
\begin{equation}\label{eq:494A}
\rho(t,x,y) = \psi(t,x)\, \psi^*(t,y)\, , \quad \rho^\dagger = \rho\, .
\end{equation}
In particular, for a plane wave solution \eqref{eq:479} one has for all $t$ 
\begin{equation}\label{eq:494B}
\rho(t,x,y) = \frac{1}{L} \exp \{ \im p\, (x-y) \}\, .
\end{equation}
In the real formulation this corresponds to $\bar{q} = \tilde{q} = q$. For
$\bar{q}$ different from $\tilde{q}$, and therefore $\bar{\psi}$ different from
$\tilde{\psi}^*$, the hermiticity of the density matrix is lost. 

In the real formulation the operation of complex conjugation acts on $\rho'$ as
\begin{equation}\label{eq:494C}
\rho' \to K_c \rho' K_c\, .
\end{equation}
Correspondingly, $\rho'_{11} (x,y)$ and $\rho'_{22} (x,y)$ are real quantities,
while $\rho'_{12}(x,y)$ and $\rho'_{21} (x,y)$ are imaginary. For an arbitrary
pure state we demand that the map to a complex formulation respects the complex
multiplication
\begin{equation}\label{eq:494D}
\rho'_{\alpha\beta} (x,y) = \tilde{q}_\alpha (x) \, \bar{q}_\beta (y) \;
\rightarrow
\; \rho(x,y) = \tilde{\psi}(x) \, \bar{\psi} (y) \, .
\end{equation}
This implies
\begin{align}\label{eq:494E}
\rho(x,y) &= \tilde{q}_1(x)\, \bar{q}_1(y) + \tilde{q}_2(x)\, \bar{q}_2(y)
\notag \\
& \quad + \im \tilde{q}_2 (x)\, \bar{q}_1(y) - \im \tilde{q}_1(x)\, \bar{q}_2(y)
\, .
\end{align}
This relates the complex hermitian density matrix $\rho$ to the matrix elements
of the real symmetric classical density matrix $\rho'$,
\begin{equation}\label{eq:494F}
\rho (x,y) = \rho'_{11} (x,y) + \rho'_{22} (x,y) + \im \big( \rho'_{21} (x,y) - 
\rho'_{12}(x,y) \big)\,.
\end{equation}
The hermitian part of $\rho$ corresponds to the symmetric part of $\rho'$,
\begin{align}\label{eq:494G}
& \frac{1}{2} \big( \rho(x,y) + \rho^*(y,x) \big) \notag \\
& \; = \frac{1}{2} \Big\{ 
\rho'_{11} (x,y) + \rho'_{11}(y,x) + \rho'_{22}(x,y) + \rho'_{22} (y,x) \notag
\\
& \quad + \im \big[ \rho'_{21}(x,y) + \rho'_{12} (y,x) - \rho'_{21}(y,x) - 
\rho'_{12} (x,y) \big] \Big\}\, ,
\end{align}
while the antihermitian part of $\rho$ obtains from the antisymmetric part of
$\rho'$.
Employing symmetric boundary conditions, $\rho'$ is symmetric and the complex
density matrix $\rho$ is hermitian.

We emphasize that the construction of the complex density matrix $\rho$ differs
from the construction of complex operators $A$ for observables. Both $\rho$ and
$A$ are hermitian complex matrices, but the map from the real to the complex
formulation differs. Arbitrary symmetric $\rho'$ are mapped to complex $\rho$,
while the map from real $\hat A$ to complex $A$ requires compatibility with the
complex structure.

Complex density matrices for mixed states obtain as sums of hermitian pure state
density matrices with real coefficients $\bar{p}_\alpha$, $0 \leq \bar{p}_\alpha
\leq 1$, $\sum_\alpha \bar{p}_\alpha = 1$. As a result the mixed state density
matrix is a hermitian 
positive matrix with real eigenvalues in the range $0 \leq \lambda_i \leq 1$. We
observe that a complex hermitian $\rho$ has $(N/2)^2$ real entries,
less than for a real symmetric matrix $\rho'$ with $N(N+1)/2$ real entries.
While for pure classical states there is a one-to-one
map between the real and complex basis for the wave function, the map from the
real classical density matrix $\rho'$ to the 
complex density matrix is not invertible. The relation \eqref{eq:494G} continues
to hold and defines the map $\rho' \to \rho$.

\paragraph*{Alternative complex structures}

The Weyl complex structure~\eqref{eq:463} extends to multi-particle states. We
discuss this in more detail in
appendix~\ref{app:complex_structure_for_two-particle_wave_function}. This
extension is rather straightforward, except for the states for which more than
one particle of a given species is present at a given position $x$. For example,
the wave function for the two-particle state where one particle of species one
and one particle of species two are present at the same position $x$ is odd
under the transformation~\eqref{eq:463}. In a complex picture this part of the
wave function would be purely imaginary, while a corresponding real part is not
present. 

A different complex structure uses the exchange of $\tilde
q_1\leftrightarrow\tilde q_2$ for the definition of complex conjugation. The
complex one-particle wave function (in a discrete formulation) is now given by
\begin{align}
\label{NCSA1}
\tilde\psi(t,x)=\frac{1+i}{\sqrt{2}}\tilde q_1(t,x)+\frac{1-i}{\sqrt{2}}\tilde
q_2(t,x)\ ,\nonumber\\
\bar\psi(t,x)=\frac{1-i}{\sqrt{2}}\bar q_1(t,x)+\frac{1+i}{\sqrt{2}}\bar
q_2(t,x)\ .
\end{align}
Correspondingly, the discrete maps $K_c$ and $I$ are given in the
basis~\eqref{eq:464} by
\bel{NCSA2}
K_c=\begin{pmatrix}0&1\\1&0\end{pmatrix}\ ,\quad
I=\begin{pmatrix}0&1\\-1&0\end{pmatrix}\ .
\ee
Eqs.~\eqref{eq:467},~\eqref{eq:474} are easily verified. The pair of
matrices~\eqref{NCSA2} is related to the pair of
matrices~\eqref{eq:471},~\eqref{eq:472} by a similarity transformation.

The complex conjugation based on the exchange of species can be extended to
general configurations with arbitrary particle numbers. We write a general wave
function,
\bel{NCSA3}
\tilde q(t)=\tilde q_{\tau\rho}(t)Q_\tau^{(1)}\otimes Q_\rho^{(2)}\ ,
\ee
in terms of basis vectors $Q_\tau^{(\gamma)}$. For a given species $\gamma$ the
components of $Q_\tau^{(\gamma)}$ are given by
\bel{NCSA4}
\gl Q_\tau\gr_\sigma=\delta_{\tau\sigma}\ ,
\ee
such that the product $Q_\tau^{(1)}\otimes Q_\rho^{(2)}$ differs from zero
precisely for the configuration $\tau$ for species $1$ and configuration $\rho$
for species $2$. The wave function $\tilde q_{\tau\rho}(t)$ can be considered as
a $2^N\times2^N$-matrix. Then the complex conjugation~\eqref{NCSA2} acts in the
matrix picture as a transposition
\bel{NCSA5}
K_c(\tilde q)=\tilde q^T\ .
\ee

We decompose $\tilde q$ as
\bel{NCSA6}
\tilde q=\tilde q^d+\tilde q^s+\tilde q^a\ ,\quad \gl\tilde q^s\gr^T=\tilde q^s\
,\quad \gl\tilde q^a\gr^T=-\tilde q^a\ ,
\ee
where $\tilde q^d$ is a diagonal matrix, while $\tilde q^s$ is a symmetric
matrix with vanishing diagonal elements and $\tilde q^a$ is antisymmetric. The
number of elements of $\tilde q^s$ and $\tilde q^a$ is equal and we can define a
complex structure by defining the complex wave function as a hermitian matrix
$\tilde\vp$ with vanishing diagonal elements,
\bel{NCSA7}
\tilde\vp=\tilde q^s+i\tilde q^a\ ,\quad \tilde\vp^\dagger=\tilde\vp\ .
\ee
The map $I$ is realized by
\bel{NCSA8}
I\gl\tilde q^s\gr=-\tilde q^a\ ,\quad I\gl\tilde q^a\gr=\tilde q^s\ ,
\ee
such that
\bel{NCSA9}
I\begin{pmatrix}\tilde q^s\\ \tilde
q^a\end{pmatrix}=\begin{pmatrix}0&-1\\1&0\end{pmatrix}\begin{pmatrix}\tilde
q^s\\ \tilde q^a\end{pmatrix}
\ee
realizes for the off-diagonal part of $\tilde q$
\bel{NCSA10}
\tilde\vp\gl I\tilde q\gr=i\tilde\vp\ .
\ee
One recognizes the same structure as for eqs.~\eqref{eq:471},~\eqref{eq:472}.
For the particular case of the one-particle wave function one has
\bel{NCSA11}
\tilde q^s=\frac{1}{\sqrt{2}}\gl q_1+q_2\gr\ ,\quad \tilde
q^a=\frac{1}{\sqrt{2}}\gl q_1-q_2\gr\ ,
\ee
and
\bel{NCSA12}
\tilde\vp=\tilde q^s+i\tilde
q^a=\frac{1+i}{\sqrt{2}}q_1+\frac{1-i}{\sqrt{2}}q_2\ .
\ee
In this case all quantities $\tilde q^s$, $\tilde q^a$, $q_1$, $q_2$ and
$\tilde\vp$ are functions of the position variable $j$.

The diagonal elements $\tilde q^d$ are particular. They correspond to states for
which the configuration for species $1$ equals precisely the configuration of
species $2$. An example are two-particle states with two particles of different
species on the same site. For this subclass of states no complex structure is
defined. These $2^N$ states are described by a real wave function $\tilde q^d$
with components $\tilde q^d_\tau=\tilde q_{\tau\tau}$. We may consider them as
particular composite particle states. In practice they often play no role since
in the continuum limit of large $\mathcal{M}_2$ they form only a very small
subclass of the total number of $2^{2\mathcal{M}_2+2}$ configurations.

Finally, for the freely propagating right movers we can divide the square
lattice into two sublattices which are not connected by the evolution. The two
species of Ising spins or fermions can be identified with Ising spins or
fermions on the two sublattices. We describe this more geometric view on
fermionic species and the associated complex structure in
appendix~\ref{app:complex_structure_based_on_sublattices}.

\subsection{Particles and holes}\label{sec:particles_and_holes}

Antiparticles play an important role in quantum field theory. In two dimensions they are present both for Weyl and Dirac fermions. In contrast, Majorana-Weyl fermions are their own antiparticles. It seems therefore natural to link the complex wave function for Weyl fermions with the notion of antiparticles. In solid state physics half-filled ground states admit both particles and holes as excitations. This is close to particles and antiparticles. In this section we identify the map between particles and holes with complex conjugation. This induces a complex structure which is different from the preceding subsection. For a particle-hole symmetric vacuum we address one-particle and multi-particle states. We describe free massless Dirac fermions in two dimensions by a generalized Ising model. 

The concept of antiparticles emerges naturally from our classical statistical setting. In sect.~\ref{sec:Particles_and_antiparticles} we will show that the presence of some complex structure compatible with the evolution, together with a Fourier transform of annihilation and creation operators, is sufficient to construct a half-filled vacuum for which the excitations are described by particles and antiparticles with positive energy. Correspondingly, we will modify the notion of particles by basing them on excitations of the half-filled vacuum.

\paragraph*{Particle-hole conjugation}

The action~\eqref{eq:FP1} is invariant under a change of sign of all Ising spins, $s(m_1,m_2)\to-s(m_1,m_2)$. This involution corresponds to particle-hole conjugation. For a given $t$ or $m_1$ a spin configuration with $\Np$ spins up and $M-\Np$ spins down corresponds to $\Np$ fermions present. It is mapped to a configuration with $\Np$ spins down and $M-\Np$ spins up. For $M$ even there exist half-filled states with $M/2$ spins up and $M/2$ spins down. For these configurations the particle-hole conjugation changes only the position of the spins up or down.

We introduce a modified concept of particles which takes into account the particle-hole conjugation. For this purpose we define the charge $Q$
\begin{equation}\label{PH1}
Q=\Np-\frac M2\ .
\end{equation}
For positive $Q>0$ we may say that $Q$ particles are present, while for negative $Q<0$ there are holes. Particle-hole conjugation switches the sign of $Q$. It transforms particles to holes and vice versa, without changing the position of the particles or holes. $Q$-particle states are mapped to $Q$-hole states. The maximal charge is $Q=M/2$, and similar for the maximal number of holes. We will later extend the particle concept to account for situations with pairs of widely separated particles and holes. Particles or holes will be considered as local excitations of some half filled ground state or vacuum.

We can divide the states $\{\tau\}$ into two classes $\{\tau'\}$ and $\{\tau^c\}$, which are mapped into each other by particle-hole conjugation. All states with $Q>0$ belong to the set $\{\tau'\}$ and all states with $Q<0$ to the set $\{\tau^c\}$. For even $\mathcal{M}_2$ or odd $M$ this classification is sufficient since all states have either positive or negative $Q$ different from zero. We can identify the observable $A$ in sect.~\ref{sec:complex_structure} with $\theta(Q)$. The map $\bar\tau_1(\tau_2)$ corresponds to the map $\tau'\leftrightarrow\tau^c$, such that $q_{\tau'}^c=q_{\tau^c}$.

One can define a similar split for even $M$. In this case we still have to distribute the configurations with $Q=0$ into the sets $\{\tau'\}$ and $\{\tau^c\}$. This distribution is to some extent arbitrary. We choose it such that configurations related by translations in $x$ belong to the same set. This is possible for almost all configurations, with only a few exceptions. A notable exception are the two states for which each spin up has two neighbors down, and each spin down has two neighbors up. A translation by $\epsilon$ moves these states into the particle-hole conjugate states. Again, each one of the sets $\{\tau'\}$ or $\{\tau^c\}$ corresponds to $2^{M-1}$ spin configurations.

According to the split we can write the wave function for a pure state (with $\tilde q=\bar q=q$) as
\begin{equation}\label{PH2}
q_\tau(t)=\begin{pmatrix}q'_{\tau'}(t)\\q^c_{\tau'}(t)\end{pmatrix}\ ,
\end{equation}
with $\tau'=1\hdots2^{M-1}$ and $q_{\tau'}^c$ denoting the component for the configuration that obtains from $\tau'$ by particle-hole conjugation. The particle-hole conjugation can be taken over as a map acting on the wave function
\begin{equation}\label{PH3}
C_{ph}\begin{pmatrix}q'\\q^c\end{pmatrix}=\begin{pmatrix}q^c\\q'\end{pmatrix}=\begin{pmatrix}0&1\\1&0\end{pmatrix}\begin{pmatrix}q'\\q^c\end{pmatrix}\ .
\end{equation}
While the action is invariant under particle-hole conjugation the boundary terms typically violate this symmetry. In consequence, a typical solution $q(t)$ is not invariant under the particle-hole transformation. For example, states with positive $Q$ may be more likely than states with negative $Q$.

\paragraph*{Complex structure relating particles and holes}

Let us use the involution $C_{ph}$ for the definition of the complex structure~\cite{FQFTPCA}. Complex conjugation is identified with particle-hole conjugation. In the basis~\eqref{PH2} the involution $K_c$ is given by the Pauli matrix $\tau_1$. The complex wave function $\varphi_\tau(t)$ is defined by
\begin{align}
\label{PH4}
\varphi_\tau(t)=&\frac{1+i}{\sqrt{2}}q'_\tau(t)+\frac{1-i}{\sqrt{2}}q^c_\tau(t)\nonumber\\
=&e^{i\pi/4}\big(q'_\tau-iq^c_\tau\big)\ ,
\end{align}
with $\tau$ restricted to the set $\{\tau'\}$. The map $q'\to q^c$ indeed corresponds to $\varphi\to\varphi^{*}$. Multiplication with $i$ in the complex picture is realized in the real picture by multiplication with $I=i\tau_2$. With $K_c^2=1$, $I^2=-1$, the choice
\begin{equation}\label{PH5}
K_c=\tau_1\ ,\quad I=i\tau_2\ ,\quad \{K_c,I\}=0\ ,
\end{equation}
indeed obeys the condition~\eqref{eq:CC2} for a complex structure. The complex density matrix is defined for pure states by
\begin{equation}\label{PH6}
\rho_{\tau\rho}=\varphi_\tau\varphi^*_{\rho}\ ,
\end{equation}
with a suitable generalization to mixed states.

The complex structure~\eqref{PH5} is compatible with the evolution. Consider first the sectors with $Q\neq0$. The step evolution operator moves both the configuration $\tau'$ and the particle-hole associated configuration $\tau^c$ by one position to the right. It is therefore block diagonal in the form
\begin{equation}\label{PH7}
\hat S=\begin{pmatrix}\hat S'&0\\0&\hat S'\end{pmatrix}\ .
\end{equation}
The matrix $\hat S'$ is a real $2^{M-1}\times2^{M-1}$-matrix. It is the same right-transport operator as before, with the restriction to states with $Q>0$. For odd $M$ this covers all states. For even $M$ we need to specify $\hat S$ in the sector with $Q=0$. The form~\eqref{PH7} is also valid for most states in the $Q=0$ sector, since for a state $\tau'$ the shifted state after an evolution step belongs to the same set $\{\tau'\}$, according to the rule for the distribution of configurations into the two sets. For the exceptional states a shift by $\epsilon$ switches from $\{\tau'\}$ to $\{\tau^c\}$. For these exceptional vacuum states the step evolution operator does not take the simple form~\eqref{PH7} and has to be discussed separately.

The evolution law in the complex picture is simply given by
\begin{equation}\label{PH8}
\varphi_\tau(t+\epsilon)=\hat S'_{\tau\rho}\varphi_\rho(t)\ ,
\end{equation}
with real orthogonal $\hat S'$ shifting each configuration one place to the right. This evolution is unitary. The normalization
\begin{equation}\label{PH9}
\varphi^*_\tau(t)\varphi_\tau(t)=1
\end{equation}
is preserved. More precisely, for even $M$ eq.~\eqref{PH8} holds except for the exceptional states. The evolution does not mix exceptional states with the other states, and the product $\varphi^\dagger\varphi$ is preserved in both sectors separately, summing up to one.

\paragraph*{Particle-hole symmetric vacua}

For possible vacuum states we require here translation invariance in space and time. (This is not the most general concept of a vacuum state.) We actually only assume invariance under translations by $2\epsilon$. This has the advantage that after two time steps the possible $x$-positions remain the same. We have seen that the Ising spins or fermions on the even and odd sublattices evolve completely independently. We will remove the odd sublattice completely, avoiding an unnecessary doubling. The remaining lattice points on the even sublattice obey $m_1+m_2=0\,\text{mod}\, 2$. The restriction to the even sublattice modifies the relation between $M$ and $\cM_2$, which becomes now $M=(\cM_2+1)/2$. The sums over space points now only sum over even $m_2$ for even $m_1$, and odd $m_2$ for odd $m_1$. With lattice distance $2\epsilon$ in the $x$-direction the integral~\eqref{eq:FP17B} becomes for even $m_1$,
\begin{equation}\label{PHA1}
\int_x=2\epsilon\sum_{m_2\,\text{even}}\ ,
\end{equation}
where the range of even $m_2$ extends from $0$ to $\cM_2-1$.

In the present section we investigate a few simple particle-hole invariant vacuum states. The discussion of half-filled vacuum states that admit particle and antiparticle excitations with positive energy is postponed to sect.~\ref{sec:Particles_and_antiparticles}. We first investigate even $M$ and half-filled vacua with $Q=0$.

One possible vacuum configuration with $Q=0$ is given (for even $m_1$) by $s(m_2)=(-1)^{m_2/2}$. The spins alter their direction from one site to the next, e.g. $s(x)=1$, $s(x+2\epsilon)=-1$, $s(x+4\epsilon)=1$ etc. We call this configuration $\tau_A$. A similar configuration $\tau_B$ is given for $s(m_2)=-(-1)^{m_2/2}$. The configuration $\tau_B$ is the particle-hole conjugate of $\tau_A$, since all spins are flipped. A particle-hole invariant vacuum state is given by the wave function
\begin{equation}\label{PHA2}
q_{\tau_A}=q_{\tau_B}=\frac{1}{\sqrt{2}}\ .
\end{equation}
Similarly, a $C_{ph}$-odd state obeys
\begin{equation}\label{PHA3}
q_{\tau_A}=-q_{\tau_B}=\frac{1}{\sqrt{2}}\ .
\end{equation}
(The overall sign of the wave function plays no role.) For both states the probability to find at any given site $m_2$ the spin up equals one half.

The configurations $\tau_A$ and $\tau_B$ are exceptional configurations since a translation by $2\epsilon$ exchanges $\tau_A\leftrightarrow\tau_B$. Distributing $\tau_A$ to $\{\tau'\}$ and $\tau_B$ to $\{\tau^c\}$ the complex wave function component reads in this vacuum sector
\begin{equation}\label{PHA4}
\varphi_0=\frac{1}{\sqrt{2}}\big(q_{\tau_A}+q_{\tau_B}\big)+\frac{i}{\sqrt{2}}\big(q_{\tau_A}-q_{\tau_B}\big)\ ,
\end{equation}
with
\begin{equation}\label{PHA5}
\varphi_0^*\varphi_0=q_{\tau_A}^2+q_{\tau_B}^2\ .
\end{equation}
The evolution operator after two time steps $\hat S_2(t)=\hat S(t+\epsilon)\hat S(t)$ acts as $\hat S_2\varphi_0=\varphi_0^*$. The particle-hole invariant vacuum state corresponds to $\varphi_0=1$. It is left invariant by the evolution. The switch of sign of the imaginary part of $\varphi_0$ cannot be accounted for by multiplication with a complex number. In this sector the evolution is not compatible with the complex structure. In practice, this plays no role. We could also consider four time steps such that arbitrary complex $\varphi_0$ remains invariant. The particle-hole invariant half-filled vacuum state $\varphi_\tau=\delta_{\tau,0}$ is invariant by translations in space and time by $2\epsilon$, has charge $Q=0$, and is invariant under particle-hole conjugation. It has maximal anti-correlation between two neighboring spins, while the spin expectation value vanishes
\begin{equation}\label{PHA6}
\langle s(x)s(x+2\epsilon)\rangle=-1\ ,\quad \langle s(x)\rangle=0\ .
\end{equation}

There exist many other states with translation invariance and particle-hole symmetry. They can be constructed in the $Q=0$ sector by superpositions of sharp wave functions for configurations that are obtained from a given configuration by applying the particle-hole conjugation and space-translations. Such wave functions will have a weaker anti-correlation of neighboring spins. An example is the half-filled equipartition vacuum. We denote by $\tau_E$ the states with $Q=0$. These states have $M/2$ occupied sites and $M/2$ empty sites. They can be characterized by specifying the positions of the occupied sites. The number of different states $\tau_E$ is denotes by $N_E$. In the real formulation the equipartition vacuum state is given by
\bel{NPHA1}
q_E=\sum_{\tau_E}\frac{1}{\sqrt{N_E}}Q_{\tau_E}\ ,
\ee
where the unit vectors $Q_{\tau_E}$ have components $\gl Q_{\tau_E}\gr_{\rho}=\delta_{\tau_E,\rho}$. This vacuum state is invariant under particle-hole conjugation since the particle-hole conjugated state $\tau_E^c$ appears with the same weight as $\tau_E$. It is also translation invariant in space. The step evolution operator leaves $q_E$ invariant. One can implement a suitable complex structure on the subspace spanned by $\tau_E$.

For $M$ odd there are no configurations with exactly half-filling. Half-filling can be reached only in the average by superposition of wave functions with positive and negative $Q$, such that $\langle Q\rangle=0$. For example, we can construct a half-filled equipartition vacuum by taking for $\tau_E$ all states with either $Q=1/2$ or $Q=-1/2$. They correspond to configurations with either $(M+1)/2$ or $(M-1)/2$ occupied sites, which can be labeled by the positions of the occupied sites. The superposition with equal weights for all $N_E$ states is again given by eq.~\eqref{NPHA1}. This vacuum wave function is invariant under particle-hole conjugation, as well as translations in space or time. The complex structure based on particle-hole symmetry can be implemented directly for the states with $|Q|=1/2$. The vacuum wave function $q_E$ is real in the complex picture.

There is no need to restrict wave functions to superpositions of basis functions with minimal $|Q|'$, either $Q=0$ for $M$ even or $|Q|=1/2$ for $M$ odd. One can admit larger ranges of $Q$ and still achieve $\langle Q\rangle=0$ and particle-hole invariance. Also equal weights for all configurations contributing to the vacuum wave function are not necessary. It is sufficient that configurations which are related by the particle-hole conjugation and by translation in space have equal weight. The ``quantum field theory vacuum'' with positive energy for all excitations that we discuss in sect.~\ref{sec:Particles_and_antiparticles} will be of this generalized type.

\paragraph*{One-particle states}

For one-particle configurations one adds either a particle or a hole to the vacuum configuration. One-particle wave functions are superpositions of sharp wave functions or basis wave functions in position space where the single particle or the single hole is added precisely at a fixed position $m_2$. Obviously, the one-particle excitations depend on the vacuum, as familiar in quantum field theory.

For a given vacuum the addition of a particle at $j$ is only possible for those components for which $n(j)=0$. All other vacuum components are projected out by the operation of adding a particle at the site $j$. This may be demonstrated for the simple particle-hole invariant vacuum~\eqref{PHA3} for even $M$. For even $m_1$ we can add a particle at $m_2=2\,\text{mod}\, 4$ by flipping the spin of the configuration $\tau_A$ at this position, while for $m_2=0\,\text{mod}\, 4$ one flips the corresponding spin of $\tau_B$. Both these configurations have three neighboring spins up, while all other spins are alternating. The one-particle wave function is characterized by the position $m_2$ of the flipped spin, i.e. the center of the three neighboring spins up, as $q_1'(m_2)=q_1'(x)$. Similarly, one-hole wave functions flip one positive spin at $m_2$ to negative, resulting in three neighboring negative spins with center at $m_2$. The one-hole wave function is $q_1^c(m_2)=q_1^c(x)$. Particle-hole conjugation maps $q_1'(m_2)\leftrightarrow q_1^c(m_2)$. For even $m_1$ and $m_2=2\,\text{mod}4$ the spin flip for the one hole configuration concerns the configuration $\tau_B$, while for $m_2=0\,\text{mod}4$ the spin in $\tau_A$ is flipped. Particle-hole conjugation not only flips the three neighboring spins with equal signs, but also the whole chain of alternating spins.

We combine the one-particle and one-hole states into a new definition of a particle which can be either an additional spin up or down. This type of particle combines both particles and holes, similar to particle physics where the same field describes both particles and antiparticles. The wave functions $q_1'(x)$ and $q_1^c(x)$ are combined into a complex wave function $\psi(x)=\varphi_1(x)$ according to eq.~\eqref{PH4}, with $\tau$ standing for $m_2$ in the one-particle sector.

In the complex picture a time step of the discrete evolution equation moves the particle position one position to the right. This is the same as for the different complex structure discussed in sect.~\ref{sec:complex_structure}. Thus for the complex picture the discussion of sect.~\ref{sec:complex_structure} can be taken over for arbitrary particle-hole invariant and space-translation invariant vacuum states, including the continuum normalization~\eqref{eq:466}, for which we have now $\bar\psi(x)=\psi^*(x)$ since we consider boundary conditions with $\tilde q_\tau(t)=\bar q_\tau(t)=q_\tau(t)$. We have the same Schrödinger equation and Fourier representation.

What is different from sect.~\ref{sec:complex_structure} is the meaning of the real and imaginary part of $\psi(t,x)$. We no longer employ two different types of spins as in eq.~\eqref{eq:459}. A single type of spin with action given by eq.~\eqref{eq:FP1} for $\beta\to\infty$ is sufficient for the description of a complex Weyl spinor. The real and imaginary parts of $\psi$ correspond to linear combinations which are even or odd under particle-hole conjugation,
\begin{align}
\label{PHA7}
\psi_{\text{r}}(t,x)&=\frac{1}{\sqrt{2}}\big(q_1'(t,x)+q_1^c(t,x)\big)\ ,\nonumber\\
\psi_{\text{i}}(t,x)&=\frac{1}{\sqrt{2}}\big(q_1'(t,x)-q_1^c(t,x)\big)\ .
\end{align}
The wave function $\psi_{\text{r}}(t,x)$ describes a Majorana-Weyl fermion, and similar for $\psi_{\text{i}}(t,x)$. Both evolve independently, according to the complex Schrödinger equation
\begin{equation}
\label{3.4.139A}
i\partial_{t}\psi=H\psi\;,\quad H=\widehat{P}=-i\partial_{x}\,.
\end{equation}

We can extract from the complex wave function $\psi$ the parts $q_1'$ for an additional spin up and $q_1^c$ for an additional spin down,
\begin{align}
\label{PHA8}
q_1'=&\frac{1}{2\sqrt{2}}\big[(1-i)\psi+(1+i)\psi^*\big]=\frac{1}{\sqrt{2}}(\psi_{\text{r}}+\psi_{\text{i}})\ ,\nonumber\\
q_1^c=&\frac{1}{2\sqrt{2}}\big[(1+i)\psi+(1-i)\psi^*\big]=\frac{1}{\sqrt{2}}(\psi_{\text{r}}-\psi_{\text{i}})\ .
\end{align}
The part with only one additional spin up obeys
\begin{equation}\label{PHA9}
q_1^c=0\ ,\quad \psi_{\text{i}}=\psi_{\text{r}}\ ,\quad \psi^*=i\psi\ ,
\end{equation}
while for only one additional spin down one has $q_1'=0$, $\psi^*=-i\psi$. As it should be, complex conjugation of $\psi$ switches between these two states. Momentum eigenstates describe an oscillation between $\psi_{\text{r}}$ and $\psi_{\text{i}}$. These oscillations constitute again a clock system.

\paragraph*{Multi-particle states}

For the multi-particle states the setting with particles as excitations of the half-filled vacuum and complex structure based on particle-hole conjugation shows some differences as compared to the Weyl complex structure with two fermion types in sect.~\ref{sec:complex_structure}. A particle and a hole cannot be placed on the same position, in contrast to two fermions of different type. For even $M$ and vacuum with $Q=0$ two-particle states can occur in the sectors with $|Q|=2$ or $Q=0$. The treatment of the sector with $|Q|=2$ is straightforward. It describes two additional particles or two additional holes at $m_2$ and $m_2'$, where $m_2'$ has to be different from $m_2$. The complex two-particle wave function $\varphi_2^{(2)}(t;x,y)$ can be taken antisymmetric in $x\leftrightarrow y$.

Two-particle states in the sector with $Q=0$ have an additional particle at $m_2$ and a hole at $m_2'$. The configuration with $(m_2,m_2')$ differs from the one with $(m_2',m_2)$, for which the hole is at $m_2$ and the additional particle at $m_2'$. The two configurations are mapped onto each other by particle-hole conjugation. The complex two-particle wave function in the $Q=0$ sector, $\varphi_2^{(0)}(t;x,y)$, combines both types of configurations into a single complex quantity. It obeys
\begin{equation}\label{PHA10}
\varphi_2^{(0)}(t;y,x)=\big(\varphi_2^{(0)}(t;x,y)\big)^*\ .
\end{equation}
Since $(m_2,m_2')$ differs from $(m_2',m_2)$ for arbitrary positions $m_2$ and $m_2'$ the complex structure is well defined for arbitrary two-particle states.

Depending on the properties of the vacuum the two-particle states can comprise states of two solitons. This may be demonstrated again for the vacuum~\eqref{PHA3}. Besides the two-particle states that obtain from the configurations $\tau_A$ and $\tau_B$ by flipping two of the spins, the $Q=0$ sector also admits ``solitonic two particle states''. For the ``normal two-particle states'' one has two clusters with three neighboring spins up and three neighboring spins down. The region between the two clusters are alternating spins, with the same type of vacuum $\tau_A$ or $\tau_B$ as outside the cluster. For the ``solitonic two-particle states'' the two clusters comprise only two neighboring spins with equal sign. The chain of alternating spins between the two clusters is now of a different type as outside, i.e. $\tau_B$ in between the clusters and $\tau_A$ outside. These solitonic states do not obtain from vacuum states by local spin flips. All spins in the string of spins between the two clusters have to be flipped compared to the vacuum.

The discussion extends to more than two particles. For example, three-particle states occur in the sectors with $|Q|=3$ and $|Q|=1$. This simply accounts for the fact that a pair of an additional particle and a hole at different positions increases the number of particles by two while it does not change the charge. This is similar to a state with a well separated electron and position. One counts it as two particles while the total charge vanishes. The complex structure remains well defined for all multi-particle states.

\paragraph*{Extended particle-hole conjugation}

The particle-hole conjugation can be extended by combining it with some other discrete transformation. For an example with two species of occupation numbers we may combine the particle-hole conjugation for each species with an exchange of the two species. The extended particle-hole conjugation maps each occupied bit of species $1$ to an empty bit of species $2$ and vice versa. Empty bits of species $1$ are mapped to occupied bits of species $2$. We can write a general wave function similar to eq.~\eqref{NCSA3}
\bel{EPC1}
\tilde q(t)=\tilde q_{\tau\rho}(t)Q_\tau^{(1)}\otimes Q_\rho^{(2)c}\ ,
\ee
where the basis vectors $Q_\rho^{(2)c}$ are the particle-hole conjugate of $Q_\rho^{(2)}$. The particle-hole conjugation of the wave function $\tilde q_{\tau\rho}(t)$ amounts to transposition as in eq.~\eqref{NCSA5}. We can proceed again to the decomposition~\eqref{NCSA6} and implement the complex structure based on $\tilde q^s$ and $\tilde q^a$.

An interesting difference to the discussion in the preceding subsection is the interpretation of the diagonal elements $\tilde q^d$ which do not admit a complex structure. The corresponding part of the wave function describes states for which each configuration $\tau$ of species $1$ is matched by the particle-hole conjugated configuration $\tau^c$ of species $2$. As a result, each occupied bit of species $1$ is accompanied by an empty bit of species $2$, and each empty bit of species $1$ is matched by an occupied bit of species $2$. Thus the states described by $\tilde q^d$ are half-filled states for which at every site $j$ the occupation numbers obey
\bel{EPC2}
n_1(j)+n_2(j)=1\ .
\ee
The different states of this type can be labeled by the configurations of occupation numbers $n_1(j)$ of species $1$, e.g. by $\tau$. One may introduce a complex structure for these particular states by employing the particle-hole conjugation without an exchange of species.

\paragraph*{Antiparticles and Dirac fermions}

The concept of antiparticles is closely related to complex conjugation. For a momentum eigenstate in the single particle sector one has
\begin{equation}\label{PHA11}
\psi(t,x)=\frac{1}{\sqrt{L}}e^{i(px-\omega t)}\ ,\quad \omega =p\ .
\end{equation}
A particle corresponds to $\omega >0$ and therefore $p>0$, as appropriate for a right mover. An antiparticle wave function obtains from the particle wave function by complex conjugation
\begin{equation}\label{PHA12}
\psi_{a}(t,x)=\psi_{p}(t,x)^*=\frac{1}{\sqrt{L}}e^{-i(|p|x-|\omega |t)}\ .
\end{equation}
This corresponds for $\psi_{a}$ to the momentum eigenstate with negative $p$ and $\omega $
\begin{align}
\label{PHA13}
\psi_{a}(t,x)&=\frac{1}{\sqrt{L}}e^{-i(|p|x-|\omega |t)}=\frac{1}{\sqrt{L}}e^{i(px-\omega t)}\ ,\nonumber\\
\omega &=p\ ,\quad \omega <0\ .
\end{align}
General pure particle wave functions are Fourier sums restricted to positive $p$, while for pure antiparticle states only negative $p$ contribute. A general wave function comprises both particle and antiparticle contributions.

At this point one-particle excitations exist with both positive and negative energy $\omega$, corresponding to positive and negative $p$. The introduction of the anti-particle wave function $\psi_a$ seems to be a pure convention. In sect.~\ref{sec:Particles_and_antiparticles} we will discuss half-filled vacua for which all particle excitations have positive energy. This will require a better understanding of the Fourier transform to a momentum basis.

Two-dimensional massless fermions admit two different types of Weyl fermions, namely right-movers and left-movers. In our setting left-movers can be implemented easily by changing the step evolution operator such that for a time step all spin configurations are moved one position to the left. A Dirac fermion consists of two different Weyl fermions, one right-mover and one left-mover. For the description of free massless Dirac fermions in two dimensions one simply adds a second type of Ising spin for which the step evolution operator accounts for the motion to the left. One may either place the left-movers on the same sites of the even sublattice as the right-movers. Alternatively, one could place the left-movers on the odd sublattice, using now the complete spacetime lattice. One needs in this case only one type of Ising spin with different evolution laws for the spin configurations on the even and odd sublattices. Since the evolution does not mix the left- and right-movers or the even and odd sublattices the two formulations are equivalent. We recall that the concept of antiparticles is formulated on the level of single Weyl fermions and does not need the extension to Dirac fermions.

In conclusion of this subsection we have found that many properties of quantum mechanics appear very naturally for our simple diagonal two-dimensional Ising model. This includes a complex wave function and the
associated concept of antiparticles related to complex conjugation. The formulation for general one-particle states
involves for appropriate boundary conditions a positive complex hermitian density matrix $\rho(x,y)$. In the continuum limit its evolution obeys the von 
Neumann equation
\begin{equation}
\partial_t \rho = -i \left[ H, \rho \right]\,.
\end{equation}
The Hamilton operator $H$ is given by the momentum operator (omitting unit operators in positions space)
\begin{equation}
H = \hat{P} = -i \partial_x\,.
\end{equation}
The position operator $\hat{X}$ in eq.~\eqref{eq:FP35} does not commute with the momentum operator. With units $\hbar = 1$ it obeys the
standard commutation relation of quantum mechanics
\begin{equation}
\left[ \hat{X}, \hat{P} \right] = i\,.
\end{equation}
In the next subsection we will see how many further concepts familiar from quantum mechanics carry over to the two-dimensional diagonal Ising model and allow for a simple description of its properties.

We have discussed different complex structures and different types of vacua. In general, the properties of single-particle excitations depend on the vacuum. For the free massless fermions discussed so far the behavior of single particles is found to be very similar for the different vacua. This will no longer be the case once interactions are included. They may introduce new phenomena as spontaneous symmetry breaking and the generation of mass terms for single particles.
\subsection[Conserved quantities and symmetries]{Conserved quantities and \\
symmetries}\label{sec:conserved_quantities_and_symmetries}

Conserved quantities play an important role for an understanding of evolution in
dynamical systems. They are often derived as properties of evolution laws which
are formulated as differential equations. Our concept of time as a property of
probabilistic models for Ising spins allows a systematic treatment of conserved
quantities in a formalism similar to quantum mechanics. Conserved quantities are
often related to symmetries, similar to classical mechanics or quantum
mechanics.
We find several conserved quantities which are associated with new types of
observables. We discuss this in detail for the momentum observable. These new
observables do no longer take fixed values in every state of the overall
probabilistic system. They rather measure properties of the probability
distribution as periodicity. The quantum formulation for classical statistical
systems is very useful for unveiling the presence of conserved quantities. Many
simple properties are not seen easily otherwise.

The quantum formalism also helps to detect symmetries which are not immediately
visible for generalized Ising models. Some of these symmetries emerge in the
continuum limit. In this limit the two-dimensional diagonal Weyl-Ising model
realizes Lorentz symmetry. This symmetry is characteristic for a model of
massless Weyl fermions, while there is no direct implementation as a symmetry of
the action for Ising spins.

\paragraph*{Conserved quantities and step evolution operator}

An observable $A$ that is represented by an operator $\hat{A}$ which commutes
with the step evolution operator $\hat{S}(t)$ is a conserved quantity: This
means that the expectation value does not depend on time,
\begin{equation}\label{eq:CQ1}
[\hat{A},\, \hat{S}(t) ] = 0 \; \Rightarrow \; \p_t \langle A(t) \rangle = 0\, .
\end{equation}
More precisely, we consider observables that do not depend explicitly on time.
This means that the values $A_\tau$ that the observable takes in a state $\tau$
at given $t$ does not depend on $t$. In turn, the associated operator $\hat{A}$
is a matrix that does not depend on $t$. From the expression \eqref{eq:DM34} for
the expectation value in terms of the classical density matrix $\rho'$, and the
evolution law \eqref{eq:DM38} for the density matrix in terms of the step
evolution operator,
\begin{equation}\label{eq:CQ2}
\langle A(t) \rangle = \tr \{ \rho'(t)\, \hat{A}\}\, , \quad
\rho'(t+\varepsilon)
= \hat{S}(t)\, \rho'(t)\, \hat{S}^{-1}(t)\, ,
\end{equation}
we infer
\begin{align}\label{eq:CQ3}
& \langle A(t+\varepsilon) \rangle = \tr \{ \rho'(t+\varepsilon)\, \hat{A} \} =
\tr \{ \hat{S}(t)\, \rho'(t)\, \hat{S}^{-1} (t) \, \hat{A} \} \notag \\
& \quad = \tr \{ \rho'(t) \, \hat{S}^{-1}(t) \, \hat{A} \, \hat{S}(t) \} \, .  
\end{align}
If $\hat{S}(t)$ commutes with $\hat{A}$ this implies
\begin{equation}\label{eq:CQ4}
\langle A(t+\varepsilon) \rangle = \langle A(t) \rangle\, ,
\end{equation}
such that $A$ is indeed a conserved quantity. The formulation in terms of the
density matrix underlines the local character of this law: only the local values
of $\hat{S}(t)$ and $\rho'(t)$ are needed. We may also derive it from the more
global view of eq.~\eqref{eq:LO5}. Since the difference between $\langle
A(t+\varepsilon) \rangle$ and $\langle A(t) \rangle$ only results from the
different position of $\hat{A}$ in the chain \eqref{eq:LO5}, with $\hat{T} =
\hat{S}$ and $Z = 1$, it vanishes if $[\hat{A},\, \hat{S}(t)] = 0$.

We have already encountered several examples of conserved quantities. For the
Ising chain the particle number operator counts the numbers of positive spins,
\begin{equation}\label{eq:CQ5}
\hat{N} = \begin{pmatrix}
1 & 0 \\
0 & 0
\end{pmatrix}\, .
\end{equation}
It commutes with the step evolution operator $\hat{S}_-$ for the attractive
Ising chain \eqref{eq:SE13A} only in the limit $\beta \to \infty$. Only in this
(trivial) limit the particle number is a conserved quantity. In contrast, for
the two-dimensional diagonal Ising model of
sect.~\ref{sec:free_particles_in_two_dimensions} the particle number is
conserved, cf. eq.~\eqref{eq:FP13}. This extends to the Weyl-Ising model of
sect.~\ref{sec:complex_structure} where the particle numbers $\Np_1$ and $\Np_2$
for the two species are conserved separately.

\paragraph*{Constrained Ising chains}

A simple way to implement a conserved quantity is to forbid in $\hat{S}$ all
matrix elements that change the value of the conserved quantity. For a diagonal
operator $\hat{A}$ one may have blocks with equal values $A_\tau$ in each
block. A block-diagonal step evolution operator commutes with $\hat{A}$ such
that $A$ is a conserved quantity. For an example with two different values
$A_\tau$, say $A_1 = A_2 = \dots = A_k = a$, $A_{k+1} = A_{k+2} = \dots = A_N =
b$, a block-diagonal $\hat{S}$ has $\hat{S}_{\tau\rho} = 0$ for $\tau \leq k$,
$\rho > k$ and $\tau > k$, $\rho \leq k$. Enforcing that these non-diagonal
elements of $\hat{S}$ vanish guarantees that $A$ is a conserved quantity.

For generalized Ising chains vanishing elements of the step evolution operator
are realized by a term in the action
\begin{equation}\label{eq:CQ6}
\cL_{\text{cons}} (m) = \kappa\, f_\tau(m+1)\, C_{\tau\rho}(m)\, f_\rho (m)\, ,
\end{equation}
with $C_{\tau\rho} > 0$ for $\tau \leq k$, $\rho > k$ and $\tau > k$, $\rho \leq
k$. Taking the limit $\kappa \to \infty$, this term will dominate over all
finite terms in the sector where $C_{\tau\rho} \neq 0$, inducing
\begin{equation}\label{eq:CQ7}
\hat{S}_{\tau\rho} = \exp \big( - \kappa\, C_{\tau\rho} \big) = 0
\end{equation}
for all elements for which $C_{\tau\rho}$ differs from zero. Generalized Ising
chains with a ``constraint term'' \eqref{eq:CQ6} are called ``constrained Ising
models''\,\cite{CWFIM,CWIT}.
The evolution of such models allows only transitions within blocks for which
$A_\tau$ has the same value. The probability distribution vanishes if for two
neighboring states at $m+1$ and $m$ the values of $A$ are different.

In the presence of a conserved quantity the evolution can be followed separately
in each block with given $A_\tau$. This is what we have done for the
two-dimensional diagonal Ising model or the Weyl-Ising model. Since the total
particle number is conserved, the sectors with fixed given particle number $\Np$
can be treated separately. The evolution in a given sector or block only
involves the step evolution operator in this sector or block. Formally the
latter can be defined by the use of appropriate projectors.

More precisely, for the wave functions $\tilde{q}$ and $\bar{q}$ characterizing
the evolution of a pure classical state the eigenfunctions to different values
of $A$ define sectors, one sector for each different value $A_\tau$ in the
spectrum of $A$. For a block-diagonal step evolution operator the evolution in a
given sector is not influenced by the other sectors. The total wave functions
are superpositions of wave functions for different sectors. This extends to the
``block-diagonal part'' of the classical density matrix, consisting of blocks
$\rho'_{\tau\rho}$ for which both indices belong to the same sector. The
evolution of each block does not depend on the other parts of the density
matrix. In this sense the evolution of each block can be discussed as a separate
model. If the initial density matrix is block diagonal, it stays so during the
evolution. In this case the total density matrix is a weighted sum over the
block density matrices, with weights independent of $t$. In general, the total
density matrix also involves elements $\rho'_{\tau\rho}$ for which $\tau$ and
$\rho$ belong to different sectors. For the evolution of those parts the step
evolution operators for both sectors involved are needed.

\paragraph*{Conserved momentum}

The diagonal operators corresponding to local observables that are functions of
occupation numbers are not the only operators that commute with $\hat{S}$. A
prominent example is the momentum operator $\hat{P}$ for the two-dimensional
Weyl-Ising model of sect.~\ref{sec:particles_and_holes} in the one-particle
sector. It is a symmetric operator with real eigenvalues given by functions of
$p$. For finite $L$ and $\varepsilon \neq 0$ the discrete spectrum of $\hat{P}$
is determined by eqs~\eqref{eq:487}, \eqref{eq:488}. If $\hat{P}$ can be
associated to an observable, with possible measurement values given by its
spectrum, it constitutes a conserved quantity similar to the diagonal operators
discussed above.

The commutator relation,
\begin{equation}\label{eq:CQ8}
[\hat{S}, \hat{P}] = 0\,,
\end{equation}
may be demonstrated for four lattice sites, $x/\varepsilon = 0, 1, 2, 3$. The
operator $\varepsilon\, \p_x$ is an antisymmetric $(4\times 4)$-matrix
\begin{equation}\label{eq:CQ9}
\varepsilon\, \p_x = \frac{1}{2} \begin{pmatrix}
0 & 1 & 0 & -1 \\
-1 & 0 & 1 & 0 \\
0 & -1 & 0 & 1 \\
1 & 0 & -1 & 0 
\end{pmatrix}\, ,
\end{equation}
where the elements on the upper right and lower left corner reflect the
periodicity in $x$. The step evolution operator for a single particle species
reads
\begin{equation}\label{eq:CQ10}
\hat{S}_1 = \begin{pmatrix}
0 & 0 & 0 & 1 \\
1 & 0 & 0 & 0 \\
0 & 1 & 0 & 0 \\
0 & 0 & 1 & 0
\end{pmatrix} \, .
\end{equation}

It is easily verified that $\hat{S}_1$ commutes with $\varepsilon\, \p_x$. The
generalization of the structure of these matrices to a larger number of lattice
points is straightforward. The operator~$\p_x$ is antisymmetric. Due to the
antisymmetry of $I$ the momentum operator $\widehat{P}=-I\p_x$ is a symmetric
matrix. For the Weyl-Ising model the operators $\hat{P}$ and $\hat{S}$ are given
in the basis with real wave functions $(q' , q^{c})$ as
\begin{equation}\label{eq:CQ11}
\hat{P} = \begin{pmatrix}
0 & -\p_x \\
 \p_x & 0
\end{pmatrix}\, , \quad
\hat{S} = \begin{pmatrix}
\hat{S}_1 & 0 \\
0 & \hat{S}_1
\end{pmatrix}\, ,
\end{equation}
where the entries are matrices in the space of a single species. It is easy to
verify that eq.~\eqref{eq:CQ8} indeed holds.
In the complex picture $\hat{P}$ is given by $-i\p_x$ and $\hat{S}_1$ remains a
real matrix of the type~\eqref{eq:CQ10}.

\paragraph*{Momentum observable}

In a fermionic quantum field theory momentum is a key observable. It is
conserved by virtue of translation symmetry in space. One may ask what
corresponds to this observable for the diagonal generalized Ising model. The
associated operator $\hat{P}$ does not commute with the position observable. The
position observable is represented by a diagonal operator in the occupation
number basis. As easily visible from eq.~\eqref{eq:CQ9} the momentum observable
has off-diagonal elements. It can therefore not be a time-local observable in
the strict sense, since these observables all correspond to diagonal operators.

We will see that the momentum observable does not take a fixed value in any
state $\omega$ of the overall probability distribution for all times. It is a
new type of observable that measures properties of the probability distribution,
in our case periodicity properties. Its status is somewhat analogous to
temperature, which can well be measured but does not take a fixed value in
anyone of the microstates of a thermal ensemble. Also temperature reflects
properties of the probability distribution. The existence of this type of
``statistical observables'' has an important implication. The classical
correlation function for the pair of observables position and momentum does not
exist. For this reason Bell's inequalities do not apply.

\paragraph*{Measurement of momentum}

For $\hat{P}$ to be associated to an observable quantity we first require that
in an eigenstate to $\hat{P}$ the eigenvalue can be measured. The momentum
eigenstates are the Fourier modes or ``plane waves'' characterized by $p$. For
the discrete derivative the eigenvalues of $\hat{P}$ are related to $p$ by
\begin{equation}\label{eq:CQ12}
\lambda(p) = \frac{\sin(p\varepsilon)}{\varepsilon}\, .
\end{equation}
Indeed, one has
\begin{align}\label{eq:CQ13}
& -\im \p_x \exp(\im p x) = - \frac{\im}{2\varepsilon} \big[ \exp\big( \im p\,
(x+\varepsilon)
\big) - \exp \big( \im p\, (x-\varepsilon ) \big) \big] \notag \\
& \quad = -\frac{\im}{2\varepsilon} \big[ \exp (ip\varepsilon) - \exp(-\im p
\varepsilon)\big]
\, \exp(\im p x) \notag \\
& \quad = \frac{\sin(p\varepsilon)}{\varepsilon} \exp(\im p x) \, .
\end{align}
If $p$ can be measured, $\lambda (p)$ can be measured. In the continuum limit
$|p\varepsilon| \to 0$ one has $\lambda(p) \to p$.


For the eigenstate~\eqref{PHA11} of the momentum operator we can compute the
probability density to find an additional particle at $x$ from eq.~\eqref{PHA8}
\begin{equation}\label{PHA14}
\big( q'_{1}(x)\big)^{2} =\frac{1}{2L}\big[ 1+\sin\big( 2p(x-t)\big)\big]\,.
\end{equation}
The probability density to find at $x$ a hole amounts to 
\begin{equation}\label{PHA15}
\big( q_{1}^{c}(x)\big)^{2} =\frac{1}{2L}\big[ 1-\sin\big( 2p(x-t)\big)\big]\,.
\end{equation}
The probabilities to find an additional particle or hole oscillate, with
\begin{equation}\label{PHA16}
\big( q'_{1}(x)\big)^{2} +\big( q_{1}^{c}(x)\big)^{2}=\frac{1}{L}\,.
\end{equation}
Momentum eigenstates describe oscillations between particles and holes. Using
the different normalization of the wave function for a discrete setting on
replaces $L$ by $M$. As it should be for a one-particle state the total
probability to find either an additional particle or a hole at anyone of the
positions $x$ equals one.

The probability to find a particle or a hole at $x$ is directly related to the
expectation value $\langle s(t,x)\rangle$ of the Ising spin,
\begin{align}\label{PHA17}
\langle s(t,x)\rangle &= \langle s (m_{1}, m_{2})\rangle \notag \\
&= 2\big[ \big( q_{1}'(m_{1},m_{2})\big)^2-\big(
q_{1}^{c}(m_{1},m_{2})\big)^2\big{]}\nonumber\\
&=\frac{2}{M}\sin\big[2p\varepsilon(m_{2}-m_{1})\big]\nonumber\\
&=\frac{2}{M}\sin\big[ 2p (x-t)\big]\,.
\end{align}
If a measurement can measure the expectation values $\langle s (x) \rangle$, it
can also measure the period of the oscillation $\Delta x$, and therefore
determine
\begin{equation}\label{eq:CQ15}
|p| = \frac{\pi}{\Delta x}\, .
\end{equation}
More precisely, a measurement counts the number of $x$-steps $\Delta m_2$
necessary for $\langle s(x) \rangle$ to turn back to the original value, and
thus finds
\begin{equation}\label{eq:CQ16}
|\varepsilon\, p| = \frac{\pi}{\Delta m_2}\, .
\end{equation}
The sign of $p$ can be determined by the direction of motion of the maximum of
$\langle s(t,x)\rangle$ with increasing $t$. We conclude that $p$ and therefore
$\lambda(p)$ is a measurable quantity in the eigenstates of $\hat{P}$.

For particle physics a more suitable operator $\tilde{P}$ has eigenvalues $p$
and can be directly related to the Fourier transform. We continue here to
discuss $\hat{P}$ because of its simple explicit form in position space.

\paragraph*{Expectation value of momentum}

We cannot assign a fixed value of the momentum observable $P$ to a given state
$\tau$ -- the quantity $P_\tau$ does not exist. 
(For a possible realization of the momentum observable as a classical observable
in a different type of ``classical'' probabilistic system see
ref.\,\cite{CWNC}.)
We may, nevertheless, assign an expectation value $\langle P \rangle$ to an
arbitrary local state defined by the classical density matrix $\rho(t)$. This is
done by generalizing the rule \eqref{eq:DM34} to arbitrary symmetric operators
\begin{equation}\label{eq:CQ17}
\langle P(t) \rangle = \tr \big\{ \hat{P}\, \rho(t) \big\} \, .
\end{equation}
For the plane-wave density matrix \eqref{eq:494B} the expectation value equals
the eigenvalue, as it should be for an eigenvector of $\hat{P}$,
\begin{align}\label{CQ18}
\langle P \rangle &= \tr \bigg\{ -\im \p_x \frac{\exp [\im p \,(x-y)]}{L}
\bigg\} \notag \\
&=\frac{\sin (\varepsilon p)}{\varepsilon\, L} \, \tr \big\{ \exp [ \im p\,
(x-y)] 
\big\} \notag \\
&= \frac{\sin (\varepsilon p)}{\varepsilon}\, ,
\end{align}
where we employ $\tr = \int_{x,y} \delta (x-y)$.
Any meaningful definition of an expectation value for an observable with real
possible measurement values $\lambda(p)$ needs $\langle P(t) \rangle$ to be a
real quantity.
This holds since $\hat{P}$ and $\rho$ are hermitean matrices.

We can express the complex density matrix $\rho$ in terms of the real symmetric
classical density matrix $\rho'$ by use of eq.~\eqref{PH4},~\eqref{PH6} for pure
classical states
\begin{equation}\label{PHA18}
\rho_{\tau\rho}'=\begin{pmatrix}
q'_{\tau}q'_{\rho} &\phantom{\Big{|}}& q'_{\tau}q_{\rho}^{c}\\
q_{\tau}^{c}q'_{\rho} &\phantom{\Big{|}}& q_{\tau}^{c}q_{\rho}^{c}
\end{pmatrix}\,.
\end{equation}
In the real picture we replace $\rho$ by $\rho'$ in eq.~\eqref{eq:CQ17}, with
real symmetric operator $\widehat{P}$ of the type~\eqref{eq:CQ11}.
Mixed state density matrices $\rho'$ are given as weighted sums of pure state
density matrices $\rho'^{(\alpha)}$, where the positive weights $w_{\alpha}>0$
are the same as for $\rho$.

We may diagonalize $\hat{P}$ by a similarity transformation
\begin{equation}\label{eq:CQ20}
\hat{P}_d = D\, \hat{P}\, D^{-1} = \text{diag}(\lambda_\tau(p))\, , \quad
\rho'_{d} = 
D\, \rho' D^{-1}\, .
\end{equation}
We numerate the eigenvalues of $\hat{P}$ by $\lambda_\tau(p)$, recalling that
these are not values of $\hat{P}$ in a given basis
state $\tau$.
Similarity transformations leave the trace in eq.~\eqref{eq:CQ17} invariant,
resulting in 
\begin{equation}\label{eq:CQ21}
\langle P(t) \rangle = \sum_\tau \lambda_\tau (p) \big( \rho'_{d}
\big)_{\tau\tau}(t)\, ,
\end{equation}
with sum over all eigenvalues of $\hat{P}$. If all diagonal elements of
$\rho'_{d}$ obey
\begin{equation}\label{eq:CQ22}
\big(\rho'_{d} \big)_{\tau\tau} (t) \geq 0\, ,
\end{equation}
the expectation value admits a classical probabilistic interpretation: the
diagonal elements $(\rho'_{d})_{\tau\tau}(t)$ can be associated with
probabilities $p_\tau^{(p)}(t)$ to find the possible measurement value
$\lambda_\tau (p)$. With eq.~\eqref{PHA18} this is indeed the case for pure
states, and generalizes to mixed states due to the positive weight factors
$w_{\alpha}$.
 Since $\tr(\rho')$ is invariant under similarity transformations the
normalization of the probability distribution $\{p_\tau^{(p)}\}$ is guaranteed,
\begin{equation}\label{eq:CQ23}
\sum_\tau p_\tau^{(p)}(t) = \tr \big( \rho'_{d} (t) \big) = 1\, .
\end{equation}

We can indeed identify the diagonal elements of $\rho'_{d}$ with probabilities
$p_{\tau}^{(p)}$ to find the value
$\lambda_\tau(p)$,
\begin{equation}
\label{NCO1}
p_{\tau}^{(p)}(t) = \left( \rho'_{d} \right)_{\tau \tau}(t)\,,
\end{equation}
and eq.\eqref{eq:CQ21} has a standard probabilistic interpretation
\begin{equation}
\label{NCO2}
\braket{P(t)} = \sum_\tau \lambda_\tau(p) p_{\tau}^{(p)}(t)\,.
\end{equation}

We can further define observables $f(P)$, as $P^2$ or $P^3$. They have the
possible measurement values
\begin{equation}
\label{NCO3}
f_\tau = f(\lambda_\tau(p))\,.
\end{equation}
We recall that $f_\tau$ does not refer to the value of $f(P)$ in one of the
basis states $\tau$ of the overall ensemble. It is
the value in a particular probabilistic state, characterized as an eigenstate of
the operator $\hat{P}$. The probabilistic interpretation \eqref{NCO2} extends to 
\begin{equation}
\label{NCO4}
\braket{f(P)(t)} = \sum_\tau f(\lambda_\tau(p)) p_{\tau}^{(p)}(t)\,.
\end{equation}
Insertion of the relation \eqref{eq:CQ20} yields
\begin{equation}
\label{NCO5}
\braket{f(P)(t)} = \tr \left\{ f(\hat{P}) \rho'(t) \right\}\,,
\end{equation}
where $f(\hat{P})$ is a matrix function or operator function. (For example,
$\hat{P}^2$ is the matrix product of $\hat{P}$ with itself.)

In summary, the operator $\hat{P}$ can be associated to an observable, namely
momentum. It is a conserved quantity of a similar status as the particle
numbers. Momentum is represented by a non-diagonal operator that does not
commute with all diagonal operators. 
Nevertheless, it admits a probabilistic interpretation \eqref{NCO2},
\eqref{NCO5} in a standard way. The only difference to the ``classical
observable'' is that it does not have a well defined or ``sharp'' value in a
given basis state or Ising spin configuration 
$\left\{ s(t,x) \right\}$. It has a sharp value only for probabilistic states
where the probability to find at a given $t$ a
single spin up is distributed over $x$ in a well defined way. This type of
observables measures properties of the 
probability distribution, in our case related to periodicity. It is a type of
``statistical observable'' as discussed in ref.\,\cite{CWQP,CWQPCG}.
The presence of a conserved momentum observable is not immediately visible from
the action of the generalized Ising model. It requires the formulation of the
time-local probabilistic information in terms of a density matrix. We emphasize
the importance of a change of basis by similarity transformations for revealing
the probabilistic properties of the momentum observable. Such basis
transformations are a very useful tool in a formulation with wave functions or a
density matrix. They are not available on the level of the time-local
probability distribution. For our very simple model the presence of a conserved
momentum may seem perhaps somewhat trivial. Conserved quantities remain,
however, central quantities for an understanding of the dynamics even for much
more complex systems. 

\paragraph*{Uncertainty relation}

It is intuitively clear that for a smooth periodic wave function the position of
the particle cannot have a sharp fixed value $y$. For such wave functions the
probability to find a particle is non-zero for many positions $x$. Inversely,
for a particle with sharp position, $\psi(x)\sim\delta(x-y)$, there cannot be
any well defined periodicity of the wave function (except the trivial one due to
the periodic boundary condition). Quantitatively, these properties are reflected
by the Heisenberg uncertainty relation.

From the non-vanishing commutator in the continuum limit,
\bel{UR1}
\left[\hat X,\hat P\right]=i\ ,
\ee
one concludes in the standard way the uncertainty relation
\bel{UR2}
\sigma_X\sigma_P\geq\frac12\ .
\ee
Here
\bel{UR3}
\sigma_X^2=\langle X^2\rangle-\langle X\rangle^2\ ,\quad \sigma_P^2=\langle
P^2\rangle-\langle P\rangle^2
\ee
are well defined in terms of the squared operators,
\bel{UR4}
\langle X^2\rangle=\langle\psi|\hat X^2|\psi\rangle\ ,\quad \langle
P^2\rangle=\langle\psi|\hat P^2|\psi\rangle\ .
\ee
For $\sigma_P^2$ this follows from the probabilistic interpretation of the
momentum observable with positive probabilities $p_\tau^{(p)}$. The uncertainty
relation~\eqref{UR2} is not restricted to quantum systems. It holds exactly if
the condition~\eqref{eq:CQ22} is obeyed. If not, one expects similar, perhaps
somewhat weaker relations forbidding simultaneous sharp values of $X$ and $P$.

No observer is implied for the validity of the uncertainty relation. This
relation results from the possibility to determine expectation values of an
extended set of observables from the time-local probabilistic information
contained in the density matrix. On the other hand, this probabilistic
information is insufficient to define/determine simultaneous values for $X$ and
$P$. We discuss this issue in more detail in
sect.~\ref{sec:observables_and_operators}.

\paragraph*{Non-diagonal operators for local observables}

The momentum operator $\hat{P}$ for the one-particle state of the
two-dimensional Weyl-Ising model is a first example for a local observable that
is not represented by a diagonal operator in the occupation number basis. We
will encounter in sect.~\ref{sec:time_local_subsystems} and
\ref{sec:local_observables_and_non_commuting_operators} further examples of
local observables for which the associated operators are not diagonal. This will
include time derivatives of local observables that can be expressed as functions
of occupation numbers. 
Since the expectation values of observables corresponding to non-diagonal
operators $\widehat{A}(t)$ can still be computed from the time-local
probabilistic information in the density matrix $\rho'(t)$ one needs a
generalization of the concept of local observables. In contrast to the narrow
definition of classical time-local observables as functions of the occupation
numbers $n_{\gamma}(t)$ at a given time $t$ we extend the concept of time-local
observables to suitable statistical observables. We briefly sketch here the
notion of generalized local observables of ref.~\cite{CWPT,CWIT}.

We postulate that generalized local observables $A(t)$ have an associated
symmetric operator $\hat{A} (t) = \hat{A}^\tp (t)$. The possible measurement
values are the eigenvalues of the operator $\hat{A}(t)$, and the expectation
value obeys
\begin{equation}\label{eq:CQ28}
\langle A(t) \rangle = \tr \big\{ \hat{A}(t)\,\rho' (t) \big\}\, .
\end{equation}
We use eq.~\eqref{eq:CQ28} now as a general definition of the expectation value
of generalized time-local observables that do not necessarily have fixed values
for the configurations of Ising spins at $t$. There is no longer a restriction
to one-particle states or a particular complex structure.
Similar to the expectation value \eqref{I1} in classical statistics the
expectation value \eqref{eq:CQ28} is first of all a definition. One has
to investigate the consistency of this definition and its consequences. We want
to know under which conditions this definition of the
expectation value is compatible with standard probabilistic properties of
observables.

We will encounter many examples where $A$ is an observable that can be
constructed from the Ising spins at different $t$. Its possible measurement
values are then determined by the values in the overall state given by spin
configurations for all times, and the expectation value is given by the
classical statistical rule in terms of the overall probability distribution. If
expectation values of such observables can also be computed from the local
probabilistic information contained in $\rho'(t)$, one can show that there
exists an associated symmetric operator $\hat{A}(t)$ and eq.~\eqref{eq:CQ28}
holds for the expectation value. For all these cases the rule \eqref{eq:CQ28}
indeed follows from the three basic axioms for classical statistical systems. We
are interested here in the question if eq.~\eqref{eq:CQ28} has to hold on
general grounds.

We first have to define the notion of generalized local
observables\,\cite{CWIT,CWPT}.
For a generalized local observable $A(t)$ the expectation value $\langle A(t)
\rangle$ can be determined from the local probabilistic information contained in
$\rho'(t)$. Thus $\langle A(t) \rangle$ has to be some function of the elements
$\rho'_{\tau\rho} (t)$ of the classical density matrix at time $t$, denoted by
$\langle A \rangle (\rho')$. We restrict the notion of generalized local
observables to those that have at most $N$ different possible measurement values
that we denote by $\mu_\tau(t)$, $\tau = 1, \dots, N$, where some values
$\mu_\tau (t)$ may be degenerate such that the spectrum of distinct possible
measurement values may contain less than $N$ values. Here $N$ is given by the
number of local states such that $\rho'(t)$ is an $(N\times N)$-matrix. For a
probabilistic setting the expectation value has to obey
\begin{equation}\label{eq:CQ29}
\langle A (t) \rangle = \sum_\tau p_\tau^{(A)} (t)\, \mu_\tau (t)\, .
\end{equation}
For non-degenerate measurement values the probabilities $p_\tau^{(A)}(t)$ to
find $\mu_\tau (t)$ have to obey
\begin{equation}\label{eq:CQ30}
p_\tau^{(A)}(t) \geq 0\, , \quad \sum_\tau p_\tau^{(A)} (t) = 1\, .
\end{equation}
For degenerate $\mu_\tau(t)$ only the sum of $p_\tau^{(A)}$ contributing to a
given $\mu_\tau(t)$ has to be positive. The probabilities $p_\tau^{(A)}$ must be
functions of $\rho'$. The functions $p_\tau^{(A)}(\rho')$ determine $\langle A
\rangle (\rho')$. 

We further require that there are local states for which the probability to find
a given $\mu_\tau (t)$ equals one. This implies that there should exist
particular density matrices $\rho'(t;\, \mu_\tau)$ for which $\langle A
\rangle(\rho'(\mu_\tau)) = \mu_\tau$. These features defining the notion of a
generalized local observable have to hold in an arbitrary basis. The
probabilities $p_\tau^{(A)}$ should be independent of the choice of basis. In
turn, the functions $p_\tau^{(A)}(\rho')$ will typically depend on the basis. It
is clear that observables $A$ that admit an associated symmetric operator
$\widehat{A}$ with spectrum $\lbrace\mu_{\tau}\rbrace$ obey all these
requirements if $\rho'$ is a symmetric positive matrix. The probabilities
$p_{\tau}^{(A)}(\rho')$ are the diagonal elements $\rho'^{(A)}_{\tau\tau}$ in a
basis where $\widehat{A}$ is diagonal. In this case eq.~\eqref{eq:CQ28} holds.
We have not investigated if there exists other possibilities to meet our
criteria for generalized local observables.


\paragraph*{Angular momentum}

As another example for a conserved quantity we may discuss angular momentum. For
rotation symmetry in a plane the angular momentum perpendicular to the plane is
expected to be a conserved quantity. In its simplest form such a rotation can be
represented as a shift of an angle $\alpha$, with periodicity identifying
$\alpha + 2\pi$ and $\alpha$. 
As a very simple example we implement the angular momentum operator in a simple
generalization of the clock system in sect.\,\ref{sec:clock_systems}. A simple
approximation to our model of Weyl fermions and possible generalizations will be
discussed subsequently.

For our simplest example we add at every position of the chain in
sect.~\ref{sec:clock_systems} a further single Ising spin $s_c(m)$. 
Together with
the $M$ Ising spins used to realize the $N=2^M$ positions of the pointer
there are now $M+1$ Ising spins at each $m$. 
The number of local states is now $2^{M+1}=2N$. We denote these states by
$(\tau,+)$ if $s_\mathrm{c} = +1$, and $(\tau,-)$ for $s_\mathrm{c}=-1$. The
additional spin induces a complex structure in a simple way. With
\begin{equation}
\tilde{q}_{(\tau,\sigma)} = 
	\begin{pmatrix} 
	\tilde{q}_{(\tau,+)} \\ \tilde{q}_{(\tau,-)}	
	\end{pmatrix}
	=
	\begin{pmatrix}
	\tilde{q}_{\mathrm{R},\tau} \\ \tilde{q}_{\mathrm{I},\tau}
	\end{pmatrix}
\label{eq:AM1}
\end{equation}
we define the complex $N$-component wave function by
\begin{equation}
\psi_\tau = \tilde{q}_{\mathrm{R},\tau} + i \tilde{q}_{\mathrm{I},\tau}.
\label{eq:AM2}
\end{equation}
The evolution equation \eqref{eq:CS8}, \eqref{eq:CS15} acts in the same way on
$\tilde{q}_\mathrm{R}$ and $\tilde{q}_\mathrm{I}$. Multiplication with the
matrix $I$ yields
\begin{equation}
I \partial_t \tilde{q} = I\omega \partial_\alpha \tilde{q}.
\label{eq:AM3}
\end{equation}
In the complex language this is the Schrödinger equation for the complex clock
system
\begin{equation}
i\partial_t \psi = H\psi,\quad H=i\omega \partial_\alpha
\label{eq:AM4}
\end{equation}

We define the angular momentum operator
\begin{equation}
\hat{L} = -i \partial_\alpha,
\label{eq:AM5}
\end{equation}
in close analogy to the momentum operator $\hat{P}$ in eq.\,\eqref{eq:CQ11}.
Similar to momentum, $\partial_\alpha$ is antisymmetric such that
$I\partial_\alpha$ is a symmetric operator. The Hamilton operator is
proportional to angular momentum
\begin{equation}
H = -\omega \hat{L}.
\label{eq:AM6}
\end{equation}
This implies $[\hat{L}, \hat{H}] = 0$ and angular momentum is a conserved
observable. Eigenstates of angular momentum are given by
\begin{equation}
\psi = e^{il\alpha},\quad \hat{L}\psi = l\psi,
\label{eq:AM7}
\end{equation}
with integer $l$ the possible momentum values for the angular momentum
observable.

The measurement procedure counts again the number of oscillations between
$\psi_\mathrm{R}$ and $\psi_\mathrm{I}$ during one period $2\pi$ for the pointer
position $\alpha$. Indeed, the wave function \eqref{eq:AM7} corresponds to
\begin{equation}
\tilde{q}_\mathrm{R}(\alpha) = \cos(l\alpha),\quad \tilde{q}_\mathrm{I}(\alpha)
= \sin(l\alpha),
\label{eq:AM8}
\end{equation}
such that for given $\tau$ or associated $\alpha$ one has
\begin{equation}
\tilde{q}_{(\tau,+)}^2 + \tilde{q}_{(\tau,-)}^2 = 1,\quad \psi^*(\alpha)
\psi(\alpha) = 1.
\label{eq:AM9}
\end{equation}
The rotation of the pointer with $\alpha$ for a fixed type $R$ or $I$ translates
to an integer number of rotations in the $(R,I)$-plane. We observe that
$\tilde{q}$ and $-\tilde{q}$ yield the same classical density matrix $\rho'$. On
the level of the density matrix the possible measurement values of angular
momentum $l$ can also be half-integer. This accounts for a well known property
of spin in quantum mechanics, namely that angular momentum is quantized with
half-integer values. (In our convention one has $\hbar=1$.) While the operator
$\partial_{\alpha}$ exists for the clock in sect.~\ref{sec:clock_systems}, the
presence of the conserved angular momentum observable involves the complex
structure. The factor $i$ or the matrix $I$ is needed for the construction of a
symmetric operator that can correspond to an observable.

For the two-dimensional diagonal Ising model there is no symmetry of space
rotations. Nevertheless, one finds a certain analogy with the rotation symmetry
of the clock system. We have seen that the 
one-particle states for the diagonal two-dimensional Ising model, doted with a
complex structure for the description of Weyl fermions, are an example for a
complex clock system. For periodic boundary conditions the variable $m_2$ is
analogous to the discrete angle $\alpha$. Up to a normalization factor $\hat{P}$
can be identified with $\hat{L}$ in this case.

\paragraph*{Symmetry generators}

With momentum and angular momentum $P$ and $L$ we have encountered two
observables directly related to symmetries. The associated local-observable
operators $\hat{P}$ and $\hat{L}$ are the symmetry generators for translations
in position or angular position. For any system realizing rotations in three
dimensions there will be three independent generators $\hat{L}_k$ of angular
momentum in three Cartesian directions $k$. The generators of the rotation group
$SO(3)$ do not commute
\begin{equation}
[\hat{L}_i, \hat{L}_j] = i \varepsilon_{ijk} \hat{L}_k,
\label{eq:AM10}
\end{equation}
with totally antisymmetric tensor $\varepsilon_{ijk}$. We can associate a local
observable $L_k$ to every $\hat{L}_k$ in complete analogy to the rotations in a
plane. The non-commuting properties of rotations around different axes require
that the operators associated to $L_k$ cannot commute and obey
eq.\,\eqref{eq:AM10}. 

In conclusion, classical statistical systems as generalized Ising models admit
observables that are associated to the generators of symmetry transformations.
These observables are not classical statistical observables with a fixed value
in every state of the overall probabilistic ensemble. They rather measure
properties of local probability distributions or density matrices. These
probabilistic observables can have sharp values in certain probabilistic states,
but not in all states. In the local-time subsystem they are represented by
local-observable operators that often do not commute.

Observables associated to symmetry generators are familiar in quantum mechanics.
The observation that such observables also exist in classical statistical
systems points to new features. In particular, classical correlation functions
for these probabilistic observables are typically not defined, and therefore
cannot be used for measurements. We will encounter new forms of correlation
functions in later parts of this work. We should emphasize that the
probabilistic observables $P$ or $L_k$ are not the classical observables in
systems which contain variables for both position and momentum. We have found
these observables without modifying or extending the variables or introducing
any ``hidden variables".

\subsection[Reference frames and Lorentz symmetry]{Reference frames and Lorentz\\symmetry}
\label{sec:reference_frames_and_lorentz_symmetry}

Since Einstein's theory of special relativity the modern view considers time as a relative quantity. If physical time is defined by a system of clocks one has to specify the relations between these clocks. Within general relativity physical time should be invariant under general coordinate transformations. It should also be independent of the choice of the metric field or the ``metric frame"~\cite{Rubakov:2022fqk,Wetterich:2014zta}. Our setting of probabilistic time shows all these features in a very natural way. 

We have already discussed in sect.~\ref{sec:physical_time} the direct connection between different reference frames and the presence of different possible time structures. In flat space the different reference frames of special relativity are related by Lorentz transformations. We will show here that this happens precisely for the continuum limit of our simple model for free massless fermions. A frame-invariant concept of physical time~\cite{Rubakov:2022fqk, Wetterich:2014zta} employs directly the periodic evolution of clocks, since the number of oscillations or ticks does not depend on the choice of the metric. This coincides with our approach to physical time.

\paragraph*{Lorentz symmetry}

The continuum limit of massless free fermions exhibits Lorentz symmetry. This symmetry is not directly visible for the generalized Ising model of sect.~\ref{sec:step_evolution_operator}. It is revealed by the Schrödinger equation~\eqref{3.4.139A} for the continuum limit of one-particle state in the complex formulation,
\begin{equation}\label{PHA19}
i(\partial_{t}+\partial_{x})\psi_{R}=0\,.
\end{equation}
This can be extended to a Dirac fermion by adding a left-mover with one-particle wave function $\psi_L$. For the two-component complex wave function $\psi$ one has
\begin{equation}
\label{PHA20}
\psi=\begin{pmatrix}
\psi_{R}\\ \psi_{L}
\end{pmatrix}\;,\quad (\partial_{t}+\tau_{3}\partial_{x})\psi=0\,.
\end{equation}
Using Dirac matrices $\gamma^{\mu}=(\gamma^{0},\gamma^{1})$,
\begin{equation}\label{PHA21}
\gamma^{0}=-i\tau_{2}\;,\quad \gamma^{1}=\tau_{1}\;,\quad \gamma^{0}\gamma^{1}=-\tau_{3}\,,
\end{equation}
eq.~\eqref{PHA20} can be written as a familiar Lorentz-covariant Dirac equation ($\partial_{0}=\partial_{t}$, $\partial_{1}=\partial_{x}$)
\begin{equation}
\label{PHA22}
\gamma^{\mu}\partial_{\mu}\psi=0\,.
\end{equation}

There exists a general bit-fermion map which maps the ``functional integral" or configuration sum of generalized Ising models to a Grassmann functional integral~\cite{CWFIM}. The model~\eqref{eq:FP1} is mapped to a discretization of a fermion model for free massless Weyl fermions, with possible extension to Dirac fermions. In the continuum limit the Lorentz-symmetry of this quantum field theory is manifest as a symmetry of the action for the Grassmann functional integral. While we postpone the discussion of the bit-fermion map to later parts of this work we may already investigate certain consequences of Lorentz-symmetry at the present level of the discussion.

\paragraph*{Reference frames}

In our discussion of physical time we have already pointed out the possibility of different reference frames by choosing different clock systems, see Fig.~\ref{figure:PT1}. In the presence of Lorentz symmetry one expects that different inertial systems are related by Lorentz transformations. We will demonstrate that this is indeed the case for our generalized Ising models formulated on a two-dimensional square lattice. This reveals the deep rooting of the relativistic concept of reference frames in our basic formulation of relativistic time.

Let us consider the continuum limit of the diagonal two-dimensional Ising model in sect.\,\ref{sec:free_particles_in_two_dimensions}. We can choose different hypersurfaces in the $x$-$t$-plane in order to define different time structures. For example, a family of hypersurfaces can be parametrized by $t'$ according to
\begin{equation}\label{PT4}
t=\frac{t'}{\cosh{\beta}}+ x\tanh\beta ,
\end{equation}
as shown in fig.\,\ref{figure:PT1}. For any given $\beta$ we can use these hypersurfaces to define a separate time structure.

The order of the observables is now given by $t'$.
Observables at the same $t'$ belong to the same equivalence class, and $t'_{1}>t'_{2}$ defines the notion that $A(t'_{2})$ is before $A(t'_{1})$. Two different $\beta$ define two different time structures. For two different time structures the ordering relations differ.
We have indicated in fig.~\ref{figure:PT1} the location of two observables $A_{1}(t_{1}, \vec{x_{1}})$ and $A_{2}(t_{2}, \vec{x_{2}})$, for which $A_{2}$ is after $A_{1}$ for the time structure labeled by $t$, and before $A_{1}$ for the time structure labeled by $t'$.

Instead of a space structure given by $x$ we can also define a different space structure according to hypersurfaces labeled by $x'$, with
\begin{equation}\label{PT5}
x=\frac{x'}{\cosh\beta}+ t\tanh\beta .
\end{equation}
This space structure is also represented in fig.\,\ref{figure:PT1}.
For the relations between $(t, x)$ and $(t', x')$ we observe the identity
\begin{equation}\label{PT6}
(t-t_{0})^{2}-(x-x_{0})^{2}=(t'-t'_{0})^{2}-(x'-x'_{0})^{2},
\end{equation}
where
\begin{align}\label{PT7}
t'_{0}=\cosh\beta \;t_{0}-\sinh\beta \;x_{0}
\nonumber\\
x'_{0}=\cosh\beta \;x_{0}-\sinh\beta \;t_{0}.
\end{align}
We recognize that the two space-time structures are related by a Lorentz transformation
\begin{equation}\label{PT8}
t'=\gamma(t-vx), \quad x'=\gamma(x-vt),
\end{equation}
where
\begin{align}\label{PT9}
\gamma=\dfrac{1}{\sqrt{1-v^2}}=\cosh\beta ,
\nonumber\\
v\gamma=\sinh\beta ,
\end{align}
and $v$ is the relative velocity between two inertial systems.

This demonstrates in a simple way that the time variables for different inertial systems correspond to different time structures rather than to different clocks of a given time structure. This explains why time dilatation between inertial systems is relative. If we take the clock system associated to an inertial system moving with $\vec{v_1}$, and compare it with clocks in an inertial system with velocity $\vec{v_2}$, the clocks in the inertial system with $\vec{v_2}$ tick slower, according to eq.\,\eqref{PT1} with $\vec{v}_i$ replaced by $\vec{v}_i'$, where $\vec{v}_1'=0 , \vec{v}_2'=\vec{v}_2-\vec{v}_1$. In contrast, in the clock systems of the inertial system with velocity $\vec{v}_2$ the clocks of the system with $\vec{v}_1$ tick slower, now following eq.\,\eqref{PT1} with $\vec{v}_1'=\vec{v}_1-\vec{v}_2 , \vec{v}_2'=0$.

There is no contradiction since the two cases use different ordering structures for observables corresponding to distinct clock systems. In this language eq.\,\eqref{PT1} corresponds still to another time structure, namely the one ``at rest'' given by $(t,x)$, for which both $\vec{v}_1$ and $\vec{v}_2$ differ from zero. We conclude that the statements which standard clocks tick slower or faster depends on the time structure that is chosen. Here we understand by standard clocks that the same physical process as oscillations of photons for a given atomic transition line is used in all inertial systems. In a formulation of probabilistic time Einstein's special relativity is rooted in the choice of different time structures.

The non-diagonal two-dimensional Ising model of sect.\,\ref{sec:free_particles_in_two_dimensions}
shows features of Lorentz symmetry. The trajectories of ``particles" with $x=x_{0}+t$ are the same in a different inertial frame, $x'=x_{0}'+t'$. This holds similarly for trajectories of left-movers $x=x_{0}-t$. 
This simple probabilistic system also admits the possibility to define different time structures that are related by Lorentz transformations.
The model actually describes a system that is invariant under Lorentz transformations. This is most easily seen in an equivalent fermionic formulation based on Grassmann variables\,\cite{CWFIM}.
In this formulation the continuous Lorentz transformations are realized as explicit symmetry transformations. They require the continuum limit. We may also realize the presence of Lorentz symmetry for the continuum limit of the expression of the step evolution operator in terms of fermionic creation and annihilation operators which will be discussed in chapter~\ref{sec:quantum_field_theory}.

\paragraph*{Diffeomorphisms and frame changes}
In sect.\,\ref{sec:probabilistic_description_of_nature} we have briefly discussed space-time structures where observables are labeled by space-time points ~$x=(t,\vec{x})$. We have argued that the choice of coordinates should not matter for physical observables and a good choice of the overall probability distribution should be invariant under general coordinate transformations or diffeomorphisms
\begin{align}
\begin{split}
x\rightarrow x'&=f(x),\\
t\rightarrow t'=f^0 (t, \vec{x}),&\quad  x^{k}\rightarrow x'^{k}=f^{k}(t,\vec{x}).
\end{split}
\label{PT10}
\end{align}
If $t$ and $t'$ label different families of hypersurfaces in $\mathbb{R}^d$ we can build different clock systems based on $t$ or $t'$ as an ordering variable. The clock systems related by diffeomorphisms correspond in this case to different time structures.

Invariance under general coordinate transformations is often realized by the introduction of some metric field. This metric field may be a composite of more fundamental fields, as the fermions in spinor gravity~\cite{Wetterich:2011yf,Hebecker:2003iw,Wetterich:2003wr}. Even more generally we may infer a metric field from the two-point correlation function of an appropriate observables~\cite{CWGEO}. From the metric one can derive geodesics and quantities as proper time along a given geodesic. In ref.~\cite{Rubakov:2022fqk} it has been shown how a frame invariant version of geodesics can be inferred from the oscillatory behavior of suitable clocks.

It is remarkable how the basic concepts of special and general relativity are already present in the formulation of probabilistic time as an ordering structure among observables.

\section[Probabilistic and deterministic evolution]{Probabilistic
and\\deterministic\\evolution}
\label{sec:probabilistic_and_deterministic_evolution}

Different types of evolution are distinguished by the way how the probabilistic
information propagates. This is reflected by the question how boundary
properties influence the behavior in the bulk. One prototype are probabilistic
systems with a finite correlation length, as the one-dimensional Ising model
with finite $\beta$. For a location inside the bulk, with distance many
correlation lengths away from the boundary, the local probabilistic system is
characterized by a unique equilibrium state. The details of the boundary
condition, or the ``boundary information'', are eventually lost. Another
prototype are oscillating systems, for which the boundary information is
transported inside the bulk without loss. The properties far inside the bulk
respond to a change of boundary conditions.

Orthogonal step evolution operators preserve the boundary information
completely. They are discussed in
sect.\,\ref{sec:orthogonal_and_unitary_step_evolution_operators}. In
sect.\,\ref{sec:probabilistic_celluar_automata} we turn to unique jump step
evolution operators which describe probabilistic automata. For such systems the
evolution is deterministic. Each given initial state at the boundary is
transported according to the deterministic rule of the cellular automaton inside
the bulk. The probabilistic aspects of this type of system arise uniquely from
the probability distribution of initial conditions. Cellular automata are
perhaps the simplest systems for which information is transported without loss.
They therefore play an important role. In
sect.~\ref{sec:Fermionic_quantum_field_theory_with_interactions} we present an
automaton whose dynamics describes a two-dimensional interacting quantum field
theory for fermions, namely a particular type of discretized Thirring model.

As an example for a probabilistic evolution we discuss static memory materials
in sect.\,\ref{sec:static_memory_materials}. For this model the
evolution is in space rather than time. 
Nevertheless, if the correlation length exceeds the separation between the
boundaries some features of an (approximately) deterministic evolution become
visible.
Interference of probabilistic boundary information from two sides shows some
similarity to interference in quantum mechanics. Such systems may offer
interesting possibilities for memory storage and computing. In parameter space
this system has a limit where it becomes a cellular automaton. It is well suited
for an understanding of connections and transitions between probabilistic and
deterministic evolution.

In sect.\,\ref{sec:partial_loss_of_memory_and_emergence_of_quantum_mechanics} we
discuss systems with a partial loss of memory of boundary information more
systematically. We argue that for an overall probability distribution with
boundaries in the infinite past and future the time-local subsystem for
``present time" in the bulk obeys a quantum evolution. Unless this evolution is
trivial, as for equilibrium states, our setting predicts that quantum mechanics
is an exact description of the world.

Markov chains are often discussed in the context of evolution of probability
distributions. For Markov chains, only the local probability distribution
$\{p_\tau(t)\}$ is needed for the formulation of an evolution law. If this
evolution law involves positive transition probabilities, the state far inside
the bulk approaches an unique equilibrium state, except for the limiting case of
cellular automata, or cellular automata for subsystems. We show in
sect.\,\ref{sec:markov_chains} that the generic evolution of the probabilistic
information is not given by Markov chains. 
A simple evolution law requires the time-local probabilistic information stored
in the classical density matrix, which exceeds the time-local probability
distribution.
We discuss how Markov chains can arise as approximations.

Even for a system with deterministic evolution as cellular automata the
evolution of subsystems is typically probabilistic. The ``coarse graining" to a
subsystem looses part of the probabilistic information which would allow a
deterministic evolution. A given state of the subsystem can evolve with certain
probabilities to different states of the subsystem. If the coarse graining to
the subsystem is compatible with the preservation of part of the boundary
information the subsystem shows typically a type of probabilistic information
whose evolution is not described by a Markov chain. It rather exhibits features
familiar from quantum mechanics. We postpone a more general discussion of
subsystems to chapter~\ref{sec:subsystems}.

\subsection[Orthogonal and unitary step evolution operators]{Orthogonal and unitary step\\evolution operators}
\label{sec:orthogonal_and_unitary_step_evolution_operators}

The clock systems encountered so far have unique jump step evolution operators. Unique jump operators are orthogonal matrices. For orthogonal $\hat{S}$ the bounary information is preserved. All eigenvalues of $\hat{S}$ obey 
$| \lambda_i | = 1$. We have
seen the particular role of orthogonal step evolution operators $\hat{S}$ in the evolution of the classical wave functions. For orthogonal
$\hat{S}$ the wave function and the conjugate wave function follow the same evolution law. For suitable initial conditions they
can be identified, such that one deals with a single classical wave function $q(t)$. In the presence of a complex structure this associates
the conjugate wave function $\bar{\psi}$ to the complex conjugate of the wave function $\psi$, $\bar{\psi} = \psi^*$, similar to 
quantum mechanics.
An orthogonal step evolution operator which is compatible with a complex structure is represented as a unitary matrix in the complex picture.

An orthogonal evolution is closely related to the oscillatory behavior that we require for physical time. In general, it will be 
sufficient that a subsector follows an orthogonal evolution. We may, however, first investigate the condition for $\hat{S}$ to be itself
an orthogonal matrix. For generalized Ising models with finite $M$ this restricts the possibilities to unique jump chains.

\paragraph*{Non-negative orthogonal matrices}

Let $\hat{S}$ be an orthogonal $N\times N$ matrix
\begin{equation}
\label{OS1}
\hat{S}^T \hat{S} = 1\,.
\end{equation}
Furthermore, we require that $\hat{S}$ is a non-negative matrix for which all elements are positive semidefinite
\begin{equation}
\label{OS2}
\hat{S}_{\tau \rho} \geq 0\,.
\end{equation}
This is appropriate for generalized Ising models. We want to establish that  eqs.~\eqref{OS1} and~\eqref{OS2} imply that $\hat{S}$ is a unique jump operator.

The orthogonality condition,
\begin{equation}
\label{OS3}
\sum_\sigma \hat{S}_{\tau \sigma} \hat{S}_{\rho \sigma} = \hat{S}_{\sigma \tau} \hat{S}_{\sigma \rho} = \delta_{\tau \rho}\,,
\end{equation}
requires for $\tau \neq \rho$ that all terms in the sum vanish, such that for each $\sigma$ (no sum here)
\begin{equation}
\label{OS4}
\hat{S}_{\tau \sigma} \hat{S}_{\rho \sigma} = \hat{S}_{\sigma \tau} \hat{S}_{\sigma \rho} = 0\,, \quad \text{for } \tau \neq \rho\,.
\end{equation}
This follows from the positivity condition \eqref{OS2} since a vanishing sum of positive terms requires each term to vanish. Furthermore,
$\hat{S}$ is invertible such that each row and each column has at least one nonzero element. We show below that only a single
element in each row and each column can differ from zero. Only this element contributes in the sum \eqref{OS3} for $\tau = \rho$.
It can therefore only take the value one. Thus $\hat{S}$ is a unique jump operator with precisely one element equal to one in
each row and column.

The proof that only a single element in each row and column can differ from zero proceeds by contradiction. Assume that in the row
$\sigma$ there are two nonzero elements, $\hat{S}_{\sigma \tau} > 0$, $\hat{S}_{\sigma \rho} > 0$, $\tau \neq \rho$.
This implies $\hat{S}_{\sigma \tau} \hat{S}_{\sigma \rho} > 0$ and contradicts the second equation \eqref{OS4}. Similarly, two 
nonzero elements in the column $\sigma$, $\hat{S}_{\tau \sigma} > 0$, $\hat{S}_{\rho \sigma} > 0$, $\tau \neq \rho$ imply
$\hat{S}_{\tau \sigma} \hat{S}_{\rho \sigma} > 0$, contradicting the first equation \eqref{OS4}.
We conclude that the only non-negative orthogonal step evolution operators are unique jump operators. 

\paragraph*{General orthogonal step evolution operators}

All orthogonal step evolution operators that are not unique jump operators have some negative elements. In particular, infinitesimal
rotations are not described by unique jump operators.  They indeed involve negative matrix elements. One may generate negative elements
of $\hat{S}$ by a change of basis. In this case $\hat{S}$ remains a unique jump operator only in the basis where it is non-negative, as the
occupation number basis for generalize Ising models. This may not remain easily visible in a different basis. 

For subsystems of generalized Ising models the step evolution operator does not need to remain a non-negative matrix. We will discuss
several examples of this type in sect.~\ref{sec:subsystems}. We will find subsystems with orthogonal step 
evolution operators different from unique jump operators. This includes infinitesimal rotations. 

Finally, we observe that positive step evolution operators are not a necessary requirement for defining a positive overall probability
distribution. We will discuss this issue in the appendix~\ref{app:positivity_of_overall_probability_distribution}.

\paragraph*{Unitary evolution operators}

The unitary transformations $U(N/2)$ are a subgroup of the rotations $SO(N)$. They are realized if a complex structure is compatible
with an orthogonal step evolution operator. This requires that $\hat{S}$ takes the form \eqref{eq:CC9}, while $\hat{S}^T \hat{S} = 1$.
A unitary evolution is one of the important characteristics of quantum mechanics. Pure classical states and unitary $\hat{S}$ behave
indeed as pure quantum systems. For subsystems we have to require, in addition, the positivity of the density matrix. 
\subsection{Probabilistic cellular automata}
\label{sec:probabilistic_celluar_automata}


Cellular automata, introduced by von Neumann~\cite{NEU}, Ulam~\cite{ULA} and Zuse~\cite{ZUS}, have found wide applications in many areas of sciences~\cite{GAR, LIRO, TOO, TDK, WOLZ, VIC, PREDU, TOMAR, LOU, HED, RIC}. The states of the system at a given time $t$ can be characterized by a configuration of bits. At the next time step from $t$ to $t+\epsilon$ each configuration is updated to precisely one new configuration. This process is deterministic, similar to a calculation step of a classical computer. For cellular automata the bits at any given time are organized in cells, that we denote by $x$ for our purpose. The updating of each cell $x$ is only influenced by the bits of a few neighboring cells at the previous time layer. In turn, each cell influences for the next time step only a few neighboring cells. We focus here on reversible or invertible automata for which the inverse of the updating map exists~\cite{TOMAR, AMPA, HPP, CRE}.

For deterministic cellular automata one starts at some initial time $t_{\text{in}}$ with precisely one sharp bit configuration. We consider here probabilistic cellular automata which are characterized by a probability distribution over initial bit configurations. In our setting the updating remains deterministic, the probabilistic aspects appearing only through the initial probability distribution. This setting should be distinguished from automata with probabilistic updating steps which are sometimes referred to with the same name.

The diagonal generalized Ising model in sect.~\ref{sec:free_particles_in_two_dimensions} is a probabilistic cellular automaton. Each cell $x$ has exactly one bit $n(x)$. Its updating is only influenced by the state of the cell $x-\epsilon$, namely if $n(x-\epsilon)$ equals zero or one. The updating is a simple copy of the state of the preceding cell. In turn, the state of the cell $x$ influences in the next time step only the cell $x+\epsilon$. Extending the model to a combination of right- and left-movers appropriate for a free massless Dirac fermion the past and future neighbors of the cell $x$ in the sense of the cellular automaton are the geometric neighbors $x-\epsilon$ and $x+\epsilon$. This automaton property induces directly a causal structure as familiar in quantum field theories. In the next section we extend this discussion to a two-dimensional generalized Ising model that describes a two-dimensional quantum field theory for Dirac fermions with interaction. In the present section we address a general formalism for probabilistic automata. It applies, in particular, to cellular automata.

For unique jump chains the evolution is the same as for automata. For a sharp initial state (and associated final state), where
the probability $p_\tau(t_\text{in})$ equals one for one particular state $\tau_0$, the unique jump chain describes indeed the deterministic 
evolution of an automaton. Each configuration $\rho$ is updated to precisely one new configuration $\tau$. The updating procedure is encoded in the invertible map $\tau(\rho)$. For more general boundary conditions the probabilistic system amounts to a probability distribution
over different initial states.
In this case we deal with probabilistic automata.

Probabilistic automata are discrete pure quantum systems if the boundary conditions factorize and are chosen such that the conjugate classical wave function coincides with the classical wave function, $\bar{q}(t) = \tilde{q}(t) = q(t)$. In the presence of a complex structure one finds the usual complex quantum mechanics. The evolution is linear and unitary, with unitary step evolution operator acting on complex wave functions that form a Hilbert space. Observables are represented by symmetric operators, which correspond to hermitean operators in the complex formulation. This generalizes to mixed state boundary conditions. If the boundary conditions are a weighted sum (with positive weights $w_\alpha$) of pure state boundary conditions with $\bar{q}^{(\alpha)}(t) = \tilde{q}^{(\alpha)}(t) = q^{(\alpha)}(t)$, the (classical) density matrix is symmetric and has only positive eigenvalues. In the complex formulation the density matrix is a positive hermitean matrix, as required for quantum mechanics. If a continuum limit exists, one obtains the usual continuous quantum mechanics with a Schrödinger- or von-Neumann-equation.

We treat probabilistic automata within our general framework for probabilistic systems. This has the advantage that powerful methods as numerical simulations or coarse graining can be applied for their investigation. 

\paragraph*{Probabilistic automata as generalized Ising models}

For a probabilistic treatment of automata we may start with eq.~\eqref{eq:TS36} for a local chain
\begin{equation}
\label{OS5}
\cL(m) = -\beta \left\{ \sum_{\tau,\rho} \left(h_\tau(m+1) \hat{S}_{\tau \rho}(m) h_\rho(m) \right)-1\right\}\,,
\end{equation}
with $\beta \to \infty$. For a configuration $\rho$ at $m$ and a configuration $\tau(\rho)$ at $m+1$ one has 
$\hat{S}_{\tau \rho}(m) = 1$, $h_\rho(m) = 1$, $h_\tau(m+1) = 1$ and therefore $\cL(m) = 0$, $\cK(m) = 1$. For all other
configurations at $m+1$, with $\tau \neq \tau(\rho)$, the corresponding elements of the step evolution operator vanishes, resulting
in $\cL(m) = \beta$ and $\cK(m) = 0$. Those sequences of configurations have zero probability, such that the unique configuration
$\tau(\rho)$ at $m+1$ is selected, in accordance with a deterministic jump $\rho \to \tau(\rho)$. 
This holds for arbitrary configurations $\rho(m)$. For every configuration $\rho(m)$ the sum over $\tau$ and $\rho$ in eq.~\eqref{OS5}
contains only a single non-zero term.

The basis function $h_\rho(m)$ is a function of the time-local occupation numbers or bits $n_\gamma(m)$, or equivalent Ising spins $s_\gamma(m)$,
while $h_\tau(m+1)$ involves $s_\gamma(m+1)$. The partition function and overall probability distribution for this setting is a 
generalized Ising model. The limit $\beta \to \infty$ can be associated with a zero temperature limit. Indeed, for finite $\beta$
the system can be seen as a local chain in thermal equilibrium at temperature $T$, $\beta = \left( k_B T \right)^{-1}$, with
Boltzmann factor in the probability distribution
\begin{equation}
\label{OS6}
e^{-S} = e^{-\beta H_\text{cl}} = e^{-\beta \sum_m \tilde{\cL}(m)}\,,
\end{equation}
where the classical Hamiltonian $H_\text{cl}$ involves next neighbor interactions in the time direction
\begin{align}
\label{OS7}
H_\text{cl} &= \sum_m \tilde{\cL}(m) \\
&= - \sum_m \left[ \sum_{\tau, \rho} \left( h_\tau(m+1) \hat{S}_{\tau \rho}(m) h_\rho(m) \right) -1 \right]\,.
\end{align}
The limit $\beta \to \infty$ corresponds to the ground state.

Without boundary terms the ground state is the minimum of $H_\text{cl}$. This minimum is highly degenerate. For every possible
configuration of ``initial spins" $s_\gamma(t_\text{in})$, or corresponding $\tau_\text{in}$, there is a ``trajectory in configuration space"
dictated by the sequence of maps $\rho \to \tau(\rho)$. For this trajectory one has $H_\text{cl} = 0$, while for all other configurations
with given $\tau_\text{in}$ one finds $H_\text{cl} > 0.$ For these other trajectories one has $H_\text{cl} = n_E >0$, where $n_E$ is the ``number of errors", 
e.g. the number of points $m$ for which $\tilde{\cL} = 1$. 
For the ground state one has $H_\mathrm{cl}=0$. The
number of degenerate ground states amounts to $2^{M}$, given by
the number of different configurations $\tau_\text{in}$.

In the presence of boundary conditions in the form of boundary terms the probabilistic system becomes more complex. Due to the 
degeneracy of the minima of $H_\text{cl}$ the boundary information propagates into the bulk. Conceptually, the solution is straightforward.
The different trajectories corresponding to different $\tau_\text{in}$ are weighed with probabilities corresponding to the boundary conditions.
One can solve the Schr\"odinger type evolution equation for the classical wave function, using the orthogonal step evolution operator.
In practice, such a solution can be rather involved, as we demonstrate by a simple example below. 

One may in this case try to solve
important aspects of the problem, as the computation of correlation functions, by other methods of statistical physics, as numerical
simulations or renormalization group techniques. Our formulation of probabilistic automata in terms of a classical statistical partition function opens rather promising avenues. This concerns, in particular the continuum limit for a very large number of cells. While functional integral methods are very well developed for this purpose, there is only little known for solutions of Schrödinger type evolution equations.

\paragraph*{Quantum formalism for probabilistic automata}

Let us concentrate on pure classical states with factorizing boundary conditions. As we have discussed in sect.\,\ref{sec:classical_density_matrix}, we can choose boundary conditions for which $\bar{q}(t) = \tilde{q}(t) = q(t)$, cf.\ eq.\,\eqref{eq:DM49}. We will assume this choice here. It leads to a symmetric classical density matrix. A pure state symmetric classical density matrix is positive, with eigenvalues one and zero. Observables are represented by symmetric operators. The evolution is orthogonal, $\hat{S}^\mathrm{T} = \hat{S}^{-1}$. A symmetric initial density matrix remains symmetric for all $t$.

The wave functions are normalized, $q_\tau(t) q_\tau(t) = 1$. As we have discussed in sect.\,\ref{sec:free_particles_in_two_dimensions}, there is no restriction on their sign. For finite $N$ the space of possible wave functions, corresponding to the space of possible boundary conditions, is a finite dimensional Hilbert space. It is at this point that the probabilistic character of the cellular automaton enters crucially. Every element of the Hilbert space can be realized by a suitable probability distribution for the initial conditions. For any arbitrary initial wave function $q(0)$ one can infer the required initial probabilities $p_\tau(0) = q_\tau^2(0)$. The limit $N\to\infty$ can lead to an infinite-dimensional Hilbert space, and the limit $\varepsilon\to 0$ to a continuum limit for the evolution.

In the presence of a complex structure the real $N$-component wave function $q(t)$ is mapped to an $N/2$-component complex wave function $\psi(t)$. The orthogonal evolution of $q(t)$ is mapped to an unitary evolution of $\psi(t)$, provided that it is compatible with the complex structure. This requires that $\psi(t+\varepsilon) = \hat{S}_\mathrm{c}(t) \psi(t)$ with $\hat{S}_\mathrm{c}$ a complex matrix. The unitarity follows from the simple observation that the conserved normalization $q^\mathrm{T}q=1$ is mapped to $\psi^\dagger \psi =1$. Complex linear transformations preserving $\psi^\dagger \psi=1$ are unitary transformations. In the complex language the step evolution operator becomes a unitary matrix. Conservation of $\psi^\dagger\psi$ implies $\hat{S}_\mathrm{c}^\dagger \hat{S}_\mathrm{c}=1$. A probabilistic cellular automaton with complex structure realizes all properties of finite-dimensional quantum mechanics with discrete unitary evolution steps.

We also have seen that operators are associated to observables. The possible measurement values of the observables correspond to the spectrum of eigenvalues of the associated operators. The expectation values of the observables are obtained from the quantum rule in terms of the operators. No new axioms have been used to derive these quantum laws. We have found non-commutative structures for operators, as for the example of the position and momentum operators. The usual Heisenberg uncertainty relations for the corresponding observables follow directly from the commutator and the quantum laws for expectation values. As for general quantum systems there is a question if for every hermitian operator a measurement procedure can be found for a suitable associated observable. We see no conceptual or formal difference between probabilistic automata and quantum systems and conclude that probabilistic automata \textit{are} particular quantum systems.

\subsection{Static memory materials}
\label{sec:static_memory_materials}
 
 In this section we discuss an example for a truly probabilistic evolution. This contrasts with the deterministic evolution of a probabilistic automaton for which the probabilistic aspects enter only by a probability distribution for initial configurations. For a truly probabilistic evolution the step evolution operator is not a unique jump matrix. Probabilistic step evolution operators map a given state at $t$ to several (or many) different states at $t+\varepsilon$ with certain probabilistic transition amplitudes. This is the generic behavior found for subsystems or the continuum limit. We discuss limits for which the step evolution operator becomes a unique jump matrix. The evolution between hypersurfaces needs not to be associated with physical time. It can equally well concern different hypersurfaces in space. As an example we investigate static memory materials.

Memory materials are probabilistic systems that can remain in different states for a sufficiently long time, such that they keep memory
of the initial state. An example are systems that admit two different possible magnetization states that keep the memory of the initial 
magnetization. The two possible magnetization states can be associated with an Ising spin or a bit of information. We consider here
the limit that the memory can be kept for infinite time. Time plays then no role anymore, and we discuss ``static memory materials''. 
A simple condition for a static memory material requires that the probabilistic system does not have a single equilibrium state.
Otherwise, the system would typically be found in the equilibrium state after sufficiently long time, such that all memory of the initial
state would be lost. For many practical memory materials it is sufficient that equilibrium is approached sufficiently slowly. We will
not be concerned with this case here and stick to static memory materials for which the initial information is kept for an infinite time.

There remains the interesting question which type of information can be stored in an extended system. We investigate here systems where the ``memory system" is in contact with external input at some boundaries. The question how this boundary information manifests itself inside the material, where we imagine that it can possibly be read out, can be investigated by following the information transport from the boundary to the interior. This generalizes the very simple boundary problem for the Ising chain in sect.~\ref{sec:influence_of_boundary_conditions}. The present section follows largely ref.\,\cite{CWIT,SEW}.
An experimental realization of static memory materials may use spin based information technology\,\cite{KWC,LBL,MMW,BRH}.

\paragraph*{Evolution in space}

We discuss static memory materials consisting of a bulk and its surface. For the bulk we consider a fixed probabilistic system in the
form of a local chain. The information to be memorized is provided by boundary conditions on the surface. Different states of the system
correspond to different boundary conditions. For multi-dimensional systems we specify the boundary conditions on those hypersurfaces
that correspond to the ``initial'' and ``final'' layer of the local chain. The probability distribution is then given by eq.\eqref{eq:LC4},
with $m$ labeling sites on the chain and not associated to time here. The integer $m$ typically denotes a position in space, but it also
can label more abstract structures, as layers in artificial neural networks. 

We consider systems for which the relevant information is stored in the bulk. The key question for this type of static memory materials concerns 
therefore the propagation of the boundary information into the bulk. We are concerned again with evolution, but this time rather with 
evolution in space. The formalism remains the same as for evolution in time. This underlines that in our probabilistic approach time is not
a special ``a priori'' concept. The same type of structure appears in certain static problems that are unrelated to any time evolution.

One of the simplest static memory materials is the Ising chain with large $\beta$ and finite volume. For $\beta \to \infty$ the
bulk has two degenerate ground states, one with spin up and the other with spin down, in which one bit of information can be stored. 
For finite $\beta$ the bulk only admits a single equilibrium state if the correlation length is small as compared to the size of the bulk. In this case the bulk cannot store information. If $\beta$ is large enough, such that the correlation
length exceeds the length of the chain, the boundary information is only partially lost in the bulk. In this case the boundary information
can be stored in the bulk. The quantitative solution of the boundary problem can be done by use of the classical wave function. We have discussed this in sect.~\ref{sec:influence_of_boundary_conditions}.

\paragraph*{Imperfect unique jump chains}

\begin{figure}[h!]
\includegraphics[width=0.5\textwidth]{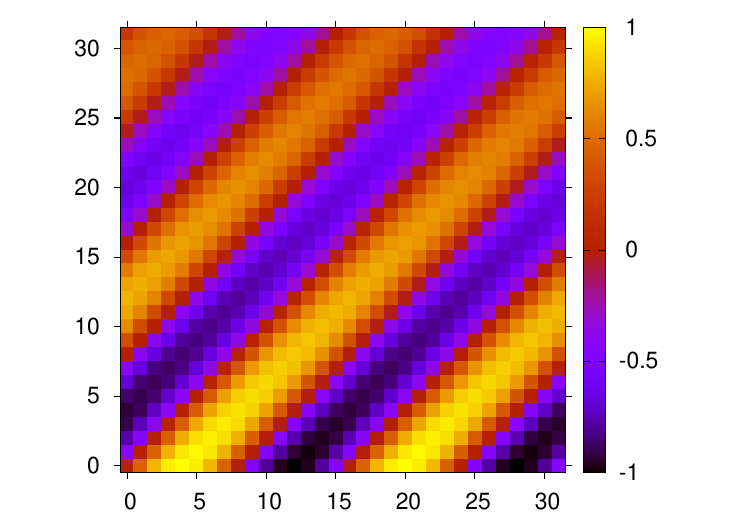}
\caption{Static memory material with open final boundary condition and uncorrelated initial boundary condition. The color code indicates the average value $\braket{s(t,x)}$. The figure is taken from ref.\,\cite{SEW}.}
\label{fig:SMA1}
\end{figure}

Unique jump chains are perfect static memory materials. The boundary information is transported to the bulk without any loss. For
$M$ Ising spins on each site $m$ the information of $M$ bits is kept by such a memory. For example, the diagonal Ising model
in two dimensions discussed in sect.~\ref{sec:free_particles_in_two_dimensions} can store $\cM_2 + 1$ bits of information if we 
consider $\cM_2 + 1$ positions $x$ and periodicity in $x$, $s(\cM_2+1) = s(0)$. This model involves a two dimensional lattice in space
with particular interactions among the Ising spins. If it can be realized in practice it constitutes a simulator of a quantum field
theory by a probabilistic state. The time evolution of the quantum field theory is represented by a particular space-pattern in a 
static system, cf. Fig.\,\ref{fig:SMA1}.

In practice, the situation with $\beta \to \infty$ will often be difficult to realize, in particular if one wants to construct small
systems for a large number of bits. We therefore discuss here the effect of a finite value of $\beta$ and also admit small additional
interactions that do not connect spins on the given diagonals. Our simple model will reveal interesting memory structures as selective
boundary conditions. We extend the diagonal two-dimensional Ising model of sect.~\ref{sec:free_particles_in_two_dimensions} by 
considering the action.
\begin{equation}
\label{SMA1}
S = -\frac{\beta}{2} \sum_{t,x} s(t,x) \left[ s(t+1,x+1) + \sigma s(t+1,x-1) \right]\,,
\end{equation}
with two integer space coordinates $(t,x)$ denoting sites on a square lattice. For $\sigma = 1$ one recovers a version of the 
next-neighbor two-dimensional Ising model while for $\sigma = 0$, $\beta \to \infty$ one has the diagonal two-dimensional Ising
model of sect.~\ref{sec:free_particles_in_two_dimensions}. For $\sigma=0$ the bulk describes independent Ising chains on the diagonals.
The even and odd sublattices are not mixed by the action \eqref{SMA1}, even for $\sigma \neq 0$. We consider boundary conditions
which factorize into an initial boundary condition $f_\text{in}$ and a final boundary condition $\bar{f}_f$.
The weight function takes the form of a local chain
\begin{equation}
\label{SMA2}
w[s] = \bar{f}_f e^{-S} f_\text{in}\,,
\end{equation}
with $f_\text{in}(s_\text{in})$, $s_\text{in} = s(0,x)$, involving spins at the "initial hyperface $t_\text{in} = 0$, and
$\bar{f}_f(s_f)$, $s_f = s(N_t, x)$, depending on the spins at the final hypersurface $t_f = N_t$.

The limit $\beta\to\infty$, $\sigma\to0$ is a probabilistic cellular automaton. For boundary conditions which are consistent with a single wave function $\tilde q=\bar q=q$ all probabilistic information is stored at the boundary. This is a realization of the holographic principle. Away from this limit the fluctuations inside the bulk acquire more independence, ending in the usual proportionality of entropy and volume for the Ising case at $\sigma=1$ and finite $\beta$.

\paragraph*{Selective boundary conditions}

This model has been investigated by Monte-Carlo simulations for various boundary conditions in ref.\,\cite{SEW}. We briefly report some of the
findings of this investigation for a $32\times 32$ lattice, with parameters $\beta = 4$, $\sigma = 0.01$. The first boundary condition
takes an open final boundary condition $\bar{f}_f = 1$. For the initial boundary condition one assumes
\begin{equation}
\label{SMA3}
f_\text{in}(s_\text{in}) = \prod_x \left[ \bar{p}_+(x) h_+\left(s_\text{in}(x)\right) + \bar{p}_-(x) h_-\left(s_\text{in}(x)\right) \right]\,,
\end{equation}
where
\begin{equation}
\label{SMA4}
h_\pm(s(x)) = \frac{1}{2} \left( 1 \pm s(x)) \right)\,, \quad \bar{p}_\pm(x) = \frac{1}{2} \left( 1 \pm \bar{s}(x) \right)\,.
\end{equation}
Due to the product structure of $f_\text{in}$ the boundary conditions for the different initial spins $s_\text{in}(x)$ are
uncorrelated. With open final boundary conditions $\bar{p}_+(x)$ amounts to the probability to find $s_\text{in}(x) = 1$,
and $\bar{p}_-(x)$ is the probability to find $s_\text{in}(x) = -1$, $\bar{p}_+(x)$ + $\bar{p}_-(x) = 1$. Thus
$\bar{s}(x)$ is the expectation value of $s_\text{in}(x)$, $\bar{s}(x) = \braket{s_\text{in}}$.

For $\bar{s}(x)$ we choose a ``wave boundary" in the form of a periodic function ($N_x = \cM_2+1 = 32$)
\begin{equation}
\label{SMA5}
\bar{s}(x) = \sin \left( \frac{2 \pi m_x x}{N_x} \right)\,, \quad m_x \in \mathbb{Z}\,.
\end{equation}
The period $N_x/m_x$ in $x$ for the initial wave function $f_\text{in}$ translates to a periodic evolution in $t$ at given $x$.
We plot the average values of the spins $\braket{s(t,x)}$ for $m_x=2$ in Fig.~\ref{fig:SMA1}. One clearly sees the periodic pattern 
in $t$. We observe that the contrast gets weaker at $t_f$ (upper part of the figure) as compared to $t_\text{in}$ (lower part of the figure).
This damping of the amplitude corresponds to a partial loss of memory due to finite $\beta$ and $\sigma \neq 0$.

\begin{figure}[h!]
\includegraphics[width=0.5\textwidth]{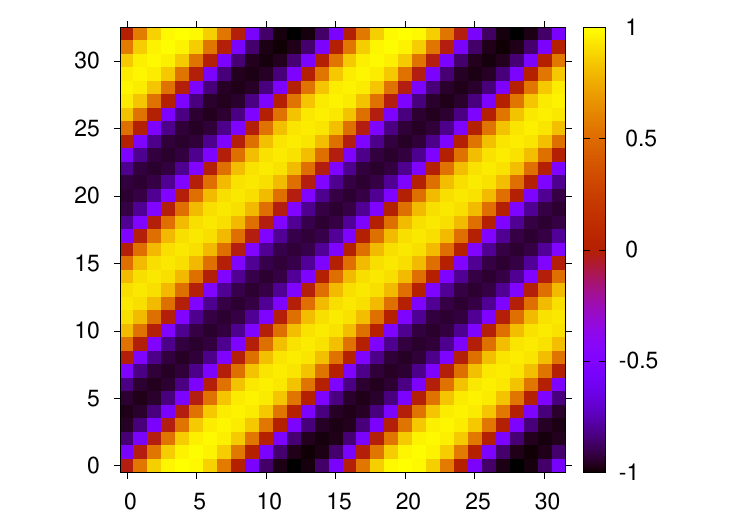}
\includegraphics[width=0.5\textwidth]{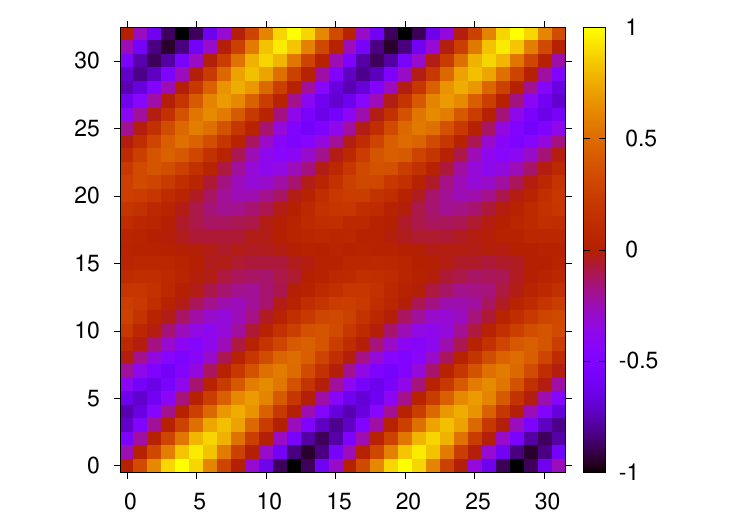}
\caption{Static memory material with uncorrelated initial and final boundary conditions. The ``phase'' of the periodic dependence on the boundary condition is shifted between the initial boundary ($m_x=2$) and the final boundary, with constructive boundary ($m_x=2$) for the upper part, and destructive boundary ($m_x=-2$) for the lower part. The color code indicates the average value $\braket{s(t,x)}$. The figure is taken from ref.\,\cite{SEW}.}
\label{fig:SMA2}
\end{figure}

We next impose non-trivial boundary conditions both at $t_\text{in}$ and $t_f$. For the final boundary term $\bar{f}_f$ we take
the same type of wave boundary \eqref{SMA3} as for $f_\text{in}$, now in dependence on $s_f$ We compare two different values for $m_x$ for the final boundary for a given $m_x$ at the initial boundary.
In Fig.~\ref{fig:SMA2} we show $\braket{s(t,x)}$ for the two different final boundary conditions, one with $m_x = 2$ (upper part),
and the other with $m_x = -2$ (lower part), keeping $m_x=2$ for the initial boundary condition. (For the plots we take units $\epsilon=1$ and $x$ on the horizontal, $t$ on the vertical axis.) As compared to Fig.~\ref{fig:SMA1}
the upper part of Fig.~\ref{fig:SMA2} shows a stronger contrast. This amounts to ``constructive boundary conditions''.

\begin{figure}[h!]
\includegraphics[width=0.5\textwidth]{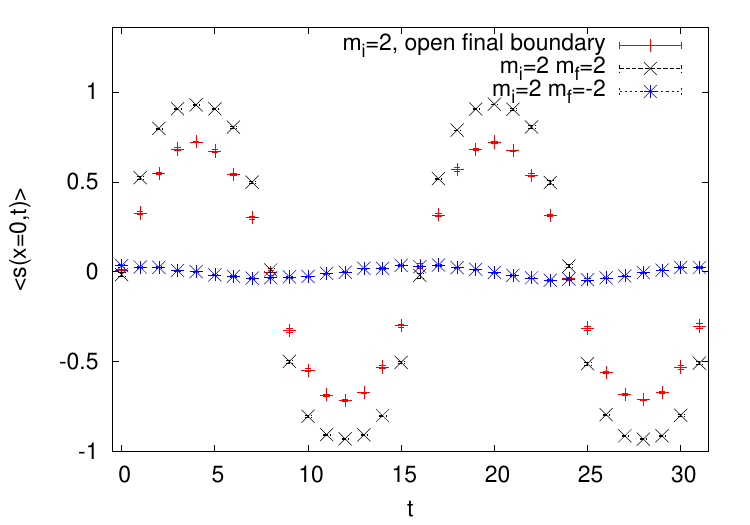}
\caption{Average spin $\braket{s(x=0,t)}$ at fixed $x=0$ in dependence on $t$, for the three boundary conditions in Figs.\,\ref{fig:SMA1} and \ref{fig:SMA2}. The figure is taken from ref.\,\cite{SEW}.}
\label{fig:SMA3}
\end{figure}

In contrast, in the lower part of Fig.~\ref{fig:SMA2} one observes a washing out of the contrast in the middle region. This indicates destructive boundary conditions.
For a more quantitative assessment of the selective boundary effect we plot in Fig.~\ref{fig:SMA3} the expectation values of the spins
at $m_t = 16$, $\braket{s(16,x)}$.
As compared to open final boundary conditions constructive boundary conditions enhance the amplitude of the oscillations, while for destructive
boundary conditions the amplitude (almost) vanishes.
Within our formalism with classical wave functions we can understand the constructive and destructive boundary conditions by following the pair of wave functions $\tilde{q}$ and $\overline{q}$ from the boundaries to the bulk. The local probabilities $p(t,x)$ are then obtained as the product $p(t,x)=\overline{q}(t,x)\tilde{q}(t,x)$. A quantitative computation needs the transfer matrix for the model~\eqref{SMA1}.

The mechanism for constructive and destructive boundary conditions can be understood by taking the limit $\sigma = 0$. In this
limit the model consists of independent Ising chains on the diagonals. The solution of the boundary problem for this case can be
inferred from sect.~\ref{sec:influence_of_boundary_conditions}. For destructive boundary conditions the spins at the two ends of the diagonal are opposite, such that the expectation value of the spin in the center vanishes. In contrast, for constructive boundaries
the spins at the two endpoints of the diagonal have the same sign, leading to an enlargement of the expectation value at the center.

Despite of its simple conceptual origin, the selective boundary effect may offer interesting computational possibilities. Details
of two boundary conditions, like a relative shift in $x$ between the initial and final boundary condition, are reflected in rather
different spatial patterns in the bulk. A readout of the patterns in the bulk can therefore ``detect" specific details of differences
of bit sequences at the initial and final boundaries. Many different geometric patterns in the bulk can be produced by appropriate 
boundary conditions. 
For example, one could identify $m$ with layers in an artificial neural network with two input layers at $t_\mathrm{in}$ and $t_\mathrm{f}$. The boundary conditions may be associated with two pictures to be compared.

\paragraph*{Coherent boundary conditions}

If one wants to use the static memory material as an imperfect simulator for the time evolution of a quantum system one is typically interested in coherent boundary conditions. For coherent boundary conditions the spins at different positions $x$ for the initial and final factors $f_{\text{in}}$ and $\overline{f}_{\text{f}}$ are correlated. For the example of an initial one-particle state the probability that two or more spins are up vanishes. This induces a strong correlation between the spins at different sites. If a a spin is up at $x$, all spins for $y\neq x$ have to be down. 

We have studied the limiting case of this model for $\sigma =0$, $\beta\to\infty $ in detail in sect.~\ref{sec:free_fermions_in_two_dimensions}. For the imperfect unique jump chain for $\sigma \neq 0 $ or finite $\beta$ one expects that part of the correlations at the boundary are lost in the bulk. In particular, one does not expect an exact conservation of particle number.

\paragraph*{Particle production}

For the diagonal two-dimensional Ising model in sect.~\ref{sec:free_particles_in_two_dimensions} particle number is conserved. We demonstrate here that this does
not hold any longer for finite $\beta$ or $\sigma \neq 0$. We can visualize particle production for our static memory material by
the choice of appropriate boundary conditions. The particle number on a given layer $t$ is given by
\begin{equation}
\label{SMA6}
N_p(t) = \sum_x n(t,x)\,, \quad n(t,x) = \frac{1}{2}(s(t,x)+1)\,.
\end{equation}
For $\beta \to \infty$, $\sigma \to 0$, it is the same for all $t$. A one particle initial boundary condition has precisely one
particle at $t_\text{in}$, $N_p(t_\text{in})=1$. It is specified by the one-particle wave function $\tilde{q}_1(t_\text{in})$,
\begin{equation}
\label{SMA7}
f_\text{in} = \sum_x \tilde{q}_1(t_\text{in},x) h_1(x)\,,
\end{equation}
with
\begin{equation}
\label{SMA8}
h_1(x) = n(x) \prod_{y \neq x} (1-n(y))\,.
\end{equation}
This type of wave function is indeed highly correlated between the different spins $s_\text{in}(x)$. Whenever one spin $s_\text{in}(x)$
is up,  $s_\text{in}(x) = 1$, $n_\text{in}(x)=1$, all other spins at $y\neq x$ must be down, $s_\text{in}(y) = -1$, $n_\text{in}(y)=0$.
The one particle basis functions $h_1(x)$ assure this correlation.

\begin{figure}[h!]
\includegraphics[width=0.5\textwidth]{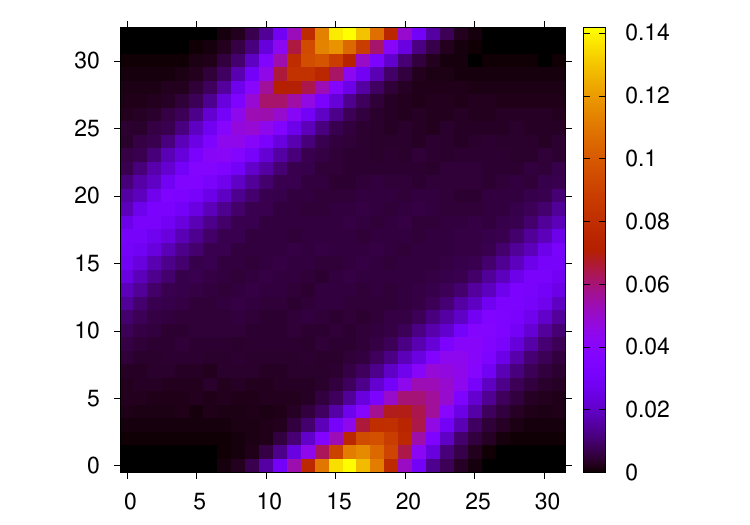}
\includegraphics[width=0.5\textwidth]{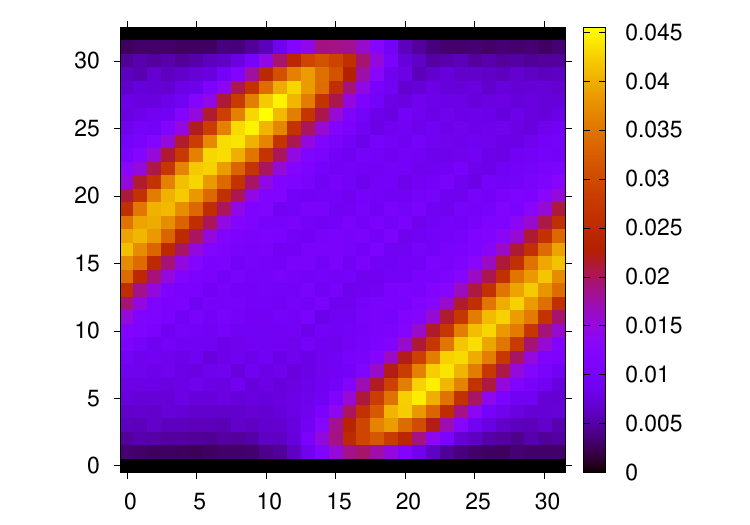}
\caption{``Particle decay'' in a static memory material. We employ correlated one-particle wave-function initial and final boundary conditions. For the upper plot the color coding shows the one-particle occupation number $n_1(t,x)$, while the lower part displays the two-particle occupation number $n_2(t,x)$. As $t$ moves towards the center the occupation shifts from the one-particle to the two-particle state. The figure is taken from ref.\,\cite{SEW}.}
\label{fig:SMA4}
\end{figure}

For the initial one particle wave function $\tilde{q}_1(t_\text{in},x)$ we choose a wave packet
\begin{equation}
\label{SMA9}
\tilde{q}_1(t_\text{in},x) \sim \exp \left\{ - \frac{(x-x_0)^2}{2 \delta^2} \right\}\,.
\end{equation}
For the final boundary condition $\bar{f}_f$ we choose the same wave packet for a one particle state, with $\tilde{q}_1(t_\text{in},x)$ 
replaced by the conjugate wave function $\bar{q}_1(t_f,x)$ and $f_\text{in}$ replaced by $\bar{f}_f$. Useful observables are the 
occupation numbers in $k$-particle states
\begin{equation}
\label{SMA10}
n_k(t,x) = \begin{cases}
n(t,x) & \text{if } N_p(t) = k \\
0 & \text{otherwise.}
\end{cases}
\end{equation}
Thus $n_2(t,x) =1$ means that the state contains two particles, for which one of the particles is located at $x$. The position of the second particle is not resolved by this observable - it can be at any position $y\neq x$.
Fig.~\ref{fig:SMA4} plots the expectation values $\braket{n_1(t,x)}$ (upper part) and $\braket{n_2(t,x)}$ (lower part), for 
$\delta = 3$ and $x_0 = 16$. Towards the middle part of the figure around $t=16$ the one particle wave function decreases, while the two-particle wave function 
reaches a maximum. This can be viewed as a decay of one particle into two particles. The decay is almost collinear since the additional
spreading of $\braket{n_2(t,x)}$ is small.

The formalism for the evolution of the classical wave functions $\tilde{q}$ and $\bar{q}$ is described in sect.~\ref{sec:classical_wave_functions}--\ref{sec:influence_of_boundary_conditions}. Since for finite $\beta$ and/or $\sigma \neq 0$ the 
particle number is not conserved the step evolution operator mixes sectors with different particle numbers. The restriction of the step evolution operator
to the one-particle sector can be found explicitely in ref.\,\cite{SEW}.

\paragraph*{Simulation of time evolution in quantum field \\theories}

As mentioned already, the static memory materials can be used for a simulation of the time evolution in certain quantum field theories.
This concerns those quantum field theories that admit a discretization equivalent to a cellular automaton.
The time $t$ in a quantum field 
theory is mapped to a space coordinate $t$ for the static memory material. The time evolution can be visualized as a geometric pattern in the
bulk of the static memory material. For example, the cellular automaton for the interacting fermionic model of the Thirring type in sect.~\ref{sec:probabilistic_celluar_automata} can be represented as a static memory material for finite $\beta$. In this form it
is accessible to Monte Carlo simulations based, for example, on the Metropolis algorithm. 

The figures of the present section can be viewed as a space-time diagram for propagating right moving fermions. In the limit $\sigma \to 0$,  $\beta\to\infty$ the deviations from an exact quantum field theory will vanish. For more complex quantum field theories similar investigations with finite large $\beta$ may already reveal important features. This could permit an extrapolation to $\beta\to\infty$ and an investigation of the continuum limit of the classical statistical model in this way.
\subsection[Partial loss of memory and emergence of quantum mechanics]{Partial loss of memory and\\emergence of quantum\\mechanics}
\label{sec:partial_loss_of_memory_and_emergence_of_quantum_mechanics}

The evolution from the boundary into the bulk often leads to partial or complete loss of memory of the information contained in the
boundary conditions. A well known example is a unique equilibrium state in the bulk. For all practical purposes the boundary information
is lost completely if the distance to the boundaries is much larger than the correlation length. Unique jump chains or cellular 
automata are the other extreme for which no boundary information is lost. Inbetween there are interesting cases of partial loss of 
memory of the boundary conditions. 

If we associate the boundaries to the infinite past and infinite future in time, the partial loss of
memory  projects effectively onto systems with an orthogonal or unitary evolution as in quantum mechanics. This constitutes a simple explanation why observations find exact quantum mechanics without deviations from it. Possible deviations from the unitary evolution of quantum mechanics have simply ``died out" due to the infinite time distance between the present and the far distant past or future.

\paragraph*{Information loss for the classical wave function}

The issue of information loss is rather apparent in the evolution of the classical wave function $\tilde{q}(t)$ with
increasing $t$, as we have already discussed in sect.~\ref{sec:influence_of_boundary_conditions}. Let us assume that the step evolution
operator has $n$ eigenvalues $\lambda_\alpha$ with $|\lambda_\alpha| = 1$, $\alpha = 1 \ldots n$, and $N-n$ eigenvalues
$\lambda_\gamma$ with $|\lambda_\gamma| \leq 1-g$, $\gamma = n+1, \ldots, N$, with finite gap $0 < g \leq 1$. For simplicity, we take
$\hat{S}$ to be independent of $t$. We can bring $\hat{S}$ to a block diagonal form
\begin{equation}
\label{PL1}
D \hat{S} D^{-1} = \hat{S}^{(bd)} = \begin{pmatrix}
\hat{S}^{(s)} & 0 \\
0 & \hat{S}^{(e)}
\end{pmatrix},
\end{equation}
where $\hat{S}^{(s)}$ is an $n\times n$-matrix with eigenvalues $|\lambda_\alpha| = 1$, and $\hat{S}^{(e)}$ is an $(N-n)\times (N-n)$-matrix with eigenvalues $|\lambda_\gamma| \leq 1-g$. We also define 
\begin{equation}
\label{PL2}
\tilde{q}^{(bd)}(t) = D \tilde{q}(t) = \begin{pmatrix}
\tilde{q}^{(s)}(t) \\ \tilde{q}^{(e)}(t)
\end{pmatrix}.
\end{equation}
Since $D$ may be a complex matrix, $\tilde{q}^{(bd)}(t)$ can be complex as well.

The evolution of $\tilde{q}^{(s)}$ and $\tilde{q}^{(e)}$ can be considered separately,
\begin{equation}
\label{PL3}
\tilde{q}^{(s)}(t+\epsilon) = \hat{S}^{(s)} \tilde{q}^{(s)}(t)\,, \quad \tilde{q}^{(e)}(t+\epsilon) = \hat{S}^{(e)} \tilde{q}^{(e)}(t)\,.
\end{equation}
Starting at $t_\text{in} = 0$ the length of the $N-n$-component vector $\tilde{q}^{(e)}$ decreases with increasing $t$ towards zero.
Diagonalizing $\hat{S}^{(e)}$,
\begin{equation}
\label{PL4}
\tilde{D}^{(e)} \hat{S}^{(e)} ( \tilde{D}^{(e)})^{-1} = \hat{S}^{(de)} = \diag (\lambda_\gamma)\,,
\end{equation}
we can write
\begin{equation}
\label{PL5}
\tilde{q}^{(e)}(t) = (\tilde{D}^{(e)})^{-1} \tilde{q}^{(de)}(t)\,.
\end{equation}
With
\begin{equation}
\label{PL6}
\tilde{q}_\gamma^{(de)}(t+m \epsilon) = (\lambda_\gamma)^m \tilde{q}^{(de)}_\gamma(t)
\end{equation}
one has
\begin{equation}
\label{PL7}
|\tilde{q}_\gamma^{(de)}(t+m \epsilon)| \leq (1-g)^m | \tilde{q}^{(de)}_\gamma(t) |\,.
\end{equation}
For $m \to \infty$ the r.h.s. of eq.~\eqref{PL7} vanishes, and we infer from eq.~\eqref{PL5} $\tilde{q}^{(e)}(t+m \epsilon) \to 0$. 
The memory of boundary information contained $\tilde{q}^{(e)}(t_\text{in})$ is lost for $m \to \infty$. This applies, in particular, 
to the continuum limit for finite $\Delta t = t - t_\text{in}$, $\epsilon \to 0$.

The only boundary information remaining for $m \to \infty$ is the one related to $\tilde{q}^{(s)}(t_\text{in})$. This information 
is never lost. Often one has $(\lambda_\alpha)^P = 1$, such that the evolution of $\tilde{q}^{(s)}$ is periodic with period $P \epsilon$.
We can repeat the steps \eqref{PL4}--\eqref{PL6}, with $(e)$ replaced by $(s)$, and eq.~\eqref{PL7} replaced by
\begin{equation}
\label{PL8}
|\tilde{q}_\alpha^{(ds)}(t+m \epsilon)| = |\tilde{q}_\alpha^{(ds)}(t)|\,.
\end{equation}
This demonstrates that $\tilde{q}^{(s)}(t)$ will never vanish.

We may associate $\tilde{q}^{(s)}$ with a subsystem, as indicated by $(s)$, and $\tilde{q}^{(e)}$ with its environment, denoted
by $(e)$. The evolution of the subsystem is independent of the environment. The step evolution operator for the subsystem $\hat{S}^{(s)}$ has all eigenvalues of the same absolute magnitude, $| \lambda_\alpha| = 1$. It is, however, in general not a positive matrix and 
therefore does not need to be a unique jump matrix. 
The time evolution of the total system projects on the subsystem if the number of time steps goes to infinity. We may therefore associate
$\tilde{q}^{(s)}$ with the wave function of the "asymptotic subsystems". 

The loss of memory is particularly simple for a finite gap $g>0$. Eq.~\eqref{PL8} indicates that this loss proceeds at least exponentially. One may encounter continuous spectra of eigenvalues for which $g$ approaches zero arbitrarily closely. This will not change the behavior of the subsystem with $|\lambda_{\alpha}|=1$. However, the rate of information loss gets modified. For a powerlike decrease of $|\tilde{q}^{(\text{de})  }_{\gamma}(t+m\varepsilon)|$ the qualitative conclusions remain similar. The sequence of limits to be taken is first an investigation of the time evolution of the wave function in the limit where the spectrum of eigenvalues becomes continuous and the gap $g$ vanishes. Subsequently one extrapolates to $t-t_{\text{in}}\to\infty$, $t_{\text{f}}-t\to\infty$. Whenever the information pertaining to the environment dies out the asymptotic subsystem undergoes the unitary evolution of quantum mechanics.

\paragraph*{Unitary basis for the asymptotic subsystem}

The unitary character of the evolution of the asymptotic subsystem is most easily seen by constructing a basis for which the evolution 
is manifestly unitary. The matrix $D$ in eq.~\eqref{PL1} is not defined uniquely. There exists a class of choices of $D$ for which the matrix $\hat{S}^{(s)}$ becomes
a unitary matrix. The existence of such matrices can be seen by diagonalizing $\hat{S}^{(s)}$,
\begin{equation}
\label{PL9}
D^{(s)} \hat{S}^{(bd)} (D^{(s)})^{-1} = \begin{pmatrix}
\hat{S}^{(ds)} & 0 \\
0 & \hat{S}^{(e)}
\end{pmatrix},
\end{equation}
with 
\begin{equation}
\label{PL10}
\hat{S}^{(ds)} = \diag (\lambda_\alpha) = \diag (e^{i\beta_\alpha})\,.
\end{equation}
The matrix $\hat{S}^{(ds)}$ is unitary. This property does not change if we transform $\hat{S}^{(ds)}$ by a further block diagonal
matrix $E$ whose $n\times n$-block is unitary. The result,
\begin{equation}
\label{PL11}
\tilde{D} \hat{S} \tilde{D}^{-1} = \begin{pmatrix}
U^{(s)} & 0 \\
0 & \hat{S}'^{(e)}
\end{pmatrix}, \quad (U^{(s)})^\dagger U^{(s)} = 1\,,
\end{equation}
with 
\begin{equation}
\label{PL12}
\tilde{D} = E D^{(s)} D\,,
\end{equation}
corresponds to a change of basis for the wave function for which $U^{(s)}$ is a unitary evolution operator for the subsystem. This
proves the existence of a basis for which the asymptotic subsystem follows a unitary evolution, similar to quantum mechanics. The choice
of this basis, or the choice of $\tilde{D}$, is not unique.

These statements about a split into an asymptotic subsystem and an environment which vanishes asymptotically are rather general. The 
unitary evolution may be trivial, however. This happens, in particular, in case of a unique equilibrium state in the bulk. For $n=1$
one has $\lambda_\alpha = 1$ and therefore $U^{(s)} = 1$. More general unit $n\times n$-matrices $U^{(s)}$ can occur as well,
and correspond to a static asymptotic subsystem.

\paragraph*{Density matrix and local probability distribution}

The loss of memory is often not directly visible in the evolution of the classical density matrix or the local probability distribution. 
For the conjugate classical wave function $\bar{q}(t)$ we can apply the same arguments as for the wave function $\tilde{q}(t)$, but now
with decreasing $t$. For $t$ decreasing from the final boundary $t_f$ to some value $t$ inside the bulk, the boundary information
can be partially lost according to eq.~\eqref{eq:SE5}. In contrast, for increasing $t$ we learn from eq.~\eqref{eq:SE7} that the 
environment follows the opposite evolution. The eigenvalues of $( \hat{S}^T)^{-1}$ have still a set $\lambda_\alpha^{-1} = \exp \left( - i \beta_{\alpha} \right)$ with magnitude equal to one. In the unitary basis $\bar{q}^{(s)}$ has the same unitary evolution
as $(\tilde{q}^{(s)})^*$. The matrix $(\hat{S}^{(e)T})^{-1}$ has, however, eigenvalues $\lambda_\gamma^{-1}$ with magnitude
$|\lambda_\gamma^{-1}| >1 $. The decrease of $\bar{q}^{(e)}$ for decreasing time corresponds to an increase of $\bar{q}^{(e)}$
for increasing time.

Pure state classical density matrices are products of $\tilde{q}$ and $\bar{q}$, given by eq.~\eqref{eq:DM1}. The evolution \eqref{eq:DM38} for increasing $t$ involves both $\hat{S}$ and $\hat{S}^{-1}$. The loss of information in $\tilde{q}$ can be compensated
by the opposite behavior of $\bar{q}$. For the asymptotic subsystem in the unitary basis the density matrix evolves according to the 
von-Neumann equation, with no difference to quantum mechanics. The parts involving the environment may typically not reflect a loss
of the boundary information. For this reason the evolution of the classical density matrix is not the appropriate tool for an investigation of the partial loss of boundary information. The same holds for the evolution of the time-local probability distribution,
which corresponds to the diagonal elements of the classical density matrix.

The reason for this limitation of the information contained in the evolution law for the classical density matrix is connected to the fact that the partial loss of information from the boundaries occurs already in the construction of the allowed values of the density matrix $\rho'(t)$. As we have discussed in sect.\ref{sec:independence_from_the_future} there are important constraints for the allowed values of $\rho'(t)$ if the step evolution operator is not an orthogonal matrix.

\paragraph*{Evolving boundary matrix}

A convenient tool for the partial loss of boundary information is the concept of an evolving boundary matrix. General boundary conditions can be formulated in terms of the boundary term $\cB$ in eqs.~\eqref{eq:DM14}, \eqref{eq:DM15}. It depends on the variables
at the final and initial time $t_f$ and $t_\text{in}$. We can expand the boundary term in terms of basis functions at $t_\text{in}$ and
$t_f$
\begin{equation}
\label{PL13}
\cB = h_\tau(t_\text{in}) \hat{\cB}_{\tau \rho}(t_\text{in},t_f) h_\rho(t_f)\,.
\end{equation}
The boundary matrix $\hat{\cB}(t_\text{in},t_f)$ involves the initial and final wave functions
\begin{equation}
\label{PL14}
\hat{\cB}_{\tau \rho}(t_\text{in},t_f) = \sum_\alpha \bar{w}_\alpha \tilde{q}_\tau(t_\text{in}) \bar{q}_\rho(t_f)\,.
\end{equation}

We can evolve the boundary matrix for increasing $t_\text{in}$ and decreasing $t_f$,
\begin{align}
\label{PL15}
\hat{\cB}(t_\text{in}+\epsilon,t_f) &= \hat{S}(t_\text{in}) \hat{\cB}(t_\text{in},t_f)\,, \nonumber \\
\hat{\cB}(t_\text{in},t_f-\epsilon) &= \hat{\cB}(t_\text{in},t_f) \hat{S}(t_f - \epsilon) \,.
\end{align}
This moves the boundary matrix farther inside the bulk, by integrating out variables at $t_\text{in}$ or $t_f$. We can repeat 
these steps until a given $t$ in the bulk. For $t_f = t_\text{in} = t$ all variables at $t' < t$ and $t' > t$ are integrated out, and
the evolving boundary matrix becomes the classical density matrix
\begin{equation}
\label{PL16}
\hat{\cB}(t,t) = \rho'(t)\,.
\end{equation}
This is a way to compute the classical density matrix and the associated local probability distribution for arbitrary boundaries.

The evolution of the boundary matrix \eqref{PL15} involves only $\hat{S}$, and not $\hat{S}^{-1}$. It therefore reflects directly the loss of memory in the environment sector. 
In the block diagonal basis \eqref{PL1} the evolving boundary matrix takes an asymptotic form
\begin{equation}
\hat{\cB}(t_1,t_2) = \begin{pmatrix}
\hat{\cB}^{(s)} & 0 \\
0 & 0
\end{pmatrix},
\end{equation}
if the number of time steps $t_1 - t_\text{in} = m_1 \epsilon$,  $t_f - t_2 = m_2 \epsilon$, goes to infinity ($m_1 \to \infty$, 
$m_2 \to \infty$) and the gap is finite. As mentioned above, this often generalizes to a continuous spectrum of eigenvalues of the step evolution operator. In the asymptotic limit all boundary information is lost in the environment sector in this case. Only boundary information in the 
asymptotic subsystem remains available.

\paragraph*{Asymptotic quantum evolution}

These findings have an important consequence for the concept of probabilistic time. If one describes the world by an overall probability
distribution for all times, and defines boundary conditions only in the infinite past $t_\text{in} \to -\infty$ and infinite future
$t_f \to \infty$, the evolution at any finite time $t$ becomes a quantum evolution. The boundary information relating to the
environment of the asymptotic subsystems is forgotten. The physics at finite time is projected onto the asymptotic subsystem. This 
projection occurs by virtue of the time evolution and does not need any observer. For the asymptotic subsystem there exists a basis
for which the density matrix follows the unitary evolution of the von-Neumann equation. This is precisely the evolution of quantum 
mechanics. The concept of probabilistic time and boundary conditions at infinity projects in a natural way on subsystems following a 
quantum evolution.

The concepts and evolution of quantum mechanics emerge in a compulsory way in this setting. This explains why quantum mechanics is
universal for all observations, without any ``corrections". In order to find deviations from quantum mechanics one either needs a 
finite distance to the past or the future (finite $m_1$ or $m_2$), or the gap $g$ has to go to zero. For a continuous spectrum 
of the step evolution operator with $g \to 0$ the double limits,
\begin{equation}
\label{PL18}
A_\text{in} = (1-g)^{m_1}\,, \quad A_f = (1-g)^{m_2}\,,
\end{equation}
for $g \to 0$, $m_{1,2} \to \infty$ matter. Only for nonzero $A_\text{in}$ or $A_f$ deviations from quantum mechanics are possible if the number
of steps to the infinite past or future is infinite.

We observe that for
\begin{equation}
\label{PL19}
m_1 = \frac{\Delta t_1}{\epsilon}\,, \quad m_2 = \frac{\Delta t_2}{\epsilon}\,,
\end{equation}
the divergence of $m_1$, $m_2$ is twofold in the continuum limit. They diverge because $\Delta t_1$ and $\Delta t_2$ diverge, 
and furthermore because of $\epsilon \to 0$. We may write
\begin{equation}
\label{PL20}
A_{\text{in}} = \left( A_\epsilon \right)^{\Delta t_{\text{in}}}\,,
\end{equation}
with
\begin{equation}
\label{PL21}
A_\epsilon = \left( 1-g \right)^{1/\epsilon}\,.
\end{equation}
For $g$ vanishing proportional to $\epsilon$ one finds a value $A_\epsilon < 1$ that is separated by a gap from one,
\begin{equation}
\label{PL22}
g = a \epsilon\,, \quad \lim_{\epsilon \to 0} A_\epsilon = e^{-a}\,.
\end{equation}
In consequence, $A_{\text{in}}$ vanishes and no corrections to quantum mechanics occur for this case. This generalizes to more complex cases where $A_{\text{in}}$ vanishes with inverse powers of $\Delta t$ or similar.

While the dynamical projection to a quantum evolution is universal in this setting, the quantum evolution could still be the trivial one with vanishing Hamiltonian. This corresponds to $\lambda_\alpha=1$ for all $\alpha$. A non-trivial unitary evolution requires that some of the eigenvalues $\lambda_\alpha = e^{i\beta_\alpha}$ have non-trivial phases $\beta_\alpha$. Whenever the evolution of the asymptotic subsystem at finite $t$ is non-trivial, and the boundary information in the environment has faded away, our concept of probabilistic time not only leads to the quantum formalism for the description of classical probabilistic systems. It predicts that one observes precise quantum mechanics!
\subsection{Markov chains}\label{sec:markov_chains}

The time evolution of probability distributions is often described by Markov
chains. For Markov chains the probabilities $p_{\tau}(t+\varepsilon)$ can be
computed from the probabilities $p_{\rho}(t)$, with positive transition
probabilities $W_{\tau\rho}(t)$ according to eqs.\,\eqref{eq:EV8},
\eqref{eq:EV9}. 
Markov chains constitute a particular evolution law which is often postulated ad
hoc. Our approach of an overall probability distribution for all times, together
with the concept of probabilistic time and the ordering of hypersurfaces, allows
for a general evolution law for the time-local probabilistic information. We can
therefore investigate under which circumstances Markov chains become a valid
approximation. We have already seen in sect.~\ref{sec:step_evolution_operator}
that with the exception of the limiting case of unique jump step evolution
operators the generic evolution of the time-local probabilistic information is
not given by a Markov chain. We will argue in the present section that Markov
chains nevertheless become often valid approximations. They typically become
appropriate for a description of the loss of initial boundary information for
the environment subsystem of the preceding section. This holds if the final
boundary is far enough away.

\paragraph*{Proper Markov chains and equilibration}

For a given $t$ the states can be divided into a subsector of states $\rho$ for
which the transition probability $W_{\tau\rho}$ differs from zero only for one
particular $\tau=\bar{\tau}(\rho)$, and the other states $\rho$ for which at
least two different $W_{\tau\rho}$ differ from zero. For the first part one has
$W_{\bar{\tau}(\rho),\rho}=1$ and the state $\rho$ changes with probability one
to the state $\bar{\tau}(\rho)$. This is the behavior of a unique jump chain or
cellular automaton. A closed subset of states that are mapped uniquely to
another state in the subset forms a unique jump chain as discussed in
sect.\,\ref{sec:probabilistic_celluar_automata}. Since we have discussed the
special case of unique jump chains extensively we subtract here closed sets of
states for which the evolution is described by unique jump step evolution
operator different from unity. We will assume that all possible subsectors
forming unit jump chains with periods $\geq 2$ are subtracted. What remains is
called here ``proper Markov chains".

We will be concerned in this section with proper Markov chains for which
typically at least two elements $W_{\tau\rho}$ differ from zero for a given
$\rho$. We concentrate on time independent $W_{\tau\rho}$. Extensions for more
general proper Markov chains are possible but nod needed for our purpose here.

For proper Markov chains the probability distribution typically approaches an
equilibrium distribution as the number of time steps increases to infinity.
Probability distributions that are eigenstates of $W$ can only occur for real
and positive eigenvalues $\lambda$ of $W$,
\begin{equation}
W_{\tau\rho} p_\rho^{(\lambda)} = \lambda p_\tau^{(\lambda)} ~\Rightarrow~
\lambda \geq 0.
\label{eq:MCB1}
\end{equation}
This follows directly from the fact that $p_\rho^{(\lambda)}$,
$p_\tau^{(\lambda)}$ and $W_{\tau\rho}$ are all positive. On the other hand, one
infers $\lambda\leq 1$ since otherwise some probabilities would become larger
than one after a sufficient number of time steps. We can extend the discussion
more generally to the eigenvalues of $W$ for which eigenstates are sums of
probabilities with possibly complex coefficients. 

The equilibrium state corresponds to the eigenvalue $\lambda=1$. Aperiodic and
irreducible Markov chains have a unique equilibrium state and therefore a unique
largest eigenvalue $\lambda=1$. If $\sum_\rho W_{\tau\rho} = 1$, the equilibrium
state is given by equipartition, $p_\tau^{(\mathrm{eq})} = 1/N$. For symmetric
$W_{\tau\rho} = W_{\rho\tau}$ (detailed balance), all eigenvalues of $W$ are
real, and the equilibrium state is equipartition. Eigenvalues corresponding to
eigenvalues $|\lambda|<1$ shrink to zero for infinitely many time steps. We
conclude that for proper Markov chains the probability distribution typically
approaches equilibrium as time goes to infinity.

\paragraph*{Markov chains for general probabilistic states of\\local chains}

We have already discussed in sect.\,\ref{sec:step_evolution_operator} that the
evolution of general probabilistic states of a local chain does not obey the
condition for Markov chains, except for the special case of unique jump chains
that are not discussed in this section. A general probabilistic state at time
$t$ is described by the classical density matrix $\rho'(t)$, whose diagonal
elements are the local probabilities $p_\tau(t)$. A Markov chain requires
$p_\tau(t+\varepsilon)$ to be computable from $p_\tau(t)$, which results for
unconstrained $\rho'$ in the rather restrictive condition \eqref{eq:SE9}.

The general evolution of the density matrix is given by the matrix equation
\eqref{eq:DM38}. We may perform a change of basis such that the step evolution
operator is diagonalized,
\begin{equation}
\tilde{S}_{\tau\rho}(t) = \lambda_\tau \delta_{\tau\rho}.
\label{eq:MCB2}
\end{equation}
In this basis one has (no index sum here)
\begin{equation}
\tilde{\rho}_{\tau\rho}(t+\varepsilon) = \lambda_\tau \lambda_\rho^{-1}
\tilde{\rho}_{\tau\rho}(t),
\label{eq:MCB3}
\end{equation}
such that the diagonal elements $\tilde{\rho}_{\tau\tau}$ are invariant. For all
$t$ they keep their boundary values at $t_\mathrm{in}$. This reflects the
discussion in
sect.\,\ref{sec:partial_loss_of_memory_and_emergence_of_quantum_mechanics} that
the loss of boundary information is often not easily visible in the evolution of
the density matrix or the local probability distribution. For a proper Markov
chain the local probability distribution typically converges towards a unique
equilibrium probability distribution, for which the memory of the detailed
initial probability distribution is lost. This contradicts the preserved
diagonal values $\tilde{\rho}_{\tau\tau}$ found from eq.\,\eqref{eq:MCB3} unless
$\rho'(t_\mathrm{in})$ has particular properties. For arbitrary
$\rho'(t_\mathrm{in})$ the step evolution operator cannot realize a Markov chain
with a unique asymptotic equilibrium state.

\paragraph*{Markov chains for particular classes of probabilistic states}

While for many forms of step evolution operators Markov chains are not found for
general probabilistic states, they will be realized for particular classes of
probabilistic states, or particular classes of density matrices $\rho'(t)$.
These particular classes are often approached asymptotically after a sufficient
number of time steps. For constrained density matrices $\rho'(t)$ the condition
\eqref{eq:SE9} does not need to hold.

As an example we take the Ising chain in
sect.\,\ref{sec:influence_of_boundary_conditions}. If the correlation length is
small as compared to the length of the chain, the conjugate wave function has
reached the equilibrium value $\bar{q}_\mathrm{eq}$ for all $t$ in the region
around $t_\mathrm{in}$. The density matrix takes in this region a form 
independent of the second index $\rho$,
\begin{equation}
\rho'_{\tau\rho}(t) = \tilde{q}_\tau(t) \left( \bar{q}_\mathrm{eq} \right)_\rho
= \rho'_{\tau\tau}(t) = p_\tau(t).
\label{eq:MCB4}
\end{equation}
The inverse of the step evolution operator \eqref{eq:BC4} reads
\begin{equation}
\hat{S}^{-1} = \frac{1}{2\sinh\beta} \begin{pmatrix}
e^\beta & -e^{-\beta} \\ 
-e^{-\beta} & e^\beta
\end{pmatrix},
\label{eq:MCB5}
\end{equation}
and one has
\begin{equation}
(\bar{q}_\mathrm{eq})^\mathrm{T} \hat{S}^{-1} =
(\bar{q}_\mathrm{eq})^\mathrm{T}.
\label{eq:MCB6}
\end{equation}

For this type of density matrix the evolution equation becomes 
\begin{equation}
\rho'(t+\varepsilon) = \hat{S} \rho'(t).
\label{eq:MCB7}
\end{equation}
One obtains for the local probabilities
\begin{align}
\begin{split}
p_\tau(t+\varepsilon) &= \rho'_{\tau\tau}(t+\varepsilon) = \sum_\rho
\hat{S}_{\tau\rho} \rho'_{\rho\tau}(t) \\
&= \sum_\rho \hat{S}_{\tau\rho} \rho'_{\rho\rho}(t) = \sum_\rho
\hat{S}_{\tau\rho} p_\rho(t).
\end{split}
\label{eq:MCB8}
\end{align}
This is a Markov chain with $W=\hat{S}$. The probabilities $p_\tau(t)$ follow
the same evolution as the components of the wave function $\tilde{q}_\tau(t)$,
approaching the equilibrium state as $t$ increases.

This example generalizes to all local chains for which the equilibrium state is
equipartition, with $(\bar{q}_\mathrm{eq})_\tau$ independent of $\tau$.
Eqs.\,\eqref{eq:MCB4}, \eqref{eq:MCB6} imply eqs.\,\eqref{eq:MCB7},
\eqref{eq:MCB8} without further information about the specific form of the step
evolution operator. Equipartition is actually not crucial for this type of
realization of Markov chains. It is sufficient that the conjugate wave function
has reached an equilibrium state for which eq.\,\eqref{eq:MCB6}, and therefore
also eq.\,\eqref{eq:MCB7}, holds. This results in a Markov chain with transition
probabilities
\begin{equation}
W_{\tau\rho} = \hat{S}_{\tau\rho}
\frac{(\bar{q}_\mathrm{eq})_\tau}{(\bar{q}_\mathrm{eq})_\rho}.
\label{eq:MCB9}
\end{equation}

\paragraph*{General Markov chains}

The reason why Markov chains can be realized for constrained classical density
matrices is rather simple. If all matrix elements $\rho'_{\tau\rho}(t)$ can be
expressed as linear combinations of the local probabilities $p_\tau(t)$, the
evolution equation for the density matrix can be rewritten as a linear evolution
equation for the probabilities. A Markov chain is realized if the coefficients
$W_{\tau\rho}$ of this linear evolution equation are all positive,
$W_{\tau\rho}\geq 0$. This setting is realized for an equilibrium conjugate wave
function. For a given $\bar{q}_\mathrm{eq}$ both the local probabilities and the
elements of the classical density matrix can be written in terms of the wave
function $\tilde{q}(t)$, 
\begin{equation}
p_\tau(t) = \tilde{q}_\tau(t) (\bar{q}_\mathrm{eq})_\tau,\quad 
\rho'_{\tau\rho}(t) = \tilde{q}_\tau(t) (\bar{q}_\mathrm{eq})_\rho,
\label{eq:MCB10}
\end{equation}
resulting in the linear relation
\begin{equation}
\rho'_{\tau\rho}(t) =
\frac{(\bar{q}_\mathrm{eq})_\rho}{(\bar{q}_\mathrm{eq})_\tau}\, p_\tau(t).
\label{eq:MCB11}
\end{equation}

This reasoning reveals the general mechanism how proper Markov chains can emerge
in our setting. If the elements of a class of density matrices depend
effectively on a number of parameters equal to or smaller than $N$, the
off-diagonal elements of $\rho'$ may become computable in terms of the diagonal
ones. In this case the evolution law only involves the probabilities. If linear
with positive coefficients it becomes a Markov chain. We can also understand why
the evolution of general probabilistic states cannot be described by Markov
chains. In the general case the classical density matrix contains boundary
information from both the initial and the final boundary. This probabilistic
information exceeds the information in the local probability distribution. Our
solution of the boundary problem for the Ising chain in
sect.\,\ref{sec:influence_of_boundary_conditions} demonstrates this clearly. 

Markov chains can arise effectively in many situations for which a large enough
part of the boundary information is effectively lost, such that the effective
evolution law needs a number of parameters smaller or equal to the effective
number of states for the description of the local probabilistic information at a
given time $t$. We have discussed explicitly only two extreme cases of Markov
chains, namely unique jump chains and chains that converge to a unique
equilibrium state. In between, many different types of evolution can be realized
by Markov chains. This includes periodic behavior of local probabilities for
chains that are not unique jump chains.

\section{Quantum field theory}
\label{sec:quantum_field_theory}

Quantum field theory emerges in a very natural and straightforward way from our
general probabilistic setting. We have seen that already very simple unique jump
chains or probabilistic cellular automata, as the transport automata, describe
two-dimensional quantum field theories for free massless fermions. Actually, it
is directly the quantum field theory which arises at the most basic level.
Single-particle quantum mechanics obtains as a special case or approximation,
rather than being the starting point as in the historical development of quantum
field theory. The generic situation for interesting probabilistic systems
describes states with many particles, with a continuum limit admitting an
infinite number.

In this chapter we extend the discussion of the simplest system. In
sect.~\ref{sec:Fermionic_quantum_field_theory_with_interactions} we modify the
updating rule for two species of right-moving and left-moving bits such that the
color of the bits can change when different particles meet on a site. This will
describe a two-dimensional quantum field theory for fermions with interactions.
We express the step evolution operator for this probabilistic automaton in terms
of creation and annihilation operators for fermions. This underlines the
observation that bits or Ising spins can be associated to fermions. It is
possible to construct probabilistic automata for rather rich classes of
fermionic field theories~\cite{FPCA, CWCA, Wetterich:2022kif,
Wetterich:2022zql}. We focus here, however, on a particularly simple case which
can be considered as a discretization of a type of Thirring~\cite{THI, KLA, AAR,
FAIV} or Gross-Neveu~\cite{GN, WWE, RSHA, RWP, SZKSR} models.

The possibility of basis transformations is a crucial feature of quantum field
theories. They permit to switch from position space to momentum space. Classical
statistical systems described by probability distributions for time-local
subsystems provide no room for basis transformations. Only the description of
the time-local probabilistic information in terms of wave functions or the
density matrix opens the door for implementing this powerful tool. In
sect.~\ref{sec:Change_of_basis_and_similarity_transformations} we develop the
systematic setting for basis transformations within the time-local subsystem. In
sect.~\ref{sec:fourier_transform_for_cellular_automata} we focus on the Fourier
transform from position- to momentum-space. On this basis we construct in
sect.~\ref{sec:Particles_and_antiparticles} the ``quantum field theory vacuum'',
which corresponds to a half-filled state in direct analogy to the ``Dirac sea''.
The presence of antiparticles follows for a wide variety of systems. ``Particle
excitations'' of this vacuum cover both particles and antiparticles. They are
related by a discrete symmetry. For the quantum field theory vacuum the
Hamiltonian is positive for all multi-particle states. This boundedness of the
Hamiltonian is another key feature familiar from quantum field theories.

\subsection{Fermionic quantum field theory with interactions}
\label{sec:Fermionic_quantum_field_theory_with_interactions}

As we have shown in sect.~\ref{sec:probabilistic_celluar_automata}, all local probabilistic cellular automata describe quantum systems. Examples of simple probabilistic cellular automata discussed so far are the clock system in sect.~\ref{sec:clock_systems} or the two-dimensional diagonal Ising models in sect.~\ref{sec:free_particles_in_two_dimensions}, \ref{sec:complex_structure}. In the inverse direction, the question arises which interesting quantum models can find a representation as a probabilistic cellular automaton. Free massless fermions are too simple for an implementation of many interesting properties of quantum field theories. One would like to have at least a model with interactions. In this section we present a probabilistic cellular automaton that describes an interacting fermionic quantum field theory in two dimensions. We discuss a model with Lorentz symmetry in the naive continuum limit. It corresponds to a particular type of discretized Thirring model.

We first formulate this model as a generalized Ising model. This approach opens the possibility to investigate its properties by standard classical statistical methods for Ising type models, as numerical simulations or renormalization group approaches. Second, we express its step evolution operator in terms of fermionic annihilation and creation operators~\cite{FQFTPCA}. This results in an equivalent picture as a discrete fermionic quantum field theory. The bit-fermion map which represents the model by a Grassmann functional integral becomes plausible, even though we postpone an explicit construction to later parts of this work.

\addtocontents{toc}{\protect\newpage}
\addtocontents{toc}{\vspace*{5.1em}} 

\subsubsection{Interacting Dirac automaton}
\label{sec:interacting_dirac_automaton}

With a suitable complex structure the action \eqref{eq:459} describes two free massless Weyl fermions. The index $\alpha=1,2$ denotes two different flavors or colors. This can be extended to free massless Dirac fermions by adding two more Ising spins per site
corresponding to left-movers.
Two-dimensional Dirac chains for two colors have at every site $(t,x)$ or $(m,j)\mathrel{\widehat{=}} (m_{1},m_{2}) $ four species of Ising spins, $s_{R \alpha}$, $s_{L \alpha}$, $\alpha = 1,2$. Free massless Dirac fermions are described by
the limit $\beta\to \infty$ of the generalized Ising model given by
\begin{multline}
\label{OS8}
\cL(m) = -\beta \sum_{\alpha = 1,2} \Big\{ s_{R \alpha}(m+1,j+1) s_{R \alpha}(m,j) \\
+ s_{L \alpha}(m+1,j-1) s_{L \alpha}(m,j) -2 \Big\}\,.
\end{multline}
At this point the four spins evolve independently. We associate $\alpha$ with colors, red for $\alpha = 1$ and green for 
$\alpha = 2$. Thus there are red and green particles moving to the right (increasing $x$ as $t$ increases), and different red and green
particles moving to the left (decreasing $x$ as $t$ increases). The free Dirac automaton is very simple. At each time step the right
movers move one position to the right, and the left movers one position to the left. 

An interaction can be introduced by the following prescription for the cellular automaton:
Whenever a red left mover and a green right mover arrive at the same location $x$, they interchange color provided that no further 
particles are present at $x$. 
In the subsequent time step the left mover will be green, and the right mover red.
Similarly, whenever a red right mover encounters a green left mover at the same $x$, colors are exchanged in the 
absence of other particles. 
We illustrate this automaton in Fig.\,\ref{fig:OS1}.
This type of interaction preserves the numbers of left movers and right movers separately. Also
the numbers of red and green particles are preserved separately. The numbers of green right movers and so on are, however, no longer
preserved once this type of interaction is added. We will later add a further interaction for which two red particles meeting at $x$ become green, and two green particles turn to red. For the sake of simplicity of the presentation we first omit this additional piece of the interaction.

The step evolution operator for this cellular automaton can be written as as matrix product of two pieces
\begin{equation}
\label{OS9}
\hat{S}(m) = \hat{S}_\text{int}(m) \hat{S}_\text{free}(m)\,.
\end{equation}
Here $\hat{S}_\text{free}(m)$ is the step evolution operator for free Dirac fermions corresponding to eq.~\eqref{OS8}, while the 
``interaction part" $\hat{S}_\text{int}(m)$ describes an interaction between the fermions which can be associated to some type of scattering.
The matrix $\hat{S}_\text{int}$ is a product of ``local exchange matrices" $\hat{E}(j)$ or $\hat{E}(x)$,
\begin{equation}
\label{OS10}
\hat{S}_\text{int}(m) = \prod_j \hat{E}(j) = \prod_x \hat{E}(x)\,.
\end{equation}

The matrix $\hat{E}(x)$ acts only on configurations of the occupation numbers $n_\gamma(t+\varepsilon,x)$ or $n_\gamma(m+1,j)$, while
it does not depend on configurations $n_\gamma(m+1,j')$ for $j' \neq j$ or on $n_\gamma(m,j)$. It can be seen as a direct product 
matrix, with one factor at position $j$ acting on the ``internal space" of configurations for the occupation numbers $n_\gamma(m+1,j)$,
and the other factors being unit matrices,
\begin{equation}
\label{OS10A}
\hat{E}(j) = 1 \otimes 1 \otimes 1 \ldots \otimes \tilde{E}(j) \otimes 1 \otimes \ldots \otimes 1\,.
\end{equation}
Equivalently, one could realize eq.~\eqref{OS9} by two consecutive time steps~\cite{FPCA}. The first time step by $\tilde\epsilon=\epsilon/2$ implements the free propagation, the second by $\tilde\epsilon=\epsilon/2$ the scattering. In this case the scattering part $\tilde{E}(j)$ involves occupation numbers $n_\gamma(m,j)$ and $n_\gamma(m+1,j)$. The net result is the same, with $\hat S(m)$ in eq.~\eqref{OS9} now combining to updatings by $\tilde\epsilon$.

In our case $\gamma = (\eta, \alpha)$ is a double index, with $\eta = (R,L) = 1, 2$ distinguishing between right and left movers
and $\alpha = 1, 2$ specifying the color. Corresponding to the four values of $\gamma$ the number of internal 
states (at a given $x$) equals $N_\text{int} = 2^4 = 16$, labeled by $\nu = 1 \ldots 16$. The matrix $\tilde{E}(j)$ is a 
$16 \times 16$-matrix. It is a unique jump operator in internal space. 

The internal states $\nu$ can be identified with sequences of the four occupation numbers for the four combinations of internal
indices,
\begin{align}
\label{OS11}
\gamma &= 1: (R1)\,,\,\,(Rr)\,,\,\,(\eta = 1, \alpha = 1)\ ,\\
\gamma &= 2: (R2)\,,\,\,(Rg)\,,\,\,(\eta = 1, \alpha = 2)\ ,\nn\\
\gamma &= 3: (L1)\,,\,\,(Lr)\,,\,\,(\eta = 2, \alpha = 1)\ ,\nn\\
\gamma &= 4: (L2)\,,\,\,(Lg)\,,\,\,(\eta = 2, \alpha = 2)\ .\nn
\end{align}
For example, the state $(1,0,0,1)$ corresponds to the presence of a red right mover and a green left mover, while no other particles
are present. In terms of Ising spins this describes the configuration $(1,-1,-1,1)$. In these terms the exchange matrix finds a very simple description: At every $x$ it exchanges two internal states with two particles,
\begin{equation}
\label{OS12}
(1,0,0,1) \leftrightarrow (0,1,1,0)\,.
\end{equation}
All other internal configurations are left invariant.

The overall feature of this automaton looks still rather simple. Left movers move left, right movers move right, and occasionally 
they change color. If we ask the simple question, what is the expectation value for a green left mover at a given $x$ and $t$, e.g.
$\braket{s_{22}(m,j)}$, we need to follow the colors in the different trajectories. This is a rather complex task, since for
the color-trajectory of a given left moving particle the trajectories of the other particles matter. We have depicted a characteristic
multi-particle trajectory in Fig.~\ref{fig:OS1}. It is clearly visible that the trajectory of a given green or red particle no longer follows
a diagonal line
and depends on the presence or absence of other particles. 

\begin{figure}[h!]
\resizebox{.4\textwidth}{!}{
\begin{tikzpicture}
\draw [help lines] (0,0) grid (12,12);
\draw [red, line width=4, rounded corners] (4,0) -- (5,1) -- (3,3) -- (4,4) -- (2,6) -- (3,7) -- (1,9) -- (4,12);
\draw [red, line width=4, rounded corners] (8,0) -- (9,1) -- (5,5) -- (10,10) -- (8,12) -- (7,11) -- (11,7) -- (6,2) -- (8,0);
\draw [green, line width=4, rounded corners] (0,0) -- (3,3) -- (0,6);
\draw [green, line width=4, rounded corners] (0,4) -- (2,6) -- (0,8) -- (1,9) -- (0,10);
\draw [green, line width=4, rounded corners] (6,0) -- (5,1) -- (6,2) -- (4,4) -- (5,5) -- (3,7) -- (7,11) -- (6,12);
\draw [green, line width=4, rounded corners] (10,0) -- (9,1) -- (12,4);
\draw [green, line width=4, rounded corners] (12,6) -- (11,7) -- (12,8) -- (10,10) -- (12,12);
\draw [green, line width=4, rounded corners] (12,10) -- (10,12);
\end{tikzpicture}
}
\caption{Trajectory for interacting Dirac automaton. A six particle state is distributed over $12$ positions $x$. Boundary conditions in $x$ are periodic.}
\label{fig:OS1}
\end{figure}

Consider for definiteness $64$ points $x$. For a cellular automaton with a fixed initial spin configuration one has to follow the time 
evolution of $256$ spins or occupation numbers by stepwise updating. This remains a relatively simple task. For a probabilistic cellular
automaton the initial condition is a probability distribution over initial spin configurations. This translates to a probability 
distribution over the associated trajectories. At each time point one has $2^{256}$ local states $\tau$, and correspondingly 
$2^{256}$ probabilities $p_\tau(t)$. 
Updating $2^{256}$ 
spin configurations and the associated probabilities
goes far beyond any computational capacity. One needs other
methods for a computation of expectation values or correlations for this type of probabilistic initial conditions. The reduction of 
complexity by investigating sectors with fixed particle numbers is only of limited help if the number of particles is large.
The problem gets even worse if the number of sites $x$ and the number of time steps $t$ increases to very large values such that a continuum limit applies. An analytic understanding of the evolution of the wave function in the continuum limit, or an approximate solution of the associated quantum field theory may become very helpful tools.

\paragraph*{Interacting Dirac chains}

One possibility uses the formulation of cellular automata as generalized Ising models. For this purpose we have to compute
$\cL(m)$ in dependence on the spins $s_\gamma(m,j)$ and $s_\gamma(m+1,j)$. We start by investigating $\cL(m)$ for a step 
evolution operator $\hat{S}_\text{int}$, adding later the effect of $\hat{S}_\text{free}$. For this task $\cL(m) = \sum_j \cL(m,j)$
decays into independent terms for the different positions $j$ or $x$. In the occupation number basis we employ the basis functions
\begin{align}
\nonumber
h_{(1001)} &= n_1 n_4 (1-n_2) (1-n_3)\,,\\
h_{(0110)} &= n_2 n_3 (1-n_1) (1-n_4)\,.
\end{align}
Here $n_\gamma$ stands for $n_\gamma(m,j)$. Similarly, we denote by $h'_{(1001)}$ and $h'_{(0110)}$ the same basis functions in
terms of $n'_\gamma$, which stands for $n_\gamma(m+1,j)$.
The product
\begin{equation}
\label{OS14}
X_1 = h'_{(1001)} h_{(0110)} + h'_{(0110)} h_{(1001)}
\end{equation}
equals one if $\tau'=(1001)$, $\tau=(0110)$ or $\tau'=(0110)$, $\tau=(1001)$, and vanishes for all other combinations. 

We can write, for $\beta \to \infty$,
\begin{equation}
\label{OS15}
\cL(m,j) = -\beta \left\{ \sum_\gamma s'_\gamma s_\gamma - 4 + X \right\}\,,
\end{equation}
with 
\begin{align}
\nonumber
X &= 8(X_1 - X_2)\,,\\
\nonumber
X_1 &= h'_{(1001)} h_{(0110)} + h'_{(0110)} h_{(1001)}\,,\\
\label{OS16}
X_2 &= h'_{(1001)} h_{(1001)} + h'_{(0110)} h_{(0110)}\,.
\end{align}
The first part without $X$ corresponds to a unit step evolution operator where every configuration $\tau$ is mapped to the same
configuration $\tau'$. This part of the bracket vanishes for $\tau'=(1001)$, $\tau=(1001)$ and $\tau'=(0110)$, $\tau=(0110)$.
For the combination $\tau'=(1001)$, $\tau=(0110)$ it takes the value $-8$, such that this sequence is suppressed.
Adding $8X_1$ erases this suppression -- the bracket vanishes now for the exchange of configurations \eqref{OS12}. Subtracting $8X_2$
the bracket takes the value $-8$ for the ``forbidden pairs" $\tau'=(1001)$, $\tau=(1001)$ and $\tau'=(0110)$, $\tau=(0110)$,
suppressing them as it should be. 

The part $X$ in eqs.~\eqref{OS15}, \eqref{OS16} reflects the interaction, 
\begin{align}
\label{OS17}
\nonumber
X &= 8 ( h'_{(1001)} h_{(0110)} + h'_{(0110)} h_{(1001)} \\
  & \quad - h'_{(1001)} h_{(1001)} - h'_{(0110)} h_{(0110)} ) \nonumber \\
  &= -8 A' A\,,
\end{align}
where
\begin{equation}
\label{OS18}
A = h_{(1001)} - h_{(0110)}\,, \quad A' = h'_{(1001)} - h'_{(0110)}\,.
\end{equation}
In terms of occupation numbers one has
\begin{align}
\label{OS19}
A &= n_1 n_4 (1-n_2) (1-n_3) - n_2 n_3 (1-n_1) (1-n_4) \nonumber \\
  &= n_1 n_4 - n_2 n_3 + (n_2 - n_1)n_3 n_4 + (n_3 - n_4)n_1 n_2\,.
\end{align}
Expressed in terms of Ising spins this yields
\begin{align}
\label{OS20}
A &= \frac{1}{8} \left[(s_1 + s_4)(1+s_2 s_3)-(s_2 + s_3)(1+ s_1 s_4) \right] \nonumber \\
&=\frac{1}{8} \left[(s_{L1} + s_{R2})(1+s_{L2} s_{R1})-(s_{L2} + s_{R1})(1+ s_{L1} s_{R2}) \right]\,.
\end{align}

Adding finally the effect of $\hat{S}_\text{free}$ in eq.~\eqref{OS9} replaces $s_{R\alpha}(m,j)$ by $s_{R\alpha}(m,j-1)$  and
$s_{L\alpha}(m,j)$ by $s_{L\alpha}(m,j+1)$. In conclusion, one finds for the interacting Ising chain
\begin{equation}
\label{OS21}
\cL(m) = \cL_\text{free}(m) + \cL_\text{int}(m)\,,
\end{equation}
where $\cL_\text{free}(m)$ is given by eq.~\eqref{OS8}. The interaction piece involves up to six Ising spins 
\begin{align}
\nonumber
\cL_\text{int}(m) = \frac{\beta}{8} &\sum_j \Big\{ \Big[ \big( s_{L1}(m+1,j)+ s_{R2}(m+1,j) \big) \\
\nonumber
&\times \big( 1 +  s_{L2}(m+1,j)  s_{R1}(m+1,j) \big) \\
\nonumber
&- \big(  s_{L2}(m+1,j) +  s_{R1}(m+1,j) \big) \\
\nonumber
&\times \big( 1 +  s_{L1}(m+1,j)  s_{R2}(m+1,j) \big) \Big] \\
\nonumber
&\times \Big[ \big ( s_{L1}(m,j+1) +  s_{R2}(m,j-1) \big) \\
\nonumber
&\times \big( 1 +  s_{L2}(m,j+1)  s_{R1}(m,j-1)\big) \\
\nonumber
&-\big(  s_{L2}(m,j+1) +  s_{R1}(m,j-1) \big) \\
\label{OS22}
&\times \big( 1 +  s_{L1}(m,j+1)  s_{R2}(m,j-1) \big) \Big] \Big\}\,.
\end{align}
The model \eqref{OS21}, \eqref{OS22} is a well behaved statistical system to which boundary terms can be added without
major problems. For the computation of expectation values as $\braket{s_{R1}(m,j)}$ or correlations as 
$\braket{s_{R1}(m,j) s_{L1}(m',j')}$ one may proceed to a Monte Carlo simulation for finite large $\beta$ and take
the limit $\beta \to \infty$. Alternatively one could employ approximative analytic techniques as block spinning or 
functional renormalization.

\paragraph*{Continuum limit}

A key question concerns the continuum limit of this type of generalized Ising model. More precisely, one is interested in the simultaneous limit $\beta\to\infty$, $\epsilon\to0$. Here some characteristic ``physical'' scale in the probability distribution or density matrix, for example momentum, is kept fixed. The number of sites on the chain and the number of time steps diverges in the continuum limit. The continuum limit of this type of model is widely unexplored. The new feature as compared to a similar model with finite $\beta$ or many other classical statistical models is the unitary evolution if one moves from one time-hypersurface to the next. The boundary information is not lost.

In the naive continuum limit, where one takes formally $\epsilon=0$ in the equivalent Grassmann functional integral for fermions, leads to a two-dimensional continuum quantum field theory for interacting fermions, as we will discuss briefly below. This naive continuum limit shows Lorentz symmetry. It needs to be established, however, if for the renormalization flow towards $\epsilon\to0$ the Lorentz symmetry violating terms of the discrete formulation are indeed ``irrelevant operators'' that vanish in the continuum limit.

\paragraph*{Propagation of local probabilistic information}

One may investigate the evolution of the local probabilistic information and hope to find some properties helping to gain insights into the behavior of the model. Unique jump chains permit for a closed update of the local probability distribution,
\begin{equation}
\label{OS23}
p_\tau(t+\epsilon) = \hat{S}_{\tau \rho}(t) p_\rho(t)\,,
\end{equation}
where $\hat{S}_{\tau \rho} = (\hat{S}_\text{int})_{\tau \sigma} (\hat{S}_\text{free})_{\sigma \rho}$ is independent of $t$ in our case.
Starting with some initial probability distribution $\{p_\tau(t_\text{in})\}$, one can use eq.~\eqref{OS23} to obtain
$\{p_\tau(t)\}$ for the time $t$, for which expectation values or correlations are to be computed. Similarly, one may follow the evolution
of the classical wave function or the classical density matrix, which reads for orthogonal step evolution operators
\begin{equation}
\label{OS24}
\rho'_{\tau \rho}(t+\epsilon) = \hat{S}_{\tau \alpha}(t)  \hat{S}_{\rho \beta}(t) \rho'_{\alpha \beta}(t)\,.
\end{equation}
For unique jump chains the diagonal elements of $\rho'$ evolve in a closed fashion due to the identity for a given $\tau$ (no sum)
\begin{equation}
\label{OS25}
 \hat{S}_{\tau \rho}  \hat{S}_{\tau \sigma} =  \hat{S}_{\tau \rho} \delta_{\rho \sigma}\,.
\end{equation}
Indeed, one has (no sum over $\tau$)
\begin{align}
\label{OS26}
p_\tau(t+\epsilon) &= \rho'_{\tau \tau}(t+\epsilon) 
= \sum_{\alpha \beta} \hat{S}_{\tau \alpha}(t) \hat{S}_{\tau \beta}(t) \rho'_{\alpha \beta}(t) \nonumber \\
&= \sum_\alpha \hat{S}_{\tau \alpha} \rho'_{\alpha \alpha}(t) = \sum_\alpha \hat{S}_{\tau \alpha} p_\alpha(t)\,.
\end{align}
In particular, a diagonal classical density matrix remains diagonal.

For the interacting Dirac automaton the step evolution operator is orthogonal, and we observe the identities
\begin{align}
\label{OS27}
\hat{S} &= \hat{S}_\text{int}\hat{S}_\text{free}\,, \quad \hat{S}^2_\text{int} = 1\,, \quad \hat{S}^T_\text{int} 
= \hat{S}_\text{int}\,,\nonumber\\
\hat{S}^{-1}_\text{free} &= \hat{S}^{T}_\text{free}\,, \quad \hat{S}^{T}\hat{S} = \hat{S}^T_\text{free} \hat{S}_\text{free} = 1\,.
\end{align}
The interaction part and the free part do not commute,
\begin{equation}
\label{OS28}
\left[ \hat{S}_\text{int}, \hat{S}_\text{free} \right] \neq 0\,.
\end{equation}
The evolution is non-trivial only because of this lack of commutativity. This can be seen from the property
\begin{align}
\label{OS29}
\hat{S}^2 &= \hat{S}_\text{int} \hat{S}_\text{free} \hat{S}_\text{int} \hat{S}_\text{free} \nonumber \\
&= \hat{S}^2_\text{free} + \left[ \hat{S}_\text{int}, \hat{S}_\text{free} \right] \hat{S}_\text{int} \hat{S}_\text{free}\,.
\end{align}
For a vanishing commutator the evolution after two time steps would be the same as for free Dirac fermions.

\subsubsection{Creation and annihilation operators}
\label{sec:creation_and_annihilation_operators}

The operator formalism for fermionic quantum field theories is based on creation and annihilation operators. We can express the step evolution operator in terms of these creation and annihilation operators. The explicit expression of the updating rule for automata in terms of the fermionic operators makes the correspondence to a fermionic model very apparent.

\paragraph*{Algebra of fermionic operators}

We represent the creation and annihilation operators explicitly by their action on the wave function for arbitrary bit configurations. This explicit implementation makes the role of these operators for the evolution of the time-local probabilistic information very concrete. Our starting point here are not abstract commutator relations. We rather express given evolution operators in terms of fermionic operators in order to get additional insight in characteristic structures of a given updating rule.

Let us first discuss a single spin, $M=1$, with two classical states. The two component wave function can be written in terms of
occupation number basis states
\begin{equation}
\label{OS30}
\ket{1} = \begin{pmatrix}
1\\0
\end{pmatrix}\,, \quad \ket{0} = \begin{pmatrix}
0\\1
\end{pmatrix}\,.
\end{equation}
The state $\ket{1}$ is occupied, $n=1$, $s=1$ and the state $\ket{0}$ is the empty state $n=0$, $s=-1$. The annihilation operator
$a$ and the creation operator $a^\dagger$ are given by the real matrices
\begin{equation}
\label{OS31}
a = \begin{pmatrix}
0 & 0 \\ 1 & 0
\end{pmatrix}\,, \quad 
a^\dagger = \begin{pmatrix}
0 & 1 \\ 0 & 0
\end{pmatrix} = a^T\,.
\end{equation}
They obey the relations 
\begin{equation}
\label{OS32}
a \ket{1} = \ket{0}\,, \quad a \ket{0} = 0\,, \quad a^\dagger \ket{1} = 0\,, \quad a^\dagger \ket{0} = \ket{1}\,,
\end{equation}
where the state $\ket{0}$ has to be distinguished from zero. The creation and annihilation operators obey the anticommutation relation
for fermions,
\begin{equation}
\label{OS33}
\left\{ a^\dagger, a \right\} = 1\,.
\end{equation}
The occupation number operator $\hat{n}$ is given by
\begin{equation}
\label{OS34}
\hat{n} = a^\dagger a = \begin{pmatrix}
1 & 0 \\ 0 & 0
\end{pmatrix}\,, \quad a a^\dagger = \begin{pmatrix}
0 & 0 \\ 0 & 1
\end{pmatrix}\,.
\end{equation}

For more than one Ising spin $(M>1)$ we define the annihilation operator for the particle $i$ as
\begin{align}
\label{OS35}
a_1 &= (a\otimes 1 \otimes 1 \otimes 1 \cdots)\,, \quad a_2 = (\tau_3 \otimes a\otimes 1 \otimes 1 \cdots)\,,\nonumber\\
a_3 &= (\tau_3 \otimes \tau_3 \otimes a \otimes 1 \cdots)\,, \quad a_4 = (\tau_3 \otimes \tau_3 \otimes \tau_3 \otimes a \cdots)\,.
\end{align}
The factors $\tau_3$ induce minus signs in appropriate places. The creation operators have the same insertion of chains of 
$\tau_3$-matrices,
\begin{equation}
\label{OS36}
a_i^\dagger = a_i^\mathrm{T}\,.
\end{equation}
We have introduced the $\tau_3$-factors in order to realize simple anticommutation relations
\begin{equation}
\label{OS37}
 \left\{ a_i, a_j \right\} = \left\{ a_i^\dagger, a_j^\dagger \right\} = 0\,, \quad \left\{ a_i, a_j^\dagger \right\} = \delta_{ij}\,.
\end{equation}
With this choice the antisymmetry of wave functions obtained by successive applications of creation operators to the vacuum state is
automatic. The particle number operators,
\begin{equation}
\label{OS38}
\hat{n}_i = a_i^\dagger a_i = 1 \otimes 1 \otimes \ldots \otimes \hat{n} \otimes 1 \otimes 1 \ldots\,,
\end{equation}
are diagonal, with appropriate eigenvalues 1 or 0 dictated by $\hat{n}$ at the place $i$.

\paragraph*{Two-bit example}

Consider the case of two particles, $M=2$, where the two annihilation and creation operators read
\begin{multline}
\label{OS39}
a_1 = a \otimes 1 = \begin{pmatrix}
0 & 0 & 0 & 0 \\ 0 & 0 & 0 & 0 \\ 1 & 0 & 0 & 0 \\ 0 & 1 & 0 & 0
\end{pmatrix},\, a_2 = \tau_3 \otimes a = \begin{pmatrix}
0 & 0 & 0 & 0 \\ 1 & 0 & 0 & 0 \\ 0 & 0 & 0 & 0 \\ 0 & 0 & -1 & 0
\end{pmatrix},\\
a_1^\dagger = a^\dagger \otimes 1 = \begin{pmatrix}
0 & 0 & 1 & 0 \\ 0 & 0 & 0 & 1 \\ 0 & 0 & 0 & 0 \\ 0 & 0 & 0 & 0
\end{pmatrix},\, a_2^\dagger = \tau_3 \otimes a = \begin{pmatrix}
0 & 1 & 0 & 0 \\ 0 & 0 & 0 & 0 \\ 0 & 0 & 0 & -1 \\ 0 & 0 & 0 & 0
\end{pmatrix}.
\end{multline}
With 
\begin{equation}
\label{OS40}
q = \begin{pmatrix}
q_1 \\ q_2 \\ q_3 \\ q_4
\end{pmatrix} = q_1 \ket{11} + q_2 \ket{10} + q_3 \ket{01} + q_4 \ket{00}\,,
\end{equation}
we may define a ``vacuum state"
\begin{equation}
\label{OS41}
\ket{0} = \ket{00} = \begin{pmatrix}
0 \\0 \\0 \\1
\end{pmatrix}\,.
\end{equation}
The one-particle states obtain by applying creation operators to the vacuum state
\begin{equation}
\label{OS42}
\ket{10} = a_1^\dagger \ket{0}\,, \quad \ket{01} = -a_2^\dagger \ket{0}\,,
\end{equation}
while for the two-particle state one has
\begin{align}
\label{OS43}
\ket{11} &= - a_1^\dagger a_2^\dagger \ket{0} = a_2^\dagger a_1^\dagger \ket{0} \nonumber\\
&= a_2^\dagger \ket{10} = a_1^\dagger \ket{01}\,.
\end{align}

\paragraph*{Exchange operator}

We may define the ``exchange operator" $\hat{t}_{12}$, which exchanges the two species of particles
\begin{align}
\label{OS44}
\hat{t}_{12} \ket{10} &= \ket{01}\,, \quad \hat{t}_{12} \ket{01} = \ket{10}\,, \\
\hat{t}_{12} \ket{11} &= \ket{11}\,, \quad \hat{t}_{12} \ket{00} = \ket{00}\,.
\end{align}
It is given by the matrix
\begin{equation}
\label{OS45}
\hat{t}_{12} = \begin{pmatrix}
1 &0 & 0 &0\\
0&0&1&0 \\
0&1&0&0\\
0&0&0&1
\end{pmatrix}, \left( \hat{t}_{12} \right)^2 = 1\,,
\end{equation}
which may be expressed in terms of annihilation and creation opertaors as
\begin{equation}
\label{OS45A}
\hat{t}_{12} = \widehat{n}_1 \widehat{n}_2 + (1-\widehat{n}_1)(1-\widehat{n}_2) - a_2^\dagger a_1 - a_1^\dagger a_2\,.
\end{equation}
We can express it in an exponential form
\begin{equation}
\label{OS46}
\hat{t}_{12} = \exp \left\{ -i \frac{\pi}{2} \hat{D}_{12} \right\}\,,
\end{equation}
where
\begin{equation}
\label{OSS42}
\hat{D}_{12}=
\begin{pmatrix}
0&0&0&0\\
0&1&-1&0\\
0&-1&1&0\\
0&0&0&0
\end{pmatrix}.
\end{equation}

In turn, we express $\hat{D}_{12}$ in terms of the annihilation and creation operators
\begin{equation}
\label{OSS43}
\hat{D}_{12} = a_1^\dagger a_2 + a_2^\dagger a_1 + \hat{n}_1(1-\hat{n}_2) + \hat{n}_2(1-\hat{n}_1)\,.
\end{equation}
We observe the identities 
\begin{equation}
\label{OSS44}
\hat{t}_{12} = 1 - \hat{D}_{12}\,, \quad \left(\hat{D}_{12}\right)^2 = 2 \hat{D}_{12}\,.
\end{equation}
Eq.~\eqref{OSS43} involves the projector $\hat{P}_1$ on the one-particle states
\begin{equation}
\label{OSS44A}
\hat{P}_1 =  \hat{n}_1(1-\hat{n}_2) + \hat{n}_2(1-\hat{n}_1)\,,
\end{equation}
which obeys
\begin{equation}
\label{OSS44B}
\hat{P}_1^2 =  \hat{P}_1\,, \quad \hat{P}_1 \hat{D}_{12} = \hat{D}_{12} \hat{P}_1 = \hat{D}_{12}\,.
\end{equation}
We can therefore write $\hat{D}_{12}$ in the form
\begin{equation}
\label{OSS44C}
\hat{D}_{12} = \left( 1 + a_1^\dagger a_2 + a_2^\dagger a_1\right) \hat{P}_1\,.
\end{equation}
The exchange operator obeys the simple relation
\bel{5.6.51A}
\hat t_{12}\hat P_1=\hat P_1\hat t_{12}=-\gl a_1^\dagger a_2+a_2^\dagger a_1\gr\ .
\ee

\paragraph*{Fermions in two dimensions}

For a single bit on each space position $j$ we need annihilation and creation operators for every space position $j$. We may order the positions and label them by $i(j)$, $i=1,\dots,\cM_2+1$. With the construction of $a_i$ above one identifies $a(j)=a_{i(j)}$, $a^\dagger(j)=a^\dagger_{i(j)}$. For several bits on each position $j$, labeled by $\gamma=1,\dots,N_c$, we may choose a suitable $i(j,\gamma)$, as $i=(j+\cM_2/2)N_c+\gamma$. This representation can be used for an arbitrary number of species of bits at a given position $j$.

For two species, $N_c=2$, we can also order the wave function in form of a matrix $q$, where the first index of the matrix element $q_{\tau\rho}$ refers to the configuration $\tau$ of the first species, and the second index to the configuration $\rho$ of the second species. With $2^{\cM_2+1}\times2^{\cM_2+1}$ matrices $a(j)$, $a^\dagger(j)$ defined as for a single bit at each position $j$ we can express the action of annihilation and creation operators $a_\gamma(j)$, $a_\gamma^\dagger(j)$ for the type $\gamma$ by matrix multiplication from the left or right,
\begin{align}
\label{NFD1}
a_1(j)(q)=&\,a(j)q\ ,\quad a_2(j)(q)=T_3qa^\dagger(j)\ ,\nonumber\\
a_1^\dagger(j)(q)=&\,a^\dagger(j)q\ ,\quad a_2^\dagger(j)(q)=T_3qa(j)\ .
\end{align}
The matrix
\bel{NFD2}
T_3=\tau_3\otimes\tau_3\otimes\tau_3\otimes\dots\otimes\tau_3
\ee
serves to implement the anticommutation of $a_1(j)$ and $a_2(j)$ etc. through the relation
\bel{NFD3}
\big\{T_3,a(j)\big\}=\big\{T_3,a^\dagger(j)\big\}=0\ .
\ee
The appearance of $a^\dagger(j)$ in the second relation~\eqref{NFD1} is due to the identity
\bel{NFD3A}
\gl a_2(j)(q)\gr_{\tau\rho}\sim a(j)_{\rho\rho'}q_{\tau\rho'}=q_{\tau\rho'}a^\dagger_{\rho'\rho}\ .
\ee

As an example we take a possible vacuum state for which $q_0$ is proportional to the unit matrix. In a direct product basis for $q$ it is given by
\bel{NFD4}
q_0=\mathcal{N}\cdot1\otimes1\otimes1\otimes\dots\otimes1\ ,\quad \mathcal N=2^{-\cM_2}\ .
\ee
With particle number operators
\bel{NFD5}
\hat n_1(j)(q)=a^\dagger(j)a(j)q\ ,\quad \hat n_2(j)(q)=qa^\dagger(j)a(j)\ ,
\ee
one finds
\bel{NFD6}
\langle n_1(j)\rangle=\langle n_2(j)\rangle=\frac12\ .
\ee

\subsubsection{Interaction part of the step evolution operator}
\label{sec:interaction_part_of_the_step_evolution_operator}

Our next task is the expression of the step evolution operator for interacting Dirac chains in terms of the fermionic annihilation and creation operators. This is particularly simple for the interaction part since only the bits on a given position $j$ or $x$ are involved. The propagation part will be reported somewhat later. We will express the interaction part of the step evolution operator in a complex exponential form. This allows for a direct extraction of a corresponding piece of the Hamiltonian.

The interaction part of the step evolution operator $\hat{S}_\text{int}$ can be expressed for every $x$ by a simultaneous exchange
for two pairs of states. We denote
\begin{align}
a_{R1} &= a\otimes 1 \otimes 1 \otimes 1 \,, \quad a_{R2} = \tau_3 \otimes a\otimes 1 \otimes 1\,,\nonumber\\
a_{L1} &= \tau_3 \otimes \tau_3 \otimes a \otimes 1\,, \quad a_{L2} = \tau_3 \otimes \tau_3 \otimes \tau_3 \otimes a \,,
\end{align}
and similarly for the creation operators. We observe that the operator
\begin{equation}
\label{OS46A}
a_{R2}^\dagger a_{L1}^\dagger a_{R1} a_{L2} = -a \otimes a^\dagger \otimes a^\dagger \otimes a
\end{equation}
yields zero when applied to any basis state different from $(1, 0, 0, 1)$. It maps
\begin{equation}
\label{OS47}
a_{R2}^\dagger a_{L1}^\dagger a_{R1} a_{L2}\, (1, 0, 0, 1) = - (0, 1, 1, 0)\,.
\end{equation}
Similarly, its transpose,
\begin{equation}
a_{L2}^\dagger a_{R1}^\dagger a_{L1} a_{R2}  = - a^\dagger \otimes a \otimes a \otimes a^\dagger\,,
\end{equation}
transforms
\begin{equation}
\label{OS48}
a_{L2}^\dagger a_{R1}^\dagger a_{L1} a_{R2}\, (0, 1, 1, 0) = -(1, 0, 0, 1)\,.
\end{equation}

We define the projector $\hat{P}_{11}$ which maps
\begin{align}
\label{OS50}
\hat{P}_{11} \, (1,0,0,1) &= (1,0,0,1)\,,\\
\hat{P}_{11} \, (0,1,1,0) &= (0,1,1,0)\,,
\end{align}
while the action of $\hat{P}_{11}$ produces zero for all other states. Thus $\hat{P}_{11}$ projects on two-particles states with one right mover and one left mover of different color.
In terms of the particle number operator one has 
\begin{equation}
\label{OS51}
\hat{P}_{11} = \hat{n}_{R1} (1-\hat{n}_{R2}) (1-\hat{n}_{L1}) \hat{n}_{L2} + (1-\hat{n}_{R1}) \hat{n}_{R2} \hat{n}_{L1} (1-\hat{n}_{L2})\,,
\end{equation}

The exchange matrix in internal space can be written in the form
\begin{equation}
\label{OS52}
\tilde{E}(j) = 1 - \tilde{D}(j)\,,
\end{equation}
where the $16\times 16$-matrix 
\begin{align}
\label{OS53}
\tilde{D}(j) &= \hat{P}_{11} + a_{R2}^\dagger a_{L1}^\dagger a_{R1} a_{L2} + a_{L2}^\dagger a_{R1}^\dagger a_{L1} a_{R2} \nonumber\\
&= \left( 1 + a_{R2}^\dagger a_{L1}^\dagger a_{R1} a_{L2} + a_{L2}^\dagger a_{R1}^\dagger a_{L1} a_{R2} \right) \hat{P}_{11}
\end{align}
involves operators at $j$ or $x$ and plays a role similar to $\hat{D}_{12}$ in eq.~\eqref{OSS44}. The exponential form reads
\begin{equation}
\label{OS54}
\tilde{E}(j) = \exp \left\{ -i \frac{\pi}{2} \tilde{D}(j) \right\}\,.
\end{equation}
This uses the property
\begin{equation}
\label{CA1}
\left( \frac{\tilde{D}(j) }{2}\right)^2 =\frac{\tilde{D}(j) }{2}\,.
\end{equation}

The interaction part of the step evolution operator is a direct product
\begin{equation}
\label{OS55}
\hat{S}_\text{int} = \tilde{E}(0) \otimes \tilde{E}(1) \cdots \otimes \tilde{E}(j) \cdots \,.
\end{equation}

We may define 
\begin{equation}
\label{OS56}
\hat{D}(j) = 1 \otimes 1 \otimes \cdots \otimes \tilde{D}(j) \otimes 1 \otimes \cdots\,,
\end{equation}
with $\tilde{D}(j)$ standing at position $j$ in the direct product. The product
\begin{equation}
\label{OS57}
\hat{D}(j) \hat{D}(j') = 1 \otimes 1 \otimes \cdots \tilde{D}(j) \otimes 1 \otimes \cdots \otimes \tilde{D}(j') \otimes 1 \cdots
\end{equation}
is commutative
\begin{equation}
\label{OS58}
\left[ \hat{D}(j), \hat{D}(j') \right] = 0\,.
\end{equation}
With the conventions of the type~\eqref{OS35} the annihilation operator $a_{R1}(j)$ has a string of $T_3$ factors at all sites $j'<j$, where $T_3=\tau_3\otimes\tau_3\otimes\tau_3\otimes\tau_3$, and a string of unit factors for all $j'>j$. With $T_3^2=1$ the operator $\hat D(j)$ is given by eq.~\eqref{OS53} with $a_\gamma$ replaced by $a_\gamma(j)$ etc.

In terms of $\hat{E}(j)$ or $\hat{D}(j)$ the interaction part of the step evolution operator has a product form
\begin{align}
\nonumber
\hat{S}_\text{int} &= \prod_j \hat{E}(j) = \prod_j \left(1-\hat{D}(j)\right) \\
\label{OS59}
&= \exp \left\{ -i \frac{\pi}{2} \sum_j \hat{D}(j) \right\}\,.
\end{align}
We can write $\hat S_{\text{int}}$ as unitary time evolution operator for a discrete time step $\epsilon$
\begin{equation}\label{CA2}
\hat S_{\text{int}}=\exp\big(-i\epsilon H_{\text{int}}\big)\ ,
\end{equation}
with hermitian interaction part of the Hamiltonian
\begin{equation}\label{CA3}
H_{\text{int}}=\frac{\pi}{2\epsilon}\sum_j\hat D(j)\ .
\end{equation}

\subsubsection{Time evolution}
\label{sec:time_evolution}

We aim at an expression of the time-evolution law for the wave function of the interacting Dirac automaton in terms of creation and annihilation operators. This makes the map to the evolution for a fermionic quantum field theory very direct. For this purpose we will need to express the propagation part of the step evolution operator in terms of the fermionic operators. Before doing so we briefly discuss some general aspects of the evolution.

\paragraph*{Perturbative expansion}

In a quantum field theory perturbation theory often plays an important role for the understanding of the behavior of particles. Perturbation theory is typically cast into the language of graphs with propagators and vertices. The basic structure can be seen very directly in our setting, even though there is no small coupling controlling the expansion.

Using the first expression in the product~\eqref{OS59} we may expand the full step evolution operator as
\begin{equation}
\label{OS60}
\hat{S} = \hat{S}_\text{free} - \sum_j \hat{D}(j) \hat{S}_\text{free} + \sum_{j, j'} \hat{D}(j) \hat{D}(j') \hat{S}_\text{free} + \cdots\,,
\end{equation}
A term $\hat{D}(j) \hat{S}_\text{free}$ describes a single two-particle interaction or scattering at position $j$. It differs from zero
only for configurations at time $m$ where a single right mover is present at $j-1$, and a single left mover at $j+1$. Furthermore, 
these two particles must belong to different species. These configurations are mapped at $m+1$ to the difference between the scattered
state and the free propagation state, according to eq.~\eqref{OS53}. Graphically, this is represented in Fig.~\ref{fig:OS2}.

\begin{figure}[h!]
\begin{tikzpicture}
\draw[red,ultra thick] (1,0) node[black,above left] {$R1$} --
(2,2) node[black,above] {$(R2,L1)-(R1,L2)$}; 
\draw[green,ultra thick] (2,2) -- (3,0) node[black,above right] {$L2$};
\draw [fill] (1,0) circle [radius=0.07];
\draw [fill] (2,2) circle [radius=0.07];
\draw [fill] (3,0) circle [radius=0.07];
\node at (-0.5,0) {$m$};
\node at (-0.5,2) {$m+1$};
\node [below] at (1,0) {$j-1$};
\node [below] at (2,-0.07) {$j$};
\node [below] at (3,0) {$j+1$};
\draw[green,ultra thick] (5,0) node[black,above left] {$R2$} --
(6,2) node[black,above] {$(R1,L2)-(R2,L1)$};
\draw[red,ultra thick] (6,2) -- (7,0) node[black,above right] {$L1$};
\draw [fill] (5,0) circle [radius=0.07];
\draw [fill] (6,2) circle [radius=0.07];
\draw [fill] (7,0) circle [radius=0.07];
\node [below] at (5,0) {$j-1$};
\node [below] at (6,-0.07) {$j$};
\node [below] at (7,0) {$j+1$};
\end{tikzpicture}
\caption{Basic two-particle scattering. The two graphs represent the two contributions to $-\hat{D}(n) \hat{S}_\text{free}$.}
\label{fig:OS2}
\end{figure}
Formally, the product $\hat{D}(j) \hat{S}_\text{free}$ applies first at $m$ a projector on one of the states $R1(j-1)L2(j+1)$ or 
$R2(j-1)L1(j+1)$, subsequently propagates these states at $m+1$ to $R1(j)L2(j)$ or $R2(j)L1(j)$, and finally takes the difference
between the unit operator and the simultaneous exchange operator. 
The operator $\hat{D}$ represents the vertex for the interaction in a perturbative expansion, while $\hat S_{\text{free}}$ accounts for the propagators.

We can next apply the sequence of evolution operators $\hat S(m)$ for the different time steps from $m$ to $m+1$ or from $t$ to $t+\eps$. If the density of particles is low, for most positions $x$ there will be no scattering at a given $t$. In this case only the first term in eq.~\eqref{OS60} contributes. One can collect all pieces with free propagation into a free propagator from $(t,x)$ to $(t+\Delta t, x+\Delta x)$, with a scattering at the endpoint of the free propagation. Furthermore, situations where the third term matters are very rare. In the first approximation this term can be neglected. This structure is analogous to perturbation theory in quantum field theory, with the difference that the couping governing the interaction is typically not small.

\paragraph*{Discrete differential evolution equation}

Using discrete lattice derivatives we can express the evolution law for the probabilistic automaton in the form of a discrete Schrödinger-type equation. This can be the basis for a continuum limit. In the presence of a complex structure this introduces the concept of a Hamiltonian in a discrete setting.

For the discrete time derivative
\begin{equation}
\label{OS61}
\partial_t \tilde{q} = \frac{1}{2\epsilon} \left( \tilde{q}(t+\epsilon) - \tilde{q}(t-\epsilon)\right) = W \tilde{q}(t)\,,
\end{equation}
we can exploit that the unique jump operator $\hat{S}$ is an orthogonal matrix. In the real formulation $W$ is given by
\begin{equation}
\label{OS62}
W = \hat{F} = \frac{1}{2\epsilon} \left( \hat{S}(t) - \hat{S}^{-1}(t) \right)\,,
\end{equation}
where $\hat{S}^{-1} = \hat{S}^T$. In terms of the free and interaction contribution this reads
\begin{align}
\label{OS63}
2 \epsilon W &= \hat{S}_\text{int} \hat{S}_\text{free} - \hat{S}^{-1}_\text{free} \hat{S}_\text{int} \nonumber \\
&= \prod_j \left[ \left( 1 - \hat{D}(j) \right) \hat{S}_\text{free} \right] - \prod_j \left[ \hat{S}_\text{free}^{-1} \left( 1 - \hat{D}(j) \right) \right]\,.
\end{align}

We can define the interaction part by 
\begin{equation}
\label{OS64}
\hat{D}_\text{int} = 1 - \prod_j \left( 1 - \hat{D}(j) \right)\,,
\end{equation}
resulting in 
\begin{equation}
\label{OS65}
W = W_\text{fee} + W_\text{int}\,,
\end{equation}
with 
\begin{equation}
\label{OS66}
W_\text{free} = \frac{1}{2\epsilon} \left( \hat{S}_\text{free} - \hat{S}^{-1}_\text{free} \right)\,,
\end{equation}
and 
\begin{equation}
\label{OS67}
W_\text{int} = -\frac{1}{2\epsilon} \left( \hat{D}_\text{int} \hat{S}_\text{free} - \hat{S}^{-1}_\text{free} \hat{D}_\text{int} \right)\,.
\end{equation}
The part $W_\text{int}$ describes particle scattering.  It differs from zero only for states with two or more particles. 
Products of the from $\hat{D}(j) \hat{D}(j')$, $j \neq j'$ can differ from zero only for states with four or more particles and so on
for products with several factors. In particular, for two-particle states the higher products do not contribute and one has
\begin{equation}
\label{OS68}
\hat{D}_\text{int} = \sum_j \hat{D}(j)\,.
\end{equation}

Eqs.~\eqref{OS61},~\eqref{OS65}-~\eqref{OS68} describe the general setting for the evolution equation for rather arbitrary local chains. Up to a factor $i$ the quantity $W$ plays the role of a Hamiltonian. In the following we will exploit that simple automata admit an explicit expression for the step evolution operator as complex exponentials of a suitable Hamiltonian. In the presence of a complex structure this will provide for a more direct access to a discrete Schrödinger equation for the time evolution.

\paragraph*{Propagation part of the step evolution operator}

Our next task is the expression of the step evolution operator for the right-transport or left-transport automata in terms of the fermionic creation and annihilation operators. This will define the free or kinetic part of the Hamiltonian. The explicit derivation encounters issues of ordering that we will not describe here explicitly. A detailed discussion and derivation can be found in ref.~\cite{FQFTPCA}.

The free part of the step evolution operator $\hat S_{\text{free}}$ consists of four independent factors for each of the right- and left-movers and each color
\begin{equation}\label{CA4}
\hat S_{\text{free}}=\hat S_1^{(R)}\otimes\hat S_2^{(R)}\otimes\hat S_1^{(L)}\otimes\hat S_2^{(L)}\ .
\end{equation}
The factor $\hat S_1^{(R)}$ involves only the operators $a_{R1}(j)$ and $a_{R1}^\dagger(j)$ for the different sites $j$, and similar for $\hat S_2^{(R)}$, $\hat S_1^{(L)}$, $\hat S_2^{(L)}$. Without writing the color index $\alpha=1,2$ explicitly one finds~\cite{FQFTPCA}
\begin{align}
\label{CA5}
\hat S^{(R)}=&N\bigg[\exp\bigg\{\sum_ja_R^\dagger(j+1)\big[a_R(j)-a_R(j+1)\big]\bigg\}\bigg]\ ,\nonumber\\
\hat S^{(L)}=&N\bigg[\exp\bigg\{\sum_ja_L^\dagger(j-1)\big[a_L(j)-a_L(j-1)\big]\bigg\}\bigg]\ .
\end{align}
This form is rather suggestive for a standard kinetic term based on a lattice derivative. It involves, however, an ordering operator $N$ which reflects the unique jump property of the automaton.

For a series expansion of the exponential the ordering operation $N$ puts all creation operators $a_\gamma^\dagger$ to the left of the annihilation operators $a_\gamma$, without modifying their respective order. Furthermore, it produces a factor $(-1)^{l-1}$ for $l\geq2$ for the $l$-th term in the exponential expansion $\exp(x)=\sum_l\big(x^l/l!\big)$. The ordering operation $N$ is important for maintaining the unique jump property and orthogonality of the operator $\hat S_{\text{free}}$. We emphasize that eq.~\eqref{CA5} is an operator identity which does not involve any model properties.

We can perform a partial continuum limit for which the ordering operator is omitted while orthogonality of $\hat S_{\text{free}}$ is maintained
\begin{align}
\label{CA6}
\hat S_R=&\exp\bigg\{-\frac12\sum_ja_R^\dagger(j)\big[a_R(j+1)-a_R(j-1)\big]\bigg\}\ ,\nonumber\\
\hat S_L=&\exp\bigg\{\frac12\sum_ja_L^\dagger(j)\big[a_L(j+1)-a_L(j-1)\big]\bigg\}\ .
\end{align}
The difference between~\eqref{CA6} and~\eqref{CA5} only concerns products of annihilation and creation operators at the same site $n$. These effects become small for smooth enough wave functions which vary only over characteristic distance scales $l$. We will discuss the suppression factor $\sim\epsilon/l$ for the omitted ordering effects below. The fact that the unique jump property is not maintained by eq.~\eqref{CA6} is, nevertheless, an important conceptual point.

With the partial continuum limit~\eqref{CA6} one obtains a simple expression for the free step evolution operator,
\begin{equation}\label{CA7}
\hat S_{\text{free}}=\exp\big(-i\epsilon H_{\text{free}}\big)\ ,
\end{equation}
with hermitian Hamiltonian $H_{\text{free}}$,
\begin{align}
\label{CA8}
H_{\text{free}}=&-\frac{i}{2\epsilon}\sum_x\sum_\alpha\Big\{a_{R\alpha}^\dagger(x)\big[a_{R\alpha}(x+\epsilon)-a_{R\alpha}(x-\epsilon)\big]\nonumber\\
&-a_{L\alpha}^\dagger(x)\big[a_{L\alpha}(x+\epsilon)-a_{L\alpha}(x-\epsilon)\big]\Big\}\ .
\end{align}
For a suitable lattice derivative
\begin{equation}\label{CA9}
\partial_xa(x)=\frac{1}{2\epsilon}\big[a(x+\epsilon)-a(x-\epsilon)\big]\ ,
\end{equation}
this yields the expression
\begin{equation}\label{CA10}
H_{\text{free}}=\sum_x\sum_\alpha\Big\{a_{R\alpha}^\dagger(x)\big(-i\partial_x\big)a_{R\alpha}-a_{L\alpha}^\dagger(x)\big(-i\partial_x\big)a_{L\alpha}\Big\}\ ,
\end{equation}
where one recognizes a type of multi-particle momentum operator. The omitted ordering effects yield corrections to $H_{\text{free}}$ that do not spoil its hermiticity.

\paragraph*{Multi-particle Schrödinger equation}

The expression of the step evolution operator in terms of fermionic annihilation and creation operators allows us to formulate a continuous time evolution equation for general states with arbitrary particle numbers. The step evolution operator can be written in the form
\begin{equation}\label{VP13}
\hat S=\exp\big(-i\epsilon H\big)\ ,
\end{equation}
where the Hamiltonian is defined by the relation
\begin{equation}\label{VP14}
\exp\big(-i\epsilon H\big)=\exp\big(-i\epsilon H_{\text{int}}\big)\exp\big(-i\epsilon H_{\text{free}}\big)\ .
\end{equation}
Since the step evolution operator $\hat S$ is a unitary (more precisely orthogonal real) matrix, the Hamilton operator is hermitian, $H^\dagger=H$. This is guaranteed even though its precise form may be rather complex, in particular of we use the exact ordered form~\eqref{CA5} for $\hat S_{\text{free}}=\exp\big(-i\epsilon H_{\text{free}}\big)$.

In the presence of an arbitrary complex structure consistent with the evolution, the time evolution of the complex wave function obeys the continuous Schrödinger equation
\begin{equation}\label{VP15}
i\partial_t\varphi_\tau(t)=H_{\tau\rho}\varphi_\rho(t)\ .
\end{equation}
Indeed, the solution of eq.~\eqref{VP15} is given by
\begin{equation}\label{VP16}
\varphi_\tau(t+\epsilon)=\exp\big(-i\epsilon H\big)_{\tau\rho}\varphi_\rho(t)=\hat S_{\tau\rho}\varphi_\rho(t)\ .
\end{equation}
For all discrete time steps by $\epsilon$ this solution coincides precisely with the discrete evolution steps of the automaton. The solution inbetween the discrete time points can be considered as a smooth interpolation. If one wishes one can rewrite eq.~\eqref{VP15} as a real differential equation for the real wave function $q'_\tau$ and $q^c_\tau$. We conclude that on very general grounds the time evolution of the wave function for automata takes the form of a Schrödinger equation, with all the important consequences known from quantum mechanics, as the superposition principle for solutions. The extension to the von-Neumann equation for the density matrix is straightforward.

The relations~\eqref{VP13},~\eqref{VP14} are complex identities that do not depend on a particular complex structure used for the wave function. A given complex structure defines the expression of the complex wave function $\vp$ in terms of the real wave function $q$. The evolution equation~\eqref{VP15} is valid for an arbitrary choice of the complex structure, provided that the step evolution operator is compatible with it.

\paragraph*{Quantum field theory}

The Schrödinger equation~\eqref{VP15} defines a quantum field theory for interacting fermions. Using our explicit representation of the creation and annihilation operators one obtains directly the action of the Hamiltonian on the complex wave function in the occupation number basis. Translating to the real picture one can infer the time-local probability distribution for the bit configurations of the automaton at any given time. One may wonder if all this formalism is necessary or useful for the description of an automaton with a rather simple evolution law. The strength of this formulation as a quantum field theory will become visible once one investigates in the continuum limit for a very large number of bits the behavior of local deviations from some vacuum state with high complexity. This holds, in particular, for more complex forms of the interaction for which an explicit following of trajectories for a high number of bits becomes impossible.

The annihilation and creation operators act on wave functions for arbitrary multi-particle states. In terms of annihilation and creation operators the particle number operators for the different species are given by
\begin{equation}\label{CA11}
\hat N_\gamma=\sum_j\hat n_\gamma(j)\ .
\end{equation}
One verifies the commutator relation for $H_{\text{free}}$ in eq.~\eqref{CA8},
\begin{equation}\label{CA12}
\big[\hat N_\gamma,H_{\text{free}}\big]=0\ .
\end{equation}
With
\begin{equation}\label{CA13}
a\hat n=a\ ,\quad \hat na=0\ ,\quad a^\dagger\hat n=0\ ,\quad \hat na^\dagger=a^\dagger
\end{equation}
one infers
\begin{align}
\label{CA14}
\big[\hat N_{R2},H_{\text{int}}\big]=\big[&\hat N_{L1},H_{\text{int}}\big]=-\big[\hat N_{R1},H_{\text{int}}\big]=-\big[\hat N_{L2},H_{\text{int}}\big]\nonumber\\
=\frac{\pi}{2\epsilon}\sum_j\Big(&a_{R2}^\dagger(j)a_{L1}^\dagger(j)a_{R1}(j)a_{L2}(j)\nonumber\\
-&a_{L2}^\dagger(j)a_{R1}^\dagger(j)a_{L1}(j)a_{R2}(j)\Big)\ .
\end{align}
The combinations $\hat N_{R1}+\hat N_{R2}$, $\hat N_{L1}+\hat N_{L2}$, $\hat N_{R1}+\hat N_{L1}$, $\hat N_{R2}+\hat N_{L2}$ commute with both $H_{\text{free}}$ and $H_{\text{int}}$ and correspond to conserved quantities.

For an investigation of the action of $H_{\text{free}}$ on a general wave function $q$ we employ a decomposition into basis vectors $Q_\tau$
\begin{equation}\label{CA15}
q(t)=q_{\tau} (t) Q_\tau\ ,\quad (Q_\tau)_\rho=\delta_{\tau\rho}\ .
\end{equation}
We use the property
\begin{equation}\label{CA16}
a_\gamma^\dagger(j)a_\gamma(j')Q_\tau=\begin{cases}\pm Q_{\tau'}&\ \text{if $\tau$ has $n_\gamma(j')=1$, $n_\gamma(j)=0$}\\ 0&\ \text{otherwise.}\end{cases}
\end{equation}
Here $\tau'$ obtains from $\tau$ by the switch from $n_\gamma(j')=1$, $n_\gamma(j)=0$ to $n_\gamma(j')=0$, $n_\gamma(j)=1$. The sign accounts for the $\tau_3$-factors in the definition of the creation and annihilation operators. For neighboring $j'$ and $j$ there are four $\tau_3$-factors inbetween the positions of $a_\gamma^\dagger(j'+1)$ and $a_\gamma(j')$, or $a_\gamma^\dagger(j'-1)$ and $a_\gamma(j')$, including the position of $a^\dagger$ or $a$ for the lowest $j'$ or $j$. A minus sign occurs if the number of spins at the corresponding sites is odd. The operator $a_\gamma^\dagger(j)a_\gamma(j')$ moves a particle of type $\gamma$ from $j'$ to $j$, provided the site $j$ is empty in the sense that there is no $\gamma$-particle.

For $H_{\text{int}}$ the operator $S_{\alpha\beta\gamma\delta}=a_\alpha^\dagger(j)a_\beta^\dagger(j)a_\gamma(j)a_\delta(j)$ with $\alpha$, $\beta$, $\gamma$, $\delta$ all different switches the particle type of two spins of type $\gamma$ and $\delta$ present at the site $j$ into the new particle types $\alpha$ and $\beta$. The result $S_{\alpha\beta\gamma\delta}Q_\tau$ differs from zero only for configurations $\tau$ with $n_\gamma(j)=n_\delta(j)=1$ and $n_\alpha(j)=n_\beta(j)=0$. There is an additional minus sign if $\alpha$ or $\beta$ correspond to a green hole, with no minus sign if both are green.

At least conceptually one can express the fully ordered form~\eqref{CA5} of the free step evolution operator in the Hamiltonian form~\eqref{CA7} and solve the operator equation~\eqref{VP14}. This demonstrates the emergence of a quantum field theory in the operator formulation. The Schrödinger equation~\eqref{VP15} acts on arbitrary states. As emphasized already, the formulation in terms of a continuous Schrödinger equation is expected to become useful if a continuum limit exists for a sufficiently smooth wave function.

The formulation of the evolution equation for the local probabilistic information in terms of fermionic creation and annihilation operators demonstrates that the concepts of quantum theories for fermions arise in a natural way for our generalized Ising model. Already
for the very simple cellular automaton discussed here the dynamics of many particle states can be quite involved. For smooth wave
function for only a few particles the continuum limit may lead to important simplifications. The corresponding Schr\"odinger equation
for a $n$-particle state can be solved by employing perturbative concepts that realize that particles move essentially freely, 
with only occasional scattering. The bit-fermion map to a quantum field 
theory for fermions represented by a Grassmann functional integral \cite{CWFIM,CWFCS} is a powerful tool for exact or approximate solutions of such systems. It holds for generalized Ising models in arbitrary dimensions. (For different maps between Ising spins or bosons to fermions which are specific to two-dimensional models see refs.~\cite{PLECH, BER1, BER2, SAM, ITS, PLE1, FUR, NAO, COL, DNS}

\paragraph*{Particular Thirring model}

A standard formulation of fermionic quantum field theories employs a functional integral for Grassmann variables. It is indeed possible to translate a given step evolution operator into the language of Grassmann functional integrals. We will not discuss here the general bit-fermion map which realizes this translation. We only refer to some of the results. In particular, one may ask if known simple two-dimensional quantum field theories for fermions with interactions can find a useful formulation as probabilistic cellular automata. We provide an example that this is indeed possible for certain types of Thirring- or Gross-Neveu models.

The Grassmann functional integral formulation for a particular discretized Thirring model\,\cite{THI,KLA} can be realized by a cellular automaton with a very simple interaction\,\cite{CWCA,FQFTPCA} which differs only slightly from the one discussed so far. For this purpose we extend
the color exchange rule for precisely one right mover and one left mover meeting at a single point $x$. We take away the restriction
that the two particles have different colors. The new rule for the automaton states that all particles change color if a single right
mover meets a single left mover at a given site. This adds the process where a green right mover and a green left mover are changed to a
red right mover and a red left mover, and similarly with the two colors exchanged. The new cellular automaton 
has a type of crossing symmetry. It
is now invariant under a 
$\pi/2$ rotation in the $(t-x)$-plane. 
There is no longer any difference between the directions $t$ and $x$ on the lattice, except for the implementation of boundary terms.

\begin{figure}[t!]
\resizebox{0.48\textwidth}{!}{%
\begin{tikzpicture}
\draw[red!85, line width=2pt] (0,0) -- (4.5,4.5) -- (6.5,2.5) -- (4,0) 
(0,6) -- (1.5,7.5) -- (3.5,5.5) --(5.5,7.5) -- (3.5,9.5) -- (4,10)
(10,10) -- (6.5,6.5) -- (8.5,4.5) -- (13.5,9.5) -- (13,10)
(12,0) -- (12.5,0.5) -- (10.5,2.5) -- (14,6)
(8,0) -- (8.5,0.5) -- (9,0);
\draw[black!60!green, line width=3pt](0,9) -- (1.5,7.5) -- ( 3.5,9.5) -- (3,10)
(0,2) -- (3.5,5.5) -- (4.5,4.5) -- (6.5,6.5) -- (5.5,7.5) -- (8,10)
(8.5,0.5) -- (6.5,2.5) -- (8.5,4.5) -- (10.5,2.5) -- (8.5,0.5)
(13,0) -- (12.5,0.5) -- (14,2)
(14,9) -- (13.5,9.5) -- (14,10);
\draw [blue] (1,7) rectangle (2,8);
\draw [blue] (3,9) rectangle (4,10);
\draw [blue] (3,5) rectangle (4,6);
\draw [blue] (4,4) rectangle (5,5);
\draw [blue] (5,7) rectangle (6,8);
\draw [blue] (6,6) rectangle (7,7);
\draw [blue] (6,2) rectangle (7,3);
\draw [blue] (8,4) rectangle (9,5);
\draw [blue] (8,0) rectangle (9,1);
\draw [blue] (10,2) rectangle (11,3);
\draw [blue] (12,0) rectangle (13,1);
\draw [blue] (13,9) rectangle (14,10);
\draw[dashed, blue]	(11,1)rectangle(12,2);
\draw[dashed, blue]	(9,1)rectangle(10,2);
\draw[dashed, blue]	(12,2)rectangle(13,3);
\draw[dashed, blue]	(10,0)--(10,1);
\draw[dashed, blue]	(11,0)--(11,1);
\draw[dashed, blue]	(13,1)--(14,1);
\draw[dashed, blue]	(13,2)--(14,2);
\draw[dashed, blue]	(13,1)--(13,2);
\draw[dashed, blue]	(10,1)--(11,1); 
\draw [blue] (9.5,1.5)  circle (1mm);
\draw [blue] (10.5,0.5)  circle (1mm);
\draw [blue] (11.5,1.5)  circle (1mm);
\draw [blue] (12.5,2.5)  circle (1mm);
\draw [blue] (13.5,1.5)  circle (1mm);
\draw [blue] (-0.2,9.4) -- (0,9.7)--(0.2,9.4);
\draw [blue] (13.4,-0.2) -- (13.7,0)--(13.4,0.2);
\draw [blue] (0,0) -- (13.5,0);
\draw [blue] (13.5,0) -- (14,0) 
node[below, font=\huge] at (13.7,-0.2){$x$};
\draw [blue] (0,0) -- (0,9.5);
\draw [blue] (0,9.5) -- (0,10)
node[left, font=\huge] at (-0.2, 9.7) {$t$};
\draw [blue] (0,10) -- (14,10);
\draw [blue] (14,0) -- (14,10);
\draw [blue] (-0.1,0) -- (0,0)
++(-0.1,1) --++(0.1,0)
++(-0.1,1) --++(0.1,0)
++(-0.1,1) --++(0.1,0)
++(-0.1,1) --++(0.1,0)
++(-0.1,1) --++(0.1,0)
++(-0.1,1) --++(0.1,0)
++(-0.1,1) --++(0.1,0)
++(-0.1,1) --++(0.1,0)
++(-0.1,1) --++(0.1,0);
\draw [blue] (14,0) -- (14.1,0)
++(-0.1,1) --++(0.1,0)
++(-0.1,1) --++(0.1,0)
++(-0.1,1) --++(0.1,0)
++(-0.1,1) --++(0.1,0)
++(-0.1,1) --++(0.1,0)
++(-0.1,1) --++(0.1,0)
++(-0.1,1) --++(0.1,0)
++(-0.1,1) --++(0.1,0)
++(-0.1,1) --++(0.1,0);
\draw [blue](0,0) -- (0,-0.1)
++(1,0.1) --++(0,-0.1)
++(1,0.1) --++(0,-0.1)
++(1,0.1) --++(0,-0.1)
++(1,0.1) --++(0,-0.1)
++(1,0.1) --++(0,-0.1)
++(1,0.1) --++(0,-0.1)
++(1,0.1) --++(0,-0.1)
++(1,0.1) --++(0,-0.1)
++(1,0.1) --++(0,-0.1)
++(1,0.1) --++(0,-0.1)
++(1,0.1) --++(0,-0.1)
++(1,0.1) --++(0,-0.1)
++(1,0.1) --++(0,-0.1)
++(1,0.1) --++(0,-0.1);
\draw [blue](0,10.1) -- (0,10)
++(1,0.1) --++(0,-0.1)
++(1,0.1) --++(0,-0.1)
++(1,0.1) --++(0,-0.1)
++(1,0.1) --++(0,-0.1)
++(1,0.1) --++(0,-0.1)
++(1,0.1) --++(0,-0.1)
++(1,0.1) --++(0,-0.1)
++(1,0.1) --++(0,-0.1)
++(1,0.1) --++(0,-0.1)
++(1,0.1) --++(0,-0.1)
++(1,0.1) --++(0,-0.1)
++(1,0.1) --++(0,-0.1)
++(1,0.1) --++(0,-0.1)
++(1,0.1) --++(0,-0.1);
\end{tikzpicture}}
	\caption{Cellular automaton for a particular Thirring model. We show single particle lines for red and green particles. Blocks for which interactions take place are indicated by solid squares. In the lower right part we also indicate by dashed boundaries and marked by a small circle a few blocks in which no interaction takes place. We do not indicate additional two-particle lines where a red and a green particle are on the same site and move together on diagonals without scattering. This figure is taken from ref.\,\cite{CWCA}.}
	\label{fig:06366} 
\end{figure}

We display in Fig.\,\ref{fig:06366} typical trajectories for red and green particles. We have surrounded the sites where a color exchange takes place by boxes. This demonstrates the local character of the interaction or the local character of the updating rule for the cellular automaton. Following the green lines one observes characteristic features of trajectories in a quantum field theory. There can be loops where green particles are created and subsequently annihilated, or trajectories going backwards in time. In the forward direction in time the numbers of red and green particles are not conserved separately, in contrast to the cellular automaton depicted in Fig.\,\ref{fig:OS1}. Other characteristic features for a quantum field theory, as vacua with spontaneous symmetry breaking of a chiral symmetry and associated solitons as defects at the boundaries between different vacua are described in ref.\,\cite{CWCA}.


The ``Thirring automaton'' has a rather simple structure relating the bit configuration at final time $t_\text{f}$ to a given bit configuration at initial time $t_\text{in}$. If at initial time $t_\text{in}$ a site $j$ is occupied by a single right-moving bit we will find at any later time $t$ a single right moving bit at the site $j+(t-t_\text{in})/\eps$. The trajectory for a single right moving bit is a straight line. The color may change along this trajectory, however. Whenever the right-moving bit encounters a single left-moving bit its color switches from red to green and vice versa. For determining the color at time $t$ one only has to count the number of switches. This is given by the number of trajectories for single left-moving bits that are crossed between $t_\text{in}$ and $t$. In turn, this counts the number of sites occupied by a single left-mover in the backward light-cone at $t_\text{in}$, e.g. in the interval between $j$ and $j+2(t-t_\text{in})/\eps$. The propagation of pairs of right-movers at a given site (one red, one green) follows straight lines, and similarly for pairs of holes (no right-movers on a given site $j$). The situation is similar for left-movers, with the opposite direction of the straight line trajectories.

Even with these very simple properties of the evolution of fixed bit configurations the behavior of probabilistic automata with a large number of bits can become a problem for which the combinatorics becomes rather demanding. This may concern single-particle excitations of a vacuum with a non-trivial structure. For these types of problems it may be useful to translate the automaton into the equivalent fermionic quantum field theory.

The extended interaction modifies the interaction part of the step evolution operator. It replaces~\cite{FQFTPCA} in eq.~\eqref{CA3} $\hat D(j)$ by $\hat H_i(j)$. One has now
\begin{equation}\label{FCA1}
H_{\text{int}}=\frac{\pi}{2\epsilon}\sum_j\hat H_i(j)\ ,
\end{equation}
with $\hat H_i^\dagger=\hat H_i$ given by
\begin{equation}
\label{4.3.93A}
\widehat{H}_{i}=\widehat{H}_{i}^{(1)}+\widehat{H}_{i}^{(2)}+\widehat{\Delta}_{i}^{(1)}+\widehat{\Delta}_{i}^{(2)}\,,
\end{equation}
where
\begin{alignat}{2}
\label{FCA2}
\widehat{H}_{i}^{(1)}(j)=& &&a_{R2}^\dagger(j)a_{L1}^\dagger(j)a_{L2}(j)a_{R1}(j)\nonumber\\
&+\,&&a_{R1}^\dagger(j)a_{L2}^\dagger(j)a_{L1}(j)a_{R2}(j)\, ,\nonumber\\
\widehat{H}_{i}^{(2)}=&\,&&a_{R2}^\dagger(j)a_{L2}^\dagger(j)a_{R1}(j)a_{L1}(j)\nonumber\\
&+\,&&a_{L1}^\dagger(j)a_{R1}^\dagger(j)a_{L2}(j)a_{R2}(j)\ ,
\end{alignat}
and
\begin{align}
\label{4.3-94A}
\widehat{\Delta}_{i}^{(1)}=&\,\widehat{n}_{R1}(1-\widehat{n}_{R2})\widehat{n}_{L1}(1-\widehat{n}_{L2})\nonumber\\
&+(1-\widehat{n}_{R1})\widehat{n}_{R2}(1-\widehat{n}_{L1})\widehat{n}_{L2}\,,\notag\\
\widehat{\Delta}_{i}^{(2)}=&-\widehat{n}_{R1}(1-\widehat{n}_{R2})(1-\widehat{n}_{L1})\widehat{n}_{L2}\nonumber\\
&-(1-\widehat{n}_{R1})\widehat{n}_{R2}\widehat{n}_{L1}(1-\widehat{n}_{L2})\,.
\end{align}
All occupation numbers one evaluated at the site $j$ and we observe
\begin{equation}
\label{4.3.94B}
\widehat{\Delta}_{i}(n)=\widehat{\Delta}_{i}^{(1)}+\widehat{\Delta}_{i}^{(2)}
=(\widehat{n}_{R1}-\widehat{n}_{R2})(\widehat{n}_{L1}-\widehat{n}_{L2})\ .
\end{equation}
(The term $\widehat{\Delta}_{i}$ is omitted in ref.~\cite{FQFTPCA}, which induces additional phases that we avoid here.)

In the product form,
\begin{align}
\label{FCA3}
\hat H_i(j)=&-\big[a_{R1}^\dagger(j)a_{R2}(j)-a_{R2}^\dagger(j)a_{R1}(j)\big]\nonumber\\
&\times\big[a_{L1}^\dagger(j)a_{L2}(j)-a_{L2}^\dagger(j)a_{L1}(j)\big]\nonumber\\
&+(\widehat{n}_{R1}-\widehat{n}_{R2})(\widehat{n}_{L1}-\widehat{n}_{L2})\,,
\end{align}
one sees directly the symmetries $L\leftrightarrow R$ and $1\leftrightarrow2$. The interaction step evolution operator obeys
\begin{equation}\label{FCA4}
\hat S_{\text{int}}=\prod_j\exp\left\{-\frac{i\pi}{2}\hat H_i(j)\right\}\ .
\end{equation}
We can equivalently write
\begin{align}
\label{NC1*}
\hat S_i(j)=&\,\exp\left\{-\frac{i\pi}{2}\hat H_i(j)\right\}\\
=&\,1-P_1^{(R)}(j)P_1^{(L)}(j)+\hat S_{12}P_1^{(R)}(j)P_1^{(L)}(j)\ ,\nn
\end{align}
with $P_1^{(R)}(j)$ projecting on one-particle states for the right movers at site $j$
\bel{NC2*}
P_1^{(R)}(j)=\hat n_{1R}(j)\gl1-\hat n_{2R}(j)\gr+\hat n_{2R}(j)\gl1-\hat n_{1R}(j)\gr\ ,
\ee
and similarly for $P_1^{(L)}(j)$ for the left-movers. The operator $\hat S_{12}(j)$ interchanges the two colors for all states,
\bel{NC3*}
\hat S_{12}(j)=t_{12}^{(R)}(j)t_{12}^{(L)}(j)\ ,
\ee
with switch operators $t_{12}^{(R,L)}(j)$ defined by eq.~\eqref{OS45A} for the right-movers and left-movers at site $j$. With eq.~\eqref{5.6.51A} this yields
\begin{align}
\label{NC4*}
\hat S_i(j)=&\,1-P_1^{(R)}(j)P_1^{(L)}(j)\nn\\
&+\gl a_{R1}^\dagger(j)a_{R2}(j)+a_{R2}^\dagger(j)a_{R1}(j)\gr\nn\\
&\times\gl a_{L1}^\dagger(j)a_{L2}(j)+a_{L2}^\dagger(j)a_{L1}(j)\gr\ .
\end{align}
The relation between eqs.~\eqref{NC1*} and~\eqref{FCA4} proceeds by the identities
\bel{NC5*}
\left(\frac{H_i}{2}\right)^2=\frac12\gl1-\hat S_{12}\gr P_1^{(R)}P_1^{(L)}\ ,\quad \left(\frac{H_i}{2}\right)^{3}=\frac{H_i}{2}\ ,
\ee
which imply
\bel{NC6*}
\hat S_i(j)=1-\frac12 H_i^2(j)\ ,\quad \hat S_i^2(j)=1\ .
\ee

For a proof of eq.~\eqref{FCA4} we first note that for all configurations for which at the site $j$ either $0$, $1$, $3$ or $4$ spins are up (fermions present) one has $\hat{H}_i(j)=0$. This is due to the presence of two creation and annihilation operators, or corresponding occupation numbers in $\widehat{\Delta}_{i}(j)$.
This extends to the sector where two right-movers are present due to the presence of $a_{R2}^\dagger a_{R1}$ or $a_{R1}^\dagger a_{R2}$, and similar for two left-movers. In all these sectors $\exp\big(-i\pi\hat{H}_i(j)/2\big)$ equals unity. For the remaining sectors of one right-mover and one left-mover present $\hat H_i(j)$ performs a map within this sector.

We denote the four configurations in this sector by 
\begin{align}
\label{4.3.96A}
1: \begin{pmatrix}
1 &0&0&1
\end{pmatrix}\,,&\quad 2: \begin{pmatrix}
0 &1&1&0
\end{pmatrix}\notag\\
3: \begin{pmatrix}
1 &0&1&0
\end{pmatrix}\,,&\quad 4: \begin{pmatrix}
0 &1&0&1
\end{pmatrix}\,.
\end{align}
For the action of $\widehat{H}_{i}$ on the four component subsector of the wave function $(\varphi_{1},\varphi_{2},\varphi_{3},\varphi_{4})$ one employs
\begin{align}
\label{4.3.96B}
\widehat{H}_{i}^{(1)}+\widehat{\Delta}_{i}^{(1)}&=\begin{pmatrix}
\tau_{1}&0\\ 0&1
\end{pmatrix}
\,,\notag\\
\widehat{H}_{i}^{(2)}+\widehat{\Delta}_{i}^{(2)}&=-\begin{pmatrix}
1&0\\0&\tau_{1}
\end{pmatrix}
\;.
\end{align}
with
\begin{align}
\label{4.3.96C}
&\left(H_{i}^{(a)}+\Delta_{i}^{(a)}\right)^{2}=1\,,\notag\\ 
&\left[(\widehat{H}_{i}^{(1)}+\widehat{\Delta}_{i}^{(1)}),(\widehat{H}_{i}^{(2)}+\widehat{\Delta}_{i}^{(2)})\right]=0\,.
\end{align}
The action of the step evolution operator~\eqref{FCA4} obeys for each $j$ 
\begin{align}
\label{4.3.96D}
\exp (-i\frac{\pi}{2}\widehat{H}_{i})=&\,\exp \left\lbrace -i\frac{\pi}{2}(\widehat{H}_{i}^{(1)}+\widehat{\Delta}_{i}^{(1)})\right\rbrace\nonumber\\
&\times\exp \left\lbrace -i\frac{\pi}{2}(\widehat{H}_{i}^{(2)}+\widehat{\Delta}_{i}^{(2)})\right\rbrace\notag\\
=&-i\begin{pmatrix}
\tau_{1}&0\\0&1
\end{pmatrix}\cdot i \begin{pmatrix}
1&0\\0&\tau_{1}
\end{pmatrix} =\begin{pmatrix}
\tau_{1}&0\\0&\tau_{1}
\end{pmatrix}\;.
\end{align}
This realizes precisely the scattering rule for the cellular automaton.

With the expression of the step evolution operator in terms of fermionic creation and annihilation operators it is manifest that this cellular automaton represents a unitary two-dimensional quantum field theory for fermions with interactions. 

The Thirring automaton can be mapped by the general bit-fermion map \cite{CWFCS,CWQFFT,CWFIM} to a Grassmann functional integral.
In the fermionic language the action in the Grassmann functional which represents this 
automaton reads in the naive continuum limit\,\cite{CWCA,FQFTPCA}
\begin{align}
\label{OS69}
S = -& \int_{t,x}\Big{\lbrace}  \bar{\psi}_{\alpha} \gamma^\mu \partial_\mu \psi_{\alpha} + \frac{1}{2} \left( \bar{\psi}_{\alpha} \gamma^\mu \psi_{\alpha} \right)
 \left( \bar{\psi}_{\beta} \gamma_\mu \psi_{\beta}  \right)\nonumber\\
& -\left(\bar{\psi}_{\alpha}\gamma^{\mu}\psi_{\beta}\right)\varepsilon^{\alpha\beta}\left(\bar{\psi}_{\gamma}\gamma_{\mu}\psi_{\delta}\right)\varepsilon^{\gamma\delta}\Big{\rbrace} 
 \, .
\end{align}
Here $\psi_{\alpha}(x)$, $\alpha=1,2$, are two complex two component Grassmann variables at every site $x$ where $\alpha$ is the color index. With $\epsilon^{12}=-\epsilon^{21}=1$, $\epsilon^{11}=\epsilon^{22}=0$ the action is invariant under $SO(2)$-rotations in color space. The two-dimensional Dirac matrices are given by the Pauli matrices
\begin{equation}
\label{OS70}
\gamma^0 = -i \tau_2\,, \quad \gamma^1 = \tau_1\,,
\end{equation}
with 
\begin{equation}
\label{OS71}
\bar{\psi} = \psi^\dagger \gamma^0\,.
\end{equation}
Indices are raised and lowered with the Lorentz metric, $\gamma_\mu = \eta_{\mu \nu} \gamma^\nu$, with
$\eta_{00}=-1$, $\eta_{11} = 1$, $\eta_{01} = \eta_{10} = 0$, and $\partial_0 = \partial_t$, $\partial_1 = \partial_x$. The
action \eqref{OS69} describes a two-dimensional Lorentz-invariant quantum field theory for interacting fermions. Similarly to free
Dirac fermions, the signature of Minkowski space and Lorentz symmetry are compatible with the formulation of a generalized Ising
model on a square lattice.

This naive continuum limit obtains by neglecting in the action of the equivalent Grassmann functional integral terms with more than four Grassmann variables that are formally of higher order in $\eps$. The naive continuum limit is not defined uniquely. A different version is based on omitting the non-vanishing commutator between $H_{\text{free}}$ and $H_{\text{int}}$ which appears in the combined step evolution operator. We will discuss this issue below. This yields a Hamiltonian with kinetic and interaction terms, expressed in terms of the fermionic creation and annihilation operators. The naive continuum limit translates these operators to Grassmann variables in the standard way~\cite{Zinn-Justin}.

Indeed a standard translation from fermionic annihilation and creation operators to a Grassmann functional integral yields
\bel{GI1}
S=-iS_M=\int_t\left\{i\mathcal{H}[\psi]+\int_x\psi_\alpha^*(t,x)\partial_t\psi_\alpha(t,x)\right\}\ ,
\ee
where $\mathcal{H}[\psi]$ obtains from $H[a_\gamma^\dagger,a_\gamma]$ by the replacement $a_\gamma(x)\to\psi_\gamma(t,x)$, $a_\gamma^\dagger(x)\to\psi_\gamma^*(t,x)$ and $e^{-S}=e^{iS_M}$. Insertion of eq.~\eqref{FCA3} yields
\bel{GI2}
S=\int_{t,x}\left\{\psi^*_{R \alpha}(\partial_t+\partial_x)\psi_{R\alpha}+\psi_{L\alpha}^*(\partial_t-\partial_x)\psi_{L\alpha}+L_i\right\}\ ,
\ee
with
\begin{align}
\label{GI3}
L_i=\,-i\pi\Big\{&\gl\psi_{R1}^*\psi_{L1}+\psi_{R2}^*\psi_{L2}\gr\gl\psi_{L1}^*\psi_{R1}+\psi_{L2}^*\psi_{R2}\gr\nn\\
-&\gl\psi_{R1}^*\psi_{L2}+\psi_{R2}^*\psi_{L1}\gr\gl\psi_{L1}^*\psi_{R2}+\psi_{L2}^*\psi_{R1}\gr\Big\}\ .
\end{align}
We observe the discrete symmetries $L\leftrightarrow R$ and $1\leftrightarrow2$ and note $L_i^*=L_i^\dagger=-L_i$, $S^\dagger=-S$, $S_M^\dagger=S_M$, where transposition involves a complete reordering of the Grassmann variables. The interaction term $L_i$ is invariant under Lorentz transformations and this naive continuum limit describes again a type of Thirring or Gross-Neveu model. It differs, however, from eq.~\eqref{OS69}.

This difference underlines that finding the true continuum limit of our model is not a simple task. The formal steps leading to the naive continuum limit can typically be justified at best for wave functions that are ``sufficiently smooth'' in a sense that needs to be specified. For the second naive continuum limit the omission of the commutator between $H_\text{free}$ and $H_\text{int}$ needs justification. Also the direct translation between fermionic operators and Grassmann variables involves an issue of ordering whose consequences are poorly understood. For the first naive continuum limit there exists an exact translation between the probabilistic automaton and the Grassmann functional integral~\cite{CWFCS, CWQFFT, CWFIM, FQFTPCA, CWCA}. The terms omitted for the naive continuum limit guarantee the unique jump property and the consequences of their omission are poorly understood.  For understanding the true continuum limit one probably will have to follow the renormalization flow, starting microscopically from a discrete Grassmann functional integral.

\subsubsection{Particles}

Applying fermionic creation and annihilation operators to some vacuum wave
function is widely used for the construction of one-particle states in quantum
field theory. We employ here this construction for the Thirring automaton.
Depending on the properties of the vacuum we either remain with massless free
fermions despite the presence of interactions. For other vacua phenomena chiral
symmetry breaking could influence the propagation even for single particles.

\paragraph*{Vacuum and one particle states}

Already for the free theory we have found that the properties of one-particle
excitations depend on the vacuum state. This becomes even more important in the
presence of interactions. For an example we focus first on particular
half-filled particle-hole symmetric vacuum states. They should be invariant
under time- and space translations by $\epsilon$.
We demonstrate the dependence of particle properties on the vacuum by comparing
two different simple possible vacuum states. In
sect.~\ref{sec:Particles_and_antiparticles} we will extend this to vacua which
lead to positive energies for particle and antiparticle excitations.

For the first vacuum state we take the ``totally empty'' spin configuration
$\tau_-$ for which all Ising spins are down, and the ``totally filled''
configuration $\tau_+$ with all spins up. The vacuum state
$|0\rangle_{\text{ef}}$ corresponds to the wave function
\begin{equation}\label{VP0}
|0\rangle_{\text{ef}}=q^{(\text{ef})}=\frac{1}{\sqrt{2}}\big(Q_{\tau_+}+Q_{\tau_-}\big)\
.
\end{equation}
Particle-hole conjugation maps $\tau_+\leftrightarrow\tau_-$, such that
$q^{(\text{ef})}$ is invariant. Translation invariance in space and time are
obvious.

We define one-particle states by a single spin up or a single spin down
\begin{equation}\label{VP1}
q^{(1)}=\sqrt{2}\sum_x\sum_\gamma\big(q'_\gamma(t,x)a_\gamma^\dagger(x)+q_\gamma^c(t,x)a_\gamma(x)\big)|0\rangle_{\text{ef}}\
.
\end{equation}
(Note
$\sqrt{2}a_\gamma^\dagger(x)|0\rangle_{\text{ef}}=a_\gamma^\dagger(x)Q_{\tau_-}$
and $\sqrt{2}a_\gamma(x)|0\rangle_{\text{ef}}=a_\gamma(x)Q_{\tau_+}$.) For the
complex structure based on particle-hole conjugation the complex two-component
one-particle wave function reads
\begin{equation}\label{VP2}
\varphi_\alpha(t,x)=\begin{pmatrix}\varphi_{R\alpha}(t,x)\\
\varphi_{L\alpha}(t,x)\end{pmatrix}\ ,
\end{equation}
with ($\eta=R,L$)
\begin{equation}\label{VP3}
\varphi_{\eta\alpha}(t,x)=\frac{1+i}{\sqrt{2}}q'_{\eta\alpha}(t,x)+\frac{1-i}{\sqrt{2}}q_{\eta\alpha}^c(t,x)\
.
\end{equation}
The time evolution reflects the motion of two free massless Dirac fermions
\begin{align}
\label{VP4}
\varphi_{R\alpha}(t+\epsilon,x)=&\varphi_{R\alpha}(t,x-\epsilon)\ ,\nonumber\\
\varphi_{L\alpha}(t+\epsilon,x)=&\varphi_{L\alpha}(t,x+\epsilon)\ .
\end{align}
For these one particle states there is no scattering, since a single-particle
trajectory never crosses another single particle trajectory. Correspondingly,
applying $H_\text{int}$ in eq.~\eqref{FCA1} to this one-particle state yields
zero.

\paragraph*{Continuum limit for one-particle states}

We may compare this with the partial continuum limit using $H_{\text{free}}$
according to eq.~\eqref{CA8}. One has
$H_{\text{free}}Q_{\tau_-}=H_{\text{free}}Q_{\tau_+}=0$, such that according to
eq.~\eqref{CA7} one concludes $\hat
S_{\text{free}}|0\rangle_{\text{ef}}=|0\rangle_{\text{ef}}$. The basis vector
$a_\gamma^\dagger(x)Q_{\tau_-}$ corresponds to a configuration with
$s_\gamma(x)=1$, while all other spins are down. One has
\begin{align}
\label{VP5}
H_{\text{free}}a_{R\alpha}^\dagger(x)Q_{\tau_-}=&\frac{i}{2\epsilon}\big(a_{R\alpha}^\dagger(x+\epsilon)-a_{R\alpha}^\dagger(x-\epsilon)\big)Q_{\tau_-}\
,\nonumber\\
H_{\text{free}}a_{L\alpha}^\dagger(x)Q_{\tau_-}=&-\frac{i}{2\epsilon}\big(a_{L\alpha}^\dagger(x+\epsilon)-a_{L\alpha}^\dagger(x-\epsilon)\big)Q_{\tau_-}\
,
\end{align}
and similar for the particle-hole conjugated states. This yields
\begin{align}
\label{VP6}
-i\epsilon&H_{\text{free}}q^{(1)}\nonumber\\
=&\frac12\sum_x\sum_\alpha\bigg\{q'_{R\alpha}(x)\big(a_{R\alpha}^\dagger(x+\epsilon)-a_{R\alpha}^\dagger(x-\epsilon)\big)Q_{\tau_-}\nonumber\\
&-q'_{L\alpha}(x)\big(a_{L\alpha}^\dagger(x+\epsilon)-a_{L\alpha}^\dagger(x-\epsilon)\big)Q_{\tau_-}\nonumber\\
&+q^c_{R\alpha}(x)\big(a_{R\alpha}(x+\epsilon)-a_{R\alpha}(x-\epsilon)\big)Q_{\tau_+}\nonumber\\
&-q^c_{L\alpha}(x)\big(a_{L\alpha}(x+\epsilon)-a_{L\alpha}(x-\epsilon)Q_{\tau_+}\bigg\}\
.
\end{align}
For the approximation $\hat S_{\text{free}}=1-i\epsilon H_{\text{free}}$ the
evolution of the wave function obeys
\begin{equation}\label{VP7}
\varphi_{R\alpha}(t+\epsilon,x)=\varphi_{R\alpha}(t,x-\epsilon)+\delta^{(1)}_{R\alpha}(t,x)\
,
\end{equation}
with
\begin{equation}\label{VP8}
\delta^{(1)}_{R\alpha}(t,x)=\varphi_{R\alpha}(t,x)-\frac12\big(\varphi_{R\alpha}(t,x+\epsilon)+\varphi_{R\alpha}(t,x-\epsilon)\big)\
.
\end{equation}

If $\varphi_{R\alpha}(t,x)$ is a differentiable function one finds approximately
for small $\epsilon$
\begin{equation}\label{VP9}
\delta^{(1)}_{R\alpha}(t,x)=-\frac{\epsilon^2}{2}\partial_x^2\varphi_{R\alpha}(t,x)\
,
\end{equation}
which vanishes in the continuum limit $\epsilon\to0$. The situation is similar
for $\varphi_{L\alpha}$. Higher order terms in the expansion of the
exponential~\eqref{CA7} vanish in this limit as well, according to
\begin{align}
\label{VP10}
&\left(-\frac{1}{2}\epsilon^2H_{\text{free}}^2\right)a_{R\alpha}^\dagger(x)Q_{\tau_-}\nonumber\\
&=\frac18\big[a_{R\alpha}^\dagger(x+2\epsilon)+a_{R\alpha}^\dagger(x-2\epsilon)-2a_{R\alpha}^\dagger(x)\big]Q_{\tau_-}\
.
\end{align}
For the continuum limit of small $\epsilon$ for differentiable
$\varphi_{R\alpha}(t,x)$ the leading contribution of eq.~\eqref{VP10} to
$\varphi_{R\alpha}(t+\epsilon,x)$ cancels $\delta_{R\alpha}^{(1)}(t,x)$ in
eq.~\eqref{VP9}. Summing all the terms in the expansion of $\hat
S_{\text{free}}=\exp\big(-i\epsilon H_{\text{free}}\big)$ replaces in
eq.~\eqref{VP7} $\delta_{R\alpha}^{(1)}(t,x)$ by a similar correction term
$\delta_{R\alpha}(t,x)$. For small enough $\epsilon$ and differentiable wave
functions we can neglect $\delta_{R\alpha}(t,x)$ and recover the exact
result~\eqref{VP4}.

Nevertheless, the unique jump property of an automaton is no longer valid for
$\delta_{R\alpha}(t,x)\neq0$. For the partial continuum limit the probability
for a spin configuration at $t+\epsilon$ is influenced by many neighboring spin
configurations at $t$. In turn, a given spin configuration influences the
probabilities for many neighboring spin configurations in the future, e.g. for
$t+2\epsilon$ etc. This conceptual change is a general property of the continuum
limit for automata. The probabilistic character of the cellular automaton is
crucial for the possible simplifications of a continuum limit. Smooth wave
functions are a key ingredient. No continuum limit exists for the sharp wave
function of a deterministic automaton.

\paragraph*{Half-filled equipartition vacuum}

As a second possible particle-hole invariant vacuum state we discuss the
half-filled equipartition state~\cite{FQFTPCA}. We denote by $\tau_i^{(E)}$ the
spin configurations which have on each site precisely one right-mover and one
left-mover. On a given site there are four possibilities, $s_{R1}=s_{L1}=1$,
$s_{R1}=s_{L2}=1$, $s_{R2}=s_{L1}=1$, $s_{R2}=s_{L2}=1$, while the Ising spins
not indicated are negative. The total number of configurations $\tau_i^{(E)}$ is
given by $4^{M_x}$. The wave function for the half-filled equipartition state is
given by
\begin{equation}\label{VP11}
|0\rangle_E=q^{(E)}=2^{-M_x}Q_{\tau_i^{(E)}}\ ,
\end{equation}
with equal probability $2^{-2M_x}$ for each configuration $\tau_i^{(E)}$.

One-particle states are again obtained by eq.~\eqref{VP1}. We observe that
$a_\gamma^\dagger(x)Q_{\tau_i^{(E)}}$ vanishes for half of the configurations
$\tau_i^{(E)}$, namely for all those for which $s_\gamma(x)=1$. For the other
half of the configurations with $s_\gamma(x)=-1$ the action of the creation
operator $a_\gamma^\dagger(x)$ flips the sign of $s_\gamma(x)$, corresponding to
adding a particle of type $\gamma$ at the site $x$. As a result, the site $x$ is
occupied by two right-movers or two left-movers, i.e. $s_{R1}(x)=s_{R2}(x)=1$ or
$s_{L1}(x)=s_{L2}(x)=1$. The pair of two right-movers or left-movers is color
neutral, as reflected by the relation
\begin{align}
\label{VP12}
a_{R1}^\dagger(x)|0\rangle_E=&a_{R2}^\dagger(x)|0\rangle_E\ ,\nonumber\\
a_{L1}^\dagger(x)|0\rangle_E=&a_{L2}^\dagger(x)|0\rangle_E\ .
\end{align}
Doubly occupied trajectories are straight lines without scattering. Similar
relations hold for the annihilation operator. In order to avoid double counting
we replace in eq.~\eqref{VP1} the sum over $\gamma$ by a sum over $\eta=(R,L)$,
and we omit contributions from $a_{\eta2}^\dagger(x)$ or $a_{\eta2}(x)$. All
one-particle excitations have the straight trajectories of massless particles.

The time evolution of the one-particle state is identical to the one for the
first vacuum combining the totally empty and filled states. There is one
important difference, however. The one-particle state describes now a single
massless Dirac fermion, in contrast to the two colored Dirac fermions for
excitations of the vacuum $|0\rangle_{\text{ef}}$. For the multi-particle states
the excitations of the vacuum $|0\rangle_{\text{ef}}$ undergo scattering by
color exchange. No such scattering occurs for the excitations of the vacuum
$|0\rangle_E$. This reflects the automaton property that double right-moving or
left-moving particles or holes do not scatter.

Already these two simple examples demonstrate the quantum field theory property
that the behavior of particles depends on the vacuum. There are many more
possibilities for particle-hole invariant vacua and particle-like excitations,
including solitons. Some of them can be found in ref.~\cite{FQFTPCA}.

\paragraph*{Particle mass and spontaneous symmetry breaking}

A typical particle mass term adds to the Hamiltonian a piece $H_M=H_M^\dagger$
\bel{MM1}
H_M=\sum_xM_{\alpha\beta}a_{L\alpha}^\dagger(x)a_{R\beta}(x)+M_{\alpha\beta}^\dagger
a_{R\alpha}^\dagger(x)a_{L\beta}(x)\ .
\ee
It changes the direction of the motion. The expression~\eqref{MM1} is invariant
under the discrete symmetry exchanging right- and left-movers if the matrix $M$
is hermitian. It violates, however, the discrete chiral symmetry
$a_{L\alpha}(x)\to-a_{L\alpha}(x)$,
$a_{L\alpha}^\dagger(x)\to-a_{L\alpha}^\dagger(x)$ with fixed $a_{R\alpha}(x)$
and $a_{R\alpha}^\dagger(x)$. One may wonder if such a term can be generated
effectively by the interaction. This requires a vacuum state that breaks
spontaneously the discrete chiral symmetry.

In particular, one may consider possible vacua that are half-filled only in the
average. The contributing configurations may not have precisely one right-mover
and one left-mover one each site, as for our second example. Then the state
$a_{R1}^\dagger|0\rangle$ can involve bit configurations with one or two
right-moving bits on the site $j$. They may be accompanied by zero, one or two
left-moving bits at $j$. The interaction part of the step evolution operator
acts non-trivially on the components with one right-mover and one left-mover.
Due to the interaction, the propagation of the single particle state can be
modified.

\paragraph*{Direct product form for vacuum states}

Let us consider a class of wave functions that take in a direct product basis
the form
\bel{DP1}
\psi=\vp(x_\text{in})\otimes\dots\otimes\vp(x)\otimes\dots\otimes\vp(x_\text{f})\
.
\ee
For every position $x$ (or $j$) the sixteen components of $\vp(x)$ are written
in a direct product form
\bel{DP2}
\vp(x)=\sum_\rho\vp_\rho(x)Q_{n_1}^{(R1)}\otimes Q_{n_2}^{(R2)}\otimes
Q_{n_3}^{(L1)}\otimes Q_{n_4}^{(L2)}\ ,
\ee
where $\rho=(n_1,n_2,n_3,n_4)$ denotes the sixteen configurations of occupation
numbers of the four bits at the position $x$, $n_\gamma=(1,0)$. The basis vector
$Q_1^{(R1)}=\begin{pmatrix}1\\0\end{pmatrix}$ stands for the bit $R1$ (at
position $x$) occupied, while $Q_0^{(R1)}=\begin{pmatrix}0\\1\end{pmatrix}$
corresponds to this bit being empty. The basis vectors $Q_\tau$ in the
occupation number basis correspond to $\vp_\rho(x)=\delta_{\rho\bar\rho(x)}$ for
suitable basis configurations $\bar\rho(x)$. We admit a possible complex
structure, such that $\vp_\rho(x)$ are complex vectors, normalized for every $x$
by
\bel{DP3}
\sum_\rho|\vp_\rho(x)|^2=1\ .
\ee

A set of candidate vacuum states has the same $\vp(x)=\vp$ at every position $x$
\bel{DP4}
\psi_0=\vp\otimes\vp\otimes\dots\otimes\vp\otimes\dots\otimes\vp\ .
\ee
This guaranteed translation symmetry in space. We further require
\bel{DP5}
\hat S_\text{free}\psi_0=\psi_0\ ,\quad \hat S_\text{int}\psi_0=\psi_0\ ,
\ee
which ensures translation symmetry in time, as appropriate for a static vacuum.
An example is the totally empty vacuum $|0\rangle_e$, for which
\bel{DP6}
\vp^{(e)}=\begin{pmatrix}0\\1\end{pmatrix}\otimes\begin{pmatrix}0\\1\end{pmatrix}\otimes\begin{pmatrix}0\\1\end{pmatrix}\otimes\begin{pmatrix}0\\1\end{pmatrix}\
,
\ee
or the totally filled vacuum
\bel{DP7}
\vp^{(f)}=\begin{pmatrix}1\\0\end{pmatrix}\otimes\begin{pmatrix}1\\0\end{pmatrix}\otimes\begin{pmatrix}1\\0\end{pmatrix}\otimes\begin{pmatrix}1\\0\end{pmatrix}\
.
\ee
A wide set of vacuum states can be obtained as linear superpositions of
candidate vacuum states, as the particle-hole symmetric vacuum
\bel{DP8}
|0\rangle_{ef}=\frac{1}{\sqrt{2}}\gl|0\rangle_e+|0\rangle_f\gr\ .
\ee
For the second vacuum discussed in this section, the half-filled equipartition
states $|0\rangle_E$, one has
\begin{align}
\label{DP9}
\vp^{(E)}=\frac12\bigg\{&\begin{pmatrix}1\\0\end{pmatrix}\otimes\begin{pmatrix}0\\1\end{pmatrix}\otimes\begin{pmatrix}1\\0\end{pmatrix}\otimes\begin{pmatrix}0\\1\end{pmatrix}\nn\\
+&\begin{pmatrix}1\\0\end{pmatrix}\otimes\begin{pmatrix}0\\1\end{pmatrix}\otimes\begin{pmatrix}0\\1\end{pmatrix}\otimes\begin{pmatrix}1\\0\end{pmatrix}\nn\\
+&\begin{pmatrix}0\\1\end{pmatrix}\otimes\begin{pmatrix}1\\0\end{pmatrix}\otimes\begin{pmatrix}1\\0\end{pmatrix}\otimes\begin{pmatrix}0\\1\end{pmatrix}\nn\\
+&\begin{pmatrix}0\\1\end{pmatrix}\otimes\begin{pmatrix}1\\0\end{pmatrix}\otimes\begin{pmatrix}0\\1\end{pmatrix}\otimes\begin{pmatrix}1\\0\end{pmatrix}\bigg\}\
.
\end{align}

\paragraph*{Total equilibrium vacuum and chiral condensates}

An interesting vacuum state with spontaneous chiral symmetry breaking is the
total equilibrium state $|0\rangle_{eq}$ with
\bel{DP10}
\vp^{(eq)}=\frac14\begin{pmatrix}1\\1\end{pmatrix}\otimes\begin{pmatrix}1\\1\end{pmatrix}\otimes\begin{pmatrix}1\\1\end{pmatrix}\otimes\begin{pmatrix}1\\1\end{pmatrix}\
.
\ee
For this state each of the sixteen configurations $\rho$ has the same
probability $1/16$. In turn, each configuration $\tau$ (at a given time $t$) has
the same probability, with
\bel{DP11}
\psi_\tau^{(eq)}=2^{-2M_x}\ .
\ee
Under the action of $\hat S_\text{free}$ the right-movers and left-movers move
to different neighboring positions. Each configuration $\tau$ changes to a new
configuration $\tau'(\tau)$. This yields for the new wave function
\bel{DP12}
\gl\hat S_\text{free}\psi^{(eq)}\gr_{\tau'}=\psi_\tau^{(eq)}=2^{-2M_x}\ .
\ee
Thus $\hat S_\text{free}\psi^{(eq)}$ is again the total equilibrium state, which
therefore obeys the first equation~\eqref{DP5}. The interaction term acts
locally, with
\bel{DP13}
\hat S_\text{int}\psi=\hat S_i\vp\otimes\hat S_i\vp\otimes\dots\otimes\hat
S_i\vp\ .
\ee
Since $\hat S_i$ only interchanges colors for the configurations with a single
right-mover and a single left-mover, and all these configurations have the same
$\vp_\rho^{(eq)}=1/4$, we conclude
\bel{DP14}
\hat S_i\vp^{(eq)}=\vp^{(eq)}\ ,
\ee
such that the second equation~\eqref{DP5} is obeyed as well. Particle-hole
conjugation acts separately on each position $x$ by exchanging occupied and
empty bits. Since $\vp_\rho^{(eq)}=1/4$ is the same for a configuration and its
particle-hole conjugate configuration we conclude that the total equilibrium
vacuum $|0\rangle_{eq}=\psi^{(eq)}$ is invariant under the particle-hole
conjugation.

The total equilibrium state does not have fixed particle numbers. At every $x$
the values $n_\gamma(x)=1$ or $0$ occur with equal probabilities. For each
individual occupation number the mean value is $1/2$,
\bel{DP15}
\langle n_\gamma(x)\rangle_{eq}=\frac12\ .
\ee
This follows from the fact that half of the configurations $\rho$ have
$n_\gamma=1$, the other half $n_\gamma=0$. In the average sense the total
equilibrium vacuum is half-filled. The mean value,
\bel{DP16}
\langle n_R(x)\rangle=\langle n_{R1}(x)+n_{R2}(x)\rangle=1\ ,
\ee
does not mean that every contribution to $\vp^{(eq)}$ has one right handed
particle. This distinguishes this state from the half-filled equilibrium state
$\vp^{(E)}$.

For the action of $a_{L1}(x)$ on the total equilibrium vacuum $\psi^{(eq)}$ one
replaces at the position $x$ the factor $\vp$ by
\begin{align}
\label{DP16*}
a_{L1}\vp=&\,\gl\tau_3\otimes\tau_3\otimes a\otimes1\gr\vp\nn\\
=&\,\frac14\begin{pmatrix}1\\-1\end{pmatrix}\otimes\begin{pmatrix}1\\-1\end{pmatrix}\otimes\begin{pmatrix}0\\1\end{pmatrix}\otimes\begin{pmatrix}1\\1\end{pmatrix}\
,
\end{align}
and multiplies all sites preceding $x$ by a factor
$\tilde\tau_3=\tau_3\otimes\tau_3\otimes\tau_3\otimes\tau_3$. Similarly, for
$a_{R2}^\dagger(x)a_{L1}(x)\psi^{(eq)}$ one replaces $\vp$ at position $x$ by
\begin{align}
\label{DP17}
a_{R2}^\dagger a_{L1}\vp=&\,\gl1\otimes a^\dagger\tau_3\otimes
a\otimes1\gr\vp\nn\\
=&\,\frac14\begin{pmatrix}1\\1\end{pmatrix}\otimes\begin{pmatrix}-1\\0\end{pmatrix}\otimes\begin{pmatrix}1\\0\end{pmatrix}\otimes\begin{pmatrix}1\\1\end{pmatrix}\
.
\end{align}
No $\tilde\tau_3$ factors for other sites are present. One concludes for the
expectation value
\bel{DP18}
\langle a_{R2}^\dagger(x)a_{L1}(x)\rangle_{eq}=\vp^\dagger a_{R2}^\dagger
a_{L1}\vp=-\frac14\ .
\ee
Similarly, one finds
\bel{DP19}
\langle a_{L1}^\dagger(x)a_{R2}(x)\rangle=-\frac14\ .
\ee
The expectation values~\eqref{DP18},~\eqref{DP19} indicate the spontaneous
breaking of the discrete chiral symmetry by the total equilibrium vacuum.

Similarly, one has
\bel{DP20}
a_{R1}^\dagger
a_{L1}\vp=\frac14\begin{pmatrix}-1\\0\end{pmatrix}\otimes\begin{pmatrix}1\\-1\end{pmatrix}\otimes\begin{pmatrix}0\\1\end{pmatrix}\otimes\begin{pmatrix}1\\1\end{pmatrix}\
,
\ee
and therefore
\begin{align}
\label{DP21}
\langle a_{R1}^\dagger(x)a_{L1}(x)\rangle_{eq}=&\,0\ ,\nn\\
\langle(2\hat n_{R2}(x)-1)a_{R1}^\dagger(x)a_{L1}(x)\rangle_{eq}=&\,-\frac14\ .
\end{align}
Again, the second expectation value~\eqref{DP21} indicates spontaneous chiral
symmetry breaking. The appearance of the operator $2\hat
n_{R2}-1=1\otimes1\otimes\tau_3\otimes1$ is due to our choice of creation and
annihilation operators. If we would choose $a_i=(a_{R1},a_{L1},a_{R2},a_{L2})$
instead of the present choice $a_i=(a_{R1},a_{R2},a_{L1},a_{L2})$ one would find
$\langle a_{R1}^\dagger(x)a_{L1}(x)\rangle_{eq}=-1/4$. The evolution of
probabilities and expectation values does not depend on the convention for the
fermionic operators.

\paragraph*{Modified particle propagation}

We conclude that the total equilibrium vacuum $\vp^{(eq)}$ leads to a
spontaneous breaking of the discrete chiral symmetry. In contrast, for the
half-filled equipartition state $\vp^{(E)}$ all expectation
values~\eqref{DP18},~\eqref{DP19},~\eqref{DP20},~\eqref{DP21} vanish. The
breaking of the chiral symmetry influences the propagation of particles. As an
example, we may consider a wave function for red right-movers
\bel{DP22}
\tilde\vp_{R1}=\sum_xq_{R1}(x)a_{R1}^\dagger(x)|0\rangle_{eq}\ .
\ee
The free part of the step evolution operator shifts the wave function one
position to the right. The subsequent interaction modifies the wave function.
This can be seen by inferring from eq.~\eqref{NC4*} the identity
\bel{DP23}
S_ia_{R1}^\dagger=\bar P_{11}a_{R1}^\dagger+\Delta\ .
\ee
The first part proportional to
\bel{DP24}
\bar P_{11}=1-P_1^{(R)}P_1^{(L)}
\ee
concerns all components of the wave function which do not involve precisely one
right-mover and one left-mover at the given position $x$. The propagation of
this component is not affected by $S_i$. The second part, 
\bel{DP25}
\Delta=\gl a_{L1}^\dagger a_{L2}+a_{L2}^\dagger a_{L1}\gr
a_{R2}^\dagger\gl1-\hat n_{R1}\gr\ ,
\ee
concerns the remaining component, $\Delta P_1^{(R)}P_1^{(L)}=\Delta$. The part
$\Delta(x)|0\rangle_{eq}$ has contributions with precisely one left-mover and
one green right-mover at the site $x$, according to
\bel{DP26}
\Delta\vp=\frac14\left\{\begin{pmatrix}0\\1\end{pmatrix}\otimes\begin{pmatrix}1\\0\end{pmatrix}\otimes\left[\begin{pmatrix}1\\0\end{pmatrix}\otimes\begin{pmatrix}0\\1\end{pmatrix}+\begin{pmatrix}0\\1\end{pmatrix}\otimes\begin{pmatrix}1\\0\end{pmatrix}\right]\right\}\
.
\ee
For the part $\Delta(x)|0\rangle_{eq}$ there is no longer a red right-mover at
the position $x$. In particular, $\Delta\vp$ has overlap with
$a_{L\alpha}^\dagger\vp$, such that we may partly interpret the interaction as a
change of direction. Without going into further details it becomes clear that
the behavior of single particles is more complex than the free propagation
without color change.

The superposition of two candidate vacuum states is again a possible vacuum
state. For example, one may consider
\bel{DP27}
|0\rangle_m=\alpha|0\rangle_E+\beta|0\rangle_{eq}\ ,
\ee
with $\alpha$ and $\beta$ restricted by normalization. For small $\beta$ the
expectation values indicating spontaneous chiral symmetry breaking are small
$\sim\beta$. This results in almost free massless propagation, with only a small
effect of the interaction. Following these issues in the operator formalism or
by combinatorics becomes rather involved. It seems more promising to apply the
powerful methods of quantum field theory.

\paragraph*{Naive continuum limit for Hamiltonian}

The naive continuum limit approximates
\begin{equation}\label{VP17}
H=H_{\text{free}}+H_{\text{int}}\ ,
\end{equation}
with $H_{\text{free}}$ given by eq.~\eqref{CA10} and $H_{\text{int}}$ given by
eqs.~\eqref{FCA1},~\eqref{FCA2}, with $j$ replaced by $x$. We can use a
continuum normalization for the annihilation and creation operators
\begin{equation}\label{VP18}
a_\gamma^{(c)}(x)=\frac{1}{\sqrt{2\epsilon}}a_\gamma(x)\ ,\quad
a_\gamma^{(c)\dagger}(x)=\frac{1}{\sqrt{2\epsilon}}a_\gamma^\dagger(x)\ ,
\end{equation}
with anticommutator
\begin{equation}\label{VP19}
\big\{a_\gamma^{(c)\dagger}(x),a_\delta^{(c)}(y)\big\}=\delta(x-y)\delta_{\gamma\delta}=\frac{1}{2\epsilon}\delta_{x,y}\delta_{\gamma\delta}\
,
\end{equation}
such that
\begin{equation}\label{VP20}
\int_x\delta(x-y)=\sum_x\delta_{x,y}=1\ .
\end{equation}
With $\sum_x=(2\epsilon)^{-1}\int_x$ this yields
\begin{align}
\label{VP21}
H=\int_x\bigg\{&-i\sum_\alpha\big(a_{R\alpha}^\dagger\partial_xa_{R\alpha}-a_{L\alpha}^\dagger\partial_xa_{L\alpha}\big)\nonumber\\
-\pi\Big[\big(&a_{R1}^\dagger a_{R2}-a_{R2}^\dagger
a_{R1}\big)\big(a_{L1}^\dagger a_{L2}-a_{L2}^\dagger a_{L1}\big)\nonumber\\
-\big(&a_{R1}^\dagger a_{R1}-a_{R2}^\dagger a_{R2}\big)\big(a_{L1}^\dagger
a_{L1}-a_{L2}^\dagger a_{L2}\big)\Big]\bigg\}\ .
\end{align}
Here the annihilation operators $a_\gamma$ are shorthands for
$a_\gamma^{(c)}(x)$, and similar for the creation operators. The
result~\eqref{VP21} constitutes the continuum Hamiltonian for a fermionic
quantum field theory, more precisely the two-color Thirring type model
corresponding to eq.~\eqref{OS66}. We emphasize that the coupling in front of
the interaction term is not a free parameter, rather taking a particular value
$\pi$.

The naive continuum neglects terms that are formally of higher order in
$\epsilon$. Comparing eqs.~\eqref{VP14} and~\eqref{VP17} the most important
neglection concerns the terms involving the non-vanishing commutator
$\big[H_{\text{int}},H_{\text{free}}\big]$ in the Baker-Campbell-Haussdorf
expansion of the product of two exponentials of matrices. A second effect arises
from the use of the partial continuum limit for $H_{\text{free}}$. Finally,
there are also corrections from the translation of the lattice derivative
$\partial_x$ to the continuum derivative. All these effects vanish formally in
the limit $\epsilon\to0$. This formal property can hold, at best, for
sufficiently smooth wave functions. For example, a correction
$~\sim\epsilon\big[H_{\text{int}},H_{\text{free}}\big]$ vanishes only if
$\big[H_{\text{int}},H_{\text{free}}\big]\psi$ does not produce a term
$\sim\epsilon^{-1}$. For general quantum field theories the true continuum limit
differs from the naive continuum limit. For example, there are important effects
of renormalization. Before a dedicated investigation of the true continuum limit
becomes available we leave it as an open question to which extent the naive
continuum limit reproduces important features of the true continuum limit.

\paragraph*{Fermion picture for cellular automata}

The ``Thirring automaton'' discussed in this section constitutes an explicit
example that a probabilistic automaton is equivalent to a discretized quantum
field theory for fermions. By construction it realizes a unitary evolution in
discrete time steps. The observables of the quantum field theory find their
correspondence as observables for the classical statistical system of the
probabilistic automaton. This includes probabilistic observables as momentum.
The expectation values of all observables are the same for the probabilistic
automaton and the fermionic quantum field theory. This includes all
correlations. We can therefore identify the probabilistic automaton and the
discrete fermionic quantum field theory. The fermionic formulation may be called
the ``fermion picture'' of the probabilistic automaton. This equivalence is
further underlined by the bit-fermion map~\cite{CWFCS, CWQFFT, CWFIM} to a
Grassmann functional integral for the fermionic quantum field theory.

By expressing the step evolution operator in terms of fermionic annihilation and
creation operators rather arbitrary probabilistic cellular automata can be
translated to an equivalent fermion picture. What is not guaranteed, however, is
a simple form of the fermionic picture. For many cases simplicity may only be
realized in the continuum limit.

One may ask the opposite question which interesting fermionic quantum field
theories admit a discretization that corresponds to a probabilistic automaton. A
general discretization will not yield an automaton. For a given discretization
one can compute the step evolution operator for suitably chosen discrete time
steps. Only if this step evolution operator is a unique jump matrix the discrete
fermion model is equivalent to an automaton. The property of a cellular
automaton follows then typically from the locality of the quantum field theory.

The unique jump property of the step evolution operator for a discrete fermionic
quantum field theory seems a priori rather restrictive. Nevertheless, rather
efficient methods have been developed for constructing discrete fermionic
quantum field theories with a unique jump step evolution operator~\cite{FPCA}.
They comprise sequences of propagation and interaction steps, or the use of
shifted blocks for the automaton. Examples of discretizations of fermionic
quantum field theories realizing the automaton property include models with
continuous abelian or non-abelian symmetries. If is possible to realize local
gauge symmetries. In particular, exact local Lorentz symmetry for fermions can
be implemented, and there are models for which the naive continuum limit is
invariant under general coordinate transformations. It is possible to construct
models with stochastically distributed fixed disorder points for which the naive
continuum limit describes a non-relativistic quantum particle in a potential in
one space dimension~\cite{Wetterich:2022kif}. So far most models are in two
dimensions~\cite{FPCA}, but a first model of four-dimensional spinor gravity has
been proposed~\cite{Wetterich:2022zql}. It shares exact local Lorentz symmetry
and diffeomorphism invariance in the naive continuum limit, as appropriate for
quantum gravity.

For all these models the interesting fermionic quantum field theories are
realized in the naive continuum limit. It is a crucial question if the true
continuum limit belongs to the same universality class as suggested by the naive
continuum limit. In particular, this concerns the continuous spacetime
symmetries as rotations or invariance under Lorentz boosts in four-dimensional
settings. If this is realized one may speculate that it could be possible to
construct a probabilistic automaton that describes the time evolution of our
whole universe.

\subsection{Change of basis and similarity\\transformations}
\label{sec:Change_of_basis_and_similarity_transformations}

The formulation of evolution in classical probabilistic systems in terms of wave functions, density matrices and step evolution operators permits us to employ changes of basis or similarity transformations. As well known from quantum mechanics these transformations are a powerful tool. We have already employed a change of basis as a Fourier transform in sect.\,\ref{sec:conserved_quantities_and_symmetries}, or for the discussion of the loss of memory in sect.\,\ref{sec:partial_loss_of_memory_and_emergence_of_quantum_mechanics}. 
The change of basis is not available on the level of the time-local probability distribution. The formulation of a linear evolution law, together with the ``quantum rule" for expectation values in terms of wave functions and operators, are crucial ingredients.

In this section we discuss the change of basis more generally. We introduce equivalence classes of overall weight distributions or probability distributions which are related by basis changes. They contain the same information about observables if operators are transformed correspondingly. The weight distributions of these equivalence classes need no longer to be real and positive, bringing our formulation even closer to complex functional integrals for quantum field theories in Minkowski space. We discuss symmetries as particular basis transformations leaving the weight distribution invariant. We also connect to the change of basis functions which parameterize the configurations of occupation numbers.
The discussion of the present section largely follows ref.\,\cite{CWQF}, where an extended discussion and more details can be found.

\paragraph*{Time-local similarity transformations}

Rather general transformations that preserve the structure of probabilistic time are the time-local similarity transformations. The step evolution operator transforms with general regular matrices $D(t)$, which can depend on $t$, as $\hat{S}(t) \to \hat{S}'(t)$,
\begin{equation}
\hat{S}'(t) = D(t+\varepsilon) \hat{S}(t) D^{-1}(t).
\label{eq:CB1}
\end{equation}
For this transformation a chain of step evolution operators is transformed only at the boundary sites
\begin{align}
\begin{split}
&\hat{S}'(t+m\varepsilon) \hat{S}'(t+(m-1)\varepsilon)...\hat{S}'(t+\varepsilon) \hat{S}'(t) \\
&\quad = D(t+(m+1)\varepsilon) \hat{S}(t+m\varepsilon) \hat{S}(t+(m-1)\varepsilon)... \\
&\qquad \times \hat{S}(t+\varepsilon) \hat{S}(t) D^{-1}(t).
\end{split}
\label{eq:CB2}
\end{align}
The partition function $Z$ remains invariant if the boundary terms transform as
\begin{align}
\begin{split}
\tilde{q}'(t_\mathrm{in}) &= D(t_\mathrm{in}) \tilde{q}(t_\mathrm{in}), \\
\bar{q}'(t_\mathrm{f}) &= \left( D^\mathrm{T}(t_\mathrm{f}) \right)^{-1} \bar{q}(t_\mathrm{f}).
\end{split}
\label{eq:CB3}
\end{align}

The wave functions at arbitrary $t$ transform as
\begin{align}
\begin{split}
\tilde{q}'(t) &= D(t) \tilde{q}(t), \\
\bar{q}'(t) &= \left( D^\mathrm{T}(t) \right)^{-1} \bar{q}(t),
\end{split}
\label{eq:CB4}
\end{align}
and the classical density matrix obeys $\rho'(t) \to \rho'_D(t)$,
\begin{equation}
\rho'_D(t) = D(t) \rho'(t) D^{-1}(t).
\label{eq:CB5}
\end{equation}
This preserves the form of the evolution equation. If we also transform the operators for local observables as $\hat{A}(t) \to \hat{A}'(t)$,
\begin{equation}
\hat{A}'(t) = D(t) \hat{A}(t) D^{-1}(t),
\label{eq:CB6}
\end{equation}
the expectation values $\braket{A(t)}$ remain invariant. The transformations of $\rho'(t)$ and $\hat{A}(t)$ act locally as similarity transformations, such that the expression for the expectation value $\tr\left\{ \rho'(t) \hat{A}(t) \right\}$ remains invariant. This will generalize to other local observables and correlations that we discuss in sect.\,\ref{sec:local_observables_and_non_commuting_operators}.

The normalization of the step evolution operator and the wave functions is not important in the present context. We can perform the transformation \eqref{eq:CB1} analogously for the transfer matrices. The matrices $D(t)$ only need to be regular. They can be complex. In this case also the transformed wave functions, density matrices and transfer matrix are complex. For the special case of orthogonal $D(t)$ the conjugate wave function $\bar{q}(t)$ transforms in the same way as $\tilde{q}(t)$. For unitary $D(t)$ the transformation of $\bar{q}(t)$ is the same as for $\tilde{q}^*(t)$. This appearance of complex quantities is a priori not related to the complex structure discussed in sects.~\ref{sec:complex_structure},~\ref{sec:particles_and_holes}. If we have already chosen a complex structure we may restrict the basis transformations to the ones that are compatible with this complex structure. They act then on complex wave functions, density matrices, step evolution operators and operators for observables, as familiar from quantum mechanics.

\paragraph*{Equivalence classes of weight distributions}

While the partition function is invariant under local similarity transformations, this does not hold for the overall weight distribution $w[n]$. In the expression of $w[n]$ in terms of products of transfer matrices by eqs.\,\eqref{eq:TS46}, \eqref{eq:TS47} the similarity transformation introduces factors $D^{-1}_{\tau\sigma}(t) D_{\sigma\rho}(t)$ that are not summed over $\sigma$. We can define an equivalence class of weight distributions for which all members can be obtained from each other by local similarity transformations. For all members of a given equivalence class all expectation values of local observables are the same. Again this will extend to a larger class of observables for which the expectation values can be computed by the ``quantum rule'' \eqref{eq:CW19}, \eqref{eq:DM34}.

From the point of view of observations, different members of an equivalence class cannot be distinguished since all observations are based on suitable expectation values of observables, including correlations. The local similarity transformations are a very large transformation group, leaving us with a very large family of equivalent, but not identical, weight distributions. Since the operators representing a given observable have to be transformed according to eq.\,\eqref{eq:CB6}, a member of an equivalence class can be seen as a pair of a weight distribution $w[n]$ and an associated family of operators $\{ \hat{A}(t) \}$. Both transform simultaneously if we switch from one member to another.

Starting from a real positive weight distribution $w[n]$, the transformed weight distribution $w'[n]$ needs no longer to be real and positive. For complex $D$ the weight distribution $w'[n]$ is, in general, complex as well. An arbitrary quantum system corresponds to a complex weight distribution, since the transfer matrix in eq.\,\eqref{eq:TS47} is replaced by the unitary step evolution operator of quantum mechanics. The question arises which quantum systems can be transformed by local similarity transformations to a probabilistic system with real positive weight distribution. If this is possible, there exists a classical statistical system for which all predictions for expectation values are identical to the ones of the quantum system. In ref.\,\cite{CWQF} it is shown that such a transformation exists for arbitrary quantum systems. In the corresponding classical statistical system the operators associated to simple quantum observables may get complicated, however. Nevertheless, the mere existence of an equivalent classical statistical system indicates that there cannot be any no-go theorem forbidding a ``classical statistical" formulation of quantum mechanics.

\paragraph*{Symmetries}

Transformations that leave the weight distribution $w[n]$ invariant are symmetries. Among the local similarity transformations all diagonal matrices $D(t)$ correspond to symmetry transformations, independently of the specific form of $\hat{S}$. (There are other possible symmetry transformations for particular step evolution operators.) The symmetries include the local sign transformation in sect.\,\ref{sec:free_particles_in_two_dimensions}. They correspond to diagonal $D(t)$ with diagonal elements $\pm 1$.

In general, symmetry transformations do not leave the step evolution operator invariant. We can use symmetry transformations to bring ``rescaled unique jump operators'' to a standard form with positive $\hat{S}$. Rescaled unique jump operators have in each column and each row precisely one element different from zero. This element is an arbitrary complex number. By the use of an appropriate symmetry transformation this element can be ``gauged'' to one. Diagonal operators $\hat{A}(t)$ are invariant under symmetry transformations with diagonal $D(t)$. Only non-diagonal operators $\hat{A}(t)$ can transform non-trivially.

\paragraph*{Change of basis functions}

We typically use basis functions $h_\tau(t) = h_\tau[n(t)]$ in the occupation number basis. There are many other possible choices of basis functions. For example, we may use linear combinations $h'_\tau(t)$, given by
\begin{equation}
h_\tau(t) = h'_\rho(t) V_{\rho\tau}(t),
\label{eq:CB7}
\end{equation}
with arbitrary regular matrices $V(t)$. If we combine the change of basis functions \eqref{eq:CB7} with a local similarity transformation one finds that the wave function $\tilde{f}(t) = \tilde{f}[n(t)] = h^\mathrm{T}(t) \tilde{q}(t)$ in eq.\,\eqref{eq:CWF3}, \eqref{eq:CWF6} transforms as
\begin{equation}
\tilde{f}'(t) = h^{\prime\mathrm{T}}(t) \tilde{q}'(t) = h^\mathrm{T}(t) V^{-1}(t) D(t) \tilde{q}(t).
\label{eq:CB8}
\end{equation}
For the choice $V(t) = D(t)$ it remains invariant. For the conjugate wave function $\bar{f}(t)$ we use a different set of basis functions $\bar{h}'_\tau(t)$
\begin{equation}
\bar{h}'_\tau(t) = V_{\tau\rho}(t) h_\rho(t).
\label{eq:CB9}
\end{equation}
Combination with the local similarity transformation \eqref{eq:CB4} yield
\begin{equation}
\bar{f}'(t) = \bar{q}^{\prime \mathrm{T}}(t) \bar{h}'(t) = \bar{q}^\mathrm{T}(t) D^{-1}(t) V(t) h(t).
\label{eq:CB10}
\end{equation}
It is again invariant for the choice $V(t) = D(t)$. For the particular case of orthogonal matrices, $V^{-1}(t) = V^\mathrm{T}(t)$, the basis functions $\bar{h}'(t)$ coincide with $h'(t)$, such that one can continue to use the same basis functions for $\tilde{f}$ and $\bar{f}$.

For the expansion of the local factors $\cK(t)$ in terms of the step evolution operator we use both sets of basis functions
\begin{align}
\begin{split}
\cK'(t) &= h^{\prime\mathrm{T}}(t+\varepsilon) \hat{S}'(t) \bar{h}'(t) \\
	&= h^\mathrm{T}(t+\varepsilon) V^{-1}(t+\varepsilon) D(t+\varepsilon) \hat{S}(t) D^{-1}(t) V(t) h(t).
\end{split}
\label{eq:CB11}
\end{align}
For $V(t) = D(t)$ the local factors remain invariant. Changing simultaneously the basis functions and the step evolution operator with $V(t) = D(t)$ the weight distribution and the partition function are invariant.

The general relation between local observables and associated local operators is given by
\begin{align}
\begin{split}
A'(t) &= h^{\prime\mathrm{T}}(t) \hat{A}'(t) \bar{h}'(t) \\
	&= h^\mathrm{T}(t) V^{-1}(t) D(t) \hat{A}(t) D^{-1}(t) V(t) h(t).
\end{split}
\label{eq:CB12}
\end{align}
Again, for $V(t) = D(t)$ the local observables are invariant, $A'(t) = A(t)$. The expectation values do not change under simultaneous basis and local similarity transformations. We conclude that the simultaneous change of basis functions and local similarity transformations with $V(t)=D(t)$ leaves the whole setting invariant.

As a consequence, we can view a local similarity transformation with $D(t)$ and fixed basis functions equivalently as a change of basis with $V(t) = D^{-1}(t)$, keeping wave functions, step evolution operator and local operators fixed. Both result in the same $f'(t)$, $\bar{f}'(t)$, $K'(t)$, $A'(t)$. All structural relations remain the same under these transformations. The explicit form of the evolution equation changes, however, since $\tilde{q}(t)$, $\bar{q}(t)$ and $\hat{S}(t)$ are transformed non-trivially. In particular, we note that the diagonal form of the local operators $\hat{A}(t)$ for local observables $A(t)$ is a property of the occupation number basis. In a different basis the operator $A'(t)$ is given by eq.\,\eqref{eq:CB6}. In general, it is no longer diagonal.

For a given overall probability distribution and a given set of observables we may treat the wave functions $f[n(t)]$, $\bar f[n(t)]$ as fixed expressions that can be computed from the probability distribution. (Uniqueness only holds up to normalization factors that can be distributed on $f$ and $\bar f$ such that $\bar ff$ remains invariant.) At this level one is free to use arbitrary basis function $h'(t)$ and $\bar h'(t)$. The wave functions and the step evolution operator depend on the choice of basis functions. In the occupation number basis we can associate to a basis function $h_\tau$ a given bit configuration $\tau$ for which the occupation numbers are fixed. This does no longer hold for a different set of basis functions $\big\{h'_\tau(t)\big\}$. It is possible that the new basis functions $h'_\tau$ are associated with other observables as momentum taking a fixed value. This is the case for the Fourier transform to be discussed below.

If we choose complex basis functions (e.g. complex $V(t)$) both the wave functions and the step evolution operator will become complex quantities. This should be distinguished from the possibility to introduce complex basis functions in the occupation number basis by use of an appropriate complex structure. Indeed, we can combine both the real components $\tilde q_\tau$ of the wave function and the real basis functions $h_\tau$ into complex quantities such that $f$ is left invariant,
\begin{align}\label{4.7.7}
f&=(\tilde q_R)_{\bar\tau}(h_R)_{\bar\tau}+(\tilde q_I)_{\bar\tau}(h_I)_{\bar\tau}=\Re\big(\psi_{\bar\tau}h_{\bar\tau}^{(c)}\big)\ ,\nonumber\\
\psi_{\bar\tau}&=(\tilde q_R)_{\bar\tau}+i(\tilde q_I)_{\bar\tau}\ ,\quad h^{(c)}_{\bar\tau}=(h_R)_{\bar\tau}-i(h_I)_{\bar\tau}\ .
\end{align}
In this picture we may still perform complex basis transformations, now acting on $N/2$-component complex vectors.

\paragraph*{Positive weight distributions}

Performing either a local similarity transformation or a change of basis functions (but not both simultaneously) the overall weight distribution $w[n]$ changes. The general expression \eqref{eq:TS46}, \eqref{eq:TS47} in terms of the transfer matrix and the step evolution operator remains the same under these transformations. The concrete expression for the step evolution operator changes, however. In particular, it needs no longer to be a positive matrix. It is an interesting question under which conditions the overall weight distribution is positive, such that a probability distribution $p[n]$ can be defined by an appropriate normalization.

We have not performed yet a systematic investigation on the general form of positive weight distributions. Some first criteria are developed in the appendix~\ref{app:positivity_of_overall_probability_distribution}. At present, we concentrate on positive local factors $\cK(t)$, for which the step evolution operator $\hat{S}(t)$ in the occupation number basis is a positive matrix. The positivity of the weight distribution for this setting (with an appropriate positive boundary term) is maintained by local similarity transformations which are symmetries, e.\,g.\ by diagonal $D(t)$. We include this obvious generalization for positive weight functions. The local factors do no longer need to be positive for the generalization.
\subsection{Fourier transform for cellular automata}
\label{sec:fourier_transform_for_cellular_automata}

In quantum mechanics or in quantum field theories the Fourier transform to a
momentum basis plays a central role for the understanding of many phenomena.
This tool is, in general, not employed for the investigation of generalized
Ising models or for cellular automata. Our formulation of the local
probabilistic information in terms of wave functions or the density matrix comes
in pair with the possibility of similarity or basis transformations. We will see
that the Fourier transform indeed opens new avenues for a simple understanding
of important aspects of classical statistical systems. We demonstrate this for
two simple probabilistic automata. One is the clock system equivalent to a
simple quantum system, the second is an automaton with transport of bit
configurations equivalent to a two-dimensional quantum field theory for free
massless fermions. For the free fermions we will find a simple expression for
the Hamiltonian in momentum space. This can later be combined with interaction
terms.

\subsubsection{Fourier basis for clock systems}

We start from a clock system with $N$ states, labeled by integers $j$ with $j+N$
and $j$ identified. As in sect.~\ref{sec:conserved_quantities_and_symmetries} we
double the states and introduce a complex structure by
eqs.~\eqref{UR4},~\eqref{eq:CQ28}. The complex wave function $\psi(t,j)$ can be
written in terms of Fourier components $\psi(t,k)$,
\bel{FT1}
\psi(t,j)=\sum_kD^{-1}(j,k)\psi(t,k)\ ,
\ee
with
\bel{FT2}
D^{-1}(j,k)=\frac{1}{\sqrt{N}}\exp\left\{\frac{2\pi i}{N}jk\right\}\ .
\ee
Here $k$ are integers with the same periodicity, with $k+N$ identified with $k$.
A suitable range for $j$ and $k$ may be the interval $[-N/2,N/2]$. For even $N$
the boundaries are identified. The transformation matrix is unitary
\bel{FT3}
D(l,j)=\frac{1}{\sqrt{N}}\exp\left\{-\frac{2\pi i}{N}jl\right\}\ ,\quad
D^\dagger D=1\ .
\ee
This employs the identity
\bel{FT4}
\frac1N\sum_j\exp\left\{\frac{2\pi i}{N}j(k-l)\right\}=\tilde\delta_{k,l}\ ,
\ee
where $\tilde\delta_{k,l}$ equals one for $l=k\ \text{mod}\ N$ and vanishes
otherwise. For a real wave function $\psi(t,j)$ one has
$\psi(t,-k)=\psi^*(t,k)$.

In the position basis the step evolution operator is given by
\bel{FT5}
\hat S(j,j')=\tilde\delta_{j,j'+1}\ .
\ee
It is diagonalized by the Fourier transform
\begin{align}\label{FT6}
\hat S(k,l)=&\sum_{j,j'}D(k,j)\hat S(j,j')D^{-1}(j',l)\nonumber\\
=&\,\exp\left(-\frac{2\pi il}{N}\right)\tilde\delta_{k,l}\ .
\end{align}
Similarly, the difference operator
\bel{FT7}
\Delta(j,j')=\frac12\big(\tilde\delta_{j,j'-1}-\delta_{j,j'+1}\Big)
\ee
reads in the Fourier basis
\bel{FT8}
\Delta(k,l)=i\sin\left(\frac{2\pi l}{N}\right)=\tilde\delta_{k,l}\ .
\ee

If we identify $j$ with an angle,
\bel{FT9}
\alpha=\frac{2\pi j}{N}\ ,
\ee
we can identify $l$ with angular momentum
\bel{FT10}
D(\alpha,l)=\exp(-i\alpha l)\ .
\ee
The angular momentum operator
\bel{FT11}
\hat L=-i\partial_\alpha=-\frac{iN\Delta}{2\pi}\ ,
\ee
is diagonal in the Fourier basis
\bel{FT12}
\hat L(k,l)=\frac{N}{2\pi}\sin\left(\frac{2\pi l}{N}\right)\tilde\delta_{k,l}\ .
\ee

We can also consider a single right-moving particle in periodic space, with
\bel{FT13}
x=\eps j\ ,\quad p=\frac{2\pi l}{\eps N}\ ,
\ee
and
\bel{FT14}
D(x,p)=\exp(-ipx)\ .
\ee
The momentum operator based on the lattice derivative,
\bel{FT15}
\hat P=-i\partial_x=-\frac i\eps\Delta
\ee
yields in the Fourier space the relation~\eqref{eq:CQ12}
\bel{FT16}
\hat P(k,l)=\frac{\sin(\eps p)}{\eps}\tilde\delta_{k,l}\ .
\ee

A momentum operator more suitable for quantum field theories can be defined in
Fourier space by
\bel{6.3.16A}
\tilde{P}(k,l) = p\tilde\delta_{k,l}\ .
\ee
It can be translated to position space by the inverse Fourier transform.

\subsubsection{Transport automata in momentum space}

As a second system we discuss an automaton which is equivalent to a fermionic
quantum field theory. The variables at fixed $t$ are occupation numbers $n(j)$.
The states $\tau$ are now the bit configurations $\{n(j)\}$. We will show that
the step evolution operator for a transport automaton finds a simple expression
in terms of creation and annihilation operators in momentum space. We focus on
the Fourier transform of the fermionic annihilation and creation operators. This
does not yet change the basis for the wave function. The Fourier transform for
the fermionic operators corresponds to linear combinations of operators. The
Fourier-transformed operators may still act on wave functions in the position
basis.

For complex wave functions one employs a complex structure. The map from the
real to the complex wave function defines how the creation and annihilation
operators in the real formulation are transformed to the complex picture. We
will focus on complex structures for which the real operators $a(j)$,
$a^\dagger(j)$ are mapped to complex matrices $A(j)$ and $A^\dagger(j)$,
preserving the anti-commutation relations. Different possible complex structures
share this requirement. We want to remain general and do not specify the
selection of the complex structure. The algebraic relations below also hold for
real $A(j)$, $A^\dagger(j)$, in particular for $A(j)=a(j)$,
$A^\dagger(j)=a^\dagger(j)$. The specification of the complex structure is not
needed for establishing our result for the Hamiltonian realizing the step
evolution operator for the transport automaton.

Let us consider at each site $j$ a pair of possibly complex annihilation and
creation operators $A(j)$, $A^\dagger(j)$ obeying the fermionic anti-commutation
relations
\begin{align}
\label{4.8.17}
\big\{A^\dagger(j),A(j')\big\}=&\,\tilde\delta_{j,j'}\ ,\nonumber\\
\big\{A(j),A(j')\big\}=&\,\big\{A^\dagger(j),A^\dagger(j')\big\}=0\ .
\end{align}
There are different possibilities how the operators $A(j)$ and $A^\dagger(j)$
are related to the annihilation and creation operators $a(j)$ and $a^\dagger(j)$
in the real formulation.

A general way to relate the complex $N/2\times N/2$-matrices $A(j)$,
$A^\dagger(j)$ to the real $N\times N$-matrices $a(j)$, $a^\dagger(j)$ employs
first a similarity transformation in the real picture
\bel{CAM1}
\hat A(j)=\tilde Da(j)\tilde D^{-1}\ ,\quad \hat A^\dagger(j)=\tilde
Da^\dagger(j)\tilde D^{-1}\ .
\ee
This preserves the anti-commutation relations. For orthogonal $\tilde D$ one has
$\hat A^\dagger(j)=\hat A^T(j)$. Compatibility with a chosen complex structure
requires
\bel{CAM2}
\big[\hat A(j),I\big]=0\ ,\quad \big[\hat A^\dagger(j),I\big]=0\ .
\ee
Then we can express the $N\times N$-matrices $\hat A(j)$ in terms of complex
$N/2\times N/2$-matrices $A(j)$,
\bel{CAM3}
\hat A(j)=1\otimes A_R(j)+\tilde I\otimes A_I(j)\ ,\quad A(j)=A_R(j)+iA_I(j)\ ,
\ee
and similar for $\hat A^\dagger(j)$,
 \bel{CAM4}
\hat A^\dagger(j)=1\otimes A_R^\dagger(j)-\tilde I\otimes A_I^\dagger(j)\ ,\quad
A^\dagger(j)=A_R^\dagger(j)-iA_I^\dagger(j)\ .
\ee
The anti-commutation relations for $\hat A(j)$, $\hat A^\dagger(j)$ translate
into the anti-commutation relations~\eqref{4.8.17} for $A(j)$, $A^\dagger(j)$.
The task consists in finding the similarity transformation $\tilde D$ which
realizes eq.~\eqref{CAM2}.

The construction of $\hat A(j)$, $\hat A^\dagger(j)$ by similarity
transformations is not the only way. Any definition of operators commuting with
$I$ and obeying the anti-commutation relations leads to creation and annihilation
operators that are compatible with the complex structure. In turn, the set $\hat
A(j)I$ and $-\hat A^\dagger(j)I$ obeys the same conditions. It is mapped in the
complex picture to the set $iA(j)$ and $-iA^\dagger(j)$. We discuss different
possibilities for complex annihilation and creation operators $A(j)$,
$A^\dagger(j)$ in the
appendix~\ref{app:particle-hole_conjugation_for_neutral_states}.

We take the integer position variables $j$ in the range $-J\leq j\leq J$. For an
odd number of positions $\cM_2+1$ we take $J=\cM_2/2$. Periodicity identifies
$J+1$ with $-J$. For an even number $\cM_2+1$ we choose $J=(\cM_2+1)/2$. Now
$-J$ is identified with $J$ and no longer an independent position.

We define creation and annihilation operators in a Fourier basis by
\begin{align}\label{FT17}
a(k)=&\sum_jD(k,j)A(j)\ ,\nonumber\\
a^\dagger(k)=&\sum_jA^\dagger(j)D^{-1}(j,k)\ ,
\end{align}
Here $D(k,j)$ is given by eq.~\eqref{FT3}, with the replacement $N\to\cM_2+1$
according to the number of positions $j$. The range of $k$ is
given by $-J\leq k\leq J$, with periodic $k$ through the identification $k+N=k$.
For odd $N$ one has $J=(N-1)/2$, while for even $N$ one takes $J=N/2$, with $-J$
and $J$ identified. From the anti-commutation relation,~\eqref{4.8.17} one infers
a similar relation in Fourier space,
\bel{FT19}
\big\{a^\dagger(k),a(l)\big\}=\tilde\delta_{k,l}\ ,\quad
\big\{a(k),a(l)\big\}=\big\{a^\dagger(k),a^\dagger(l)\big\}=0\ .
\ee
Occupation number operators for Fourier modes are given by
\bel{FT20}
\hat n_k=a^\dagger(k)a(k)\ .
\ee

It is our aim to express the step evolution operators for right-transport and
left-transport in terms of the operators $a(k)$ and $a^\dagger(k)$. We will find
that this leads to a rather simple expression which can be directly translated
to the Hamiltonian for free fermions, or more generally to the ``kinetic part''
of the Hamiltonian for a fermionic quantum field theory. This simple form leads
to the concept of the ``quantum field theory vacuum'' which will be discussed in
the following section.

For a right-moving particle the step evolution operator moves the whole bit
configuration one position to the right
\bel{FT21}
\psi\big(t+\eps;\{n_j\}\big)=\psi\big(t;\{n_{j+1}\}\big)=\hat
S\psi\big(t;\{n_j\}\big)\ .
\ee
This notation is understood in the sense that for periodic $j$ one has
$\psi(t+\eps;n_1,n_2,n_3,n_4)=\psi(t;n_2,n_3,n_4,n_1)$. We assume that for every
$t$ an arbitrary wave function can be expressed by applying a suitable linear
combination of products of operators $A(j)$ and $A^\dagger(j)$ on some reference
wave function. The wave function at $t+\eps$ is then obtained by the replacement
$A(j)\to A(j+1)$, $A^\dagger(j)\to A^\dagger(j+1)$. We first show that the
relation~\eqref{FT21} is achieved if the step evolution operator obeys the
relations ($A_j=A(j)$)
\begin{align}\label{FT22}
\hat SA_j=A_{j+1}\hat S\ ,\nonumber\\
\hat SA_j^\dagger=A_{j+1}^\dagger\hat S\ .
\end{align}

We start with simple one-particle states. For this purpose we first assume the
existence of some vacuum state $|0\rangle$ that is invariant under the
evolution,
\bel{FT23}
\hat S|0\rangle=|0\rangle\ .
\ee
A sharp one-particle state at the position $j$ obtains as
$A_j^\dagger|0\rangle$. After an evolution step one has $\hat
SA_j^\dagger|0\rangle=A_{j+1}^\dagger|0\rangle$, such that the particle has
moved to the position $j+1$. Correspondingly, a general one-particle wave
function
\bel{FT24}
\psi^{(1)}(t)=\sum_j\psi^{(1)}(t,j)A_j^\dagger|0\rangle\ ,
\ee
evolves to
\begin{align}\label{FT25}
\psi^{(1)}(t&+\eps)=\,\hat
S\psi^{(1)}(t)=\sum_j\psi^{(1)}(t,j)A_{j+1}^\dagger|0\rangle\nonumber\\
=&\sum_j\psi^{(1)}(t,j-1)A_j^\dagger|0\rangle=\sum_j\psi^{(1)}(t+\eps,j)A_j^\dagger|0\rangle\
.
\end{align}
This realizes eq.~\eqref{FT21} in the one-particle sector,
\bel{FT25A}
\psi^{(1)}(t+\eps,j)=\psi^{(1)}(t,j-1)\ .
\ee

Similar relations hold for one-hole wave functions $\sim A_j|0\rangle$. The
precise properties of the assumed vacuum state play no role for this shift in
position of particles or holes. For multi-particle wave functions one replaces
$A_j^\dagger$ by products of creation or annihilation operators. The
relation~\eqref{FT22} moves the whole product to the new position. This entails
eq.~\eqref{FT21}. Actually, the existence of a vacuum state is not necessary in
order to show that eq.~\eqref{FT22} is a necessary and sufficient condition for
the right-transport operator $\hat S$. For this purpose one only needs that the
sharp wave functions for all bit configurations can be obtained from a given
``starting configuration'' by applying a product of suitable creation and
annihilation operators $A_j^\dagger$ and $A_j$. The step evolution operator
obeying eq.~\eqref{FT22} displaces all occupation numbers $\hat
n_j=A^\dagger(j)A(j)$ by one position. The extension from sharp to smooth
probabilistic wave functions is straightforward. The relations~\eqref{FT22} also
hold in the real formulation with $A_j$, $A_j^\dagger$ replaced by $a_j$,
$a_j^\dagger$.

We next express the first relation~\eqref{FT22} in terms of annihilation
operators for Fourier modes
\bel{FT26}
\sum_k\hat SD^{-1}(j,k)a(k)=\sum_kD^{-1}(j+1,k)a(k)\hat S\ .
\ee
Using eq.~\eqref{FT2} with $M=\cM_2+1=N$,
\bel{FT27}
D^{-1}(j+1,k)=\exp\left(\frac{2\pi ik}{M}\right)D^{-1}(j,k)\ ,
\ee
this yields
\bel{FT28}
\sum_kD^{-1}(j,k)\Big[\hat Sa(k)-\exp\left(\frac{2\pi ik}{M}\right)a(k)\hat
S\Big]=0\ .
\ee
Thus a step evolution operator obeying the relations
\begin{align}\label{FT29}
\hat Sa(k)&=\exp\left(\frac{2\pi ik}{M}\right)a(k)\hat S\ ,\nonumber\\
\hat Sa^\dagger(k)&=\exp\left(-\frac{2\pi ik}{M}\right)a^\dagger(k)\hat S\ ,
\end{align}
achieves the right-transport by one unit. (For the second relation we express
the second relation~\eqref{FT22} in terms of $a^\dagger(k)$.)

For $\hat S$ we can make an ansatz in a product form
\bel{FT30}
\hat S=\gamma\prod_l\hat S(l)\ ,
\ee
with $\hat S(l)$ involving an even number of operators $a^\dagger(l)$ or $a(l)$,
such that $[\hat S(l),a(k)]=0$, $[\hat S(l),a^\dagger(k)]=0$ for $k\neq l$. The
condition~\eqref{FT29} is obeyed if we replace in this relation $\hat S$ by
$\hat S(k)$. We can therefore deal with each $k$-mode separately. Using the
relations~\eqref{CA13} for $a(k)$, $a^\dagger(k)$,
\begin{align}
\label{4.8.31A}
a(k)\hat n_k=&\,a(k)\ ,\quad \hat n_ka(k)=0\ ,\nn\\
a^\dagger(k)\hat n_k=&\,0\ ,\quad \hat n_ka^\dagger(k)=a^\dagger(k)\ ,
\end{align}
the condition~\eqref{FT29} is solved by
\begin{align}\label{FT31}
\hat S(k)&=\exp\left(\frac{i\pi k}{M}\right)(1-\hat n_k)+\exp\left(-\frac{i\pi
k}{M}\right)\hat n_k\nonumber\\
&=\cos\left(\frac{\pi k}{M}\right)-i\sin\left(\frac{\pi k}{M}\right)(2\hat
n_k-1)\nonumber\\
&=\exp\left[-\frac{2\pi i k}{M}\left(\hat n_k-\frac12\right)\right]\ .
\end{align}
Here we employ for the last identity the relation $(2\hat n_k-1)^2=1$.

For even $M$ a particular case is $k=M/2$. In this case eq.~\eqref{FT29} reads
\bel{FT31A}
\left\{\hat S\left(\frac M2\right),a(k)\right\}=\left\{\hat S\left(\frac
M2\right),a^\dagger(k)\right\}=0\ .
\ee
Eq.~\eqref{FT31} yields
\bel{FT31B}
\hat S\left(\frac M2\right)=-i(2\hat n_k-1)\ .
\ee
We could use equivalently
\bel{FT31C}
\hat S'\left(\frac M2\right)=i(2\hat n_k-1)=\hat S\left(-\frac M2\right)\ ,
\ee
which results in the same relation~\eqref{FT22}.

We conclude that the step evolution operator for right-transport takes the
simple exponential form
\bel{FT32}
\hat S^{(R)}=\gamma^{(R)}\exp\left\{-i\sum_k\frac{2\pi
k}{M}\left[a^\dagger(k)a(k)-\frac12\right]\right\}\ .
\ee
We can write it in Hamiltonian form
\begin{align}
\label{FT33}
\hat S^{(R)}=&\,\gamma^{(R)}\exp\big(-i\eps\tilde{H}^{(R)}\big)\ ,\nonumber\\
\tilde{H}^{(R)}=&\,\sum_k\frac{2\pi
k}{M\eps}\left[a^{\dagger}(k)a(k)-\frac12\right]\ .
\end{align}
The Hamiltonian $\tilde{H}^{(R)}$ describes independent fermionic harmonic
oscillators, including their ground state energy $-\pi k/(M\eps)$. The
eigenvalues of $\tilde H^{(R)}$ are $\pm\pi k/(M\eps)$, reflecting the
simplicity of a map between occupied and empty momentum modes.

In this form we still have an undetermined overall phase $\gamma^{(R)}$. We
choose it such that $\hat S^{(R)}$ equals unity when applied to the totally
empty state $|0\rangle_p$. For this state one has $a(k)|0\rangle_p=0$ for all
$k$. For even $M$ this implies
\begin{align}
\label{APH1}
\hat S^{(R)}=&\,\gamma^{(R)}\exp\left\{i\sum_k\frac{\pi
k}{M}\right\}=\gamma^{(R)}\exp\left(\frac{i\pi}{2}\right)\nonumber\\
=&\,i\gamma^{(R)}=1\ .
\end{align}
The choice of $\gamma^{(R)}=-i$ absorbs the the constant part in $\tilde
H^{(R)}$. For odd $M$ one finds $\gamma^{(R)}=1$. The constant part in $\tilde
H^{(R)}$ vanishes by virtue of the opposite contributions from $k$ and $-k$.

We may extend the discussion to settings where also $A^*(j)$ and $A^T(j)$ are
needed to generate arbitrary states from the same vacuum. In this case the
relation~\eqref{FT22} has to be extended by the condition
\bel{4.8.38A}
\hat SA_j^*=A_{j+1}^*\hat S\ ,\quad \hat SA_j^T=A_{j+1}^T\hat S\ .
\ee
For real $\hat S$ this follows directly by taking the complex conjugate of
eq.~\eqref{FT22}. In this case on concludes that the step evolution
operator~\eqref{FT33} obeys eq.~\eqref{FT21} as well.

\paragraph*{Hamiltonian for free massless particles}

With the appropriate value for $\gamma^{(R)}$ we can immediately write the
right-transport operator in a Hamiltonian form,
\bel{APH2}
\hat S^{(R)}=\exp\big\{-i\eps H^{(R)}\big)\ ,\quad H^{(R)}=\sum_k\frac{2\pi
k}{M\eps}a^\dagger(k)a(k)\ .
\ee
(For $M$ even $H^{(R)}$ and $\tilde{H}^{(R)}$ are related by an additive shift,
for $M$ odd they are identical.)
In this form the periodicity properties of the time evolution for momentum
eigenstates are directly visible. For example, we may consider the sector of
particular one-particle states with fixed $k=\tilde p$, defined by
\bel{APH3}
\psi_l(k)\sim a^\dagger(l)|0\rangle_p\tilde\delta_{k,l}\ .
\ee
For these states the step evolution operator takes the form
\bel{APH4}
\hat S^{(R)}(k,l)=\exp\left(-\frac{2\pi il}{M}\right)\tilde\delta_{k,l}\ ,
\ee
since $\hat n_k\psi_l=a^\dagger(k)a(k)\psi_l=\delta_{kl}\psi_l$. This coincides
with the simple clock system~\eqref{FT6}.
With
\bel{APH5}
\big(\hat S^{(R)}\big)^{M/l}=1\ ,
\ee
the evolution is periodic in time with period given by $N_t=sM/l$, where $s$ is
the smallest integer such that $N_t$ is integer.

It is instructive to compare the form~\eqref{APH2} of the step evolution
operator with the expression~\eqref{CA5}. This comparison reveals two important
simplifications. First, the different $k$-modes decouple and $\hat S^{(R)}$
becomes block diagonal. Second, no ordering operation of the type $N[\,]$ is
needed for guaranteeing the unique jump property. We encounter a situation well
known from particle physics. The kinetic part of the Hamiltonian, which
describes free particles, becomes simple in momentum space. We observe the
linear dispersion relation for the Fourier modes $H_k\sim ka^\dagger(k)a(k)$,
rather than a formula $H_k\sim\sin\gl2\pi k/M\gr$ which often characterizes
typical lattice Hamiltonians for relativistic fermions. The latter produces
``lattice doublers'' since $H_k$ vanishes both for $k=0$ and $k=M/2$. These
lattice doublers are absent in our setting.

For the left-transport by one position one replaces in
eqs.~\eqref{FT21},~\eqref{FT22} $j+1$ by $j-1$. This replaces in
eqs.~\eqref{FT28},~\eqref{FT29} the factor $\exp(2\pi ik/M)$ by $\exp(-2\pi
ik/M)$. As a result the sign of the Hamiltonian is reversed
\bel{FT34}
\hat S^{(L)}=\exp\big(-i\eps H^{(L)}\big)\ ,\quad H^{(L)}=-\sum_k\frac{2\pi
k}{M\eps}a^\dagger(k)a(k)\ .
\ee
For left-movers the normalization factor for even $M$ is $\gamma^{(L)}=i$, while
$\gamma^{(L)}=1$ for odd $M$. As a consequence, for a model with both
right-movers and left-movers (e.g. Dirac fermions) the normalization factor
drops out, $\tilde H^{(R)}+\tilde H^{(L)}=H^{(R)}+H^{(L)}$.

For the continuum limit $M\to\infty$ with fixed $L=M\eps$ we employ the
continuous momentum variable
\bel{FT35}
q=\frac{2\pi k}{M\eps}=\frac{2\pi k}{L}\ ,\quad \frac
1M\sum_k=\eps\int_q=\eps\int\frac{\text{d}q}{2\pi}\ ,
\ee
where the interval of $q$ is given by $[-\pi/\eps,\pi/\eps]$, with boundaries
identified. The continuum creation and annihilation operators are defined by
\bel{FT36}
a(q)=L^{1/2}a(k)\ ,\quad a^{\dagger}(q)=L^{1/2}a^{\dagger}(k)\ ,
\ee
such that
\bel{FT37}
\big\{a^\dagger(p),a(q)\big\}=2\pi\delta(q-p)\ ,
\ee
and
\bel{FT38}
\frac{2\pi}{M\eps}\sum_kka^\dagger(k)a(k)=\int_qqa^\dagger(q)a(q)\ .
\ee
With a continuous Fourier transform,
\bel{FT39}
a(x)=\int_qe^{iqx}a(q)\ ,
\ee
one recovers for $H^{(R)}$ the continuum limit
\bel{FT40}
H^{(R)}=\int_xa^\dagger(x)(-i\partial_x)a(x)\ ,
\ee
in accordance with eq.~\eqref{VP21}. For left-movers one replaces
$\partial_x\to-\partial_x$.

For a direct comparison with the discrete formulation~\eqref{CA5} or~\eqref{CA8}
we can translate eq.~\eqref{APH2} to position space,
\begin{align}
\label{FT41}
H^{(R)}=&\,\frac{2\pi}{M\eps}\sum_kka^\dagger(k)a(k)\nonumber\\
=&\,-\frac{i}{2\eps}\sum_j\sum_mB(m)A^\dagger(j)A(j+m)\ ,
\end{align}
with real
\bel{FT42}
B(m)=\frac{4\pi i}{M^2}\sum_kk\exp\left(-\frac{2\pi ikm}{M}\right) =
-\frac{2\pi(-1)^m}{M\sin\left(\frac{\pi m}{M}\right)}\ .
\ee
For $M\to\infty$ this yields
\bel{FT42A}
B(m) = -\frac{2(-1)^m}{m}\ .
\ee
The expression~\eqref{FT41} becomes the SLAC lattice
derivative~\cite{PhysRevD.14.1627}. The Hamiltonian operator~\eqref{FT41} is
non-local. Nevertheless, the corresponding step evolution operator only
displaces all bits one position to the right. In the discrete sense this is a
local operation. The Hamiltonians~\eqref{CA8} and~\eqref{FT41} are rather
different. Nevertheless, the expressions~\eqref{CA5} and~\eqref{APH2} both
produce the same step evolution operator. The choice of a Hamiltonian for a
given step evolution operator is not unique.

\subsection[Particles and antiparticles]{Particles and antiparticles}
\label{sec:Particles_and_antiparticles}

In momentum space one can find a vacuum state for which the Hamiltonian for a
model of free massless fermions is positive for all excitations. For this
half-filled vacuum all momentum modes with negative $k$ are occupied, and all
modes with positive $k$ are empty. This corresponds to the ``Dirac sea'' in
particle physics. We may call it the ``quantum field theory vacuum''. The
excitations of this vacuum are particles and antiparticles. The presence of
antiparticles is directly related to the implementation of a complex structure
which permits a Fourier transform, and the possibility of a vacuum with minimal
energy.

\paragraph*{Antiparticles}

For a general vacuum state $|0\rangle$ we require that it is invariant under
time evolution and therefore obeys
\bel{FT43}
H|0\rangle=0\ ,\quad \sum_kka^\dagger(k)a(k)|0\rangle=0\ .
\ee
We want to construct a particular half-filled vacuum for which all excitations
have positive $H$. This leads directly to the concept of antiparticles. We
assume that a particle-antiparticle transformation maps
\bel{FT44}
C_{\text{pa}}:\, A(j)\leftrightarrow A^\dagger(j)\ ,\quad a(k)\leftrightarrow
a^\dagger(-k)\ .
\ee
The second relation follows from the first by use of eq.~\eqref{FT17} and
$D(k,j)=D^{-1}(j,-k)$. We introduce for $k>0$ the creation and annihilation
operators for antiparticles
\bel{FT45}
b(k)=a^\dagger(-k)\ ,\quad b^\dagger(k)=a(-k)\ .
\ee
The operators $b(k)$, $b^\dagger(k)$ anticommute with $a(k)$, $a^\dagger(k)$ and
obey $\big\{b^\dagger(k),b(l)\big\}=\delta_{kl}$. Particle-antiparticle
conjugation maps for $k\geq0$
\bel{FT46}
C_{\text{pa}}:\, a(k)\leftrightarrow b(k)\ ,\quad a^\dagger(k)\leftrightarrow
b^\dagger(k)\ ,\quad k>0.
\ee
This corresponds indeed to a map between particles and antiparticles.

The Hamiltonian $H=H^{(R)}$ in eq.~\eqref{APH2} can be written in the form
\bel{FT47}
H=\frac{2\pi}{L}\sum_{k=1}^{J'}\big(h_k^{(p)}+h_k^{(a)}\gr+\Delta H\ ,
\ee
with
\bel{FT48}
h_k^{(p)}=k\big(a^\dagger(k)a(k)-\frac12\big)\ ,\quad
h_k^{(a)}=k\big(b^\dagger(k)b(k)-\frac12\big)\ .
\ee
This uses for $k\neq0$ the identity
\begin{align}\label{Ft49}
k\big(b^\dagger(k)b(k)-\frac12\big)&=k\big(a(-k)a^\dagger(-k)-\frac12\big)\nonumber\\
&=-k\big(a^\dagger(-k)a(-k)-\frac12\big)\ .
\end{align}
The detailed definition depends slightly on even or odd $M$. For odd $M$ one has
$J'=J$ and $\Delta H=0$. For even $M$ one takes $J'=J-1$, and the remaining
contribution for $k=M/2$ results in
\bel{4.8.58A}
\Delta H=\frac\pi\eps a^\dagger\left(\frac M2\right)a\left(\frac M2\right)\ .
\ee

We split off the ground state energy $H_0$
\begin{align}\label{FT50}
H=&\,\hat H+H_0\ ,\nonumber\\
\hat
H=&\,\frac{2\pi}{L}\sum_{k=1}^{J'}k\big[a^\dagger(k)a(k)+b^\dagger(k)b(k)\big]\
.
\end{align}
For even $M$ one has for $H_0$
\bel{4.8.58B}
H_0=\frac\pi\eps\left(a^\dagger\left(\frac M2\right)a\left(\frac
M2\right)-\frac14(M-2)\right]\ ,
\ee
while for odd $M$ it reads
\bel{4.8.58C}
H_0=-\frac{(M^2-1)\pi}{4M\eps}\ .
\ee
The part $\hat H$ is positive,
\bel{FT51}
\hat H\geq0\ ,
\ee
and invariant under the exchange of particles and antiparticles. In order to see
that the particle-hole invariance of the step evolution operator is also
preserved by $H_0$ for $M$ even we recall that the replacement
$a^\dagger(M/2)a(M/2)-1/2$ by $-\big(a^\dagger(M/2)a(M/2)-1/2\big)$ does not
change the step evolution operator according to eq.~\eqref{FT31C}. The step
evolution operator is also left invariant by a shift $H\to H+2\pi/\eps$.

\paragraph*{Half-filled vacua}

We consider half-filled ground states which obey
\bel{4.8.60A}
a(k)|0\rangle=0\ ,\quad b(k)|0\rangle=0\ \text{for}\ 0<k<J'\ .
\ee
For odd $M$ we further require $\langle a^\dagger(0)a(0)\rangle=1/2$. For even
$M$ we consider $k=M/2$ as special and not included in the Fourier modes for
$k\leq J'$. We require
\bel{FT52}
\left(a^\dagger\left(\frac M2\right)a\left(\frac
M2\right)+a^\dagger(0)a(0)\right)|0\rangle=|0\rangle\ .
\ee
These conditions are invariant under the particle-antiparticle transformation.
For these ground states one has
\begin{align}\label{FT55}
a^\dagger(k)a(k)|0\rangle&=0\quad \text{for}\ 0<k\leq J'\nonumber\\
a^\dagger(k)a(k)|0\rangle&=1\quad \text{for}\ -J'\leq k<0\ .
\end{align}

The interpretation in Fourier space is simple: All modes with negative $k$ are
maximally occupied, and no mode with positive $k$ is occupied. For even $M$ the
vacuum is an eigenstate of the total particle number
\bel{FT54}
\hat
N|0\rangle=\sum_jA^\dagger(j)A(j)|0\rangle=\sum_ka^\dagger(k)a(k)|0\rangle=\frac
M2|0\rangle\ .
\ee
For odd $M$ and $M$ large this is still the case approximately
\bel{4.8.63A}
\hat N|0\rangle=\frac M2|0\rangle+\gl a^\dagger(0)a(0)-\frac12\gr|0\rangle\ .
\ee
For $\langle a^\dagger(0)a(0)\rangle=1/2$ one has $\langle\hat N\rangle=M/2$.
Translations by one unit in space are realized by
\bel{FT56}
A(j)\to A(j+1)\ ,\quad a(k)\to\exp\left\{\frac{2\pi ik}{M}\right\}a(k)|0\rangle\
,
\ee
and the vacuum remains invariant. These are the characteristics of a half-filled
ground state. 

For the mean particle number per site $j$ one has
\begin{align}\label{FT53}
\langle n(j)\rangle&=\langle0|A^\dagger(j)A(j)|0\rangle\nonumber\\
&=\frac1M\sum_{k,l}\exp\left\{\frac{2\pi
ij(l-k)}{M}\right\}\langle0|a^\dagger(k)a(l)|0\rangle\nonumber\\
&=\frac1M\bigg\{\sum_{k<0}\sum_{l<0}\exp\left\{\frac{2\pi
ij(l-k)}{M}\langle0|a^\dagger(k)a(l)|0\rangle+1\right\}\ .
\end{align}
Here we use for even $M$ the observation that for $|l-k|=M/2$ the exponential
yields $-1$ such that there is no contribution from
$\big\{a^\dagger(M/2),a(0)\big\}$. For negative $k$ and $l$ and $k\neq l$ we
employ
\bel{4.8.61B}
\langle0|a^\dagger(k)a(l)|0\rangle=-\langle0|a(l)a^\dagger(k)|0\rangle=0\ .
\ee
As a result, one obtains for the mean occupation number per site
\bel{4.8.61C}
\langle
n(j)\rangle=\frac1M\bigg\{\sum_{k<0}\langle0|a^\dagger(k)a(k)|0\rangle+c\bigg\}=\frac12\
,
\ee
with $c=1$ for $M$ even and $c=\langle a^\dagger(0)a(0)\rangle=1/2$ for $M$ odd.

The conditions
\bel{PA1}
a(k>0)|0\rangle=a^\dagger(k<0)|0\rangle=0\ ,
\ee
can be translated to position space by the Fourier transform~\eqref{FT17}. For a
complete characterization of the vacuum wave function for even $M$ one needs an
additional condition in the sector $k=0\ ,\ M/2$ beyond the
condition~\eqref{FT52}. The possibility to impose eq.~\eqref{PA1} on a
non-trivial state requires $a^\dagger(k>0)|0\rangle\neq0$,
$a(k<0)|0\rangle\neq0$.

We can obtain the half-filled vacua from the totally empty vacuum $|0\rangle_p$
by
\bel{4.8.66A}
|0\rangle=f\prod_{k<0}a^\dagger(k)|0\rangle_p\ .
\ee
The totally empty vacuum obeys for all $k$ and $j$
\bel{4.8.66B}
a(k)|0\rangle_p=0\ ,\quad A(j)|0\rangle_p=0\ .
\ee
By virtue of equation~\eqref{FT52} the factor $f$ is given for even $M$ by
\bel{4.8.66C}
f=ca^\dagger(0)+\bar ca^\dagger\left(\frac M2\right)\ .
\ee
The normalization of the vacuum wave function requires
\bel{4.8.66D}
|c|^2+|\bar c|^2=1\ .
\ee
Equivalently, we can obtain the half filled vacuum from the totally filled
vacuum $|0\rangle_a$ which obeys for all $k$,
\bel{4.8.66E}
a^\dagger(k)|0\rangle_a=0\ .
\ee
It obeys the relation
\bel{4.8.66F}
|0\rangle=\bar f\prod_{k=1}^{M/2-1}a(k)|0\rangle_a\ .
\ee
with
\bel{4.8.66G}
\bar f=ca(0)+\bar ca\left(\frac M2\right)\ .
\ee
As an example we may take $c=1$, $\bar c=0$. For odd $M$ we take
\bel{4.8.66GA}
f=\frac{1}{\sqrt{2}}\gl1+a^\dagger(0)\gr\ .
\ee

With eq.~\eqref{4.8.66A} we can express the half-filled vacuum in position space
for even $M$ and $\bar c=0$ as
\begin{align}
\label{4.8.66H}
|0\rangle=&\,\prod_{k\geq0}a^\dagger(k)|0\rangle_p\nonumber\\
=&\,\prod_{k\geq0}\left(\sum_{j_k}\frac{1}{\sqrt{M}}\exp\left\{\frac{2\pi
ij_kk}{M}\right\}A^\dagger(j_k)\right)|0\rangle_p\nonumber\\
=&\,M^{-\frac M2}\sum_{j_0}\sum_{j_1}\sum_{j_2}\dots\sum_{j_{M/2-1}}\nonumber\\
&\times\exp\left\{\frac{2\pi i}{M}\gl j_1+2j_2+\dots+\left(\frac
M2-1\right)j_{M/2-1}\gr\right\}\nonumber\\
&\times A^\dagger(j_0)A^\dagger(j_1)A^\dagger(j_2)\dots
A^\dagger(j_{M/2-1})|0\rangle_p\ .
\end{align}
Each term in the sum involves $M/2$ factors $A^\dagger(j_k)$. Thus the
half-filled vacuum $|0\rangle$ is a superposition of a very large number of
configurations with $M/2$ different occupied site
$(j_0,j_1,j_2,\dots,j_{M/2-1})$. The other $M/2$ sites are empty. Configurations
with an occupied site $j_k$ obey $A^\dagger(j_k)A(j_k)\vp_\tau=\vp_\tau$, and
for an empty site one has $A(j_k)A^\dagger(j_k)\vp_{\tau'}=\vp_{\tau'}$. The
particular phase factors for the different terms single out the particular
vacuum with positive energies for particles and antiparticles from more general
half-filled states with arbitrary coefficients for the different half-filled
configurations. The simple structure of the half-filled vacuum in momentum space
corresponds to a rather complex structure in position space!

For odd $M$ one obtains an expression very similar to eq.~\eqref{4.8.66H}, now
with another series of terms for which the contribution from $k=0$ is omitted.
Also for even $M$ $\bar c\neq 0$ there is a second series of terms. We also
emphasize that for this discussion the notion of occupied or empty sites relates
to the eigenvalues of $A^\dagger(j)A(j)=(1,0)$. They do not necessarily coincide
with the original occupation numbers which are given by the eigenvalues of
$a^\dagger(j)a(j)$.

\paragraph*{One-particle wave function}

The complex one-particle wave function describes both a single particle or a
single antiparticle. In momentum space we define it as
\bel{PA2}
\psi^{(1)}(t)=\left[\sum_{k>0}\tilde\psi(t,k)a^\dagger(k)+\sum_{k<0}\tilde\psi(t,k)a(k)\right]|0\rangle\
,
\ee
where for even $M$ the value $M/2$ is not included in the sums over $k$.
Expressed in terms of one-particle basis functions $b_k$,
\bel{PA3}
\psi^{(1)}(t)=\sum_k\tilde\psi(t,k)b_k
\ee
one has for $k\neq0\ ,\ M/2$
\bel{PA4}
b_{k>0}=a^\dagger(k)|0\rangle\ ,\quad b_{k<0}=a(k)|0\rangle\ .
\ee
In the remaining space of basis functions corresponding to $k=0$ or $k=M/2$ for
even $M$ the Fourier components $\tilde\psi(t,k)$ vanish. We can equivalently
write eq.~\eqref{PA2} in terms of particles and holes as
\bel{PA5}
\psi^{(1)}(t)=\sum_{k>0}\left[\tilde\psi(t,k)a^\dagger(k)+\tilde\psi(t,-k)b^\dagger(k)\right]|0\rangle\
.
\ee
As appropriate for a right-moving Weyl fermion the sum is only over positive
$k$.

Despite their simple formal structure the one-particle excitations of this type
of particle-hole invariant half-filled vacuum have a rather complex wave
function once expressed in terms of configurations of Ising spins. Even for a
plane wave solution with $\tilde\psi(t,k)=\vp(t)\delta_{k,\bar k}$ one has
\begin{align}
\label{4.8.70A}
\psi_{\bar k}^{(1)}(t)=&\,\vp(t)a^\dagger(\bar k)|0\rangle\nonumber\\
=&\,\frac{\vp(t)}{\sqrt{M}}\sum_j\exp\left(\frac{2\pi ij\bar
k}{M}\right)A^\dagger(j)|0\rangle\ ,
\end{align}
where we recall the rather complex structure~\eqref{4.8.66H} for the half-filled
vacuum in position space.

If we repeat the same arguments for left-movers the vacuum conditions in this
sector are $a_L(k)|0\rangle=0$, $a_L^\dagger(-k)|0\rangle=0$ for $k<0$.
Correspondingly, the one-particle wave function for left-movers replaces in
eq.~\eqref{PA5} the summation range to negative $k$. The general condition for
this type of vacuum is that all states with negative eigenvalues of $H$ are
completely filled, and those with positive eigenvalues completely empty.

The organization of the wave function in terms of particles and antiparticles is
well known in quantum field theory. Revealing the corresponding structures for
probabilistic cellular automata relies crucially on a formulation of the
probabilistic information in terms of wave functions rather than probability
distributions. Only for wave functions the Fourier transform is possible, and
only in the Fourier basis the characterization~\eqref{FT55} of the half-filled
vacuum is simple. And only in the Fourier basis the discrete formulation of the
step evolution operator or Hamiltonian in terms of fermionic operators becomes
simple. A discrete description of the half-filled ground state and its
one-particle excitations in terms of the time-local probability distribution in
position space seems extremely complicated. Only the quantum formalism for
cellular automata allows for a simple description of these rather complicated
states. The precise form of the quantum field theory vacuum in position space is
not needed for the description of particle and antiparticle states.

\paragraph*{Dirac equation for particles in quantum field theory vacuum}

The evolution equation for the one-particle excitations of the half-filled
ground state are given by
\bel{PA6}
i\partial_t\tilde\psi^{(1)}=\hat H^{(1)}\tilde\psi^{(1)}\ ,
\ee
with $\hat H^{(1)}$ the restriction of $\hat H$ in eq.\eqref{FT50} to the
one-particle states. For free massless fermions the Hamiltonian is diagonal in
momentum space
\bel{PA7}
i\partial_t\tilde\psi(k)=\frac{2\pi}{L}\big[k\theta(k)-k\theta(-k)\big]\tilde\psi(k)=\frac{2\pi|k|}{L}\tilde\psi(k)\
.
\ee
This form guarantees that $\hat H^{(1)}$ is a positive operator. For theories
with interactions the boundedness of the Hamiltonian from below, combined with
$H$ being a conserved observable, constitutes strong restrictions for the
evolution.

The physical content of eq.~\eqref{PA7} is very intuitive, with right-movers
having positive momentum given by $|q|=2\pi|k|/L>0$ and positive energy $H=|q|$.
On the other hand, the $\theta$-functions in eq.~\eqref{PA7} do not admit a
simple transformation to position space even in the continuum limit. We can
introduce a related ``Weyl wave function'' with a simpler form of the evolution
equation. For this we split $\psi^{(1)}$ into parts for particle and
antiparticle excitations
\bel{PA8}
\tilde\psi^{(1)}=\tilde\psi^{(1)}_p+\tilde\psi^{(1)}_a\ ,
\ee
where
\bel{PA9}
\tilde\psi^{(1)}_p=\sum_{k>0}\tilde\psi(k)a^\dagger(k)|0\rangle\ ,\quad
\tilde\psi^{(1)}_a=\sum_{k<0}\tilde\psi(k)a(k)|0\rangle\ .
\ee
The Weyl wave function is defined by using the complex conjugate of the
antiparticle part of the wave function
\bel{PA10}
\psi^{(1)}=\tilde\psi_p^{(1)}+\gl\tilde\psi_a^{(1)}\gr^*=\psi_p^{(1)}+\psi_a^{(1)}\
,
\ee
where the Fourier modes are given by
\bel{PA11}
\psi(k<0)=\tilde\psi^*(k<0)\ ,\quad \psi(k>0)=\tilde\psi(k>0)\ .
\ee
With
\bel{PA12}
i\partial_t\psi(k<0)=\frac{2\pi k}{L}\psi(k<0)\ ,
\ee
we obtain for arbitrary signs of $k$ or $q$
\bel{PA13}
i\partial_t\psi(q)=q\psi(q)\ .
\ee

In the continuum limit this permits a straightforward Fourier transform
\bel{PA14}
i\partial_t\psi(x)=-i\partial_x\psi(x)\ .
\ee
In this basis the evolution equation for excitations of the half-filled vacuum
takes the same form as for the empty vacuum. The difference is that one-particle
states are described by a complex wave function, instead of the real wave
function for the totally empty vacuum. This reflects that the one-particle wave
function describes both particles and antiparticles.

For left-movers the sign of $k$ and $q$ is reversed. For an automaton describing
free massless right-movers and left-movers we can combine the one-particle wave
function for either left- or right-movers in a two-component complex Dirac wave
function
\bel{PA15}
\psi=\begin{pmatrix}\psi_R\\ \psi_L\end{pmatrix}\ ,\quad
i\partial_t\psi(q)=q\tau_3\psi(q)\ .
\ee
In position space this results in the two-dimensional Dirac equation
\bel{PA16}
i\partial_t\psi(x)=-i\tau_3\partial_x\psi(x)\ ,\quad
\gamma^\mu\partial_\mu\psi=0\ .
\ee
The second eq.~\eqref{PA16} uses the Majorana representation~\eqref{PHA21} of
the Dirac matrices. By a change of basis one can switch to any other
representation of these matrices.

\paragraph*{Interactions}

So far we have discussed the quantum field theory vacuum only for free fermions.
In the presence of interactions the corresponding generalization is more
involved. One can still proceed to a division between particles and
antiparticles by dividing eigenfunctions of the Hamiltonian into those with
positive and negative eigenvalues. For the negative eigenvalues one may again
occupy all states in order to define the quantum field theory vacuum (Dirac
sea). As a result, all excitations of the vacuum have positive energies. While
conceptually simple, this procedure is in practice a formidable task. For a
field theory with interactions the precise spectrum of the Hamiltonian is not
available. The quantum field theory vacuum is known to be a rather complicated
object. Quantum field theory methods based on a functional integral are usually
much more efficient than the use of the operator formalism. This includes, for
example, the Thirring automaton. Constructing the quantum field theory vacuum
and following one-particle excitations is most likely too difficult for
combinatoric procedures despite the extremely simple updating rule.

\section{Subsystems}
\label{sec:subsystems}

Subsystems are a central ingredient for a probabilistic description of the
world.
For most practical purposes one does not want to deal with the overall
probability 
distribution for all events in the universe from the past to the future. One
rather
concentrates on subsystems.

Let us take four examples. The solar system is a rather well isolated subsystem
located in space. At least for the recent cosmological epoch the evolution
within the solar system would be almost the same if we would discard somehow the
rest of the universe -- neglecting here a few astronomers and physicists and
their
cultural impact. The earth can also be treated as a subsystem. The evolution is,
however, not independent of its environment. Day and night, the seasons or the
tides
reflect its correlation with the sun or the moon. An isolated atom is a third
subsystem -- paradigmatic for the microworld. Finally, we may take the present
as a fourth
subsystem. It is described by probabilistic information at the present time,
rather
than by a probability distribution for all times.

What is common to all subsystems is that only part of the probabilistic
information for the overall system is used -- often a very small part. This
information should be sufficient to determine the expectation values of
(sub)system observables and the time-evolution of the subsystem without the
explicit need of probabilistic information for the ``environment'' of the
subsystem. This does not mean that the subsystem is disconnected from the
environment in the sense that no correlations exist between the subsystem and
its environment. We mainly encounter subsystems that are correlated with the
environment and therefore cannot be separated completely from it.

We will see how new probabilistic features emerge for subsystems. This concerns
probabilistic observables that do not have sharp values for every state of the
subsystem, or incomplete statistics where the expectation values of two
observables $A$ and $B$ can be computed from the probabilistic information of
the subsystem, while the classical correlation function $\langle
AB\rangle_{\text{cl}}$ is not available for the subsystem.

Often subsystems are obtained by some averaging or coarse graining, thereby
discarding part of the information contained in the overall probability
distribution. For the coarse graining of a unique jump chain or automaton the
subsystem typically looses the unique jump property. For example, we may
consider an automaton for which a particle at $x$ moves to the left, while at
the position $x+\epsilon$ it moves to the right. Combining $x$ and $x+\epsilon$
to a coarse grained cell a particle in this cell may move either to the right or
to the left with certain probabilities. If no information is lost by the
evolution on the coarse grained level the coarse grained step evolution operator
will continue to be an orthogonal matrix. It is no longer a unique jump matrix,
with several non-zero entries in a given row corresponding to the different
possibilities for the evolution. A finite dimensional orthogonal matrix that is
not a unique jump matrix must have elements with different signs. The continuum
limit corresponds to a coarse graining.

Quantum systems are subsystems of ``classical'' overall probabilistic systems.
We discuss here subsystems of a classical statistical overall system more
generally. This will show that many typical ``quantum properties'' as
observables represented by non-commuting operators and the corresponding
uncertainty relations are actually present for many subsystems and not a
particular property of quantum systems.

An important type of subsystem that we have already encountered is the
time-local subsystem. The appropriate description of observables and time
evolution of this subsystem is in terms of the (classical) density matrix,
operators associated to observables and the step evolution operator. A crucial
feature is that many different observables $B_i$ of the overall system are
mapped to the same operator $\hat B$. While their expectation values $\langle
B_i\rangle$ are therefore the same, the classical correlation function with
another observable, $\langle AB_i\rangle$, differs for different $B_i$. These
classical correlation functions can no longer be reconstructed from the
time-local subsystem, in particular if some averaging in time is performed
towards the continuum limit. The non-availability of the classical correlation
functions enforces new definitions of correlations for the time-local subsystem.
These correlations are built on operator products. Since they do not correspond
to classical correlations they do not have to obey Bell's inequalities.

By the map to the density matrix and from observables to operators much of the
detailed information contained in the overall probability distribution is lost.
This gives to the time-local subsystem a high degree of robustness, making it
insensitive to many ``microscopic details''. It is precisely this information
loss which is responsible for the new probabilistic features of subsystems.

\subsection{Subsystems and environment}\label{sec:subsystems_and_environment}

This section gives a general definition of subsystems of the overall
probabilistic system. It emphasizes the importance of correlations with the
environment.

\subsubsection{Subsystems and correlation with\\environment}
\label{sec:subsystems_and_correlation_with_environment}

Defining a subsystem divides somehow the probabilistic information into a ``system'' and
its environment. The ``system'' is often used as a shorthand for the subsystem. The division
depends on the particular choice of a subsystem and we need to understand the
underlying formal concept.

\paragraph*{Subsystems}
A general subsystem is characterized by a number of ``system variables'' $\rho_z$. These
are $N_S$ real numbers, where $N_S$ may be infinite. For simplicity we stay here with
finite $N_S$, and the limit $N_S \rightarrow \infty$ can be taken at the end if needed.
The ``state of the subsystem'', $\rho = (\rho_1, \dotsc, \rho_{N_S})$, is a point in
$\mathbb{R}^{N_S}$. This point characterizes the probabilistic information in the
subsystem. Expectation values of ``system observables'' and the other properties of the
subsystem can be computed as a function of $\rho$.

For a subsystem $\rho$ must be computable for any given overall probability distribution
$\{p_{\tau}\}$, $\rho = \rho[p_\tau]$. A subsystem with $N_S$ variables is therefore a map from the
space $\mathscr{P}$ of all overall probability distributions $\{p_\tau\}$ to $\mathbb{R}^{N_S}$,
\begin{equation}
  \label{SUB1}
  \mathscr{P} \to \mathbb{R}^{N_S},\quad \{p_\tau\} \mapsto \rho[p_\tau]\,.
\end{equation}
To every point in the space of probability distributions $\{p_\tau\}$ it associates values of the
system variables. A given overall probability distribution $\{p_\tau\}$ contains always all the
probabilistic information that is available for the subsystem. The space of general subsystems are
the maps from $\mathscr{P}$ to $\mathbb{R}^{N_S}$ with arbitrary $\mathbb{R}^{N_S}$.

For a genuine subsystem the map $\mathscr{P} \to \mathbb{R}^{N_S}$ is not injective. Different
probability distributions $\{p_\tau\}$ and  $\{{p'}_\tau\}$ are mapped to the same state of the
subsystem $\rho$. The subsystem contains therefore less probabilistic information than the overall
probabilistic system. Parts of the overall probabilistic information are lost if one concentrates on
the subsystem. This is precisely the purpose of dealing with subsystems: the amount of necessary
information is reduced for all system observables for which $\rho$ is sufficient to compute
their probabilistic properties. The probabilistic information contained in the overall probability
distribution, but not in $\rho$, may be called the ``environment''. In this sense the environment
is the complement of the system within the overall probabilistic system. For some subsystems the
system variables $\rho_z$ are linear combinations of the probabilities $p_\tau$,
\begin{equation}
  \label{SUB2}
  \rho_z = a_z^\tau  p_\tau\,.
\end{equation}
In this case the environment consists of the other linear combinations independent of $\rho_z$.
(Linear independence has to be defined in presence of the constraint $\sum_\tau p_\tau = 1$.)

We emphasize that the system variables are real numbers that can be negative. Appropriate
probabilities $w_i$ within the subsystem are functions of $\rho$. They have to obey the positivity
and normalization constraints for probabilities. We will find that often the probabilistic
information in $\rho$ is sufficient to define different probability distributions, e.g.
$\{w_i^{(1)}\}$ and $\{w_i^{(2)}\}$, for which one is not a subsystem of the other. For this reason
$\rho$ cannot simply be associated with a probability distribution for the subsystem.

\paragraph*{Correlation with environment}

Different types of subsystems can be characterized by their correlation with the environment. The
solar system can be treated to a good approximation as a subsystem that is not correlated with its
environment. In the absence of correlations the overall probability distribution can be written
as a direct product of independent probability distributions, one for the system and the other for
the environment. The subsystem earth is correlated with its environment. Changes in the
environment, for example removing the sun, would impact the evolution on earth. The earth is an
open system. To a good approximation one could parametrize the influence of the environment by
additional parameters, for example the time-dependent photon flux from the sun and the varying
gravitational field produced by masses outside the earth. These parameters may be included in the
state of the earth and could provide for a closed description of the subsystem earth.
Nevertheless, we should recall that the subsystem is correlated with its environment. The overall
probability distribution is no longer a direct product of a probability distribution for the earth
and a probability distribution for the environment -- otherwise the correlation with the sun would be
absent. The correlation with the environment has to be taken into account when one ``embeds'' the
subsystem into the overall probability distribution.

Atoms are subsystems correlated with their environment. In quantum field theory isolated atoms are
particular excitations in a surrounding vacuum. The probabilistic information of the atom system
permits a closed description, for example by the Schr\"odinger equation. Nevertheless, the
embedding of the atom subsystem into the overall probability distribution is not of the direct
product type for uncorrelated subsystems. The properties of the atom depend on the properties of
the vacuum. For example, if one changes the expectation value of the Higgs field, the electron mass
changes correspondingly and a central parameter in the Schr\"odinger equation is modified. Even though
in practice the expectation value of the Higgs field is static and we can well describe the atom
with a fixed electron mass, the influence of the vacuum properties clearly indicates correlations
of the isolated atom with its environment. Without such a correlation no response of the atom to a
change of the Higgs field would be possible. This argument is simple, but we will see that it has far
reaching consequences. Too often atoms are treated as subsystems without correlation with the
environment.

The subsystem ``present'' is a correlated subsystem as well. It may often be possible to describe
it effectively as a closed system, if we take ``present'' in a wider sense of a finite time interval
$\Delta t$ around the present time $t_0$. Often one can formulate evolution equations that are
time-local in the sense that properties at some time $t_2$ within the time interval $\Delta t$ can be
computed from the probabilistic information for a neighbouring time $t_1$ within the time interval
$\Delta t$. Nevertheless, the subsystem ``present'' is correlated with its environment. The present
is influenced by the past, and we can use present probabilistic information for making predictions
for the future. This is only possible in the presence of correlations between the subsystem and its
environment. For a direct product structure the environment would not be influenced by the system,
and there would be no influence of the environment on the subsystem. We will investigate in the next
section the ``local time subsystem'' extensively.

\paragraph*{Uncorrelated subsystems}

In the absence of correlations between the system and its environment we may formulate separate
probability distributions $p_\alpha^{(S)}$ for the system and  $p_a^{(E)}$ for the environment.
The overall probability distribution takes a direct product form, with $\tau = (\alpha, a)$ a double
index,
\begin{equation}
  \label{SUB3}
  p_\tau = p_{\alpha a} = p_\alpha^{(S)} p_a^{(E)}\,.
\end{equation}
The probability distribution of the system is obtained from the overall probability distribution by
``integrating out'' the environment
\begin{equation}
  \label{SUB4}
  p_\alpha^{(S)} = \sum_a p_{\alpha a} = \sum_a p_\alpha^{(S)} p_a^{(E)} =  p_\alpha^{(S)} \sum_a p_a^{(E)}\,,
\end{equation}
where we use the normalization $\sum_a  p_a^{(E)} = 1$. Inversely, a direct product form of the
overall probability distribution~\eqref{SUB3} implies that the system and the environment are
uncorrelated. Any change in the probability distribution for the environment $p_a^{(E)}$ does not
affect the probability distribution of the system $ p_\alpha^{(S)}$ and vice versa.

System observables are those for which the probabilities for the possible measurement values
$A_\alpha^{(S)}$ are those of the system. The possible measurement values do not depend on the
environment,
\begin{equation}
  \label{SUB5}
  A_\tau^{(S)} = A_{\alpha a}^{(S)} = A_\alpha^{(S)}\,.
\end{equation}
The general classical statistical rule for expectation values directly translates to the
subsystem,
\begin{align}
  \label{SUB6}
  \braket{A^{(S)}} &= \sum_\tau p_\tau A_\tau^{(S)} = \sum_{\alpha a} p_{\alpha a} A_{\alpha a}^{(S)} \nonumber \\
  &= \sum_\alpha p_\alpha^{(S)} A_\alpha^{(S)} \sum_a p_a^{(E)} = \sum_\alpha p_\alpha^{(S)} A_\alpha^{(S)}\,.
\end{align}
All correlation functions of system observables can be computed from $\{p_\alpha^{(S)}\}$, e.g.
\begin{equation}
\label{SUB7}
  \braket{A^{(S)} B^{(S)}} = \sum_{\alpha a} p_{\alpha a} A_\alpha^{(S)} B_\alpha^{(S)} 
  = \sum_{\alpha} p_{\alpha}^{(S)} A_\alpha^{(S)} B_\alpha^{(S)}\,.
\end{equation}
We deal with ``complete statistics'' in this sense. The parameter $\rho$ for the subsystem can be
identified with the probabilities $\{p_\alpha^{(S)}\}$. Thus the system probability distribution
$\{p_\alpha^{(S)}\}$ defines the probabilistic state of the subsystem. We can also identify $\alpha$ with
the microstates of the subsystem. Every system observable $A^{(S)}$ has a fixed value $A_\alpha^{(S)}$
in any given microstate $\alpha$.

We may similarly define ``environment observables'' $B^{(E)}$ for which the probabilities to find
possible measurement values are given by the $p_a^{(E)}$,
\begin{equation}
  \label{SUB8}
  B_{\alpha a}^{(E)} = B_a^{(E)}\,,\quad \braket{B^{(E)}} = \sum_a p_a^{(E)} B_a^{(E)}\,.
\end{equation}
All connected correlation functions for system observables and environment observables vanish,
\begin{align}
\nonumber
&\braket{A^{(S)} B^{(E)}} - \braket{A^{(S)}} \braket{B^{(E)}} \\
\label{SUB9}
&\hspace*{-1.5mm}= \sum_{\alpha a} p_{\alpha a} A_\alpha^{(S)} B_a^{(E)} - \left( \sum_\alpha p_\alpha^{(S)} A_\alpha^{(S)} \right) \left( \sum_a p_a^{(E)} B_a^{(E)} \right) = 0.
\end{align}
This is a formal statement of the absence of correlations between the system and its environment.
It is directly related to the direct product form of the overall probability distribution.

\paragraph*{Correlated subsystems}

Many subsystems of practical importance, as atoms or the local time subsystem, are correlated with their
environment. For such ``correlated subsystems'' a direct product form~\eqref{SUB3} of the overall probability
distribution is not possible. We will concentrate in this section on correlated subsystems and discuss
specific important subsystems in the following: Time local subsystems in sect.~\ref{sec:time_local_subsystems}, correlation subsystems in sect.~\ref{sec:correlation_subsystems}, local chains in sect.~\ref{sec:matrix_chains_as_subsystems_of_local_chains}, subtraces of density matrices in sect.~\ref{sec:subtraces} and general subsystems in~\ref{sec:general_local_subsystems}.

Correlated subsystems show specific probabilistic features that differ from the classical statistics for the
overall probability distribution or for uncorrelated subsystems. One important feature are ``probabilistic
observables''. In general, correlated subsystems are characterized by system variables $\rho$ without the
presence of microstates $\alpha$. For probabilistic system observables the probabilities $w_i(\rho)$ for
finding a possible measurement value $\lambda_i$ can be computed for every probabilistic state $\rho$.
There are, however, no microstates $\alpha$ for which all system observables take fixed values. One
typically finds restrictions of the type of uncertainty relations which forbid that all system obervables
can take simultaneously sharp values for a given $\rho$.

A second characteristic feature for many correlated subsystems is ``incomplete statistics''. There are
system observables $A$ and $B$ for which the expectation values $\braket{A}_\rho$ and  $\braket{B}_\rho$
can be computed from $\rho$. The classical correlation function $\braket{AB}_\text{cl}$, which is well
defined in the overall system, can no longer be computed from $\rho$. This is closely related to the fact
that system observables often represent equivalence classes. Two observables which differ in the overall
probabilistic system can lead to the same system observable. We will discuss these features of
correlated subsystems in sects.~\ref{sec:local_observables_and_non_commuting_operators} and~\ref{sec:classical_correlations_and_continuum_limit}.
\subsubsection{Time-local subsystems}
\label{sec:time_local_subsystems}

An important class of subsystems are time-local subsystems. They only use
probabilistic information at a
given time $t$ or site $m$ on a chain. This information is sufficient for the
computation of expectation values of local
observables. The time-local information may consist of a local probability
distribution or a classical density
matrix. Time-local subsystems are a crucial tool for a physicist to make
predictions. Given the local probability
information at time $t_1$, she will attempt to predict probabilities for
observations at a later time $t_2 > t_1$.
We often use the wording ``local subsystem'' if it is clear from the context
that locality relates to time. Parts
of the material of this section can be found in refs. \cite{CWIT,CWPT}.

\paragraph*{Local probability distribution}

A simple, albeit not general, form of local probabilistic information is the
local probability distribution.
It obtains from the overall probability distribution by ``summation'' or
``integration'' over the variables
at time different from the given time $t$. For a one-bit local chain the overall
probability distribution
$p[s(m')]$ depends on the Ising spins $s(m')$ for all times or positions on the
chain $m'$. The local probability
distribution at time $m$ is given by eq.~\eqref{eq:TOS10b}
\begin{equation}
 \label{LS1}
 p(m) = p\left(s(m)\right) = \prod_{m' \neq m} \sum_{s(m') = \pm 1} p[s(m')]\,.
\end{equation}
It depends on a single local Ising spin $s(m)$.

This definition extends directly to generalized Ising models with an arbitrary
number of local Ising spins
$s_\gamma(m')$. The local probability distribution is a function of the local
spins $s_\gamma(m)$ at the
position $m$ on the chain. Generalizations to the limiting cases of continuous
time $t$ or continuous
variables $\phi(m)$ or $\phi(t)$ are straightforward.

\paragraph*{Expectation values of strictly local observables}

Local observables in a narrow sense are functions of local Ising spins
$s_\gamma(m)$. We will call them
``strictly local observables'', in distinction to a more general notion of local
observables to be discussed
below. The expectation values of strictly local observables $A(m)$ can be
computed from the local probability
distribution,
\begin{equation}
 \label{LS2}
 \braket{A(m)} = \int \dif s(m)\, A(m)\,p(m)\,.
\end{equation}
This follows directly from the general law for expectation values~\eqref{I1},
\begin{equation}
 \label{LS3}
 \braket{A(m)} = \int \mathcal{D}s\, A(s(m))\, p[s(m')]\,,
\end{equation}
by performing first the integration over $s(m' \neq m)$ according to
eq.~\eqref{LS1}. The symbols $\int \dif s(m)$
and $\int \mathcal{D}s$ imply summations over all local Ising spins
$s_\gamma(m)$ or over all Ising spins at
arbitrary $m'$, respectively.

From the point of view of the time-local subsystem at $m$ an observable
$B(s(\hat m))$, which is a function of the Ising spins at a different time $\hat
m\neq m$, is an environment observable. The overall probability distribution
allows for the computation of the connected correlation function $\langle
A(m)B(\hat m)\rangle_c$. In general, this ``unequal time correlation'' will not
vanish, indicating that the system is correlated with its environment. Neither
$\langle B(\hat m)\rangle$ not the correlation $\langle A(m)B(\hat m)\rangle_c$
can be computed from the local probabilities $\{p(m)\}$.

\paragraph*{Classical density matrix}

We have already seen in
sects.~\ref{sec:step_evolution_operator}-~\ref{sec:classical_density_matrix},~\ref{sec:markov_chains}
that in general the local probabilistic information contained in the local
probability distribution is insufficient for the computation of the evolution of
the local probability distribution. The local object that permits the
formulation of a simple evolution law is the classical density matrix $\rho'(m)$
introduced in sect.~\ref{sec:classical_density_matrix}. The local probabilistic
information contained in the classical density matrix at time $m$ or $t$ is
sufficient for a computation of the classical density matrix at a neighboring
time $t+\epsilon$ or $m+1$. For a given model this evolution law involves the
transfer matrix or step evolution operator~\eqref{eq:SE1}. The local
probabilities are the diagonal elements of the classical density matrix.

The expectation values of strictly local classical observables can be computed
form the local probabilistic information in the classical density matrix by the
quantum rule~\eqref{eq:DM34}. Here the strictly local observables are
represented in the occupation numbers basis by diagonal
operators~\eqref{eq:CW24}. The operators for strictly local observables all
commute. In the following, we define the time-local subsystems by the classical
density matrix $\rho'(t)$. The variables $\rho$ of this subsystem are the
independent elements of $\rho'$, or suitable combinations thereof.

The local probabilistic information contained in the density matrix exceeds the
one in the time-local probability distribution which corresponds to the diagonal
elements of the density matrix. The information stored in the off-diagonal
elements of the density matrix permits the computation of expectation values of
further observables, extending the notion of local observables beyond the family
of strictly local observables. For example, if the step evolution operator $\hat
S(m)$ is known we can find operators for the observables $B(m+1)$ which involve
neighboring spins $s(m+1)$, such that $\langle B(m+1)\rangle$ can be computed
from $\rho'(m)$. From the density matrix we can also compute the expectation
values of statistical observables as momentum in the preceding section.

\subsection{Observables and operators}\label{sec:observables_and_operators}

In this section we address the connection between observables and associated operators in more detail. We will find observables that correspond to operators with non-diagonal elements. These operators do not commute with the diagonal operators for the strictly local observables. The computation of their expectation values from the probabilistic information of the time-local subsystem involves the off-diagonal elements of the density matrix $\rho'(m)$. We will establish the generalization of the Heisenberg picture for operators in quantum mechanics to general classical statistical systems.

\subsubsection{Local observables and non-commuting operators}
\label{sec:local_observables_and_non_commuting_operators}

The local probabilities are the diagonal elements of the classical density matrix. The off-diagonal elements of
the classical density matrix contain additional local probabilistic information. This information is necessary,
in general, in order to formulate a local evolution law. Furthermore, the additional local probabilistic information
in the off-diagonal elements of $\rho'$ permits the computation of expectation values for an extended set of
local observables, beyond the strictly local observables.

\paragraph*{Neighboring observables}

In particular, for a given model with given step evolution operators $\hat{S}(t)$, the classical density matrix
$\rho'(m)$ permits the computation of expectation values for observables $A(m+1)$ or $A(m-1)$ \cite{CWIT,CWQF}.
These expectation values can be computed from the quantum rule
\begin{equation}
  \label{LS4}
  \braket{A(m+1)} = \tr \left\{ \rho'(m)\, \hat{A}(m+1, m)\right\}\,.
\end{equation}
Here $\hat{A}(m+1, m)$ is a suitable operator associated to the observable $A(m+1) = A(s(m+1))$, where the second index $m$ indicates that this operator refers to the time-local subsystem at $m$. Typically, this
operator is not diagonal in the occupation number basis. The trace~\eqref{LS4} therefore involves the off-diagonal
elements of $\rho'(m)$.

With respect to the local subsystem at time $m+1$ the observable $A(m+1) = A(s(m+1))$ is a strictly local observable,
with
\begin{equation}
  \label{LS5}
    \braket{A(m+1)} = \tr \left\{ \rho'(m+1)\, \hat{A}(m+1, m+1)\right\}
  \end{equation}
and diagonal operator $\hat{A}(m+1)\equiv\hat{A}(m+1, m+1)$. We label operators here with two different times $\hat{A}(m_1; m_2)$.
The first time label $m_1$ indicates that the associated observable $A(m_1)$  depends only on Ising spins $s_\gamma(m_1)$ at time
$m_1$. The second time label $m_2$ indicates with respect to which classical density matrix $\rho'(m_2)$ the operator refers, such
that the quantum rule applies at $m_2$, as in eq.~\eqref{LS4} or \eqref{LS5}. Using the evolution law for the classical density matrix
eq.~\eqref{LS5} yields
\begin{align}
  \label{LS6}
  \braket{A(m+1)} &= \tr \left\{ \hat{S}(m)\, \rho'(m)\, \hat{S}^{-1}(m)\, \hat{A}(m+1, m+1)\right\} \nonumber \\
  &= \tr \left\{ \rho'(m)\, \hat{S}^{-1}(m)\, \hat{A}(m+1, m+1)\, \hat{S}(m)\right\}\,.
\end{align}
Comparing with eq.~\eqref{LS4} this determines
\begin{equation}
  \label{LS7}
  \hat{A}(m+1,m) = \hat{S}^{-1}(m)\, \hat{A}(m+1, m+1)\, \hat{S}(m)\,.
\end{equation}
In general, the step evolution operator $\hat{S}(m)$ does not commute with the diagonal operator $\hat{A}(m+1,m+1)$
and the operator $\hat{A}(m+1,m)$ is not diagonal.

The observable $A(m+1)$ is a first example for an extended set of local observables for which expectation values can be computed
from the local probabilistic information contained in the classical density matrix.
By a similar construction other such observables are $A(m-1)$ or $A(m+2)$ etc.. As long as the step evolution operators
$\hat{S}(m')$ are known for a certain range of $m'$ around $m$, the expectation values of all observables $A(m')$ can be computed
from $\rho'(m)$. As long as $m'$ remains in some sense in the vicinity of $m$ we may include $A(m')$ in the extended set of
local observables.

\paragraph*{Heisenberg operators}

The operators $\hat{A}(n, m)$ or $\hat{A}(t_1,t)$ are Heisenberg operators. Heisenberg operators are a familiar concept
in quantum mechanics. We argue here that for classical statistics with local information encoded in the classical density
matrix the concept of Heisenberg operators is very useful as well. The definition of Heisenberg operators in classical
statistics is analogous to quantum mechanics.

For $t_b > t_a$ we define the evolution operator $U(t_b,t_a)$ as the ordered product of step evolution operators,
\begin{equation}
  \label{LS7A}
  U(t_b,t_a) = \hat{S}(t_b - \epsilon)\,\hat{S}(t_b - 2\epsilon) \cdots \hat{S}(t_a + \epsilon)\, \hat{S}(t_a)\,.
\end{equation}
We also define
\begin{equation}
  \label{LS7B}
   U(t_a,t_b) =  U^{-1}(t_b,t_a)\,,
 \end{equation}
 and observe the relation
 \begin{equation}
   \label{LS7C}
   U(t_c,t_b)\,U(t_b,t_a) = U(t_c,t_a)\,.
 \end{equation}
 As an important possible difference to quantum mechanics the evolution operator does, in general, not need to be
 a unitary or orthogonal matrix.
 
 The Heisenberg operator associated to $A(t_1)$ is given by
 \begin{equation}
   \label{LS7D}
   \hat{A}_H(t_1,t) = U(t,t_1)\, \hat{A}(t_1,t_1) U^{-1}(t,t_1)\,.
 \end{equation}
 We recognize in eq.~\eqref{LS7} the operator $\hat{A}(t+\epsilon) = \hat{A}_H(t+\epsilon,t)$ to be the Heisenberg
 operator associated to $A(t+\epsilon)$, with $U(t,t+\epsilon) = \hat{S}^{-1}(t)$. For any observable $A(t_1)$ which
 only depends on variables $s(t_1)$, the associated local operator at $t$ is precisely the Heisenberg operator at $t$,
 \begin{equation}
   \label{LS7E}
   \braket{A(t_1)} = \tr \left\{ \rho'(t)\, \hat{A}_H(t_1,t) \right\}\,.
 \end{equation}

 \paragraph*{Local observables and local operators}
 
 At this point one realizes that the concept of local observables should be extended beyond the strictly local observables. Roughly speaking, from the point of view of the time-local subsystem a generalized setting for local observables should meet two conditions. First the expectation value should be computable from the probabilistic information of the time-local subsystem. Second, the observable should somehow refer to the vicinity of $t$. While the first criterion is mathematically exact, requiring that for a local observable an operator exists such that its expectation value can be computed from the quantum rule, the second criterion is more vague.

 We require for local observables $A$ the following two properties:
 \begin{enumerate}
 \item The observable $A$ has a certain number (possibly infinite) of real possible measurement values $\lambda_i^{(A)}$
   that can be found by some type of ``local measurement''.
 \item The real and positive probabilities $w_i^{(A)}$ to find the value $\lambda_i^{(A)}$ can be computed from the classical
   density matrix $\rho'(t)$.
 \end{enumerate}

 These two conditions imply that the expectation value of $A$ can be computed from the classical density matrix
 \begin{equation}
   \label{LS7F}
   \braket{A} = \sum_i w_i^{(A)} \lambda_i^{(A)}\,.
 \end{equation}
 Furthermore, the observable $A^p$ with integer $p$ is again a local observable. The possible measurement values are given
 by $\left(\lambda_i^{(A)}\right)^p$, and the probabilities to find $\left(\lambda_i^{(A)}\right)^p$ for $A^p$ are the same as
 for $\lambda_i^{(A)}$ for $A$ in case of a non-degenerate spectrum of $A^p$. (If for $\lambda_i^{(A)} \neq  \lambda_j^{(A)}$ one
 has  $\left(\lambda_i^{(A)}\right)^p = \left(\lambda_j^{(A)}\right)^p$, the probabilities for $i$ and $j$ add for $A^p$.) The expectation value of $A^p$ obeys
 \begin{equation}
   \label{LS7G}
   \braket{A^p} = \sum_i w_i^{(A)}  \left(\lambda_i^{(A)}\right)^p\,.
 \end{equation}

 A useful concept are local operators that correspond to local observables. For such ``local-observable operators''
 we require four conditions.
 \begin{enumerate}[label=(\arabic*),labelsep=5mm]
 \item $\braket{A} = \braket{\hat{A}(t)} = \tr\left\{ \rho'(t)\hat{A}(t) \right\}\,,$
 \item $\spec \left(\hat{A}(t)\right) = \left\{ \lambda_i^{(A)} \right\}\,,\quad \lambda_i^{(A)} \in \mathbb{R}\,,$
 \item $w_i^{(A)}\left(\rho, \hat{A}\right) \geq 0\,,$
 \item $A^p \rightarrow \hat{A}^p\,. \hfill \refstepcounter{equation}(\theequation)\label{LS8}$
 \end{enumerate}
The first condition specifies the rule for the computation of $\braket{A}$. The second condition identifies the eigenvalues
of $\hat{A}(t)$ with the possible measurement values of $A$. It requires that all eigenvalues of the operator $\hat A$ are real.

The third condition states that the probabilities $w_i^{(A)}$ to find
$\lambda_i^{(A)}$ must be computable from $\hat{A}(t)$ and $\rho'(t)$.
We can diagonalize the operator $\hat{A}(t)$ by
\begin{equation}
  \label{LS9}
  D(t) \hat{A}(t) D^{-1}(t) = \hat{A}_D(t) = \diag \left(\lambda_i^{(A)}\right)\,.
\end{equation}
Correspondingly, we may apply the same similarity transformation to the classical density matrix,
\begin{equation}
  \label{LS10}
  \rho'_{(D)}(t) = D(t) \rho'(t) D^{-1}(t)\,,
\end{equation}
with
\begin{equation}
  \label{LS11}
  \braket{A} = \tr \left\{ \rho'_{(D)}(t) \hat{A}_{(D)}(t) \right\} = \sum_\tau \left( \hat{A}_{(D)}(t) \right)_{\tau \tau} \left(  \rho'_{(D)}(t) \right)_{\tau \tau}\,.
\end{equation}
The diagonal elements $\left( \hat{A}_{(D)}(t) \right)_{\tau \tau}$ are given by the possible measurement values $\lambda_i^{(A)}$. For a
non-degenerate spectrum we can associate the diagonal elements of $\rho'_{(D)}$ with the probabilities $w_i^{(A)}$ for the
possible measurements values  $\lambda_i^{(A)}$, provided that for all $\tau$ they are real and positive
\begin{equation}
  \label{LS12}
   \left(  \rho'_{(D)}(t) \right)_{\tau \tau} \geq 0\,.
 \end{equation}

 Eq.~\eqref{LS12} is a central condition for an operator to represent a local observable. The elements
 $\left(  \rho'_{(D)}(t) \right)_{\tau \tau}$ depend on $\rho'(t)$ as well as on $\hat{A}$ via $D(t)$ which
 diagonalizes $\hat{A}(t)$. For a degenerate spectrum the probabilities $w_i^{(A)}$ obtain by summing over
 all $\left(  \rho'_{(D)}(t) \right)_{\tau \tau}$ which  ``belong'' to a given $\lambda_i^{(A)}$. Degenerate
 spectra can be considered a limiting case of non-degenerate spectra. Finally, the fourth condition states that
 the operator associated to $A^p$ should be given by the corresponding matrix product $\hat{A}^p$.

 The four conditions are not independent. Conditions $(1), (2), (3)$ imply $(4)$, while conditions
 $(1), (3), (4)$ imply $(2)$. We can therefore either use condition $(2)$ or $(4)$ equivalently. Inversely, for
 all $D(t)$ for which all diagonal elements of $\rho'_{(D)}(t)$ are real and positive, we can define local
 operators
 \begin{equation}
   \label{LS13}
   \hat{A}(t) = D^{-1}(t) \hat{A}_d(t) D(t)\,,
 \end{equation}
for arbitrary real diagonal matrices $\hat{A}_d(t)$. They obey all criteria~\eqref{LS8} and therefore are
local-observable operators. If there exist appropriate local measurement procedures with possible outcome
given by the diagonal elements of $\hat{A}_d(t)$, these operators can be associated to local observables.
We will discuss the general relations between local observables and local operators in more detail below.

The neighboring observables described above obey all criteria for local observables. The corresponding
Heisenberg operators are the associated local-observable operators. For $A(m+1)$ the matrix $D(t)$ is given by
the step evolution operator $\hat{S}(t)$ in eq.~\eqref{LS7}. The transformed classical density matrix
$\rho'_{(D)}(t)$ equals the classical density matrix $\rho'(t+\epsilon)$ at $t+\epsilon$. The diagonal elements
of $\rho'_{(D)}(t)$ are therefore the local probabilities at $t+\epsilon$, such that the condition \eqref{LS12} is
obeyed.

For a positive density matrix all eigenvalues are positive. There exists a matrix $\bar D$ in eq.~\eqref{LS10} leading to eq.~\eqref{LS12} in the basis where $\rho'$ is diagonal. If the density matrix is symmetric as for unique jump chains $\bar D$ is an orthogonal matrix. The diagonal elements of $\rho'$ are then positive in an arbitrary basis, such that eq.~\eqref{LS12} holds for arbitrary $D$. In turn, all operators with real eigenvalues obey the condition~\eqref{LS12}.

\paragraph*{Classical correlations}

The local probabilistic information in the classical density matrix is sufficient for the computation of
classical correlation functions for local observables $A(t_1)$ and $B(t_2)$ at different times $t_1 \neq t_2$.
The operator associated to the classical product observable $A(t_1) B(t_2)$ is the time ordered product of
Heisenberg operators $\hat{A}(t_1,t)$ and $\hat{B}(t_2,t)$, with classical correlation functions given by
\begin{equation}
  \label{LS14}
  \braket{A(t_1)B(t_2)} = \tr \left\{ \rho'(t) \TO \left( \hat{A}_H(t_1,t) \hat{B}_H(t_2,t) \right) \right\}\,.
\end{equation}
Here we define the time ordered operator product as
\begin{equation}
  \label{LS15}
  \TO \left( \hat{A}(t_1,t) \hat{B}(t_2,t) \right) =
  \begin{cases}
    \hat{A}(t_1,t) \hat{B}(t_2,t), & \text{for } t_1 > t_2 \\
     \hat{B}(t_2,t) \hat{A}(t_1,t) & \text{for } t_1 < t_2
  \end{cases}
\end{equation}
such that the operator with the larger time argument stands on the left in the matrix product. (Recall that
the second time label $t$ designates the reference point  and is not used for the time ordering.) The time ordered
operator product is commutative
\begin{equation}
  \label{LS16}
   \TO \left( \hat{A}(t_1,t) \hat{B}(t_2,t) \right) =  \TO \left(  \hat{B}(t_2,t) \hat{A}(t_1,t) \right)\,,
\end{equation}
as appropriate for the classical correlation which has no concept of ordering for the factors $A(t_1)$ and
$B(t_2)$,
\begin{equation}
  \braket{A(t_1)B(t_2)} = \braket{B(t_2)A(t_1)}\,.
\end{equation}

The proof of eq.\,\eqref{LS14} proceeds in analogy to eq.\,\eqref{eq:LO5}. For a local chain we start from the definition
of the classical correlation function
\begin{align}
  \nonumber
  &\braket{A(t_1)B(t_2)} = \int \mathcal{D}s\, p[s] A(t_1)B(t_2) = \\
  \label{LS17}
  &\int \mathcal{D}s\, \prod_{t' \geq t_2} \cK(t') B(t_2) \prod_{t_1 \leq t''< t_2}\cK(t'') A(t_1) \prod_{t'''< t_1} \cK(t''') B.
\end{align}
Here we take without loss of generality $t_2 > t_1$, with $B(t_2)$ depending only on the variables $s(t_2)$, and $A(t_1)$
depending on $s(t_1)$. We assume that the local factors $\cK(t')$ are normalized such that the associated transfer matrix is the
step evolution operator. In the occupation number basis this yields
\begin{align}
  \label{LS18}
  &\braket{A(t_1)B(t_2)} = \\ 
\nonumber  
  &\quad\tr \left\{  U(t_f,t_2) \hat{B}(t_2,t_2) U(t_2,t_1) \hat{A}(t_1,t_1) U(t_1, t_\text{in}) \hat{\cB} \right\}\,.
\end{align}
(The same expression is valid for the matrix chains discussed in appendix~\ref{app:matrix chains}.) The classical density matrix $\rho'(t)$ reads
\begin{equation}
  \label{LS19}
  \rho'(t) = U(t, t_\text{in}) \hat{\cB} U(t_f,t)\,,
\end{equation}
with $\hat{\cB}$ the boundary matrix.

Inverting eq.~\eqref{LS19} one has
\begin{equation}
  \label{LS20}
  \hat{\cB} = U(t_\text{in},t) \rho'(t) U(t,t_\mathrm{f})\,,
\end{equation}
and insertion into eq.~\eqref{LS18} yields with eq.~\eqref{LS7C}
\begin{align}
  \label{LS21}
  &\braket{A(t_1)B(t_2)} = \\ 
  \nonumber
  &\quad\tr \left\{ U(t,t_2) \hat{B}(t_2,t_2) U(t_2,t_1) \hat{A}(t_1,t_1) U(t_1,t) \rho'(t) \right\}\,.
\end{align}
Using $U(t_2,t_1) = U(t_2,t) U(t,t_1)$, and the definition \eqref{LS7D} of the Heisenberg operators, this establishes indeed
eq.~\eqref{LS14}.

\paragraph*{Time-ordered operator product}

The classical correlation function $\langle A(t_1)B(t_2)\rangle$ has to obey Bell's inequalities. We can use the overall probability distribution in order to show that all conditions for their validity are met. In particular, $A(t_1)$ and $B(t_2)$ have both well defined values for the states $\omega$ of the overall classical statistical ensemble. We will argue next that this classical correlation is not an observable accessible for the time-local subsystem, however. It does not obey the criteria for time-local observables. The time ordered operator product $T\big(\hat A_H(t_1,t)\hat B_H(t_2,t)\big)$ is well defined and its expectation value can be evaluated by the quantum rule. Nevertheless, there is no subsystem-observable associated to it. While both $A(t_1)$ and $B(t_2)$ are local observables, the ``classical product'' $A(t_1)B(t_2)$ is not. The classical correlation does not measure only properties of the time-local subsystem. It involves properties of the environment. This has important implications for quantum mechanics formulated as a time-local subsystem. If one wants to define some type of ``subsystem correlation'' for the local observables $A(t_1)$ and $B(t_2)$  this has to be compatible with pure subsystem properties not involving the environment. It therefore cannot be the classical correlation. In turn, such a ``subsystem correlation'' has not to obey Bell's inequalities.

We want to show that the time ordered operator product does not define a local observable. Let us associate to two operators $\hat{A}_H(t_1,t)$ and $\hat{B}_H(t_2,t)$ a new operator $\hat{C}(t)$ by the time ordered
operator product
\begin{equation}
  \label{LS22}
  \hat{C}(t) = \TO \left( \hat{A}_H(t_1,t) \hat{B}_H(t_2,t) \right)\,.
\end{equation}
The classical correlation function $\braket{A(t_1)B(t_2)}$ is the expectation value of $\hat{C}$ according to the first
eq.~\eqref{LS8},
\begin{equation}
  \label{LS23}
  \braket{A(t_1)B(t_2)} = \braket{\hat{C}(t)} = \tr \left\{ \rho'(t) \hat{C}(t) \right\}\,.
\end{equation}

Nevertheless, the operator $\hat{C}(t)$ is in general not a local-observable operator associated with a local observable.
Indeed, the operator product of two non-commuting operators for local observables is typically no longer
a local-observable operator corresponding to a local observable. As an example, we may consider a single Ising spin at each $m$ and take
\begin{align}
  \label{LS24}
  \hat{A}_H &= \begin{pmatrix} 1 & 0 \\ 0 & -1 \end{pmatrix}\,, \quad
  \hat{B}_H = \begin{pmatrix} \cos \theta & \sin \theta \\ \sin \theta & - \cos \theta \end{pmatrix}\,, \notag \\
  \hat{A}_H \hat{B}_H &= \begin{pmatrix} \cos \theta & \sin \theta \\ - \sin \theta &  \cos \theta \end{pmatrix}\,.
\end{align}
Both $\hat{A}_H$ and $\hat{B}_H$ are possible operators for two-level observables with eigenvalues $\lambda_i^{(A)} = \pm 1$
,  $\lambda_i^{(B)} = \pm 1$. The eigenvalues of the operator product $\hat{A}_H\hat{B}_H$ are given by
$\lambda_i = \cos \theta \pm i \sin \theta$. For $\sin \theta \neq 0$ the eigenvalues are not real, contradicting the
condition $(2)$ in eq.~\eqref{LS8}. The operator $\hat{A}_H\hat{B}_H$ can therefore not correspond to a local observable.

An operator with real eigenvalues may be constructed by the anticommutator
\begin{equation}\label{5.2.29A}
\frac12\big\{\hat A_H,\hat B_H\big\}=\begin{pmatrix}\cos\theta&0\\0&\cos\theta\end{pmatrix}\ .
\end{equation}
The eigenvalue $\cos\theta$ does, however, not correspond to one of the products of eigenvalues for $\hat A_H$ and $\hat B_H$, which are the possible measurement values $\pm1$ of the associated observables. The classical product observable has possible measurement values $\pm1$.

As another indication, the time ordering may not be compatible with condition $(4)$ in eq.~\eqref{LS8}.
If $A(t_1)$ and $B(t_2)$ are
two-level observables, the product $C = A(t_1)B(t_2)$ is again a two-level observable, $C^2 = 1$. The time ordered operator
associated to $C^2$ is for $t_1 > t_2$
\begin{equation}
  \label{LS25}
  \TO \left( \left( \hat{A}_H(t_1,t) \hat{B}_H(t_2,t) \right)^2 \right) = \hat{A}_H^2(t_1) \hat{B}_H^2(t_2) = 1\,,
\end{equation}
while
\begin{align}
  \label{LS26}
  \hat{C}^2 &= \left( \TO \left( \hat{A}_H(t_1,t) \hat{B}_H(t_2,t) \right) \right)^2 \nonumber \\
  &= \hat{A}_H(t_1,t) \hat{B}_H(t_2,t) \hat{A}_H(t_1,t) \hat{B}_H(t_2,t) \nonumber \\
  &= 1 + \hat{A}_H(t_1,t) \left[\hat{B}_H(t_2,t),\hat{A}_H(t_1,t) \right] \hat{B}_H(t_2,t)
\end{align}
can differ from one if $\hat{B}_H$ and $\hat{A}_H$ do not commute. In this case $\hat{C}$ is again
not a local operator for a local observable. While the expectation value of the classical product observable
$A(t_1)B(t_2)$ is computable in the local subsystem, there is, in general, no local-observable operator
obeying the conditions \eqref{LS8} associated to this observable.

This formal mismatch between time ordered operator products and local observables has a simple reason. A given operator $\hat A$ represents a whole equivalence class of different observables $A_i$ that are all mapped to $\hat A$. Similarly, different $B_j$ are all mapped to the operator $\hat B$. The classical correlation $\langle A_iB_j\rangle$ depends on the individual observables $A_i$ and $B_j$. On the other side, operator products depend only on the equivalence class and are therefore independent of $i$ and $j$. This simple argument demonstrates directly that the classical correlation involves, in general, properties of the environment. In turn, it is not surprising that expressions based on operator products may not correspond to local observables. We will discuss later observables associated to operator products which do not correspond to classical correlations.
\subsubsection{Algebras of local observables and\\operators}
\label{sec:algebras_of_local_observables_and_operators}

From the point of view of the local subsystem there are different algebraic structures for the local observables and local operators. The first type of structure are the classical linear combinations and products of observables. The second type is based on linear combinations and products of local operators. By analogy to quantum mechanics it may be called the ``quantum algebra''. We briefly address the relations between these structures.

The classical algebra and the associated classical correlations can, in principle, be accessible from the probabilistic information in the time-local subsystem. In practice, however, these structures are not relevant for observations. We demonstrate this by a discussion of time-derivative observables. The classical algebra conflicts with the continuum limit. Different definitions of time-derivative observables can differ by microscopic details on a time scale $\varepsilon$. Measuring their classical correlations would need a time resolution of the order $\varepsilon$. Different microscopic definitions of time-derivative observables lead to different classical correlation functions. In contrast, the quantum algebra based on the operator product is robust and compatible with the continuum limit. A given local-observable operator for a time-derivative observable represents a whole equivalence class of different microscopic time-derivative observables. The operator product preserves the structure of equivalence classes and is insensitive to microscopic details of the precise definition of the time-derivative. These findings generalize to arbitrary averaged observables that are relevant for the continuum limit. We conclude that the quantum algebra is robust and well suited for measurements in the continuum limit where the microscopic details should not matter. In contrast, the classical algebra is inappropriate for the continuum limit.

\paragraph*{Classical algebra}

The classical linear combinations and products of observables are defined by the overall probability distribution. The question arises to which extend the local probabilistic information of the subsystem remains sufficient to compute sums and products of local observables. Consider two local observables $A$ and $B$ with possible measurement values $\lambda_i^{(A)}$ and $\lambda_j^{(B)}$, and probabilities $w_i^{(A)}$ and $w_j^{(B)}$ to find these values. (We omit time arguments for simplicity of the notation.) The classical product observable $C=AB$ has possible measurement values $\lambda_k^{(C)} = \lambda_{ij}^{(C)} = \lambda_i^{(A)}\lambda_j^{(B)}$. The expectation value of $C$,
\begin{equation}
\braket{C} = \sum_k w_k^{(C)} \lambda_k^{(C)} = \sum_{i,j} w_{ij}^{(AB)} \lambda_i^{(A)} \lambda_j^{(B)},
\label{eq:LS27}
\end{equation} 
involves the \textit{simultaneous probabilities} $w_{ij}^{(AB)}$ to find for $A$ the value $\lambda_i^{(A)}$ and for $B$ the value $\lambda_j^{(B)}$. In the presence of correlations $w_{ij}^{(AB)}$ differs from the product $w_i^{(A)} w_j^{(B)}$ and the question arises if $w_{ij}^{(AB)}$ remains computable from the information within the subsystem.

In general, the simultaneous probabilities $w_{ij}^{(AB)}$ cannot be computed from the classical density matrix $\rho'$ alone. The local information in the classical density matrix is sufficient to compute $w_i^{(A)}$ and $w_j^{(B)}$, but in general does not give access to $w_{ij}^{(AB)}$. One can, however, often reconstruct $w_{ij}^{(AB)}$ from a sufficient set of classical correlation functions. This involves beyond $\rho'$ knowledge of the step evolution operator $\hat{S}$.

Let us take for $A$ and $B$ Ising spins with $\lambda_i^{(A)}=\pm 1$, $\lambda_j^{(B)}=\pm 1$. For $A$ we consider a two-level observable at time $t+\varepsilon$, with associated Heisenberg operator
\begin{equation}
\hat{A}_H(t) = \hat{S}^{-1}(t) \hat{A}_\mathrm{d} \hat{S}(t),
\label{eq:LS28}
\end{equation}
while for $B$ we take a two-level observable at $t$ represented by the diagonal operator $\hat{B}_\mathrm{d}$. Both $\hat{A}_\mathrm{d}$ and $\hat{B}_\mathrm{d}$ are diagonal matrices with eigenvalues $\pm 1$, but $\hat{A}_H$ does not commute with $\hat{B}_\mathrm{d}$ for a general step evolution operator $\hat{S}$. From the expectation values and classical correlation, 
\begin{align}
\begin{split}
\braket{A} &= \tr (\rho' \hat{A}_H),\quad 
\braket{B} = \tr (\rho' \hat{B}_\mathrm{d}), \\
\braket{AB} &= \tr(\rho'\hat{A}_H \hat{B}_\mathrm{d}),
\end{split}
\label{eq:LS29}
\end{align}
we can construct the simultaneous probabilities $w_{ij}^{(AB)}$. One has
\begin{align}
\begin{split}
\braket{A} &= \sum_{i,j} w_{ij}^{(AB)} \lambda_i^{(A)},\quad \braket{B} = \sum_{i,j} w_{ij}^{(AB)} \lambda_j^{(B)}, \\
\braket{AB} &= \sum_{i,j} w_{ij}^{(AB)} \lambda_i^{(A)} \lambda_j^{(B)},\quad 1 = \sum_{i,j} w_{ij}^{(AB)},
\end{split}
\label{eq:LS30}
\end{align}
where the indices $i,j$ take the values $(+,-)$ with $\lambda_+^{(A)}=1$, $\lambda_-^{(A)}=-1$, $\lambda_+^{(B)}=1$, $\lambda_-^{(B)}=-1$, 
\begin{align}
\begin{split}
\braket{A} &= w_{++}^{(AB)} + w_{+-}^{(AB)} - w_{-+}^{(AB)} - w_{--}^{(AB)}, \\
\braket{B} &= w_{++}^{(AB)} - w_{+-}^{(AB)} + w_{-+}^{(AB)} - w_{--}^{(AB)}, \\
\braket{AB} &= w_{++}^{(AB)} - w_{+-}^{(AB)} - w_{-+}^{(AB)} + w_{--}^{(AB)}, \\
1 &= w_{++}^{(AB)} + w_{+-}^{(AB)} + w_{-+}^{(AB)} + w_{--}^{(AB)}.
\end{split}
\label{eq:LS31}
\end{align}
The simultaneous probabilities are found as
\begin{align}
\begin{split}
w_{++}^{(AB)} &= \frac{1}{4} ( 1 + \braket{A} + \braket{B} + \braket{AB} ), \\
w_{+-}^{(AB)} &= \frac{1}{4} ( 1 + \braket{A} - \braket{B} - \braket{AB} ), \\
w_{-+}^{(AB)} &= \frac{1}{4} ( 1 - \braket{A} + \braket{B} - \braket{AB} ), \\
w_{--}^{(AB)} &= \frac{1}{4} ( 1 - \braket{A} - \braket{B} + \braket{AB} ).
\end{split}
\label{eq:LS32}
\end{align}

Since the local probabilistic information is sufficient for the computation of $\braket{A}$, $\braket{B}$ and $\braket{AB}$, it is also sufficient for the computation of the simultaneous probabilities $w_{ij}^{(AB)}$. One needs, however, knowledge of the step evolution operator $\hat{S}$ in order to compute the Heisenberg operator $\hat{A}_H$. Thus only the knowledge of both $\rho'$ and $\hat{S}$ permits the computation of the simultaneous probabilities from the information of the subsystem. This extends to products of observables at different times $t_1$ and $t_2$, provided the step evolution operators $\hat{S}(t')$ are known in the range $t_1 \geq t' < t_2$, and the classical density matrix is known at $t_1$, $t_2$, or at some time between $t_1$ and $t_2$. For observables with more than two possible measurements values also higher correlations, corresponding to classical products with more than two factors, are needed for the reconstruction of the simultaneous probabilities from the local probabilistic information of the subsystem.

In summary, the simultaneous probabilities $w_{ij}^{(AB)}$ are defined by the overall probability distribution as for a classical statistical system. For many circumstances they remain accessible, in principle, for the time-local subsystems defined by the classical density matrix $\rho'(t)$. The relation between $\rho'$ and the simultaneous probabilities is not direct, however. While the simultaneous probabilities are linear combinations of the matrix elements of $\rho'$, the coefficients of this relation depend on the step evolution operator $\hat{S}$. 

Once the simultaneous probabilities $w_{ij}^{(AB)}$ are known, one can also compute the expectation values of linear combinations of two local observables $A$ and $B$. Defining
\begin{equation}
D = \alpha A + \beta B,
\label{eq:LS33}
\end{equation}
the possible measurement values of $D$ are
\begin{equation}
\lambda_k^{(D)} = \lambda_{ij}^{(D)} = \alpha \lambda_i^{(A)} + \beta \lambda_j^{(B)}.
\label{eq:LS34}
\end{equation}
The probabilities to find $\lambda_{ij}^{(D)}$ are given by the simultaneous probabilities $w_{ij}^{(AB)}$. We have not discussed in detail the cases of degenerate spectra. They can be seen as limiting cases of non-degenerate spectra.

\paragraph*{Quantum algebra}

A different algebraic structure is related to sums and products of operators. We are interested in the operator algebra for local-observable operators that obey the conditions (1)--(4) in eq.\,\eqref{LS8}. Let $\hat{A}$ and $\hat{B}$ be two local-observable operators obeying the conditions \eqref{LS8}. The question arises if the product operator,
\begin{equation}
\hat{C} = \frac12\big(\hat{A}\hat{B}+\hat B\hat A\big)\ ,
\label{eq:LS35}
\end{equation}
obeys again the conditions \eqref{LS8}. We take here the anticommutator in order to ensure real eigenvalues of $\hat C$ if $\hat A$ and $\hat B$ are symmetric matrices. If $\hat C$ obeys the conditions~\eqref{LS8} we may associate a local observable $C$ to this operator.

The question arises what are the relations between the local observable $C$ and the classical product observable $AB$, if $A$ and $B$ are local observables associated to $\hat{A}$ and $\hat{B}$, respectively. As we have seen, the operator product (matrix product) of $\hat{A}_H$ and $\hat{B}_\mathrm{d}$ appears in the computation of the classical correlation function. In general, this operator product does not represent a local observable, however. We demonstrate this by the case where $A$ corresponds to a local observable at time $t_2$, and $B$ to a local observable at time $t_1$, with $t_2>t_1$. One has
\bel{OLA1}
\langle A(t_2)B(t_1)\rangle_{\text{cl}}=\langle\TO\big[\hat A_H(t_2)\hat B_H(t_2)\big]\rangle=\langle\hat A_H(t_2)\hat B(t_1)\rangle\ ,
\ee
where we omit the explicit notation of the wave function or density matrix. We assume that $\hat A_H(t_2)$ and $\hat B_H(t_1)$ do not commute. In this case the operator $\TO\big[\hat A_H(t_2)\hat B_H(t_1)\big]=\hat A_H(t_2)\hat B_H(t_1)$ associated to the classical product $A(t_2)B(t_1)$ is not hermitian. Its eigenvalues are not real. (The classical correlation $\langle A(t_2)B(t_1)\rangle_{\text{cl}}$ is real, nevertheless.) Thus $\TO\big[\hat A_H(t_2)\hat B_H(t_1)\big]$ cannot be associated with a local observable. On the other hand, the quantum product
\bel{OLA2}
\hat C=\big\{\hat A_H(t_2),\hat B_H(t_1)\big\}
\ee
is a hermitian operator with real eigenvalues. It is possible that it represents a local observable.

For $\hat{C}$ to obey the condition (3) in eq.\,\eqref{LS8} one needs the condition \eqref{LS12}. For local chains that are not unique jump chains this is not obeyed, in general. Let us now consider the case where $\hat C$ obeys all criteria~\eqref{LS12}. In this case we may associate a local observable $C$ to it. In general, this observable will differ from the classical product variable $AB$. The possible measurement values of $C$, as given by the eigenvalues of $\hat C$, do not match with the possible measurement values of $AB$. The classical observable algebra and the operator algebra are not simply matched to each other for local subsystems.

An important exception is the case of commuting operators, $[\hat{A},\hat{B}]=0$. Commuting operators can be diagonalized simultaneously by the same matrix $D$. This implies that the eigenvalues of $\hat{A}\hat{B}$ are indeed given by the products $\lambda_i^{(A)} \lambda_j^{(B)}$ and therefore real. If eq.\,\eqref{LS12} holds for the matrix $D$ that diagonalizes simultaneously $\hat{A}$ and $\hat{B}$, the simultaneous probabilities $w_{ij}^{(AB)}$ can be extracted from the diagonal elements $(\rho'_{(D)})_{\tau\tau}$. Then the product operator is a local-observable operator, obeying the condition \eqref{LS8}.
\subsubsection{Classical correlations and continuum limit}
\label{sec:classical_correlations_and_continuum_limit}

In this subsection we argue that the classical correlations are often not useful quantities for the time-local subsystem. Classical correlations should primarily be seen as properties of observables for the overall system for all times. Their meaning for the time-local subsystem suffers from two shortcomings. First, different classical observables with different correlations to a third observable are represented by the same operator in the subsystem. Second, observables that lead to the same continuum limit have different classical correlations. In this case the classical correlations are related to the microscopic structure which is washed out in the continuum limit of observables that correspond to the time derivative of another observable.

For a quantum subsystem the ``quantum correlations'' are typically expressed in terms of operator products of the type $\hat C$. Also the classical correlation $AB$ is expressed in terms of $\hat C$. This does not imply that the quantum correlations are identical to the classical correlations. For the example of derivative observables the classical correlations fail to be useful for a time-local subsystem associated with some type of continuum limit. This gives also an explanation why the classical correlations play no role for quantum systems, even though they are sometimes formally accessible by the probabilistic information of the time-local subsystem. In short, a measurement of quantum correlations, or more generally correlations in a time-local subsystem should not be associated to a measurement of classical correlations. The error to be avoided assigns a single unique observable to an operator, neglecting that different observables are mapped to the same operator.

\paragraph*{Derivative observables}

Time derivatives of observables are often important observables themselves. An example is the velocity observable as the time derivative of the position observable. Consider classical local observables $A(t)$ and the derivative observable
\begin{equation}
B(t) = \dot{A}(t) = \frac{1}{2\varepsilon} ( A(t+\varepsilon) - A(t-\varepsilon) ),
\label{eq:LS36}
\end{equation}
with expectation value
\begin{equation}
\braket{\dot{A}(t)} = \frac{1}{2\varepsilon} ( \braket{A(t+\varepsilon)} - \braket{A(t-\varepsilon)} ).
\label{eq:LS37}
\end{equation}
The expectation value of the derivative observable $\dot{A}(t)$ can be computed from the local probabilistic information contained in the classical density matrix. The associated operator to $B=\dot{A}$ is expressed in terms of Heisenberg operators as
\begin{equation}
\hat{B}(t) = \dot{\hat{A}}(t) = \frac{1}{2\varepsilon} \left[ \hat{A}_H(t+\varepsilon,t) - \hat{A}_H(t-\varepsilon,t) \right].
\label{eq:LS38}
\end{equation}
We concentrate on observables which do not depend explicitly on time, e.\,g.\ $\hat{A}_H(t,t) = \hat{A}$, and on time independent step evolution operators $\hat{S}$. In this case one has
\begin{equation}
\hat{B} = \frac{1}{2\varepsilon} \{ \hat{S}^{-1}, [\hat{A},\hat{S}] \}.
\label{eq:LS39}
\end{equation}
In general, the operator $\hat{A}$ and the associated time derivative operator $\dot{\hat{A}} = \hat{B}$ dot not commute
\begin{equation}
[\hat{A},\hat{B}] = \frac{1}{2\varepsilon} \left( \{ \hat{A},\hat{S}^{-1} \} [\hat{A},\hat{S}] - \{ \hat{A},\hat{S} \} [\hat{A},\hat{S}^{-1}] \right).
\label{eq:LS40}
\end{equation}

We can express $\hat{B}$ in terms of the $W$-operator given by eq.\,\eqref{eq:ct2},
\begin{equation}
\hat{B} = \hat{S}\hat{A}W - W\hat{A}\hat{S}.
\label{eq:LS41}
\end{equation}
In the continuum limit of smooth enough wave functions or density matrix, $\hat{S}\tilde{q} = \tilde{q} + \mathcal{O}(\varepsilon)$ etc., we can neglect the factor $\hat{S}$ and replace eq.\,\eqref{eq:LS41} effectively by
\begin{equation}
\hat{B} = [\hat{A},W].
\label{eq:LS42}
\end{equation}
For $\hat{A}$ commuting with $W$ the observable $A$ is a conserved quantity. This is the continuum version of eq.\,\eqref{eq:CQ1}. In the presence of a complex structure eq.\,\eqref{eq:LS42} translates to the operator identity
\begin{equation}
\hat{B} = \dot{\hat{A}} = i[G,\hat{A}] = i[H,\hat{A}] - [J,\hat{A}].
\label{eq:LS43}
\end{equation}
For $J=0$ this is the standard expression of quantum mechanics expressing the time derivative operator $\hat{B}$ as the commutator of $\hat{A}$ and the Hamiltonian.

\paragraph*{Unique operator for different derivative\\observables}

The choice of a derivative observable $B$ is not unique\,\cite{CWPT}. For example, we could use as a different derivative observable
\begin{equation}
B_+(t) = \dot{A}_+(t) = \frac{1}{\varepsilon} ( A(t+\varepsilon) - A(t) ).
\label{eq:LS44}
\end{equation}
The corresponding local operator reads
\begin{equation}
\hat{B}_+(t) = \frac{1}{\varepsilon} \hat{S}^{-1}(t) [ \hat{A}_H(t,t), \hat{S}(t) ].
\label{eq:LS45}
\end{equation}
For constant $\hat{A}_H(t,t)$ and $\hat{S}(t)$ one finds
\begin{equation}
\frac{1}{2} ( \hat{B}_+ + \hat{S}\hat{B}_+\hat{S} ) = [\hat{A},W]\hat{S}\ .
\label{eq:LS46}
\end{equation}
For smooth enough wave functions or density matrices this yields in the continuum limit the same effective replacement as for $\hat{B}$,
\begin{equation}
\hat{B}_+ = [\hat{A},W].
\label{eq:LS47}
\end{equation}
One infers that in the continuum limit the two derivative observables $B$ and $B_+$ are represented by the same local operator $\hat{B}$. They therefore have the same expectation value.

Nevertheless, $B(t)$ or $B_+(t)$ are different classical observables. If $A(t)$ are Ising spins the possible measurement values for $B(t)$ are $(2,0,-2)/(2\varepsilon)$, while for $B_+(t)$ the spectrum of possible measurement values is given by $(2,0,-2)/\varepsilon$. For both observables the spectrum diverges for $\varepsilon\to 0$. In contrast, the operator $\hat{B}$ has a spectrum of eigenvalues that is independent of $\varepsilon$ if $W$ is independent of $\varepsilon$. If $\hat{B}$ is a local-observable operator, it cannot be associated to the classical observables $B(t)$ or $B_+(t)$. 

This is also apparent by the violation of condition (4) in eq.\,\eqref{LS8}. The squared observable
\begin{equation}
B^2(t) = \frac{1}{4\varepsilon^2} ( A^2(t+\varepsilon) + A^2(t-\varepsilon) - 2A(t+\varepsilon)A(t-\varepsilon) )
\label{eq:LS48}
\end{equation}
has for Ising spins possible measurement values $1/\varepsilon^2$ and $0$. Its expectation value can be computed from the classical density matrix. It diverges for $\varepsilon\to 0$ unless $A(t)$ is a conserved Ising spin for which the probability to find different values at $t$ and $t+\varepsilon$ is of measure zero\,\cite{CWPT}. This contrasts with the squared operator $\hat{B}^2(t)$ which has finite spectrum and a finite value for $\tr \{ \rho'(t) \hat{B}^2(t) \}$. The expectation value of $B_+^2(t)$ also diverges for $\varepsilon\to 0$ and differs from the one for $B^2(t)$.

\paragraph*{Failure of classical correlations}

The classical correlation functions $\braket{B(t)A(t)}$ and $\braket{B_+(t)A(t)}$ can be computed from the classical density matrix. They differ, however, and are no longer accessible in the continuum limit. Consider the classical product observable
\begin{equation}
B(t)A(t) = \frac{1}{2\varepsilon} ( A(t+\varepsilon)A(t) - A(t-\varepsilon)A(t) )
\label{eq:LS49}
\end{equation}
and the associated classical correlation function
\begin{align}
\nonumber
\braket{B(t)A(t)} = \frac{1}{2\varepsilon} \tr &\left\{ \rho'(t) \left( \hat{A}_H(t+\varepsilon,t) \hat{A}_H(t,t) \right.\right. \\
\label{eq:LS50}
	&\left.\left.\quad-\hat{A}_H(t,t) \hat{A}_H(t-\varepsilon,t) \right) \right\}.
\end{align}
Taking $\hat{A}_H(t,t) = \hat{A}$, $\hat{S}(t)=\hat{S}$ both independent of $t$, this results in
\begin{align}
\nonumber
&\braket{B(t)A(t)} = \frac{1}{2\varepsilon} \tr \left\{ \rho'(t) (\hat{S}^{-1} \hat{A} \hat{S} \hat{A} - \hat{A} \hat{S} \hat{A} \hat{S}^{-1} ) \right\} \\
\label{eq:LS51}
&\qquad=\frac{1}{2\varepsilon} \tr \left\{ \big(\rho'(t+\varepsilon) - \rho'(t)\big) \hat{A}\hat{S}\hat{A}\hat{S}^{-1} \right\}.
\end{align}
Similarly, one obtains
\begin{align}
\nonumber
\braket{B_+(t)A(t)} &= \frac{1}{\varepsilon} \tr \left\{ \rho'(t+\varepsilon) \hat{A}\hat{S}\hat{A}\hat{S}^{-1} - \rho'(t) \hat{A}^2 \right\} \phantom{ZZZZZZZ}\\
\label{eq:LS52}
&\hspace*{-16mm}= 2\braket{B(t)A(t)} + \frac{1}{\varepsilon} \tr \left\{ \rho'(t) (\hat{A}\hat{S}\hat{A}\hat{S}^{-1} - \hat{A}^2) \right\}.
\end{align}
The operator $\hat A\hat S\hat A\hat S^{-1}$ is typically not close to $\hat A^2$. For a smooth density matrix the classical correlation $\langle B(t)A(t)\rangle$ remains finite for $\epsilon\to0$, involving $\partial_t\rho'$. In contrast, $\langle B_+(t)A(t)\rangle$ diverges for $\epsilon\to0$. The two classical correlation functions are rather different.

We can express the classical correlation functions~\eqref{eq:LS51} and~\eqref{eq:LS52} as expectation values of time ordered Heisenberg operators. The different results for $\hat B$ and $\hat B_+$ indicates directly that time ordered operator products are much less robust than simple operator products. This arises from the discontinuity in the order of operators, $T[\hat A_H(t+\epsilon)\hat A_H(t)]=\hat A_H(t+\epsilon)\hat A_H(t)$, $T[\hat A_H(t-\epsilon)\hat A_H(t)]=\hat A_H(t)\hat A_H(t-\epsilon)$. One concludes that the classical correlations involving derivative operators exist and are even accessible from the local probabilistic information with knowledge of $\hat{S}$. They are not very robust quantities, depending on the precise definition of the derivatives. This contrasts with the robust operator structure. The classical correlations typically do not admit a smooth continuum limit.

Observations that differentiate between the derivative-observables $B$ or $B_+$ would require observations with a time resolution of $\varepsilon$. In many circumstances where a continuum limit applies this does not correspond to reality. Observations of time-derivatives are typically done with a resolution $\Delta t$ that stays finite for $\varepsilon\to 0$. The precise observable corresponding to $\dot{A}$ on the microscopic level is a linear combination of many different classical derivative-observables as $B$ and $B_+$. The coefficients of the linear combination may differ from one to another experiment. At best probabilities for these coefficients will be known. All this does not matter. All the different possible derivative-observables lead to the same operator $\hat{B}$, such that for a given classical density matrix the expectation value of $\dot{A}$ can be computed as
\begin{equation}
\braket{\dot{A}} = \tr \{ \rho'(t) \hat{B}(t) \}.
\label{eq:LS53}
\end{equation}

The operator $\hat{B}(t)$ represents an \textit{equivalence class} of classical derivative observables that all lead to the same expectation value $\braket{\dot{A}}$. This concept of equivalence classes is crucial for an understanding of subsystems. We will discuss it in more detail below. Under many circumstances the time-derivative operator $\hat{B}$ can be associated to a local observable $\bar{B}(t)$, with possible measurement values given by the spectrum of $\hat{B}$. This local observable is, however, not a particular classical observable of the type $B(t)$ or $B_+(t)$. It rather represents a whole equivalence class of classical observables that lead to identical results for measurements on a coarse grained level that averages over microscopic time and has no longer resolution $\varepsilon$.
\subsubsection{Averaged observables}
\label{sec:averaged_observables}

We have seen that many different observables are mapped to the same operator. Structures among operators, as the definition of the derivative operator $\dot{\hat A}=[\hat A,W]$, are robust in the sense that they do not resolve which precise observable is mapped to $\hat A$ and $\dot{\hat A}$. The observables mapped to a given operator $\hat A$ need not to obey the criteria for local observables defined in sect.~\ref{sec:local_observables_and_non_commuting_operators}. There exist useful observables whose expectation values can be computed from the probabilistic information of the time-local subsystem which do not obey all criteria~\eqref{LS8}. We discuss as an example average observables which are typically needed on the way to some continuum limit.

\paragraph*{Average Ising spins}

In the continuum limit for time one can no longer resolve the difference between local observables $A(t)$ and $A(t+\varepsilon)$. Observables that are adapted to the continuum limit are time averaged observables. For this purpose the interval $\Delta t$ for the time averaging is much larger than $\varepsilon$, but still small as compared to a typical time scale for the evolution of the classical density matrix. The outcome of measurements for suitably averaged observables is expected to be independent (within small errors) of the precise form of the averaging procedure.

Let us consider Ising spins $s(t)$ and the averaged observable
\begin{equation}
\sigma(\bar{t}) = \sum_{t'} a^{(\sigma)} (\bar{t} + t') s(\bar{t}+t').
\label{eq:LS54}
\end{equation}
Here $\bar{t}$ denotes the central time of the average, and the precise averaging is encoded in the function $a^{(\sigma)}(\bar{t}+t')$. This function should vanish rapidly for $|t'|\gg \Delta t$ and we normalize it according to
\begin{equation}
\sum_{t'} a^{(\sigma)} (\bar{t}+t') = 1.
\label{eq:LS55}
\end{equation}
An example could be Gaussians characterized by $\Delta t$,
\begin{equation}
a^{(\sigma)}(\bar{t}+t') = N_\sigma \exp \left\{ -\frac{t'^2}{\Delta t^2} \right\},\quad N_\sigma = \frac{\varepsilon}{\sqrt{\pi}\Delta t},
\label{eq:LS56}
\end{equation}
where the last expression uses $\sum_{t'} = \int \d t' /\varepsilon$. Many other shape functions $a^{(\sigma)}$ are possible. The possible measurement values of the average observable $\sigma(\bar t)$ are in the interval between $-1$ and $1$. If a large number of sites $t'$ contributes one effectively finds a continuous spectrum within this interval.

The expectation value of the average spin $\sigma(\bar{t})$ can be computed from the overall probability distribution of a local chain since $\braket{s(t)}$ is computable for all $t$,
\begin{equation}
\braket{\sigma(\bar{t})} = \sum_{t'} a^{(\sigma)} (\bar{t}+t') \braket{s(\bar{t}+t')}.
\label{eq:LS57}
\end{equation}
It can also be obtained from the time-local subsystem by use of the classical density matrix,
\begin{equation}
\braket{\sigma(\bar{t})} = \tr \{ \rho'(\bar{t}) \hat{\sigma}(\bar{t},\bar{t}) \}.
\label{eq:LS58}
\end{equation}
Here the operator $\hat{\sigma}(\bar{t},\bar{t})$ involves the Heisenberg operators $\hat{s}_H(\bar{t}+t',\bar{t})$ associated to the spins $s(\bar{t}+t')$,
\begin{align}
\label{eq:LS59}
\hat{\sigma}(\bar{t},\bar{t}) &= \sum_{t'} a^{(\sigma)}(\bar{t}+t') \hat{s}_H(\bar{t}+t',\bar{t}), \\
\nonumber
\hat{s}_H(\bar{t}+t',\bar{t}) &= U(\bar{t},\bar{t}+t') \hat{s}(\bar{t}+t',\bar{t}+t') U^{-1}(\bar{t},\bar{t}+t').
\end{align}
The operator $\hat{s}(\bar{t}+t',\bar{t}+t')$ does not depend on $\bar{t}+t'$. (For a single spin local chain one may take $\hat{s}(\bar{t}+t',\bar{t}+t') = \tau_3$.) Since $\hat{\sigma}$ is a linear combination of Heisenberg operators it can be transported to arbitrary $t$,
\begin{equation}
\hat{\sigma}(\bar{t},t) = U(t,\bar{t}) \hat{\sigma}(\bar{t},\bar{t}) U^{-1}(t,\bar{t}),
\label{eq:LS60}
\end{equation}
with 
\begin{equation}
\braket{\sigma(\bar{t})} = \tr \{ \rho'(t) \hat{\sigma}(\bar{t},t) \}.
\label{eq:LS61}
\end{equation}
We conclude that the expectation value of the average spin can be computed from the probabilistic information of the time-local subsystem. The operator associated to this observable is well defined. This extends trivially to the average occupation number.

\paragraph*{Local observables and operators for averaged\\observables}

We next ask if the average spin is a local observable in the sense of the criteria in subsection~\ref{sec:local_observables_and_non_commuting_operators}. This issue is important for the continuum limit. Since for realistic observations a time resolution of the order $\varepsilon$ is not possible, the practically relevant observables are all averaged or coarse grained observables. Representing them by local observables greatly simplifies the discussion since no explicit averaging has to be performed.

We choose the averaging procedure such that $\hat{\sigma}(\bar{t},\bar{t})$ is diagonal. The Heisenberg operators in the sum~\eqref{eq:LS59} have both diagonal and off-diagonal elements. Our choice of $a^{(\sigma)}(\bar{t}+t')$ should be such that the off-diagonal elements cancel. We can write this requirement in the form that for $\tau\neq \rho$
\begin{align}
\begin{split}
\sum_{t'>0} &\left\{ a^{(\sigma)}(\bar{t}+t') [\hat{s}_H(\bar{t}+t',\bar{t})]_{\tau\rho} \right. \\
&\quad + \left. a^{(\sigma)}(\bar{t}-t') [\hat{s}_H(\bar{t}-t',\bar{t})]_{\tau\rho} \right\} = 0.
\end{split}
\label{eq:LS62}
\end{align}
For the example of symmetric shape functions,
\begin{equation}
a^{(\sigma)}(\bar{t}-t') = a^{(\sigma)}(\bar{t}+t'),
\label{eq:LS63}
\end{equation}
the requirement \eqref{eq:LS62} is obeyed if for $\tau\neq\rho$
\begin{align}
\begin{split}
&\sum_\alpha \left\{ U_{\tau\alpha}(\bar{t},\bar{t}+t') [\hat{s}_H(\bar{t},\bar{t})]_{\alpha\alpha} U_{\alpha\rho}^{-1}(\bar{t},\bar{t}+t') \right. \\
&+ \left. U_{\tau\alpha}(\bar{t},\bar{t}-t')[\hat{s}_H(\bar{t},\bar{t})]_{\alpha\alpha} U_{\alpha\rho}^{-1}(\bar{t},\bar{t}-t') \right\} =0.
\end{split}
\label{eq:LS64}
\end{align}
Here we employ the property that $\hat{s}_H(\bar{t}+t',\bar{t}+t')$ is given by the same diagonal operator for all $\bar{t}+t'$, which is associated to the spin $s(\bar{t}+t')$. With evolution operators $U(t_1,t_2)$ only depending on the difference $t_1-t_2$ one has
\begin{equation}
U(\bar{t},\bar{t}-t') = U^{-1}(\bar{t}-t',\bar{t}) = U^{-1}(\bar{t},\bar{t}+t'),
\label{eq:LS65}
\end{equation}
such that the second term in eq.\,\eqref{eq:LS64} corresponds to the inverse evolution of the first term. The l.\,h.\,s.\ of eq.\,\eqref{eq:LS64} is of the form
\begin{equation}
U\hat{s}U^{-1} + U^{-1}\hat{s}U = 2\hat{s} + [ [U,\hat{s}],U^{-1} ].
\label{eq:LS66}
\end{equation}
The term $2\hat{s}$ is diagonal, such that only the double commutator in the second term on the l.\,h.\,s.\ of eq.\,\eqref{eq:LS66} can contribute to off-diagonal elements. This contribution vanishes for a large class of evolution operators. In particular, if $U$ is a unique jump operator one finds diagonal $U\hat{s}U^{-1}$ for diagonal $\hat{s}$, such that $\sigma(\bar{t},\bar{t})$ is indeed a diagonal operator for symmetric shape functions. 

A diagonal operator $\hat{\sigma}(\bar{t},\bar{t})$ is uniquely characterized by its diagonal elements $[\hat{\sigma}(\bar{t},\bar{t})]_{\tau\tau}$. Many different shape functions $a^{(\sigma)}(\bar{t}+t')$ can lead to the same $\hat{\sigma}(\bar{t},\bar{t})$. Different shape functions correspond to different averaging procedures and therefore to different observables in the overall probabilistic system. If they correspond to the same $\hat{\sigma}(\bar{t},\bar{t})$ these observables cannot be distinguished in the continuum limit of the local time subsystem. This is precisely what one aims for in the continuum limit. The continuum limit becomes independent of the microscopic details -- only ``macroscopic quantities'' as the diagonal elements of $\hat{\sigma}(\bar{t},\bar{t})$ remain observable.

For constant $\rho'(t)$ one has $\braket{s(\bar{t}+t')} = \braket{s(\bar{t})}$ and therefore
\begin{equation}
\braket{\sigma(\bar{t})} = \braket{s(\bar{t})}.
\label{eq:LS67}
\end{equation}
For a slowly varying classical density matrix the relation \eqref{eq:LS67} remains a good approximation. One may therefore use the local spin $s(\bar{t})$ as a good proxy for the averaged spin $\sigma(\bar{t})$. This is the reason why one can continue to employ strictly local observables as $s(\bar{t})$ in the continuum limit for many purposes, even though conceptually average observables as $\sigma(\bar{t})$ are more appropriate.

A given operator $\hat{\sigma}(\bar{t},\bar{t})$ represents many different microscopic observables. The question arises if there are observables for which $\hat{\sigma}(\bar{t},\bar{t})$ is the associated local-observable operator. Such an observable requires that its possible measurement values equal the eigenvalues of the operator $\hat{\sigma}(\bar{t},\bar{t})$. In other words, one asks of there are measurement procedures whose possible outcomes are only the eigenvalues of $\hat{\sigma}(\bar{t},\bar{t})$. Suitable observables exist on the microscopic level since a diagonal operator $\hat{\sigma}(\bar{t},\bar{t})$ can be obtained by an appropriate linear combination of strictly local observables $A(\bar{t})$. In practice, however, the question concerns the issue if one can find a suitable ``macroscopic'' measurement prescription.

The spectrum of $\sigma$ is continuous. One may also consider a discrete averaged spin observable represented by the operator
\bel{5.2.73A}
\hat s_{\text{av}}(\bar t)=\theta\gl\hat\sigma(\bar t,\bar t)-\theta(-\hat\sigma(\bar t,\bar t)\gr\ .
\ee
Here the $\theta$-functions acting on the operator $\hat\sigma$ are defined by first diagonalizing $\hat\sigma$ with $\hat\sigma_D=D\hat\sigma D^{-1}$, replacing all positive eigenvalues of $\hat\sigma_D$ by one and all negative eigenvalues by minus one, and transforming back with $D^{-1}$. The spectrum of $\hat s_{\text{av}}$ has the values $\pm1$, corresponding to the yes/no-decision if the averaged spin $\sigma(\bar t)$ is positive or negative. For diagonal $\hat\sigma$ the matrix $D$ is not needed. One may imagine some measurement device with a suitable threshold that gives a signal if $\sigma(\bar t)>0$ and no signal if $\sigma(\bar t)<0$.

The averaged observables $\sigma(\bar{t})$ can also be employed to define averaged derivative observables as
\begin{equation}
\partial_t \sigma(\bar{t}) = \frac{\sigma(\bar{t}+\Delta t) - \sigma(\bar{t}-\Delta t)}{2\Delta t}.
\label{eq:LS68}
\end{equation}
They are represented by the operators
\begin{equation}
\partial_t \hat{\sigma}(\bar{t},\bar{t}) = \frac{\hat{\sigma}(\bar{t}+\Delta t,\bar{t}) - \hat{\sigma}(\bar{t}-\Delta t,\bar{t})}{2\Delta t} .
\label{eq:LS69}
\end{equation}
This type of derivative observable no longer needs a resolution $\varepsilon$. It is compatible with the continuum limit.
\subsubsection{Probabilistic observables and \\incomplete statistics}
\label{sec:probabilistic_observables_and_incomplete_statistics}

A characteristic feature of subsystems are probabilistic observables. They do not have fixed
values in a given state of the subsystem. A state of the subsystem provides only probabilities
to find a possible measurement value of a probabilistic observable. We follow the here partly
the discussions in ref.\,\cite{CWPO,CWQM,CWB}.
The notion of probabilistic observables or ``fuzzy observables''\,\cite{HOL,ALP,SISU,BEBU,BEBU2,BUG} is well known in measurement theory\,\cite{BINE,VONE1} and used for ``classical extensions'' of quantum mechanics\,\cite{MIS,STUBU}.

\paragraph*{Probabilistic observables}

Consider time-local subsystems. A state of the subsystem is given by a classical density matrix
$\rho'(t)$. The classical density matrix specifies the state of the subsystem completely --
no probabilistic information beyond the one contained in $\rho'(t)$ is available for the subsystem.
We have seen that for local observables represented by local-observable operators one can compute
the probabilities $w_i$ to find a given possible measurement value $\lambda_i$ from the classical
density matrix. These probabilities obtain as the diagonal elements of a suitably transformed matrix
$\left( \rho'_D \right)_{\tau \tau}$. The local observable has a definite value in a state of the
subsystem only if the probability equals one for a particular value $\lambda_a$, $w_i = \delta_{ai}$.
Otherwise only a probability distribution for the possible measurement values is available in the
subsystem. In this respect subsystems share important properties with quantum systems. In a given
state of the subsystem only a subclass of observables can have definite values, while for the others
only probabilities are known.

Let us discuss the concept of probabilistic observables on a more general level. We assume that the
state of the subsystem can be described by a number of real variables $\rho_z$. These variables contain
all the probabilistic information of the subsystem. The possible measurement values of a given
probabilistic observable are denoted by real numbers $\lambda_i$. In the subsystem a probabilistic
observable is characterized by the spectrum of possible measurements values $\lambda_i$, together
with the probabilities $w_i\left(\rho_z\right)$ to find these values. The possible measurement
values $\lambda_i$ are independent of the state of the subsystem, while the probabilities
$w_i(\rho)$ are computable from the probabilistic information of the subsystem. They are functions
of the variables $\rho_z$ characterizing the state of the subsystem, $\rho = \left\{ \rho_z \right\}$.
The probabilities $w_i$ have to obey the usual rules for probabilities
\begin{equation}
  \label{PO1}
  w_i(\rho) \geq 0\,, \quad \sum_i w_i(\rho) = 1\,.
\end{equation}
The expectation value of a probabilistic observable $A$ is given by the standard rule of classical
statistics
\begin{equation}
  \label{PO2}
  \braket{A} = \sum_i \lambda_i^{(A)} w_i^{(A)}(\rho)\,.
\end{equation}
It depends on the state of the subsystem via the dependence of the probabilities $w_i^{(A)}(\rho)$ on
$\rho$.

Taking the example of a time-local subsystem the state-variables $\rho_z$ could be associated with the
elements $\rho'_{\tau \sigma}(t)$ of the classical density matrix. It will often be more convenient to
choose some ``generator basis'' with
\begin{equation}
  \label{PO3}
  \rho'_{\tau \sigma}(t) = \sum_z \rho_z(t) \left( L_z \right)_{\tau \sigma}\,,
\end{equation}
with a number of linearly independent generators $L_z$ given by the number of independent elements of
$\rho'$, and $\rho_z(t)$ specifying the particular density matrix. For the probabilistic observable
we could take an Ising spin with $\lambda_1 = 1$, $\lambda_2 = -1$. The expectation value in the state
$\rho = \{\rho_z(t)\}$ of the subsystem,
\begin{equation}
  \label{PO4}
  \braket{A}_\rho = \sum_i w_i(\rho) \lambda_i = w_1(\rho) - w_2(\rho)\,,
\end{equation}
is directly related to the probabilities
\begin{equation}
  w_1(\rho) = \frac{1}{2} \left( 1 +\braket{A}_\rho \right)\,, \quad w_2 = \frac{1}{2} \left( 1 -\braket{A}_\rho \right)\,.
\end{equation}
Only for $\braket{A}_\rho = \pm 1$ the observable $A$ has a fixed value in the corresponding state
$\rho$. For $\braket{A}_\rho \neq \pm 1$ only probabilities to find $A = +1$ or $A=-1$ are available.

\paragraph*{Uncertainty relations for subsystems}

We call ``system observables'' those probabilistic observables for which the probabilities $w_i(\rho)$
can be computed from the variables $\rho_z$ characterizing the state of the subsystem. Typically, not
all system observables can have fixed values in a given state of the subsystem. This generates
``uncertainty relations'' for classical statistical subsystems in close analogy to quantum mechanics.

As one example, we take again the time local subsystem. Consider two local observables $A(t)$ and
$B(t)$ that are represented by local-observable operators $\hat{A}(t)$ and $\hat{B}(t)$ that
do not commute. For example, $A(t)$ may be a strictly local observable represented by a diagonal
operator $\hat{A}(t)$, while $B(t)$ is a neighbouring local observable represented by an operator
$\hat{B}(t)$ with off-diagonal elements. We further assume that $A(t)$ and $B(t)$ have a non-degenerate
spectrum. For example, both could be two-level observables or Ising spins in a one-bit local
chain, e.g. $A(t) = s(t)$ and $B(t) = s(t+\epsilon)$. The classical density matrix $\rho'(t)$ is then a
$2\times 2$-matrix and the operator associated to $A(t)$ and $B(t)$ are
\begin{equation}
  \label{PO5}
  \hat{A}(t) = \tau_3\,, \quad \hat{B}(t) = \hat{S}^{-1} \tau_3 \hat{S}\,,
\end{equation}
with $\left[ \hat{S}, \tau_3 \right] \neq 0$.

A state of the subsystem for which $A(t)$ has the sharp value $+1$ corresponds to a classical density
matrix
\begin{equation}
  \label{PO6}
  \rho'(t) = 
  \begin{pmatrix}
    1 & \rho'_{12} \\
    \rho'_{21} & 0 \\
  \end{pmatrix}
  \,.
\end{equation}
In this state $B(t)$ typically does not have a sharp value. Indeed, we may diagonalize $\hat{B}$ by
a similarity transformation $D=\hat{S}$ and compute
\begin{equation}
  \label{PO7}
  \rho'_D = \hat{S} \rho'(t) \hat{S}^{-1}\,.
\end{equation}
The probability to find $B(t) = s(t+\epsilon) = 1$ is given by
\begin{align}
  \label{PO8}
  \left( \rho'_D \right)_{11} &= \hat{S}_{1 \alpha} \rho'_{\alpha \beta} \hat{S}^{-1}_{\beta 1} \nonumber \\
                              &= \hat{S}_{11} \hat{S}^{-1}_{11} + \hat{S}_{11}\rho'_{12} \hat{S}^{-1}_{21} + \hat{S}_{12}\rho'_{21} \hat{S}^{-1}_{11} \nonumber \\
                              &= \left( \det \hat{S} \right)^{-1} \left( \hat{S}_{11} \hat{S}_{22} + \rho'_{12}\hat{S}_{11} \hat{S}_{21} + \rho'_{21}\hat{S}_{12} \hat{S}_{22}\right) \nonumber \\
                              &= 1 - \left( \rho'_D \right)_{22}\,,
\end{align}
with
\begin{equation}
  \label{PO9}
  \left( \rho'_D \right)_{22} =  \left( \det \hat{S} \right)^{-1} \left( \hat{S}_{12} \hat{S}_{21} + \rho'_{12}\hat{S}_{21} \hat{S}_{11} - \rho'_{21}\hat{S}_{12} \hat{S}_{22}\right)\,.
\end{equation}
Whenever the r.h.s. in eq.~\eqref{PO9} differs from zero we know that $\left( \rho'_D \right)_{11}$ is
smaller than one,  $\left( \rho'_D \right)_{22} \geq 0$. Thus $B(t)$ could ony have a sharp value
for $\left( \rho'_D \right)_{22} = 1$.

More generally, one may use the quantum rule \eqref{LS8} for the expectation values of $A,A^2,B,B^2$ in order to establish uncertainty relations. Whenever $\braket{A^2} > \braket{A}^2$ or  $\braket{B^2} > \braket{B}^2$ the corresponding observable cannot have a sharp value. For the special case of
symmetric operators $\hat{A}(t)$ and $\hat{B}(t)$ and symmetric density matrices $\rho'(t)$ these are
precisely the Heisenberg uncertainty relations for quantum systems.

The presence of uncertainty relations obstructs the construction of microstates for the subsystem.
Microstates can be associated to specific states of the subsystem $\rho $ for which all system
observables take sharp values. Parameterizing by $\alpha$ all those states $\rho_\alpha$ for which
at least the values of two system observables differ, we could call $\alpha $ the microstates of
the subsystem.
In such a microstate all system observables have given sharp values $A_\alpha$, which have to
belong to the spectrum of possible measurement values of the observable $A$. We could then try to
construct probabilities $w_\alpha(\rho) \geq 0, \sum_\alpha w_\alpha(\rho) = 1$, such that
\begin{equation}
  \label{PO9A}
  \braket{A}= \sum_\alpha A_\alpha w_\alpha(\rho)\,.
\end{equation}
Such a construction is impossible if not all system observables can take a sharp value
simultaneously.

\paragraph*{Equivalence classes of classical observables}

Many different classical observables of the overall probabilistic system are mapped to the same
probabilistic observable in a subsystem. We have seen simple examples as the derivative observables
or the averaged observables in the continuum limit for time-local subsystems. Consider two
different classical observables $A$ and $A'$ with the same spectrum of possible measurement values $\lambda_i$.
An example are two different Ising spins. If both $A$ and $A'$ are system observables the probabilities
$w_i$ and $w'_i$ are sufficient for the computation of the expectation values $\braket{A^p}$ and
$\braket{A'^p}$ for arbitrary powers $p$. If for all $i$ one has $w_i(\rho) = w'_i(\rho)$ for arbitrary
states of the subsystem, the two observables have the same expectation values,
\begin{equation}
  \label{PO10}
  \braket{A^p}_\rho = \braket{A'^p}_\rho\,,
\end{equation}
for all possible states of the subsystem. It is impossible to distinguish between $A$ and $A'$ by
any measurement or observation that only uses probabilistic information of the subsystem. From the
point of view of the subsystem the two observables $A$ and $A'$ are identical -- they are described
by the same probabilistic observable. All classical observables that correspond to the same
probabilistic observable in a subsystem form an equivalence class. Only the equivalence class
matters for the subsystem. Different members of the equivalence class are indistinguishable on the
level of the subsystem.

Inversely, two observables with an identical finite spectrum of possible measurements values
(finite number of different $\lambda_i$) correspond to the same probabilistic system observable if
the relation \eqref{PO10} holds for arbitrary $p$ and $\rho$. For a given $\rho$ the relation \eqref{PO10}
implies for the probabilities $w_i(\rho) =w'_i(\rho)$. If this holds for all $\rho$ both observables
correspond to the same probabilistic observable characterized by $\left\{ \lambda_i, w_i(\rho) \right\}$.
For Ising spins it is actually sufficient that $\braket{A}_\rho = \braket{A'}_\rho$  and
$A^2 = A'^2 = 1$.

In general, an equivalence class has more than a unique member. There are different observables in
the overall probabilistic system that lead to the same probabilistic observable in a subsystem. The
map from the observables in the overall system to probabilistic observables in the subsystem is not
invertible. As a simple example we may consider two Ising spins $A$ and $A'$. We select a family
of overall probability distributions for which the expectation values are equal, $\braket{A} = \braket{A'}$.
We may then define a subsystem for which part of the probabilistic information is given precisely by the expectation
value of $A$ or $A'$
\begin{equation}
  \label{PO11}
  \rho_z = \braket{A} = \braket{A'}\,.
\end{equation}
In this subsystem the probabilistic observables associated to $A$ and $A'$ are identical. With
eigenvalues $\pm 1$ the probabilities $w_\pm $ to find the value $+1$ or $-1$ are obviously the same
\begin{equation}
  \label{PO12}
  w_\pm(\rho) = w'_\pm (\rho) = \frac{1}{2}(1 \pm \rho_z)\,.
\end{equation}

Nevertheless, in the overall system $A$ and $A'$ are different observables. There can be states
$\omega $ for which the values of $A$ and $A'$ differ, $A_\omega \neq A'_\omega $.
The constraint \eqref{PO12} constrains only certain combinations of probabilities $p_\omega $. In
particular, the classical correlation function with another observable may be different for $A$ and
$A'$. We may compare $\braket{AA}$ and $\braket{A' A}$. While $\braket{AA} = 1$, the correlation
$\braket{A' A}$ can be smaller than one. To be very concrete, denote by $p_{++}, p_{+-}, p_{-+}, p_{--}$
the probabilities to find in the overall system for $(A, A')$ the values $(1,1), (1,-1), (-1,1)$ and $(-1,-1)$. The
subsystem can be defined if
\begin{equation}
  \label{PO13}
  p_{+-} = p_{-+}\,, \quad \rho_z = p_{++} - p_{--}\ .
\end{equation}
For the classical correlation $\braket{A A'}$ one has
\begin{equation}
  \label{PO14}
  \braket{A A'} = 1 - 4 p_{+-}\,,
\end{equation}
which differs from one for $p_{+-} > 0$. This demonstrates that $A$ and $A'$ are different observables for the overall system.

Let us focus on local-time subsystems and consider system observables for which an associated
local-observable operator exists. If the operator associated to $A$ and $A'$ is the same
$\hat{A}(t)$, the condition \eqref{PO10} is automatically obeyed
\begin{equation}
  \label{PO15}
  \braket{A^p} = \braket{A'^p} = \tr \left\{ \hat{A}^p \rho' \right\}\,.
\end{equation}
In the other direction, a local-observable operator obeying the conditions \eqref{LS8} defines
a probabilistic observable uniquely. The eigenvalues of $\hat{A}$ are the possible measurements
values $\lambda_i$, and the probabilities $w_i(\rho)$ can be computed for every state $\rho$
according to the conditions \eqref{PO3}. For time local subsystems the local-observable operators
define equivalence classes of observables. All observables in the overall system that are
represented in the subsystem by the same local-observable operator belong to the same equivalence
class.

It is not difficult to find different observables in the overall system that are represented in the
local-time subsystem by the same operator. Local-observable operators are actually a very economical
way to define probabilistic observables for the local-time subsystem. For $\rho'(t)$ and $\hat{A}(t)$
being $N \times N$-matrices, the specification of the probabilistic observable needs at most the $N^2$
real elements of the matrix $\hat{A}$. They define both the possible measurement values $\lambda_i$ and
the probabilities $w_i(\rho)$ for each state characterized by the classical density matrix
$\rho'(t)$. The functions $w_i(\rho_z)$ are linear in $\rho_z$, with coefficients depending on the
matrix elements of $\hat{A}$. For symmetric  operators $\hat{A}$ the number $N(N+1)/2$ of elements is even
smaller.

\paragraph*{Incomplete statistics}

Incomplete statistics is a characteristic feature of many subsystems. This means that for two
probabilistic observables $A$ and $B$ the probabilistic information of the subsystem is sufficient
for the computation of $\braket{A^p}$ and $\braket{B^p}$, but not for all classical correlations as $\braket{AB}$.
Indeed, the probabilistic observables are characterized by $w_i^{(A)}(\rho)$ and $w_j^{(B)}(\rho)$, but
the joint probabilities $w_{ij}^{(AB)}$ to find $\lambda_i^{(A)}$ for $A$ and $\lambda_j^{(B)}$ for $B$
may not be part of the probabilistic information in the subsystem. As we have seen, probabilistic
observables correspond to equivalence classes of observables rather than to a specific local
observable of the overall system. For incomplete statistics the classical correlation function $\braket{AB}$
may not find a formulation in terms of equivalence classes. If $A$ and $A'$ belong to the same
equivalence class, the correlation $\braket{AB}$ may nevertheless differ from $\braket{A'B}$. Is is
then not an object that is consistent with the concept of equivalence classes.

Incomplete statistics will be a crucial feature for many later developments, in particular the
quantum subsystems. Overlooking the incompleteness of statistics for subsystems often leads to
paradoxes or ``no go theorems'' that are actually not valid for subsystems. We will
therefor demonstrate the issue in a very simple example.

Consider three Ising spins $A, A'$ and $B$ and an overall probability distribution characterized
by $p_{\sigma_1 \sigma_2 \sigma_3}$ for the states $(\sigma_1, \sigma_2 ,\sigma_3)$ for which the triple
$(A,A',B)$ has values $(\sigma_1, \sigma_2 ,\sigma_3)$, $\sigma_k = \pm 1$. One has
\begin{equation}
  \label{PO16}
\begin{split}
  \braket{A}& = p_{+++} +  p_{++-} +  p_{+-+} +  p_{+--} \\
  & \quad-  p_{-++} - p_{-+-} -  p_{--+} -  p_{---} \\
  \braket{A'}& = p_{+++} + p_{++-} - p_{+-+} -  p_{+--} \\
  & \quad + p_{-++} + p_{-+-} -  p_{--+} -  p_{---} \\
  \braket{B}& = p_{+++} -  p_{++-} +  p_{+-+} -  p_{+--} \\
  & \quad + p_{-++} - p_{-+-} +  p_{--+} -  p_{---} 
\end{split}
\end{equation}
Consider a family of probability distributions subject to the constraint
\begin{equation}
\label{PO17}
p_{+-+} + p_{+--} = p_{-++} + p_{-+-}\,.
\end{equation}
This implies 
\begin{equation}
\label{PO18}
\braket{A} = \braket{A'} = \rho_1 = p_{+++} + p_{++-} - p_{--+} - p_{---}\,.
\end{equation}

We define a subsystem by $\rho_1$ and $\rho_2 = \braket{B}$. The observables $A$ and $A'$ correspond
to the same probabilistic observable for the subsystem and therefore belong to the same equivalence
class. The probabilistic information in the subsystem is sufficient for the computation of the expectation values $\braket{A} = \braket{A'}$ and $\braket{B}$. For the classical correlations one has
\begin{equation}
\label{PO19}
\begin{split}
\braket{AB}& = p_{+++} -  p_{++-} +  p_{+-+} -  p_{+--} \\
  & \quad-  p_{-++} + p_{-+-} -  p_{--+} + p_{---} \\
\braket{A'B}& = p_{+++} -  p_{++-} -  p_{+-+} +  p_{+--} \\
  & \quad+  p_{-++} - p_{-+-} -  p_{--+} +  p_{---} \\
\end{split}
\end{equation}
They cannot be expressed in terms of the local probabilistic information of the subsystem. Instead
of being functions of $\rho_1, \rho_2$, these correlations require probabilistic information about
the ``environment". In general, the correlation functions for the two different members $A, A'$ of
the equivalence class are different,
\begin{equation}
\label{PO20}
\braket{AB}-\braket{A'B} = 2 (p_{+-+} - p_{+--} - p_{-++} + p_{-+-})\,.
\end{equation}
The subsystem specified by $(\rho_1, \rho_2)$ is characterized by incomplete statistics. The 
simultaneous probabilities for finding for $(A, B)$ the values $\lambda_i^{(A)}$ and $\lambda_j^{(B)}$
are not available in the subsystem. They can only be computed in the overall system, e.g.
\begin{equation}
\label{PO21}
w_{++}^{(AB)} = p_{+++} + p_{+-+}\,,\quad w_{++}^{(A'B)} = p_{+++} + p_{-++}\,,
\end{equation}
but these combinations cannot be expressed in terms of $\rho_1$ and $\rho_2$.

We will see that incomplete statistics arises in many practical subsystems. Many examples have already been 
found for local time subsystems, as the derivative observables in the continuum limit. This extends to correlations of occupation numbers at different times. Even though classical correlation functions for observables at different $t$
may be, in principle, accessible from time ordered operator products, the simultaneous probabilities cannot 
be expressed in terms of the state of the subsystem, i.e. the classical density matrix $\rho'(t)$. They need, in
addition, knowledge to which operator $\hat{C}$ the correlation function $\braket{AB}$ is associated, involving
in turn control over a sequence of step evolution operators. For two equivalent local observables $A$ and $A'$, 
associated to the same local-observable operator $\hat{A}$, the correlation functions $\braket{AB}$ and $\braket{A'B}$
are typically associated to two different operators $\hat{C}$ and $\hat{C'}$ and are therefore different. We may want to
formulate the local time subsystem uniquely in terms of the state of the subsystem $\rho$ and the probabilistic observables
$\{ \lambda_i, w_i(\rho) \}$. The simultaneous probabilities for observables represented by non-commuting operators are not available in this setting.

We will next discuss correlation subsystems for which the incompleteness of the subsystem is a central characteristic. Incomplete
statistics is generic for many types of subsystems. This has perhaps often been overlooked because the simplest subsystems obtained from 
direct product systems have complete statistics. This is, however, a special case where all correlations with the environment are absent.

\subsection{Classes of subsystems}\label{sec:classes_of_subsystems}

The general definition of subsystems in
sect.~\ref{sec:subsystems_and_environment} allows for a large variety of
subsystems. In this section we present a few simple examples.

\subsubsection{Correlation subsystems}
\label{sec:correlation_subsystems}

The information contained in the probability distribution for $N$ Ising spins is equivalent to the $2^N$ correlation functions. This
involves up to $N$-point correlations, which are the expectation values of the correlation basis observables. We have discussed this in
sect.~\ref{sec:correlations}. For a large number $N$ the very high correlation functions cannot be resolved in 
practice. One typically proceeds to subsystems with correlation functions of a moderate order.

\paragraph*{Correlation functions of finite order}

For a correlation subsystem one considers only a finite subset of correlation functions, say up to four point functions. A well
known example is the approach to thermal equilibrium. The time-local probability distribution typically does not converge to the 
equilibrium distribution -- there are obstructions as infinitely many conserved quantities\,\cite{CWQMTE}. Nevertheless, 
all the low correlation functions often approach their equilibrium values for asymptotic time\,\cite{ABW1,BER}. The subsystem of low correlation
functions apparently decouples from the high correlation functions where the obstructions to the equilibrium of the full probability
distribution are located. 

By definition, a correlation subsystem is characterized by incomplete statistics. Only the low correlation functions included in the 
correlation subsystem can be computed from the probabilistic information of the subsystem. They typically constitute themselves the
probabilistic information of the correlation subsystem. The high correlation functions are not computable from the probabilistic 
information of the subsystem. The subsystem has incomplete statistics which does not permit the computation of all correlation functions.

From the point of view of embedding the correlation subsystem into the total time-local subsystem the high correlation functions are 
the environment. The subsystem is not some separated physical part. It is a subsystem in the space of correlation functions. The 
subsystem and the environment are correlated. Typically, the evolution of the high correlation functions depends on the values of the
low correlation functions. On the other hand, effective evolution equations for the low correlation functions involve terms that depend on the higher correlation functions. Correlation subsystems are a very simple example for probability distributions that are not direct products
of separate probability distribution for the subsystem and the environment. 

\paragraph*{Closed time evolution}

For a useful concept of a correlation subsystem it is important that its dynamics can be computed from the probabilistic information
of the subsystem. For models with interactions, the evolution of the low correlation functions typically depends on higher correlation
functions, as encoded in the BBGKY-hierarchy~\cite{YVO, BOG, BOGR, KIR}. For the example of a classical scalar field theory with quartic interaction the evolution of the two-point function
depends on the four-point function. In turn, the evolution law for the four-point function involves the six-point function, and so on.
The system of evolution equations for correlation functions is, in general, not closed. The appearance of ``high" correlation
functions belonging to the environment in the evolution equations for the ``low" correlation functions belonging to the correlation
subsystem is a consequence of the fact that the subsystem and its environment are correlated. 

Closed evolution equations for the subsystem become possible, nevertheless, if the effect of higher correlation functions can be 
expressed in terms of the values of correlation functions in the subsystem. For example, an expansion in one-particle-irreducible 
correlation functions\,\cite{CWTE,CWNEQFT,ABW1,ABW2, Schmied_2019, Pr_fer_2020, PhysRevX.10.011020, ott2022equaltime} expresses the six-point correlation functions in terms of four-point and two-point functions, plus
an irreducible part. If the irreducible part can be neglected, the evolution of the correlation subsystem becomes closed. Often 
the closed evolution equations for the correlation subsystem are only a good approximation. Some cases are known, however, where the 
evolution for particular two-point functions is closed even in the presence of interactions\,\cite{CWQMTE}. Some quantum subsystems can be viewed as correlation subsystems with a closed evolution.
\subsubsection{Integrating out variables}\label{sec:matrix_chains_as_subsystems_of_local_chains}

There are various ways of obtaining subsystems by ``integrating out'' variables. Such subsystems can have structures that are much richer than for the uncorrelated direct product case in sect.\,\ref{sec:subsystems_and_correlation_with_environment}. We first demonstrate that integrating out variables leads in general to matrix chains for the remaining variables. These subsystems may be reduced further to smaller subsystems, for example by taking subtraces of the matrices. We will discuss examples where the subsystem of a local chain is again a local chain.

\paragraph*{Integrating out variables in a two-bit chain}

A simple form of subsystems arises if one ``integrates out'' variables. We demonstrate the general consequences first for a two-bit chain, $M=2$, for which there are only two occupation numbers $n_\gamma(m)$, $\gamma=1, 2$, on each site $m$. Assume that we are interested in observables that only depend on the first bit $n_1(m)$, and not on the second bit $n_2(m)$. One would like to discuss a subsystem for which the weight function and the overall probability distribution depends only on the variables $n_1(m)$. This obtains by integrating out the other variables $n_2(m)$. We define
\begin{equation}\label{eq:MCS1}
w[n_1(m)] = \int \cD n_2\, w[n_1(m),\, n_2(m)]\, ,
\end{equation}
where the ``integral'' $\int \cD n_2$ is the sum over all configurations of occupation numbers $n_2(m)$, taken for fixed configurations of $n_1(m)$. For the partition function one has
\begin{equation}\label{eq:MCS1A}
Z = \int \cD n_1\, \cD n_2\, w[n_1,\, n_2] = \int \cD n_1\, w[n_1]\, ,
\end{equation}
and the probability distribution for the subsystem obtains as
\begin{equation}\label{eq:MCS2}
p[n_1] = Z^{-1}\, w[n_1]\, .
\end{equation}
The expectation values of all observables that depend only on $\{ n_1 (m)\}$ can be computed from $p[n_1]$ in the usual way
\begin{align}\label{eq:MCS3}
\langle A \rangle &= Z^{-1}\, \int \cD n_1\, \cD n_2\, A[n_1]\, w[n_1,\, n_2] \notag \\
&= Z^{-1}\, \int \cD n_1\, A[n_1]\, w[n_1] \notag \\
&= \int \cD n_1\, A[n_1]\, p[n_1]\, .
\end{align}

\paragraph*{Matrix chains for the subsystem}

The question arises how the local chain structure of the two-bit chain translates to the subsystem. We will show that the generic result is a matrix chain for the subsystem, with the rank of the matrix $n=2$. For a general discussion of matrix chains we refer to appendix~\ref{app:matrix chains}. A general method for focusing on the variables $n_1$ expands the local factors in the occupation number basis for $n_2$, keeping the $n_1$-dependence without expansion,
\begin{align}\label{eq:MCS4}
& \cK\big( n_1(m+1),\, n_2(m+1),\, n_1(m)\, n_2(m) \big) \notag \\
& \quad = \hat{\cK}_{\alpha\beta} \big( n_1(m+1),\, n_1(m)\big) \, h_\alpha 
\big( n_2 (m+1)\big)\, h_\beta \big( n_2 (m)\big)\, .
\end{align}
Here, $\alpha, \beta = 1, 2$ denotes the two basis states for $n_2$, and $\hat{\cK}_{\alpha\beta}$ are the elements of a $(2\times 2)$-matrix $\hat{\cK}$. They depend on $n_1(m+1)$ and $n_1(m)$.

We next consider the product of two neighboring local factors $\cK(m+1)$ and $\cK(m)$ and integrate over $n_2(m+1)$,
\begin{align}\label{eq:MCS5}
& \int \dif n_2(m+1) \notag \\
& \; \times \cK \big( m+1;\, n_1(m+2),\, n_2(m+2),\, n_1(m+1),\, n_2(m+1) \big) \notag \\
& \; \times\cK \big( m;\, n_1(m+1),\, n_2(m+1),\, n_1(m),\, n_2(m+1)\big) \notag \\
& \; = \int \dif n_2 (m+1)\, \hat{\cK}_{\alpha\beta} 
\big( m+1;\, n_1(m+2),\, n_1(m+1)\big) \notag \\
& \qquad \times h_\alpha \big( n_2(m+2)\big) \, h_\beta \big( n_2 (m+1)\big) \notag \\
& \qquad \times \hat{\cK}_{\gamma\delta}\big( m;\, n_1(m+1),\, n_1(m)\big) \notag \\
& \qquad \times h_\gamma\big( n_2(m+1)\big)\, h_\delta \big( n_2(m)\big) \notag \\
& \; = \hat{\cK}_{\alpha\beta} \big( m+1;\, n_1 (m+2),\, n_1(m+1)\, \big) \notag \\
& \qquad \times \hat{\cK}_{\beta\delta} \big( m;\, n_1(m+1),\, n_1(m)\big) \notag \\
& \qquad \times h_\alpha \big( n_2(m+2)\big)\, h_\delta \big( n_2(m)\big)\, .
\end{align}
Here we have indicated explicitly the arguments of $\cK(m+1)$ and $\cK(m)$ after the semicolon, and similar for $\hat{\cK}$. The last line uses the orthogonality relation \eqref{eq:TS12} for the basis functions for $n_2$ at the site $m+1$. The result contains a matrix multiplication $\hat{\cK}_{\alpha\beta}(m+1)\, \hat{\cK}_{\beta\delta}(m)$. For $\hat{\cK}(m+1)$ different from $\hat{\cK}(m)$ the order of the two matrices is such that the matrix for higher $m'$ stands to the left of the one for lower $m'$.

This procedure can be repeated for multiple products along the chain, resulting in
\begin{align}\label{eq:MCS6}
& \prod_{m' = 1}^{\cM - 1}\int \dif n_2(m') \prod_{m' = 0}^{\cM - 1} \cK(m') \notag \\
& = h_\alpha \big(n_2(\cM)\big) \big( \hat{\cK}(\cM - 1)\, \hat{\cK}(\cM - 2)
\dots \hat{\cK}(0)\big)_{\alpha\beta} h_\beta \big( n_2(0)\big)\, .    
\end{align}
The matrices $\hat{\cK}(m)$ depend on the occupation numbers $n_1(m+1)$ and $n_1(m)$. For the overall weight distribution $w[n_1,\, n_2]$ appearing in eq.\,\eqref{eq:MCS1} we have to multiply by the boundary term $\cB \big( n_1(\cM),\, n_2(\cM),\, n_1(0),\, n_2(0)\big)$, that we expand as
\begin{equation}\label{eq:MCS7}
\cB = \hat{\cB}_{\gamma\delta}\big( n_1(\cM),\, n_1(0)\big)\, 
h_\gamma \big( n_2(0)\big)\, h_\delta \big( n_2(\cM)\big)\, .
\end{equation}
The weight function for the subsystem obtains then by performing the remaining integrations over $n_2(0)$ and $n_2(\cM)$,
\begin{align}\label{eq:MCS8}
w[n_1] &= \int \cD n_2 \prod_{m' = 0}^{\cM - 1} \cK (m')\, \cB \notag \\
&= \int \dif n_2 (0) \, \int \dif n_2(\cM) \notag \\
& \quad \times h_\alpha\big( n_2(\cM)\big)\, 
\big( \hat{\cK}(\cM-1)\cdots\, \hat{\cK}(0)\big)_{\alpha\beta} \, 
h_\beta \big( n_2(0)\big) \notag \\
& \quad \times h_\gamma \big( n_2(0)\big)\, \hat{\cB}_{\gamma\delta}\, 
h_\delta \big( n_2(\cM)\big)\, .
\end{align}
Using the orthogonality relation at sites $m' = 0$, $\cM$, one arrives at the final expression
\begin{equation}\label{eq:MCS9}
w[n_1] = \tr \big\{ \hat{\cK}(\cM - 1)\cdots \, \hat{\cK}(0)\, \hat{\cB} \big\}\, . 
\end{equation}
This is precisely the defining formula for a matrix chain, with $(2\times 2)$-matrices $\hat{\cK}(m)$ depending on neighboring variables in the subsystem $n_1(m+1)$ and $n_1(m)$.

\paragraph*{Local chain as subsystem of local chain}

The general reduction of a local chain to a subsystem results in a matrix chain. Only for certain systems the subsystem can again be described by a local chain. As a simple case we consider a system where both $\hat{\cK}$ and $\hat{\cB}$ are proportional to unit matrices,
\begin{equation}\label{eq:MCS10}
\hat{\cK} = \begin{pmatrix}
\cK_1[n_1] & 0 \\
0 & \cK_1[n_1]
\end{pmatrix}\, , \quad
\hat{\cB} = \begin{pmatrix}
\cB_1[n_1] & 0 \\
0 & \cB_1[n_1]
\end{pmatrix}\, .
\end{equation}
The order of the matrices in eq.~\eqref{eq:MCS9} plays no longer a role, and the trace results in a factor $2$. One obtains the weight function for a local chain depending on  only one bit,
\begin{equation}\label{eq:MCS11}
w[n_1] = 2\prod_{m' = 0}^{\cM - 1} \cK_1(m')\, \cB_1\, .
\end{equation}
Matrices
\begin{equation}\label{eq:MCS12}
\hat{\cK}_{\alpha\beta} = \cK_1[n_1]\, \delta_{\alpha\beta}
\end{equation}
correspond to an $n_2$-dependence of the local factors in the two-bit chain of the form
\begin{align}\label{eq:MCS13}
& \cK(m) = \cK_1[n_1]\, h_\alpha\big( n_2(m+1)\big)\, h_\alpha \big( n_2(m)\big) 
\notag \\
& \; = \cK_1[n_1]\big\{ n_2(m+1) n_2(m) + [ 1 - n_2(m+1)] [ 1 - n_2(m)]\big\}\, .
\end{align}
We deal with two independent equal chains, one for $n_2 = 1$, the other for $n_2 = 0$. The resulting factor $2$ in $w_1$ can be absorbed by renormalization.

Another, trivial, case for a subsystem described by a local chain occurs if the local factors and boundary term in the two-bit local chain are all independent of $n_2$. This corresponds to
\begin{equation}\label{eq:MCS14}
\hat{\cK}(m) = \cK_1(m)\, \begin{pmatrix}
1 & 1 \\
1 & 1
\end{pmatrix}\, , \quad \hat{\cB} = \cB_1\, \begin{pmatrix}
1 & 1 \\
1 & 1
\end{pmatrix} \, .
\end{equation}
With
\begin{equation}\label{eq:MCS15}
\hat{F} = \begin{pmatrix}
1 & 1  \\
1 & 1
\end{pmatrix}\, , \quad \hat{F}^{\cM + 1} = 2^{\cM + 1}\, \hat{F}
\end{equation}
one obtains
\begin{equation}\label{eq:MCS16}
w[n_1] = 2^{\cM + 1} \prod_{m'=0}^{\cM - 1} \cK_1(m')\, \cB_1\, .
\end{equation}
The configuration sum of $n_2$ produces a factor $2$ for each site, resulting in the normalization $2^{\cM + 1}$.

We can generalize this to arbitrary $(2\times 2)$-matrices $\hat{F}(m)$, with
\begin{equation}\label{eq:MCS17}
\hat{\cK}(m) = \cK_1 (m)\, \hat{F}(m)\, , \quad \hat{\cB} = \cB_1\, \hat{F}_{\cB}\, ,
\end{equation}
where $\cK_1(m)$ depends on $n_1(m+1)$ and $n_1(m)$, while $\hat{F}$ is independent of $n_1$. The expression \eqref{eq:MCS9} factorizes
\begin{equation}\label{eq:MCS18}
w[n_1] = \cN \prod_{n' = 0}^{\cM - 1} \cK_1(m')\, \cB_1\, ,
\end{equation}
with
\begin{equation}\label{eq:MCS19}
\cN = \tr \big\{ \hat{F}(\cM - 1)\cdots \, \hat{F}(0)\, \hat{F}_{\cB} \big\}
\end{equation}
a normalization constant independent of $n_1$. Again, the normalization factor $\cN$ can be absorbed by multiplicative renormalization of $w[n_1]$.

A factorization of the matrices \eqref{eq:MCS17} corresponds to a product structure of the local factors in the two-bit chain,
\begin{align}\label{eq:MCS20}
\cK(m) &= \cK_1\big( m;\, n_1(m+1), n_1(m)\big) \notag \\
& \quad \times \cK_2\big( m;\, n_2(m+1), n_2(m)\big)\, ,
\end{align}
with $\hat{F}(m)$ obtained from the expression of $\cK_2$ in the occupation number basis for $n_2$, and similar for the boundary term $\cB$. It is obvious that the integration over the variables $n_2$ only results in an overall normalization factor $\cN$ in this case,
\begin{equation}\label{eq:MCS21}
\cN = \int \cD n_2 \prod_{m'} \cK_2 (m')\, \cB_2\, .
\end{equation}

Such a product structure is, however, not the general case. We denote by $n'_1$, $n_1$, $n'_2$, $n_2$ the occupation numbers $n_1(m+1)$, $n_1(m)$, $n_2(m+1)$, $n_2(m)$, respectively. If the dependence of the local factor $\cK(n'_1,\, n'_2,\, n_1,\, n_2)$ on the occupation numbers $n'_1$ and $n_1$ is not the same for all possible configurations of $n'_2$ and $n_2$, the factorized form \eqref{eq:MCS20} is not given. The matrix elements $\hat{\cK}_{\alpha\beta}$ cannot have the same dependence on $n'_1$ and $n_1$ in this case. The subsystem is a matrix chain that cannot be written as a local chain. We conclude that matrix chains are the generic case for subsystems where some variables are integrated out.

\paragraph*{Observables}

The computation of expectation values of observables $A[n_1]$ that only involve the occupation numbers $n_1(m)$ is straightforward according to eq.~\eqref{eq:MCS3}. If $w[n_1]$ is formulated as a matrix chain one may use
\begin{equation}\label{eq:MCS22}
\langle A[n_1] \rangle = Z^{-1} \, \int \cD n_1\, \tr \bigg\{ \hat{A}[n_1] 
\prod_{m'=0}^{\cM - 1} \hat{\cK}(m')\, \hat{\cB} \bigg\}\, ,
\end{equation}
with operator $\hat{A}$ proportional to the unit matrix,
\begin{equation}\label{eq:MCS23}
\big( \hat{A}[n_1] \big)_{\alpha\beta} = A[n_1]\, \delta_{\alpha\beta}\, .
\end{equation}
Since $\hat{A}[n_1]$ commutes with all matrices, its position under the trace plays no role.

The information contained in the chain of matrices $\hat\cK(m)$ is still sufficient to compute the expectation value of a local observable $A(m;\, n_1(m),\, n_2(m))$ that depends both on $n_1$ and $n_2$. This information is only lost by performing matrix multiplications and the trace. If we admit generalized local observables represented by operators that are not proportional to $\delta_{\alpha\beta}$ (cf. appendix~\ref{app:matrix chains}) the expectation values of local observables depending on $n_2(m)$ can be recovered. This can be seen most directly from the fact that according to appendix~\ref{app:matrix chains} the classical wave functions remain four-component vectors $\tilde{q}_{\alpha,\tau}(m)$, $\bar{q}_{\alpha,\tau}(m)$. Using the expression \eqref{eq:CW25} for the expectation value of a local observable the local probabilistic information contained in the classical wave functions is sufficient for the computation of $\langle A (m;\, n_1(m),\, n_2(m)) \rangle$. For example, one has the association of operators to observables according to
\begin{equation}\label{eq:MCS24}
n_1 \, \to \, \hat{N}_1 = \begin{pmatrix}
1 & 0 & 0 & 0 \\
0 & 1 & 0 & 0 \\
0 & 0 & 0 & 0 \\
0 & 0 & 0 & 0
\end{pmatrix} , \;
n_2 \, \to \, \hat{N}_2 = \begin{pmatrix}
1 & 0 & 0 & 0 \\
0 & 0 & 0 & 0 \\
0 & 0 & 1 & 0 \\
0 & 0 & 0 & 0 
\end{pmatrix}\, .
\end{equation}

In fact, for the matrix chain no information is lost by the integration over the variables $n_2(m)$ if we admit observables represented by operators that are not proportional to $\delta_{\alpha\beta}$. The information is only reshuffled to information contained in the matrices $\hat{\cK}$. If we expand the matrix $\hat{\cK}$ in basis functions for $n_1$,
\begin{equation}\label{eq:MCS25}
\hat{\cK}_{\alpha\beta} [ n_1(m) ] = \hat{T}_{\tau\alpha,\, \rho\beta} (m)\,
h_\tau [ n_1 (m+1) ]\, h_\rho [n_1 (m) ]\, ,
\end{equation}
we obtain the same transfer matrix as for the original two-bit local chain. (We observe that the assignment of indices to the transfer matrix differs from eq.~\eqref{eq:TS41}. For the conventions it matters which variable is integrated out. If we integrated out $n_1$ instead of $n_2$, the index argument \eqref{eq:TS41} would be valid. Since we can express the weight distribution for the two-bit local chain by a product of transfer matrices, eq.~\eqref{eq:TS52} with exchanged indices, the full information is available for both procedures of integrating out variables.) The expectation values of arbitrary observables built from both $n_1$ and $n_2$ can be computed from the matrix chain.

We conclude that in this view the matrix chain is a reformulation rather than a true subsystem. A true subsystem for which the information about $n_2$ is lost follows only once the trace~\eqref{eq:MCS9} is performed explicitly without memory of the underlying ordered chain of matrices. For example, true subsystems for which the full information of the two-bit local chain is no longer available are given for the cases where the subsystem is again a local chain. Once all probabilistic information about the occupation numbers $n_2$ is absorbed in the normalization $\cN$ given by eq.~\eqref{eq:MCS19} or \eqref{eq:MCS21}, it is no longer available for the subsystem. Expectation values for observables involving $n_2$ can no longer be computed from the probabilistic information based on the local chain built with $\cK_1$.

\paragraph*{General subsystems from integrating out variables}

The construction of matrix chains from local chains can be generalized for arbitrary local chains. First, for a total number of $M = M_1 + 1$ bits on each site of the local chain we may ``integrate'' one bit and leave the other $M_1$ bits without integration in eq.~\eqref{eq:MCS4}. The results in $(2\times 2)$-matrices depend on $M_1$ occupation numbers. We may also split $M$ differently, $M = M_1 + M_2$, and integrate $M_2$ bits. There are now $2^{M_2}$ basis functions, and the matrices are $(n\times n)$-matrices, $n = 2^{M_2}$, with elements depending on $M_1$ bits at sites $m+1$ and $m$. Integrating out a large number $M_2$ of bits results in very large matrices.

True subsystems, for which only part of the probabilistic information of the original local chain is kept, obtain if the chain of $(n\times n)$-matrices can be reduced to a matrix chain with $(n'\times n'$)-matrices, $n' < n$. (For local chains obtained from two-bit local chains, one has $n=2$, $n'=1$.) For given matrices smaller matrices can be obtained by various forms of taking subtraces. For such subtraces one can compute the values of all observables that have an operator expression in terms of $(n'\times n')$-matrices. This is no longer the set of all possible observables of the original $M$-bit local chain.

Useful subsystems obtained by integrating out variables and subsequent partial traces for the resulting matrix chains are those for which the time evolution is closed. This means that the probabilistic time-local information characterizing the subsystem at $m$ can be obtained from the one characterizing this system at $m-1$. (Often it is sufficient that the evolution is approximately closed.) Typical subsystems obtained by integrating out variables are correlated with their environment. The connected correlation functions do not vanish for pairs of observables where one observable depends on the system variables $n$ and the other ``environment observable'' depends on variables $n'$ that are integrated out. This is visible directly from the matrix chains before reducing the matrices by subtraces. Correlated subsystems are the generic case.
\subsubsection{Subtraces}\label{sec:subtraces}

In quantum mechanics subtraces of the density matrix are an important way to define relevant subsystems. The same holds for ``classical'' probabilistic systems. Our formulation of time-local subsystems in terms of the classical density matrix offers the appropriate starting point for defining subsystems by subtraces. We discuss in this subsection subsystems for local chains. A generalization to matrix chains, possibly obtained from integrating out variables, is straightforward, but not considered here. Large classes of interesting subsystems for local chains can be obtained by subtraces. They are only found in a formulation of the time-local probability information in terms of classical wave functions or the classical density matrix, while they are not accessible from the time-local probability distribution. The possibility to find relevant subsystems is an important advantage of our formulation of evolution in classical statistics. A crucial criterion for useful subsystems requires that they are closed with respect to the time evolution. The probabilistic information of the subsystem, encoded in the density matrix of the subsystem, should be sufficient for the computation of the time evolution of the subsystem. No probabilistic information related to its environment should play a role.

\paragraph*{Properties of substraced subsystems}

Subsystems defined by subtraces show important new properties. The time evolution of the subsystem is still governed by a step evolution operator $\tilde{S}$ for the subsystem, with density matrix $\tilde{\rho}$ of the subsystem evolving similar to eq.\,\eqref{eq:DM38},
\begin{equation}
\tilde{\rho}(m+1) = \tilde{S}(m) \tilde{\rho}(m) \tilde{S}^{-1}(m).
\label{SUB01}
\end{equation}
What is new is that the step evolution operator $\tilde{S}$ of the subsystem is generically no longer a real positive matrix, even if the step evolution operator $\hat{S}$ of the total time-local subsystem is real and positive. In general, $\tilde{S}$ will be found to be a complex matrix. For certain types of subsystems it is necessarily complex. Correspondingly, also the density matrix $\tilde{\rho}$ of the subsystem is a complex matrix. Furthermore, the unique jump property of automata is typically lost for subsystems. If $\hat S$ is a unique jump step evolution operator the reduced evolution operator $\tilde S$ has no longer this property.

We can define a weight distribution $\tilde{w}$ for the subsystem by replacing in eqs.\,\eqref{eq:TS46}, \eqref{eq:TS47} the transfer matrix $\hat{T}$ by the step evolution operator $\tilde{S}$ of the subsystem, and similar for the boundary conditions, $\hat{B}\to \tilde{B}$. The weight distribution $\tilde{w}$ for the subsystem is, in general, no longer real and positive. For complex $\tilde{S}$ it will be a complex quantity. This constitutes an important bridge to the path integral in quantum mechanics or the complex functional integral for quantum field theories in Minkowski space. For these quantum systems the weight distribution is complex, similar to the one for subsystems of ``classical'' probabilistic systems. For the step evolution operator of subsystems the restrictions discussed in appendix~\ref{app:positivity_of_overall_probability_distribution} do no longer apply.

Another new property of subsystems is that a pure classical state of the total time-local subsystem generically leads to a mixed state for the subsystem defined by a subtrace. This property is familiar from quantum mechanics. In the appendix~\ref{app: subtraces for the two-bit local chain} we will discuss all these new properties in the context of a rather simple example of a total time-local system with four states ($N=4$) and a subsystem with two states.

For the general setting of subtraces we consider step evolution operators $\hat{S}_{\tau\alpha,\, \rho\beta}$ and classical density matrices $\rho'_{\tau\alpha,\,\rho\beta}$ for which the indices of the matrix elements are written as double indices, e.g. $(\tau,\,\alpha)$ or $(\rho,\, \beta)$. We may define a subsystem by taking for the density matrix a subtrace over the indices $\alpha$, $\beta$,
\begin{equation}\label{eq:ST1}
\rho^{(s)}_{\tau\rho} = \delta^{\alpha\beta} \rho'_{\tau\alpha,\,\rho\beta} = 
\rho'_{\tau\alpha,\,\rho\alpha}\, .
\end{equation}
Due to the sum, the density matrix for the subsystem $\rho^{(s)}$ contains, in general, less local probabilistic information than the original density matrix $\rho'_{\tau\alpha,\, \rho\beta}$. We note that the density matrix for the subsystem is properly normalized, 
\begin{equation}\label{eq:ST10}
\tr\, \rho^{(s)} = \rho^{(s)}_{\tau\tau} = \rho'_{\tau\alpha,\,\tau\alpha} = \tr \,\rho' = 1\, 
.
\end{equation}
in appendix~\ref{app: subtraces for the two-bit local chain} we discuss a generalization to modified subtraces which use in eq.~\eqref{eq:ST1} a metric $g^{\alpha\beta}$ instead of $\delta^{\alpha\beta}$.
 
Once the subtrace is taken, the probabilistic information about the environment of the subsystem is lost. The expectation values of arbitrary observables of the total system can no longer be computed. Still, observables with associated operators of the form
\begin{equation}
\hat{A}_{\tau\alpha,\rho\beta} = \hat{A}_{\tau\rho}^\mathrm{(s)} \delta_{\alpha\beta}
\label{eq:SUBA}
\end{equation}
remain compatible with the subsystem. Their expectation values can be computed from the density matrix of the subsystem
\begin{equation}
\braket{A} = \hat{A}_{\tau\alpha,\rho\beta} \rho'_{\rho\beta\,\tau\alpha} = \hat{A}_{\tau\rho}^\mathrm{(s)} \rho_{\rho\tau}^\mathrm{(s)}.
\label{eq:SUBB}
\end{equation}
We can associate to such an observable $A$ the subsystem operator $\hat{A}^\mathrm{(s)}$, such that the standard ``quantum rule'' for the expectation value holds in the subsystem,
\begin{equation}
\braket{A} = \tr\{ \hat{A}^\mathrm{(s)}\rho^\mathrm{(s)} \}.
\label{eq:SUBC}
\end{equation}

Subtraces can be taken in an arbitrary basis for the classical density matrix. In such a basis the operators may be non-diagonal even for strictly local observables that are diagonal in the occupation number basis. As a result, strictly local observables for the overall time-local system often result in non-diagonal operators $\hat A^{(s)}$ for the subsystem. This is an important general origin for non-diagonal operators representing observables in classical statistics. We emphasize the direct connection with the incomplete statistics for subsystems discussed in sect.~\ref{sec:probabilistic_observables_and_incomplete_statistics}.

\paragraph*{Evolution of the subsystem}

A subsystem obtained by the subtrace \eqref{eq:ST1} is useful if its time evolution can be described in terms of the probabilistic information of the subsystem. The time evolution has to be ``closed'' in this sense. This will impose ``compatibility conditions'' on the step evolution operator $\hat{S}_{\tau\alpha,\rho\beta}$.

The evolution law for the density matrix $\rho^{(s)}$ is given by
\begin{align}\label{eq:ST2}
\rho^{(s)}_{\tau\rho} (m+1) &= \rho'_{\tau\alpha,\, \rho\alpha} (m+1) \notag \\
& = \hat{S}_{\tau\alpha,\, \sigma\gamma}(m)\, \rho'_{\sigma\gamma,\, \mu\delta} (m)
\, \big( \hat{S}^{-1} \big)_{\mu\delta,\, \rho\alpha} (m)\, .
\end{align}
In general, the r.h.s. of eq.~\eqref{eq:ST2} cannot be written in terms of $\rho^{(s)}$ alone. An evolution law that only involves the local statistical information in the subsystem requires for the step evolution operator the condition
\begin{equation}\label{eq:ST3}
\hat{S}_{\tau\alpha,\, \sigma\gamma}\, \big( \hat{S}^{-1} \big)_{\mu\delta,\, \rho\alpha}
= A_{\tau\sigma\mu\delta}\, \delta_{\gamma\delta}\, .
\end{equation}
In this case one has
\begin{equation}\label{eq:ST4}
\rho^{(s)}_{\tau\rho}(m+1) = A_{\tau\sigma\mu\rho}(m)\, \rho^{(s)}_{\sigma\mu}(m)\, .
\end{equation}
If, furthermore,
\begin{equation}\label{eq:ST5}
A_{\tau\sigma\mu\rho} = \hat{S}^{(s)}_{\tau\sigma}\, \big(\hat{S}^{(s)} \big)^{-1}_{\mu\rho}
\, ,
\end{equation}
the evolution law takes for the subsystem the same form as for the original system
\begin{equation}\label{eq:ST6}
\rho^{(s)}_{\tau\rho} (m+1) = \hat{S}^{(s)}_{\tau\sigma}(m)\, \rho^{(s)}_{\sigma\mu}\, 
\big( \hat{S}^{(s)} \big)^{-1}_{\mu\rho}\, .
\end{equation}
Only in this case the subtrace produces a subsystem that is compatible with the standard evolution. We will require that subsystems which are compatible with the local evolution in time obey the condition \eqref{eq:ST6}.
The evolution of the density matrix of a subsystem will show new features, similar to the ones for subsystems in quantum mechanics\,\cite{KOS,LIN,ZOL}.

\paragraph*{Factorized step evolution operator}

A simple case for such subtraces can be realized if the step evolution operator has a direct product form
\begin{equation}\label{eq:ST7}
\hat{S}_{\tau\alpha,\,\rho\beta} = \big( \hat{S}_1 \big)_{\tau\rho} 
\big(\hat{S}_2\big)_{\alpha\beta}\, .
\end{equation}
The inverse of the step evolution operator takes the form
\begin{equation}\label{eq:ST8}
\big( \hat{S}^{-1} \big)_{\mu\delta,\, \rho\beta} = \big( \hat{S}_1^{-1} \big)_{\mu\rho}
\big( \hat{S}_2^{-1} \big)_{\delta\beta}\, ,
\end{equation}
such that eq.~\eqref{eq:ST3} reads
\begin{equation}\label{eq:ST9}
\hat{S}_{\tau\alpha,\, \sigma\gamma}\, \hat{S}^{-1}_{\mu\delta,\,\rho\alpha} = 
\big( \hat{S}_1 \big)_{\tau\sigma} \big( \hat{S}_1^{-1} \big)_{\mu\rho} \, 
\delta_{\gamma\delta}\, ,
\end{equation}
and $\rho^{(s)}$ obeys equation \eqref{eq:ST6}, with $\hat{S}^{(s)}$ identified with $\hat{S}_1$. For the factorized form \eqref{eq:ST7} the matrix $\hat{S}_1$ accounts for the evolution of the subsystem, while $\hat{S}_2$ describes the evolution of its ``environment''. Once the environment is integrated out by the subtrace \eqref{eq:ST1}, $\hat{S}_2$ does not matter for the evolution of the subsystem.

For a factorized step evolution operator \eqref{eq:ST7} the eigenvalues $\lambda_i$ of $\hat{S}$ are products of eigenvalues $\lambda^{(1)}_k$ of $\hat{S}_1$ and eigenvalues $\lambda_l^{(2)}$ of $\hat{S}_2$, 
\begin{equation}\label{eq:ST11}
\lambda_i = \lambda_{kl} = \lambda^{(1)}_k\, \lambda_l^{(2)}\, .
\end{equation}
Indeed, the product of eigenfunctions to $\hat{S}_1$ and $\hat{S}_2$ is an eigenfunction of $\hat{S}$, since from
\begin{equation}\label{eq:ST12}
\big( \hat{S}_1 \big)_{\tau\rho}\, q^{(1)}_\rho = \lambda^{(1)}_k\, q_\tau^{(1)}\, , \quad
\big( \hat{S}_2\big)_{\alpha\beta}\, q^{(2)}_\beta = \lambda^{(2)}_l\, q^{(2)}_\alpha
\end{equation}
one infers
\begin{equation}\label{eq:ST13}
\hat{S}_{\tau\alpha,\, \rho\beta}\, q^{(1)}_\rho\, q^{(2)}_\beta = \lambda^{(1)}_k\,
\lambda^{(2)}_l\, q^{(1)}_\tau\, q^{(2)}_\alpha\, .
\end{equation}
The relation
\begin{equation}\label{eq:ST14}
\tr \,\hat{S} = \hat{S}_{\tau\alpha,\,\tau\alpha} = \big( \hat{S}_1\big)_{\tau\tau} 
\big( \hat{S}_2 \big)_{\alpha\alpha} = \tr \,\hat{S}_1\; \tr\, \hat{S}_2
\end{equation}
implies consistently
\begin{align}\label{eq:ST15}
\sum_i \lambda_i = \Big( \sum_k \lambda_k^{(1)} \Big)\, \Big( \sum_l \lambda_l^{(2)} \Big)
= \sum_{kl} \lambda_k^{(1)} \lambda_l^{(2)} = \sum_{kl} \lambda_{kl}\, .
\end{align}

\paragraph*{General subtraces}

The direct product form of the step evolution operator \eqref{eq:ST7} is not the only possibility for defining subsystems by subtraces, such that the evolution law for the subsystem only requires local probabilistic information contained in the density matrix of the subsystem. 
There are many interesting possibilities for embeddings of subsystems into a total system. They can be realized by suitable modifications of the definition of subtraces. One rather straightforward way uses similarity transformations.

Indeed, we can use a similarity transformation
\begin{equation}\label{eq:ST16}
\tilde{S} = D\, \hat{S}\, D^{-1}
\end{equation}
in order to bring $\tilde{S}$ to a factorized form \eqref{eq:ST7},
\begin{equation}\label{eq:ST17}
\tilde{S} = \tilde{S}_1 \otimes \tilde{S}_2\, .
\end{equation}
Similarly, we transform the density matrix
\begin{equation}\label{eq:ST18}
\tilde{\rho} = D\, \rho'\, D^{-1}\, .
\end{equation}
In the new basis the evolution law keeps its form
\begin{equation}\label{eq:ST19}
\tilde{\rho}(m+1) = \tilde{S}(m)\, \tilde{\rho}(m)\, \tilde{S}^{-1}\, .
\end{equation}

We can now define the subsystem by taking a subtrace in the new basis. Let us consider a basis for which the step evolution operator factorizes
\begin{equation}\label{eq:ST20}
\tilde{S}_{\tilde{\tau}\tilde{\alpha},\, \tilde{\rho}\tilde{\beta}} =
\big( \tilde{S}_1 \big)_{\tilde{\tau}\tilde{\rho}} 
\big( \tilde{S}_2 \big)_{\tilde{\alpha}\tilde{\beta}}\, .
\end{equation} 
In this basis the density matrix has elements $\tilde{\rho}_{\tilde{\tau}\tilde{\alpha},\,\tilde{\rho}\tilde{\beta}}$, and we define the density matrix of the subsystem by
\begin{equation}\label{eq:ST21}
\tilde{\rho}^{(s)}_{\tilde{\tau}\tilde{\rho}} = 
\tilde{\rho}_{\tilde{\tau}\tilde{\alpha},\, \tilde{\rho}\tilde{\alpha}}\, .
\end{equation}
We can take over the preceding discussion on factorized step evolution operators \eqref{eq:ST7} to the new basis. Thus $\tilde{\rho}^{(s)}$ defines a subsystem whose evolution obeys the standard evolution law
\begin{equation}\label{eq:ST22}
\tilde{\rho}^{(s)} (m+1) = \tilde{S}_1 (m)\, \tilde{\rho}^{(s)}(m)\, \tilde{S}_1^{-1}(m)\, 
.
\end{equation}

The factorized form \eqref{eq:ST17} always exists provided one can find sets of eigenvalues $\tilde{\lambda}^{(1)}_k$ and $\tilde{\lambda}_k^{(2)}$ such that the eigenvalues $\lambda_i$ of $\hat{S}$ can be written as
\begin{equation}\label{eq:ST23}
\lambda_i = \lambda_{kl} = \tilde{\lambda}^{(1)}_k\, \tilde{\lambda}^{(2)}_l\, .
\end{equation}
This condition is necessary since the eigenvalues of $\hat{S}$ and $\tilde{S}$ are the same. It is also sufficient. For example, we can use for $D$ the matrix that diagonalizes $\hat{S}$. In this diagonal form, $\tilde{S}$ can be written as a product of diagonal matrices $\tilde{S}_1$ and $\tilde{S}_2$ if the relation \eqref{eq:ST23} is obeyed. We emphasize, however, that this is not the only possible choice of $D$. For example, further similarity transformations respecting the factorized form can be applied to the diagonal submatrices $\tilde{S}_1$ and $\tilde{S}_2$, such that $\tilde{S}_1$ and $\tilde{S}_2$ remain no longer diagonal.

For any given matrix which achieves eq.\,\eqref{eq:ST20} we can define the density matrix of the subsystem by the modified subtrace
\begin{equation}
\tilde{\rho}_{\tau\rho}^\mathrm{(s)} = (D\hat{S}D^{-1})_{\tau\alpha,\rho\alpha}.
\label{eq:SUBD}
\end{equation}
Many interesting subsystems can be obtained in this way. For the time-local system the physical properties are independent of the choice of basis. This does no longer hold for subtraced subsystems. It matters in which basis the subtrace is taken. Different choices of the basis result in different subsystems, for which different parts of the time-local probabilistic information are neglected, as being associated to the environment. Subsystems obtained by subtraces in position or momentum space can be rather different.

\subsubsection{General local subsystems}
\label{sec:general_local_subsystems}

In this section we look at the probabilistic information which characterizes a subsystem from a somewhat different point of view. We focus on expectation values of observables. They can be computed from the overall probability distribution and define the probabilistic information which specifies the subsystem. General local subsystems are subsystems of the time-local subsystem that we have discussed in sect.~\ref{sec:time_local_subsystems}. They
only use part of the probabilistic information contained in the classical density matrix $\rho'(t)$. Examples that we have
already encountered are correlation subsystems in sect.~\ref{sec:correlation_subsystems}, asymptotic subsystems for the continuum limit in sect.~\ref{sec:markov_chains} or subtraces in sect.~\ref{sec:subtraces}.
In the present section we review these subsystems by putting the focus on expectation values of observables that are used as variables for the subsystem.

\paragraph*{Reduced density matrix}

We associate the probabilistic information in the general local subsystem to expectation values of local observables
\begin{equation}
\label{GLS1}
\rho_z(t) = \braket{A_z(t)} = \tr \left( \hat{A}_z  \rho'(t) \right)\,.
\end{equation}
For given operators $\hat{A}_z$ associated to the local observables $A_z$ the subsystem variables $\rho_z(t)$ are linear combinations
of the elements of the classical density matrix $\rho'(t)$. If the number of variables $\rho_z(t)$ is smaller than the number of
independent elements of $\rho'(t)$, we have a true subsystem that only uses part of the probabilistic information in the time-local
subsystem.

We next employ $R \times R$-matrices $L_z$ with the following properties:
\renewcommand{\labelenumi}{(\roman{enumi})}
\begin{enumerate}
\item The matrix $L_0 = 1$ is the unit matrix
\item The matrices $L_z$ for $z \neq 0$ are traceless, $\tr(L_z) = 0$
\item We choose the normalization \begin{equation}
\label{GLS2}
\tr \left( L_y L_z \right) = R \delta_{yz}\,.
\end{equation}
\end{enumerate}
We define the reduced density matrix by 
\begin{equation}
\label{GLS3}
\rho(t) = \frac{1}{R} \rho_z(t) L_z\,, \quad \rho_0(t) = 1\,.
\end{equation}
It is normalized as a density matrix
\begin{equation}
\label{GLS4}
\tr(\rho(t)) = 1\,.
\end{equation}
The reduced density matrix is hermitian if we employ hermitian matrices $L_z^\dagger = L_z$. In many respects the reduced density matrix
will play a role similar to the classical density matrix or the quantum density matrix. We do, however, not necessarily require positivity
of the matrix $\rho(t)$.

The properties (i)--(iii) are sufficient to extract from the reduced density matrix the expectation values of observables that we
have used as an input 
\begin{equation}
\label{GLS5}
\braket{A_z(t)} = \rho_z(t) = \tr \left( L_z \rho(t) \right)\,.
\end{equation}
There is no need that the matrices $L_z$ obey the same commutation relations as the operators $\hat{A}_z$ used in the time-local
subsystem in eq.~\eqref{GLS1}. In particular, if for two different observables $A_z$ and $A_y$ the correlation function 
$\braket{A_z(T) A_y(t)}$ is not available in the general local subsystem, the matrices $L_z$ and $L_y$ typically do not commute.

In the most general case the possible measurement values of $A_z$ need not to coincide with the eigenvalues of $L_z$. We are mainly interested, however, in the case where the
possible measurement values of $A_z$ are indeed given by the eigenvalues of $L_z$ and powers of $A_z$ are mapped to powers of $L_z$. If $\rho$ is a positive hermitian matrix, the operators
$L_z$ define then the same probabilistic observables $A_z$ as for the time-local subsystem. For a non-degenerate spectrum, the probability
to find an eigenvalue $\lambda_\alpha^{(z)}$ of $L_z$ is given by the diagonal element $\tilde{\rho}_{\alpha \alpha}$, with 
$\tilde{\rho}$ the reduced density matrix in the basis where $L_z$ is diagonal.

The normalization with the factor $R$ in eqs.~\eqref{GLS2},
\eqref{GLS3} is adapted to $A_z$ being Ising spins. Other normalizations of $L_z$ which keep the relations \eqref{GLS4}, \eqref{GLS5}
unchanged, are possible. We could also add to $L_z$ a part proportional to the unit matrix. We prefer here to subtract from the observable $A_z$ a constant part which makes $L_z$ traceless.

For a positive density matrix the probabilistic information of the subsystem is sufficient for the computation of expectation values of arbitrary functions $f(A_z)$. This can be seen by diagonalizing the density matrix. In general, classical correlations between two different observables $A_z$ and $A_{z'}$ are not available for the subsystem. Exceptions are only possible if $L_z$ and $L_{z'}$ commute. If at least two of the generators $L_z$ do not commute the incomplete statistics of the general local subsystem is directly visible. It is an interesting question which other types of observables become computable in the case of subsystems realized by non-commuting $L_z$.

\paragraph*{Subtraces}

The subtraces discussed in sect.~\ref{sec:subtraces} are a particular case of this setting. Employing for $z$ a double index notation,
$z = (ab)$, we define observables associated to the operators
\begin{equation}
\label{GLS6}
\left( \hat{A}_{ab} \right)_{\tau \alpha,\rho \beta} = \delta_{a \rho} \delta_{b \tau} \delta_{\alpha \beta}\ .
\end{equation}
These are projectors with possible measurement values one or zero. Their expectation values obey
\begin{align}
\label{GLS7}
\bar\rho_{ab} &= \tr \left( \hat{A}_{ab} \rho' \right) = \left( \hat{A}_{ab} \right)_{\tau \alpha,\rho \beta} \rho'_{\rho \beta,\tau \alpha} \nonumber \\
&= \delta_{\alpha \beta} \rho'_{a \beta, b \alpha}\,.
\end{align}
We define the matrices $\bar{L}_z = L_{ab}$ 
\begin{equation}
\label{GLS8}
(L_{ab})_{\tau \rho} = \delta_{a \tau} \delta_{b \rho}\,,
\end{equation}
such that
\begin{equation}
\label{GLS9}
\rho = L_{ab} \bar\rho_{ab}
\end{equation}
is indeed the density matrix of the subsystem formed with a subtrace \eqref{eq:ST1},
\begin{equation}
\label{GLS10}
\rho_{\tau \rho} = \delta_{\alpha \beta} \rho'_{\tau \alpha,\rho \beta}\,.
\end{equation}
The matrices $L_{ab}$ do not all commute. For example for $a\neq b$ one finds (no sum over $b$)
\bel{5.3.96A}
\big[L_{ab},L_{bb}\big]_{\tau\rho}=\delta_{a\tau}\delta_{b\rho}\ .
\ee

The matrices $\bar L_{z}=L_{ab}$ are not yet traceless normalized matrices. The matrices $L_z$ can be formed as linear combinations of $\bar{L}_z$. This demonstrates that subtraces can be cast into the general framework of reduced density matrices defined by eqs.~\eqref{GLS1}-~\eqref{GLS4}. From a general point of view they are only a very particular
case of the general local subsystems. Many other subsystems can be found by other choices of the operator $\hat{A}_z$ and $L_z$.

\paragraph*{Evolution of general local subsystems}

The evolution equation for a general local subsystem can be inferred from the evolution equation for the time local subsystem. 
Writing
\begin{equation}
\label{GLS11}
\rho(t) = \frac{1}{R} L_z \tr \left\{ \hat{A}_z \rho'(t) \right\},
\end{equation}
we infer from eq.~\eqref{eq:DM38}
\begin{align}
\label{GLS12}
\rho(t+\epsilon) &= \frac{1}{R} L_z \tr \left\{ \hat{A}_z \hat{S}(t) \rho'(t) \hat{S}^{-1} \right\} \nonumber \\
&= \frac{1}{R} L_z \tr \left\{ \hat{B}_z(t) \rho'(t) \right\}\,,
\end{align}
with 
\begin{equation}
\label{GLS13}
\hat{B}_z(t) = \hat{S}^{-1}(t) \hat{A}_z \hat{S}(t)\,.
\end{equation}
This evolution equation is closed if we can express the expectation value of $B_z$ in terms of $A_z$,
\begin{equation}
\label{GLS14}
\braket{B_z(t)} = c_{zy}(t) \braket{A_y(t)}\,.
\end{equation}
or
\begin{align}
\label{GLS15}
\rho_z(t+\epsilon) &= \tr \left\{ \hat{B}_z(t) \rho'(t) \right\} = c_{zy}(t) \tr \left\{ \hat{A}_y \rho'(t) \right\} \nonumber \\
&= c_{zy}(t) \rho_y(t)\,.
\end{align}
In this case one has
\begin{equation}
\label{GLS16}
\rho(t+\epsilon) = \frac{1}{R} L_z c_{zy}(t) \rho_y(t)\,.
\end{equation}

If, furthermore, a regular $R \times R$ matrix $\bar{S}(t)$ exists which obeys
\begin{equation}
\label{GLS17}
L_z c_{zy}(t) = \bar{S}(t) L_y \bar{S}^{-1}(t)\,,
\end{equation}
one finds a von-Neumann type evolution equation for the reduced density matrix
\begin{equation}
\label{GLS18}
\rho(t+\epsilon) = \bar{S}(t) \rho(t) \bar{S}^{-1}(t)\,.
\end{equation}
Here we have normalized $\bar{S}(t)$ such that the largest eigenvalues obey $| \lambda_i | = 1$. The matrix $\bar{S}(t)$ is therefore
the step evolution operator for the reduced density matrix of the general local subsystem.

We emphasize that the matrix $\bar{S}(t)$ needs not to be a non-negative real matrix even if $\hat{S}(t)$ has this property. The step
evolution operator for the reduced density matrix can be a complex matrix. In particular, if the reduced density matrix is a positive
matrix and $\hat{S}(t)$ is a unitary matrix, eq.\,\eqref{GLS18} is precisely the von Neumann equation for quantum mechanics.

As far as subtraces are concerned the approach in the present section is somewhat complementary to the discussion in sect.\,\ref{sec:subtraces}. The subtraces focus on sets of eigenvalues of the step evolution operator $\hat{S}$, defining the similarity transformation $D$ accordingly. One has then to find out which are the observables that are compatible with the subsystem. The general local subsystems of the present section focus on expectation values of observables. For a suitable choice of $L_z$ they define probabilistic observables of the subsystem. One has then to find out if the evolution is compatible with such a system, i.\,e.\ if eq.\,\eqref{GLS17} is obeyed.
\subsubsection{Incomplete statistics for subsystems}
\label{sec:incomplete_statistics_for_subsystems}

In summary, probabilistic subsystems show characteristic features that differ from the overall probabilistic system. For the overall 
probabilistic system the observables have fixed values $A_\tau$ in every state $\tau$. From the probabilities $p_\tau$ for the states
$\tau$ the expectation values of the observables can be computed. Products of observables are again observables. The expectation values
of such product observables are the classical correlation functions. All classical correlation functions can, in principle, be computed
from the overall probability distribution. The overall probabilistic system is characterized by ``complete statistics". In particular,
all joint probabilities to find for $A$ the value $\lambda_i^{(A)}$, and for $B$ the value $\lambda_j^{(B)}$, are available for the
overall probabilistic system.

In contrast, the probabilistic information for a subsystem is typically given by expectation values of observables. A state of the 
subsystem can be specified by a certain number of such expectation values. Observables have typically no fixed values in a given state
$\rho$ of the subsystem. This is the basic reason why classical correlation functions are often not available from the probabilistic
information of the subsystem. The subsystem is then characterized by ``incomplete statistics''\,\cite{CWICS}. For incomplete statistics at most
a part of the correlation functions is accessible for the subsystem. 

General observables in subsystems are probabilistic observables. For a given state $\rho$ one can often compute the probabilities $w_i^{(A)}$
to find the value $\lambda_i^{(A)}$ of an observable $A$. The probabilistic information of the subsystem in terms of expectation values
is typically sufficient for this purpose. For Ising spin observables with two possible measurement values $\lambda_+^{(A)} = 1$,
$\lambda_-^{(A)} = -1$, the expectation value $\braket{A}$ fixes $w_+^{(A)}$ and  $w_-^{(A)}$. This extends to many observables with a 
larger set of possible measurement values. For a second probabilistic observable $B$ of the subsystem one can determine the probabilities
$w_j^{(B)}$ to find $\lambda_j^{(B)}$. Both $w_i^{(A)}$ and $w_j^{(B)}$ are functions of $\rho$. What is not available, in general,
are joint probabilities to find for $A$ the value $\lambda_i^{(A)}$ and for $B$ the value $\lambda_j^{(B)}$. Even for $A$ and $B$ 
being Ising spins the joint probabilities cannot be inferred from $\braket{A}$ and $\braket{B}$ alone.

Many subsystems do not contain the probabilistic information necessary to determine the classical correlation $\braket{AB}$ for \emph{all} 
probabilistic observables of the subsystem. 
For example, the information may be sufficient for the two-point function $\braket{AB}$ of two basis observables.
The product $AB$ may then be considered itself as one of the probabilistic system 
observables. The expectation value of the product of this ``composite observable'' with a third basis observable $C$, $\braket{ABC}$, is a
three point function that may not be accessible from the subsystem. Complete statistics requires that all expectation values of 
classical product observables are accessible. Otherwise, we deal with incomplete statistics.

This setting is most easily visible for correlation subsystems. By definition, only a certain number of correlations are included in the
subsystem. If the other correlations cannot all be computed from the subsystem correlations, the statistics is incomplete. For the
time-local subsystem, the probabilistic information is encoded in the classical density matrix $\rho'(t)$. The elements of $\rho'(t)$
can be obtained from expectation values of observables. For the diagonal elements one can use the expectation values of strictly local
observables, while for the off-diagonal elements one employs the expectation values of observables at $t' \neq t$ or derivative 
observables. Again, the state of the time-local subsystem can be associated with a certain number of expectation values. System observables
are probabilistic observables. Not all classical correlations for system observables are accessible and the statistics of the time-local
subsystem is incomplete. 

An example 	for a subsystem with complete statistics is the uncorrelated subsystem discussed in sect.~\ref{sec:subsystems_and_correlation_with_environment}. For this special case, where the subsystem and its environment are uncorrelated, 
the overall probability distribution can be written in a direct product form. Even though the uncorrelated subsystems are most often 
discussed in the literature on probabilistic subsystems, they do not cover many relevant subsystems in the real world. Concentration 
on this special case hides the important particular features of the probabilistic setting for subsystems. Their incomplete statistics is closely connected to the correlations between the subsystem and its environment.

\section[Discussion and conclusions]{Discussion and \\
conclusions}\label{sec:Discussion}

This work proposes a probabilistic formulation of the fundamental laws of
Nature. 
On a fundamental level, the description of the Universe
is entirely based on the concepts of classical statistics: Observables that take
definite values in the states of the system, probabilities for these states, and
expectation values of observables computed from these probabilities. The overall
probability distribution covers everything in the world, all times, all
locations, all possible events.

Any efficient description of natural phenomena concentrates on suitable
subsystems of the overall probabilistic system for the Universe. Such subsystems
are generically correlated with their environment. For example, a simple atom is
a particular ``excitation'' of the vacuum of a quantum field theory. The
evolution of the atom-subsystem depends on the properties of the vacuum. It is
correlated with its environment.

Subsystems that are correlated with their environment exhibit probabilistic
structures that differ from the overall probabilistic system for the world.
Observables in subsystems are typically probabilistic observables that do not
take definite values in all states of the subsystem. Only the probabilities to
find a given value of a probabilistic observable are available in the subsystem.
Many subsystems are characterized by incomplete statistics. Not all classical
correlation functions for the observables of the subsystem are computable from
the probabilistic information in the subsystem, in contrast to the complete
statistics of the overall probabilistic system. All probabilistic laws and
properties of the subsystems follow from the basic ``classical'' probabilistic
formulation of the overall system. They reflect the particular embedding of
subsystems in the overall system.

An important structure in the overall probabilistic system is probabilistic
time. Time emerges as an ordering relation between observables, rather than
being a concept postulated ``a priori''. The structure of time induces an
important type of subsystems -- the time-local subsystem. Time-local subsystems
contain the relevant probabilistic information for the ``present'' at a given
time, while the past and the future are the environment. The present is
correlated with the past and the future, such that time-local subsystems are
correlated with their environment.

Evolution describes the laws how the time local subsystem at a neighboring time
$t+\varepsilon$ is connected to the time-local subsystem at $t$. The formulation
of this rather simple issue for classical statistical systems reveals the
importance of probabilistic concepts that are familiar from quantum mechanics:
wave functions, density matrix and operators representing observables.
Formulated in terms of wave functions or the density matrix the evolution law
becomes a linear equation, generalizing the Schrödinger or von-Neumann equation.
In general, no simple evolution law can be formulated in terms of the time-local
probabilities at a given time $t$ alone. Similar to quantum mechanics, the
time-local probabilities are bilinears of the wave functions. They correspond to
the diagonal elements of the density matrix. All these concepts follow from the
basic laws of classical statistics for the overall system without further
assumptions. They reflect the embedding of the time-local subsystem into the
overall probabilistic system. Similarly, the quantum rules for the expectation
value of observables in terms of operators acting on wave functions or the
density matrix follow from the laws of classical statistics.

These findings open the possibility to employ the quantum formalism with
operators and wave functions or density matrix to classical statistical systems.
As compared to the traditional approach of concentrating on the time-local
probability distribution at a given time $t$, the use of wave functions or
probability amplitudes offers three important advantages:
\begin{inlinelist}
\item A simple general time evolution law follows from the overall probability
distribution if the latter has suitable locality properties in time. This
includes settings with discrete variables for which it is often not obvious
otherwise how to formulate time evolution laws.
A structurally simple linear evolution law holds for the wave functions or the
density matrix. Such a law cannot be formulated, in general, for the time local
probability distribution.
\item Wave functions or the density matrix permit the use of similarity
transformations or basis changes as the Fourier transform. This introduces
important observables as momentum for the description of Ising type systems for
which this possibility is usually not considered. 
\item New symmetries can be found in the quantum formalism. One can exploit the
quantum relation between symmetries and conserved quantities, realized by the
vanishing commutator between symmetry generators and the Hamiltonian. In
particular, in the presence of time-translation symmetry the energy is
conserved, as expressed by a time-independent Hamiltonian in the quantum
formalism. Again, the concept of a time-local energy is usually not considered
in Ising-type classical statistical systems.
\end{inlinelist}

Quantum mechanics corresponds to a particular type of time-local subsystems. It
is characterized by an evolution law for which no probabilistic information is
lost as time increases. The evolution of wave functions or the density matrix is
unitary or, more general, orthogonal in a setting without a complex structure.
Setting boundary conditions in the infinite past, the quantum subsystems are
naturally selected by the evolution. While the environment of a quantum
subsystem typically approaches an equilibrium state as time increases, the
boundary information is preserved in the quantum subsystem. 
This dynamical selection of quantum subsystems predicts that quantum mechanics
is precisely valid for the ``Universe at present time''. Within time-local
physics possible modifications of quantum mechanics can only arise for the
treatment of subsystems that are not well isolated from their
environment~\cite{KOS,LIN,ZOL, Sieberer_2016}. In principle, these modifications
are computable from the precise quantum mechanics for the ``present Universe".

We have discussed several classical statistical systems that show explicitly all
the properties of quantum systems. This demonstrates that quantum mechanics does
not need any new postulates. It is conceptually a subfield of classical
probabilistic theories, realized by subsystems with particular properties. 
We emphasize that our embedding of quantum systems as subsystems of classical
probabilistic systems realizes both the often discrete measurement values of
observables, encoded in the spectrum of the associated operators, and the
quantum law for the continuous expectation values. This holds without any
additional axioms. The discreteness of possible measurement values is usually
not accounted for if one associates the wave function with a continuous
observable, rather than a carrier of the probabilistic information.
All the ``paradoxes'' of quantum mechanics find a natural explanation. The ``no
go theorems'' for an embedding of quantum mechanics into classical statistics
\cite{BELL2},~\cite{CHSH,BELL,PER,MER,KOSP,CLHO},~\cite{CLSH, CWA, STRA} do not
apply because their assumptions are not met by the quantum subsystems.
Incomplete statistics plays an important role in this context.

The perhaps most direct way of realizing quantum systems in a ``classical"
probabilistic setting employs probabilistic cellular automata. Automata or
unique jump chains have a deterministic updating from a bit configuration at $t$
to a neighboring bit configuration at the next time layer $t+\varepsilon$. The
crucial probabilistic aspect enters through a probability distribution for the
initial bit configurations. Unique jump chains or probabilistic automata obey an
evolution law for which the probabilistic information is not lost from one time
layer to the next one. The time-local subsystems of unique jump chains are
discrete quantum systems. Their time evolution is described by a unitary
evolution operator for each discrete time step. In the presence of a locality
property for the cells of a cellular automaton these quantum systems are quantum
field theories for fermions. 

In the opposite direction discrete quantum mechanics with a discrete spectrum of
the Hamiltonian can be thought as a collection of clocks, provided the evolution
of the system is periodic. One can construct unique jump chains that precisely
reproduce the unitary evolution operator of these quantum systems. This does not
imply that unique jump chains are the only possibility for classical systems to
realize quantum systems as subsystems. It suggests, however, that unique jump
chains are a rather simple and very versatile way to implement quantum
subsystems. If, in contrast to our view, one interprets probabilities not as
fundamental but rather as a lack of knowledge (observer probabilities) the
evolution of automata opens the road for a deterministic view of the
world~\cite{TH, TH2, HOOFT2, HOOFT3, HOOFT4, hooft2023ontological}.

Our investigation of probabilistic cellular automata has revealed the presence
and usefulness of statistical local observables. A prime example is the momentum
observable. Statistical observables measure properties of the probabilistic
information, similar to temperature in classical statistical thermal equilibrium
systems. They do not take fixed values for the bit configurations of the
time-local subsystem. One may ask if there is a fundamental distinction between
``ontological observables" that take sharp values for the bit-configurations at
a given time, and statistical observables for which this is not the case. For
our simple examples the fermionic occupation numbers in position space are
ontological observables, while occupation numbers in momentum space are
statistical observables. In view of the simple mapping by a change of basis one
may have doubts concerning a fundamental distinction between the two sets of
occupation numbers. In any case, for a given choice of variables for the
time-local subsystem the simultaneous presence of ``ontological" and statistical
observables is one of the main sources of non-commuting operators and possible
violations of Bell's inequalities. 

In a certain sense our approach brings quantum physics close to classical
physics. Many connections and analogies between the quantum and classical
concepts have been discussed in the
past~\cite{KOP,VNE,WIG,MOJ,GORE,MAMA,MAU,GOMAS,NKO,VOL, volovich2009time,
Nikoli__2006, Nikolic__2007, Gozzi_2009}. 
In our setting both classical physics and quantum physics are formulated on a
fundamental level by the same basic concept based on fundamental probabilism.
This only employs classical probabilities. Classical and quantum systems differ
only by describing different types of time-local subsystems. For quantum
subsystems the classical limits are the same as usual, both for decoherence or
the measurement process~\cite{ZEH,JZ,ZUR,JZKG} and for the limit where the
action is large as compared to $\hbar$. (Recall that $\hbar$ is only a
conversion factor for units and set to one in our conventions.)

While the present work settles the conceptual foundations of quantum mechanics,
important steps remain to be done. So far we have only discussed the naive
continuum limit of the discrete fermionic quantum field theories which are
equivalent to probabilistic cellular automata. This reflects the true time
continuum limit for the case of massless free fermions. In the presence of
interactions the establishment of the true continuum limit remains an important
task. This concerns, in particular, the emergence of the continuous Lorentz
symmetry and the issue of spontaneous symmetry breaking with the possible
generation of particle masses. The true continuum limit requires the
renormalization program for lattice quantum field theories.

We have only given very simple examples for quantum subsystems or quantum field
theories. What remains is the construction of an overall probability
distribution that accounts for realistic quantum field theories as the standard
model of particle physics or its extension by quantum gravity. The concept of
time discussed here has to be extended to spacetime, and the concept of
particles needs to be extended to four-dimensional models with interactions. No
direct conceptual obstacles are visible for these tasks. For example, it has
been demonstrated that local gauge symmetries can be realized for probabilistic
cellular automata with suitable updating rules~\cite{FPCA}. Bosons can be
described as composites of fermions, similar to mesons in the theory of strong
interactions. A probabilistic cellular automaton for spinor gravity in four
dimensions~\cite{Wetterich:2022zql} suggests that diffeomorphism symmetry can
arise in the continuum limit of a discrete setting, starting from lattice
diffeomorphism invariance~\cite{Wetterich:2012np}. Still, only an explicit
construction can settle the question if a fundamental theory of the world can be
formulated along the concepts discussed in the present work. It is conceivable
that the requirement of existence of an overall probability distribution
constitutes a selection criterion among many possible quantum field theories.

\textbf{Acknowledgement:}\newline
Part of this work was supported by the DFG collaborative research center SFB
1225 ISOQUANT and by the DFG excellence cluster ``STRUCTURES''. The author
thanks D.\ Sexty, C.\ Pehle, K.\ Meier and M.\ Oberthaler for collaboration on
particular topics. He thanks T.\ Budde, M.\ Corona, B.\ Kellers, H.\ Koeper, A.\
Simon and K.\ Wolz for typing the manuscript.

\numberwithin{equation}{section}

\begin{appendices}
\section{Matrix chains}\label{app:matrix chains}

Local chains are not the only possibility for probability distributions with a locality property. In this appendix we consider ``matrix chains''.  In this case the local factors $\cK(m)$ of the local chain are replaced by real $n\times n$-matrices $\hat{\cK}(m)$, with elements depending on spins at sites $m$ and $m+1$. For the sake of clarity we often use hats for matrices. We do, however, not employ this notation systematically. We define
\begin{equation}\label{eq:MC1}
\hat{\cW} = \hat{\cK} (\cM - 1)\, \hat{\cK}(\cM - 2) \dots\,  \hat{\cK}(1)\, 
\hat{\cK}(0)\, \hat{\cB}
\end{equation}
by matrix multiplication, where $\hat{\cK} (m)$ is a function of the two neighboring spins $s(m+1)$ and $s(m)$. The order of the matrices is such that larger $m$ is on the left. The boundary matrix $\hat{\cB}$ depends on initial and final spins $s_\text{in}$ and $s_\text{f}$. The weight function is a scalar quantity as $\tr(\hat{A})$ or $\det(\hat{A})$. We will see in sect.\,\ref{sec:matrix_chains_as_subsystems_of_local_chains} that matrix chains arise as subsystems of local chains.
Matrix chains are also an example how complex numbers can appear in classical statistics.

\paragraph*{Probability distribution}

For the definition of a probability distribution,
\begin{equation}\label{eq:MC2}
p[s] = Z^{-1}\, w[s] \, , \quad Z = \int \cD s \, w[s]
\end{equation}
we need to construct a scalar $w[s]$ from $\hat{\cW}$, with $w[s] \geq 0$ for all spin configurations. We use the trace and define
\begin{equation}\label{eq:MC4}
w[s] = \tr\, \hat{\cW} [s]\, .
\end{equation}
The positivity of $w[s]$ imposes restrictions on $\hat{\cW}[s]$ and therefore on the possible choices for $\hat{\cK}(m)$. We further require that for at least one configuration $w[s]$ differs from zero, such that $Z > 0$. Even though formulated in terms of matrices
as an intermediate step, the weight distribution is at the end only a function of the Ising spins $s_\gamma(m)$. As such, it does not 
differ from other possible general functions of Ising spins.

The locality property of the probability distribution follows from the fact that each matrix $\hat{\cK}(m)$ only connects spins at two neighboring sites on the chain, e.g. $s_\gamma (m+1)$ and $s_\gamma (m)$. It is the analogue of next-neighbor interactions for the generalized Ising models. From the point of view of $(n\times n)$-matrices $\hat{\cK}(m)$, the local chains in sect. \ref{sec:local_chains} correspond to $n=1$. The particular organization of the locality property in terms of more general matrices generates new possibilities for local observables. Expectation values for such an extended set of observables may be defined by inserting matrices $\hat\cA(m)=\hat\cA(s_\gamma(m))$ into the claim~\eqref{eq:MC1} between the factors $\hat\cK(m)$ and $\hat\cK(m-1)$. This may at first sight not seem very natural from the point of view of a classical probability distribution. We will see in sect.~\ref{sec:matrix_chains_as_subsystems_of_local_chains} how this structure arises naturally if matrix chains appear as subsystems of local chains.

\paragraph*{Complex structure}

A particular form of matrix chains can be mapped to local chains with complex local factors. Let us consider a particular form of $(2\times 2)$-matrices $\hat{\cK}(m)$, given by
\begin{equation}\label{eq:MC5}
\hat{\cK}(m) = \begin{pmatrix}
\cK_R (m) & - \cK_I(m) \\
\cK_I (m) & \cK_R (m)
\end{pmatrix}
= \cK_R (m)\, \Id + \cK_I (m)\, I\, ,
\end{equation}
with 
\begin{equation}\label{eq:MC6}
I = \begin{pmatrix}
0 & -1 \\ 1 & 0
\end{pmatrix}
\, , \quad I^2 = - \Id\, .
\end{equation}
Similarly, we take for the boundary matrix
\begin{equation}\label{eq:MC7}
\cB = \cB_R \, \Id + \cB_I\, I\, .
\end{equation}

The $(2\times 2)$-matrices \eqref{eq:MC5}, \eqref{eq:MC7} can be mapped to complex numbers $\tilde{\cK}(m)$ and $\tilde{\cB}(m)$,
\begin{align}\label{eq:MC7A}
\tilde{\cK}(m) = \cK_R (m) + \im\, \cK_I (m)\, , \\ \notag
\tilde{\cB} (m) = \cB_R (m) + \im\, \cB_I (m)\, .
\end{align}
Consider two matrices $\hat{\cK}_1$ and $\hat{\cK}_2$ that are mapped to the complex numbers $\tilde{\cK}_1$ and $\tilde{\cK}_2$, respectively. The matrix multiplication
\begin{equation}\label{eq:MC8}
\hat{\cK}_1 \hat{\cK}_2 = \hat{\cK}_2 \hat{\cK}_1 = \hat{\cK}_3
\end{equation}
is mapped to the multiplication of complex numbers
\begin{equation}\label{eq:MC9}
\tilde{\cK}_3 = \tilde{\cK}_1 \tilde{\cK}_2\, .
\end{equation}
Indeed, one has
\begin{equation}\label{eq:MC10}
\hat{\cK}_3 = \begin{pmatrix}
\cK_{1R} \cK_{2R} - \cK_{1I} \cK_{2I} & -(\cK_{1R} \cK_{2I} + \cK_{1I} \cK_{2R}) \\
\cK_{1R} \cK_{2I} + \cK_{1I} \cK_{2R} & \cK_{1R} \cK_{2R} - \cK_{1I} \cK_{2I}
\end{pmatrix}
\end{equation}
and
\begin{equation}\label{eq:MC11}
\tilde{\cK}_3 = \cK_{1R} \cK_{2R} - \cK_{1I} \cK_{2I} + \im\, (\cK_{1R} \cK_{2I} + \cK_{1I} \cK_{2R})\, ,
\end{equation}
such that $\hat{\cK}_3$ is mapped to $\tilde{\cK}_3$.

On the level of the $2\times 2$-matrices we can define a complex structure by the matrix
\begin{equation}\label{eq:MC11A}
K_c = \begin{pmatrix}
1 & 0 \\ 0 & -1
\end{pmatrix}
\, ,
\end{equation}
together with the matrix $I$ in eq.~\eqref{eq:MC2}. They obey
\begin{equation}\label{eq:MC12}
K_c^2 = 1\, , \quad I^2 = -1\, , \quad \{ K_c, \, I \} = 0\, .
\end{equation} 
Matrices of the type \eqref{eq:MC5}, \eqref{eq:MC7} are compatible with this complex structure. The map to complex numbers $\hat{\cK} \to \tilde{\cK}$ translates $K_c$ to complex conjugation and $I$ to multiplication by $\im$, according to
\begin{align}\label{eq:MC13}
K_c \, \hat{\cK}\, K_c & \rightarrow \tilde{\cK}^*\, , \notag \\
\hat{\cK}\, I = I \,\hat{\cK} & \rightarrow \im\, \tilde{\cK}\, .
\end{align}

\paragraph*{Complex local chains}

We can write  the complex number $\tilde{\cK}(m)$ as
\begin{equation}\label{eq:MC18}
\tilde{\cK} (m) = \exp \{  -\cL (m) \}\, ,
\end{equation}
with complex $\cL(m)$. Here $\tilde{\cK} (m) = 0$ corresponds to the limit where the real part of $\cL (m)$ goes to positive infinity, similar to the unique jump chains in sect.~\ref{sec:unique_jump_chains}. With a complex action $\cS$,
\begin{equation}\label{eq:MC19}
\cS = \sum_{m=0}^{\cM - 1} \cL (m)\, ,
\end{equation}
we can define a complex chain by
\begin{equation}\label{eq:MC20}
w[s] = 2\,\text{Re} \big[ \text{e}^{-\cS} \, \tilde{\cB} \big] \, .
\end{equation}
This corresponds to the real matrix chain with $n=2$ and matrix structure \eqref{eq:MC5}, \eqref{eq:MC7}. The definition \eqref{eq:MC4}, $w = \tr (\hat{\cW})$, projects indeed on the real part
\begin{equation}\label{eq:MC21}
\tr \begin{pmatrix}
\cW_R & - \cW_I \\ \cW_I &  \cW_R
\end{pmatrix}
= 2\,\cW_R\,.
\end{equation}

We conclude that the weight function can be written in terms of complex numbers, with a complex action $\cS$. This amounts to
a matrix chain with real $2 \times 2$ matrices that have the particular structure (3.1.69).
More generally, real matrix chains for even $n$ and $\hat{\cK} = \cK_R + \cK_I \, I$, $\hat{\cB} = \cB_R + \cB_I\, I$ can be reformulated as complex matrix chains, with complex $(n/2 \times n/2)$-matrices $\tilde{\cK} = \cK_R + \im\, \cK_I$. (Here $\cK_R$ and $\cK_I$ are real $(n/2\times n/2)$-matrices, similar for $\cB_R$ and $\cB_I$, and $I$ involves in eq.~\eqref{eq:MC6} the unit $(n/2 \times n/2)$-matrix at the place of $1$.)

An interesting case could be purely imaginary $\cS = -\im\, \cS_M$ with real $\cS_M$,
\begin{equation}\label{eq:MC22}
w[s] = 2\,\text{Re}\big[ \text{e}^{\im\, \cS_M} \, \cB \big]\, .
\end{equation}
This resembles Feynman's path integral for quantum mechanics. Care has to be taken, however, for the positivity of $w[s]$. A general $\cS_M$ will not correspond to positive $w[s]$. For the complex chain~\eqref{eq:MC22} one has $\tilde{\cK}(m) = \exp ( \im\, \tilde{\cL}_M(m))$ or
\begin{equation}\label{eq:MC23}
\cK_R (m) = \cos \left( \tilde{\cL}_M (m) \right) \, ,\quad \cK_I (m) = \sin 
\left( \tilde{\cL}_M(m) \right) \, .
\end{equation}
Thus the real $2\times2$-matrix $\hat\cK(m)$ typically has both positive and negative elements.

\paragraph*{Determinant chains}

Instead of the definition \eqref{eq:MC4} we could also define the weight distribution by 
\begin{equation}\label{eq:MC3}
w_{\det} [s] = \det \hat{\cA} [s] = \prod_{m=0}^{\cM - 1} \det \hat{\cK} (m)\, \det\hat{\cB}\, .
\end{equation}
This is a product of scalars similar to eq.~\eqref{eq:LC4}, with $\cK(m)$ and $\cB$ replaced by $\det \hat{\cK}(m)$ and $\det \hat{\cB}$. It is therefore no new structure. For $\det \hat{\cK} (m) \geq 0$ we recover generalized Ising models. For the special case $n=2$ and matrices $\hat{\cK}(m)$ of the type \eqref{eq:MC5}, the determinant $\det \hat{\cK}$ maps to the absolute square of the associated complex number $\tilde{\cK}$
\begin{equation}\label{eq:MC14}
\det \hat{\cK} \; \rightarrow\; |\tilde{\cK}|^2 = \cK_R^2 + \cK_I^2 \, .
\end{equation}
With
\begin{equation}\label{eq:MC15}
w_{\det} = \prod_{m=0}^{\cM - 1} |\tilde{\cK}(m)|^2 \, |\tilde{\cB}|^2
\end{equation}
the positivity of $w_{\det}$ is obvious. Similarly, one maps
\begin{equation}\label{eq:MC16}
\prod_{m=0}^{\cM - 1} \hat{\cK} (m)\, \hat{\cB} \;\rightarrow\; \prod_{m=0}^{\cM - 1} \tilde{\cK}(m) \, \tilde{B}\, ,
\end{equation}
and $w_{\det}$ is the absolute square of a complex number
\begin{equation}\label{eq:MC17}
w_{\det} = \Big\vert \prod_{m=0}^{\cM - 1} \tilde{\cK}(m)\, \cB \Big\vert^2\, .
\end{equation}
While $w_{\det}$ has the structure of a generalized Ising model, it is sometimes convenient to use a formulation with matrices for which the sign of $\cK_R$ and $\cK_I$ is not restricted. Since determinant chains are a particular form of local chains, we subsume them to this category and reserve the name ``matrix chain'' to the definition \eqref{eq:MC4}.

\paragraph*{Transfer matrix for matrix chains}

The general strategy for the computation of transfer matrices in sect.~\ref{sec:transfer_matrix} can be extended to matrix chains. Each element in the matrix $\hat{\cK} (m)$ is a function of the occupation numbers at sites $m+1$ and $m$, for which we can employ the double expansion in terms of the basis functions in the occupation number basis. For the example \eqref{eq:MC5} the local $(2\times 2)$-matrix $\hat{\cK} (m)$ reads
\begin{equation}\label{eq:TS37}
\hat{\cK} (m) = \begin{pmatrix}
(\hat{T}_R (m))_{\tau\rho}  & - (\hat{T}_I (m) )_{\tau\rho} \\
(\hat{T}_I (m) )_{\tau\rho} & (\hat{T}_R (m) )_{\tau\rho}
\end{pmatrix}
h_\tau(m+1)\, h_\rho (m) \, ,
\end{equation}
where $(\hat{T}_R (m) )_{\tau\rho}$ and $(\hat{T}_I (m) )_{\tau\rho}$ are the expansion coefficients for $\cK_R(m)$ and $\cK_I (m)$, respectively. This extends to arbitrary $(n\times n)$-matrices.

We may define an overall transfer matrix $\hat{T}$ such that
\begin{equation}\label{eq:TS38}
\hat{\cK} (m) = \hat{T}_{\tau\rho} (m)\, h_\tau (m+1) \, h_\rho (m)\, .
\end{equation}
For the particular setting \eqref{eq:TS37} one has
\begin{equation}\label{eq:TS39}
\hat{T} = \begin{pmatrix}
\hat{T}_R & - \hat{T}_I \\
\hat{T}_I & \hat{T}_R
\end{pmatrix}\, .
\end{equation}
More in detail, writing the indices $\alpha$, $\beta$ of the matrix $\hat{\cK} (m)$ explicitly, one has
\begin{equation}\label{eq:TS40}
\hat{\cK}_{\alpha\beta} (m) = \hat{T}_{\alpha\tau,\, \beta\rho} (m) \, 
h_\tau(m+1)\, h_\rho (m)\, ,
\end{equation}
with
\begin{equation}\label{eq:TS41}
\hat{T}_{\alpha\tau,\, \beta\rho} = \left( \hat{T}_{\alpha\beta} \right)_{\tau\rho}
\end{equation}
the expansion coefficients of the matrix element $\hat{\cK}_{\alpha\beta}$. If $\tau$, $\rho$ are indices of $(2^M\times 2^M)$-matrices and $\alpha$, $\beta$ indices of $(n\times n)$-matrices, we can interpret the double indices $(\alpha\tau)$ and $(\beta\rho)$ as indices of a $(n\cdot 2^M \times n\cdot 2^M)$-matrix.

The multiplication rule \eqref{eq:TS16} extends to the matrix chains as
\begin{align}\label{eq:TS42}
&\int \cD n(m+1)\, \cK_{\alpha\gamma}(m+1)\, \cK_{\gamma\beta} (m) \notag \\
& \quad = \hat{T}_{\alpha\tau,\, \gamma\sigma} (m+1)\, 
\hat{T}_{\gamma\sigma,\, \beta \rho} (m) \, h_\tau (m+2) \, h_\rho (m)\, .
\end{align}
One encounters a matrix multiplication of the extended transfer matrices. For $(2\times 2)$-matrices admitting a complex structure of the type \eqref{eq:MC5} the multiplication rule \eqref{eq:TS42} is compatible with the complex structure. In this case one has
\begin{align}\label{eq:TS43}
& \hat{T}_{1\tau,\, 1\rho} = \hat{T}_{2\tau,\, 2\rho} = (\hat{T}_R)_{\tau\rho}\, ,   
\notag \\
& \hat{T}_{1\tau,\, 2\rho} = - \hat{T}_{2\tau,\, 1\rho} = - (\hat{T}_I)_{\tau\rho}\, , 
\end{align}
which can be mapped to a complex $(2^{\cM} \times 2^{\cM})$-matrix
\begin{equation}\label{eq:TS44}
\tilde{T}_{\tau\rho} = (\hat{T}_R)_{\tau\rho} + \im\,(\hat{T}_I)_{\tau\rho}\, .
\end{equation}
The matrix multiplication in eq.~\eqref{eq:TS42} is mapped to the complex matrix multiplication $\tilde{T}(m+1)\, \tilde{T}(m)$. The formula \eqref{eq:TS21} for the partition function remains valid for matrix chains, involving now the extended transfer matrices or complex transfer matrices.


We can  generalize the expressions~\eqref{eq:TS46},~\eqref{eq:TS47} to matrix chains for which we employ the product
\begin{align}\label{eq:TS51}
& \hat{\cK}_{\alpha\gamma} (m+1)\, \hat{\cK}_{\gamma\beta} (m) = \notag \\
& \quad\quad \sum_{\tau, \rho, \sigma} \,  \sum_\gamma \, \hat{T}_{\alpha\tau,\,
	\gamma\sigma} (m+1)\, \hat{T}_{\gamma \sigma,\, \beta \rho} (m) \notag \\
& \quad\quad\qquad \times h_\tau (m+2)\, h_\sigma (m+1) \, h_\rho (m)\, .
\end{align}
Extension to multiple factors yields for the weight function $w[n]$ the formula \eqref{eq:TS46}. The coefficients are now given by
\begin{align}\label{eq:TS52}
& w_{\rho_0 \rho_1 \cdots \rho_{\cM}} = \sum_{\alpha_0,\, \dots,\, \alpha_{\cM}} \, 
\hat{T}_{\alpha_{\cM}\, \rho_{\cM}\, \alpha_{\cM - 1}\, \rho_{\cM - 1}} (\cM -1) \cdots \notag \\
& \qquad \times
\hat{T}_{\alpha_2 \rho_2 ,\,\alpha_1 \rho_1} (1) \, 
\hat{T}_{\alpha_1 \rho_1,\, \alpha_0 \rho_0} (0) \, 
\hat{B}_{\alpha_0 \rho_0,\, \alpha_{\cM} \rho_{\cM}}\, .
\end{align}
This amounts to matrix multiplication and a trace in the indices $\alpha_m$, but not for the indices $\rho_m$.


\paragraph*{Classical wave functions for matrix chains}

The construction of classical wave functions for matrix chains proceeds in complete analogy to local chains. The pure state boundary conditions is now given by
\begin{equation}\label{eq:CW12}
\hat{\cB}_{\alpha \beta} (n_{in},\, n_f) = \tilde{f}_{in,\, \alpha} (n_{in})\, 
\bar{f}_{f,\, \beta} (n_f)\, ,
\end{equation}
with initial and final boundary conditions given by $n$-component vectors $\tilde{f}_{in}$ and $\bar{f}_f$. We choose again a normalization $Z=1$. The classical wave function is now an $n$-component vector, with index $\alpha \equiv \alpha_m$,
\begin{align}\label{eq:CW13}
& \tilde{f}_\alpha (m) = \prod_{m' = 0}^{m-1} \,\int \cD n(m') \, \hat{\cK}_{\alpha,\, \alpha_{m-1}} (m-1) \notag \\
& \qquad \times \hat{\cK}_{\alpha_{m-1},\, \alpha_{m-2}} (m-2) \cdots \, \hat{\cK}_{\alpha_1,\, \alpha_0}(0)\, \tilde{f}_{in,\, \alpha_0} \, ,
\end{align}
and summation over all double indices $\alpha_0,\, \dots,\, \alpha_{m-1}$, such that eq.~\eqref{eq:CW13} involves an ordered matrix multiplication. 

Similarly, the conjugate wave function is an $n$-component vector
\begin{align}\label{eq:CW14}
\bar{f}_\alpha (m) = \prod_{m' = m+1}^{\cM} \, \int & \cD n(m') \, 
\bar{f}_{f,\, \alpha_{\cM}} \, \hat{\cK}_{\alpha_{\cM},\, 
	\alpha_{\cM - 1}} (\cM - 1) \notag \\ 
& \times \cdots \, \hat{\cK}_{\alpha_{m+1},\, \alpha} (m)\, .
\end{align}
The local probability distribution involves a sum over $\alpha$,
\begin{equation}\label{eq:CW15}
p_1 (m) = \bar{f}_\alpha (m)\, \tilde{f}_\alpha (m)\, .
\end{equation}

With local factors $\hat{\cK} (m)$ being $(n\times n)$-matrices, the evolution law involves the multiplication of a vector by a matrix
\begin{align}\label{eq:CW16}
& \tilde{f}_\alpha (m+1) = \int \cD n(m) \, \hat{\cK}_{\alpha\beta} (m) 
\tilde{f}_\beta (m)\, , \notag \\
& \bar{f}_\alpha (m-1) = \int \cD n(m) \, \bar{f}_\beta (m) \, 
\hat{\cK}_{\beta\alpha} (m-1) \, .
\end{align}
Here $\tilde f_\alpha(m+1)$ depends on the occupation numbers $n_\gamma(m+1)$, and $\bar f_\alpha(m-1)$ is a function of $n_\gamma(m-1)$.

Similar to sect.~\ref{sec:classical_wave_functions} we can express the classical wave function in the occupation number basis. In this basis the classical wave functions $\tilde{q}_{\alpha\tau}$ and $\bar{q}_{\alpha\tau}$ carry an additional index $\alpha$,
\begin{equation}\label{eq:CW17}
f_\alpha (m) = \tilde{q}_{\alpha\tau} (m)\, h_\tau\, ,\quad \bar{f}_\alpha (m) = 
\bar{q}_{\alpha\tau} (m)\, h_\tau (m)\, .
\end{equation}
The local probabilities involve a sum over $\alpha$, but not over $\tau$,
\begin{equation}\label{eq:CW18}
p_\tau (m) = \sum_\alpha \bar{q}_{\alpha\tau} (m) \, \tilde{q}_{\alpha\tau} (m)\, .
\end{equation}
We conclude that the classical wave functions for matrix chains are very similar to simple local chains $(n=1)$, except for the additional indices of all objects.

\paragraph*{Operators for observables}

For matrix chains we can compute the expectation value of a generalized local observable by a rule analogous to quantum mechanics,
\begin{equation}\label{eq:CW25}
\langle A(m) \rangle = \bar{q}_{\alpha,\, \tau} (m) \, 
\hat{A}_{\alpha\tau,\,\beta\rho} (m)\, \tilde{q}_{\beta,\,\rho} (m)\, .
\end{equation}
The operator $\hat{A}(m)$ associated to a function $A[n(m)]$ of the occupation numbers $n_\gamma(m)$ is given by
\begin{equation}\label{eq:CW26}
\left( \hat{A}(m) \right)_{\alpha\tau,\,\beta\rho} = A_\tau (m) \, \delta_{\tau\rho}
\delta_{\alpha\beta}\,.
\end{equation}
Generalized local observables are represented by more general operators.

\paragraph*{Density matrix}

The classical density matrix for matrix chains is constructed in complete analogy to local chains. For pure-state boundary conditions the density matrix is again a product of the classical wave function and the conjugate wave function
\begin{equation}\label{eq:DM53}
\rho'_{\alpha\tau,\,\beta\rho} (m) = \tilde{q}_{\alpha\tau} (m) \bar{q}_{\beta\rho} 
(m)\, .
\end{equation}
The only difference is the larger dimension of the matrix -- $\rho'$ is now an $(N\times N)$-matrix with $N = n\cdot 2^M$. For matrix chains negative signs of components of wave functions are possible, such that also some elements of $\rho'$ may be negative. For general boundary conditions the classical density matrix is constructed as a weighted sum of pure-state density matrices according to eq.~\eqref{eq:DM23}. The quantum rule for expectation values of generalized local observables is given by eq.~\eqref{eq:DM34}, now with matrices $\hat A$ and $\rho'$ in an extended space.


\paragraph*{Step evolution operator}

As compared to simple local chains the step evolution operators for matrix chains extend the possibilities. This holds, in 
particular, for unique jump chains and for Markov chains. Since matrix chains can be obtained as subsystems of local chains with
more degrees of freedom, cf. sect. \ref{sec:matrix_chains_as_subsystems_of_local_chains}, this demonstrates that subsystems permit a 
richer structure for their evolution. 

For matrix chains we choose again a normalization where $Z=1$. Also the local matrices $\hat{\cK} (m)$ are normalized such that the largest eigenvalue of the transfer matrix obeys $|\lambda| = 1$. The corresponding step evolution  operators have many things in common with the ones for local chains. They are, however, larger matrices with ``internal'' indices $\alpha,\, \beta = 1,\, \dots,\, n$, such that the elements are labeled by double-indices $\hat{S}_{\alpha\tau,\, \beta\rho}$. The evolution of the classical wave functions is the same as for local chains, except for the higher number of components of the wave function $n\cdot 2^M$. Unique jump operators have again one element equal to $1$ in each column and row, and all other elements $0$.

For the evolution of the local probability distribution the unique jump operators of matrix chains offer additional possibilities. This can be seen for $(2\times 2)$-matrices depending on a single bit or occupation number, with local matrix factors $\hat{\cK}_{\alpha\beta} (n(m+1),\, n(m))$. Consider the unique jump operator
\begin{equation}\label{eq:SE28A}
\hat{S} = \begin{pmatrix}
0 & 1 & 0 & 0 \\
0 & 0 & 0 & 1 \\
1 & 0 & 0 & 0 \\
0 & 0 & 1 & 0 \\
\end{pmatrix}\, .
\end{equation}
In our basis the wave functions read $\tilde{q}_1 = \tilde{q}_{11}$, $\tilde{q}_2 = \tilde{q}_{12}$, $\tilde{q}_3 = \tilde{q}_{21}$, $\tilde{q}_4 = \tilde{q}_{22}$, where the notation with two indices refers to $\tilde{q}_{\alpha\tau}$. The local probability for finding the occupation number $n=1$, is given at each $m$ by
\begin{equation}\label{eq:SE28B}
p_1 (m) = \bar{q}_1 \tilde{q}_1 + \bar{q}_3 \tilde{q}_3\, .
\end{equation}

With shorthands $\tilde{q}'_\tau = q_\tau (m+1)$, $q_\tau = q_\tau (m)$, the step evolution operator \eqref{eq:SE28A} induces the map
\begin{equation}\label{eq:SE28C}
\tilde{q}'_1 = \tilde{q}_2\, , \quad \tilde{q}'_2 = \tilde{q}_4\, , \quad 
\tilde{q}'_3 = \tilde{q}_1\, , \quad \tilde{q}'_4 = \tilde{q}_3\, ,
\end{equation}
and similar for $\bar{q}$. This implies
\begin{align}\label{eq:SE28D}
& p_1 (m+1) = \bar{q}_1 \tilde{q}_1 + \bar{q}_2 \tilde{q}_2\, ,\quad  p_1 (m+2) = 
\bar{q}_2 \tilde{q}_2 + \bar{q}_4 \tilde{q}_4\, , \notag \\
& p_1 (m+3) = \bar{q}_3 \tilde{q}_3 + \tilde{q}_4 \tilde{q}_4\, , \quad
p_1 (m+4) = p_1 (m)\, .    
\end{align}
If we take as an example $\bar{q}_1 \tilde{q}_1 = 5/16$, $\bar{q}_2 \tilde{q}_2 = 3/16$, $\bar{q}_3 \tilde{q}_3 = 7/16$, $\bar{q}_4 \tilde{q}_4 = 1/16$ we find for the local probabilities 
\begin{align}\label{eq:SE28E}
p_1 (m) =&\,\frac34\, , \quad p_1 (m+1) =\frac12\, , \notag \\
p_1 (m+2) =&\,\frac14\, , \quad p_1 (m+3) =\frac12\, .
\end{align}

For a second example with $\bar q_1\tilde q_1=1/4$, $\bar q_2\tilde q_2=1/8$, $\bar q_3\tilde q_3=1/2$, $\bar q_4\tilde q_4=1/8$ one has
\begin{align}
\label{A48A}
p_1(m)=&\,\frac34\ ,\quad p_1(m+1)=\frac38\ ,\nonumber\\
p_1(m+2)=&\,\frac14\ ,\quad p_1(m+3)=\frac58\ .
\end{align}
There is no rule how the probability $p_1(m+1)$ can be computed from the probability $p_1(m)$. The two examples have the same $p_1(m)$, but different $p_1(m+1)$. In particular, the evolution of the time local probabilities $p_1(m)$ and $(1-p_1(m))$ to find $n(m)=1$ or $n(m)=0$ is no longer given by a unique jump chain.

The evolution~\eqref{eq:SE28C} has period four, corresponding to $\hat{S}^4 = 1$. In contrast, for a local chain with a single bit the only unique jump operators are the identity and the exchange $p_1 \leftrightarrow p_2 = 1 - p_1$, which has period two. For matrix chains with large $n$ and matrix elements depending on a single bit $n(m+1)$ and $n(m)$, the maximal period for unique jump operators is $2n$. Longer periods can be achieved if the matrices differ for different sites $m$.

\paragraph*{Evolution of time-local probabilities}

For the evolution of the local probabilities $p_\tau(m)$ the difference between matrix chains and local chains concerns the relations of the step evolution operators to the overall probability distribution and the local probability distribution. They involve additional summations over the index $\alpha$, as in eq.~\eqref{eq:CW18}. The evolution law for the local probabilities reads for matrix chains (no sum over $\tau$)
\begin{align}\label{eq:SE29}
p_\tau (m+1) &= \sum_\alpha \bar{q}_{\alpha\tau} (m+1)\, \tilde{q}_{\alpha\tau} (m+1) \, \notag \\
&= \sum_{\alpha,\beta, \gamma} \, \sum_{\sigma, \rho} \bar{q}_{\beta\sigma} (m) 
\hat{S}^{-1}_{\beta\sigma,\, \alpha\tau} (m) \, \hat{S}_{\alpha\tau,\, \gamma\rho}
(m)\, \tilde{q}_{\gamma\rho} (m)\, .
\end{align}

In particular, Markov chains are realized if the step evolution operator obeys for all $\tau$, $\rho$, $\sigma$ and $\beta$, $\gamma$ the condition
\begin{align}\label{eq:SE30}
\sum_\alpha \hat{S}^{-1}_{\beta\sigma,\, \alpha\tau} \,
\hat{S}_{\alpha\tau,\, \gamma\rho} = W_{\tau\rho} \delta_{\rho\sigma}
\delta_{\beta\gamma}\, .
\end{align}
The coefficients
\begin{equation}\label{eq:SE31}
W_{\tau\rho} = \sum_\alpha \hat{S}^{-1}_{\gamma\rho,\, \alpha\tau}\, 
\hat{S}_{\alpha\tau,\, \gamma\rho}\, ,
\end{equation}
must be the same for all $\gamma$ according to the condition \eqref{eq:SE30}. In this case they are the transition probabilities in the Markov chain. Eq.~\eqref{eq:SE31} is a weaker condition as compared to eq.\,\eqref{eq:SE9} since an additional summation over $\alpha$ is performed.
The possibilities for Markov chains are extended beyond the rather restricted possibilities \eqref{eq:SE19}. This extends to situations
where the matrix elements depend on more than one local spin, $M \geq 2$.
\section[Positivity of overall probability distribution]{Positivity of overall probability\\distribution}
\label{app:positivity_of_overall_probability_distribution}

The condition that all probabilities $p[n]$ are positive or zero is a  central cornerstone of classical statistics and the probabilistic approach to physics that we take here. In turn, this implies that the corresponding weight function $w[n]$ has to be positive or zero for any configuration of occupation numbers. For generalized Ising chains and unique jump chains this is obviously obeyed since the factors $\mathscr{K}(m)$ are positive for arbitrary $n(m)$ and $n(m+1)$ by definition. It is then sufficient that the boundary term $\mathscr{B}$ is also positive. For unique jump chains the local factors are $\delta$-functions that only take the values one and zero.

Positivity of all local factors $\mathscr{K}(m)$, or all matrix elements of the local matrices $\hat{\mathscr{K}}(m)$ for matrix chains, is sufficient for establishing the positivity of the weight function $w[n]$ (for the appropriate boundary conditions). This property is, however, not necessary. It is very simple to construct examples of positive $w[n]$ for which not all $\mathscr{K}(m)$ are positive. For example, one could multiply an even number of local factors by $-1$. We would like to have some general criteria for a positive weight function $w[n]$. The transfer matrix in the occupation number basis will be a useful tool for this purpose.
In turn, the requirement of a positive weight distribution places important restrictions on the properties of the transfer matrix.

Consider a particular site $m$ and define
\begin{align}
&f_{\rho_0\rho_1 \dots \rho_{m-1}\rho_{m+1}\dots \rho_\cM} \\
&\qquad= \hat{T}_{\rho_\cM\rho_{\cM-1}}(\cM-1) \dotsb \hat{T}_{\rho_{m+2}\rho_{m+1}}(m+1) \nonumber \\
\nonumber
&\qquad \times\ \hat{T}_{\rho_{m-1}\rho_{m-2}}(m-2)\dotsb \hat{T}_{\rho_1\rho_0}(0) \hat{B}_{\rho_0\rho_\cM},
\end{align}
and
\begin{equation}
g_{\rho_{m}-1\rho_m\rho_{m+1}} = \hat{T}_{\rho_{m+1}\rho_{m}}(m)\hat{T}_{\rho_m\rho_{m-1}}(m-1),
\end{equation}
such that the coefficients of the weight distribution \eqref{eq:TS47} can be written as a product of two factors
\begin{equation}
w_{\rho_0\rho_1\dots\rho_\cM} = f_{\rho_0\dots\rho_{m-1}\rho_{m+1} ... \rho_{\cM}} g_{\rho_{m-1}\rho_{m}\rho_{m+1}}.
\end{equation}
For any given values of the indices except $\rho_m$, i.e. $(\rho_0,\dots,\rho_{m-1},\rho_{m+1},\rho_\cM),$ the factor $f$ is independent of $\rho_m$. It can be positive or negative or zero. For $f>0$ one needs $g\ge 0$ for all values of $\rho_m$, and $f<0$ requires $g\le 0$ for all values of $\rho_m$.
Consider now a pair $(\rho_{m-1},\rho_{m+1})$ for which $f$ differs from zero for at least one combination of the indices except $\rho_{m-1}$ and $\rho_{m+1}$, and $g$ differs from zero for at least one $\rho_m$. Then $g_{\rho_{m-1}\rho_m\rho_{m+1}}$ needs to have the same sign for all values of $\rho_m$ for which it does not vanish. 
Otherwise one would obtain two non-zero elements $w_{\rho_0...\rho_\cM}$ with opposite sign, which contradicts positivity.
For all $\rho_m$ for which both $\hat{T}_{\rho_{m+1}\rho_{m}}(m)$ and $\hat{T}_{\rho_{m}\rho_{m-1}}(m-1)$ differ from zero, the sign of $\hat{T}_{\rho_{m+1}\rho_m}(m)$ has to be either the same  as for $\hat{T}_{\rho_{m}\rho_{m-1}}(m-1)$ for all $\rho_m$, or the signs must be opposite for all $\rho_m$.

This condition places severe restrictions on transfer matrices for which not all elements are positive semidefinite. Consider $2\times 2$ matrices that are rotations
\begin{align}
\hat{T}(m) = 
\begin{pmatrix}
\cos \psi & -\sin \psi\\ 
\sin \psi & \cos \psi
\end{pmatrix}, 
\nonumber \\
\hat{T}(m-1) = 
\begin{pmatrix}
\cos \varphi & -\sin \varphi\\ 
\sin \varphi & \cos \varphi
\end{pmatrix}.
\end{align}
One finds
\begin{align}
g_{111} = g_{222} &= \cos \psi \cos \varphi
\nonumber \\
g_{121} = g_{212} &= -\sin \psi \sin \varphi
\nonumber \\
g_{112} = -g_{221} &= -\cos \psi \sin \varphi
\nonumber \\
g_{122} = -g_{211} &= -\sin \psi \cos \varphi.
\end{align}
For $\psi = \varphi$ the same sign for $g_{111}$ and $g_{121}$ requires $\sin \varphi =0$ or $\cos \varphi =0$. (We include the value zero in the ``same sign".) Up to an overall possible minus sign these are only the unique jump operations or $\pi/2$-rotations. More possibilities could open up if $\psi$ differs from $\varphi$. For both $\sin \varphi \neq 0$, $\cos \varphi \neq 0$ the only possibility consistent with our condition is $\sin \psi = 0$ or $\cos \psi =0$.

We conclude that the requirement of positivity of the overall probability or weight distribution imposes important constraints on the transfer matrix or step evolution operator. These constraints are, however, not strong enough to impose that all elements of the step evolution operator are positive, or that a similarity transformation exists for which all elements of the transformed step evolution operator are positive. A step evolution operator realizing a $\pi/2$-rotation in the two-component system is a $2\times 2$-matrix with eigenvalues $\pm i$. This cannot be realized by a positive $2\times 2$-matrix. Since eigenvalues do not change under similarity transformations, this extends to step evolution operators that can be mapped by a similarity transformation to a positive matrix.

The constraints on the step evolution operator concern the step evolution operator of the overall probabilistic system for which the weight distribution has to be positive. They do not apply to subsystems. We will see in sect.\,\ref{sec:subtraces} that the step evolution operator for subsystems is often complex, with a complex weight distribution for the subsystems. Also real weight distributions for subsystems can occur, but they are no longer restricted to being positive. This opens many new possibilities for the step operators of subsystems. In particular, they can realize infinitesimal rotations or unitary transformations. 
\section[Weyl complex structure for two-particle wave function]{Weyl complex structure for two-particle wave function}\label{app:complex_structure_for_two-particle_wave_function}

The Weyl complex structure for two-particle wave functions is somewhat more involved. We consider first the complex structure~\eqref{eq:463}. For states with two particles at different positions $x\neq y$ we take without loss of generality $x< y$. In the real formulation the general two-particle wave function has four components. The first component $q_{20}(x,y)$ accounts for states where both particles are of type one. It is antisymmetric, $q_{20}(y,x) = - q_{20}(x,y)$, similar to the case of a single species discussed above. For the second component $q_{02}(x,y) = - q_{02}(y,x)$ both particles are of type two. A further component $q_{11}(x,y) $ accounts for states where the particle at $x$ is of type one, and the particle at $y$ is of type two. Finally, for the fourth component $\bar{q}_{11}(x,y)$ the particle of type two sits at $x$, while the particle of type one is located at $y$. The states accounted for by $q_{11}$ and $\bar{q}_{11}$ are different since the two particles are distinguished by their type. The relation $q_{11}(y,x)=-\bar q_{11}(x,y)$ avoids double counting. Each of the four components accounts for $(\cM_2 + 1)\cM_2/2$ states related to the different positions $(x,y)$.

We may group the four components into two two-component real vectors
\begin{equation}\label{eq:495}
\hat{q}_1 = \begin{pmatrix}
q_{20} \\
\bar{q}_{11}
\end{pmatrix}\, , \quad \hat{q}_2 = 
\begin{pmatrix}
q_{02} \\
- q_{11}
\end{pmatrix} \, .
\end{equation}
On each vector the involution $K_c$ for the complex conjugation is realized by multiplication with a matrix $K_c$, while the operation corresponding to multiplication with $\im$ is performed by multiplication with $I$, with $K_c$ and $I$ given by eqs.~\eqref{eq:471}, \eqref{eq:472}. Correspondingly, we introduce complex wave functions
\begin{equation}\label{eq:496}
\psi_1 = q_{20} + \im \bar{q}_{11}\, , \quad \psi_2 = q_{02} - \im q_{11} \, ,
\end{equation}
with $K(\psi_a) = \psi^*_a$, $I(\psi_a) = \im \psi_a$. The general two-particle wave function for $x\neq y$ is given by
\begin{equation}\label{eq:497}
\hat{q} = \begin{pmatrix}
\hat{q}_1 \\
\hat{q}_2
\end{pmatrix}\, , \quad \psi = \begin{pmatrix}
\psi_1 \\
\psi_2
\end{pmatrix} \, ,
\end{equation}
in the real and complex formulation, respectively. Complex conjugation switches the sign of $q_{11}$ and $\bar{q}_{11}$, which indeed transforms the product wave function $\psi \to \psi^*$. Applying the transformation $I$ on the real components indeed leads to $\psi \to \im \psi$, demonstrating compatibility.

We have defined the complex structure such that it is compatible with the complex product for two-particle states that are products of one-particle states. Let us define two complex one-particle wave functions by
\begin{equation}\label{eq:498}
\varphi(x) = q_1 (x) + \im q_2(x)\, , \quad \varphi'(y) = q'_1 (y) + \im q'_2 (y)\, .
\end{equation}
In the real formulation, the product wave functions for two particles of type one or two are given by
\begin{align}\label{eq:499}
& q^{}_{20} (x,y) = q^{}_1(x)\, q'_1(y) - q^{}_1(y)\, q'_1(x) \, , \notag \\
& q^{}_{02} (x,y) = q^{}_2(x)\, q'_2(y) - q^{}_2(y) \, q'_2(x)\, ,
\end{align} 
while the components for one particle at $x$ and one at $y$ read for the product wave function
\begin{align}\label{eq:500}
& q^{}_{11} (x,y) = q^{}_1(x)\, q'_2(y) - q'_1(x)\, q_2(y)\, , \notag \\
& \bar{q}^{}_{11} (x,y) = q^{}_2 (x)\, q'_1(y) - q'_2(x)\, q_1(y)\, .
\end{align}
In the complex basis the product wave function is given by the complex product,
\begin{equation}\label{eq:501}
\psi(x,y) = \varphi(x)\, \varphi'(y) - \varphi(y)\, \varphi'(x)\, . 
\end{equation}
(We omit here normalization factors.)

We want to verify that this complex product is compatible with the real product wave function \eqref{eq:499}, \eqref{eq:500}, and the complex structure \eqref{eq:495}, \eqref{eq:497}. Insertion of eq.~\eqref{eq:498} into the complex product \eqref{eq:501} yields
\begin{align}\label{eq:502}
\psi(x,y) &= q_{20} (x,y) - q_{02}(x,y) + \im q_{11} (x,y) + \im \bar{q}_{11}(x,y) 
\notag \\
&= \psi_1(x,y) - \psi_2(x,y)\, .
\end{align}
We may also consider the product of $\varphi(x)$ and $\varphi'^*(y)$,
\begin{align}\label{eq:503}
\psi'(x,y) &= \varphi(x)\, \varphi'^*(y) - \varphi^*(y)\, \varphi'(x) \notag \\
&= \psi_1(x) + \psi_2(x)\, .
\end{align}
Again, the complex product is compatible with the real product wave function \eqref{eq:499}, \eqref{eq:500} and the complex structure \eqref{eq:495}, \eqref{eq:496}.

So far we have accounted for all two-particle states except for the $\cM_2+1$ states where both a particle of type one and a particle of type two are localized at the same position. This component of the two-particle wave function forms a separate class. According to our rule the wave function is purely imaginary. The multiplication with $\im$ is not defined. These particular features of the complex structure for the two-particle wave function can be related to the complex structure that can be introduced in the equivalent formulation of the system in terms of a Grassmann wave function \cite{CWFIM}. In this language the complex structure for other multi-particle states finds a more systematic description.

\section[Complex structure based on sublattices]{Complex structure\\based on sublattices}\label{app:complex_structure_based_on_sublattices}

The choice of a complex structure is, in general, not unique. As an alternative to the complex structure for Weyl fermions based on two types of fermions we discuss here a complex structure based on sublattices rather than doubling the degrees of freedom. The diagonal Ising model can be decomposed into two subsystems on two sublattices. This can be used for an example of a complex structure \cite{CWIT}. We split the two-dimensional square lattice with lattice points
\begin{equation}
(t,x) = (t_{in} + m\epsilon, x_{in} + j\epsilon)
\end{equation}
into an ``even sublattice" with $m+j$ even, and an ``odd sublattice" with $m+j$ odd. The propagation of particles on diagonals does not mix the sublattices. We can employ this observation for introducing a complex structure for the one-particle wave function $q(t,x) = q(m,j)$. A complex conjugation $K_c$ is an involution that changes the sign of $q$ for $m+j$ odd, e.g.
\begin{equation}\label{eq.456}
K_c(q(m,j)) = (-1)^{m+j} q(m,j).
\end{equation}
Correspondingly, we define ``real parts" or ``imaginary parts" of a wave function as the parts that are even or odd with respect to $K_c$, i.e.
\begin{align}
K_c q_R = q_R, && K_c q_I = - q_I.
\end{align}
In more detail, we define for even $m$ and even $j$ 
\begin{align}
q(t,x) = q_R(t,x), && q(t,x+\epsilon) = q_I(t,x).
\end{align} 
For odd $m$ and even $j$ we take
\begin{align}
q(t,x-\epsilon) = q_R(t,x-\epsilon), && q(t,x) = q_I(t,x-\epsilon).
\end{align}
The imaginary part $q_I$ corresponds to $q(m,j)$ with support on the odd sublattice and therefore changes sign under the complex conjugation $K_c$ given by eq.~\eqref{eq.456}.

A complex wave function is defined by 
\begin{equation}
\psi(t,x) = q_R(t,x) + i q_I(t,x).
\end{equation}
It has support on the even sublattice only -- we have eliminated the wave function on the odd sublattice in favor of the imaginary part of a complex wave function $\psi(t,x)$ on the even sublattice. The imaginary part of $\psi(m,j)$ is given by $q(m,j+1)$. Multiplication with $i$ corresponds to the discrete transformation $I$,
\begin{align}
q_R(t,x) \to q_I(t,x), && q_I(t,x) \to - q_R(t,x),
\end{align}
or
\begin{align}
\nonumber
\begin{rcases*}
q(m,j) \to q(m,j+1) \\
q(m,j+1) \to -q(m,j)
\end{rcases*} \text{for }m\text{ even, }j\text{ even}
\\
\begin{rcases*}
q(m,j) \to -q(m,j-1) \\
q(m,j-1) \to q(m,j)
\end{rcases*} \text{for }m\text{ odd, }j\text{ even}
\end{align}
Thus the transformation $I$ transports the wave equation on even lattice sites to the next neighbor in the $x$-direction on the odd lattice, while wave functions on odd lattice sites are transported to the next neighbor in the negative $x$-direction on the even sublattice, with an additional minus sign that accounts for $I^2 = -1$. One verifies the anticommutation realation $\{K_c,I\} = 0$.

We can write the evolution equation \eqref{eq:FP20} in the form
\begin{equation}
I\del_t q(t,x) = -I\del_x q(t,x).
\end{equation}
This translates in the complex language to a discrete Schrödinger equation
\begin{equation}
i\del_t \psi(t,x) = -i\del_x \psi(t,x).
\end{equation}
For an initial condition
\begin{equation}
\psi_p(t_{in},x) = c e^{ipx}
\end{equation}
the solution of the evolution equation is
\begin{equation}
\psi_p(t,x) = c e^{ip(x-t+t_{in})}.
\end{equation}
For real positive c one has for $m+j$ even
\begin{align}
\nonumber
q_p(m,j) = c \cos\{\epsilon p(j-m)\},
\\
q_p(m,j+1) = c \sin\{\epsilon p(j-m)\}.
\end{align}
This ``plane wave function" is constant on the diagonals, with jumps between neighboring diagonals on even and odd sublattices. It can take negative values. The possible values of the ``momentum" $p$ are discrete, as required by the periodicity in the $x$-direction,
\begin{align}
p=\frac{2\pi k}{\epsilon(\cM_2+1)}, && k \in \mathbb{Z}.
\end{align}
(Here $\cM_2$ is taken to be odd.) For a given fixed $j$ the evolution in $m$ is periodic, with the period given by the inverse momentum,
\begin{equation}
P=\frac{\pi}{\epsilon p}.
\end{equation}
The normalization factor is given by
\begin{equation}
c^2 = \frac{1}{\cM_2 -1},
\end{equation}
guaranteeing for every $m$
\begin{equation}
\sum_j q^2(m,j) = 1
\end{equation}
or, equivalently
\begin{equation}
\sum_{j'}  \psi^*(m,j')\psi(m,j') =1,
\end{equation}
with $j'$ even for $m$ even and $j'$ odd for $m$ odd.

An arbitrary one-particle wave function can be represented in terms of the plane wave functions
\begin{equation}
\psi(m,j) = \sum_p \tilde\psi(m,p)\psi_p(m,j).
\end{equation}
The coefficients $\tilde\psi(m,p)$ are the standard discrete Fourier transform of $\psi(m,j)$. 
\section[Complex fermionic operators]{Complex fermionic\\operators}\label{app:particle-hole_conjugation_for_neutral_states}

The Fourier transform to annihilation and creation operators in momentum space in sect.~\ref{sec:fourier_transform_for_cellular_automata} starts from general complex annihilation and creation operators $A(j)$, $A^\dagger(j)$ in position space. There are several scenarios how $A(j)$ and $A^\dagger(j)$ are connected to the real operators $a(j)$ and $a^\dagger(j)$ that act on bit configurations as described in sect.~\ref{sec:Fermionic_quantum_field_theory_with_interactions}. The simplest way takes real $A(j)=a(j)$, $A^\dagger(j)=a^\dagger(j)=a^T(j)$. For real $A$, $A^\dagger$ one has in Fourier space
\bel{CAC1}
a(-k)=a^*(k)\ .
\ee
One can implement a complex wave function on which the annihilation and creation operators act. A simple possibility is an additional bit which distinguishes between the real and imaginary part.

It is not necessary that all possible bit configurations correspond to multi-particle states. The multi-particle states may account only for a part of the most general wave function of the system, whereas the remaining part may be seen as a type of environment in which the particles propagate. In the functional integral for a quantum field theory many strongly fluctuating field configurations are often not interpreted in a multi-particle language. We demonstrate this issue by considering a scenario with two species of bits or two colors of fermions, with associated real annihilation operators $a_1(j)$ and $a_2(j)$ and creation operators $a_1^\dagger(j)$ amd $a_2^\dagger(j)$. As usual, the fermionic operators $a_2$, $a_2^\dagger$ anticommute with $a_1$, $a_1^\dagger$. We define complex annihilation and creation operators
\begin{align}
\label{CAC2}
A(j)=&\,\frac{1+i}{2}a_1(j)+\frac{1-i}{2}a_2^\dagger(j)\ ,\nonumber\\
A^\dagger(j)=&\,\frac{1-i}{2}a_1^\dagger(j)+\frac{1+i}{2}a_2(j)\ .
\end{align}
They obey the standard anticommutation relations
\begin{align}
\label{CAC3}
&\big\{A(j),A^\dagger(j')\big\}=\, \delta_{j,j'}\ , \nn\\
&\big\{A(j),A(j')\big\}=\big\{A^\dagger(j),A^\dagger(j')\big\}=0\ .
\end{align}
The same anticommutation relation holds if we exchange in eq.~\eqref{CAC2} $a_2\leftrightarrow a_2^\dagger$. The complex conjugate operators $A^*(j)$ and $\gl A^*\gr^\dagger(j)=A^T(j)$ are an independent set of annihilation and creation operators. The relation~\eqref{CAC1} no longer holds. The map $a_1(j)\leftrightarrow a_2^\dagger(j)$, $a_2(j)\leftrightarrow a_1^\dagger(j)$ induces the map $A(j)\leftrightarrow A^*(j)$, $A^\dagger(j)\leftrightarrow A^T(j)$. The particle-antiparticle transformation $A(j)\leftrightarrow A^\dagger(j)$ is realized by the color change $a_1(j)\leftrightarrow a_2(j)$.

The occupation number operator associated to the operators~\eqref{CAC2} is given by
\begin{align}
\label{CAC4}
\hat n(j)=&\,A^\dagger(j)A(j)\nonumber\\
=&\,\frac12\gl a_1^\dagger(j) a_1(j)+a_2(j)a_2^\dagger(j)\nn \\&+ia_2(j)a_1(j)-ia_1^\dagger(j)a_2^\dagger(j)\gr\nonumber\\
=&\,\frac12 +\frac12\gl\hat n_1(j)-\hat n_2(j)\gr \nn\\ 
&+\frac i2\gl a_2(j)a_1(j)-a_1^\dagger(j)a_2^\dagger(j)\gr\ .
\end{align}
With $\hat n^2(j)=\hat n(j)$ this operator has indeed the eigenvalues one and zero. The vacuum with positive energies for particles and antiparticles discussed in sect.~\ref{sec:Particles_and_antiparticles} has $\langle n(j)\rangle=1/2$. This is realized for an equal mean occupation number of the two species,
\bel{CAC5}
\langle n_1(j)\rangle=\langle n_2(j)\rangle\ ,
\ee
together with the property
\bel{CAC6}
\langle a_2(j)a_1(j)-a_1^\dagger(j)a_2^\dagger(j)\rangle=0\ .
\ee
The action of the creation operator $A^\dagger(j)$ either enhances $n_1$ by one unit, or lowers $n_2$ by one unit. As a result, the charge $Q(j)=n_1(j)-n_2(j)$ is enhanced by one unit. The relation $n(j)A^\dagger(j)=A^\dagger(j)$ involves the imaginary part of $n(j)$.

For the action of the fermionic operators on the complex wave function we choose a matrix representation $\vp$, in analogy to eq.~\eqref{NFD1}. States with different local charges $Q(j)$ are constructed by applying the complex operators $A(j)$ and $A^\dagger(j)$ to a reference ground state $\vp_0$. We choose $\vp_0$ real and diagonal, $\vp_0=q_0$, as given by eg.~\eqref{NFD4}. It coincides with the wave function in the real picture. The real operators $a_i(j)$, $a_i^\dagger(j)=a_i^T(j)$ act as specified by eq.~\eqref{NFD1}. Due to the factors of $i$ in eq.~\eqref{CAC2} the action of $A(j)$ and $A^\dagger(j)$ on $\vp_0=q_0$ produces a complex wave function. The reference state is an eigenstate of the local charge operator $Q(j)=\hat n_1(j)-\hat n_2(j)$,
\bel{CAC8}
Q(j)\gl\vp_0\gr=\gl\hat n_1(j)-\hat n_2(j)\gr\gl\vp_0\gr=0\ .
\ee
For eq.~\eqref{CAC4} one infers
\begin{align}
\label{CAC9}
\hat n(j)\gl\vp_0\gr=&\,\frac12\vp_0+\frac i2\Big(T_3a(j)\vp_0a^\dagger(j)-a^\dagger(j)T_3\vp_0a(j)\Big)\nonumber\\
=&\,\frac12\gl1+iT_3\gr\vp_0\ .
\end{align}
We conclude that $\vp_0$ is half-filled in the average with respect to $\hat n(j)=A^\dagger(j)A(j)$,
\bel{CAC10}
\langle\hat n(j)\rangle=\frac12\ ,
\ee
as for the state with positive energies for particles and antiparticles in sect.~\ref{sec:Particles_and_antiparticles}.

Applying on $\vp_0$ the complex creation operator yields the complex wave function
\begin{align}
\label{CAC11}
\vp_1(j)=&\,A^\dagger(j)\gl\vp_0\gr\nonumber\\
=&\,\frac12\gl a(j)\vp_0+T_3\vp_0a^\dagger(j)\gr+\frac i2\gl a(j)\vp_0-T_3\vp_0a^\dagger(j)\gr\ .
\end{align}
With
\bel{CAC11}
Q(j)=\hat n_1(j)-\hat n_2(j)\ ,\quad \big[Q(j),Q(j')\big]=0\ ,
\ee
and
\bel{CAC12}
\big[Q(j'),A^\dagger(j)\big]=A^\dagger(j)\delta_{j'j}\ ,\quad \big[Q(j'),A(j)\big]=-A(j)\delta_{j'j}\ ,
\ee
we conclude that $\vp_1(j)$ has charge $Q(j)=1$, $Q(j')=0$ for $j'\neq j$,
\bel{CAC13}
Q(j')\gl\vp_1(j)\gr=\delta_{j'j}\vp_1(j)\ .
\ee
We can interpret $\vp_1(j)$ as a sharp one-particle state (up to normalization). Defining the total charge,
\bel{CAC13A}
Q=\sum_jQ(j)\ ,
\ee
a general one-particle state with charge $Q=1$ is a linear superposition of wave functions $\vp_1(j)$. Similarly, the state $\vp_{-1}(j)=A(j)(\vp_0)$ has charge $Q(j)=-1$. A linear superposition of wave functions $\vp_{-1}(j)$ constitutes a one-particle state for a particle with total charge $Q=-1$. It can be interpreted as the antiparticle of the particle with $Q=1$.

One can construct from $\vp_0$ states with arbitrary values of the total charges $Q(j)=(1,0,-1)$ by applying suitable products of operators $A^\dagger(j)$ or $A(j)$ on $\vp_0$. They can be seen as multi-particle states. These multi-particle states are only a small part of the general wave function of the system. Indeed, the $3^M$ states generated in this way cover only part of the $4^M$ configurations of occupation numbers that form the basis of the real wave function $q$. The multi-particle states generated by applying operators $A^\dagger(j)$, $A(j)$ on $\vp_0$ are not eigenstates of the operators $\hat n(j)=A^\dagger(j)A(j)$. Therefore an extended set of states can be constructed by applying either $A(j)$, $A^\dagger(j)$, $\hat n(j)$ or $1$ at each site $j$ on the reference state $\vp_0$. This yields $4^M$ linearly independent wave functions. A general complex $2^M\times 2^M$ matrix $\vp$ can be constructed as a linear superposition of these states with complex coefficients. If we restrict the multi-particle states to the ones obtained by applying creation and annihilation operators $A^\dagger(j)$ and $A(j)$ on the reference state $\vp_0$, the multi-particle states are only particular linear combinations of the most general wave function.

More explicitly, we may start from a general complex wave function $\vp$ for the two colors of bits given as a complex $2^M\times 2^M$ matrix. We employ a basis in a direct product form
\bel{CAC14}
\vp=\vp_\beta B_\beta\ ,\quad \beta=\{\gamma(j)\}\ ,
\ee
where the index $\beta$ denotes configurations $\{\gamma(j)\}$ with $\gamma(j)$ taking the four values $(1,1)$, $(1,0)$, $(0,1)$ and $(0,0)$ for each $j$. The basis vectors $B_\beta$ are given by direct products ($J=(M-1)/2$ for odd $M$)
\bel{CAC15}
B_\beta=H_{\gamma(-J)}\otimes H_{\gamma(-J+1)}\otimes\dots\otimes H_{\gamma(J)}\ ,
\ee
with $2\times2$-matrices
\begin{align}
\label{CAC16}
H_{11}=&\,\begin{pmatrix}1&0\\0&0\end{pmatrix}\ ,\quad H_{10}=\begin{pmatrix}0&1\\0&0\end{pmatrix}\nn\\
H_{01}=&\,\begin{pmatrix}0&0\\1&0\end{pmatrix}\ ,\quad H_{00}=\begin{pmatrix}0&0\\0&1\end{pmatrix}\ .
\end{align}
With
\bel{CAC17}
\hat n_1(j)\gl B_\beta\gr=\begin{cases}B_\beta\ &\text{if}\ \gamma(j)=(1,1)\ \text{or}\ (1,0)\\
0\ &\text{if}\ \gamma(j)=(0,1)\ \text{or}\ (0,0)\end{cases}\ ,
\ee
and 
\bel{CAC18}
\hat n_2(j)\gl B_\beta\gr=\begin{cases}B_\beta\ &\text{if}\ \gamma(j)=(1,1)\ \text{or}\ (0,1)\\
0\ &\text{if}\ \gamma(j)=(1,0)\ \text{or}\ (0,0)\end{cases}\ ,
\ee
the basis vectors are eigenstates of $\hat n_1(j)$ and $\hat n_2(j)$ and therefore of $Q(j)$,
\bel{CAC19}
Q(j)=\begin{cases}1\ &\text{if}\ \gamma(j)=(1,0)\\
-1\ &\text{if}\ \gamma(j)=(0,1)\\
0\ &\text{if}\ \gamma(j)=(1,1)\ \text{or}\ (0,0)\end{cases}\ .
\ee

Basis vectors with $Q(j)=0$ can be distinguished by the local ``filling operator''
\bel{CAC20}
F(j)=\hat n_1(j)+\hat n_2(j)-1\ .
\ee
It takes the value $1$ if $\gamma(j)=(1,1)$ and $-1$ for $\gamma(j)=(0,0)$, while for $\gamma(j)=(1,0)$ or $(0,1)$ one has $F(j)=0$. Thus $F(j)$ measures the local deviation from half-filling. With $[F(j),Q(j')]=0$, $[Q(j),Q(j')]=0$, $[F(j),F(j')]=0$ one can characterize the basis states by the eigenvalues of $Q(j)$ and $F(j)$. We observe the relation
\bel{CAC21}
\gl Q(j)+F(j)\gr^2=1\ ,
\ee
such that for $\gl Q(j),F(j)\gr$ only the combinations of eigenvalues $(1,0)$, $(-1,0)$, $(0,1)$ and $(0,-1)$ are possible. The reference state $\vp_0$ is not an eigenstate of $F(j)$. The expectation value of $F(j)$ vanishes $\langle F(j)\rangle=0$, as appropriate for half filing, $\langle\hat n_1(j)+\hat n_2(j)\rangle=1/2$.

In our matrix representation one has
\bel{CAC22}
Q(j)\gl\vp\gr=\big[a^\dagger(j)a(j),\vp\big]\ ,\ F(j)\gl\vp\gr=\big\{a^\dagger(j)a(j),\vp\big\}-\vp\ .
\ee
From the commutation relation
\begin{align}
\label{CAC23}
\big[Q(j'),\hat n(j)\big]=&\,0\ ,\nn\\
\big[F(j'),\hat n(j)\big]=&\,-i\gl a_2(j)a_1(j)+a_1^\dagger(j)a_2^\dagger(j)\gr\delta_{jj'}
\end{align}
one infers that applying $\hat n(j)$ does not change the charge $Q(j')$ of a given eigenstate of $Q(j')$, while the filling number $F(j')$ changes. Applying $\hat n(j)$ to the reference state $\vp_0$ leads to a new state with vanishing local charges $Q_j(0)$, but different combinations of local filling numbers. The part of the wave function not described by the multi-particle states -- the environment -- can be obtained by first applying combinations of $\hat n(j)$ on $\vp_0$, and subsequently changing the local charges $Q(j)$ by applying $A^\dagger(j)$ or $A(j)$.

We have not specified the complex structure which is at the basis of the general complex wave function $\vp$. In the simplest case one uses again an additional bit. Different complex structures may be employed as well.

\section{Subtraces for the two-bit local chain}\label{app: subtraces for the two-bit local chain}

The conceptual aspects of subsystems defined by subtraces can already be seen for very simple local chains. We discuss a particular two-bit local chain with two Ising spins at each site of the chain. The number of local states is $N=4$, such that the step evolution operator and the density matrix of the time-local subsystem are $4\times4$-matrices. In order to take a subtrace which defines a subsystem which has a closed time evolution we first bring $\hat S$ to a block diagonal form by a suitable similarity transformation, $\tilde S=D\hat SD^{-1}$.

\paragraph*{Complex density matrices for subsystems}

In general, the matrix $D$ that brings $\hat{S}$ to the factorized form $\tilde{S}$ is not a real matrix. In consequence, the density matrix $\tilde{\rho}$ needs not to be real in the new basis, and the density matrix $\rho^{(s)}$ for the subsystem may be a complex matrix. As an example we take the unique jump step evolution operator
\begin{equation}\label{eq:ST24}
\hat{S} = \begin{pmatrix}
0 & 0 & 1 & 0 \\
1 & 0 & 0 & 0 \\
0 & 0 & 0 & 1 \\
0 & 1 & 0 & 0
\end{pmatrix}\, ,
\end{equation}
which corresponds to eq.\,\eqref{eq:SE28A} with indices $\alpha$ and $\tau$ exchanged. The eigenvalues of $\hat{S}$ are
\begin{equation}\label{eq:ST25}
\lambda = (\, 1\,, - 1\, ,\, \im\, ,\, - \im\, )\, .
\end{equation}

We select a basis where $\tilde{S} = D\, \hat{S}\, D^{-1}$ factorizes, $\tilde{S} = \tilde{S}_1\otimes\tilde{S}_2$, with eigenvalues of $\tilde{S}_1$ and $\tilde{S}_2$ given by
\begin{equation}\label{eq:ST26}
\tilde{\lambda}^{(1)} = (\, \im\, ,\, - \im\, )\, , \quad 
\tilde{\lambda}^{(2)} = (\, 1\, ,\, \im\, )\, .
\end{equation}
The eigenvalues \eqref{eq:ST25} obtain indeed as products $\lambda_{kl} = \tilde{\lambda}^{(1)}_k \tilde{\lambda}^{(2)}_l$. We take
\begin{equation}\label{eq:ST27}
\tilde{S}_1 = \begin{pmatrix}
0 & 1 \\
-1 & 0
\end{pmatrix}\, , \quad \tilde{S}_2 =
\begin{pmatrix}
1 & 0 \\
0 & \im
\end{pmatrix}\, ,
\end{equation}
and therefore
\begin{equation}\label{eq:ST28}
\tilde{S} = \begin{pmatrix}
0 & 0 & 1 & 0 \\
0 & 0 & 0 & \im  \\
-1 & 0 & 0 & 0 \\
0 & -\im & 0 & 0
\end{pmatrix}\, .
\end{equation}
A possible matrix $D$ is given by
\begin{equation}\label{eq:ST29}
D = \frac{1}{\sqrt{2}} \begin{pmatrix}
1 & 0 & 0 & -1 \\
0 & 1 & 1 & 0 \\
0 & -1 & 1 & 0 \\
-\im & 0 & 0 & -\im
\end{pmatrix}\, ,\ D^{-1} = \frac{1}{\sqrt{2}}
\begin{pmatrix}
1 & 0 & 0 & \im \\
0 & 1 & -1 & 0 \\
0 & 1 & 1 & 0 \\
-1 & 0 & 0 & \im
\end{pmatrix}\, .
\end{equation} 

For an arbitrary matrix $A$ with elements $A_{ij}$ one finds
\begin{equation}\label{eq:ST30}
\tilde{A} = D\, A\, D^{-1}\, ,
\end{equation}
with
\begin{align}\label{eq:ST31}
& \tilde{A}_{i1} = \frac{1}{2} \begin{pmatrix}
A_{11} - A_{41} - A_{14} + A_{44} \\
A_{21} + A_{31} - A_{24} - A_{34} \\
- A_{21} + A_{31} + A_{24} - A_{34} \\
-\im\, (A_{11} + A_{41} - A_{14} - A_{44} )
\end{pmatrix}\, , \notag \\
& \tilde{A}_{i2} = \frac{1}{2} 
\begin{pmatrix}
A_{12} - A_{42} + A_{13} - A_{43} \\
A_{22} + A_{32} + A_{23} + A_{33} \\
-A_{22} + A_{32} - A_{23} + A_{33} \\
-\im\, (A_{12} + A_{42} + A_{13} + A_{43} )
\end{pmatrix}\, , \notag \\
& \tilde{A}_{i3} = \frac{1}{2} \begin{pmatrix}
- A_{12} + A_{42} + A_{13} - A_{43} \\
- A_{22} - A_{32} + A_{23} + A_{33} \\
A_{22} - A_{32} - A_{23} + A_{33} \\
+ \im\, (A_{12} + A_{42} - A_{13} - A_{43} )
\end{pmatrix}\, , \notag \\
& \tilde{A}_{i4} = \frac{\im}{2} 
\begin{pmatrix}
A_{11} - A_{41} + A_{14} - A_{44} \\
A_{21} + A_{31} + A_{24} + A_{34} \\
-A_{21} + A_{31} - A_{24} + A_{34} \\
-\im\, (A_{11} + A_{41} + A_{14} + A_{44} )
\end{pmatrix}\, .
\end{align}
In particular, for $A = \hat{S}$, with only non-vanishing elements $\hat{S}_{13} = \hat{S}_{21} = \hat{S}_{34} = \hat{S}_{42} = 1$, one has $\tilde{A} = \tilde{S}$. For $A = \rho'$ the transformed matrix corresponds to $\tilde{A} = \tilde{\rho}$.

We write the density matrix $\tilde\rho$ in a double index form corresponding to the direct product structure $\tilde S=\tilde S_1\otimes\tilde S_2$. The subtrace is then taken over the indices of the subspace in which $\tilde S_2$ acts. In detail, the density matrix $\rho^{(s)}$ of the subsystem has the following elements,
\begin{align}\label{eq:ST32}
& \rho^{(s)}_{11} = \tilde{\rho}_{11,\,11} + \tilde{\rho}_{12,\,12} = \tilde{\rho}_{11} + 
\tilde{\rho}_{22}\, , \notag \\
& \rho^{(s)}_{12} = \tilde{\rho}_{11,\,21} + \tilde{\rho}_{12,\,22} = \tilde{\rho}_{13} + 
\tilde{\rho}_{24}\, , \notag \\
& \rho^{(s)}_{21} = \tilde{\rho}_{21,\,11} + \tilde{\rho}_{22,\,12} = \tilde{\rho}_{31} + 
\tilde{\rho}_{42}\, , \notag \\
& \rho^{(s)}_{22} = \tilde{\rho}_{21,\,21} + \tilde{\rho}_{22,\,22} = \tilde{\rho}_{33} + 
\tilde{\rho}_{44}\, ,
\end{align}
where the second expression uses the double index notation $11\,\hat{=}\,1$, $12\,\hat{=}\, 2$, $21\,\hat{=}\,3$, $22\,\hat{=}\,4$. Insertion of the expression \eqref{eq:ST31} for $\tilde{\rho}$ in terms of $\rho'$ yields
\begin{align}\label{eq:ST33}
\rho^{(s)}_{11} =&\, \frac{1}{2} \big( \rho'_{11} + \rho'_{22} + \rho'_{33} + \rho'_{44}
+ \rho'_{23} + \rho'_{32} - \rho'_{14} - \rho'_{41} \big)\, , \notag \\
\rho^{(s)}_{22} =&\, \frac{1}{2} \big( \rho'_{11} + \rho'_{22} + \rho'_{33} + \rho'_{44}
- \rho'_{23} - \rho'_{32} + \rho'_{14} + \rho'_{41} \big)\, , \notag \\
\rho^{(s)}_{12} =&\, \frac{1}{2} \big( \rho'_{13} + \rho'_{42} - \rho'_{12} - \rho'_{43}
\big)\nonumber\\
&+ \frac{\im}{2}\big( \rho'_{21} + \rho'_{31} + \rho'_{24} + \rho'_{34} \big)\, , 
\notag \\
\rho^{(s)}_{21} =&\, \frac{1}{2} \big( \rho'_{31} + \rho'_{24} - \rho'_{21} - \rho'_{34}
\big)\nonumber\\
&- \frac{\im}{2}\big( \rho'_{12} + \rho'_{13} + \rho'_{42} + \rho'_{43} \big)\, . 
\end{align}
In general, the elements $\rho_{12}^\mathrm{(s)}$ and $\rho_{21}^\mathrm{(s)}$ will have an imaginary part. The density matrix of the subsystem is complex.

\paragraph*{Hermitean density matrices with unitary evolution}

If $\rho'$ is a symmetric matrix this property is preserved by the orthogonal step evolution operator \eqref{eq:ST24}. The density matrix of the subsystem turns out to be a complex hermitean matrix, and the step evolution operator for the subsystem is a unitary matrix. This demonstrates that the complex structure characteristic for quantum mechanics can arise as a consequence of considering suitable subsystems.

In eq.\,\ref{eq:ST33} we observe that for symmetric $\rho'$, the subsystem density matrix $\rho^{(s)}$ is hermitean,
\begin{equation}\label{eq:ST34}
\rho'^\tp = \rho' \; \Rightarrow \; \big( \rho^{(s)} \big)^\dagger = \rho^{(s)}\, .
\end{equation}
As it should be, one verifies $\tr\,\rho^{(s)} = \tr\,\rho' = 1$. 
For symmetric $\rho'$, one has
\begin{equation}\label{eq:ST35}
\rho^{(s)} = \frac{1}{2} \rho_\mu \tau_\mu\, ,
\end{equation}
with $\mu = 0,\,\dots,\, 3$, $\tau_0 = 1$, $\tau_k$ the Pauli matrices, $\rho_0 = 1$, and
\begin{align}\label{eq:ST36}
& \rho_1 = \rho'_{13} + \rho'_{24} - \rho'_{12} - \rho'_{34}\, , \notag \\
& \rho_2 = - \big( \rho'_{12} + \rho'_{13} + \rho'_{24} + \rho'_{34} \big) \, , \notag \\
& \rho_3 = 2\,\rho'_{23} - 2\, \rho'_{14}\, .
\end{align}

We observe that $\hat{S}$, $\tilde{S}$, $\tilde{S}_1$, $\tilde{S}_2$ and $D$ are all unitary matrices,
\begin{align}\label{eq:ST37}
& \hat{S}^\dagger S = 1\, ,\quad \tilde{S}^\dagger \tilde{S} = 1\, , \notag \\
& \tilde{S}_1^\dagger \tilde{S}_1 = 1\, , \quad \tilde{S}_2^\dagger \tilde{S}_2 = 1\, , \quad D^\dagger D = 1\, .
\end{align}
The evolution law for the hermitean density matrix $\rho^{(s)}$ is the same as for quantum mechanics,
\begin{equation}\label{eq:ST38}
\rho^{(s)} (m+1) = \tilde{S}_1\, \rho^{(s)} (m)\, \tilde{S}_1^\dagger\, .
\end{equation}
For the particular choice \eqref{eq:ST29} for $D$ the unitary step evolution operator for the subsystem $\tilde{S}_1 = i\tau_2$ is real. We could choose different $D$ by applying an additional similarity transformation acting only in the subsystem. In this way one can obtain complex step evolution operators for the subsystem, as $\tilde{S}_1 = i\tau_1$.

\paragraph*{Mixed states from pure states}

In close analogy to quantum mechanics a subsystem obtained by a subtrace can be in a mixed state, even if the total system is described by a pure classical state. We can see this directly by the subsystem density matrix given by eq.\,\eqref{eq:ST33}.

We can define the ``purity'' $P$ of the reduced system by
\begin{align}\label{eq:ST39}
P &= \rho_k\rho_k = \rho_1^2 + \rho_2^2 + \rho_3^2 \notag \\
&= 2\, (\rho'_{12} + \rho'_{34})^2 + 2\, (\rho'_{13} + \rho'_{24} )^2 + 
4\, (\rho'_{14} - \rho'_{23})^2\, .
\end{align}
For $P=1$ the density matrix $\rho^{(s)}$ describes a pure quantum state. Positive $\rho^{(s)}$, with all eigenvalues $\lambda_\rho \geq 0$, require $P\leq 1$. Consider a pure classical system with $\bar{q} = \tilde{q} = q$ and $\rho'_{\varepsilon\eta} = q_\varepsilon\, q_\eta$. The purity for the subsystem is given by
\begin{equation}\label{eq:ST40}
P = 2\, \big\{ 1 - (q_1^4 + q_2^4 + q_3^4 + q_4^4) - (q_1^2 + q_4^2)\, (q_2^2 + q_3^2) 
\big\}\, .
\end{equation}
For a classical pure state with $q_1 = 1$, $q_2 = q_3 = q_4 = 0$, one has $P=0$ and therefore a mixed density matrix $\rho^{(s)} = \text{diag}\big( \frac{1}{2}, \, \frac{1}{2} \big).$ For $q_1 = q_2 = 1/\sqrt{2}$, $q_3 = q_4$, one finds $P=1/2$. A pure state with $P=1$ obtains for $q_1 = q_2 = q_3 = q_4 = 1/2$.

We conclude that generically subsystem are in a mixed state even if the total system is a pure classical state. For particular states of the total system the subsystem can be in a pure state, however.

\paragraph*{Complex step evolution operators}

Instead of taking a subtrace over the indices $\tilde{\alpha}$, $\tilde{\beta}$ of $\tilde{\rho}_{\tilde{\tau}\tilde{\alpha},\,\tilde{\rho}\tilde{\beta}}$ we can also take a subtrace over the indices $\tilde{\tau}$, $\tilde{\rho}$. The resulting evolution operator for the subsystem is then $\tilde{S}_2$. The reduced density matrix $\rho^{(s)}_{\tilde{\alpha}\tilde{\beta}}$ takes the same form as eq.~\eqref{eq:ST35}, now with coefficients (for symmetric $\rho'$)
\begin{align}\label{eq:ST41}
& \rho_1 = \rho'_{12} + \rho'_{13} - \rho'_{24} - \rho'_{34}\, , \notag \\
& \rho_2 = \rho'_{12} - \rho'_{13} + \rho'_{24} - \rho'_{34}\, , \notag \\
& \rho_3 = - 2\, (\rho'_{14} + \rho'_{23})\, .
\end{align}
As compared to eq.~\eqref{eq:ST36}, the contributions of the elements $\rho'_{12}$, $\rho'_{13}$, $\rho'_{24}$ have the opposite sign. The two possible subsystems are rather similar.

An interesting difference should be noted, however. For the first subsystem with eigenvalues $\pm i$ of the step evolution operator there exists a basis for which $\tilde{S}_1$ is real, as given by eq.\,\eqref{eq:ST27}. This is no longer possible for the second subsystem with step evolution operator $\tilde{S}_2$. The eigenvalues of $\tilde{S}_2$ are $(1,i)$. No real $2\times 2$ matrix with these eigenvalues exists. For a real matrix with eigenvalue $\lambda$ the complex conjugate $\lambda^*$ is also an eigenvalue. This property is not obeyed for the set of eigenvalues $(1,i)$. This has the interesting consequence that there are subsystems for which the step evolution operator is necessarily complex. 

\paragraph*{General closed subsystems}

The possibility to diagonalize $\tilde S$ by a similarity transformation brings us to a more general concept of subsystems that are not necessarily related to subtraces. General subsystems that are closed with respect to the time evolution can easily be understood in a basis where the step evolution operator $\hat{S}$ is made diagonal by a suitable similarity transformation. Combining a set of eigenvalues of $\hat{S}$, the corresponding subspace spanned by the eigenvectors evolves independently of the complement spanned by the eigenvectors of eigenvalues not in the set. This subspace forms a closed subsystem. One can now apply further similarity transformations in this subspace without changing the property of a closed subsystem. In general, the step evolution operator $\tilde{S}$ of the subsystem is no longer a real positive matrix, even if the step evolution operator $\hat{S}$ of the total system has this property. If the eigenvalues in the set used for the definition of the subsystem include a complex eigenvalue $\lambda$, but not its conjugate $\lambda^*$, the step evolution operator $\tilde{S}$ of the subsystem is necessarily complex.

\paragraph*{Modified subtraces}

So far we have formed subtraces by contracting $\rho'_{\tau\alpha,\,\rho\beta}$ with $\delta_{\alpha\beta}$, or more generally $\tilde{\rho}_{\tilde{\tau}\tilde{\alpha},\,\tilde{\rho}\tilde{\beta}}$ with $\delta_{\tilde{\alpha}\tilde{\beta}}$. A wider class of possible subsystems obtains by contraction with a general matrix $g_{\alpha\beta}$ or $\tilde{g}_{\tilde{\alpha}\tilde{\beta}}$,
\begin{equation}\label{eq:MS1}
\rho^{(s)}_{\tau\rho} = g_{\alpha\beta} \, \rho'_{\tau\alpha,\,\rho\beta}\, ,
\end{equation}
provided the transposed  matrix $g^\tp$ commutes with $\hat{S}_2$,
\begin{equation}\label{eq:MS2}
[g^\tp,\, \hat{S}_2] = 0\, .
\end{equation}
Here we assume the factorized form \eqref{eq:ST7}, with straightforward generalization to $\tilde{S}$ in eq.~\eqref{eq:ST20}. The evolution law for $\rho^{(s)}$ reads in this case
\begin{align}\label{eq:MS3}
& \rho^{(s)}_{\tau\rho}(m+1) = \rho'_{\tau\alpha,\, \rho\beta} (m+1)\, g_{\alpha\beta} 
\notag \\
& \quad = \hat{S}_{\tau\alpha,\,\sigma\gamma}\, \rho'_{\sigma\gamma,\,\mu\delta}(m)\, 
\big( \hat{S}^{-1} \big)_{\mu\delta,\,\rho\beta}\, g_{\alpha\beta} \notag \\
& \quad = \big( \hat{S}_1 \big)_{\tau\sigma}\, \rho'_{\sigma\gamma,\,\mu\delta}(m)\,
\big( \hat{S}_1^{-1} \big)_{\mu\rho}\, \big( \hat{S}_2 \big)_{\alpha\gamma}\, 
\big( \hat{S}_2^{-1} \big)_{\delta\beta}\, (g^\tp)_{\beta\alpha} \notag \\
& \quad = \big( \hat{S}_1 \big)_{\tau\sigma}\, \rho'_{\sigma\gamma,\, \mu\delta}(m)\, 
g_{\gamma\delta}\, \big( \hat{S}^{-1} \big)_{\mu\rho} \notag \\
& \quad = \big( \hat{S}_1 \big)_{\tau\sigma}\, \rho^{(s)}_{\sigma\mu}(m)\, 
\big( \hat{S}^{-1} \big)_{\mu\rho}\, ,
\end{align}
where we use eq.~\eqref{eq:MS2},
\begin{equation}\label{eq:MS4}
\hat{S}^{-1}_2\, g^\tp\, \hat{S}_2 = g^\tp\, .
\end{equation}
Thus the evolution law for the reduced density matrix $\rho^{(s)}$ is closed and keeps the standard form,
\begin{equation}\label{eq:MS5}
\rho^{(s)}(m+1) = \hat{S}_1\, \rho^{(s)}(m)\, \hat{S}_1^{-1}\, ,
\end{equation}
with evolution operator $\hat{S}_1$.

The evolution \eqref{eq:MS5} guarantees that the norm of $\rho^{(s)}$ is preserved,
\begin{equation}\label{eq:MS5A}
\tr \,\rho^{(s)} (m+1) = \tr\, \rho^{(s)} (m)\, .
\end{equation}
The contraction with a matrix $g_{\alpha\beta}$ does, however, not guarantee $\tr\, \rho^{(s)} = 1$, since
\begin{equation}\label{eq:MS5B}
\tr \,\rho^{(s)} = \delta_{\tau\rho} \,g_{\alpha\beta} \,\rho'_{\tau\alpha,\, \rho\beta}
\end{equation}
may not coincide with $\tr \rho' = \rho'_{\tau\alpha,\,\tau\beta} = 1$. For arbitrary $g$ obeying eq.~\eqref{eq:MS4}, also the probabilistic interpretation of the reduced density matrix $\rho^{(s)}$ has to be guaranteed. The diagonal elements $\rho^{(s)}_{\tau\tau}$ are not necessarily real numbers obeying $\rho_{\tau\tau}^{(s)} \geq 0$. One may define a normalized reduced density matrix by
\begin{equation}\label{eq:MS5C}
\tilde{\rho}^{(s)} = ( \tr\, \rho^{(s)} )^{-1}\, \rho^{(s)}\, .
\end{equation}

We illustrate the wider possibilities to define subsystems by the example with the evolution operator $\hat{S}$ in eq.~\eqref{eq:ST24}. Using $\tilde{S}$ in eqs~\eqref{eq:ST27}, \eqref{eq:ST28}, we observe that all diagonal matrices $\tilde{g}$ commute with $\tilde{S}_2$,
\begin{equation}\label{eq:MS6}
\tilde{g} = \begin{pmatrix}
g_1 & 0 \\
0 & g_2
\end{pmatrix}\, .
\end{equation}
With the modified subtrace the reduced density matrix for the subsystem becomes
\begin{align}\label{eq:MS7}
& \rho^{(s)}_{11} = g_1\, \tilde{\rho}_{11,\,11} + g_2\, \tilde{\rho}_{12,\,12} = 
g_1\, \tilde{\rho}_{11} + g_2\, \tilde{\rho}_{22}\, , \notag \\
& \rho^{(s)}_{12} = g_1\, \tilde{\rho}_{11,\,21} + g_2\, \tilde{\rho}_{12,\,22} = 
g_1\, \tilde{\rho}_{13} + g_2\, \tilde{\rho}_{24}\, , \notag \\
& \rho^{(s)}_{21} = g_1\, \tilde{\rho}_{21,\,11} + g_2\, \tilde{\rho}_{22,\,12} = 
g_1\, \tilde{\rho}_{31} + g_2\, \tilde{\rho}_{42}\, , \notag \\
& \rho^{(s)}_{22} = g_1\, \tilde{\rho}_{21,\,21} + g_2\, \tilde{\rho}_{22,\,22} = 
g_1\, \tilde{\rho}_{33} + g_2\, \tilde{\rho}_{44}\, .
\end{align}
Expressed in terms of the elements of $\rho'$, one finds
\begin{align}\label{eq:MS8}
\rho^{(s)}_{11} + \rho^{(s)}_{22} &= \frac{g_1 + g_2}{2} + 
\frac{g_2 - g_1}{2}\, ( \rho'_{14} + \rho'_{41} + \rho'_{23} + \rho'_{32} )\, , 
\notag \\
\rho^{(s)}_{11} - \rho^{(s)}_{22} &= \frac{g_1 - g_2}{2}\, (\rho'_{11} + \rho'_{44} - 
\rho'_{22} - \rho'_{33}) \notag \\
& \quad + \frac{g_1 + g_2}{2} \, (\rho'_{23} + \rho'_{32} - \rho'_{14} - 
\rho'_{41}) \, , \notag \\
\rho^{(s)}_{12} &= \frac{g_1}{2}\, (-\rho'_{12} + \rho'_{13} + \rho'_{42} - \rho'_{43}) 
\notag \\
& \quad + \frac{\im g_2}{2}\, (\rho'_{21} + \rho'_{24} + \rho'_{31} + \rho'_{34})\, , \notag \\
\rho_{21}^{(s)} &= \frac{g_1}{2} \, (-\rho'_{21} + \rho'_{31} + \rho'_{24} - \rho'_{34})
\notag \\
& \quad - \frac{\im g_2}{2}\, (\rho'_{12} + \rho'_{42} + \rho'_{13} + \rho'_{43})\, .
\end{align}

The trace depends on elements of $\rho'$, more precisely on the quantity
\begin{equation}\label{eq:MS9}
V = \rho'_{24} + \rho'_{41} + \rho'_{23} + \rho'_{32}\, .
\end{equation}
This quantity is preserved by the evolution with the unique jump operator $\hat{S}$ in eq.~\eqref{eq:ST24}, which transforms
\begin{align}\label{eq:MS10}
& \rho'_{14} (m+1) = \rho'_{32}(m)\, , \quad \rho'_{41} (m+1) = \rho'_{23}(m)\, , \notag \\
& \rho'_{23} (m+1) = \rho'_{14}(m)\, , \quad \rho'_{32} (m+1) = \rho'_{41}(m)\, .
\end{align}
We could use this for a $V$-dependent renormalization of the reduced density matrix according to eq.~\eqref{eq:MS5C}. For real $g_1$, $g_2$ and symmetric $\rho'$, the reduced density matrix is hermitean. The difference
\begin{equation}\label{eq:MS11}
\rho'_{11} + \rho'_{44} - \rho'_{22} + \rho'_{33} = \langle s_1\, s_2 \rangle
\end{equation}
corresponds to the correlation function for the two original spins. In the limit $g_1 = g_2$ this quantity does not influence the reduced density matrix.

We conclude that modified subtraces \eqref{eq:MS1} may open further generalizations of possible subsystems. Since we have not encountered useful applications so far we will not pursue this possibility further.

\end{appendices}

\vspace{2.0cm}\noindent

\nocite{*}

\vspace{2.0cm}\noindent

\bibliography{probabilistic_world}

\end{document}